\documentclass[a4paper,12pt]{article}%
 \usepackage[margin=1in,footskip=0.5in]{geometry}

\usepackage{setspace}
 \usepackage[symbol]{footmisc}
 \usepackage{caption}
\usepackage{subcaption}

\usepackage{amsthm}
\usepackage{amsmath}
\usepackage{amssymb,amsmath,amsthm,amsfonts,mathrsfs,xcolor}
\usepackage[authoryear,round]{natbib}
\usepackage[colorlinks,citecolor=blue,urlcolor=blue,filecolor=blue,backref=page]{hyperref}
\usepackage{graphicx}
\usepackage{enumitem}
 \usepackage{tcolorbox}
\theoremstyle{plain}
\newtheorem{theorem}{Theorem}
  
\newtheorem{assumption}{Assumption}
\renewcommand{\theassumption}{A\arabic{assumption}}
 \newtheorem{corollary}{Corollary}
 \newtheorem{lemma}{Lemma}
 
 \newtheorem*{definition*}{Definition}
\newtheorem{proposition}{Proposition}
\theoremstyle{remark}
\newtheorem{remark}{Remark}
\theoremstyle{plain}

\usepackage{times}

\usepackage{amssymb,amsmath,amsthm,amsfonts,bbm,bm,mathrsfs,xcolor}
 \usepackage{float}
 \usepackage{algorithm}
\usepackage{algpseudocode}
\usepackage{placeins}
 \usepackage{xr}
\usepackage{multirow}
\usepackage[OT1]{fontenc}
\ifdefined\directlua
\usepackage{luacolor,lua-ul}
\fi

\usepackage{longtable,booktabs,needspace}

\usepackage{bigints}
\usepackage{tikz}
\usetikzlibrary{positioning,arrows.meta,fit}
\DeclareMathOperator*{\argmin}{arg\,min}
\DeclareMathOperator*{\argmax}{arg\,max}

\def\m{\mathcal}
\def\mb{\mathbb}

\def\ind{\mathbbm{1}}

\title{Learning to Infer: Statistical Inference through Meta Learning}

\author{
Xiaolu Li\textsuperscript{$\ast$}
and
Yun Yang\textsuperscript{$\dagger$}
}

\date{\small
Department of Mathematics, University of Maryland, College Park
\\
\textsuperscript{$\ast$}\texttt{xl020216@umd.edu}
\qquad
\textsuperscript{$\dagger$}\texttt{yy84@umd.edu}
\vspace{-2em}
}

\newcommand{\icpt}{\alpha}

\begin{document}
\maketitle
\addtocontents{toc}{\protect\setcounter{tocdepth}{-1}}

\begin{abstract}
We study meta learning for statistical inference, in which neural networks learn inference rules from simulated tasks and apply them to new datasets without parameter updates, providing a form of in-context learning. We draw on statistical decision theory to design training objectives whose population minimizers are Bayes rules and exploit their connections to classical inference procedures to establish frequentist guarantees for the learned rules. The structure of classical inference procedures also motivates the design of our neural network architecture, called Residual Recurrent Deep Sets (RRDS), which repeatedly aggregates information across observations in a permutation-invariant manner and refines its internal state through residual updates. Under suitable regularity and training conditions, we establish asymptotic efficiency in parameter estimation, consistency in model selection, and variable selection guarantees in high-dimensional regression models. The RRDS architecture allows us to avoid the curse of sample size that arises in generic neural network approximation when the network takes the entire dataset as input. For fixed-dimensional models, the network size and the number of meta training tasks need only grow polynomially in the sample size; under additional analyticity conditions, these requirements improve to polylogarithmic growth in network size and nearly linear growth in the number of training tasks. Numerical experiments illustrate the finite-sample advantages of the learned inference rules over classical methods.

\end{abstract}


\section{Introduction}\label{sec:intro}

Statistical inference concerns drawing conclusions about an unknown data-generating distribution from observed data. Depending on the problem, the goal may be to estimate an unknown parameter, select a model among competing candidates, or identify relevant variables in a regression model while estimating their coefficients. Classical statistical procedures are typically designed on a case-by-case basis, relying on problem-specific knowledge to construct estimators through likelihood optimization, estimating equations, or other methods~\citep{casella2024statistical,van2000asymptotic,heyde1997quasi}. Model and variable selection further require criteria for comparing candidate models or balancing goodness of fit against model complexity, often through penalized likelihood or regularization. While such procedures can have desirable statistical properties, their construction and the verification of their optimality often require substantial problem-specific analysis. Moreover, even when a classical procedure is asymptotically optimal, it need not minimize the risk of interest at a given finite sample size.

Recent developments in machine learning have motivated an alternative, amortized computational approach in which inference procedures are learned from a population of simulated statistical tasks rather than constructed separately for each problem. This approach is closely related to \emph{meta learning}~\citep{thrun1998learning,hospedales2022meta}, where a learning procedure is trained across multiple tasks and subsequently applied to new ones. In our setting, each task consists of a dataset generated from a statistical model, together with the generating parameter, model label, or other information specifying the inferential target. By training a neural network across such tasks, we seek to learn an inference rule that maps an observed dataset directly to an estimate or decision. Once trained, the network performs inference on new datasets without further parameter updates or solving a separate estimation problem. This connects our approach to \emph{in-context learning}~\citep{brown2020language,xie2022explanation,garg2022what}, where a pretrained model uses information supplied in its context to perform a new task. Here, the context is an observed dataset, from which the network produces the corresponding statistical estimate or decision.

Statistical decision theory provides a natural framework for learning such procedures~\citep{berger2013statistical}. Although parameter estimation, model selection, and variable selection have different objectives, they can all be formulated as decision problems in which a procedure maps observed data to an estimate or decision, with its performance measured by a suitable loss function. Suppose that statistical tasks are generated from a distribution $\pi$ over the underlying parameters or model labels. Meta learning seeks to minimize the expected loss over the joint distribution of tasks and their observed datasets. For each fixed sample size, the population minimizer over all measurable rules is a Bayes decision rule under the specified task distribution and loss. The choice of training loss determines the Bayesian quantity that the network is encouraged to learn. This allows us to target finite-sample optimality directly through the training objective, rather than designing a procedure primarily based on its large-sample properties. A central question in our analysis is whether a network trained to approximate the population rule can also retain the frequentist guarantees of classical inference methods.

To illustrate this formulation, consider the classical problem of \emph{point estimation}. Let $\m P=\{\mb P_\theta:\theta\in\Theta\}$ be a parametric model, where $\Theta\subset\mb R^d$, and let $X^n=(X_1,\ldots,X_n)$ consist of $n$ independent and identically distributed (i.i.d.) observations from $\mb P_\theta$ on $\m X\subset\mb R^{d_X}$. Rather than constructing an estimator $F:\m X^n\to\mb R^d$ through a model-specific optimization problem or estimating equation, we draw $\theta\sim\pi$ from a user-specified distribution and then $X^n\sim\mb P_\theta^{(n)}$, where $\mb P_\theta^{(n)}$ is the $n$-fold product distribution of $\mb P_\theta$. Under squared-error loss, we obtain the population-level regression problem
\begin{equation}
\min_{F\in\m F}\mb E_{\theta,X^n}[\|F(X^n)-\theta\|_2^2],
\label{eqn:regression}
\end{equation}
where $\m F$ is a function approximation class, such as neural networks. The minimizer over all measurable functions is the posterior mean $F^*(X^n)=\mb E[\theta\mid X^n]$, which minimizes the mean squared error (MSE) averaged over $\pi$ at the given sample size. In practice, we approximate this rule by minimizing the empirical risk on independently simulated parameter--dataset pairs. Note that the network is trained directly on the generating parameters, not on estimates produced by classical procedures. Nevertheless, as we show later, the learned estimator can attain first-order asymptotic equivalence to the maximum likelihood estimator (MLE).

The same formulation applies to model and variable selection by suitably choosing the training loss. For model selection, cross-entropy loss targets the posterior model probabilities, whose maximizer minimizes the probability of selecting the wrong model under the task distribution. For variable selection, a combination of squared-error and binary cross-entropy losses allows us to learn the posterior mean coefficients and posterior inclusion probabilities, with thresholding of the latter at $1/2$ minimizing the expected Hamming loss. As in parameter estimation, training requires only simulated datasets and their associated parameters or labels, without explicitly computing posterior quantities or evaluating likelihoods.

The finite-sample optimality of these population rules, however, does not automatically imply frequentist guarantees for a network trained on finitely many tasks. In parameter estimation, these guarantees include consistency, asymptotic normality, and efficiency under a fixed data-generating distribution. For model and variable selection, they concern recovery of the correct model or support and control of estimation and predictive risks. To attain such properties, the network must approximate its population target sufficiently accurately, with the required accuracy generally increasing as the sample size grows. The question is whether this can be achieved while keeping the network complexity and the number of meta training tasks computationally manageable.

This challenge is particularly severe when the inference rule is approximated by a generic neural network. For a dataset $X^n=(X_1,\ldots,X_n)$ with observations $X_i\in\mb R^{d_X}$, treating the entire sample as a single vector gives a network input dimension of $nd_X$. Generic neural-network approximation bounds for arbitrary functions on $\mb R^{nd_X}$ can scale as $W=\varepsilon^{-O(nd_X)}$ in the number of parameters needed to achieve accuracy $\varepsilon>0$~\citep{yarotsky2017error}. The number of meta training tasks needed to control generalization may likewise grow exponentially with $n$. We refer to this unfavorable dependence as the \emph{curse of sample size} in meta learning. Statistical inference procedures, however, possess additional structure that is not captured by such generic bounds. For i.i.d. observations, inference rules are naturally invariant to permutations of the sample. Moreover, many quantities used in classical estimation procedures, such as empirical moments, scores, and information matrices, can be expressed as averages of functions of individual observations. Exploiting these structures may substantially reduce the complexity of neural approximation and meta training.

To this end, we propose a neural network architecture called \emph{Residual Recurrent Deep Sets} (RRDS), which combines the permutation-invariant aggregation of Deep Sets~\citep{zaheer2017deep} with recurrent residual updates motivated by classical iterative estimation. The network first constructs an initial estimate by applying a shared encoder to each observation and averaging the resulting representations. It then iteratively refines this estimate using information aggregated from the full dataset, with the same network blocks reused across observations and refinement steps. This design incorporates the additive structure of statistical estimating procedures while keeping the number of distinct trainable parameters independent of the sample size and the number of refinement steps when the block sizes are fixed.

Our theoretical analysis exploits classical connections between the Bayesian quantities targeted by the training losses and frequentist inference procedures. Specifically, for each inferential task, we identify a classical procedure that serves as an intermediate step between the population target of training and the desired frequentist properties. These connections also guide our construction of RRDS networks capable of approximating the population targets with manageable network complexity.

For parameter estimation, the posterior mean is asymptotically equivalent to the MLE in regular parametric models~\citep[Section~10.3]{van2000asymptotic}. Motivated by this connection, we construct RRDS networks that approximate the classical one-step estimator~\citep[Section~5.7]{van2000asymptotic}, which applies a Fisher scoring update to an initial method-of-moments estimator. Both steps involve empirical averages of functions of individual observations and can thus be approximated by RRDS without incurring the curse of sample size. Together with a quantitative bound on the mean-square difference between the posterior mean and the constrained MLE, our approximation and meta-generalization results show that an empirically trained network can attain first-order asymptotic equivalence to the constrained MLE on a new task drawn from the meta-distribution. For fixed-dimensional models, the network complexity and the required number of meta training tasks grow at most polynomially in $n$. Under additional analyticity conditions, the network size grows only polynomially in $\log n$, with nearly linear growth in the number of training tasks.

For model and variable selection, we follow a similar strategy, using the connections between posterior probabilities and the Bayesian information criterion (BIC)~\citep{schwarz1978dimension} and extended BIC (EBIC)~\citep{chen2008extended}, respectively. Here BIC and EBIC play a role analogous to the MLE in parameter estimation: they relate the population targets of training to classical selection procedures with desirable frequentist properties. We construct RRDS networks that approximate the corresponding posterior probabilities and establish model selection consistency, task-averaged bounds on variable selection errors, and bounds on estimation and prediction risks. For model selection, we also establish predictive Kullback--Leibler oracle inequalities for the constructed network under model misspecification. The variable selection analysis allows the number of covariates to grow with $n$. Our meta-generalization bounds further control the corresponding errors for networks trained on finitely many tasks.

The classical constructions used in our proofs, however, do not prescribe the inference rules that RRDS must learn. In practice, the network is trained end to end on simulated tasks by minimizing the corresponding estimation or selection loss, without using the MLE, BIC, or EBIC as training targets or explicitly incorporating these criteria into the training objective. Their role in our analysis is to facilitate the neural approximation of the population targets and establish the frequentist properties of the resulting inference rules. This distinction is important for finite-sample performance: the population decision rule minimizes the expected loss for the given sample size under the task distribution, whereas classical procedures are often justified by their asymptotic optimality or consistency. The learned network therefore need not reproduce the classical procedures used in the analysis and may improve upon them at finite sample sizes.

Our numerical experiments in Section~\ref{sec:numerical} illustrate this potential advantage. In logistic regression with small samples, RRDS achieves lower MSE than the MLE (Section~\ref{sec:sim-estimation}); in model selection, it outperforms BIC for small sample sizes (Section~\ref{sec:sim-model}); and in sparse linear regression, it achieves better support recovery than LASSO~\citep{tibshirani1996regression} and the minimax concave penalty (MCP)~\citep{zhang2010nearly} at moderate signal strengths (Section~\ref{sec:sim-variable}). These findings illustrate how inference rules learned under a specified task distribution and loss can improve on classical procedures at finite sample sizes.

The amortized nature of our approach also offers computational advantages when inference needs to be performed repeatedly. For bootstrap uncertainty quantification, for example, each resampled dataset can be passed through the same trained network without repeatedly solving an estimation problem. We establish bootstrap consistency under additional regularity conditions, providing a theoretical justification for this approach. We also develop a divide-and-conquer method that extends the applicability of a trained network to datasets larger than those used in meta training, while retaining first-order asymptotic efficiency under suitable conditions.

Our work is closely related to neural Bayes estimation~\citep{lenzi2023neural,sainsbury2024likelihoodfree,lenzi2025towards,rodder2025theoretical} and recent theoretical work on in-context learning. We extend the direct learning of statistical estimators to model and variable selection and study how classical inference procedures can be used to establish frequentist guarantees for learned rules without incurring the curse of sample size. A detailed discussion of the related literature is given in Appendix~\ref{sec:supp_related}.

\medskip
\noindent {\bf Notation.}
For $m\geq1$, let $[m]=\{1,\ldots,m\}$. Write $\|\cdot\|_2$ for the Euclidean norm, $I_d$ for the identity matrix, and $\mathbb S_+^d$ and $\mathbb S_{++}^d$ for the sets of symmetric positive semidefinite and positive definite $d\times d$ matrices. For a set $A$, let $\operatorname{int}(A)$, $\partial A$, and $\operatorname{diam}_2(A)$ denote its interior, boundary, and Euclidean diameter, respectively. Write $\operatorname{supp}(P)$ for the support of $P$ and $\operatorname{dist}(x,A)=\inf_{y\in A}\|x-y\|_2$. Throughout the paper, constants $c,C>0$ may change from line to line, are independent of $n$, and may depend on other quantities as specified.
We use the usual notation $C^r$ for spaces of continuously differentiable or H\"older continuous functions of smoothness $r$; precise definitions are given in Appendix~\ref{app:notation}. For probability measures $P,Q$, let $D_{\rm KL}(P\|Q)=\int\log(dP/dQ)\,dP$, taken as $+\infty$ if $P\not\ll Q$. Write $Pf=\int f\,dP$, $\mathcal L(Y)$ and $\mathcal L(Y\mid Z)$ for the law and conditional law of $Y$, and $\delta_x$ for the point mass at $x$. We use $d_{\rm BL}$ for the bounded Lipschitz distance defined in Appendix~\ref{app:notation}.

\medskip
\noindent {\bf Organization of the paper.}
Section~\ref{sec:problem-formulation} formulates the inference problems and introduces the meta learning objectives, RRDS architecture, and training procedure. Section~\ref{sec:theory-results} gives the theoretical results, and Section~\ref{sec:numerical} presents the numerical experiments and real-data analysis. Section~\ref{sec:discussion} discusses extensions and future work. The supplementary material contains the related literature, additional methodological and theoretical details, further numerical results, and proofs.

\section{Meta Learning for Statistical Inference}
\label{sec:problem-formulation}\label{sec:method-section}

We first formulate the three inference tasks and their training objectives in a common framework, then introduce the RRDS architecture, its training procedure, and extensions for larger datasets and uncertainty quantification.

\subsection{Meta learning formulation}\label{sec:preliminaries}\label{sec:meta-formulation}\label{sec:method}

Recall from Section~\ref{sec:intro} the statistical model $\mathcal P=\{\mb P_\theta:\theta\in\Theta\}$ with parameter space $\Theta$ and observation space $\mathcal X\subseteq\mathbb R^{d_X}$. We observe $n$ i.i.d. observations $X^n=(X_1,\ldots,X_n)\sim\mb P_0^{(n)}$, where $\mb P_0$ is the fixed, unknown data-generating distribution. Write $\mb E_0$ for expectation under $\mb P_0^{(n)}$, and $\mb E_\theta^{(n)}$ when the observations follow $\mb P_\theta^{(n)}$.

We assume that observations can be simulated from the statistical model at any specified parameter value. Let $\mathsf T$ denote the inference target, which may be a parameter, a model label, or a parameter--support pair, depending on the inference problem. The value of $\mathsf T$ is known for each simulated task and is used directly in the training loss. The simulation procedure induces a joint distribution $\mb Q_n$ of $(\mathsf T,X^n)$. We generate $M$ independent training tasks $\mathcal D_M=\{(\mathsf T^{(m)},X^{n,(m)}):m\in[M]\}$.

The network need not predict $\mathsf T$ directly. For example, it may output model probabilities rather than a model label. To allow for such outputs, we introduce an augmented output space $\mathcal O$. Specifically, we learn a map $F$ that takes a dataset $X^n$ as input and produces $F(X^n)\in\mathcal O$.
A task-specific map $g:\mathcal O\to\mathcal A$ then converts this output into the final estimate or selection, where $\mathcal A$ denotes the set of possible estimates or selections. The map $g$ is fixed in advance.
Given a loss function $\ell_{\rm tr}$ measuring the accuracy of $F(X^n)$ relative to $\mathsf T$, the population and empirical training risks are
\[
\mathcal R(F)=\mb E_{\mb Q_n}[\ell_{\rm tr}\{F(X^n),\mathsf T\}]
\quad \mbox{and} \quad
\widehat{\mathcal R}_M(F)=\frac1M\sum_{m=1}^M\ell_{\rm tr}\{F(X^{n,(m)}),\mathsf T^{(m)}\}.
\]
Let $F^\ast$ denote a minimizer of $\mathcal R$ over all measurable maps from datasets to $\mathcal O$, and let $\mathcal R^\ast=\mathcal R(F^\ast)=\inf_G\mathcal R(G)$, where the infimum is taken over the same class of measurable maps. Thus, $F^\ast$ is the ideal target of training, while $\mathsf T$ is the quantity to be inferred. We approximate $F^\ast$ by minimizing $\widehat{\mathcal R}_M(F)$ over a class $\mathcal F$ of neural networks introduced in the next subsection, obtaining the trained network $\widehat F$.
At inference time, $\widehat F$ is applied to a new dataset $X^n\sim\mb P_0^{(n)}$, independent of the training tasks, and $g\{\widehat F(X^n)\}$ gives the final estimate or selection. For each of the three inference problems, we specify the statistical target, the task distribution, and the training loss below.

\medskip
\noindent {\bf Parameter estimation.}
We assume correct specification, $\mb P_0=\mb P_{\theta_0}$ for some unknown $\theta_0\in\Theta\subseteq\mathbb R^d$, and seek an estimator $\widehat\theta$ with small MSE $\mb E_0[\|\widehat\theta-\theta_0\|_2^2]$. Let $\ell_\theta(x)=\log p_\theta(x)$, where $p_\theta$ is the density of $\mb P_\theta$, and let $s_\theta(x)=\nabla_\theta\ell_\theta(x)$ and $I(\theta)=\mb E_\theta[s_\theta(X_1)s_\theta(X_1)^\top]$ denote the score and Fisher information, respectively. Our benchmark is the MLE $\widehat\theta_{\rm MLE}\in\argmax_{\theta\in\Theta}\sum_{i=1}^n\ell_\theta(X_i)$, whenever it exists. Under standard regularity conditions with fixed $d$, $\sqrt n(\widehat\theta_{\rm MLE}-\theta_0)\xrightarrow{d}\mathcal N(0,I(\theta_0)^{-1})$. We seek first-order asymptotic equivalence to the constrained MLE $\widehat\theta_{0,n}$ defined in Section~\ref{sec:theory-estimation}, namely, $\sqrt n\|\widehat\theta-\widehat\theta_{0,n}\|_2\to0$ in probability, so that the learned estimator inherits the asymptotic efficiency of the MLE under suitable regularity conditions.

We generate each task by drawing $\theta\sim\pi$ from a user-specified distribution $\pi$ on $\Theta$, followed by $X^n\sim\mb P_\theta^{(n)}$. The inference target is $\mathsf T=\theta$, and its joint distribution with the dataset is
\begin{equation}
\mb Q_n(d\theta,dx^n)=\pi(d\theta)\,\mb P_\theta^{(n)}(dx^n).
\label{eq:meta-task-law}
\end{equation}
The network directly produces an estimate $F(X^n)=\widehat\theta_F(X^n)\in\mathbb R^d$, so $\mathcal O=\mathcal A=\mathbb R^d$ and $g$ is the identity map. We use the squared-error loss $\ell_{\rm tr}\{F(X^n),\theta\}=\|F(X^n)-\theta\|_2^2$, as in~\eqref{eqn:regression}. When $\mb E_\pi[\|\theta\|_2^2]<\infty$, the population minimizer is the posterior mean $F^\ast(x^n)=\mb E_{\mb Q_n}[\theta\mid X^n=x^n]$. For any estimator $F$ with finite risk, we have
\begin{equation}
\mathcal R(F)-\mathcal R^\ast
=\mb E_{\mb Q_n}\bigl[\|F(X^n)-F^\ast(X^n)\|_2^2\bigr].
\label{eq:excess-mse}
\end{equation}
Thus, training over $\m F$ aims to approximate the posterior mean. Although $\pi$ plays the mathematical role of a prior, our analysis takes a frequentist perspective, distinguishing the approximation guarantees at fixed data-generating parameters from the generalization guarantees averaged over the task distribution; see Section~\ref{sec:theory-estimation}.

\medskip
\noindent {\bf Model selection.}
For model selection, let $\m P=\bigcup_{k=1}^K\mathcal M_k$, where $\mathcal M_k=\{\mb P_{k,\theta_k}:\theta_k\in\Theta_k\subseteq\mathbb R^{d_k}\}$. Under correct specification, we seek a rule $\widehat k$ satisfying $\mb P_0^{(n)}\{\widehat k\ne k_0\}\to0$, where $k_0=\argmin_{k:\,\mb P_0\in\mathcal M_k}d_k$ is assumed unique. Under misspecification, the target is $k^*\in\argmin_k\inf_{\theta_k\in\Theta_k}D_{\rm KL}(\mb P_0\|\mb P_{k,\theta_k})$. For fitted parameters $\widetilde\theta_k$, our aim is to bound the expected predictive KL risk $\mb E_0[D_{\rm KL}(\mb P_0\|\widetilde{\mb P})]$ of the selected distribution $\widetilde{\mb P}=\mb P_{\widehat k,\widetilde\theta_{\widehat k}}$ in terms of the minimum KL divergence among the candidate models.

To generate a task, we first draw a model index $k\in[K]$ with probability $w_k$, then draw its parameter $\theta_k\sim\pi_k$ from $\Theta_k$, and finally generate $X^n\sim\mb P_{k,\theta_k}^{(n)}$. The inference target is the model index $\mathsf T=k$, with $\theta_k$ integrated out in the joint distribution $\mb Q_n$ of $(k,X^n)$. Rather than directly predicting a model index, the network outputs model probabilities $F(X^n)=q(X^n)\in\Delta_K$, where $\Delta_K=\{u\in[0,1]^K:\sum_{k=1}^K u_k=1\}$. Thus, $\mathcal O=\Delta_K$ and $\mathcal A=[K]$, with $g$ selecting the model with the largest predicted probability:
\[
g\{q(X^n)\}=\widehat k_F(X^n)
\in\operatorname*{arg\,max}_{k\in[K]}q_k(X^n),
\]
where ties are resolved by a fixed rule.

We use the cross-entropy loss $\ell_{\rm tr}\{q(X^n),k\}=-\log q_k(X^n)$, giving the population risk $\mathcal R(F)=\mb E_{\mb Q_n}[-\log q_{\mathsf T}(X^n)]$. The population minimizer is $F^\ast(x^n)=q^\ast(x^n)=(q_1^\ast(x^n),\ldots,q_K^\ast(x^n))$, where $q_k^\ast(x^n)=\mb Q_n\{\mathsf T=k\mid X^n=x^n\}$ is the posterior model probability. The corresponding excess risk satisfies
\[
\mathcal R(F)-\mathcal R^\ast
=\mb E_{\mb Q_n}\left[
\sum_{k=1}^K q_k^\ast(X^n)\log\frac{q_k^\ast(X^n)}{q_k(X^n)}
\right].
\]
The connection between posterior model probabilities and BIC~\citep{schwarz1978dimension,kassRaftery1995bayes} motivates our theoretical analysis in Section~\ref{sec:theory-model}.

\medskip
\noindent {\bf Variable selection.}
We consider variable selection in canonical generalized linear models (GLMs) with $p$ covariates. Let $X_i=(Y_i,Z_i)$ denote the $i$th observation, where $Y_i$ is the response and $Z_i\in\mathbb R^p$ is the covariate vector. The observations $(Y_i,Z_i)$ are i.i.d., with $Z_i\sim Q_Z$ independently of the model parameter $\theta=(\icpt,\beta)$, where $\icpt\in\mathbb R$ is the intercept and $\beta\in\mathbb R^p$ collects the regression coefficients subject to selection. Conditional on $Z_i=z$, the response follows
\begin{equation}
p_\theta(y\mid z)
=\exp\left\{\frac{y\eta_\theta(z)-b\{\eta_\theta(z)\}}{a}
+c(y,a)\right\},
\qquad
\eta_\theta(z)=\icpt+z^\top\beta,
\label{eq:glm-model}
\end{equation}
with respect to a fixed measure $\nu_Y$, where $\eta_\theta(z)$ is the linear predictor, $a>0$ is the known dispersion parameter, $b$ is the cumulant function, and $c(y,a)$ is the remaining term in the exponential-family representation. The joint distribution is thus
$\mb P_\theta(dy,dz)=p_\theta(y\mid z)\nu_Y(dy)Q_Z(dz)$, under which all subsequent risks and probabilities are defined.
We assume correct model specification, with $\mb P_0=\mb P_{\theta_0}$ for some $\theta_0=(\icpt_0,\beta_0)$. Let $S_0=\{j\in[p]:\beta_{0j}\ne0\}$ denote the true support, with size $s_0=|S_0|$. Our goal is to recover $S_0$ while accurately estimating $\theta_0$. Accordingly, we seek a selected support $\widehat S\subseteq[p]$ and a parameter estimate $\widehat\theta=(\widehat\icpt,\widehat\beta)$, treating the intercept as a nuisance parameter.

For variable selection, we study the probability of incorrect support recovery and Hamming loss $H=|\widehat S\setminus S_0|+|S_0\setminus\widehat S|$, averaged over the task distribution. This formulation accommodates tasks with arbitrarily weak signals, for which exact support recovery at a fixed parameter may be information-theoretically impossible~\citep{wainwright2009information}. In addition, we study the MSE $\mb E_0[\|\widehat\theta-\theta_0\|_2^2]$ and expected KL risk $\mb E_0[D_{\rm KL}(\mb P_0\|\mb P_{\widehat\theta})]$, allowing $p$ to grow with $n$.

Each task is generated by drawing a support $S\subseteq[p]$, the intercept $\icpt$, and nonzero coefficients $\beta_S$, with $\beta_j=0$ for $j\notin S$, followed by observations from the resulting model. The inference target is $\mathsf T=(\theta,S)$, where $\theta=(\icpt,\beta)$ contains the full parameter vector. The network outputs $F(X^n)=(\widehat\theta_F(X^n),q(X^n))$, where $\widehat\theta_F\in\mathbb R^{1+p}$ is the parameter estimate and $q=(q_1,\ldots,q_p)\in[0,1]^p$ contains the predicted inclusion probabilities. Thus, $\mathcal O=\mathbb R^{1+p}\times[0,1]^p$ and $\mathcal A=\mathbb R^{1+p}\times 2^{[p]}$, where $2^{[p]}$ denotes the collection of subsets of $[p]$.

We combine squared-error loss for parameter estimation with binary cross-entropy loss for variable selection:
\[
\begin{aligned}
\ell_{\rm tr}\{F(X^n),(\theta,S)\}
&=\|\widehat\theta_F(X^n)-\theta\|_2^2
+\lambda\,\ell_{\mathrm{BCE}}\{q(X^n),S\},\\
\mbox{with} \ \ \ell_{\mathrm{BCE}}(u,S)
&=-\frac1p\Big\{\sum_{j\in S}\log u_j
+\sum_{j\in[p]\setminus S}\log(1-u_j)\Big\},
\end{aligned}
\]
where $\lambda>0$ is a tuning parameter. For any $\lambda>0$, the population minimizer $F^\ast(x^n)=(\widehat\theta^\ast(x^n),q^\ast(x^n))$
consists of the posterior mean
$\widehat\theta^\ast(x^n)=\mb E_{\mb Q_n}[\theta\mid X^n=x^n]$
and the marginal posterior inclusion probabilities
$q_j^\ast(x^n)=\mb Q_n\{j\in S\mid X^n=x^n\}$.
Thus, $\lambda$, which may depend on $n$, does not affect the population minimizer but serves to balance the two loss terms when training over the restricted network class $\mathcal F$. In Theorem~\ref{thm:vs-gen}, we quantify its effect on parameter estimation and variable selection.

To obtain the final parameter estimate and selected variables, we define the map $g$ by
\[
g\{F(X^n)\}
=\bigl(\widehat\theta_F(X^n),\widehat S_F(X^n)\bigr),
\qquad
\widehat S_F(X^n)=\{j\in[p]:q_j(X^n)>t\},
\]
where $t\in(0,1)$ is a threshold. In particular, setting $t=1/2$ and applying $g$ to the population minimizer $F^\ast$ corresponds to selecting the median probability model~\citep{barbieri2004optimal}, which minimizes posterior expected Hamming loss.
Section~\ref{sec:theory-variable} studies the concentration of the posterior mean and inclusion probabilities and the corresponding guarantees for the learned rule.
Further derivations of the population risk minimizers and excess-risk identities, together with the justification for thresholding the posterior inclusion probabilities, are provided in Appendix~\ref{app:meta-bayes}.

\subsection{Residual recurrent deep sets}\label{sec:rrds}

To construct the neural network class $\mathcal F$ introduced in the preceding subsection, we start with Deep Sets~\citep{zaheer2017deep}, which take the form
\[
F_{\mathrm{DS}}(X^n)=\rho_{\mathrm{DS}}\bigg(\frac1n\sum_{i=1}^n\phi_{\mathrm{DS}}(X_i)\bigg),
\]
where $\phi_{\mathrm{DS}}$ and $\rho_{\mathrm{DS}}$ are neural networks serving as the encoder and decoder, respectively. This structure resembles the classical method of moments: the encoder $\phi_{\mathrm{DS}}$ learns summary statistics, whose empirical averages, also known as mean pooling, are transformed into an estimate by the decoder $\rho_{\mathrm{DS}}$. Mean pooling also ensures permutation invariance, as required for inference from i.i.d. observations. However, beyond exponential families, most statistical models do not admit sufficient statistics of fixed dimension. Consequently, compressing the entire dataset into a single fixed-dimensional representation may lose information needed for efficient estimation, even with highly expressive encoders and decoders. This limitation motivates us to extend the Deep Sets architecture by allowing the network to revisit the observations as its estimate evolves.

Classical iterative estimation provides a natural way to revisit the observations. Starting from an initial estimate obtained by the method of moments, one can refine it through Newton--Raphson or Fisher scoring updates. Under suitable regularity conditions, even a single update from a sufficiently accurate initial estimate can achieve first-order asymptotic equivalence to the MLE~\citep[Section~5.7]{van2000asymptotic}. Both initialization and refinement involve empirical averages, but the quantities being averaged during refinement depend on the current estimate. We use the same structure in RRDS, repeatedly aggregating information from the observations to refine the current estimate. Although a single refinement suffices in our theoretical construction, the corresponding networks may be large. Multiple refinements allow the network to approach the target gradually, so that each update only needs to make progress toward the target rather than approximate an entire Fisher scoring step.

\begin{figure}[tbp]
\centering
\includegraphics[width=\textwidth]{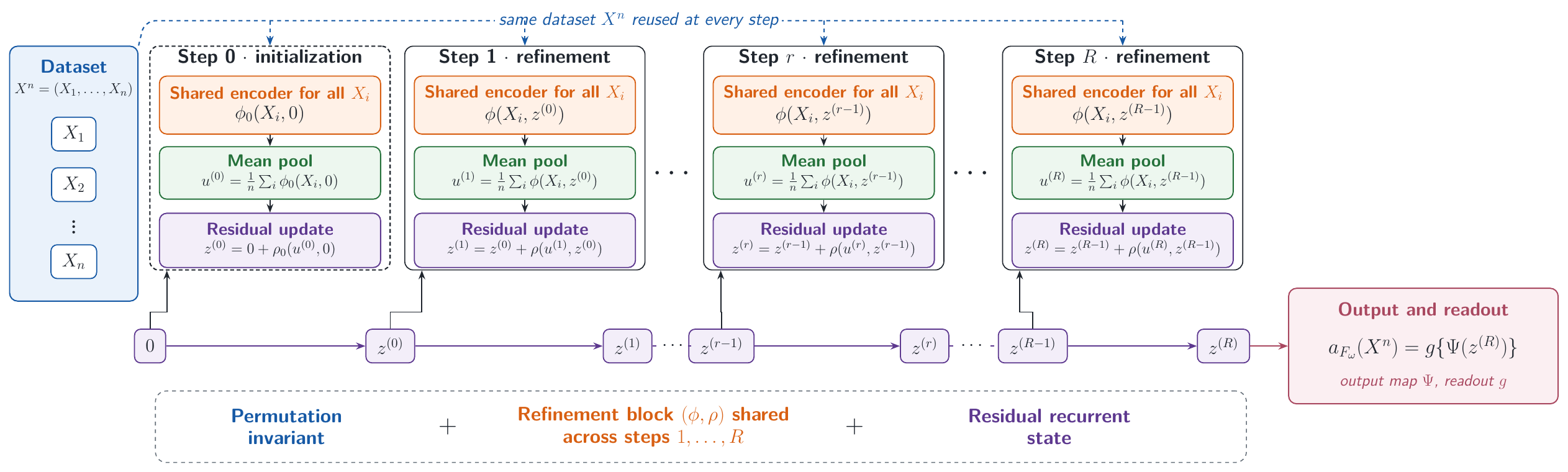}
\caption{Residual recurrent deep sets. Starting from the zero state, the initialization block encodes each observation and averages the resulting vectors to form $u^{(0)}$, which the decoder transforms into the initial state $z^{(0)}$. At each refinement step, the encoder processes the same observations together with the current state, and the decoder computes a correction that is added to that state. All refinement blocks share the same encoder $\phi$ and decoder $\rho$. After $R$ refinements, the output map $\Psi$ and the fixed map $g$ produce the final estimate or selection.}
\label{fig:rrds}
\end{figure}

Figure~\ref{fig:rrds} illustrates the RRDS architecture. Let $d_z$ and $d_u$ denote the dimensions of the state and pooled representation, respectively. We first construct an initial state using empirical averages of learned summary statistics, analogous to the method of moments. Specifically, we use two neural networks $\phi_0:\mathcal X\times\mathbb R^{d_z}\to\mathbb R^{d_u}$ and $\rho_0:\mathbb R^{d_u}\times\mathbb R^{d_z}\to\mathbb R^{d_z}$, both with rectified linear unit (ReLU) activation functions. Starting from the zero state, we compute
\begin{equation}
u^{(0)}=\frac1n\sum_{i=1}^n\phi_0(X_i,0),
\qquad
z^{(0)}=\rho_0(u^{(0)},0).
\label{eq:rrds-init}
\end{equation}
Here $\phi_0$ learns functions of the observations that play the role of summary statistics, while $\rho_0$ transforms their empirical averages into the initial state $z^{(0)}$.

We then refine the initial state through $R$ iterations, motivated by likelihood-based iterative estimation. Each refinement retains the Deep Sets structure of encoding, averaging, and decoding, but now both the encoder and decoder depend on the current state. Specifically, we introduce two ReLU networks $\phi:\mathcal X\times\mathbb R^{d_z}\to\mathbb R^{d_u}$ and $\rho:\mathbb R^{d_u}\times\mathbb R^{d_z}\to\mathbb R^{d_z}$ and update the state according to
\begin{equation}
u^{(r)}=\frac1n\sum_{i=1}^n\phi\bigl(X_i,z^{(r-1)}\bigr),
\qquad
z^{(r)}=z^{(r-1)}+\rho\bigl(u^{(r)},z^{(r-1)}\bigr),
\quad r=1,\ldots,R.
\label{eq:rrds-update}
\end{equation}
The encoder $\phi$ can approximate functions such as the likelihood score evaluated at the current estimate, while $\rho$ uses their empirical averages to compute a correction, analogous to a gradient descent, Newton--Raphson, or Fisher scoring update. The residual connection~\citep{he2016deep} allows the network to learn corrections to the current state, which may facilitate optimization with multiple refinements.

Although the method of moments initialization and likelihood-based refinement motivate different stages of RRDS, both use the same Deep Sets structure. We use separate networks $(\phi_0,\rho_0)$ for initialization and $(\phi,\rho)$ for refinement, but reuse the same $\phi$ and $\rho$ across all refinement blocks, giving RRDS its recurrent structure~\citep{elman1990finding}.
Applying the same encoder to every observation and averaging its outputs preserves permutation invariance throughout the iterations. For fixed architectures of the initialization and refinement networks, increasing either the sample size or the number of refinements does not increase the number of trainable parameters.

We use this parameter-sharing scheme throughout; alternative schemes are considered in Appendix~\ref{app:est-R}. When $R=0$, RRDS reduces to an ordinary Deep Sets network.

After $R$ refinements, we apply a fixed, task-specific output map $\Psi:\mathbb R^{d_z}\to\mathcal O$ to transform the final state $z^{(R)}$ into the augmented output space $\mathcal O$ defined in the preceding subsection. Let $\omega$ denote the trainable parameters of $\phi_0$, $\rho_0$, $\phi$, and $\rho$. The resulting network and inference rule are given by
\begin{equation}
F_\omega(X^n)=\Psi\bigl(z^{(R)}\bigr),
\qquad
a_{F_\omega}(X^n)=g\{F_\omega(X^n)\},
\label{eq:rrds-output}
\end{equation}
where $a_{F_\omega}(X^n)$ denotes the final estimate or selection, and $g:\mathcal O\to\mathcal A$ is the fixed map introduced in the preceding subsection.

The choice of $\Psi$ depends on the inference task. For parameter estimation, $\Psi$ can be the identity map or the projection onto a compact working parameter set, while $g$ is the identity map. For model selection, $\Psi$ applies the softmax function to convert the final state into model probabilities, and $g$ selects the model with the largest probability. For variable selection, the final state contains estimates of the intercept, coefficient values $b_j$, and real-valued inclusion scores $a_j$. We set the inclusion probabilities $q_j=\{1+\exp(-a_j)\}^{-1}$ and coefficient estimates $\widehat\beta_j=q_jb_j$, so that $\Psi$ produces both parameter estimates and inclusion probabilities, while $g$ selects variables satisfying $q_j>t$. For theoretical purposes, $\Psi$ may additionally incorporate clipping and projection operations to ensure boundedness of the relevant quantities, without introducing additional trainable parameters. Further details of these output maps and the boundedness requirements used in Section~\ref{sec:theory-results} are given in Appendix~\ref{app:meta-network}.

In Section~\ref{sec:theory-results}, the connection to classical iterative estimation guides our construction of RRDS networks, whereas in practice we train RRDS end to end using simulated tasks. Unlike transformer-based approaches to in-context learning, our construction does not require attention with learned weights: mean pooling suffices to approximate the iterative procedures used to establish asymptotic efficiency.

\subsection{Meta training and evaluation}\label{sec:training}

Given the simulated tasks $\mathcal D_M$, we train RRDS by approximately
minimizing the empirical risk $\widehat{\mathcal R}_M(F_\omega)$ over the
network parameters $\omega$, using minibatch Adam updates~\citep{kingma2015adam}
with a cosine learning-rate schedule~\citep{loshchilov2017sgdr}.
The loss function is chosen according to the inference task, as described
in Section~\ref{sec:meta-formulation}. Once trained, the network parameters
are held fixed, and inference on a new dataset requires only a forward
pass. Algorithm~\ref{alg:rrds} in Appendix~\ref{app:meta-training} gives the training procedure and implementation details.

The choice of the task distribution should reflect available knowledge about the parameters or models expected at inference time, as discussed in Appendix~\ref{app:meta-training}.

For the theoretical analysis, let $\mathcal F(L,W)$ denote a class of RRDS networks with depth at most $L$ and at most $W$ distinct trainable parameters. The depth counts the maximum number of successive affine transformations, including those across repeated refinement steps, and affects both approximation accuracy and the sensitivity of the network output to its parameters. The parameter count includes the trainable parameters of all network blocks, with shared parameters counted only once~\citep{bartlett2019nearly}, reflecting the actual number of parameters optimized during meta training. Let $L(f)$ and $W(f)$ denote the number of affine layers and distinct trainable parameters of a ReLU network $f$, respectively. For an RRDS network of the form~\eqref{eq:rrds-init}--\eqref{eq:rrds-output} with $R\geq1$, the depth and parameter count are
\begin{equation}
L=L(\phi_0)+L(\rho_0)+R\{L(\phi)+L(\rho)\},
\qquad
W=W(\phi_0)+W(\rho_0)+W(\phi)+W(\rho).
\label{eq:rrds-complexity}
\end{equation}
Increasing $R$ therefore increases the depth but not the number of trainable parameters, and neither quantity depends explicitly on $n$. For $R=0$, only the initialization blocks contribute to $L$ and $W$. Sample averaging, residual additions, and the fixed output map $\Psi$ introduce no additional trainable parameters. The full specification of $\mathcal F(L,W)$ also involves the number of refinement steps $R$, the dimensions $d_z$ and $d_u$, the output map $\Psi$, and the architectures of the four network blocks. These specifications are provided in the corresponding theoretical results. We simply write $\mathcal F$ when the network class is clear from the context.

To account for the optimization error in training, we define $\widehat F\in\mathcal F(L,W)$ as an approximate empirical risk minimizer satisfying
\begin{equation}
\widehat{\mathcal R}_M(\widehat F)\leq
\inf_{F\in\mathcal F(L,W)}\widehat{\mathcal R}_M(F)+\varepsilon_{\mathrm{opt}},
\label{eq:approx-erm}
\end{equation}
where $\widehat{\mathcal R}_M$ is the empirical counterpart of the task-specific population risk $\mathcal R$ introduced in Section~\ref{sec:method}, computed from $M$ independent meta training tasks, and $\varepsilon_{\mathrm{opt}}\geq0$ denotes the optimization error. Appendix~\ref{app:meta-complexity} discusses further aspects of network complexity and clarifies the distinction between approximation under a fixed data-generating distribution and generalization across tasks drawn from the meta-distribution.

We evaluate the learned inference rule on independently simulated tasks and report averages and Monte Carlo standard errors, with task-specific metrics defined in Appendix~\ref{app:meta-training}.

\subsection{Divide-and-conquer estimation and uncertainty quantification}
\label{sec:extensions}
We consider two applications of the learned estimator: inference from datasets larger than those used in meta training, and uncertainty quantification through bootstrap resampling.

\medskip
\noindent {\bf Divide-and-conquer parameter estimation.}
Although the RRDS architecture can accommodate datasets of different
sizes, a network trained at a fixed sample size need not retain
its estimation accuracy when applied directly to substantially
larger datasets. Moreover, meta training on large datasets can be
computationally expensive. Divide-and-conquer estimation addresses
both issues by allowing us to train on smaller datasets and apply
the learned estimator to larger samples without retraining.
Suppose $\widehat F$ is trained on datasets of size $n_0$.
For a larger dataset of size $n=Jn_0$, partition the observations
into $J$ disjoint blocks $X^{n_0}_{(1)},\ldots,X^{n_0}_{(J)}$
and average the block estimates,
\begin{equation}
\widehat\theta_{\mathrm{DC}}(X^n)=\frac1J\sum_{j=1}^J\widehat F\bigl(X^{n_0}_{(j)}\bigr).
\label{eq:dc-estimator}
\end{equation}
The usual divide-and-conquer approach~\citep{zhang2013communication,rosenblatt2016optimality} thus allows the network to operate at its original training sample size. We establish first-order equivalence to the full-sample MLE when $n=o(n_0^2)$ under suitable training conditions.

\medskip
\noindent {\bf Bootstrap uncertainty quantification.}
For bootstrap inference, the trained estimator can be evaluated on many resampled datasets without repeatedly solving an estimation problem. For a network trained at sample size
$n$, let $\mb P_n=n^{-1}\sum_{i=1}^n\delta_{X_i}$ be the empirical
distribution of the observations. Conditional on $X^n$, draw a
bootstrap sample $X^{*n}=(X_1^*,\ldots,X_n^*)$ i.i.d. from $\mb P_n$
and compute
\begin{equation}
\widehat\theta^*=\widehat F(X^{*n}).
\label{eq:bootstrap-estimator}
\end{equation}
Repeated evaluations on resampled datasets give the bootstrap distribution~\citep{efron1994introduction}. We establish bootstrap consistency in Corollary~\ref{cor:bootstrap}; confidence interval construction and numerical results are given in Appendices~\ref{app:meta-extensions} and~\ref{app:impl-bootstrap}.

\section{Theoretical Guarantees}\label{sec:theory-results}

For each inference task, we relate its population target to a classical inference procedure (the MLE, BIC, or EBIC), construct an RRDS network approximating that procedure, and use an excess risk bound to study an empirically trained network.

Throughout this section, $\widetilde F$ denotes a constructed RRDS network and $\widehat F$ denotes an approximate empirical risk minimizer trained on $M$ independent tasks $\mathcal D_M$, with optimization error $\varepsilon_{\rm opt}$ as in~\eqref{eq:approx-erm}. We use the notation $R$, $d_z$, $d_u$, $L$, and $W$ introduced in Section~\ref{sec:training} and~\eqref{eq:rrds-complexity}, with shared trainable parameters counted only once in $W$. Unless otherwise specified, the guarantees for trained networks concern a new task drawn from $\mb Q_n$ independently of $\mathcal D_M$, while for constructed networks we also establish guarantees under fixed data-generating distributions. Appendix~\ref{app:theory-discussion} contains the detailed network constructions and additional results, and all proofs are given in Appendix~\ref{app:proofs}.

\subsection{Parameter estimation}\label{sec:theory-estimation}

Consider $X^n\sim\mb P_\theta^{(n)}$ and fix a compact convex working parameter set $\Theta_0\subset\operatorname{int}(\Theta)$ with nonempty interior. Since the unrestricted MLE may not exist for some datasets, we use the constrained MLE $\widehat\theta_{0,n}\in\operatorname*{arg\,max}_{\theta\in\Theta_0}\sum_{i=1}^n\ell_\theta(X_i)$ as our benchmark. By compactness of $\Theta_0$ and the continuity assumption below, a maximizer always exists, and we take a fixed measurable choice if it is not unique. If the unrestricted MLE exists with probability tending to one and is consistent at $\theta_0\in\operatorname{int}(\Theta_0)$, the two estimators coincide with probability tending to one; see Appendix~\ref{app:class}. Let $\Pi_{\Theta_0}$ denote Euclidean projection onto $\Theta_0$.

We assume that $\mathcal X\subset\mb R^{d_X}$ is compact and that, for $\theta$ in an open neighborhood $U$ of $\Theta_0$, the densities $p_\theta$ are positive with respect to a common finite measure on $\mathcal X$. This includes both discrete and continuous observations.

\begin{assumption}[Regularity and nondegenerate moment function]\label{ass:regular}
For a real number $r\geq2$, the joint log density $(x,\theta)\mapsto\ell_\theta(x)$ has a $C^{r+1}$ extension to a neighborhood of $\mathcal X\times U$. There is a bounded measurable map $T:\mathcal X\to\mb R^d$ whose mean $\mu(\theta)=\mb E_\theta[T(X_1)]$ is one-to-one and has a nonsingular Jacobian on $U$.
\end{assumption}

Assumption~\ref{ass:regular} ensures that the model is identifiable and that the Fisher information satisfies $I\in C^r(U;\mathbb S_{++}^d)$. It also ensures that the method-of-moments estimator $\mu^{-1}(\bar T_n)$, with $\bar T_n=n^{-1}\sum_iT(X_i)$, is well defined with probability tending to one and is root-$n$ consistent at each $\theta_0\in\operatorname{int}(\Theta_0)$ (Lemma~\ref{lem:consequences}). Starting from this estimator, a single Fisher scoring update,
\begin{equation}
z\longmapsto z+I(z)^{-1}\frac1n\sum_{i=1}^ns_z(X_i),
\label{eq:one-step}
\end{equation}
yields an estimator that is first-order asymptotically equivalent to the MLE~\citep[Section~5.7]{van2000asymptotic}. To approximate the one-step estimator using RRDS, we use ReLU networks to approximate the inverse moment map $\mu^{-1}$, the score function, and the Fisher scoring correction $I(z)^{-1}u$. The resulting network has a single refinement ($R=1$), dimensions $d_z=d_u=d$, and output map $\Psi=\Pi_{\Theta_0}$. The four network blocks are specified in~\eqref{eq:est-blocks}--\eqref{eq:structured-class} in Appendix~\ref{app:rrds-estimation}.

The Fisher scoring update corrects the leading error of the initial method-of-moments estimator. Both the initialization and the refinement involve empirical averages of functions whose input dimensions do not depend on $n$. Thus, rather than approximating the mapping from the entire dataset $X^n$ to the MLE, we only need to approximate these individual functions. As the following theorem shows, a network whose size grows at most polynomially, rather than exponentially, in $n$ suffices to achieve the accuracy required for asymptotic efficiency.

\begin{theorem}[RRDS approximates the MLE]\label{thm:est-approx}
Under Assumption~\ref{ass:regular}, set $\kappa=(d_X+d)/r$. For each $\varepsilon\in(0,1)$, there is an RRDS network $\widetilde F\in\mathcal F(L,W)$ of the preceding form, with
\[
L\leq C\log(e/\varepsilon),\qquad
W\leq C\varepsilon^{-\kappa}\log(e/\varepsilon),
\]
such that, for every compact $\Theta_1\subset\operatorname{int}(\Theta_0)$ and all $n\geq1$,
\begin{equation}
\sup_{\theta\in\Theta_1}\mb E_\theta^{(n)}\big[
\|\widetilde F(X^n)-\widehat\theta_{0,n}\|_2^2\big]
\leq C_{\Theta_1}(\varepsilon^2+n^{-2}).
\label{eq:est-approx-mse}
\end{equation}
The constants are independent of $n$ and $\varepsilon$, and $C_{\Theta_1}$ depends in addition on $\Theta_1$. Consequently, if $\varepsilon=o(n^{-1/2})$, then, at each fixed $\theta_0\in\operatorname{int}(\Theta_0)$,
$\sqrt n(\widetilde F-\widehat\theta_{0,n})\to0$ in probability and
$\sqrt n(\widetilde F-\theta_0)\xrightarrow{d}\mathcal N\{0,I(\theta_0)^{-1}\}$.
\end{theorem}

The two terms in~\eqref{eq:est-approx-mse} correspond to the squared error in approximating the functions involved in the one-step estimator and the squared difference between the one-step estimator and the MLE, respectively. Since the same score network is applied to each observation before averaging, the exponent $\kappa$ depends on $d_X+d$, the input dimension of this network, rather than on $nd_X$, the dimension of the full sample. In particular, for a fixed approximation accuracy $\varepsilon$, the required network size is independent of $n$. Theorem~\ref{thm:theory-unbounded} in the appendix extends the result to unbounded observations by clipping the observations and controlling the error on suitable bounded sets.

\medskip
\noindent{\bf Guarantees for a trained network.}
We next study whether a network trained on finitely many tasks can attain the same asymptotic efficiency. Recall that the squared-error loss targets the posterior mean $F^*(X^n)=\mb E_{\mb Q_n}[\theta\mid X^n]$, rather than the MLE. To relate the posterior mean to the MLE, we impose the following condition on the task distribution.

\begin{assumption}[Meta training distribution]\label{ass:proposal}
The density $\pi\in C^1(\mb R^d)$ has compact support in $\operatorname{int}(\Theta_0)$ and finite prior information
$\mathcal J(\pi)=\int_{\{\pi>0\}}\|\nabla\log\pi(\theta)\|_2^2\,\pi(\theta)\,d\theta$.
\end{assumption}

Under Assumptions~\ref{ass:regular} and~\ref{ass:proposal}, Proposition~\ref{prop:bayes-mle} gives
$\mb E_{\mb Q_n}[\|F^*(X^n)-\widehat\theta_{0,n}\|_2^2]\leq C\{1+\mathcal J(\pi)\}/n^2$.
The root mean squared difference between the posterior mean and the constrained MLE is therefore $O(n^{-1})$ under the task distribution. Consequently, the RRDS network constructed in Theorem~\ref{thm:est-approx} also approximates the posterior mean. Since the excess risk under squared-error loss equals the expected squared distance from the posterior mean (see~\eqref{eq:excess-mse}), we can use an excess risk bound for the trained network to control its distance from the MLE.

Fix $\varepsilon_n\in[n^{-1},1)$ and consider the class of RRDS networks with the architecture in Theorem~\ref{thm:est-approx} at approximation accuracy $\varepsilon_n$. We train $\widehat F$ under squared-error loss over this class, subject to $\|\omega\|_2\leq\Gamma=O(n^C)$, where $\Gamma$ is sufficiently large to include the constructed network.

\begin{theorem}[Meta learning generalization for parameter estimation]\label{thm:est-gen}
Under Assumptions~\ref{ass:regular} and~\ref{ass:proposal}, for $n\geq3$, $M\geq1$, and $\delta\in(0,1)$, with probability at least $1-\delta$ over $\mathcal D_M$,
\begin{equation}
\begin{aligned}
\mb E[\|\widehat F(X^n)-\widehat\theta_{0,n}\|_2^2\mid\mathcal D_M]
&\leq C\frac{W\{(\log n)^2+\log(eM)\}+\log(3/\delta)}{M}\\
&\quad+C\varepsilon_n^2+\frac{C\{1+\mathcal J(\pi)\}}{n^2}
+4\varepsilon_{\rm opt}.
\end{aligned}
\label{eq:est-gen-bound}
\end{equation}
Here $C$ is independent of $n$, $M$, $\delta$, $\varepsilon_n$, and $\varepsilon_{\rm opt}$. If
$\varepsilon_n=o(n^{-1/2})$,
$W\{(\log n)^2+\log(eM)\}/M=o(n^{-1})$, and
$\varepsilon_{\rm opt}=o(n^{-1})$, then, for every $\gamma>0$,
$\mb P\{\sqrt n\|\widehat F-\widehat\theta_{0,n}\|_2>\gamma\mid\mathcal D_M\}\to0$
in probability over $\mathcal D_M$, and
\[
d_{\rm BL}\!\left[
\mathcal L\{\sqrt n(\widehat F-\theta)\mid\theta,\mathcal D_M\},
\mathcal N\{0,I(\theta)^{-1}\}\right]\to0
\]
in probability over $(\theta,\mathcal D_M)$ as $n\to\infty$.
\end{theorem}

While Theorem~\ref{thm:est-approx} provides approximation guarantees uniformly over compact subsets of the interior of $\Theta_0$, Theorem~\ref{thm:est-gen} establishes asymptotic efficiency on a new task drawn from the meta-distribution rather than at each fixed parameter value. Since the training risk is averaged over the task distribution, the network may have larger estimation errors in regions of the parameter space with small prior probability.

The four terms in~\eqref{eq:est-gen-bound} arise from the finite number of meta training tasks, the neural approximation error, the difference between the posterior mean and the MLE, and the optimization error, respectively. Since the third term is already $O(n^{-2})$, asymptotic efficiency can be achieved by making each of the remaining terms $o(n^{-1})$. For example, choosing $\varepsilon_n=n^{-1/2}/\log n$ gives $W=O\{n^{\kappa/2}(\log n)^{\kappa+1}\}$, and $M=\lceil n^{\kappa/2+1}(\log n)^{\kappa+4}\rceil$ meta training tasks suffice, provided $\varepsilon_{\rm opt}=o(n^{-1})$. For fixed $d$ and $d_X$, both the network size and the required number of meta training tasks thus grow at most polynomially in $n$.
Under the conditions of Corollary~\ref{cor:analytic-est-gen}, which require the log density and the inverse moment map to admit suitable analytic extensions, the bounds improve to $W=O\{(\log n)^{d_X+d+1}\}$ and $M=\lceil n(\log n)^{d_X+d+5}\rceil$. Thus, a network whose size grows only polynomially in $\log n$ suffices for asymptotic efficiency, and the required number of meta training tasks is nearly linear in $n$.

Although a single refinement suffices for asymptotic efficiency, our numerical experiments show that increasing the number of refinements can further improve estimation accuracy (Appendix~\ref{app:est-R}).

\medskip
\noindent{\bf Divide-and-conquer estimation and bootstrap inference.}
Corollary~\ref{cor:dc} establishes first-order equivalence of the divide-and-conquer estimator~\eqref{eq:dc-estimator} to the full-sample constrained MLE when $n=o(n_0^2)$, provided that the approximation, meta-generalization, and optimization errors on the squared-error scale are $o(n^{-1})$. Averaging does not remove the order-$n_0^{-1}$ bias of the block estimates, which explains the restriction on $n$.

Corollary~\ref{cor:bootstrap} establishes consistency of the bootstrap distribution for the constructed and trained networks. For the latter, additional uniform bounds on the magnitudes and Lipschitz constants of network components allow us to control the approximation error on bootstrap samples, which may contain repeated observations. The precise statements are given in Appendix~\ref{app:dc-bootstrap}.

\subsection{Model selection}\label{sec:theory-model}

Consider the $K$ candidate families $\mathcal M_k=\{\mb P_{k,\theta}:\theta\in\Theta_k\}$ introduced in Section~\ref{sec:preliminaries}, where the number of models $K$ and their dimensions $d_k$ are fixed. For each $k$, let $\Theta_{0,k}\subset\operatorname{int}(\Theta_k)$ be a compact convex polytope with nonempty interior and $U_k$ an open neighborhood of $\Theta_{0,k}$. Assume that $\mathcal X$ is compact and that, for $\theta\in U_k$, the candidate models have positive densities on $\mathcal X$ with respect to a common finite measure. Write $\ell_{k,\theta}=\log p_{k,\theta}$, $s_{k,\theta}=\nabla_\theta\ell_{k,\theta}$, $I_k(\theta)=\mb E_{k,\theta}[s_{k,\theta}(X_1)s_{k,\theta}(X_1)^\top]$, and $\ell_{n,k}(\theta)=\sum_i\ell_{k,\theta}(X_i)$. As in parameter estimation, we use a constrained MLE $\widehat\theta_k$ over $\Theta_{0,k}$. The polytope condition ensures that the projections within the network can be represented exactly by ReLU networks.

\begin{assumption}[Candidate models and meta training distribution]\label{ass:models}
\textup{(i)} Each candidate $k$ satisfies Assumption~\ref{ass:regular} on $\mathcal X\times U_k$, with moment function $T_k$, mean map $\mu_k$, and smoothness index $r_k$.

\textup{(ii)} For each $k$, $w_k>0$ and $\pi_k$ has a Lebesgue density $f_k$ supported on a rectangle $\Xi_k=\prod_{j=1}^{d_k}[a_{k,j},b_{k,j}]\subset\operatorname{int}(\Theta_{0,k})$, with $a_{k,j}<b_{k,j}$ and $c_\pi\leq f_k\leq C_\pi$ on $\Xi_k$, for constants $0<c_\pi\leq C_\pi<\infty$.
\end{assumption}

Recall that the population minimizer of the cross-entropy loss for model selection is the posterior model probability vector,
\[
q_k^*(X^n)=\frac{w_km_k(X^n)}{\sum_{l=1}^Kw_lm_l(X^n)},
\qquad
m_k(X^n)=\int e^{\ell_{n,k}(\theta)}\,\pi_k(d\theta).
\]
The Laplace approximation relates these probabilities to BIC~\citep{schwarz1978dimension,kassRaftery1995bayes}. Writing $\mathrm{BIC}_k=-2\ell_{n,k}(\widehat\theta_k)+d_k\log n$, we define $a_k^{\rm BIC}=\log w_k-\mathrm{BIC}_k/2$ and the corresponding probabilities by
\begin{equation}
a_k^{\rm BIC}=\log w_k+\ell_{n,k}(\widehat\theta_k)-\tfrac{d_k}2\log n,
\qquad q^{\rm BIC}=\operatorname{softmax}(a^{\rm BIC}).
\label{eq:bic-logits}
\end{equation}
Let $\psi_n$ denote the probability that a task fails the separation condition~\eqref{eq:weak-separation} in Appendix~\ref{app:rrds-model}, which requires competing models either to contain the generating distribution at a suitable interior point of their prior supports with larger dimension, or to be separated in KL divergence by at least $A_1\log n/n$. BIC omits terms involving the prior density and likelihood curvature in the Laplace expansion, which need not vanish as $n$ grows. While these terms can affect the posterior probabilities that the network is trained to approximate, bounded errors in the log odds have little effect on the probabilities themselves when the posterior concentrates on a single model.

Write ${\rm BE}_n=\mb E_{\mb Q_n}[1-\max_kq_k^*(X^n)]$ for the Bayes misselection probability. Proposition~\ref{prop:ms-bridge} bounds the expected KL divergence from $q^*$ to $q^{\rm BIC}$ by $C\{{\rm BE}_n+\psi_n\log n+n^{-1/2}\}$. Motivated by this connection, we construct an RRDS network that approximately implements BIC minimization using two refinement steps with shared network blocks. In the first step, the network computes the one-step parameter estimates for all candidate models. In the second, it evaluates the log likelihoods at these estimates and uses them to obtain the model probabilities. We center and clip the logits before applying softmax to ensure that all model probabilities are at least $1/(Kn)$, so the cross-entropy loss is bounded. The detailed construction is given in Appendix~\ref{app:rrds-model}. Let $d_{\max}=\max_kd_k$, $\kappa_{\max}'=\max_k(d_X+d_k)/(r_k+1)$, and $b_n=(\log n)^{d_{\max}/2}/\sqrt n$.

\begin{theorem}[RRDS approximates the posterior model probabilities]\label{thm:ms-approx}
Under Assumption~\ref{ass:models}, for every $n\geq3$ there is an RRDS network $\widetilde q\in\mathcal F(L,W)$ of the preceding form, with $R=2$,
$L\leq C\log n$, and $W\leq CKn^{\kappa_{\max}'}\log n$, such that
\begin{equation}
\begin{aligned}
\mb E_{\mb Q_n}[D_{\rm KL}(q^*\|\widetilde q)]
&\leq C\{{\rm BE}_n+\psi_n\log n+n^{-1/2}\}\leq C(b_n+\psi_n\log n),\\
\mb Q_n(\widetilde k\ne k)&\leq Cb_n+\psi_n,
\end{aligned}
\label{eq:ms-approx-kl}
\end{equation}
where $\widetilde k\in\operatorname*{arg\,max}_k\widetilde q_k$ and $C$ is independent of $n$.
For a fixed distribution $\mb P_0=\mb P_{k_0,\theta_0}$ with $\theta_0\in\operatorname{int}(\Theta_{0,k_0})$, suppose every $k\ne k_0$ either contains $\mb P_0$ at an interior point of $\Theta_{0,k}$ with $d_k>d_{k_0}$, or satisfies $\inf_{\theta\in\Theta_{0,k}}D_{\rm KL}(\mb P_0\|\mb P_{k,\theta})>0$. Then
\begin{equation}
\mb P_0^{(n)}(\widetilde k\ne k_0)\leq Cb_n,
\label{eq:ms-recovery-rate}
\end{equation}
where $C$ may also depend on $\mb P_0$.
\end{theorem}

The $n^{-1/2}$ rate in the probability of selecting the wrong model (up to logarithmic factors) arises from comparisons with larger models that also contain the generating distribution. In such cases, BIC penalizes the increase in the log likelihood by $(\log n)/2$ for each additional parameter. The term $\psi_n$ accounts for tasks that fail the separation condition, for example, because the generating distribution is close to an intersection of candidate models or because of the restrictions on the prior supports. Lemma~\ref{lem:psi-nested} gives bounds on $\psi_n$ for nested and separated model families. If, for every task satisfying the separation condition, no larger candidate model contains the generating distribution, the remainder in the KL bound of Proposition~\ref{prop:ms-bridge} improves to $\log n/n$.

We also study the predictive risk of the constructed network when the candidate models may be misspecified. Proposition~\ref{prop:ms-misspec} in Appendix~\ref{app:ms-misspecification} gives oracle inequalities for predictive KL risk in terms of the KL approximation errors of the candidate models, their dimension penalties, and the bias of the one-step estimators. The remainder is $O(n^{-1/2})$ in general. Under bounded log-likelihood ratios, it improves to $O(\log n/n)$, at the cost of a factor of two on the approximation error. The bias term arises because, under misspecification, the method-of-moments estimator need not converge to the parameter that minimizes KL divergence from the generating distribution.

\medskip
\noindent{\bf Guarantees for a trained network.}
We now consider a network $\widehat q$ trained under cross-entropy loss over the class of RRDS networks with the same architecture as $\widetilde q$, subject to $\|\omega\|_2\leq\Gamma=O(n^C)$, with $\Gamma$ sufficiently large to include the constructed network. Let $\widehat k\in\operatorname*{arg\,max}_k\widehat q_k$ denote the selected model, and define the training error term
\[
\Delta_{\rm ms}
=\frac{\log n\,[W\{(\log n)^2+\log(eM)\}+\log(3/\delta)]}{M}
+\varepsilon_{\rm opt}.
\]
For brevity, write $\mb E_M[\cdot]=\mb E[\cdot\mid\mathcal D_M]$ and $\mb P_M(\cdot)=\mb P(\cdot\mid\mathcal D_M)$ here and in the next subsection.

\begin{theorem}[Meta learning generalization for model selection]\label{thm:ms-gen}
Under Assumption~\ref{ass:models}, for $n\geq3$, $M\geq1$, and $\delta\in(0,1)$, with probability at least $1-\delta$ over $\mathcal D_M$,
\begin{align}
\mb E_M[D_{\rm KL}(q^*\|\widehat q)]
&\leq C(b_n+\psi_n\log n+\Delta_{\rm ms}),\label{eq:ms-gen-kl}\\
\mb P_M(\widehat k\ne k)
&\leq C(b_n+\psi_n\log n+\Delta_{\rm ms}).\label{eq:ms-gen-err}
\end{align}
Here $C$ is independent of $n$, $M$, $\delta$, and $\varepsilon_{\rm opt}$. If $\psi_n\log n\to0$ and $\Delta_{\rm ms}\to0$, the conditional misselection probability tends to zero in probability over $\mathcal D_M$.
\end{theorem}

As in parameter estimation, the excess risk allows us to compare the trained network with the constructed one. Under cross-entropy loss, it equals the expected KL divergence from the posterior model probabilities to the probabilities estimated by the network. To see how this controls selection errors, consider datasets for which the posterior assigns a high probability to one model. Selecting a different model then requires a substantial error in the estimated probability vector, and we can bound the probability of such an error by the expected KL divergence. The probability of observing a dataset for which the posterior is not sufficiently concentrated is controlled by ${\rm BE}_n$. This gives the linear dependence of the selection error on the expected KL divergence in the proof of Theorem~\ref{thm:ms-gen} (Appendix~\ref{app:proof-ms-gen}).

The factor $\log n$ in $\Delta_{\rm ms}$ comes from bounding the cross-entropy loss after clipping the logits. With $\delta=n^{-1}$ and $\varepsilon_{\rm opt}=O(n^{-1/2})$, $M=\lceil n^{\kappa_{\max}'+1/2}(\log n)^4\rceil$ meta training tasks suffice for $\Delta_{\rm ms}=O(n^{-1/2})$. The error due to training therefore does not worsen the approximation rate. Under analogous analyticity conditions for the candidate models, Corollary~\ref{cor:analytic-ms} improves these bounds to $W=O\{K(\log n)^{d_X+d_{\max}+1}\}$ and $M=\lceil\sqrt n(\log n)^{d_X+d_{\max}+5}\rceil$. For a fixed collection of candidate models, the network size thus grows only polylogarithmically in $n$, and even a sublinear number of meta training tasks suffices. As in parameter estimation, the guarantees for the trained network are averaged over the task distribution $\mb Q_n$, while the predictive risk bounds under model misspecification apply to the constructed network.

\subsection{Variable selection}\label{sec:theory-variable}

Consider the canonical GLM~\eqref{eq:glm-model} with intercept $\icpt$ and candidate supports $\mathcal S=\{S\subseteq[p]:1\leq|S|\leq s_{\max}\}$. We assume that the maximal support size $s_{\max}$ is fixed, but allow $p\geq2s_{\max}$ to grow with $n$. Write $\theta_S=(\icpt,\beta_S)$, $d_S=1+|S|$, and let $\theta^{(S)}\in\mb R^{1+p}$ denote the corresponding full parameter vector with coefficients outside $S$ set to zero. For each support, let $\Theta_{0,S}\subset[-B,B]^{d_S}$ be a compact working rectangle with nonempty interior. We assume that if $S\subseteq S'$, extending a parameter in $\Theta_{0,S}$ by zeros yields a parameter in $\Theta_{0,S'}$. Let $\Theta_{\mathcal S}$ denote the union of these rectangles after extending their parameters to $\mb R^{1+p}$. We also assume that $\nu_Y$ is a finite measure supported on $[-B,B]$ and that $Q_Z\{\max_j|Z_j|\leq B\}=1$, where $B$ is independent of $n,p,S$. This setting accommodates bounded responses, including binary and binomial outcomes, and bounded covariates.

Let $\ell_{n,S}$ and $\widehat\theta_S$ denote the log likelihood and constrained MLE for support $S$, respectively, and let $\widehat\theta^{(S)}$ be the corresponding full parameter vector with coefficients outside $S$ set to zero. Recall that the squared-error and binary cross-entropy components of the training loss target the posterior mean $\theta^*=\mb E[\theta\mid X^n]$ and the marginal posterior inclusion probabilities $q_j^*=\sum_{S\ni j}Q_S^*$, where
\[
Q_S^*(X^n)=\frac{w(S)m_S(X^n)}{\sum_{S'\in\mathcal S}w(S')m_{S'}(X^n)},
\qquad
m_S(X^n)=\int e^{\ell_{n,S}(\theta_S)}\,\pi_S(d\theta_S).
\]
The binary cross-entropy loss targets each marginal inclusion probability regardless of the posterior dependence among the inclusion indicators. As in model selection, we use a Laplace approximation to relate the posterior weights to EBIC. For a prior uniform over support sizes and over supports of a given size, $w(S)=\{s_{\max}\binom p{|S|}\}^{-1}$, the corresponding EBIC weights are~\citep{chen2008extended}
\begin{equation}
\begin{gathered}
Q_S^{\rm E}=\frac{\exp(a_S^{\rm E})}{\sum_{S'\in\mathcal S}\exp(a_{S'}^{\rm E})},
\qquad a_S^{\rm E}=\ell_{n,S}(\widehat\theta_S)-\operatorname{pen}(S),\\
\operatorname{pen}(S)=\frac{d_S}2\log n+\log\binom p{|S|}.
\end{gathered}
\label{eq:ebic-weights}
\end{equation}
The penalty $\log\binom p{|S|}$ arises from the prior on the supports. The corresponding inclusion probabilities and parameter estimate are
\begin{equation}
q_j^{\rm E}=\sum_{S\ni j}Q_S^{\rm E},\qquad
\theta^{\rm E}=\sum_{S\in\mathcal S}Q_S^{\rm E}\widehat\theta^{(S)}.
\label{eq:ebic-targets}
\end{equation}
In particular, each regression coefficient in $\theta^{\rm E}$ is the product of its inclusion probability and its weighted average estimate conditional on inclusion, as in the RRDS output map for variable selection.

For each support, we use $v_S(Z)=(1,Z_S^\top)^\top$ and the moment function $T_S(Y,Z)=Yv_S(Z)$ for initialization. Its mean is $\mu_S(\theta_S)=\mb E_{Q_Z}[b'\{v_S(Z)^\top\theta_S\}v_S(Z)]$, with Jacobian $aI_S(\theta_S)$, where $I_S$ is the Fisher information for support $S$. We impose the following conditions to control the estimation and neural approximation errors uniformly over the candidate supports.

\begin{assumption}[Sparse generalized linear models]\label{ass:sparse}
There are constants $m_b,M_b,\underline\lambda_Z,\overline\lambda_Z,\rho_g>0$, independent of $n,p,S$, such that:

\textup{(i)} There is a compact interval $\mathcal I_{\rm pred}$ in the natural-parameter interval such that $Q_Z\{\eta_\theta(Z)\in\mathcal I_{\rm pred}\}=1$ for every $\theta\in\Theta_{\mathcal S}$, and $m_b\leq b''(u)\leq M_b$ on $\mathcal I_{\rm pred}$.

\textup{(ii)} For every $h_{\icpt}\in\mb R$ and $h_\beta\in\mb R^p$ with at most $2s_{\max}$ nonzero entries,
\begin{equation}
\underline\lambda_Z(h_{\icpt}^2+\|h_\beta\|_2^2)
\leq\mb E_{Q_Z}[(h_{\icpt}+Z^\top h_\beta)^2]
\leq\overline\lambda_Z(h_{\icpt}^2+\|h_\beta\|_2^2).
\label{eq:glm-sparse-design}
\end{equation}

\textup{(iii)} The range of $\mu_S$ contains the closed $\rho_g$-neighborhood $K_{g,S}$ of $\mu_S(\Theta_{0,S})$ defined in~\eqref{eq:moment-inverse-domain}. The inverse $\mu_S^{-1}:K_{g,S}\to\mb R^{d_S}$ admits a holomorphic extension with constants $(\mathcal R_{g,S},\varrho_{g,n},\mathcal B_{g,n})$ in the precise sense of~\eqref{eq:extension-main} in Appendix~\ref{app:analytic-extension}, where $\mathcal R_{g,S}$ is a rectangle and $\varrho_{g,n}>1$, $\mathcal B_{g,n}<\infty$ are common to all $S$.
\end{assumption}

Conditions~(i)--(ii) ensure that the Fisher information is uniformly nonsingular, while allowing correlated covariates. Condition~(iii) allows the inverse moment map to be approximated by a network whose size grows slowly with the required accuracy. Let $L_{{\rm vs},n}$ and $V_{{\rm vs},n}$ denote the resulting depth and parameter bounds per support derived in Appendix~\ref{app:vs-setting}. If $\varrho_{g,n}$ is bounded away from one and $\mathcal B_{g,n}$ grows at most polynomially in $n$, then $L_{{\rm vs},n}=O\{\log^2(en)\}$ and $V_{{\rm vs},n}=O\{\log^{2+s_{\max}}(en)\}$.

\begin{assumption}[Meta training distribution]\label{ass:sparse-proposal}
There are constants $a_\pi,c_-,c_+,\rho>0$ and $L_\pi\geq0$, independent of $n,p,S$, such that:

\textup{(i)} $w(S)=\{s_{\max}\binom p{|S|}\}^{-1}$ for every $S\in\mathcal S$.

\textup{(ii)} On $\mathcal I_S=\mathcal I_0\times\prod_{j\in S}\mathcal I_j$, the prior has density $\pi_S(\theta_S)=g_0(\icpt)\prod_{j\in S}g_j(\beta_j)$. Each $g_j$ is a probability density supported on an interval $\mathcal I_j$, with $|\mathcal I_0|\geq2a_\pi$, $[-a_\pi,a_\pi]\subseteq\mathcal I_j$ for $j\geq1$, and
\[
c_-\leq g_j(u)\leq c_+,\qquad
|\log g_j(u)-\log g_j(u')|\leq L_\pi|u-u'|
\quad(u,u'\in\mathcal I_j).
\]

\textup{(iii)} For every $S$ and $\theta_S\in\mathcal I_S$, $\theta_S\in\operatorname{int}(\Theta_{0,S})$ and $\operatorname{dist}(\theta_S,\partial\Theta_{0,S})\geq\rho$.
\end{assumption}

The prior in Assumption~\ref{ass:sparse-proposal} has a spike-and-slab structure: coefficients outside $S$ are fixed at zero (the spike), while those in $S$ follow continuous distributions (the slab). We refer to $g_j$ and $\mathcal I_j$, for $j\geq1$, as the slab densities and their supports, respectively. Since each slab density is bounded away from zero near the origin, the task distribution includes arbitrarily weak signals. The bounds on the densities and the Lipschitz condition on their logarithms allow us to control the Laplace approximation uniformly over candidate supports. Condition~(iii) keeps the generating parameters away from the boundaries of the working rectangles, including when extended by zeros to larger supports.

The posterior inclusion probabilities also determine the optimal variable selection rule under Hamming loss, which selects variable $j$ if $q_j^*>1/2$. Write $\Lambda_n=\log(en)+\log(ep)$, and define
\[
H_n^*=\sum_{j=1}^p\mb E_{\mb Q_n}[\min\{q_j^*,1-q_j^*\}],
\qquad
h_n=s_{\max}^{7/2}\sqrt{\frac{\Lambda_n}n}.
\]
Here $H_n^*$ is the Bayes Hamming risk, which satisfies $H_n^*\leq Cs_{\max}^{5/2}\sqrt{\Lambda_n/n}$ by Proposition~\ref{prop:vs-bridge}. The same proposition, together with Proposition~\ref{prop:vs-bridge-est}, shows that the EBIC-based quantities $(\theta^{\rm E},q^{\rm E})$ approximate the posterior mean and inclusion probabilities $(\theta^*,q^*)$. In particular, the sum of the expected KL divergences for the inclusion probabilities is bounded by $Ch_n$, while the mean squared difference between the coefficient estimates is bounded by $Cs_{\max}/n$; see~\eqref{eq:vs-bridge}. These results connect the Bayesian targets of the training losses to the EBIC-based procedure used in our network construction.

We construct an RRDS network to approximate the EBIC-based quantities in~\eqref{eq:ebic-targets}. As in model selection, the construction uses two refinement steps with shared network blocks. The first computes one-step parameter estimates for all candidate supports, while the second evaluates the log likelihoods at these estimates and combines the resulting EBIC weights to obtain the coefficient estimates and inclusion probabilities. The network outputs $(\widetilde\theta,\widetilde q)$, and the final selected support is $\widetilde S=\{j:\widetilde q_j>1/2\}$. The construction is given in Appendix~\ref{app:rrds-variable}, together with an illustration in Figure~\ref{fig:vs-rrds}. For a generating parameter $\theta_0=(\icpt_0,\beta_0)$ with support $S_0$, write $s_0=|S_0|$ and $d_0=1+s_0$.

\begin{theorem}[RRDS approximation for variable selection and estimation]\label{thm:vs-approx}
Suppose Assumptions~\ref{ass:sparse} and~\ref{ass:sparse-proposal} hold and $s_{\max}^4\Lambda_n^3\leq c_0n$ for a sufficiently small $c_0>0$. There is an RRDS network $\widetilde F=(\widetilde\theta,\widetilde q)\in\mathcal F(L,W)$ with $R=2$, the architecture described in Appendix~\ref{app:rrds-variable}, and the complexity bounds in~\eqref{eq:vs-network-complexity}, satisfying:

\textup{(a)} Under the task distribution,
\[
\sum_{j=1}^p\mb E_{\mb Q_n}[D_{\rm KL}(q_j^*\|\widetilde q_j)]
\leq Ch_n,
\qquad
\mb E_{\mb Q_n}[\|\widetilde\theta-\theta\|_2^2]
\leq Cs_{\max}/n.
\]
Here $D_{\rm KL}(u\|v)$ denotes Bernoulli KL divergence.

\textup{(b)} For every $S_0\in\mathcal S$ and $\theta_0$ in the support of $\pi_{S_0}$,
\[
\mb E_0[\|\widetilde\theta-\theta_0\|_2^2]
+\mb E_0[D_{\rm KL}(\mb P_{\theta_0}\|\mb P_{\widetilde\theta})]
\leq\frac{Cs_{\max}\Lambda_n}n.
\]
The constants $c_0,C$ depend only on the response family, $B$, and the constants in the assumptions, not on $n,p,s_{\max},S_0$, or the generating coefficients.
\end{theorem}

Theorem~\ref{thm:vs-approx}(b) gives estimation and prediction risk bounds at each fixed generating parameter in the prior support. Averaging over the task distribution yields the sharper estimation rate $s_{\max}/n$ in part~(a), without the logarithmic dependence on $p$. This improvement relies on the bounded slab densities, which imply that tasks with coefficients near zero occur with small probability. For fixed $s_{\max}$, the sample size condition allows $p$ to grow as fast as $\exp(cn^{1/3})$ for a suitable constant $c>0$. Under the analyticity conditions above, the network size per support grows only polylogarithmically in $n$, although the total number of parameters also depends on the number of candidate supports. Further rate comparisons are given in Appendix~\ref{app:vs-training}.

\medskip
\noindent{\bf Guarantees for a trained network.}
For $F=(\theta_F,q_F)$, the composite training loss with weight $\lambda>0$ has excess risk
\begin{equation}
\mathcal R(F)-\mathcal R^*
=\mb E_{\mb Q_n}[\|\theta_F-\theta^*\|_2^2]
+\frac\lambda p\sum_{j=1}^p\mb E_{\mb Q_n}[D_{\rm KL}(q_j^*\|q_{F,j})].
\label{eq:vs-excess}
\end{equation}
The population minimizer is unchanged for any $\lambda>0$, but the choice of $\lambda$ determines the relative weight given to coefficient estimation and inclusion probability errors. Let $\widehat F=(\widehat\theta,\widehat q)$ be trained over the class of RRDS networks with the architecture and fixed output map specified in Theorem~\ref{thm:vs-approx}, subject to $\|\omega\|_2\leq\Gamma$, where $\Gamma\geq1$ is sufficiently large to contain $\widetilde F$ and bound the fixed coefficients. Let $\widehat S=\{j:\widehat q_j>1/2\}$ denote the selected support, and define
\begin{equation}
\Delta_{\rm vs}
=\frac{(s_{\max}+\lambda\Lambda_n)
[W\{L\log(eW\Gamma)+\log(eM)\}+\log(3/\delta)]}{M}
+\varepsilon_{\rm opt}.
\label{eq:sparse-complexity-training}
\end{equation}

\begin{theorem}[Meta learning generalization for variable selection]\label{thm:vs-gen}
In the setting of Theorem~\ref{thm:vs-approx}, for $n\geq3$, $M\geq1$, $\lambda>0$, and $\delta\in(0,1)$, with probability at least $1-\delta$ over $\mathcal D_M$,
\begin{align}
\mathcal R(\widehat F)-\mathcal R^*
&\leq C\{s_{\max}/n+\lambda h_n/p+\Delta_{\rm vs}\},
\label{eq:vs-trained-excess}\\
\mb E_M[\|\widehat\theta-\theta\|_2^2]
&\leq C\{s_{\max}/n+\lambda h_n/p+\Delta_{\rm vs}\},
\label{eq:vs-trained-mse}\\
\mb P_M(\widehat S\ne S)
&\leq\mb E_M[H]
\leq5H_n^*+\frac{8p}{\lambda}
\{\mathcal R(\widehat F)-\mathcal R^*\}\notag\\
&\leq C\{h_n+s_{\max}p/(\lambda n)
+p\Delta_{\rm vs}/\lambda\},
\label{eq:vs-trained-hamming}
\end{align}
where $H=|\widehat S\mathbin{\triangle}S|$ and the excess risk is conditional on $\mathcal D_M$. The constant $C$ is independent of $n,p,s_{\max},M,\Gamma,\lambda,\delta$, and $\varepsilon_{\rm opt}$.
\end{theorem}

The bound~\eqref{eq:vs-trained-hamming} relates the expected Hamming loss of the trained network to the Bayes Hamming risk $H_n^*$ and the excess risk. In particular, the excess risk controls the additional selection error arising from approximating the posterior inclusion probabilities. The factor $p/\lambda$ appears because the binary cross-entropy losses are weighted by $\lambda/p$ in the training objective.
The factor $s_{\max}+\lambda\Lambda_n$ in~\eqref{eq:sparse-complexity-training} bounds the training loss after projection and clipping, while the remaining factor reflects the network complexity and the number of meta training tasks. If $\Delta_{\rm vs}=O(s_{\max}/n)$, choosing $\lambda\asymp s_{\max}p/(nh_n)$ yields an estimation risk of order $s_{\max}/n$ and an expected Hamming loss of order $h_n$. The sum of the KL divergences between the posterior and learned inclusion probabilities is also of order $h_n$; see~\eqref{eq:vs-trained-kl} in the proof. Appendices~\ref{app:vs-training} and~\ref{app:lambda-calibration} give further details on the meta training requirements and the choice of $\lambda$. As in the preceding subsections, these guarantees for trained networks are averaged over the task distribution $\mb Q_n$, whereas Theorem~\ref{thm:vs-approx}(b) gives estimation and prediction risk bounds for the constructed network at fixed generating parameters in the prior support.

\section{Numerical Experiments}\label{sec:numerical}

We examine the performance of RRDS in parameter estimation, model selection, and variable selection through simulations and an application to the HTRU2 pulsar data. Appendix~\ref{app:implementation} provides implementation details and additional experiments on network size and training, recurrent refinement, bootstrap inference, and divide-and-conquer estimation.

\subsection{Simulation study}\label{sec:simulation}

In each simulation study, the generating parameter, model, or support is known, allowing us to evaluate RRDS and the classical methods against the same target on a common set of tasks drawn independently of those used for training. Unless otherwise specified, we use $R=12$ refinement steps with shared network parameters. As described in Section~\ref{sec:training}, we report Monte Carlo standard errors for the averages over evaluation tasks conditional on the trained network. These standard errors do not account for variability due to retraining. Appendix~\ref{app:impl-protocol} describes how we compute standard errors for paired comparisons and ratios.

\subsubsection{Parameter estimation}\label{sec:sim-estimation}

We first compare RRDS with the MLE and examine how estimation accuracy depends on the number of trainable parameters $W$ and the sample size $n$. The effects of the number of training tasks, recurrent refinement, and parameter sharing are examined in Appendices~\ref{app:est-M} and~\ref{app:est-R}. We consider linear and logistic regression with $d=5$ coefficients, varying $n$ from 250 to 8000 and $W$ from $10^4$ to $10^5$. The simulation designs, task distributions, and training settings are described in Appendix~\ref{app:impl-estimation}. We report the MSE of RRDS, $\bar e_{\widehat F}$, and its mean squared distance from the MLE, $\bar e_{\widehat F,\mathrm{MLE}}$, relative to the MSE of the MLE, $\bar e_{\mathrm{MLE}}$ (see Appendix~\ref{app:meta-training} for their definitions).

\begin{figure}[!htb]
\centering
\includegraphics[width=\textwidth]{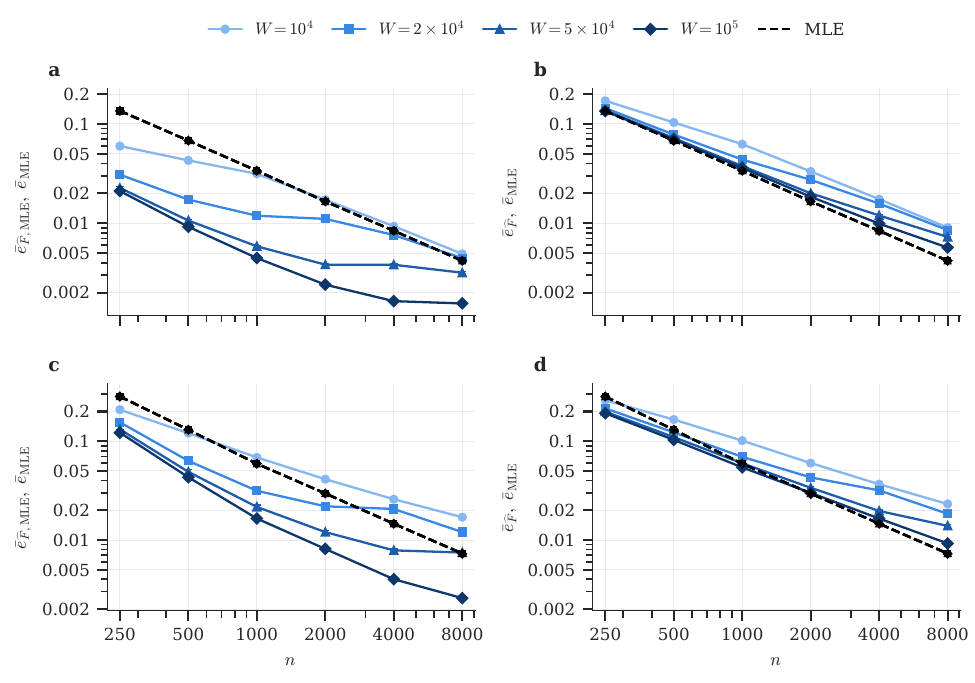}
\caption{Distance to the MLE (left) and estimated risk with respect to $\theta$ (right) in linear regression (a, b) and logistic regression (c, d).}
\label{fig:est-dist}
\end{figure}

The right panels of Figure~\ref{fig:est-dist} compare the MSEs of RRDS and the MLE. For $W=10^5$, the ratio $\bar e_{\widehat F}/\bar e_{\mathrm{MLE}}$ increases with $n$ in both regression models and exceeds one at larger sample sizes. In logistic regression with $n=250$, for example, the MSE of RRDS averaged over $\pi$ is about 32\% lower than that of the MLE. One possible explanation is that training under squared-error loss encourages RRDS to learn the posterior mean, which incorporates prior information about the parameter values and can help stabilize estimation when the sample is small. Tuning the LASSO and ridge penalties on tasks drawn from $\pi$ rather than by cross-validation within each task also improves their estimation accuracy, lending support to this interpretation (Table~\ref{tab:est-regularization} in Appendix~\ref{app:est-regularization}). As $n$ increases, the posterior mean becomes close to the MLE and the benefit of this regularization diminishes. With $W$ and $M$ fixed, the error incurred in learning the estimation rule then becomes increasingly important relative to the statistical error of the MLE.

The left panels of Figure~\ref{fig:est-dist} show how closely RRDS approximates the MLE relative to the MSE of the MLE itself. For each $n$, increasing $W$ reduces the mean squared distance $\bar e_{\widehat F,\mathrm{MLE}}$ between their estimates. Although this distance also decreases with $n$ for each $W$, it becomes larger relative to the MSE of the MLE, which decreases at the rate $n^{-1}$. With a larger $W$, the squared distance between the two estimators remains below the MSE of the MLE over a wider range of sample sizes. These results illustrate the importance of increasing $W$ with $n$, as suggested by the approximation bounds in Theorem~\ref{thm:est-approx}. Theorem~\ref{thm:est-gen} further suggests increasing the number of meta training tasks along with the network size. In logistic regression, increasing $M$ reduces the relative error (Figure~\ref{fig:app-m-sweep} in Appendix~\ref{app:est-M}). The reduction is much smaller when the number of gradient updates, rather than the number of epochs, is held fixed, suggesting that both the number of tasks and the optimization error matter.

Recurrent refinement contributes substantially to the accuracy of RRDS. Figure~\ref{fig:est-R} in Appendix~\ref{app:est-R} shows that most of the reduction in $\bar e_{\widehat F,\mathrm{MLE}}/\bar e_{\mathrm{MLE}}$ is achieved within the first few refinement steps, although further refinements can improve accuracy at larger sample sizes, particularly in logistic regression. At a fixed $W$, sharing parameters across refinements also improves accuracy at larger sample sizes (Figure~\ref{fig:app-est-unshared} in the same appendix).

In Appendices~\ref{app:impl-bootstrap} and~\ref{app:impl-dc}, we reuse the trained networks for bootstrap inference and divide-and-conquer estimation without further training. In linear regression, the marginal coverage of basic and normal 95\% bootstrap intervals approaches the nominal level and remains close to that of the MLE-based intervals as $n$ increases. In logistic regression, the RRDS intervals have lower coverage than the MLE-based intervals, but increasing $W$ improves coverage for larger $n$ (Figure~\ref{fig:boot}); similar patterns hold for simultaneous coverage (Figure~\ref{fig:app-boot}). This illustrates the importance of estimation accuracy for bootstrap inference, as discussed in Corollary~\ref{cor:bootstrap}. For divide-and-conquer estimation in logistic regression, Figure~\ref{fig:dc} shows that the mean squared distance from the full-sample MLE relative to the MSE of that MLE generally increases with the number of blocks, both for the averaged RRDS estimates and for the averaged block MLEs. This ratio exceeds one for all methods at $n=n_0^2$, illustrating the restriction $n=o(n_0^2)$ in Corollary~\ref{cor:dc}.

\subsubsection{Model selection}\label{sec:sim-model}

For model selection, we compare RRDS trained directly under the cross-entropy loss of Section~\ref{sec:meta-formulation} with BIC~\citep{schwarz1978dimension} at small and large sample sizes. We consider two nested mixture models based on the linear and logistic regression models of Section~\ref{sec:sim-estimation}. In the smaller model, the mixing probability $\nu$ is fixed at $1/2$, whereas in the larger model it is an unknown parameter. Thus, the models share the same distribution at $\nu=1/2$. We train RRDS at $n\in\{35,1000\}$ and evaluate its model selection accuracy against that of BIC, which has an explicit form in this example, for different values of $\nu$. Appendix~\ref{app:impl-model} describes the simulation design, training, and evaluation procedures.

\begin{figure}[!htb]
\centering
\includegraphics[width=\textwidth]{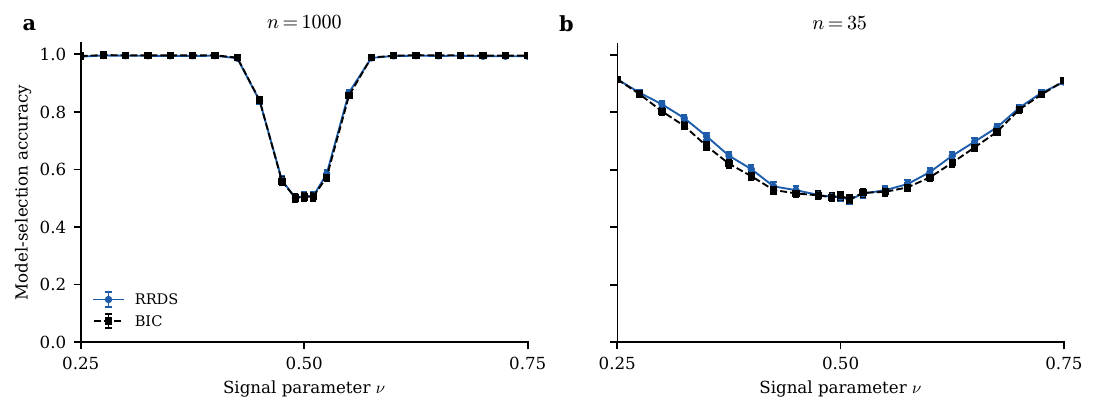}
\caption{Model selection accuracy of RRDS and BIC as a function of the mixing probability $\nu$.}
\label{fig:ms}
\end{figure}

Figure~\ref{fig:ms} compares the model selection accuracy of RRDS and BIC for $n=1000$ and $n=35$ over $0.25\leq\nu\leq0.75$. At both sample sizes, RRDS and BIC perform similarly when $\nu$ is far from $1/2$, where the evidence for the larger model is stronger. At $n=35$, the largest improvement over BIC occurs at intermediate distances from $1/2$, where model selection is difficult but the data still contain information for distinguishing the models. The difference between the two methods becomes smaller as $\nu$ approaches $1/2$, where the two models coincide and the selection accuracy approaches $1/2$ in this experiment.

The improvement at $n=35$ may be attributed to the difference between the Bayes selection rule targeted by cross-entropy training and its large-sample approximation based on BIC. The two rules nearly agree at $n=1000$ (Appendix~\ref{app:impl-model}).

\subsubsection{Variable selection}\label{sec:sim-variable}

For variable selection, we compare RRDS with classical penalized regression methods in terms of support recovery and coefficient estimation at different signal strengths. We consider a sparse Gaussian linear regression model with $p=10$ covariates, $s_0=3$ active coefficients, and $n=1000$ observations. The signal strength is controlled by a multiplier $\tau$ relative to the scale $(\log p/n)^{1/2}$, which is relevant to support recovery by LASSO~\citep{wainwright2009sharp}. The competing methods are LASSO~\citep{tibshirani1996regression} and the minimax concave penalty (MCP)~\citep{zhang2010nearly}, with the penalty parameters chosen by cross-validation. Further details of the simulation design, training, and evaluation are given in Appendix~\ref{app:impl-variable}.

\begin{figure}[!htb]
\centering
\includegraphics[width=\textwidth]{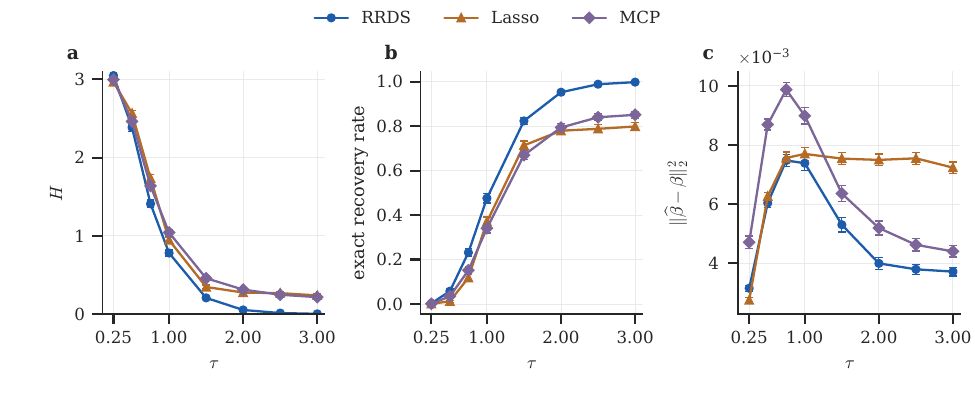}
\caption{Hamming loss (a), exact support recovery rate (b), and squared coefficient error (c) as functions of the signal multiplier $\tau$.}
\label{fig:vs}
\end{figure}

As shown in Figure~\ref{fig:vs}, RRDS recovers the true support more accurately than LASSO and MCP once the signals are moderately strong. As $\tau$ increases, the Hamming loss of RRDS decreases toward zero and its probability of exact support recovery approaches one. The coefficient estimation error initially increases with signal strength before decreasing for stronger signals. This can be explained by the fact that omitting a coefficient close to zero contributes little to the squared estimation error, even though it counts as a selection error. For $\tau\geq1$, the improved support recovery of RRDS is accompanied by a decrease in the coefficient estimation error. Among the three methods, RRDS has the smallest mean squared coefficient error for $\tau\geq0.5$.

\subsection{Real data analysis}\label{sec:real-data}

We also examine the performance of RRDS trained on simulated tasks when applied to real data, for which the assumed statistical model may be misspecified. We use the HTRU2 pulsar data~\citep{lyon2015htru2,lyon2016fifty} and model the pulsar label using logistic regression with eight continuous features and an intercept. We apply networks trained on simulated logistic regression tasks and calibrated using a held-out portion of the data to subsamples of size $n\in\{500,1000\}$ drawn from the remaining records. Since the true parameter and support are unknown, we use the unrestricted MLE $\widehat\theta_{\rm ref}$ and the BIC-selected support $\widehat S_{\rm ref}$ obtained from all the remaining records as reference values. The reference support contains six of the eight features. For parameter estimation, we compare RRDS with the MLE, LASSO, and ridge regression~\citep{lecessie1992ridge}. For joint variable selection and coefficient estimation, we compare RRDS with LASSO and BIC, where the latter is evaluated over all $2^8$ subsets and followed by maximum likelihood estimation within the selected model. Appendix~\ref{app:impl-htru} describes the data split, task distributions, training settings, and implementation of the competing methods.

\begin{figure}[!htb]
\centering
\includegraphics[width=\textwidth]{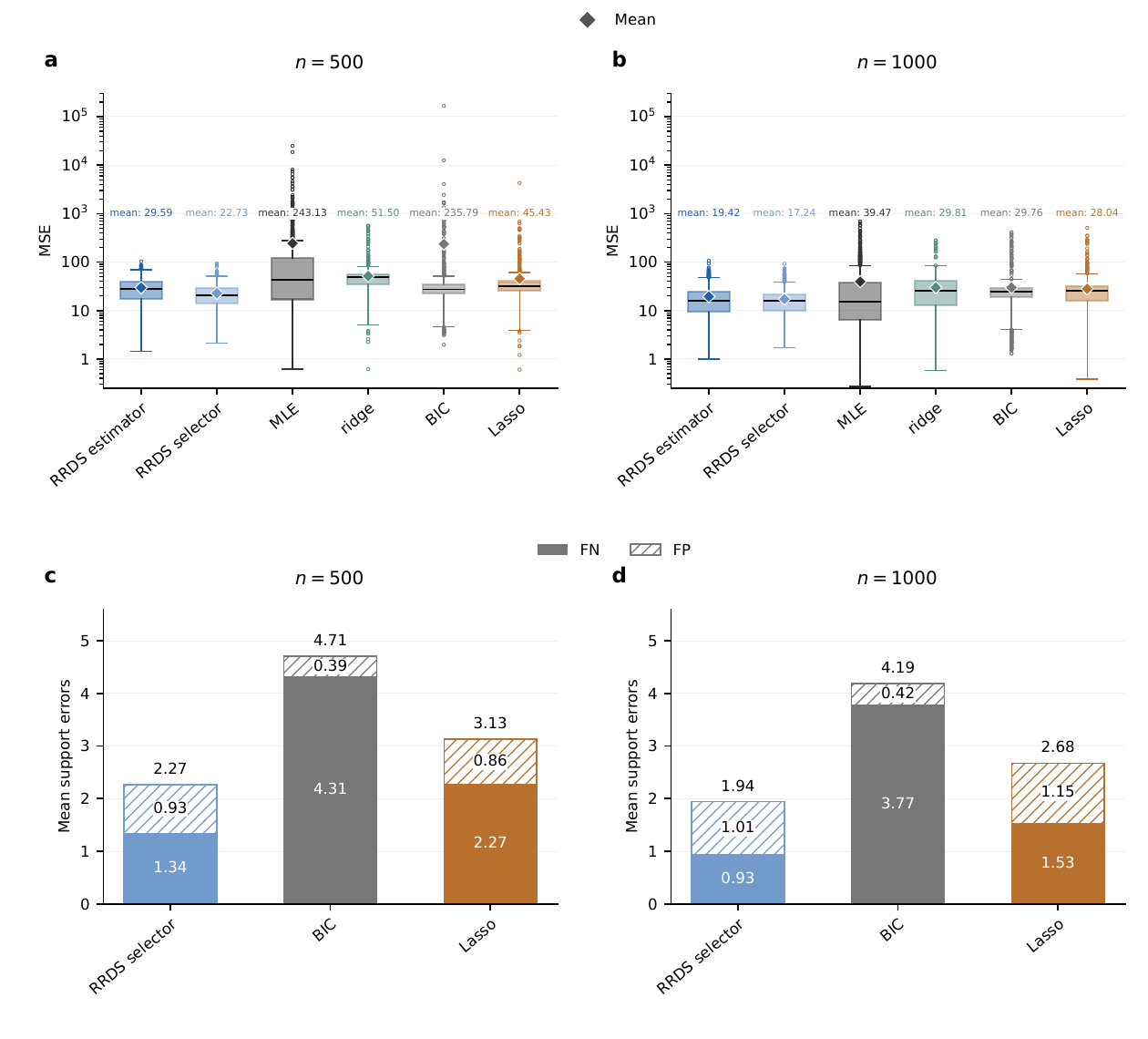}
\caption{HTRU2 coefficient errors relative to $\widehat\theta_{\rm ref}$ (a, b) and support errors relative to $\widehat S_{\rm ref}$ (c, d). Diamonds mark means. Stacked bars show false negatives and false positives.}
\label{fig:htru}
\end{figure}

We first consider parameter estimation without variable selection. Figure~\ref{fig:htru}(a, b) shows that RRDS achieves a smaller mean squared error relative to $\widehat\theta_{\rm ref}$ than the MLE, LASSO, and ridge regression at both sample sizes. The squared errors of the MLE have a long right tail, particularly at $n=500$, whereas RRDS reduces the estimation error along the direction in which the data contain the least information (Appendix~\ref{app:impl-htru}). This is consistent with the regularizing effect of training under the task distribution observed in the logistic regression simulations. As $n$ increases, the mean squared errors decrease for all four estimators and the difference between RRDS and the MLE becomes smaller.

For joint variable selection and coefficient estimation, RRDS achieves smaller mean squared coefficient error and mean Hamming loss than both BIC with an MLE refit and LASSO at both sample sizes (Figure~\ref{fig:htru}). The smaller Hamming loss is mainly due to fewer false negatives relative to the reference support. Compared with RRDS trained for parameter estimation alone, the network trained for joint selection and estimation also achieves a smaller mean squared coefficient error on the same evaluation tasks. Both its mean Hamming loss and mean squared coefficient error decrease as $n$ increases. Appendix~\ref{app:impl-htru} provides paired comparisons and results on the sizes of the selected supports.

\section{Discussion}\label{sec:discussion}

We developed a meta learning framework that uses statistical decision theory to select training objectives and the structure of classical inference procedures to design RRDS. Our results relate
the accuracy of learning a population decision rule to its
frequentist properties, and show how the structure of statistical inference problems can be exploited to reduce the complexity of neural approximation and meta training.

Several directions are worth exploring. Although our analysis
considers meta training at a fixed sample size, the averaging
operations in RRDS allow a single network to process datasets of
different sizes. A natural question is whether joint training
across sample sizes~\citep[Section~2.2.2 and Supplementary
Section~S2]{sainsbury2024likelihoodfree} can improve inference
accuracy or reduce the number of training tasks needed while
retaining first-order efficiency. Another direction is to
accommodate dependent data through aggregation operations that
account for dependence, such as aggregating features of consecutive
observations in Markov models or using attention with positional
information.
The framework may also be extended to hypothesis testing and
learning posterior distributions. For hypothesis testing, a key
question is how to design training objectives that yield high
power while controlling type I error. For learning posterior
distributions~\citep{greenberg2019automatic,radev2022bayesflow},
the Bernstein--von Mises theorem~\citep[Section~10.2]{van2000asymptotic}
suggests that, in regular parametric models, learning the posterior
center and covariance may suffice. A theoretical question is how
accurately these quantities must be learned for credible sets to
attain asymptotically correct frequentist coverage.


\bibliographystyle{apalike}
\bibliography{references}

\clearpage
\appendix
\setstretch{1}
\numberwithin{theorem}{section}
\numberwithin{lemma}{section}
\numberwithin{proposition}{section}
\numberwithin{corollary}{section}
\numberwithin{remark}{section}
\numberwithin{assumption}{section}
\renewcommand{\theassumption}{\thesection.\arabic{assumption}}
\renewcommand{\theHtheorem}{\thesection.\arabic{theorem}}
\renewcommand{\theHlemma}{\thesection.\arabic{lemma}}
\renewcommand{\theHproposition}{\thesection.\arabic{proposition}}
\renewcommand{\theHcorollary}{\thesection.\arabic{corollary}}
\renewcommand{\theHremark}{\thesection.\arabic{remark}}
\renewcommand{\theHassumption}{\thesection.\arabic{assumption}}

\begin{center}
{\LARGE\bfseries Supplementary Materials for ``Learning to Infer:\\
Statistical Inference through Meta Learning''\par}
\end{center}


Appendix~\ref{sec:supp_related} reviews additional related literature. Appendix~\ref{app:meta-details} gives further details of the meta learning framework, and Appendix~\ref{app:theory-discussion} presents the network constructions and additional theoretical results for parameter estimation, model selection, and variable selection. Appendix~\ref{app:implementation} describes the implementation and additional numerical results. The proofs of the main results and technical lemmas are given in Appendices~\ref{app:proofs} and~\ref{app:technical}, respectively, and Appendix~\ref{app:tools} collects general tools used in the proofs.

\begingroup
\renewcommand{\contentsname}{Contents}
\tableofcontents
\endgroup

\bigskip

\addtocontents{toc}{\protect\setcounter{tocdepth}{3}}

\section{Related Literature}\label{sec:lit_rev}\label{sec:supp_related}
\medskip
\noindent {\bf Neural estimators and amortized inference.}
Neural estimators provide a way to learn statistical procedures directly from simulated data without evaluating the likelihood. \citet{lenzi2023neural,lenzi2025towards} study neural estimators for statistical models in which conventional inference can be computationally difficult, while \citet{sainsbury2024likelihoodfree} develop neural Bayes estimators trained under a specified loss and use permutation-invariant architectures for replicated observations. The latter work also considers uncertainty quantification through the bootstrap. These approaches are closely related to amortized inference, which replaces repeated inference computations with a pretrained neural network.

More broadly, simulation-based inference uses neural networks to approximate posterior distributions, likelihoods, likelihood ratios, or invertible transformations~\citep{cranmer2020frontier,greenberg2019automatic,papamakarios2019snl,radev2022bayesflow,zammit2025neural}.

Our framework builds on neural point estimation and extends it to model and variable selection through suitable choices of training loss. Rather than approximating an entire posterior distribution or a likelihood for subsequent inference, the network learns the posterior quantity relevant to the decision problem.

On the theoretical side, \citet{rodder2025theoretical} develop a general framework for studying neural estimators by decomposing their risk averaged over the task distribution into Bayes risk, approximation error, generalization error, and optimization error. They establish consistency in terms of this average risk and study efficiency through its ratio to the optimal Bayes risk. Their analysis treats the full dataset as a high-dimensional network input, and the resulting sufficient conditions can require the number of meta training tasks to grow faster than exponentially with the within-task sample size.

In contrast, our approximation results exploit the statistical structure of the inference map, so that the network size and the number of meta training tasks need only grow polynomially in the sample size for fixed-dimensional models. Under suitable analyticity conditions, the required network size grows only polynomially in the logarithm of the sample size. We also establish first-order equivalence to the MLE, which gives a stronger characterization of the asymptotic behavior of the learned estimator than a comparison of average risks alone.

\medskip
\noindent {\bf Learned summary statistics and permutation-invariant networks.}
Regression from simulated datasets to parameters also arises in approximate Bayesian computation (ABC). \citet{fearnhead2012constructing} study the construction of approximately optimal summary statistics from a decision-theoretic perspective, while \citet{jiang2017learning} use neural networks to learn low-dimensional summaries by predicting parameters from simulated data. In ABC, the learned summary is typically used in a subsequent posterior approximation, whereas in our framework the learned map directly produces the inferential output.

For unordered observations, Deep Sets~\citep{zaheer2017deep} represent permutation-invariant functions through shared encoders and aggregation across observations. Related constructions include invariant and equivariant networks~\citep{bloemreddy2020probabilistic} and Set Transformers~\citep{lee2019set}, which use attention to model interactions among elements of a set. These architectures provide general representations for unordered data. RRDS retains permutation-invariant aggregation but introduces recurrent residual updates, in which the representation of each observation depends on the current estimate. The resulting architecture resembles classical iterative estimation procedures, in which the score or another estimating function is evaluated repeatedly at updated parameter values. This structure allows the network to reuse both the observations and the trainable parameters across refinement steps.

\medskip
\noindent {\bf Meta learning and in-context learning.}
Meta learning broadly refers to learning from a population of tasks so that experience from training tasks improves performance on new ones~\citep{thrun1998learning,hospedales2022meta}. Different meta learning methods achieve this through different forms of adaptation. For example, model-agnostic meta learning~\citep{finn2017model} learns an initialization that can be adapted to a new task through gradient updates. Our approach instead learns the complete mapping from a dataset to an inferential action, so no task-specific parameter updates are needed after training.

Our formulation is also closely related to in-context learning, where a pretrained model uses information supplied in its context to perform a new task without updating its parameters~\citep{brown2020language,xie2022explanation,garg2022what}. Prior-Data Fitted Networks and TabPFN~\citep{muller2022transformers,hollmann2025accurate} train transformer-based models across synthetic datasets so that a forward pass on a new dataset approximates a Bayesian prediction rule. Although these methods primarily target prediction, they share the idea of learning a reusable statistical procedure through pretraining across tasks.

A growing literature studies the statistical and computational properties of in-context learning. For linear models, \citet{akyurek2023what,vonoswald2023transformers,bai2023transformers} investigate how transformers can implement or approximate classical learning algorithms, including gradient-based and least-squares procedures.
More recent work extends this analysis to nonparametric regression. \citet{kim2024transformers} show that transformers pretrained on tasks from Besov and related function classes can attain minimax estimation rates by representing suitable basis expansions. \citet{shen2025understanding} relate attention mechanisms to kernel regression and establish near-minimax rates for H\"older regression functions on low-dimensional manifolds, where the rate depends on the intrinsic rather than ambient dimension.

For $\alpha$-H\"older regression functions with $d_{\mathrm c}$-dimensional covariates, \citet{ching2026efficient} show that a pretrained transformer with $\Theta(\log n)$ parameters and
$M\gtrsim n^{2\alpha/(2\alpha+d_{\mathrm c})}\log^3 n$
meta training sequences can attain the minimax rate
$n^{-2\alpha/(2\alpha+d_{\mathrm c})}$.
Their construction approximates a local polynomial estimator and implements the associated optimization through gradient descent. These results illustrate how the statistical structure of an estimator can be used to control the approximation complexity of a neural network and, consequently, the number of meta training tasks required for generalization.

Our analysis follows a similar approach but focuses on statistical inference rather than prediction. The network is trained to perform parameter estimation, model selection, or joint parameter estimation and variable selection, and we study the resulting inferential guarantees, including asymptotic efficiency and model or support recovery. Our approximation analysis uses classical estimation and selection procedures to connect the Bayesian population targets of the training losses to frequentist guarantees.

In particular, the parameter estimation construction starts with a method-of-moments estimator and refines it using a Fisher scoring update. Rather than representing these operations through attention, RRDS uses shared observation-wise transformations, permutation-invariant aggregation, and recurrent residual updates. These operations exploit the additive structure of classical estimating procedures. The classical procedures guide our approximation analysis rather than serve as targets for training. The same decision-theoretic formulation can also be extended to other tasks, such as hypothesis testing and prediction, by changing the inference target and training loss.

\section{Further Details of Meta Learning Framework}
\label{app:meta-details}

This section provides further details of the population training objectives, the RRDS network class, and the training and evaluation procedures introduced in Section~\ref{sec:problem-formulation}.

\subsection{Additional notation}\label{app:notation}

For sets $A,B\subset\mathbb R^m$, we write $A\Subset B$ if $A$ is compact and $A\subset B$. For a multi-index $\nu\in\mathbb N_0^m$, let $|\nu|=\sum_{j=1}^m\nu_j$ and let $D^\nu$ denote the corresponding partial derivative. For an open set $A\subset\mathbb R^m$, $C^k(A)$ denotes the space of functions with continuous derivatives up to order $k$. For $s=k+\gamma$, where $k\geq0$ is an integer and $\gamma\in(0,1)$, $C^s(A)$ further requires that all order-$k$ derivatives be locally $\gamma$-H\"older continuous. That is, for every compact $K\subset A$ and $|\nu|=k$,
\[
[D^\nu f]_{\gamma,K}
=\sup_{\substack{x,y\in K\\x\ne y}}
\frac{|D^\nu f(x)-D^\nu f(y)|}{\|x-y\|_2^\gamma}<\infty.
\]
On compact domains, we define these spaces by restriction from a neighborhood. For vector- and matrix-valued functions, these definitions apply coordinatewise. For probability measures $P$ and $Q$, let $D_{\rm KL}(P\|Q)=\int\log(dP/dQ)\,dP$, taken as $+\infty$ if $P\not\ll Q$. Write $\mb E_\theta$ and $\mb E_\theta^{(n)}$ for expectations under $\mb P_\theta$ and $\mb P_\theta^{(n)}$, respectively, and $\mb E[\cdot\mid Z]$ for conditional expectation. We write $Pf=\int f\,dP$. We also write $\mathcal L(Y)$ and $\mathcal L(Y\mid Z)$ for the law and conditional law of $Y$, and $\delta_x$ for the point mass at $x$. For a map $f$ between Euclidean spaces, let $\|f\|_\infty=\sup_x\|f(x)\|_2$ and ${\rm Lip}(f)=\sup_{x\ne y}\|f(x)-f(y)\|_2/\|x-y\|_2$. We define the bounded Lipschitz distance as
\[
d_{\rm BL}(P,Q)
=\sup_{\|f\|_\infty\leq1,\ {\rm Lip}(f)\leq1}|Pf-Qf|,
\]
where the supremum is over real-valued functions. This distance metrizes weak convergence.

\subsection{Population risks and Bayes rules}
\label{app:meta-bayes}

Recall the joint task distribution $\mb Q_n$ introduced in~\eqref{eq:meta-task-law}. Let $\Pi_n(\cdot\mid x^n)$ denote the conditional distribution of the inference target $\mathsf T$ given $X^n=x^n$. For a measurable output map $F$, we can write the population risk as
\[
\mathcal R(F)
=\mb E_{\mb Q_n}\left[
\int \ell_{\rm tr}\{F(X^n),t\}\,\Pi_n(dt\mid X^n)
\right].
\]
For the squared-error and joint losses below, we assume $\mb E_{\mb Q_n}[\|\theta\|_2^2]<\infty$, as in Section~\ref{sec:problem-formulation}. For the categorical loss, no moment condition is needed. The conditional Bayes rules derived below are measurable and have finite population risk. Consequently, every unrestricted population minimizer minimizes the conditional expected loss at almost every $x^n$.

\medskip
\noindent {\bf Parameter estimation.}
For this problem, $\mathsf T=\theta$. Under squared-error loss, the population minimizer is the posterior mean $F^\ast(x^n)=\mb E_{\mb Q_n}[\theta\mid X^n=x^n]$. For any measurable $F$ with finite squared-error risk, conditioning on $X^n$ gives
\[
\mb E_{\mb Q_n}[\|F(X^n)-\theta\|_2^2] =\mb E_{\mb Q_n}[\|F^\ast(X^n)-\theta\|_2^2] +\mb E_{\mb Q_n}[\|F(X^n)-F^\ast(X^n)\|_2^2].
\]
The cross term $\mb E_{\mb Q_n}[\langle F(X^n)-F^\ast(X^n),F^\ast(X^n)-\theta\rangle]$ vanishes because $\mb E_{\mb Q_n}[\theta\mid X^n]=F^\ast(X^n)$. Hence $F^\ast$ minimizes the risk, and the first term on the right is $\mathcal R(F^\ast)=\mathcal R^\ast$. This yields the excess-risk identity~\eqref{eq:excess-mse}.

\medskip
\noindent {\bf Model selection.}
For this problem, $\mathsf T=k$ is the generating model label. The network outputs a vector of model probabilities $q(x^n)=(q_1(x^n),\ldots,q_K(x^n))\in\Delta_K$. Denote the posterior probability of model $k$ by $q_k^\ast(x^n)=\mb Q_n\{\mathsf T=k\mid X^n=x^n\}$, and write $q^\ast=(q_1^\ast,\ldots,q_K^\ast)$. Conditional on $X^n=x^n$, the expected loss of $q$ is $-\sum_{k=1}^Kq_k^\ast(x^n)\log q_k(x^n)$, and subtracting its value at $q^\ast$ gives a Kullback--Leibler divergence, which is nonnegative and vanishes only when $q(x^n)=q^\ast(x^n)$. This property is known as strict propriety of the cross-entropy scoring rule~\citep{gneitingRaftery2007proper}. Hence the population minimizer is $q^\ast$. For any prediction map $q$ with finite risk, taking expectations over $X^n$ gives
\[
\mathcal R(q)-\mathcal R^\ast =  \mb E_{\mb Q_n}\Bigl[ \sum_{k=  1}^Kq_k^\ast(X^n) \log\frac{q_k^\ast(X^n)}{q_k(X^n)} \Bigr].
\]
Thus, the excess risk is the expected Kullback--Leibler divergence between the posterior model probabilities and their network predictions.

When we generate a task by drawing model $k$ with probability $w_k$, followed by $\theta_k\sim\pi_k$ and $X^n\sim\mb P_{k,\theta_k}^{(n)}$, Bayes' formula expresses $q^\ast$ in terms of the marginal likelihoods displayed in Section~\ref{sec:theory-model}. Selecting a model that maximizes $q_k^\ast(x^n)$ minimizes the probability of misclassifying the generating label. This differs from selecting the smallest correct model $k_0$, which is the target of frequentist model selection consistency. In particular, when candidate models overlap, the generating label need not uniquely identify the smallest correct model.

\medskip
\noindent {\bf Joint parameter estimation and variable selection.}
For this problem, $\mathsf T=(\theta,S)$. The network outputs a parameter estimate $\widehat\theta_F$ and a vector of inclusion probabilities $q=(q_1,\ldots,q_p)$. The squared-error and binary cross-entropy terms in the training loss act on separate outputs. Their unrestricted population minimizers are therefore $\widehat\theta^\ast(x^n)=\mb E_{\mb Q_n}[\theta\mid X^n=x^n]$ and $q_j^\ast(x^n)=\mb Q_n\{j\in S\mid X^n=x^n\}$ for $j\in[p]$.

For the binary cross-entropy term, condition on $X^n=x^n$. The conditional loss for coordinate $j$, minus its value at $q_j^\ast(x^n)$, is the KL divergence between Bernoulli distributions with success probabilities $q_j^\ast(x^n)$ and $q_j(x^n)$, respectively. Each divergence is nonnegative and vanishes only when the two probabilities agree, including at the endpoints under the usual conventions for the logarithm. Summing over $j$ proves the assertion for the inclusion probabilities, while the squared-error decomposition above proves it for the parameter estimate.

These minimizers do not depend on the weight $\lambda>0$. However, $\lambda$ affects the relative importance of the two losses when we restrict optimization to a neural network class.

To justify the thresholding rule, consider the weighted selection loss $t\,\mathrm{FP}+(1-t)\,\mathrm{FN}$ for $t\in(0,1)$. Conditional on $X^n=x^n$, including variable $j$ incurs expected loss $t\{1-q_j^\ast(x^n)\}$, while excluding it incurs expected loss $(1-t)q_j^\ast(x^n)$. Including $j$ incurs a smaller expected loss precisely when $t\{1-q_j^\ast(x^n)\}<(1-t)q_j^\ast(x^n)$, which is equivalent to $q_j^\ast(x^n)>t$. Thus, the optimal decision includes $j$ whenever $q_j^\ast(x^n)>t$, and either decision is admissible at equality.

\subsection{Network class and recurrent aggregation}
\label{app:meta-network}\label{app:meta-complexity}

We use feedforward ReLU networks for the encoder and decoder blocks of RRDS. A depth-$L$ feedforward network is a composition of affine maps and coordinatewise ReLU activations $\sigma(u)=\max\{u,0\}$. Following the notation of Section~\ref{sec:training}, we write $\mathcal F(L,W)$ for an RRDS class whose unrolled network blocks have depth at most $L$ and at most $W$ distinct trainable parameters, together with the specified fixed operations. We count parameters constrained to be equal only once, and we exclude fixed transformations and weights from $W$.

In each block, we apply a common encoder to all observations and average the outputs, so the average does not depend on the order of the observations. Consequently, the state is invariant under permutations of the observations at initialization and, by induction, after each residual update. Hence $z^{(r)}(X_{\varpi(1)},\ldots,X_{\varpi(n)})=z^{(r)}(X_1,\ldots,X_n)$ for $r=0,\ldots,R$ and every permutation $\varpi$ of $[n]$. The output $F_\omega=\Psi(z^{(R)})$ is therefore permutation invariant whether or not parameters are shared across refinement blocks.

Sharing parameters across refinement blocks controls network complexity rather than permutation invariance. In~\eqref{eq:rrds-complexity}, we count the parameters of $\phi$ and $\rho$ once regardless of the number of refinements, and for $R=0$ only the initialization networks contribute. For $R\geq1$, increasing $n$ or $R$ therefore leaves $W$ unchanged. Increasing either quantity raises the computational cost of evaluating the network and its gradients during meta training, while increasing $R$ also increases the depth $L$. The complexity required to achieve a prescribed approximation accuracy may nevertheless depend on $n$, as in the constructions of Section~\ref{sec:theory-results}.

Other choices of parameter sharing are also possible. For example, we could reuse $(\phi_0,\rho_0)$ for several consecutive blocks before switching to $(\phi,\rho)$, assign a separate encoder-decoder pair to each block, or divide the blocks into multiple groups with parameters shared within each group. In our numerical experiments, however, we find that assigning different parameters to each block substantially increases the number of parameters and the computational burden of meta training, without significantly improving accuracy. We therefore focus on the simpler architecture with a single initialization block followed by refinement blocks with shared parameters, both in our numerical experiments and in our theoretical analysis. When $R=0$, RRDS reduces to an ordinary Deep Sets network.

The output map $\Psi$ and the map $g$ in~\eqref{eq:rrds-output} have different roles. We apply the training loss to $F_\omega(X^n)=\Psi(z^{(R)})$, whereas $g$ converts this output into the final estimate or selection. For parameter estimation, $\Psi$ is the identity map in our numerical experiments and the fixed projection onto $\Theta_0$ in the approximation analysis. For variable selection, the output map of Section~\ref{sec:rrds} couples the coefficient estimates to the inclusion probabilities while allowing the intermediate states to remain unconstrained.

\subsection{Training and evaluation details}
\label{app:meta-training}

\begin{algorithm}[tbp]
\caption{Meta training and inference with RRDS}\label{alg:rrds}
\begin{algorithmic}[1]
\Require task distribution and simulator defining $\mb Q_n$, loss $\ell_{\rm tr}$, RRDS architecture, number of training tasks $M$
\State Simulate $\mathcal D_M=\{(\mathsf T^{(m)},X^{n,(m)}):m\in[M]\}$ from $\mb Q_n$
\State Initialize the network parameters $\omega$
\For{each training epoch}
  \For{each minibatch $\mathcal I\subseteq[M]$}
    \State Compute $F_\omega(X^{n,(m)})$ using~\eqref{eq:rrds-init}--\eqref{eq:rrds-output} for $m\in\mathcal I$
    \State Update $\omega$ using the gradient of $|\mathcal I|^{-1}\sum_{m\in\mathcal I}\ell_{\rm tr}\{F_\omega(X^{n,(m)}),\mathsf T^{(m)}\}$
  \EndFor
\EndFor
\State Set $\widehat F=F_\omega$ and apply $a_{\widehat F}(X^n)=g\{\widehat F(X^n)\}$ to new data
\end{algorithmic}
\end{algorithm}

The choice of the task distribution $\pi$ determines the parameter values encountered by the network. Ideally, $\pi$ should reflect available knowledge about the parameter values expected at inference time. For example, when the true distribution is $\mb P_0=\mb P_{\theta_0}$, it is desirable for $\pi$ to have a positive and continuous density in a neighborhood of $\theta_0$, as required for the asymptotic equivalence between the posterior mean and the MLE discussed earlier. See Assumption~\ref{ass:proposal} in Section~\ref{sec:theory-estimation} for the precise conditions on $\pi$ used in our theoretical analysis. Similar considerations apply to the model probabilities $w_k$, the within-model distributions $\pi_k$, and the distribution over candidate supports. Although $\pi$ plays the mathematical role of a prior, this Bayesian interpretation primarily serves as a theoretical bridge between population risk minimization and asymptotically efficient estimation. The learned inference rule need not be interpreted as Bayesian. In our frequentist analysis, the new dataset $X^n\sim\mb P_0^{(n)}$ is independent of the meta training tasks. The distribution $\mb P_0$ and its underlying parameter, when correctly specified, are treated as fixed and unknown.

The $M$ simulated tasks in $\mathcal D_M$ are generated before training and held fixed throughout optimization. During each epoch, we shuffle the tasks and divide them into minibatches $\mathcal I\subseteq[M]$. The minibatch objective is
\[
\widehat{\mathcal R}_{\mathcal I}(\omega) =  \frac1{|\mathcal I|}\sum_{m\in\mathcal I} \ell_{\rm tr}\{F_\omega(X^{n,(m)}),\mathsf T^{(m)}\}.
\]
We compute its gradient by backpropagation and update $\omega$ as described in Section~\ref{sec:training}. The approximate empirical risk minimization condition~\eqref{eq:approx-erm} allows for an optimization tolerance $\varepsilon_{\mathrm{opt}}$. Our theoretical results depend on this tolerance and the number of independent training tasks, rather than the convergence properties of a particular stochastic optimization algorithm.

For numerical evaluation, we keep the trained network $\widehat F$ fixed and generate $M_{\mathrm{test}}$ independent tasks from the stated evaluation law. For parameter estimation, we draw the evaluation tasks from $\mb Q_n$. For model selection and variable selection, each curve fixes the displayed difficulty parameter and averages over the remaining task variables. The evaluation criteria depend on the inferential task.

For a per-task quantity $a_m$, we write $\bar a=M_{\mathrm{test}}^{-1}\sum_{m=1}^{M_{\mathrm{test}}}a_m$. We let $\ind\{E\}$ denote the indicator of an event $E$. For parameter estimation, $\mathsf T^{(m)}=\theta^{(m)}$, $\widehat\theta_m=\widehat F(X^{n,(m)})$ and $\widehat\theta_{\mathrm{MLE},m}=\widehat\theta_{\mathrm{MLE}}(X^{n,(m)})$. We define $e_{\widehat F,m}=\|\widehat\theta_m-\theta^{(m)}\|_2^2$, $e_{\mathrm{MLE},m}=\|\widehat\theta_{\mathrm{MLE},m}-\theta^{(m)}\|_2^2$ and $e_{\widehat F,\mathrm{MLE},m}=\|\widehat\theta_m-\widehat\theta_{\mathrm{MLE},m}\|_2^2$. We report their means $\bar e_{\widehat F}$, $\bar e_{\mathrm{MLE}}$ and $\bar e_{\widehat F,\mathrm{MLE}}$. Dividing $\bar e_{\widehat F}$ and $\bar e_{\widehat F,\mathrm{MLE}}$ by $\bar e_{\mathrm{MLE}}$ compares the risk of $\widehat F$ and its discrepancy from the MLE with the estimation error of the MLE.

For model selection, $\mathsf T^{(m)}=k^{(m)}$, and we average $\ind\{\widehat k_{\widehat F}(X^{n,(m)})=k^{(m)}\}$. For variable selection, $\mathsf T^{(m)}=(\theta^{(m)},S^{(m)})$ with $\theta^{(m)}=(\icpt^{(m)},\beta^{(m)})$. We average $\ind\{\widehat S_{\widehat F}(X^{n,(m)})=S^{(m)}\}$ for exact recovery, $H^{(m)}=|\widehat S_{\widehat F}(X^{n,(m)})\setminus S^{(m)}|+|S^{(m)}\setminus\widehat S_{\widehat F}(X^{n,(m)})|$ for Hamming loss, and $\|\widehat\beta_{\widehat F}(X^{n,(m)})-\beta^{(m)}\|_2^2$ for coefficient error.

We also examine model selection at prescribed parameter values, rather than only averaging over the task-generating distribution, to assess performance across different degrees of model separation. In the real-data analysis, the generating parameters and supports are unknown, so we use estimates from the full data pool as the reference values.

We compute Monte Carlo standard errors across independently generated evaluation tasks, keeping the trained network fixed~\citep{morris2019using}. These standard errors reflect variation across evaluation tasks, conditionally on the trained network. Appendix~\ref{app:meta-extensions} gives further details on bootstrap confidence intervals and divide-and-conquer comparisons.

\subsection{Bootstrap and divide-and-conquer evaluation}
\label{app:meta-extensions}

\medskip
\noindent {\bf Bootstrap confidence intervals.}
For an evaluation dataset $X^n$, let $\mb P_n$ denote its empirical distribution. We generate $B$ independent bootstrap samples $X_1^{*n},\ldots,X_B^{*n}$ from $\mb P_n$ and compute $\widehat\theta_b^*=\widehat F(X_b^{*n})$. Let $\widehat\theta_j$ be coordinate $j$ of $\widehat F(X^n)$, and let $\widehat\theta_{b,j}^*$ be the corresponding bootstrap estimate.

Write $s_j$ for the sample standard deviation of $\widehat\theta_{1,j}^*,\ldots,\widehat\theta_{B,j}^*$ and $c_{\alpha,j}$ for their empirical $\alpha$ quantile. At nominal level $0.95$, we use the normal and basic bootstrap intervals
$[\widehat\theta_j-1.96s_j,\,\widehat\theta_j+1.96s_j]$, $[2\widehat\theta_j-c_{0.975,j},\,2\widehat\theta_j-c_{0.025,j}]$,
respectively~\citep[Chapter~5]{davison1997bootstrap}.

For simultaneous inference, we use the sup-$t$ rectangle $\prod_{j= 1}^d [\widehat\theta_j-c^*s_j,\,\widehat\theta_j+c^*s_j]$,
where $c^*$ is the empirical $0.95$ quantile, over bootstrap draws $b$, of $\max_{1\leq j\leq d} \frac{|\widehat\theta_{b,j}^*-\widehat\theta_j|}{s_j}$.
We record coordinatewise coverage, simultaneous coverage and interval width, averaged over independent evaluation tasks~\citep{montielolea2019simultaneous}. All bootstrap computations use the trained network with its parameters fixed.

\medskip
\noindent {\bf Divide-and-conquer evaluation.}
For each evaluation task $m$, we draw a parameter $\theta^{(m)}$ from the task-generating distribution and a dataset $X^{n,(m)}$ of size $n=Jn_0$ from the corresponding sampling model. We compare the divide-and-conquer estimator $\widehat\theta_{\mathrm{DC}}$ with the full-sample MLE and the average of the MLEs fitted to the $J$ blocks. Using the averaging convention of Appendix~\ref{app:meta-training}, the means $\bar e_{\mathrm{DC},\mathrm{MLE}}$ and $\bar e_{\mathrm{DC}}$ average the squared discrepancy $e_{\mathrm{DC},\mathrm{MLE},m}=\|\widehat\theta_{\mathrm{DC}}(X^{n,(m)})-\widehat\theta_{\mathrm{MLE}}(X^{n,(m)})\|_2^2$ and the squared error $e_{\mathrm{DC},m}=\|\widehat\theta_{\mathrm{DC}}(X^{n,(m)})-\theta^{(m)}\|_2^2$, respectively. The ratio $\bar e_{\mathrm{DC},\mathrm{MLE}}/\bar e_{\mathrm{MLE}}$ compares the discrepancy from the full-sample MLE with the estimation error of the MLE. The ratio $\bar e_{\mathrm{DC}}/\bar e_{\mathrm{MLE}}$ compares the estimation risk with the same error. We vary $J$ while keeping the training sample size $n_0$ fixed and report Monte Carlo standard errors across evaluation tasks. The network is not retrained as $J$ varies.

\medskip
\noindent{\bf Extensions to selection.}
Similar divide-and-conquer strategies have also been developed for model selection and variable selection~\citep{zhang2018robust,chen2014split}. These approaches can be adapted to our framework by aggregating the model probabilities or selected supports obtained from meta trained networks on individual blocks.

\section{Network Constructions and Additional Theoretical Results}\label{app:theory-discussion}

This appendix supplements Section~\ref{sec:theory-results}. For each of the three inference tasks, we describe the network construction, explain the additional assumptions, and relate the approximation bounds to the statistical guarantees. Readers interested in the improved rates for analytic models may turn directly to Appendix~\ref{app:analytic}. In particular, Appendix~\ref{app:analytic-extension} introduces the approximation condition used both there and in the sparse GLM analysis.

\subsection{Parameter estimation}

\subsubsection{Construction of the one-step network}\label{app:rrds-estimation}

For the construction with a single refinement in Theorem~\ref{thm:est-approx}, we take $d_z=d_u=d$, $R=1$, and $\Psi=\Pi_{\Theta_0}$. For the inner projection, we choose a polytope $P$ satisfying $\Theta_0\subseteq P\subset U$, and we set $P=\Theta_0$ when the working set is itself a polytope. The four blocks are
\begin{equation}
\begin{aligned}
\phi_0(x,z)&=T(x), &\qquad \rho_0(u,z)&=\Pi_P\{g_{\mathrm{init}}(u)\},\\
\phi(x,z)&=S(x,z), &\qquad \rho(u,z)&=h(u,z),
\end{aligned}
\label{eq:est-blocks}
\end{equation}
so the network is
\begin{equation}
\begin{aligned}
z^{(0)}&=\Pi_P\{g_{\rm init}(\bar T_n)\},\quad F(X^n)=\Pi_{\Theta_0}\left[z^{(0)}+h\left\{\frac1n\sum_iS(X_i,z^{(0)}),z^{(0)}\right\}\right].
\end{aligned}
\label{eq:structured-class}
\end{equation}
The ReLU networks $g_{\rm init}$, $S$, and $h$ approximate $\mu^{-1}$, $(x,z)\mapsto s_z(x)$, and $(u,z)\mapsto I(z)^{-1}u$, respectively. By Lemma~\ref{lem:relu-moment}, we can replace the moment function by a fixed ReLU function while retaining the required injectivity and nonsingular Jacobian. The moment mean and its inverse then correspond to this chosen function.

The inner projection $\Pi_P$ is represented exactly by a fixed-size ReLU network within $\rho_0$ (Lemma~\ref{lem:polytope}), whereas the outer projection $\Pi_{\Theta_0}$ is part of the fixed output map.

The same construction applies under the analyticity conditions of Appendix~\ref{app:analytic}, with smaller component networks; see Corollary~\ref{cor:analytic-est-gen}.

The construction shows that a single refinement suffices for asymptotic efficiency. Additional refinements can nevertheless improve empirical accuracy (Appendix~\ref{app:est-R}). A possible explanation is that several updates allow the network to approach its target gradually, without requiring an individual update to approximate a full Fisher scoring step.

The constrained MLE $\widehat\theta_{0,n}$ of Section~\ref{sec:theory-estimation} carries the subscript $0$ to refer to the working set $\Theta_0$ and to distinguish it from $\widehat\theta_{\mathrm{MLE}}$. We always write a true parameter as $\theta_0$ without a hat. By Lemma~\ref{lem:consequences}, for each compact $\Theta_1\subset\operatorname{int}(\Theta_0)$, there are $C,c>0$ such that, uniformly over generating parameters $\theta\in\Theta_1$, the maximizer is unique and lies in $\operatorname{int}(\Theta_0)$ with $\mb P_\theta$-probability at least $1-Ce^{-cn}$. On this event, the constrained MLE solves the likelihood equation. It therefore coincides with a consistent unrestricted MLE with probability tending to one.

The network of Theorem~\ref{thm:est-approx} contains two projections, which play different roles. The output map $\Psi=\Pi_{\Theta_0}$ acts on the final state. Like the softmax function in model selection, it is a fixed map without trainable parameters, and it is defined for every compact convex $\Theta_0$. This projection guarantees that the estimate lies in $\Theta_0$. On the event of exponentially small probability where the one-step comparison is not available, the error is therefore at most the diameter of $\Theta_0$.

The second projection acts on the initial estimate inside the block $\rho_0$, which must be a ReLU network. Its only purpose is to keep the initial state in a compact subset of $U$ on which the score network and the information network are accurate, whatever the data. Projection onto $\Theta_0$ would serve this purpose.

For a general convex set, however, it is not a piecewise linear map, so a ReLU network cannot compute it. In addition, the general approximation bound for Lipschitz maps would cost of the order of $\varepsilon^{-d}$ parameters for accuracy $\varepsilon$. Instead, we use the projection $\Pi_P$ onto a polytope $P$ with $\Theta_0\subseteq P\subset U$. Such a polytope exists because $\Theta_0$ is compact and $U$ is open (Lemma~\ref{lem:polytope}(a)). Moreover, $\Pi_P$ is continuous and piecewise linear, so a ReLU network of fixed size computes it exactly (Lemma~\ref{lem:polytope}(b)). Since $\Pi_P$ fixes every point of $\Theta_0$ and is nonexpansive, it does not change $g_{\rm init}(\bar T_n)$ when this point lies in $P$, in particular when it lies in $\Theta_0$, and it never increases its distance to the MLE. We then approximate the score and the inverse information on $\mathcal X\times P$ and on $P$ instead of $\mathcal X\times\Theta_0$ and $\Theta_0$. Since $P$ is a compact subset of $U$, this changes only the constants.

When $\Theta_0$ is a polytope, we take $P=\Theta_0$. When $\Theta_0$ is a rectangle, $\Pi_P$ is coordinatewise clipping. This is the case in Sections~\ref{sec:theory-model} and~\ref{sec:theory-variable}, where the working sets are polytopes and rectangles, respectively. In those sections, we apply the projections onto the working sets inside the refinement blocks. We compare the network logits with the criterion evaluated at the constrained MLE over the same working set, which requires the refined estimates to lie in this set (Lemma~\ref{lem:criterion}). We therefore take the working sets to be polytopes from the start. By Lemma~\ref{lem:polytope}(a), every compact convex set contained in an open set is contained in a polytope inside that open set, so we can replace a general compact convex working set by such a polytope. We treat the moment function $T$ in Lemma~\ref{lem:relu-moment}.

\subsubsection{Approximation error and the population target}\label{app:param-space}

The score network determines the exponent $\kappa$ in Theorem~\ref{thm:est-approx}. By Lemma~\ref{lem:consequences} and the inverse function theorem, $\mu(U)$ is open and $\mu^{-1}\in C^{r+1}\{\mu(U)\}$. Hence the moment inverse is a $C^{r+1}$ map of $d$ variables with approximation cost $\varepsilon^{-d/(r+1)}$. The cost $\varepsilon^{-(d_X+d)/r}$ of the score dominates this cost. The input dimension $d_X+d$ of the score does not depend on $n$.

For the trained network, the population target is the posterior mean. The following proposition, used before Theorem~\ref{thm:est-gen}, shows how far the posterior mean is from the constrained MLE.

\begin{proposition}[Posterior mean and constrained MLE]\label{prop:bayes-mle}
Under Assumptions~\ref{ass:regular} and~\ref{ass:proposal}, for all $n\geq1$, $\mb E_{\mb Q_n}[\|F^*(X^n)-\widehat\theta_{0,n}\|_2^2]\leq C\{1+\mathcal J(\pi)\}/n^2$.
Here $C$ depends on $\pi$ only through the distance of $\operatorname{supp}(\pi)$ from $\partial\Theta_0$.
\end{proposition}

Combining Proposition~\ref{prop:bayes-mle} with Theorem~\ref{thm:est-approx}, we reduce the analysis of the trained network $\widehat F$ to the comparisons
$\widehat F\ \approx\ F^*\ \approx\ \widehat\theta_{0,n}\ \approx\ \widetilde F$
in mean square under $\mb Q_n$, the first conditionally on $\mathcal D_M$. Proposition~\ref{prop:bayes-mle} controls the middle comparison. Theorem~\ref{thm:est-approx} with $\Theta_1=\operatorname{supp}(\pi)$, averaged over $\theta\sim\pi$, controls the last comparison. Theorem~\ref{thm:est-gen} controls the first difference through the excess risk.

\medskip
\noindent{\bf Effect of analyticity on network size.}\label{app:smoothness-complexity}
Under the analyticity conditions of Appendix~\ref{app:analytic}, the component approximation bounds improve substantially. If the moment inverse, the score and the inverse information satisfy the analytic extension condition of Appendix~\ref{app:analytic-extension}, Corollary~\ref{cor:analytic-est-gen} gives $W=O\{\log^{d_X+d+1}(en)\}$ at $\varepsilon=n^{-1/2}/\log n$.

Under this condition, for the class of Corollary~\ref{cor:analytic-est-gen}(b), which is built at $\varepsilon=n^{-1}$ and has $W$ of the same order, $M=\lceil n\log^{d_X+d+5}(en)\rceil$ tasks with $\varepsilon_{\rm opt}=o(n^{-1})$ suffice for first-order equivalence on a new task. The larger depth of the networks enters the covering bound through Lemma~\ref{lem:param-lipschitz}. This is why $(\log n)^2$ becomes $\log^3(en)$ in Corollary~\ref{cor:analytic-est-gen}. The extension condition on the moment inverse is the only requirement beyond an analytic log density. It asks for a rectangle that contains $\mathcal K$ and on which $\mu^{-1}$ has a bounded holomorphic extension.

Assumption~\ref{ass:proposal} allows smooth compactly supported densities that vanish inside their support, and excludes the uniform density. For task distributions $\pi_n$ that vary with $n$ and keep a fixed distance from $\partial\Theta_0$, the same bound holds with $\mathcal J(\pi_n)$. In this case, the bound gives first-order equivalence when in addition $\mathcal J(\pi_n)=o(n)$.

\subsubsection{Divide-and-conquer estimation and bootstrap inference}\label{app:dc-bootstrap}

We give the theoretical guarantees for divide-and-conquer estimation and bootstrap inference introduced in Section~\ref{sec:extensions}, with proofs in Appendices~\ref{app:proof-dc} and~\ref{app:proof-boot}. For divide-and-conquer estimation, we train $\widehat F$ as in Theorem~\ref{thm:est-gen} at sample size $n_0$ and accuracy $\varepsilon_{n_0}\in[n_0^{-1},1)$, so that $W=O\{\varepsilon_{n_0}^{-\kappa}\log(e/\varepsilon_{n_0})\}$. We then apply~\eqref{eq:dc-estimator} to a new task $\theta\sim\pi$, $X^n\mid\theta\sim\mb P_\theta^{(n)}$ with $n=Jn_0$, independent of $\mathcal D_M$.

\begin{corollary}[Divide-and-conquer efficiency]\label{cor:dc}
Under Assumptions~\ref{ass:regular} and~\ref{ass:proposal}, suppose $n_0\to\infty$, $n=o(n_0^2)$, $\varepsilon_{n_0}=o(n^{-1/2})$, $W\{(\log n_0)^2+\log(eM)\}/M=o(n^{-1})$ and $\varepsilon_{\rm opt}=o(n^{-1})$. Then $\sqrt n(\widehat\theta_{\rm DC}-\widehat\theta_{0,n})\to0$ in probability over $(\theta,X^n,\mathcal D_M)$, and
\[
d_{\rm BL}\!\left[\mathcal L\{\sqrt n(\widehat\theta_{\rm DC}-\theta)\mid\theta,\mathcal D_M\},\mathcal N\{0,I(\theta)^{-1}\}\right]\to0
\]
in probability over $(\theta,\mathcal D_M)$.
\end{corollary}

Under the remaining conditions of Corollary~\ref{cor:dc}, the condition $n=o(n_0^2)$ suffices to transfer a network trained at sample size $n_0$ to samples of size $n$. Averaging over blocks does not in general reduce the bias of the block MLEs, which is $O(n_0^{-1})$ and can therefore contribute an error of order $\sqrt n/n_0=(J/n_0)^{1/2}$ on the full-sample scale. When we apply~\eqref{eq:est-gen-bound} at $n_0$, its remainder $C\{1+\mathcal J(\pi)\}/n_0^2$ contributes $n/n_0^2=J/n_0$. Both vanish when $n=o(n_0^2)$. Similarly, conditional on $(\theta,\mathcal D_M)$, the network errors on the blocks have the same mean. Averaging reduces their conditional variance, but preserves this mean. Hence the approximation, pretraining and optimization errors must be negligible on the full-sample scale. We require $\varepsilon_{n_0}^2=o(n^{-1})$ and meta training error $o(n^{-1})$, rather than $o(n_0^{-1})$. The choice $\varepsilon_{n_0}=\max\{n_0^{-1},n^{-1/2}/\log n\}$ meets the accuracy condition whenever $n=o(n_0^2)$. It gives $W=O\{n^{\kappa/2}(\log n)^{\kappa+1}\}$, so $M=\lceil n^{\kappa/2+1}(\log n)^{\kappa+4}\rceil$ tasks of size $n_0$ suffice. These are the requirements of Theorem~\ref{thm:est-gen} at the full sample size $n$, while each training task has only $n_0$ observations. For $n=n_0^\beta$ with $\beta\in[1,2)$, the network has $O\{n_0^{\beta\kappa/2}(\log n_0)^{\kappa+1}\}$ parameters.

For bootstrap inference, we hold the same network fixed and evaluate it on $X^n$ and resamples $X^{*n}$ as in~\eqref{eq:bootstrap-estimator}.

\begin{corollary}[Bootstrap consistency]\label{cor:bootstrap}
Suppose Assumption~\ref{ass:regular} holds.
\textup{(a)} For the constructed network $\widetilde F$ of Theorem~\ref{thm:est-approx} with $\varepsilon=o(n^{-1/2})$ and every fixed $\theta_0\in{\rm int}(\Theta_0)$,
\[
d_{\rm BL}\!\left[\mathcal L\{\sqrt n(\widetilde F(X^{*n})-\widetilde F(X^n))\mid X^n\},\mathcal N\{0,I(\theta_0)^{-1}\}\right]\to0
\]
in $\mb P_{\theta_0}^{(n)}$-probability.
\textup{(b)} Suppose in addition Assumption~\ref{ass:proposal} and the rate conditions of Theorem~\ref{thm:est-gen} hold. Assume that, for a fixed stability constant $K_{\rm st}>0$ and with probability tending to one over $\mathcal D_M$, the components $g_{\rm init}$, $S$ and $h$ of the trained network $\widehat F$ satisfy
\[
{\rm Lip}(\Pi_P\circ g_{\rm init})\leq K_{\rm st},\qquad {\rm Lip}(h)\leq K_{\rm st},\quad
\sup_{x\in\mathcal X}\max\{\|S(x,\cdot)\|_\infty,{\rm Lip}(S(x,\cdot))\}\leq K_{\rm st},
\]
where $P$ is the polytope of the network class, so that $\Pi_P\circ g_{\rm init}$ is the block $\rho_0$ that is actually evaluated, the first two bounds are global and those for $S(x,\cdot)$ are on $P$. Then, for a new task $(\theta,X^n)\sim\mb Q_n$ independent of $\mathcal D_M$,
\[
d_{\rm BL}\!\left[\mathcal L\{\sqrt n(\widehat F(X^{*n})-\widehat F(X^n))\mid X^n,\mathcal D_M\},\mathcal N\{0,I(\theta)^{-1}\}\right]\to0
\]
in probability over $(\theta,X^n,\mathcal D_M)$.
\end{corollary}

Part~(b) requires a stability condition because a trained network is accurate only on average over samples drawn from tasks, whereas a resample is drawn from the empirical distribution and repeats observations of $X^n$. The stability condition transfers the accuracy of $\widehat F$ on $X^n$ to resamples through fixed-dimensional empirical averages (Appendix~\ref{app:proof-boot}).

\subsubsection{Unbounded observations}\label{app:unbounded}

We can relax compactness of $\mathcal X$ for the approximation result of Theorem~\ref{thm:est-approx}. Theorem~\ref{thm:theory-unbounded} below treats unbounded observations by clipping every observation coordinatewise to a compact rectangle $\mathcal X_n$, which increases with $n$, before it enters the moment statistic and the score network. The likelihood and the constrained MLE remain those of the original model. Suppose the rectangles have radius of order $\sqrt{\log n}$, the bound $B_n$ on the component maps defined below grows polynomially in the radius, and the smoothness $\alpha$ of the moment map below satisfies $d/\alpha\leq(d_X+d)/r$. Then the size exceeds that of Theorem~\ref{thm:est-approx} only by logarithmic factors. All other results, for trained networks, divide-and-conquer estimation, bootstrap inference and selection, are stated for compact observation spaces.

We retain the fixed parameter sets $\Theta_0\subset U$ and the original experiment. As in Appendix~\ref{app:proof-approx}, we project the initial estimate onto a polytope $P$ with $\Theta_0\subseteq P\subset U$, and the suprema and Lipschitz constants over $\Theta_0$ that enter $M_I$ and $B_n$ below are taken over $P$. We suppose
\[
 X_1,\ldots,X_n\stackrel{\rm iid}{\sim}\mathbb P_\theta,\quad
 \mathbb E_\theta[\|T(X_1)\|_2]<\infty\quad(\theta\in U),
\]
\[
 \begin{gathered}
 \mu(\theta)=\mathbb E_\theta [T(X_1)],\qquad\mu\in C^\alpha(U),\qquad\alpha\geq1,\\
 \mu(\theta)=\mu(\eta)\Longrightarrow\theta=\eta,\qquad
 \det D\mu(\theta)\ne0\qquad(\theta,\eta\in U).
 \end{gathered}
\]
We assume in addition that a fixed ReLU network computes the moment function $T$ exactly. That is, $T$ is continuous and piecewise linear, for example $T(x)=x$. Lemma~\ref{lem:relu-moment} is not available here because $\mathcal X$ is not compact, and the smoothness of $\mu$ alone does not control the cost of computing $T$. The depth and the number of parameters of this network do not depend on $n$ or $\varepsilon$ and are absorbed into the constants below. We set $g_{\rm MoM}=\mu^{-1}$ and fix a compact set $\mathcal K$ with $\mu(\Theta_0)\subset\operatorname{int}(\mathcal K)\subset\mathcal K\Subset\mu(U)$. Let $\mathcal X_n$ be increasing compact rectangles and let $\mathsf C_n$ be their coordinatewise clipping maps:
\[
 \mathcal X_n\subseteq\mathcal X_{n+1},\qquad
 \mathsf C_n:\mathbb R^{d_X}\to\mathcal X_n,\qquad
 \mathsf C_n(x)=x\quad(x\in\mathcal X_n).
\]
We apply $\mathsf C_n$ before evaluating $T$ and the score network. For $r\geq2$, we assume that extensions on open neighborhoods satisfy
\[
 \ell^{\rm ext}\in C^{r+1}(O_n),\quad
 \mathcal X_n\times U\subset O_n,\quad
 s^{\rm ext}=\nabla_\theta\ell^{\rm ext}\in C^r(O_n),
\]
\[
 \begin{gathered}
 g_{\rm MoM}^{\rm ext}\in C^\alpha(O_g),\qquad\mathcal K\subset O_g,\\
 (I^{-1})^{\rm ext}\in C^r(O_I),\qquad P\subset O_I,\qquad
 M_I:=\sup_{\Theta_0}\|I^{-1}\|_{\rm op}<\infty.
 \end{gathered}
\]
Here $I$ is the Fisher information of the original law. On $\mathcal A_n=\bigcup_{m\geq1}\mathcal X_m^n$, we fix the measurable maximizer rule of Appendix~\ref{app:class}. Its existence follows from continuity on each compact rectangle. With a fixed $\theta_{\rm fix}\in\Theta_0$, we set
\[
 \widehat\theta_{0,n}(x^n)\in\argmax_{z\in\Theta_0}\sum_i\ell_z(x_i)
 \quad(x^n\in\mathcal A_n),\qquad
 \widehat\theta_{0,n}(x^n)=\theta_{\rm fix}\quad(x^n\notin\mathcal A_n).
\]
The tail condition below implies
\[
 \mathbb P_\theta\!\left(X_1\notin\bigcup_m\mathcal X_m\right)
 =\lim_{m\to\infty}\mathbb P_\theta(X_1\notin\mathcal X_m)=0,
 \qquad\mathbb P_\theta^{(n)}(\mathcal A_n)=1.
\]

For a real number $s>0$ and a scalar function $f$ on a rectangle $R$, write $k=\lfloor s\rfloor$ and $\gamma=s-k$, and let $[D^\nu f]_{\gamma,R}$ be the H\"older seminorm defined in the notation paragraph in Section~\ref{sec:intro}. The $C^s$ norm of $f$ on $R$ is
\[
\|f\|_{C^s(R)}=\max_{|\nu|\leq k}\sup_{x\in R}|D^\nu f(x)|+\ind\{\gamma>0\}\max_{|\nu|=k}[D^\nu f]_{\gamma,R},
\]
and for vector- or matrix-valued functions it is the maximum of the coordinatewise norms.

Let $\mathcal R_{s,n}$ be a rectangle containing $\mathcal X_n\times P$, and let $A_n$ normalize it to the unit cube. Choose $B_n\geq1$ so that
\[
 \begin{aligned}
 B_n\geq\max\{&\|s^{\rm ext}\circ A_n^{-1}\|_{C^r},
 \sup_{\mathcal X_n\times\Theta_0}\|s\|_2,
 \sup_{\mathcal X_n\times\Theta_0}\|D_\theta s\|_{\rm op},
 \sup_{\mathcal X_n\times\Theta_0}\|D_\theta^2s\|_{\rm op},\\
 &\sup_{x\in\mathcal X_n}\operatorname{Lip}_{\Theta_0}\{s(x,\cdot)\},
 \operatorname{Lip}_{\Theta_0}(I),
 \|g_{\rm MoM}^{\rm ext}\|_{C^\alpha},
 \|(I^{-1})^{\rm ext}\|_{C^r},K_{{\rm geom},n}\}.
 \end{aligned}
\]
The last two norms are taken on their fixed approximation rectangles. Here $K_{{\rm geom},n}$ denotes the largest absolute value of any coefficient (slope or translation) of the affine maps $A_n$ and $A_n^{-1}$. As in the proof of Theorem~\ref{thm:est-approx}, with the original sample and likelihood, let $\bar T_n=n^{-1}\sum_{i=1}^nT(X_i)$, and set $\theta_n^{\rm m}=g_{\rm MoM}(\bar T_n)$ when $\bar T_n\in\mathcal K$ and $\theta_n^{\rm m}=\theta_{\rm fix}$ otherwise. We also write $H_n(z)=-n^{-1}\sum_{i=1}^n\nabla_z^2\ell_z(X_i)$ and define
\[
\begin{gathered}
\mathcal E_n=\{\bar T_n\in\mathcal K,\ \theta_n^{\rm m}\in\Theta_0,\ \widehat\theta_{0,n}\in\operatorname{int}(\Theta_0)\},\quad
\rho_n=\|\theta_n^{\rm m}-\widehat\theta_{0,n}\|_2,\quad
\delta_n=\|I(\theta_n^{\rm m})-H_n(\theta_n^{\rm m})\|_{\rm op}.
\end{gathered}
\]

\begin{theorem}[Approximation on expanding domains]\label{thm:theory-unbounded}
Suppose the preceding component-map regularity and bounds hold, and set $\kappa=\max\{d/\alpha,(d_X+d)/r\}$. For each $\varepsilon\in(0,1)$, there is a clipped-input network for which
\[
 L\leq C\log(eB_n/\varepsilon),\qquad
 W\leq C(B_n^2/\varepsilon)^\kappa
                         \log(eB_n/\varepsilon),
\]
and, on $\mathcal E_n\cap\{X_i\in\mathcal X_n\text{ for all }i\}$,
\[
 \|\widetilde F-\widehat\theta_{0,n}\|_2
 \leq\varepsilon+C(\varepsilon+\rho_n)
                         \{\delta_n+B_n(\varepsilon+\rho_n)\}.
\]
At a fixed $\theta\in\operatorname{int}\Theta_0$, suppose
\[
\begin{gathered}
 n\mb P_\theta(X_1\notin\mathcal X_n)\to0,\qquad
 \mb P_\theta(\mathcal E_n)\to1,\\
 \rho_n=O_P(n^{-1/2}),\qquad\delta_n=o_P(1),\qquad
 B_n=o(\sqrt n),\qquad\varepsilon=o(n^{-1/2}).
\end{gathered}
\]
Then $\sqrt n\|\widetilde F-\widehat\theta_{0,n}\|_2\xrightarrow{\mathbb P_\theta}0$. If the original-model MLE also satisfies $\sqrt n(\widehat\theta_{0,n}-\theta)\xrightarrow{d}\mathcal N\{0,I(\theta)^{-1}\}$, then $\sqrt n(\widetilde F-\theta)\xrightarrow{d}\mathcal N\{0,I(\theta)^{-1}\}$.
\end{theorem}

\subsection{Model selection}

\subsubsection{Model separation and the RRDS construction}\label{app:rrds-model}

We call a task generated from $(k,\theta)$ \emph{separated} if every $j\ne k$ satisfies one of the following conditions:
\begin{equation}
\begin{aligned}
&\text{either }d_j>d_k\text{ and there is }\theta'\in\Xi_j\text{ such that }
\mb P_{j,\theta'}=\mb P_{k,\theta},\\
&\qquad\operatorname{dist}(\theta',\partial\Xi_j)
\geq\min\{\operatorname{dist}(\theta,\partial\Xi_k),\rho\},\\
&\text{or }\inf_{\theta'\in\Theta_{0,j}}
D_{\rm KL}(\mb P_{k,\theta}\|\mb P_{j,\theta'})
\geq A_1\frac{\log n}n.
\end{aligned}
\label{eq:weak-separation}
\end{equation}
We specify $\rho>0$ and $A_1$ in Appendix~\ref{app:model-recovery}. Let $\psi_n$ be the probability that a task fails the separation condition. In the first case, we need the condition on the prior support. This is because the BIC fit uses the working set, whereas the marginal likelihood integrates only over the prior support. In the second case, the KL threshold is large enough for the likelihood comparison to overcome the BIC penalties in our analysis.

Lemma~\ref{lem:psi-nested} gives explicit bounds on $\psi_n$ under its ordered-pair conditions. In particular, $\psi_n=0$ eventually when all candidates are uniformly separated. A regular nested pair with dimension difference $b$ contributes at most $C(\log n/n)^{b/2}$ when the task is generated from the larger candidate. The familiar $n^{-1/2}$ difficulty corresponds to a one-dimensional difference. In that case, a neighborhood of radius $n^{-1/2}$ around the smaller model has prior mass of that order. For larger dimension differences, the mass of this neighborhood has a different order.

With ${\rm BE}_n=\mb E_{\mb Q_n}[1-\max_kq_k^*(X^n)]$, Proposition~\ref{prop:ms-bridge} gives
\begin{equation}
\mb E_{\mb Q_n}[D_{\rm KL}(q^*\|q^{\rm BIC})]
\leq C\Bigl\{{\rm BE}_n+\psi_n\log n+\frac1{\sqrt n}\Bigr\}.
\label{eq:ms-bridge}
\end{equation}
In general, we cannot remove the factor $\log n$ on $\psi_n$ for these BIC probabilities. Remark~\ref{rem:bic} gives an example with different prior supports. If no separated task has a competitor of the first kind in~\eqref{eq:weak-separation}, Proposition~\ref{prop:ms-bridge} improves the bound to $C\{\psi_n\log n+\log n/n\}$. Global nonnesting alone does not exclude intersections between candidates at particular distributions.

\medskip
\noindent{\bf State and initialization.}
For each candidate $k$, the model-selection network stores an initial estimate $\vartheta_k^0$ and a current estimate $\theta_k$. Together with the $K$ logits in $a$, these form the state $z=(\vartheta^0,\theta,a)$. Thus $d_z=2\sum_kd_k+K$. In the initialization, we compute the projected moment estimates $\vartheta_k^0$ for all candidates and set $\theta_k=\vartheta_k^0$.

\medskip
\noindent{\bf Shared refinements and output.}
At each refinement, the encoder evaluates the approximate scores at the fixed initial estimates $\vartheta_k^0$ and the approximate log densities $\varphi_k$ at the current estimates $\theta_k$. We choose each $\varphi_k$ to satisfy $\|\varphi_k-\ell_{k,\cdot}\|_\infty\leq n^{-1}$ on $\mathcal X\times\Theta_{0,k}$. The encoder output dimension is $d_u=\sum_kd_k+K$. The decoder computes the one-step estimates $\widetilde\theta_k$ from the averaged scores and forms the logits from the averaged likelihoods. It returns the corrections needed for the residual update~\eqref{eq:rrds-update}.

The first refinement updates the parameters but evaluates the likelihoods at the initial estimates. The second refinement evaluates them at $\widetilde\theta_k$ and gives the uncentered logits $\widetilde a_k=\log w_k+\sum_i\varphi_k(X_i,\widetilde\theta_k)-(d_k/2)\log n$. The decoder subtracts their maximum and clips them below at $-\log n$. The output map then clips the resulting logits to $[-\log n,\log n]$ and applies softmax to produce the model probabilities $\widetilde q_k$. Since the maximum centered logit is zero, the clipping preserves the maximizing indices and ensures $\widetilde q_k\geq1/(Kn)$. The selected fitted distribution is $\widetilde{\mb P}=\mb P_{\widetilde k,\widetilde\theta_{\widetilde k}}$. Further refinements leave this construction unchanged. Lemma~\ref{lem:ms-rrds} gives the exact blocks.

Two sources of error arise when we approximate the information criteria: the parameter estimates and the neural approximation of the log densities. Lemma~\ref{lem:ms-logits} gives, for every dataset,
\[
\widetilde a_k=a_k^{\rm BIC}-D_k+e_k,\qquad
D_k=\ell_{n,k}(\widehat\theta_k)-\ell_{n,k}(\widetilde\theta_k)\geq0,
\qquad |e_k|\leq1.
\]
When candidate $k$ contains the generating distribution, we can bound the likelihood deficit $D_k$ using the one-step approximation. For other candidates, the nonnegativity of $D_k$ suffices for the selection comparison, even if $\widetilde\theta_k$ does not approximate the corresponding MLE accurately.

The likelihood approximation determines the leading parameter count in Theorem~\ref{thm:ms-approx}. Its input dimension is $d_X+d_k$, and its required accuracy is $n^{-1}$. The parameter subnetworks use accuracy $n^{-1/2}/\log(en)$. Under the full analytic conditions of Corollary~\ref{cor:analytic-ms}, $W=O\{K(\log n)^{d_X+d_{\max}+1}\}$ and $M=\lceil\sqrt n(\log n)^{d_X+d_{\max}+5}\rceil$ tasks suffice to preserve the approximation rate. We give the detailed calculations in Appendix~\ref{app:ms-discussion}.

\subsubsection{Posterior model probabilities and BIC}\label{app:model-recovery}

In our recovery analysis, $\psi_n$ denotes the prior probability of \emph{non-separated} tasks, that is, generating pairs $(k,\theta)$ that fail the weak-separation condition~\eqref{eq:weak-separation} for at least one competitor $j\ne k$:
\[
\psi_n=\sum_{k=1}^Kw_k\,\pi_k\bigl\{\theta\in\Xi_k:(k,\theta)\text{ violates~\eqref{eq:weak-separation} for some }j\ne k\bigr\}.
\]
To specify the separation threshold $A_1$ in~\eqref{eq:weak-separation} explicitly, let $B$ be such that $|\ell_{k,\theta}(x)|\le B$ for all $k\in[K]$, $x\in\mathcal X$ and $\theta\in\Theta_{0,k}$, put $B_Y=\max\{1,2B\}$, and let $\lambda_Y$ be the constant of Lemma~\ref{lem:lik-process} for this $B_Y$. We set
\begin{equation}
A_1=\max\Bigl\{8,\ \frac83\Bigl(\frac{d_{\max}+1}{\lambda_Y}+\frac{d_{\max}}2\Bigr),\ 4(d_{\max}+2),\ \frac{4(d_{\max}+2)}{\lambda_Y}\Bigr\},
\label{eq:ms-A1}
\end{equation}
and fix any $\rho\in(0,\tau]$, where $\tau=\min_k\operatorname{dist}(\Xi_k,\partial\Theta_{0,k})>0$ is the buffer distance between each prior support and the boundary of its working parameter set.

We make no structural assumption on $\psi_n$ in our general error bounds, where it enters additively. Instead, the next lemma shows that $\psi_n$ vanishes automatically as $n\to\infty$ under the two standard model configurations discussed in Section~\ref{sec:theory-model}: well-separated candidates and nested candidates.

\begin{lemma}[Weak-separation probability for nested and separated candidates]\label{lem:psi-nested}
Suppose Assumption~\ref{ass:models} holds and every ordered pair $(k,j)$ with $k\ne j$ (where $k$ is the generating model and $j$ is a competitor) falls into one of the following three types, for a fixed constant $\gamma>0$:
\begin{itemize}
\item[\textup{(S)}] \emph{(Separated)} $\inf_{\theta'\in\Theta_{0,j}}D_{\rm KL}(\mb P_{k,\theta}\|\mb P_{j,\theta'})\geq\gamma$ for every $\theta\in\Xi_k$.
\item[\textup{(U)}] \emph{(Upward-nested, $d_j>d_k$)} Every $\theta\in\Xi_k$ is matched by some $\theta'\in\Xi_j$ with $\mb P_{j,\theta'}=\mb P_{k,\theta}$ and $\operatorname{dist}(\theta',\partial\Xi_j)\geq\min\{\operatorname{dist}(\theta,\partial\Xi_k),\rho\}$.
\item[\textup{(D)}] \emph{(Downward-nested, $d_j<d_k$)} There is a Lipschitz map $\iota:\Theta_{0,j}\to\Theta_{0,k}$ such that $\mb P_{k,\iota(\vartheta)}=\mb P_{j,\vartheta}$ for all $\vartheta\in\Theta_{0,j}$.
\end{itemize}
Then, for all $n$ large enough that $A_1\log n/n\leq\gamma$,
\[
\psi_n\leq C\sum_{(k,j)\text{ of type \textup{(D)}}}w_k\Bigl(\frac{\log n}n\Bigr)^{(d_k-d_j)/2}\leq C\sqrt{\frac{\log n}n},
\]
and $\psi_n=0$ if no pair is of type \textup{(D)}.
\end{lemma}

Once $A_1\log n/n\leq\gamma$, pairs of types~\textup{(S)} and~\textup{(U)} cause no violations. Instead, violations only come from type~\textup{(D)} when $\theta\in\Xi_k$ lies within distance $O(\sqrt{\log n/n})$ of $\iota(\Theta_{0,j})$. This region is a tube of codimension $d_k-d_j$, so its prior mass is $O\{(\log n/n)^{(d_k-d_j)/2}\}$. For example, suppose candidate $j$ has parameter $(\theta,\nu)$ and candidate $k$ fixes $\nu=\nu_0$. If $\Xi_j\supseteq\Xi_k\times\prod_i[\nu_{0,i}-\rho,\nu_{0,i}+\rho]$, then $(k,j)$ is of type~\textup{(U)} and $(j,k)$ is of type~\textup{(D)} with dimension gap $\dim\nu$. Here, requiring $\theta'\in\Xi_j$ instead of $\theta'\in\Theta_{0,j}$ in type~\textup{(U)} is key. This is because the BIC logit maximizes over $\Theta_{0,j}$, while the marginal likelihood $m_j$ integrates over $\Xi_j$ only. If $\theta'\notin\Xi_j$, the two differ by order $n$, so BIC fails to describe the posterior model probabilities.

\begin{proposition}[Posterior model probabilities and BIC]\label{prop:ms-bridge}
Under Assumption~\ref{ass:models}, for $n\geq3$,
\[
\mb E_{\mb Q_n}\bigl[D_{\rm KL}(q^*\|q^{\rm BIC})\bigr]
\leq C\Bigl\{{\rm BE}_n+\psi_n\log n+\frac1{\sqrt n}\Bigr\},
\qquad
{\rm BE}_n\leq C\frac{(\log n)^{d_{\max}/2}}{\sqrt n}+\psi_n .
\]
The first bound also holds for the BIC probabilities with logits centered by their maximum and clipped at $-\log n$. If no separated task has a competitor of the first kind in~\eqref{eq:weak-separation}, in particular if all pairs are of type \textup{(S)} in Lemma~\ref{lem:psi-nested}, then the term $n^{-1/2}$ can be replaced by $\log n/n$ and ${\rm BE}_n\leq C/n+\psi_n$. Here $C$ depends on the candidate families and working sets, on $\max_{j,l}w_j/w_l$, on $\rho$, and on the density bounds and the supports of the $\pi_k$.
\end{proposition}

We give the proof in Appendix~\ref{app:proof-ms-bridge}. It rests on the Laplace bounds of Lemma~\ref{lem:ms-laplace}. To compare the posterior log odds with BIC, we write $\delta_j=\log m_j-\{\ell_{n,j}(\widehat\theta_j)-(d_j/2)\log n\}$ for the remainder in the BIC approximation to $\log m_j$. Then the posterior log odds satisfy
\[
\log\frac{q_j^*}{q_l^*}=\bigl(a_j^{\rm BIC}-a_l^{\rm BIC}\bigr)+(\delta_j-\delta_l).
\]
For every candidate $j$ that contains the generating distribution at a parameter in the support $\Xi_j$ of $\pi_j$, $\delta_j$ is bounded above by a constant and below by $-C-Cn\operatorname{dist}(\widehat\theta_j,\Xi_j)^2$, with probability exponentially close to one. The lower bound is a constant unless the MLE falls outside the support of $\pi_j$. That event has non-negligible probability only for tasks within $O(n^{-1/2})$ of the boundary of the support. This result gives the precise form of the statement that BIC arises from an asymptotic approximation of the posterior model probabilities~\citep[Section~3]{schwarz1978dimension}, \citep[Section~4.1]{kassRaftery1995bayes}. It also explains why Assumption~\ref{ass:models}(ii) requires densities that are bounded and bounded away from zero. This condition differs from Assumption~\ref{ass:proposal}, which requires a continuously differentiable density on $\mb R^d$. We use Assumption~\ref{ass:proposal} to compare the posterior mean with the MLE.

\begin{remark}[The factor $\log n$ of $\psi_n$]\label{rem:bic}
Proposition~\ref{prop:ms-bridge} and Theorem~\ref{thm:ms-approx} charge non-separated tasks at the clipping level $\log n$. We cannot remove this factor without a further condition. Let $X=(X_1,X_2)\in\{-1,1\}^2$ have density
\[
p_{\mu,\nu}(x)=\frac{e^{\mu x_1+\nu x_2}}{4\cosh\mu\cosh\nu}
\]
with respect to the counting measure. Candidate~1 fixes $\nu=0$, with $\Theta_{0,1}=[-2,2]$ and $\mu$ uniform on $[-1,1]$. Candidate~2 has $\Theta_{0,2}=[-3,3]^2$ and $(\mu,\nu)$ uniform on $[-3/2,3/2]\times[-1,1]$. We set $w=(1/2,1/2)$. Assumption~\ref{ass:models} holds. The pair $(1,2)$ is of type~\textup{(U)} and the pair $(2,1)$ is of type~\textup{(D)}, so $\psi_n$ is of order $\sqrt{\log n/n}$. Consider the tasks from candidate~2 with $\mu\in[5/4,11/8]$ and $|\nu|\leq\frac14\sqrt{\log n/n}$. They have probability of order $\sqrt{\log n/n}$. Uniformly over these tasks, with probability tending to one, the BIC logits satisfy $a_2^{\rm BIC}-a_1^{\rm BIC}\leq-(\log n)/4$, and hence $q_2^{\rm BIC}\leq n^{-1/4}$. On the other hand, the marginal likelihood of candidate~1 integrates over $\mu\in[-1,1]$ only. This support is at a fixed KL distance from the generating distribution, so $q_2^*\to1$. Hence $D_{\rm KL}(q^*\|q^{\rm BIC})\geq c\log n$ on these tasks. This gives
\[
\mb E_{\mb Q_n}\bigl[D_{\rm KL}(q^*\|q^{\rm BIC})\bigr]\geq\frac{c(\log n)^{3/2}}{\sqrt n},
\]
which is the order of $\psi_n\log n$. In this example, the working-set MLE of the smaller candidate fits the common parameter $\mu$ outside the support of its task distribution. The lower bound concerns $q^{\rm BIC}$ and networks that follow our construction.
\end{remark}

\subsubsection{Predictive risk under model misspecification}\label{app:ms-misspecification}

Consider the network of Theorem~\ref{thm:ms-approx} and its fitted distribution $\widetilde{\mb P}$. Let $\mb P_0$ have a density $p_0$ with respect to the common dominating measure and satisfy $\mb E_0[|\log p_0(X_1)|]<\infty$. Suppose every candidate $k$ satisfies:

\textup{(i)} $\argmin_{\theta\in\Theta_{0,k}}D_{\rm KL}(\mb P_0\|\mb P_{k,\theta})=\{\theta_k^*\}$ for some $\theta_k^*\in\operatorname{int}(\Theta_{0,k})$;

\textup{(ii)} $-\mb E_0[\nabla_\theta^2\ell_{k,\theta}(X_1)]\big|_{\theta=\theta_k^*}\in\mathbb S_{++}^{d_k}$;

\textup{(iii)} there is $\theta_{m,k}\in\operatorname{int}(\Theta_{0,k})$ such that $\mu_k(\theta_{m,k})=\mb E_0[T_k(X_1)]$.

Define $\tau_k=\|\theta_{m,k}+I_k(\theta_{m,k})^{-1}\mb E_0[s_{k,\theta_{m,k}}(X_1)]-\theta_k^*\|_2$. This measures the discrepancy between one population Fisher scoring step from the moment target and the KL projection. It vanishes for a well-specified candidate.

\begin{proposition}[Predictive KL oracle inequalities]\label{prop:ms-misspec}
Under Assumption~\ref{ass:models} and the preceding conditions, for $n\geq3$,
\begin{equation}
\begin{aligned}
\mb E_0[D_{\rm KL}(\mb P_0\|\widetilde{\mb P})]
&\leq\min_{k\in[K]}\left\{
D_{\rm KL}(\mb P_0\|\mb P_{k,\theta_k^*})
+\frac{d_k\log n}{2n}+C\tau_k^2\right\}+C\left(\frac{\log K+d_{\max}}n\right)^{1/2}.
\end{aligned}
\label{eq:ms-oracle}
\end{equation}
If, in addition, $|\log\{p_0(x)/p_{k,\theta}(x)\}|\leq B_0<\infty$ for $\mb P_0$-almost every $x$, every $k$, and every $\theta\in\Theta_{0,k}$, then
\begin{equation}
\mb E_0[D_{\rm KL}(\mb P_0\|\widetilde{\mb P})]
\leq\min_{k\in[K]}\left\{
2D_{\rm KL}(\mb P_0\|\mb P_{k,\theta_k^*})
+\frac{d_k\log n}n+C\tau_k^2\right\}
+\frac{C\log n}n.
\label{eq:ms-oracle-fast}
\end{equation}
The constant $C$ may depend on $\mb P_0$ and, in the second bound, on $B_0$, but not on $n$.
\end{proposition}

We give the proof in Step~4 of Appendix~\ref{app:proof-ms-approx}. The bounds compare the selected fit with the best trade-off among approximation error, dimension penalty and one-step bias. Under bounded log-likelihood ratios, the KL divergence controls the variance. This allows a Bernstein bound with remainder $O(\log n/n)$, at the cost of a factor two on the approximation error. Appendix~\ref{app:ms-discussion} gives further interpretation. These guarantees are for the constructed selection rule.

\subsubsection{Selection rates, network size, and meta training}\label{app:ms-discussion}

The discussion following Theorem~\ref{thm:ms-approx} explains the sources of the rates in that theorem and the improvement when no larger candidate model contains the generating distribution. Here we discuss the oracle inequalities and the model selection error bounds, and then the network complexity and the meta training requirements.

In the oracle inequality~\eqref{eq:ms-oracle}, the discrepancy $\tau_k$ vanishes when candidate $k$ contains $\mb P_0$. It also vanishes in a regular exponential family initialized by sufficient-statistic means, because the KL projection then solves the same moment equation. The two oracle inequalities differ in the coefficient of the best approximation error and the size of the remainder. In~\eqref{eq:ms-oracle}, the best approximation error enters with constant one, and the remainder is of order $n^{-1/2}$. This remainder is the fluctuation of the empirical likelihood comparison between two misspecified candidates. It exceeds the dimension penalty, so~\eqref{eq:ms-oracle} is informative when the candidates differ in approximation error by more than $n^{-1/2}$.

In~\eqref{eq:ms-oracle-fast}, the constant is two and the remainder is of order $\log n/n$. The reason is that, for bounded log likelihood ratios, the corresponding KL divergence controls the variance of a ratio. When some candidate contains $\mb P_0$,~\eqref{eq:ms-oracle-fast} gives predictive risk of order $\log n/n$.

The recovery bound~\eqref{eq:ms-recovery-rate} collects three terms with different sources (Appendix~\ref{app:proof-ms-approx}). The first term, of order $n^{-1}$, is the fitting error of the generating candidate. It is the probability that the parameter network of this candidate loses more than a fixed amount of likelihood relative to the constrained MLE.

The second term, of order $e^{-cn}$, accounts for the candidates whose working families have a positive KL gap from $\mb P_0$. The expected log-likelihood loss for such candidates is of order $n$. The third term accounts for the larger candidates that also contain $\mb P_0$. Such a candidate must overcome the difference in the BIC penalties, which is $(\log n)/2$ per additional parameter. When it nests the generating candidate, it gains likelihood only through its redundant parameters. With one additional parameter, it overcomes the penalty with probability at most of order $n^{-1/2}$ up to a logarithmic factor. With more additional parameters, the bound is smaller. This last term determines the rate of the selection error bound. Precisely, a larger candidate $k$ that contains the generating distribution competes against the penalty difference $(d_k-d_{k_0})\log n/2$. Lemma~\ref{lem:ms-compare}(b) bounds its likelihood-gain tail at this threshold by
\[
C(\log n)^{d_k/2}n^{-(d_k-d_{k_0})/2}.
\]
The power of the logarithm comes from the covering argument in that lemma. This bound requires no nesting among the correct candidates. The same mechanism gives the bound on the Bayes error in Proposition~\ref{prop:ms-bridge}.

In Theorem~\ref{thm:ms-approx}, the networks that approximate the log densities dominate the network size. Because the criteria compare unnormalized log likelihoods, each $\varphi_k$ needs uniform accuracy $n^{-1}$. The parameter networks use accuracy $n^{-1/2}/\log(en)$, as in Section~\ref{sec:theory-estimation}. Up to logarithmic factors, their size $n^{(d_X+d_k)/(2r_k)}$ is of smaller order because $r_k\geq2$. Now suppose that, for every $k\in[K]$, the maps $\mu_k^{-1}$, $(x,\theta)\mapsto s_{k,\theta}(x)$, $\theta\mapsto I_k(\theta)^{-1}$ and $(x,\theta)\mapsto\ell_{k,\theta}(x)$ satisfy the analytic extension condition of Appendix~\ref{app:analytic-extension} on their respective approximation domains. Corollary~\ref{cor:analytic-ms}(a) then gives the conclusions of Theorem~\ref{thm:ms-approx} and Proposition~\ref{prop:ms-misspec} with
\[
L=O\{\log^2(en)\},\qquad W=O\{K\log^{d_X+d_{\max}+1}(en)\},
\]
so the polynomial factor in $n$ disappears.

We use the constructed network as a reference function to bound the approximation error in meta training. In Theorem~\ref{thm:ms-gen}, we train the network end to end with the cross-entropy loss. All blocks of the class $\mathcal F(L,W)$ are trainable, including the encoders that play the role of the likelihood networks in the construction. The sampling term of $\Delta_{\rm ms}$ in Theorem~\ref{thm:ms-gen} is $\log n$ times the corresponding term of Theorem~\ref{thm:est-gen}. The number of tasks stated after Theorem~\ref{thm:ms-gen} makes $\Delta_{\rm ms}$ of the order of the approximation terms for nested candidates. In parameter estimation, the training error is taken to be of smaller order than $n^{-1}$. Here it only needs to be of the order of the posterior uncertainty, because we compare the trained probabilities with the posterior probabilities through the excess risk.

For analytic candidates, Corollary~\ref{cor:analytic-ms}(b) gives $W=O\{K\log^{d_X+d_{\max}+1}(en)\}$ and replaces $(\log n)^2$ in the definition of $\Delta_{\rm ms}$ by $\log^3(en)$. With $\delta=n^{-1}$ and $\varepsilon_{\rm opt}=O(n^{-1/2})$, $M=\lceil\sqrt n\,\log^{d_X+d_{\max}+5}(en)\rceil$ tasks suffice for $\Delta_{\rm ms}=O(n^{-1/2})$.

\subsection{Variable selection}

\subsubsection{Regularity and analytic approximation for sparse GLMs}

As in parameter estimation, we start from a moment estimator and apply a Fisher scoring update. For variable selection, however, we need to construct such an estimator for every candidate support. It is therefore important to distinguish the dimension of each candidate model from the ambient dimension $p$, and to control the approximation errors uniformly over the supports.

The domain of the moment inverse in Assumption~\ref{ass:sparse}(iii) is
\begin{equation}
K_{g,S}=\{m\in\mathbb R^{d_S}:\operatorname{dist}\{m,\mu_S(\Theta_{0,S})\}\leq\rho_g\}.
\label{eq:moment-inverse-domain}
\end{equation}
The neighborhood ensures that a moment estimate close to the population moment can be inverted, including when the generating parameter is close to the boundary of the working set. The additional extension condition in Appendix~\ref{app:analytic-extension} concerns the approximation of this inverse by a ReLU network. In Appendix~\ref{app:vs-setting}, we show how to choose $\rho_g=\rho/\{2(1+s_{\max})\}$ after imposing $\rho\leq\rho_\mu$, where the model regularity constants determine $\rho_\mu$.

For the sparse GLM, Assumption~\ref{ass:sparse} concerns $b$ and $Q_Z$. The common term $c(y,a)$ cancels from likelihood comparisons. Bounded responses imply that $b$ is finite and smooth on $\mathbb R$, with $|b'|\leq1\vee B$, $b''\leq(1\vee B)^2/a$, and $|b'''|\leq16(1\vee B)^3/a^2$ (Lemma~\ref{lem:glm-natural-modules}). The centered log density $[y\eta_\theta(z)-b\{\eta_\theta(z)\}]/a$ and its parameter derivatives are bounded on the working domains (Appendix~\ref{app:sparse-regularity}). Bounded observations allow us to approximate the relevant functions uniformly on these domains and control the likelihood fluctuations across candidate supports. The output clipping and projection keep the training loss bounded.

Assumption~\ref{ass:sparse}(iii) allows us to approximate the inverse moment map with a network whose size grows only as a power of the logarithm of the inverse accuracy, for fixed support size and extension constants. For sufficiently small working rectangles, the sufficient conditions in Appendix~\ref{app:vs-setting} give extension constants independent of $n$ and $p$. Larger rectangles may impose a restriction, because injectivity and real smoothness do not ensure the required complex extension. The same appendix derives the analytic approximations of $b$, $b'$, and $I_S^{-1}$ and specifies $L_{{\rm vs},n}$ and $V_{{\rm vs},n}$ in~\eqref{eq:sparse-analytic-complexity}.

The product structure and density bounds in Assumption~\ref{ass:sparse-proposal} help control the marginal likelihoods. The slab supports include a neighborhood of zero, allowing small coefficients to be omitted. The Lipschitz condition on the log densities controls changes in the Laplace corrections across nearby supports. Uniform slabs satisfy this condition with $L_\pi=0$. The margin in part~(iii) makes the generating parameter interior in every containing support. The localization results show that the corresponding MLE is interior with high probability.

\subsubsection{Setting and approximation of the component functions}\label{app:vs-setting}

We give further details of the setting of Section~\ref{sec:theory-variable}, the analytic approximation used in Assumption~\ref{ass:sparse}\textup{(iii)}, and the per-support bounds $L_{{\rm vs},n}$ and $V_{{\rm vs},n}$. Throughout the appendix we write $D=1+s_{\max}$ and $\Lambda_n=\log(en)+\log(ep)$. The main text states its bounds in terms of $s_{\max}$, and $s_{\max}<D\leq2s_{\max}$. We first work with fixed $D$ and track its effect separately in Appendix~\ref{app:vs-smax}. The intercept is included in every candidate.

We fix constants $B,\rho,c_\Theta>0$ independent of $n$, $p$ and $S$. Each candidate $S$ has a working rectangle $\Theta_{0,S}\subset[-B,B]^{d_S}$ whose side lengths are at least $c_\Theta$. The rectangles are nested, in the sense that $\{\theta^{(S)}:\theta_S\in\Theta_{0,S}\}\subset\{\theta^{(S')}:\theta_{S'}\in\Theta_{0,S'}\}$ whenever $S\subset S'$. We write $\Theta_{\mathcal S}=\bigcup_{S\in\mathcal S}\{\theta^{(S)}:\theta_S\in\Theta_{0,S}\}$ for their union.

The observations lie in a compact set $\mathcal X_n\subset[-B,B]\times\mb R^p$, so the responses are bounded, as in binary and binomial regression. The conditional response densities $p_\theta(\cdot\mid z)$ for $\theta\in\Theta_{\mathcal S}$ all share a common support, on which the response measure $\nu_Y$ is finite. For $S\supseteq S_0$, we write $\theta_{0,S}=(\icpt_0,\beta_{0,S})$ for the generating parameter in the coordinates of candidate $S$. The generating parameter satisfies $\operatorname{dist}(\theta_{0,S},\partial\Theta_{0,S})\geq\rho/D$ for every $S\in\mathcal S$ with $S\supseteq S_0$, so it stays away from the boundary of every working rectangle that contains it. By Assumption~\ref{ass:sparse-proposal}(iii), this bound holds for every parameter in the support of the task distribution.

The bounds in Assumption~\ref{ass:sparse}\textup{(ii)} control the population second-moment matrix on sparse vectors, including differences between two candidate parameters, and permit correlated covariates. Lemma~\ref{lem:glm-kl} gives
$D_{\mathrm{KL}}(\mb P_{\theta_0}\|\mb P_\theta)\geq(c_Q/2)\|\theta-\theta_0\|_2^2$ for $\theta\in\Theta_{\mathcal S}$, where $c_Q=m_b\underline\lambda_Z/a$.
Thus omitting active variables incurs KL loss at least $c_Q\|\beta_{0,S_0\setminus S}\|_2^2/2$, even after refitting the intercept and retained slopes. In the setting of Section~\ref{sec:theory-variable}, bounded observations give uniform bounds on the centered log density and its parameter derivatives up to order three (see~\eqref{eq:sparse-envelopes} in Appendix~\ref{app:sparse-regularity}). These bounds control fluctuations and initialization uniformly over the supports, whose number grows with $p$. Under the curvature condition of Assumption~\ref{ass:sparse}\textup{(i)}, bounded covariates are also necessary for such bounds, because the second derivative of the log density with respect to the coefficient of $z_j$ has absolute value $a^{-1}b''\{\eta_\theta(z)\}z_j^2$.

The parameter network for $S$ uses the natural moment of the GLM. We write $z_S=(z_j)_{j\in S}$, $v_S(z)=(1,z_S^\top)^\top$, and $T_S(y,z)=y\,v_S(z)$. Under $\mb P_{S,\theta_S}$, the mean $\mu_S(\theta_S)$ of $T_S$ and the Fisher information $I_S(\theta_S)$ of candidate $S$ are
\[
\mu_S(\theta_S)=\mb E_{Q_Z}[b'\{v_S(Z)^\top\theta_S\}v_S(Z)],\quad
I_S(\theta_S)=a^{-1}\mb E_{Q_Z}[b''\{v_S(Z)^\top\theta_S\}v_S(Z)v_S(Z)^\top].
\]
The network for support $S$ approximates four maps: the moment inverse $\mu_S^{-1}$, the inverse information $I_S^{-1}$, and the response functions $b$ and $b'$. Its size is polylogarithmic in $n$ when these maps have holomorphic extensions with constants $(\mathcal R,\varrho,\mathcal B)$ in the sense of~\eqref{eq:extension-main} in Appendix~\ref{app:analytic-extension}. All rectangles below are fixed and do not depend on the data.

In the setting above and under Assumption~\ref{ass:sparse}\textup{(i)}--\textup{(ii)}, Lemma~\ref{lem:glm-natural-modules} gives holomorphic extensions with explicit constants for $I_S^{-1}$ on $\Theta_{0,S}$ and for $b$ and $b'$ on an interval that contains all linear predictors. For the moment inverse, Lemma~\ref{lem:glm-inverse-extension}\textup{(a)} shows that $\mu_S$ is injective on $\mb R^{d_S}$. By Lemma~\ref{lem:glm-natural-modules}, there is a constant $\rho_\mu>0$ such that the range of $\mu_S$ contains the set $K_{g,S}$ of~\eqref{eq:moment-inverse-domain} with radius $\rho_g=\rho/(2D)$ whenever $\rho\leq\rho_\mu$. Here $\rho_\mu$ depends only on the constants of the setting and of Assumption~\ref{ass:sparse}\textup{(i)}--\textup{(ii)}. We take $\rho\leq\rho_\mu$ from now on. Decreasing $\rho$ only weakens the margin required of the generating parameter. Thus $\mu_S^{-1}:K_{g,S}\to\mb R^{d_S}$ is well defined, and $K_{g,S}$ is the set on which the network evaluates it.

Lemma~\ref{lem:glm-inverse-extension} gives a sufficient condition for Assumption~\ref{ass:sparse}\textup{(iii)} in terms of the image of $\mu_S$. Fix $S\in\mathcal S$. Suppose there are a compact convex set $\Theta_S^+$ with $\Theta_{0,S}\subseteq\Theta_S^+\subseteq[-B_+,B_+]^{d_S}$ and a constant $m_+\in(0,m_b]$ such that $b''\{v_S(z)^\top\theta\}\geq m_+$ for every $\theta\in\Theta_S^+$ and $Q_Z$-almost every $z$. Then the lemma provides a constant $\rho_+>0$ with the following property. If a rectangle $\mathcal R_{g,S}=\prod_j[u_j,v_j]$ with $\bar h_S=\max_j(v_j-u_j)/2$ satisfies
\begin{equation}
 K_{g,S}\subseteq\mathcal R_{g,S}\subseteq
 \bigl\{m\in\mb R^{d_S}:\operatorname{dist}\{m,\mu_S(\Theta_S^+)\}\leq\rho_+/(2\sqrt D)\bigr\},
 \label{eq:moment-rectangle-condition}
\end{equation}
then $\mu_S^{-1}:K_{g,S}\to\mb R^{d_S}$ has a holomorphic extension with constants
\[
 \bigl(\mathcal R_{g,S},\ 1+\min\{1,\rho_+/(4D\bar h_S)\},\ B_++2\bigr).
\]
Here $\rho_+$ depends only on $m_+$ and the constants of the setting and of Assumption~\ref{ass:sparse}\textup{(i)}--\textup{(ii)}. When $\Theta_S^+=\Theta_{0,S}$ and $m_+=m_b$, we have $\rho_+=\rho_\mu$. In particular, take $\Theta_S^+=\Theta_{0,S}$, $m_+=m_b$ and $B_+=B$. If $\rho\leq\rho_\mu/6$ and
\begin{equation}
 \operatorname{diam}_2(\Theta_{0,S})\leq\frac{\rho_\mu}{4DM_b\overline\lambda_Z},
 \label{eq:small-working-set}
\end{equation}
then~\eqref{eq:moment-rectangle-condition} holds for the smallest rectangle $\mathcal R_{g,S}$ containing $K_{g,S}$, and the ellipse parameter is $2$. Consequently, when $\rho\leq\rho_\mu/6$ and every working rectangle satisfies~\eqref{eq:small-working-set}, Assumption~\ref{ass:sparse}\textup{(iii)} follows from Assumption~\ref{ass:sparse}\textup{(i)}--\textup{(ii)} with $\varrho_{g,n}$ and $\mathcal B_{g,n}$ independent of $n$ and $p$ (Remark~\ref{ex:glm-verified-modules}). For larger working rectangles, Assumption~\ref{ass:sparse}\textup{(iii)} is a restriction. The range $\mu_S(\mb R^{d_S})$ is a bounded open set, and $\mu_S^{-1}$ is unbounded near its boundary. When $d_S\geq2$, a rectangle that contains $K_{g,S}$ need not lie in this range because $\mu_S(\Theta_{0,S})$ is curved.

Let $\varrho_n$ be the smallest ellipse parameter and $\mathcal B_n$ the largest modulus bound among the extensions of the four maps. The resulting depth and parameter bounds $L_{{\rm vs},n},V_{{\rm vs},n}$ are given in~\eqref{eq:sparse-analytic-complexity}. The orders stated after Assumption~\ref{ass:sparse}\textup{(iii)} follow when $\varrho_n$ is bounded away from one and $\mathcal B_n$ grows at most polynomially in $n$.

\subsubsection{Posterior inclusion probabilities, EBIC and the task distribution}\label{app:sparse-criteria}

We quantify how accurately the EBIC-based estimates and inclusion probabilities approximate the population targets of meta training.

\begin{proposition}[Posterior inclusion probabilities and EBIC]\label{prop:vs-bridge}
In the setting of Section~\ref{sec:theory-variable}, suppose Assumptions~\ref{ass:sparse} and~\ref{ass:sparse-proposal} hold and $s_{\max}^4\Lambda_n^3\leq c_0n$ for a sufficiently small constant $c_0>0$. Let $q_j^{\rm E}$ be the EBIC inclusion probabilities of~\eqref{eq:ebic-weights}, clipped to $[\epsilon_n,1-\epsilon_n]$ with $\epsilon_n=(1+np)^{-1}$. Then, with the sets $\mathcal T$ and $\mathcal B$ of weak and boundary tasks of Appendix~\ref{app:vs-network},
\[
\begin{gathered}
\sum_{j=1}^p\mb E_{\mb Q_n}\bigl[\ind_{(\mathcal T\cup\mathcal B)^c}D_{\rm KL}(q_j^*\|q_j^{\rm E})\bigr]
\leq C\Bigl\{H_n^*+\frac{s_{\max}^2\Lambda_n^2}n+\frac1n\Bigr\}\leq C\Bigl\{H_n^*+s_{\max}\sqrt{\frac{\Lambda_n}n}\Bigr\},\\
\sum_{j=1}^p\mb E_{\mb Q_n}\bigl[\ind_{\mathcal T\cup\mathcal B}D_{\rm KL}(q_j^*\|q_j^{\rm E})\bigr]\leq Cs_{\max}^{7/2}\sqrt{\frac{\Lambda_n}n},\qquad
\mb Q_n(\mathcal T\cup\mathcal B)\leq Cs_{\max}^{3/2}\sqrt{\frac{\Lambda_n}n},
\end{gathered}
\]
where $c_0$ and $C$ do not depend on $n$, $p$ or $s_{\max}$. Moreover, $H_n^*\leq Cs_{\max}^{5/2}\sqrt{\Lambda_n/n}$.
\end{proposition}

We give the proof in Appendix~\ref{app:proof-vs-bridge}, together with the corresponding bound for the network of Theorem~\ref{thm:vs-approx}.

Three features of the argument explain the form of the bound. First, for a regular task, the posterior concentrates on the supports that contain the generating one. By the Lipschitz condition of Assumption~\ref{ass:sparse-proposal}\textup{(ii)}, the Laplace corrections of two such supports differ by a bounded amount per added variable (Lemma~\ref{lem:vs-laplace-diff}). The posterior support probabilities and the EBIC weights therefore differ by a tilt that is bounded on the supports with at most one added variable. This costs a constant multiple of the Bayes Hamming risk, with a constant free of $s_{\max}$ (Lemma~\ref{lem:added-tilt}). The supports with two or more added variables contribute $s_{\max}^2\Lambda_n^2/n$. Second, a task with weak signals is close to the smaller supports that omit them. Every slab covers a neighborhood of zero. When the task is not within $O(r_n)$ of the boundary of a slab, these supports are therefore slightly misspecified candidates whose pseudo-true parameters lie in the support of their own task distributions. Here $r_n=\sqrt{A_n/n}$ with $A_n=D\log(en)+s_{\max}\log(ep/s_{\max})$ (Appendix~\ref{app:vs-network}). The Laplace approximation then remains valid for them. This is why we need no minimum signal strength.

For the tasks within $O(r_n)$ of the boundary of a slab, the Laplace approximation fails on one side only. If such a task is not weak, the Laplace corrections are bounded above by a multiple of $s_{\max}$, and the lower bound for the generating support has an additional term whose expectation is of order $s_{\max}$. For the tasks that are weak or near the boundary, but not both, we use only these bounds of order $s_{\max}$ on the Laplace corrections. The tasks that are both weak and near the boundary have probability at most $Cs_{\max}^2r_n^2$, and for them we use the bound $Cs_{\max}\Lambda_n$ on the sum of the divergences, which holds for every dataset. We lose another factor $s_{\max}$ when we pass from the supports to the inclusion probabilities (Lemma~\ref{lem:sparse-contraction}). Third, these tasks have probability at most $Cs_{\max}^{3/2}\sqrt{\Lambda_n/n}$, because every coefficient has a bounded density. This gives the last term. The bound on $H_n^*$ follows from the Hamming loss of the constructed network, since no selection rule has a smaller expected Hamming loss than the Bayes rule. This loss is at most $Cs_{\max}\sqrt{\Lambda_n/n}$ in expectation for tasks that are not weak, and at most $5s_{\max}$ for the weak tasks.

The next proposition is the counterpart of Proposition~\ref{prop:vs-bridge} for the coefficients. It compares the posterior mean $\theta^*$ with the EBIC-weighted MLE $\theta^{\rm E}=\sum_{S\in\mathcal S}Q_S^{\rm E}\,\widehat\theta^{(S)}$ of~\eqref{eq:ebic-targets}. Here $\widehat\theta^{(S)}$ is the zero-padded extension of $\widehat\theta_S$, and $Q^{\rm E}$ denotes the unclipped EBIC weights~\eqref{eq:ebic-weights}. Recall that $A_n=D\log(en)+s_{\max}\log(ep/s_{\max})\leq2s_{\max}\Lambda_n$ (Appendix~\ref{app:vs-network}).

\begin{proposition}[Posterior mean and EBIC-weighted MLE]\label{prop:vs-bridge-est}
In the setting of Section~\ref{sec:theory-variable}, suppose Assumptions~\ref{ass:sparse} and~\ref{ass:sparse-proposal} hold and $s_{\max}^4\Lambda_n^3\leq c_0n$ for a sufficiently small constant $c_0>0$. In this statement, $c_0$, $C_1$ and $C$ do not depend on $n$, $p$ or $s_{\max}$.

\textup{(a)} For every $S_0\in\mathcal S$, every $\theta_0$ in the support of $\pi_{S_0}$ and every dataset,
\[
\|\theta^{\rm E}-\theta_0\|_2^2\leq\frac{4a}{m_b\underline\lambda_Z\,n}\bigl\{nZ_n+\operatorname{pen}(S_0)+\log N_{\mathcal S}\bigr\},
\]
with $Z_n$ as in Lemma~\ref{lem:vs-soft-oracle}. Consequently $\mb E_0[\|\theta^{\rm E}-\theta_0\|_2^2]\leq C_1A_n/n\leq2C_1s_{\max}\Lambda_n/n$.

\textup{(b)} Under the task distribution, $\mb E_{\mb Q_n}[\|\theta^*-\theta^{\rm E}\|_2^2]\leq\mb E_{\mb Q_n}[\|\theta^{\rm E}-\theta\|_2^2]\leq Cs_{\max}/n$.
\end{proposition}

We give the proof in Appendix~\ref{app:proof-vs-bridge-est}. Part~(a) is a deterministic oracle inequality. The squared norm is convex, so the error of the weighted average is at most the weighted average of the errors of the individual MLEs. Up to the fluctuation term $Z_n$, a support with a large error has a small likelihood and therefore a small EBIC weight. The proof does not need accurate MLEs on the supports that omit active variables. It also needs no condition on the signal strength. The dimension enters only through $A_n$, so the bound is proportional to $s_{\max}$. The network output $\widetilde\theta$ is the same weighted average with the EBIC weights and the MLEs replaced by their network counterparts. Lemma~\ref{lem:vs-soft-oracle} gives the same inequality for $\widetilde\theta$. In part~(b), the first inequality is the Pythagorean identity of Lemma~\ref{lem:ce-selection}(c).

For the second inequality, we need a bound on the expected weight of the supports that strictly contain the generating one, multiplied by their squared score statistics. For the supports with one added variable, we obtain this bound from Lemma~\ref{lem:cramer-prefactor}. Its tail bound has the factor $1/x$ of the Gaussian tail, which Bernstein's inequality does not have.

\begin{remark}[EBIC and the support prior]\label{rem:ebic}
With the constrained MLE $\widehat\theta_S$, the EBIC with $\gamma=1$ in~\citet[Section~2]{chen2008extended} is $\mathrm{EBIC}_S=-2\ell_{n,S}(\widehat\theta_S)+d_S\log n+2\log\binom p{|S|}$, so that $Q_S^{\rm E}\propto\exp(-\mathrm{EBIC}_S/2)$ in~\eqref{eq:ebic-weights}. A term of the log likelihood that is common to all supports, such as $c(y,a)$ in~\eqref{eq:glm-model}, cancels from $Q^{\rm E}$, and Appendix~\ref{app:sparse-regularity} subtracts it in the computational log likelihood. Under Assumption~\ref{ass:sparse-proposal}(i), $\log\{w(S)m_S\}=-\mathrm{EBIC}_S/2+\delta_S-\log s_{\max}$ with the Laplace error $\delta_S$ of Appendix~\ref{app:vs-network}, so the combinatorial term $2\log\binom p{|S|}$ of EBIC equals $-2\log w(S)$ up to an additive constant and is not an additional assumption. Our proofs use the size-uniform distribution through the identity
\[
\binom{p-s_0}{b}\exp\{-\operatorname{pen}(S)+\operatorname{pen}(S_0)\}=\binom{s_0+b}{b}n^{-b/2}
\]
for a support $S\supsetneq S_0$ with $b$ added variables, which cancels the number of such supports and is used to control the contributions of larger supports in the task-averaged bounds.
\end{remark}

Assumption~\ref{ass:sparse-proposal} differs from Assumption~\ref{ass:proposal}. It requires densities that are bounded and bounded away from zero on their supports and Lipschitz on the logarithmic scale, such as uniform slabs. We assume no further smoothness. This is because in Section~\ref{sec:theory-variable} we do not use the comparison of the posterior mean with the MLE through the Fisher information of the task distribution.

\subsubsection{EBIC approximation and the RRDS construction}\label{app:rrds-variable}

In the notation of Section~\ref{sec:theory-variable}, the joint distribution for support $S$ is $\mb P_{S,\theta_S}=\mb P_{\theta^{(S)}}$. The candidates have positive densities with respect to the common finite measure $\nu_Y\otimes Q_Z$. We write $\ell_{S,\theta_S}$ for the log density, $s_{S,\theta_S}=\nabla_{\theta_S}\ell_{S,\theta_S}$ for the score, and $I_S(\theta_S)$ for the Fisher information. We denote the compact observation support by $\mathcal K_n\subset[-B,B]^{1+p}$. It lies inside a compact rectangle $\mathcal X_n$ used for the approximation networks. The constrained MLEs exist by compactness and continuity. We draw each task by sampling $S\sim w$, then $\theta_S\sim\pi_S$, and then $X^n\sim\mb P_{S,\theta_S}^{(n)}$.

\medskip
\noindent{\bf Comparison with the population targets.}
Under the conditions of Theorem~\ref{thm:vs-approx}, with $H_n^*$ and $\Lambda_n$ as defined in the main text, Propositions~\ref{prop:vs-bridge} and~\ref{prop:vs-bridge-est} give
\begin{equation}
\begin{aligned}
\sum_{j=1}^p\mb E_{\mb Q_n}[D_{\rm KL}(q_j^*\|q_j^{\rm E})]
&\leq C\Bigl\{H_n^*+s_{\max}\sqrt{\frac{\Lambda_n}n}
+s_{\max}^{7/2}\sqrt{\frac{\Lambda_n}n}\Bigr\},\\
\mb E_{\mb Q_n}[\|\theta^*-\theta^{\rm E}\|_2^2]
&\leq\mb E_{\mb Q_n}[\|\theta^{\rm E}-\theta\|_2^2]
\leq\frac{Cs_{\max}}n.
\end{aligned}
\label{eq:vs-bridge}
\end{equation}
In the first line, we clip $q_j^{\rm E}$ to $[\epsilon_n,1-\epsilon_n]$, where $\epsilon_n=(1+np)^{-1}$. The coefficient estimate uses the original EBIC weights. Clipping gives uniform control of the logarithmic loss and adds at most $O(n^{-1})$ after summing over coordinates. In the first bound, the first term reflects the Bayes Hamming risk, the second bounds supports with two or more added variables, and the third accounts for weak signals and slab boundaries. The third term is absent when we restrict the expectation to tasks with all active coefficients away from zero and all generating coordinates away from the slab endpoints, as specified in the proof of Proposition~\ref{prop:vs-bridge}. We write $J_n=H_n^*+s_{\max}\sqrt{\Lambda_n/n}+s_{\max}^{7/2}\sqrt{\Lambda_n/n}$ for the sum in braces in the first line. By the bound on $H_n^*$ in Proposition~\ref{prop:vs-bridge}, $J_n\leq Ch_n$, with $h_n=s_{\max}^{7/2}\sqrt{\Lambda_n/n}$ as in the main text.

\medskip
\noindent{\bf State and initialization.}
The network uses $R=2$ and a state of dimension $d_z=2\sum_{S\in\mathcal S}d_S+2p+1$,
\begin{equation}
z=\bigl(\{\vartheta_S^0\}_{S\in\mathcal S},\;\{\theta_S\}_{S\in\mathcal S},\;b_0,\;\{(a_j,b_j)\}_{j\in[p]}\bigr).
\label{eq:vs-state}
\end{equation}
The state stores the initial moment estimates $\vartheta_S^0$, the current within-support estimates $\theta_S$, an intercept estimate $b_0$, and an inclusion log-odds $a_j$ and conditional active-coefficient estimate $b_j$ for each covariate $j\in[p]$. The initialization encoder evaluates $Y_i$ and the products $Y_iZ_{ij}$ once per observation using a shared multiplication subnetwork. From their averages, the decoder extracts $\bar T_{n,S}$, applies $g_{{\rm init},S}$ followed by exact projection onto $\Theta_{0,S}$, and sets $\vartheta_S^0=\Pi_{\Theta_{0,S}}\{g_{{\rm init},S}(\bar T_{n,S})\}$ and $\theta_S=\vartheta_S^0$. It initializes the other entries to zero.

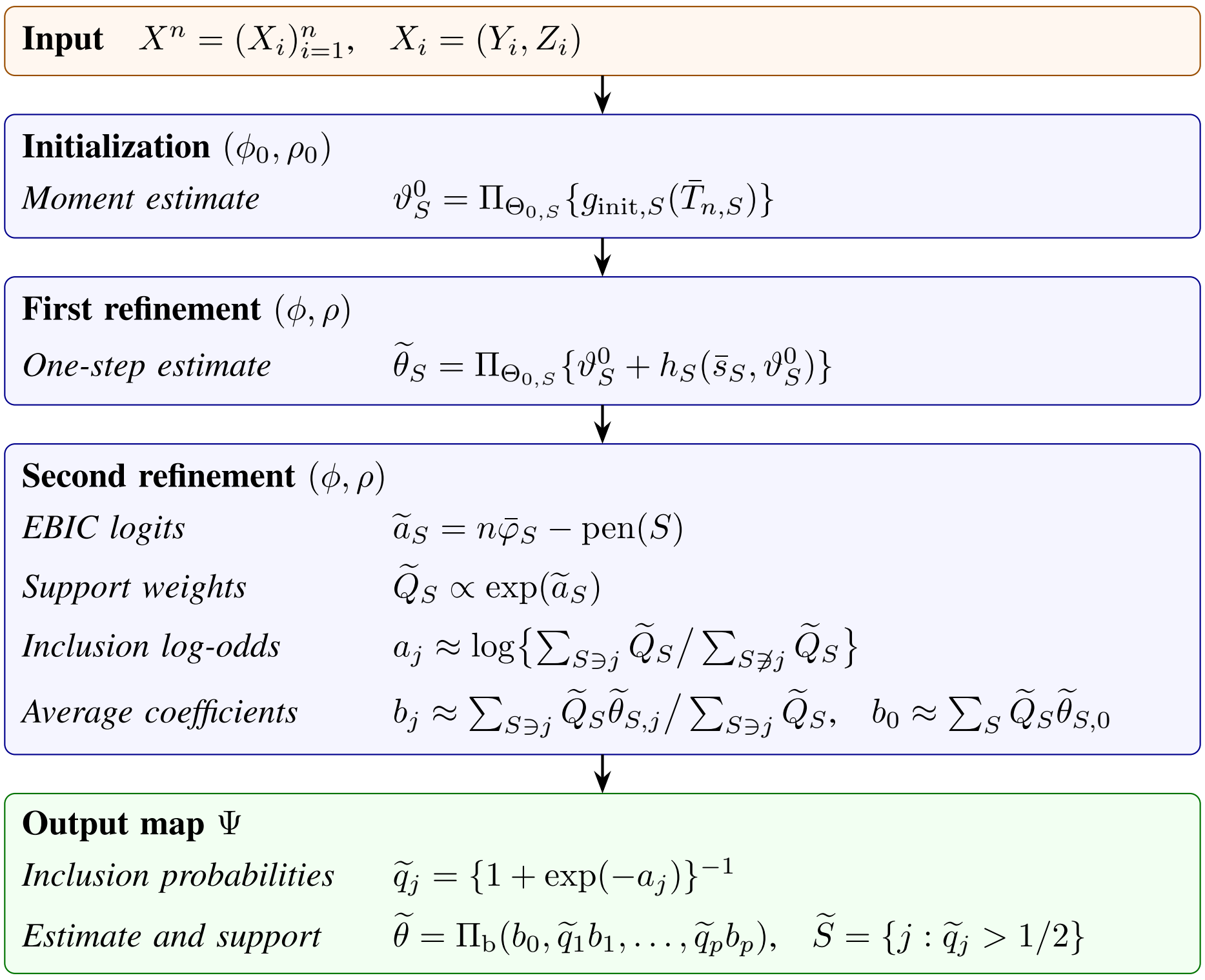
\begin{figure}[t]
\centering
\begin{tikzpicture}[
    font=\footnotesize,
    >=Stealth,
    blk/.style={rounded corners=3pt, inner sep=5pt, outer sep=0pt, align=left, text width=11.9cm},
    bluebox/.style={blk, draw=blue!55!black, fill=blue!4},
    greenbox/.style={blk, draw=green!45!black, fill=green!5},
    orangebox/.style={blk, draw=orange!60!black, fill=orange!6},
    flow/.style={->, thick}
]
\newcommand{\vsrow}[2]{\parbox[t]{3.6cm}{\raggedright\itshape #1}\hspace{0.2cm}\parbox[t]{8.0cm}{\raggedright #2}}

\node[orangebox] (sample) {
    \textbf{Input}\quad $X^n=(X_i)_{i=1}^n$,\quad $X_i=(Y_i,Z_i)$
};

\node[bluebox, below=4mm of sample] (init) {
    \textbf{Initialization $(\phi_0,\rho_0)$}\\[3pt]
    \vsrow{Moment estimate}{$\vartheta_S^0=\Pi_{\Theta_{0,S}}\{g_{{\rm init},S}(\bar T_{n,S})\}$}
};

\node[bluebox, below=4mm of init] (ref1) {
    \textbf{First refinement $(\phi,\rho)$}\\[3pt]
    \vsrow{One-step estimate}{$\widetilde\theta_S=\Pi_{\Theta_{0,S}}\{\vartheta_S^0+h_S(\bar s_S,\vartheta_S^0)\}$}
};

\node[bluebox, below=4mm of ref1] (ref2) {
    \textbf{Second refinement $(\phi,\rho)$}\\[3pt]
    \vsrow{EBIC logits}{$\widetilde a_S=n\bar\varphi_S-\operatorname{pen}(S)$}\\[4pt]
    \vsrow{Support weights}{$\widetilde Q_S\propto\exp(\widetilde a_S)$}\\[4pt]
    \vsrow{Inclusion log-odds}{$a_j\approx\log\bigl\{\sum_{S\ni j}\widetilde Q_S\big/\sum_{S\not\ni j}\widetilde Q_S\bigr\}$}\\[4pt]
    \vsrow{Average coefficients}{$b_j\approx\sum_{S\ni j}\widetilde Q_S\widetilde\theta_{S,j}\big/\sum_{S\ni j}\widetilde Q_S$,\quad $b_0\approx\sum_{S}\widetilde Q_S\widetilde\theta_{S,0}$}
};

\node[greenbox, below=4mm of ref2] (out) {
    \textbf{Output map $\Psi$}\\[3pt]
    \vsrow{Inclusion probabilities}{$\widetilde q_j=\{1+\exp(-a_j)\}^{-1}$}\\[4pt]
    \vsrow{Estimate and support}{$\widetilde\theta=\Pi_{\rm b}(b_0,\widetilde q_1b_1,\ldots,\widetilde q_pb_p)$,\quad $\widetilde S=\{j:\widetilde q_j>1/2\}$}
};

\draw[flow] (sample.south) -- (init.north);
\draw[flow] (init.south) -- (ref1.north);
\draw[flow] (ref1.south) -- (ref2.north);
\draw[flow] (ref2.south) -- (out.north);
\end{tikzpicture}
\caption{Two-refinement ($R=2$) RRDS network for sparse GLM variable selection. Quantities indexed by $S$ are computed for every support $S\in\mathcal S$, and those indexed by $j$ for every covariate $j\in[p]$. Both refinements apply the same blocks $(\phi,\rho)$.}
\label{fig:vs-rrds}
\end{figure}

\medskip
\noindent{\bf Two shared refinements.}
The encoder has output dimension $d_u=\sum_{S\in\mathcal S}d_S+|\mathcal S|$. For each support, it evaluates an approximate score at the fixed initial estimate $\vartheta_S^0$ and a likelihood subnetwork $\varphi_S$ at the current $\theta_S$. Here $\varphi_S$ approximates $\ell_{S,\theta_S}(y,z)-c(y,a)$ with uniform error $\{6n\log(en)\}^{-1}$ on $\mathcal K_n\times\Theta_{0,S}$. The omitted term $c(y,a)$ is common to all supports.

From the averaged scores and likelihoods, the decoder computes the one-step estimate $\widetilde\theta_S=\Pi_{\Theta_{0,S}}\{\vartheta_S^0+h_S(\bar s_S,\vartheta_S^0)\}$, where $h_S(u,\vartheta)$ approximates $I_S(\vartheta)^{-1}u$. It also forms the support logits and weights
\begin{equation}
\widetilde a_S=n\bar\varphi_S-\operatorname{pen}(S),\qquad
\widetilde Q_S=\frac{\exp(\widetilde a_S)}{\sum_{S'\in\mathcal S}\exp(\widetilde a_{S'})},
\qquad S\in\mathcal S.
\label{eq:vs-network-logits}
\end{equation}
ReLU subnetworks then approximate the summaries
\begin{equation}
\begin{aligned}
a_j&\approx\log\frac{\sum_{S\ni j}\widetilde Q_S}{\sum_{S\not\ni j}\widetilde Q_S},\qquad j\in[p],\\
b_0&\approx\sum_{S\in\mathcal S}\widetilde Q_S\theta_{S,0},\qquad
b_j\approx\frac{\sum_{S\ni j}\widetilde Q_S\theta_{S,j}}{\sum_{S\ni j}\widetilde Q_S},\qquad j\in[p],
\end{aligned}
\label{eq:vs-bma-targets}
\end{equation}
within $(np)^{-2}$ for every dataset (Lemma~\ref{lem:vs-rrds}). The decoder leaves $\vartheta_S^0$ unchanged and returns differences from the current state, so its updates have the residual form~\eqref{eq:rrds-update}. In the first pass, the network evaluates the likelihoods at $\vartheta_S^0$ and updates the parameter state to $\widetilde\theta_S$. In the second pass, it evaluates the likelihoods and summaries at the one-step estimates. Both passes use the same blocks.

\medskip
\noindent{\bf Output map.}
The fixed map $\Psi$ clips the final $a_j$ to $[-\log(np),\log(np)]$ and $b_k$ ($k=0,\ldots,p$) to $[-B,B]$, and returns
\[
\widetilde q_j=\{1+\exp(-a_j)\}^{-1},\qquad
\widetilde\theta=\Pi_{\rm b}(b_0,\widetilde q_1b_1,\ldots,\widetilde q_pb_p),\qquad
\widetilde S=\{j:\widetilde q_j>1/2\}.
\]
Here $\Pi_{\rm b}$ projects onto the ball of radius $2B\sqrt{s_{\max}}$ in $\mathbb R^{1+p}$, which contains $\Theta_{\mathcal S}$. This bounds the squared loss by a multiple of $s_{\max}$ throughout the network class and leaves the constructed network unchanged (Lemma~\ref{lem:vs-heads}). Since the within-support estimates are zero outside their supports, $\widetilde q_jb_j$ approximates $\sum_S\widetilde Q_S\theta_{S,j}$. The coefficient output therefore approximates the weighted average in~\eqref{eq:ebic-targets}.

\medskip
\noindent{\bf Complexity and interpretation of the rates.}
The construction in Theorem~\ref{thm:vs-approx} satisfies
\begin{equation}
\begin{aligned}
L\leq C\{L_{{\rm vs},n}+s_{\max}\Lambda_n^3\},\quad
W\leq|\mathcal S|\{V_{{\rm vs},n}+C(p+s_{\max}^2)\Lambda_n^3\}.
\end{aligned}
\label{eq:vs-network-complexity}
\end{equation}
The additional terms arise from the inclusion log-odds and weighted coefficient summaries. Appendix~\ref{app:vs-training} describes the binary-tree calculations that avoid unstable division by small support weights. The factor $|\mathcal S|\leq(ep/s_{\max})^{s_{\max}}$ comes from enumerating candidate supports. For fixed $s_{\max}$, this factor is polynomial in $p$. However, it can still be large and need not be polynomial in $n$ when $p$ grows faster than a power of $n$.

When $s_{\max}$ is bounded and $p$ grows at least polynomially in $n$, the rate $s_{\max}\Lambda_n/n$ in Theorem~\ref{thm:vs-approx}(b) has the same order as the fixed-design sparse estimation benchmarks $s_0\log(ep/s_0)/n$ discussed for linear regression and GLMs~\citep{raskutti2011minimax,abramovich2016model}. This comparison concerns the order of the rate. Coefficients near the detection threshold can be shrunk toward zero, so the pointwise bound cannot generally be replaced by a uniform $n^{-1}$ rate. The bounded slab densities make such tasks sufficiently uncommon for the integrated $s_{\max}/n$ bound to hold. Appendix~\ref{app:vs-training} explains the sample-size condition and the explicit dependence on $s_{\max}$.

For the trained network, if $\Delta_{\rm vs}=O(s_{\max}/n)$ and $\lambda\asymp s_{\max}p/(nh_n)$, both components attain the rates $s_{\max}/n$ and $h_n$ of Theorem~\ref{thm:vs-approx}(a). For bounded $s_{\max}$, this choice is $\lambda\asymp p/\sqrt{n\Lambda_n}$. Under the uniform analytic conditions stated after Assumption~\ref{ass:sparse}, we can take the radius $\Gamma$ to be polynomial in $np$. Then
$M\asymp(s_{\max}+\lambda\Lambda_n)nWs_{\max}\Lambda_n^4$
tasks suffice for $\Delta_{\rm vs}=O(s_{\max}/n)$ when $\varepsilon_{\rm opt}=O(n^{-1})$, $\delta=n^{-1}$, and $\lambda$ is at most polynomial in $np$. Appendices~\ref{app:vs-training} and~\ref{app:lambda-calibration} give the derivation. These are bounds for the specified network class and task distribution. The construction does not require the empirically learned updates to follow the EBIC calculation.

\subsubsection{Selection and estimation rates and meta training}\label{app:vs-training}

Theorem~\ref{thm:vs-approx}(b) gives estimation and prediction risk bounds at every fixed generating parameter in the prior support. The sharper estimation risk bound $s_{\max}/n$ in part~(a) follows by averaging over the task distribution. In the proof, the bounded slab densities control the probability of tasks with coefficients close to zero. For fixed $s_{\max}$, the condition $s_{\max}^4\Lambda_n^3\leq c_0n$ allows the ambient dimension $p$ to grow as fast as $\exp(cn^{1/3})$ for a suitable constant $c>0$. The cubic dependence on $\log p$ arises from controlling the likelihood expansions uniformly over the candidate supports.

Under the analyticity conditions discussed above, the network size required for each candidate support grows only polylogarithmically in $n$. The total number of parameters also depends on the number of candidate supports $|\mathcal S|$, since the construction considers each support separately. Further comparisons of these rates and their dependence on $s_{\max}$ are given below.

The estimation risk bound in Theorem~\ref{thm:vs-approx}(b) follows from a deterministic oracle inequality for the output $\widetilde\theta=(b_0,\widetilde q_1b_1,\ldots,\widetilde q_pb_p)$ (Lemma~\ref{lem:vs-soft-oracle}). Up to the approximation errors, the output is the average of the one-step estimates of the supports with the weights $\widetilde Q_S$. The squared norm is convex, so the error of the output is at most the average of their errors. The likelihood controls the error of each support, uniformly over all supports. We need no accuracy for the estimates on the supports that omit active variables.

\medskip
\noindent{\bf The sample-size condition.} Theorem~\ref{thm:vs-approx} assumes $s_{\max}^4\Lambda_n^3\leq c_0n$. For bounded $s_{\max}$ and large $n$, this reads $(\log p)^3\leq cn$. Consistent estimation does not need the cube, since $\log p=o(n)$ suffices for the bound of Proposition~\ref{prop:vs-bridge-est}\textup{(a)}. The cube arises as follows. The EBIC weights depend on the maximized log likelihoods through an exponential, and we compare them with the posterior probabilities through the Laplace approximation. The quadratic approximation of the log likelihood must therefore hold with a bounded error. Its remainder is the cubic term $\frac1{6a}\sum_{i=1}^nb'''(\cdot)\,\{v_S(Z_i)^\top\Delta\}^3$,
where $\Delta$ is the deviation of the parameter from the generating one and $b'''$ is evaluated at intermediate points. To compare the $p$ candidate covariates simultaneously, we must control deviations of size $\|\Delta\|_2\asymp\sqrt{(\log p)/n}$. At this size the cubic term is of order $n\|\Delta\|_2^3\asymp\{(\log p)^3/n\}^{1/2}$.

With the dependence on $s_{\max}$, the deviations have size $\|\Delta\|_2\leq C\sqrt{A_n/n}$ with $A_n\leq2s_{\max}\Lambda_n$, $|v_S(Z_i)^\top\Delta|\leq C\sqrt D\,\|\Delta\|_2$ and $\sum_{i=1}^n\{v_S(Z_i)^\top\Delta\}^2\leq Cn\|\Delta\|_2^2$ (Lemma~\ref{lem:vs-hessian}), so the cubic term is at most $C(DA_n^3/n)^{1/2}\leq C(s_{\max}^4\Lambda_n^3/n)^{1/2}$. This gives the factor $s_{\max}^4$.

The tail bound for the score statistics in Lemma~\ref{lem:cramer-prefactor} has the correction factor $\exp(Kx^3/\sqrt n)$, whose exponent at the level $x^2\asymp\log p$ is of the same order. This is the range in which the cubic correction in this moderate-deviation bound remains bounded. The sample-size condition keeps both cubic remainders bounded. For comparison, in the Gaussian linear model the log likelihood is exactly quadratic, so the likelihood expansion itself does not impose this restriction. \citet{chen2012extended} assume $\log p=O(n^{\kappa})$ with $\kappa<1/3$ for the consistency of EBIC in GLMs. The condition $s_{\max}^4\Lambda_n^3\leq c_0n$ is a sufficient condition in our proof.

\medskip
\noindent{\bf Dependence on the maximal support size.} The dependence on $s_{\max}$ differs between the estimation and selection bounds. In particular, the third term of $J_n$ determines the rate $h_n$ in part~\textup{(a)}. The support size is uniform on $[s_{\max}]$, so the task distribution does not adapt to the sparsity $s_0$ of a particular task. Thus $s_{\max}$ rather than $s_0$ governs parts~\textup{(a)} and~\textup{(b)}. The logarithm of the number of candidate supports and of the covering numbers of their working sets is at most a constant multiple of $A_n\leq2s_{\max}\{\log(en)+\log(ep)\}$. This gives the factor in part~\textup{(b)} and in Proposition~\ref{prop:vs-bridge-est}\textup{(a)}.

In part~\textup{(a)}, the constant in front of $H_n^*$ does not depend on $s_{\max}$. The reason is that, under Assumption~\ref{ass:sparse-proposal}\textup{(ii)}, the Laplace corrections of two supports differ by a bounded amount per variable in which the supports differ.

For the tasks that are weak or boundary tasks, but not both, the bounds on the Laplace corrections used in the proof are of order $s_{\max}$, in expectation for the boundary tasks, and we lose another factor of order $s_{\max}$ when we pass from the supports to the inclusion probabilities. These tasks have probability at most $Cs_{\max}^{3/2}\sqrt{\{\log(en)+\log(ep)\}/n}$, which gives the third term of $J_n$. The tasks that are both weak and boundary tasks have the smaller probability $Cs_{\max}^2r_n^2$, and their contribution is at most of the same order. The bound on $H_n^*$ has the factor $s_{\max}^{5/2}$, the product of the order $s_{\max}^{3/2}$ of this probability and the largest Hamming loss $5s_{\max}$ of the constructed network.

\medskip
\noindent{\bf Network size.} The additional terms in the bounds on $L$ and $W$ of Theorem~\ref{thm:vs-approx} come from the subnetworks that compute the inclusion logits and the average coefficients in~\eqref{eq:vs-bma-targets}. Dividing by $\sum_{S\ni j}\widetilde Q_S$ is unstable when this weight is small, so we compute $b_j$ by merging the supports containing $j$ two at a time along a balanced binary tree. Each node replaces two pairs $(v_1,\widetilde a_1)$ and $(v_2,\widetilde a_2)$ of a value and a logit by
\[
\bigl(v_2+\sigma_{\rm log}(\widetilde a_1-\widetilde a_2)(v_1-v_2),\ \log(e^{\widetilde a_1}+e^{\widetilde a_2})\bigr),\qquad\sigma_{\rm log}(t)=(1+e^{-t})^{-1},
\]
namely the weighted average of the two values and the logit of their total weight. The root returns $b_j$ and $\log\sum_{S\ni j}e^{\widetilde a_S}$, and $a_j$ is the difference between this logit and its analogue over the supports not containing $j$. The nodes of these trees approximate the logistic and softplus functions within a power of $(np)^{-1}$, which costs $O(\Lambda_n^3)$ parameters per node (Lemma~\ref{lem:vs-rrds}).

In Theorem~\ref{thm:vs-gen}, the network is trained end to end with the joint loss, and all blocks of the class $\mathcal F(L,W)$ are trainable. The output map $\Psi$ clips the inclusion scores and the coefficient values and then projects the coefficient vector onto the ball of radius $2B\sqrt{s_{\max}}$, both during training and at inference. We choose this radius so that the projection leaves the constructed network unchanged for every dataset. The constructed network therefore belongs to the class, and we can compare the empirical risk of the trained network with its empirical risk.

In the training error $\Delta_{\rm vs}$ in~\eqref{eq:sparse-complexity-training}, the factor $s_{\max}+\lambda\Lambda_n$ bounds the loss over the network class. The logarithmic covering scale $L\log(eW\Gamma)$ is the depth of the network times the logarithm of its number of parameters and parameter radius. The radius $\Gamma$ only needs to contain the constructed network and bound the fixed coefficients of the architecture. We assume no growth condition on it. By Lemma~\ref{lem:vs-rrds}(d), $L\leq C(L_{{\rm vs},n}+s_{\max}\Lambda_n^3)$, and we can take $\Gamma$ of order $\sqrt W(n+\Lambda_n+P_{{\rm vs},n}+K_{{\rm dom},n})$. Under the analytic conditions stated after Assumption~\ref{ass:sparse}\textup{(iii)}, $\varrho_n$ is bounded away from one and $\mathcal B_n$ is polynomial in $n$. Then $L_{{\rm vs},n}=O\{\log^2(en)\}$, $V_{{\rm vs},n}=O\{\log^{D+1}(en)\}$ and $P_{{\rm vs},n}$ is polynomial in $n$. In addition, $K_{{\rm dom},n}$ is bounded because every normalized rectangle has side lengths bounded below ($\mathcal R_{g,S}$ contains $K_{g,S}$, and the working rectangles have side lengths at least $c_\Theta$). Hence $W=O[|\mathcal S|\{\log^{D+1}(en)+(p+s_{\max}^2)\Lambda_n^3\}]$.

When $\Gamma$ is polynomial in $np$ and in the number of parameters, we obtain $\log(eW\Gamma)\leq Cs_{\max}\Lambda_n$ and therefore $L\log(eW\Gamma)\leq C(L_{{\rm vs},n}+s_{\max}\Lambda_n^3)s_{\max}\Lambda_n\leq Cs_{\max}^2\Lambda_n^4$.

With $\delta=n^{-1}$ and $\varepsilon_{\rm opt}=O(n^{-1})$, a number of tasks $M$ of order $n(s_{\max}+\lambda\Lambda_n)Ws_{\max}\Lambda_n^4$ then gives $\Delta_{\rm vs}=O(s_{\max}/n)$, provided $\lambda$ is at most polynomial in $np$. The reason is that the bound $\Delta_{\rm vs}$ contains $\log(eM)$, and $\log(eM)\leq Cs_{\max}^2\Lambda_n^4$ requires $\log(1+\lambda\Lambda_n)\leq Cs_{\max}^2\Lambda_n^4$. For general $\lambda$, we replace the factor $s_{\max}^2\Lambda_n^4$ by $s_{\max}^2\Lambda_n^4+\log(1+\lambda\Lambda_n)$.

The guarantees of Theorem~\ref{thm:vs-gen} are averages over the task distribution because a small excess risk is compatible with a large error on a set of tasks of small probability, so the excess risk does not give a bound at a fixed parameter value.

\subsubsection{The loss weight in end-to-end variable selection}
\label{app:lambda-calibration}

For the joint loss of Section~\ref{sec:meta-formulation}, the population targets are $\theta^*$ and $q_j^*$ for every $\lambda>0$, including $\lambda=\lambda_n$, and the excess risk is~\eqref{eq:vs-excess}. Lemma~\ref{lem:ce-selection}(b) and~(c) convert a bound on the excess risk $\mathcal R(F)-\mathcal R^*$ into bounds for the two components,
\[
\mb E[\|\theta_F-\theta\|_2^2]\leq\mb E[\|\theta^*-\theta\|_2^2]+\{\mathcal R(F)-\mathcal R^*\},\qquad
\mb E[|\widehat S_F\mathbin{\triangle}S|]\leq5H_n^*+\frac{8p}\lambda\,\{\mathcal R(F)-\mathcal R^*\},
\]
and these comparisons also hold conditionally on independent training data. Thus the parameter risk increases by at most $\mathcal R(F)-\mathcal R^*$, whereas the selection error depends on $p\{\mathcal R(F)-\mathcal R^*\}/\lambda$. In Theorem~\ref{thm:vs-gen}, the excess risk of the trained network is at most $C\{s_{\max}n^{-1}+(\lambda/p)J_n+\Delta_{\rm vs}\}$, with $J_n$ as in Appendix~\ref{app:rrds-variable}. The theorem holds for every $\lambda>0$, and the value used in training enters the bounds as displayed. Since $J_n\leq Ch_n$, the choice $\lambda\asymp s_{\max}p/(nh_n)$ discussed after Theorem~\ref{thm:vs-gen} gives the rates $s_{\max}/n$ and $h_n$ for the two components, and it does not depend on the unknown $H_n^*$. The same holds with $J_n$ in place of $h_n$, which gives the sharper order $J_n$ for the Hamming loss.

More precisely, suppose $\Delta_{\rm vs}=O(s_{\max}/n)$. The bound on the parameter risk is $O(s_{\max}/n)$ if and only if $\lambda=O\{s_{\max}p/(nJ_n)\}$. The bound on the Hamming loss is $O(J_n)$ if and only if $s_{\max}p/(nJ_n)=O(\lambda)$. Thus this order of $\lambda$ balances the two sufficient risk bounds at the rates of the constructed network. At this value the two components of the excess risk~\eqref{eq:vs-excess} of the constructed network are both $O(s_{\max}/n)$, by Theorem~\ref{thm:vs-approx}(a).

When $s_{\max}$ is bounded and $J_n$ is of the order $\sqrt{\Lambda_n/n}$, this is the choice $\lambda\asymp p/\sqrt{n\Lambda_n}$. Note that the sum over the $p$ covariates in $J_n$ is not $p$ times a per-coordinate error. At most $s_{\max}$ covariates are active, and the inactive ones contribute only through the supports with added variables. A larger weight lets the selection component dominate the excess risk and loosens the estimation bound. A smaller weight controls the selection error only through $s_{\max}p/(\lambda n)$.

More generally, the parameter risk tends to zero if $(\lambda/p)J_n+\Delta_{\rm vs}\to0$, and the Hamming loss if $J_n\to0$, $s_{\max}p/(\lambda n)\to0$ and $(p/\lambda)\Delta_{\rm vs}\to0$. These are sufficient conditions obtained from the excess risk of the joint loss.

\subsection{Improved approximation and training bounds under analyticity}\label{app:analytic-training-results}\label{app:analytic}

We now explain the additional regularity conditions that permit smaller networks and give the resulting approximation and meta-generalization bounds for parameter estimation and model selection. The same approximation condition is used for the sparse GLMs in Appendix~\ref{app:vs-setting}. 

\subsubsection{Analyticity and neural network approximation}\label{app:analytic-extension}

The improvement over the finite-smoothness approximation bounds comes from a classical fact about polynomial approximation: a function that extends analytically to a complex neighborhood of a real interval can be approximated by polynomials with an error that decreases geometrically in the degree. A ReLU network can evaluate such a polynomial using approximate multiplication (Appendix~\ref{app:relu}). Consequently, the number of network parameters needed to achieve an accuracy $\varepsilon$ can be bounded by a power of $\log(1/\varepsilon)$, rather than a power of $1/\varepsilon$. This distinction matters here because the approximation error must decrease as the sample size grows to retain the statistical properties of the target inference rule.

To quantify the size of the complex neighborhood, we use Bernstein ellipses. Their foci are the endpoints of the normalized real interval, and the parameter $\varrho$ controls how far the ellipse extends into the complex plane. The following definition also allows functions defined only on a subset of a rectangle, as occurs for the inverse of the moment map. The affine transformation maps the rectangle to $[-1,1]^m$, so that the same approximation result can be applied to functions on different domains.

For example, in one dimension the endpoints of the real interval remain fixed after normalization, while the semi-axes of the ellipse are $(\varrho+\varrho^{-1})/2$ and $(\varrho-\varrho^{-1})/2$. As $\varrho$ decreases to one, the ellipse shrinks to the interval. Requiring a bounded extension to a larger ellipse is a stronger condition, but allows faster decay of the polynomial approximation error. In several dimensions we use a product of these ellipses, corresponding to a separate polynomial expansion in each coordinate.

For $\varrho>1$, let $E_\varrho\subset\mathbb C$ be the closed region bounded by the Bernstein ellipse $\{(w+w^{-1})/2:|w|=\varrho\}$ with foci $\pm1$, and write $E_\varrho^m$ for its $m$-fold product. For a rectangle $\mathcal R=\prod_{j=1}^m[u_j,v_j]$, with $u_j<v_j$, define $[A_{\mathcal R}(z)]_j=(2z_j-u_j-v_j)/(v_j-u_j)$. A map $f:K\to\mathbb R^q$, with $K\subset\mathbb R^m$, admits a \emph{holomorphic extension with constants $(\mathcal R,\varrho,\mathcal B)$} if $K\subseteq\mathcal R$ and there are an open set $O\subset\mathbb C^m$ and a holomorphic map $F:O\to\mathbb C^q$ such that
\begin{equation}
\begin{gathered}
E_\varrho^m\subset O,\qquad F\{A_{\mathcal R}(x)\}=f(x)\quad(x\in K),\\
F([-1,1]^m)\subseteq\mathbb R^q,\qquad
\max_{1\leq j\leq q}\sup_{z\in E_\varrho^m}|F_j(z)|\leq\mathcal B.
\end{gathered}
\label{eq:extension-main}
\end{equation}

In this condition, $\varrho$ controls the rate of polynomial approximation, while $\mathcal B$ bounds the size of the extension. For a fixed dimension and fixed extension constants, Lemma~\ref{lem:analytic-approx} gives a network size polynomial in $\log(1/\varepsilon)$. In the sparse GLM analysis, we also need to track the dependence on the dimension and the extension constants, since these affect the cost of constructing a network for each candidate support. Real analyticity on a neighborhood of a compact rectangle suffices to obtain a complex neighborhood of that rectangle, but real analyticity on an irregular domain does not by itself guarantee an extension to a rectangle containing that domain. This is why we impose a separate condition on the inverse moment map.

\subsubsection{Analyticity of the log density and the moment inverse}

We use the analytic extension condition in~\eqref{eq:extension-main}. We denote the restriction of the extension to the real rectangle $\mathcal R$ by $f^{\mathcal R}=F\circ A_{\mathcal R}|_{\mathcal R}$.

In the setting of Section~\ref{sec:theory-estimation}, we fix a compact set $\mathcal K$ with $\mu(\Theta_0)\subset\operatorname{int}(\mathcal K)$ and $\mathcal K\subset\mu(U)$. The one-step construction uses the moment inverse $g_{\rm MoM}=\mu^{-1}:\mathcal K\to\mb R^d$, the score $s:\mathcal X\times\Theta_0\to\mb R^d$ and the inverse Fisher information $I^{-1}:\Theta_0\to\mb R^{d\times d}$. Model selection also uses the log density $\ell:\mathcal X\times\Theta_0\to\mb R$ of each candidate.

The next lemma shows that $s$, $I^{-1}$ and $\ell$ satisfy the extension condition whenever the log density is analytic. We say that the log density is analytic on the pair $(\mathcal R_X,\mathcal R_\Theta)$ if $\mathcal R_X\subset\mb R^{d_X}$ and $\mathcal R_\Theta\subset\mb R^d$ are compact rectangles with nonempty interiors and there are an open set $O_\ell\subset\mb R^{d_X}\times\mb R^d$ and a real analytic function $\bar\ell:O_\ell\to\mb R$ such that
\begin{equation}
\begin{gathered}
 \mathcal X\subseteq\mathcal R_X,\qquad \Theta_0\subseteq\mathcal R_\Theta\subset U,\qquad
 \mathcal R_X\times\mathcal R_\Theta\subset O_\ell,\\
 \bar\ell(x,\theta)=\ell_\theta(x)\quad(x\in\mathcal X,\ \theta\in\mathcal R_\Theta).
\end{gathered}
\label{eq:analytic-log-density}
\end{equation}
When $\Theta_0$ is a rectangle, we may take $\mathcal R_\Theta=\Theta_0$.

\begin{lemma}[Score and inverse information under an analytic log density]\label{lem:analytic-density}
Suppose Assumption~\ref{ass:regular} holds and the log density is analytic on $(\mathcal R_X,\mathcal R_\Theta)$ in the sense of~\eqref{eq:analytic-log-density}. Then there are constants $\varrho_\ell>1$ and $\mathcal B_\ell<\infty$, depending only on $\bar\ell$, the rectangles $\mathcal R_X$ and $\mathcal R_\Theta$, and the model $\{\mb P_\theta:\theta\in\mathcal R_\Theta\}$, such that the log density $(x,\theta)\mapsto\ell_\theta(x)$ and the score $(x,\theta)\mapsto s_\theta(x)$ on $\mathcal X\times\Theta_0$ satisfy~\eqref{eq:extension-main} with constants $(\mathcal R_X\times\mathcal R_\Theta,\varrho_\ell,\mathcal B_\ell)$, and the inverse Fisher information $\theta\mapsto I(\theta)^{-1}$ on $\Theta_0$ satisfies~\eqref{eq:extension-main} with constants $(\mathcal R_\Theta,\varrho_\ell,\mathcal B_\ell)$.
\end{lemma}

The proof is given in Appendix~\ref{app:proof-analytic}. By Lemma~\ref{lem:analytic-density}, the analytic results below assume the extension condition only for the moment inverse. In these results, the polytope $P$ of Appendix~\ref{app:proof-approx} is taken with $\Theta_0\subseteq P\subseteq\mathcal R_\Theta$, for example $P=\mathcal R_\Theta$, so that the score and the inverse information are approximated on $P$. As explained after Lemma~\ref{lem:relu-moment}, this condition refers to the moment function computed by the block $\phi_0$ of the network.

\subsubsection{Consequences for parameter estimation and model selection}

We apply the preceding lemma to the component functions in the one-step construction. The gain in approximation complexity also reduces the number of training tasks needed to control the excess risk, although the required statistical accuracy remains unchanged.

\begin{corollary}[Analytic approximation and meta training for parameter estimation]
\label{cor:analytic-est-gen}
Suppose Assumption~\ref{ass:regular} holds, the log density is analytic on $(\mathcal R_X,\mathcal R_\Theta)$ in the sense of~\eqref{eq:analytic-log-density}, and the moment inverse $g_{\rm MoM}:\mathcal K\to\mb R^d$ satisfies~\eqref{eq:extension-main} with constants $(\mathcal R_g,\varrho_g,\mathcal B_g)$. Put $m=d_X+d$.

\textup{(a)} The approximation and asymptotic conclusions of Theorem~\ref{thm:est-approx} hold with $L\leq C\log^2(e/\varepsilon)$ and $W\leq C\log^{m+1}(e/\varepsilon)$, with component maps chosen independently of $n$. The constant $C$ depends only on $d,d_X,r,\ell,T,\Theta_0$ and the fixed extension rectangles, ellipse parameters and amplitude bounds; $C_{\Theta_1}$ depends in addition on the compact $\Theta_1\Subset\operatorname{int}(\Theta_0)$.

\textup{(b)} Suppose also Assumption~\ref{ass:proposal} holds. Use part~(a) at $\varepsilon=n^{-1}$, with constructed vector $\omega_{0,n}$, and for sufficiently large fixed $c_\Gamma$ take $\|\omega_{0,n}\|_2\leq\Gamma\leq c_\Gamma n^{c_\Gamma}$ and $\mathcal F=\{F_\omega:\|\omega\|_2\leq\Gamma\}$. Thus $L_n\leq C\log^2(en)$ and $W\leq C\log^{m+1}(en)$. Under the training and independent-new-task setup of Theorem~\ref{thm:est-gen}, for $n\geq3$, $M\geq1$ and $\delta\in(0,1)$, on a training event $\mathcal G$ of probability at least $1-\delta$,
\begin{equation}
\begin{aligned}
 \mb E[\|\widehat F(X^n)-\widehat\theta_{0,n}\|_2^2\mid\mathcal D_M]
 &\leq C\frac{W\{\log^3(en)+\log(eM)\}+\log(3/\delta)}{M}\\
 &\quad+\frac{C\{1+\mathcal J(\pi)\}}{n^2}+4\varepsilon_{\rm opt}.
\end{aligned}
\label{eq:analytic-est-gen-bound}
\end{equation}
Here $C$ depends only on the quantities in part~(a), $c_\Gamma$ and $\operatorname{dist}(\operatorname{supp}\pi,\partial\Theta_0)$; $\mathcal J(\pi)$ enters through the displayed term. For fixed $\pi$, the two conditional convergence conclusions of Theorem~\ref{thm:est-gen} hold with its rate condition replaced by $W\{\log^3(en)+\log(eM)\}/M=o(n^{-1})$ and the same $\varepsilon_{\rm opt}=o(n^{-1})$. A sufficient task complexity bound is $M=\lceil n\log^{m+5}(en)\rceil$.
\end{corollary}

For model selection, we assume that for every candidate $k\in[K]$, the log density $\ell_k$ is analytic on a pair $(\mathcal R_X,\mathcal R_{\Theta,k})$ in the sense of~\eqref{eq:analytic-log-density}, with $\Theta_{0,k}\subseteq\mathcal R_{\Theta,k}\subset U_k$. We also assume that for each $k\in[K]$, there is a compact set $\mathcal K_k$ with $\mu_k(\Theta_{0,k})\subset\operatorname{int}(\mathcal K_k)$ and $\mathcal K_k\subset\mu_k(U_k)$ such that the moment inverse $\mu_k^{-1}:\mathcal K_k\to\mb R^{d_k}$ satisfies~\eqref{eq:extension-main}. The dimensions, extension constants and rectangles are fixed. By Lemma~\ref{lem:analytic-density}, the score, the inverse Fisher information and the log density of every candidate then satisfy~\eqref{eq:extension-main}. Put $m_{\rm ms}=d_X+d_{\max}$.

\begin{corollary}[Analytic approximation and meta training for model selection]
\label{cor:analytic-ms}
Under the preceding analytic conditions and Assumption~\ref{ass:models}, the following hold.

\textup{(a)} All conclusions of Theorem~\ref{thm:ms-approx} and Proposition~\ref{prop:ms-misspec} hold for an RRDS network of the same form, with $R=2$, $d_z=2\sum_kd_k+K$, $d_u=\sum_kd_k+K$, $L\leq C\log^2(en)$ and $W\leq CK\log^{m_{\rm ms}+1}(en)$.

\textup{(b)} For the corresponding class with $\|\omega\|_2\leq\Gamma=O(n^C)$, the conclusions of Theorem~\ref{thm:ms-gen} hold with $(\log n)^2$ replaced by $\log^3(en)$ in the definition of $\Delta_{\rm ms}$. With $M=\lceil\sqrt n\,\log^{m_{\rm ms}+5}(en)\rceil$, $\delta=n^{-1}$ and $\varepsilon_{\rm opt}\leq n^{-1/2}$, the training error is at most $C/\sqrt n$.

\noindent The constants depend in addition on the extension rectangles, ellipse parameters and amplitude bounds.
\end{corollary}

\section{Further Implementation Details and Additional Numerical Results}\label{app:implementation}

This appendix gives the implementation details and additional results cited in Section~\ref{sec:numerical}. The evaluation criteria and resampling constructions are defined in Appendices~\ref{app:meta-training} and~\ref{app:meta-extensions}.

\subsection{Common protocol}\label{app:impl-protocol}

The encoders and decoders have two hidden ReLU layers and linear outputs, with widths chosen to match $W$. We use Adam with a cosine learning rate schedule from $10^{-3}$ to $10^{-5}$ over 200 epochs. The effective minibatch size is 256 for parameter estimation and model selection, and 8 for variable selection. For estimation, we use microbatches of 16 with gradient accumulation. We implement the networks in PyTorch and train them on NVIDIA GPUs. Classical fits use Python numerical and statistical libraries.

For $T$ evaluation tasks, we compute the Monte Carlo standard error of an average as the sample standard deviation across tasks divided by $\sqrt T$. For paired comparisons, we apply the same formula to within-task differences. For paired task values $a_m$ and $b_m$, the ratio $r=\bar a/\bar b$ has delta method standard error $\operatorname{SD}(a_m-rb_m)/(\sqrt T\,\bar b)$. Error bars and paired comparison intervals use $1.96$ times the corresponding standard error. These calculations condition on the trained network, as in Appendix~\ref{app:meta-training}.

\subsection{Parameter estimation}\label{app:impl-estimation}

The experiments of Section~\ref{sec:sim-estimation} use linear and logistic regression with $d=5$ coefficients and no intercept. In both models, $Z_i\sim\mathcal N(0,I_5)$, and the task distribution $\pi$ draws each coordinate of $\theta$ independently from ${\rm Uniform}(-2,2)$. The linear model has Gaussian errors with variance $20/3=\mathbb E_\pi[\|\theta\|_2^2]$, so its signal to noise ratio is one on average over $\pi$. The logistic model has success probability $\{1+\exp(-Z_i^\top\theta)\}^{-1}$. We vary $n$ from 250 to 8000 and $W$ over $\{10^4,2\times10^4,5\times10^4,10^5\}$. We train each network on $M=50{,}000$ tasks and evaluate it on $M_{\mathrm{test}}=5{,}000$ new tasks. The ratios of $\bar e_{\widehat F}$ and $\bar e_{\widehat F,\mathrm{MLE}}$ to $\bar e_{\mathrm{MLE}}$ measure the estimation risk and the distance to the MLE relative to the statistical error.

\subsubsection{Regularization using the task distribution}
\label{app:est-regularization}
We examine how information in the task distribution contributes to the small-sample results in Section~\ref{sec:numerical}. We use the same linear and logistic models, task distribution $\pi$, and 5,000 test tasks at each $n\in\{250,500,1000\}$. RRDS uses $W=10^5$, $R=12$, and shared refinements.

LASSO and ridge estimates minimize the mean negative log likelihood plus $\lambda\|\theta\|_1$ and $\lambda\|\theta\|_2^2/2$, respectively. The Gaussian likelihood uses the noise variance $20/3$, for which the signal-to-noise ratio equals one. We fix the intercept at zero in both models. For each model, sample size, and penalty, we select one $\lambda$ from $\{10^{-5},10^{-4.5},\ldots,10\}$ by minimizing the average squared parameter error over 5,000 independently generated validation tasks from $\pi$. We then fix this value for all test tasks. We also include penalties selected separately within each task by cross-validation, denoted CV. The validation criterion based on the task distribution matches the coefficient loss used to train RRDS.

Table~\ref{tab:est-regularization} shows that tuning on tasks drawn from the task distribution lowers the mean error of both penalized estimators relative to their CV versions in all six settings. In logistic regression, this tuning narrows their gap to RRDS, which retains the smallest mean error. In linear regression, both estimators tuned using $\pi$ have slightly smaller mean errors than RRDS. In this model, the mean error of RRDS exceeds that of the MLE at $n=500$ and $n=1000$ ($0.0701$ against $0.0679$, and $0.0360$ against $0.0337$). In logistic regression, RRDS also has a shorter upper error tail. At $n=250$, the 99th percentile of the squared coefficient error is $1.57$ for the MLE, $0.65$ for RRDS, $0.93$ for LASSO tuned using $\pi$, and $0.80$ for ridge tuned using $\pi$. These comparisons suggest that information from the task distribution helps stabilize coefficient estimates at small sample sizes.

\begin{table}[!htb]
\centering
\caption{Mean squared coefficient errors, with Monte Carlo standard errors in parentheses.}
\label{tab:est-regularization}
\setlength{\tabcolsep}{3pt}
\renewcommand{\arraystretch}{1.3}
\small
\begin{tabular}{lrcccccc}
\hline
Model & $n$ & MLE & RRDS & LASSO CV & Ridge CV & LASSO $\pi$ & Ridge $\pi$\\
\hline
Linear & 250 & \shortstack{0.1352\\(0.0012)} & \shortstack{0.1344\\(0.0012)} & \shortstack{0.1383\\(0.0013)} & \shortstack{0.1365\\(0.0013)} & \shortstack{0.1343\\(0.0012)} & \shortstack{0.1322\\(0.0012)}\\
Linear & 500 & \shortstack{0.0679\\(0.0006)} & \shortstack{0.0701\\(0.0006)} & \shortstack{0.0693\\(0.0006)} & \shortstack{0.0683\\(0.0006)} & \shortstack{0.0677\\(0.0006)} & \shortstack{0.0673\\(0.0006)}\\
Linear & 1000 & \shortstack{0.0337\\(0.0003)} & \shortstack{0.0360\\(0.0003)} & \shortstack{0.0341\\(0.0003)} & \shortstack{0.0338\\(0.0003)} & \shortstack{0.0337\\(0.0003)} & \shortstack{0.0336\\(0.0003)}\\
Logistic & 250 & \shortstack{0.2835\\(0.0048)} & \shortstack{0.1918\\(0.0019)} & \shortstack{0.2535\\(0.0032)} & \shortstack{0.2603\\(0.0033)} & \shortstack{0.2257\\(0.0027)} & \shortstack{0.2113\\(0.0023)}\\
Logistic & 500 & \shortstack{0.1305\\(0.0019)} & \shortstack{0.1031\\(0.0011)} & \shortstack{0.1244\\(0.0016)} & \shortstack{0.1269\\(0.0016)} & \shortstack{0.1194\\(0.0016)} & \shortstack{0.1137\\(0.0014)}\\
Logistic & 1000 & \shortstack{0.0592\\(0.0008)} & \shortstack{0.0543\\(0.0006)} & \shortstack{0.0578\\(0.0007)} & \shortstack{0.0584\\(0.0007)} & \shortstack{0.0560\\(0.0007)} & \shortstack{0.0553\\(0.0006)}\\
\hline
\end{tabular}
\end{table}

\subsubsection{Number of training tasks}\label{app:est-M}

All estimation networks in Section~\ref{sec:sim-estimation} are trained on $M=50{,}000$ tasks for 200 epochs. To assess how much of the increase in $\bar e_{\widehat F,\mathrm{MLE}}/\bar e_{\mathrm{MLE}}$ at large $n$ comes from the number of training tasks and epochs, we retrain RRDS in logistic regression with $n=8000$, $W=10^5$, $R=12$ and shared refinements on $M\in\{12{,}500,\,25{,}000,\,50{,}000,\,100{,}000,\,200{,}000\}$ tasks. All other settings follow Section~\ref{sec:sim-estimation}, including 200 epochs and the cosine learning rate schedule. We evaluate every network on the same 5,000 test tasks with the same MLE.

\begin{figure}[!htb]
\centering
\includegraphics[width=\textwidth]{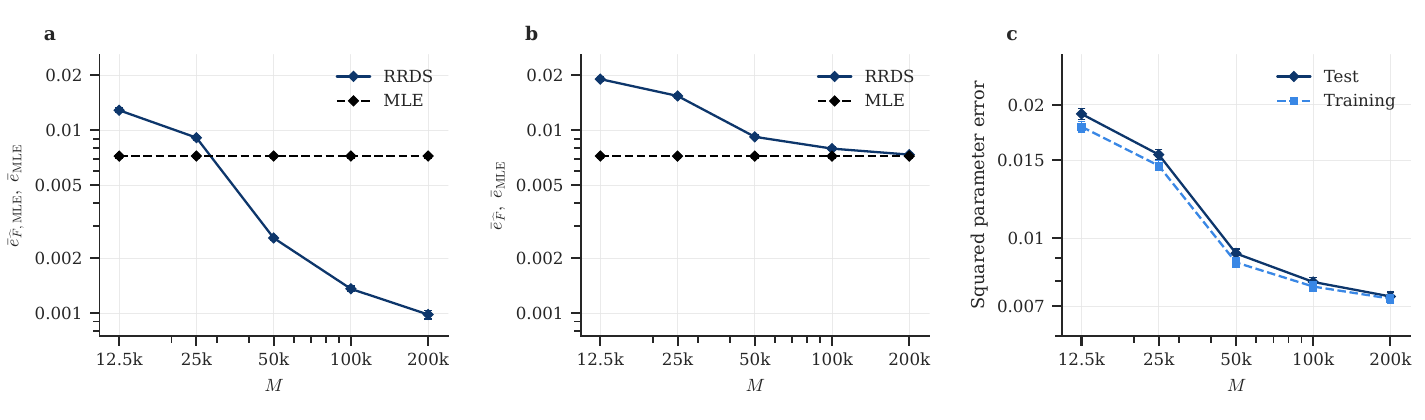}
\caption{Distance to the MLE (a), estimated risk (b), and training and test risks (c) against the number of training tasks.}
\label{fig:app-m-sweep}
\end{figure}

Panels (a) and (b) of Figure~\ref{fig:app-m-sweep} show that both the distance to the MLE and the risk with respect to $\theta$ keep decreasing as $M$ grows.

The ratio $\bar e_{\widehat F,\mathrm{MLE}}/\bar e_{\mathrm{MLE}}$ falls from 0.36 at $M=50{,}000$ to 0.14 at $M=200{,}000$, and the risk of RRDS approaches that of the MLE, with $\bar e_{\widehat F}/\bar e_{\mathrm{MLE}}=1.02$. Panel (c) shows that the training and test risks decrease together and that their gap shrinks from $1.2\times10^{-3}$ at $M=12{,}500$ to $8\times10^{-5}$ at $M=200{,}000$. Thus, larger task banks show no sign of overfitting over this range. With the number of epochs fixed, the number of gradient steps grows in proportion to $M$, so these reductions reflect both the larger bank and the longer optimization. When we instead hold the number of steps at its value for $M=50{,}000$, the ratio decreases only from 0.36 to 0.34 at $M=200{,}000$. Both the number of tasks and the number of epochs therefore matter, in line with the conditions on $W$, $M$ and the optimization error in Theorem~\ref{thm:est-gen}.

\subsubsection{Number of refinements}\label{app:est-R}

\begin{figure}[!htb]
\centering
\includegraphics[width=\textwidth]{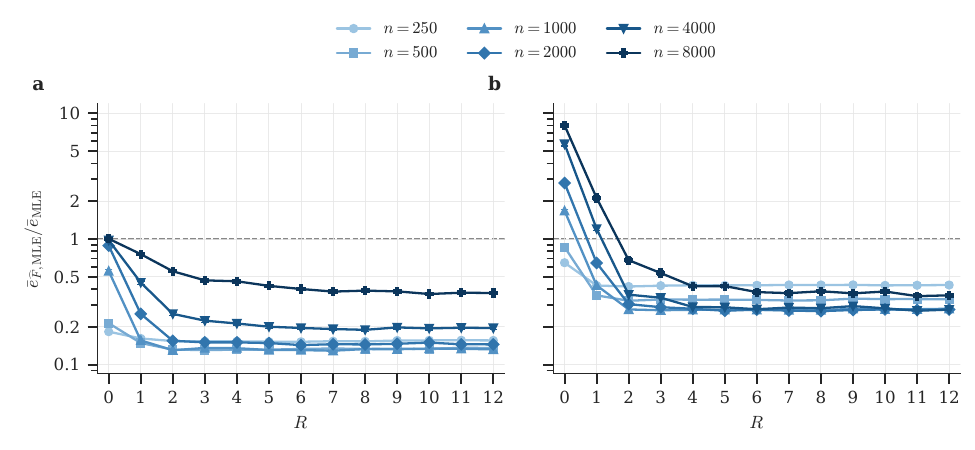}
\caption{Relative distance to the MLE against the number of shared refinements in linear (a) and logistic (b) regression.}
\label{fig:est-R}
\end{figure}

We vary $R$ from 0 to 12 with $W=10^5$ in the linear and logistic designs of Section~\ref{sec:sim-estimation}. Figure~\ref{fig:est-R} shows the ratios $\bar e_{\widehat F,\mathrm{MLE}}/\bar e_{\mathrm{MLE}}$ for shared refinements. In logistic regression at $n=8000$, for example, the ratio falls from 2.12 after one refinement to 0.54 after three.

\begin{figure}[!htb]
\centering
\includegraphics[width=\textwidth]{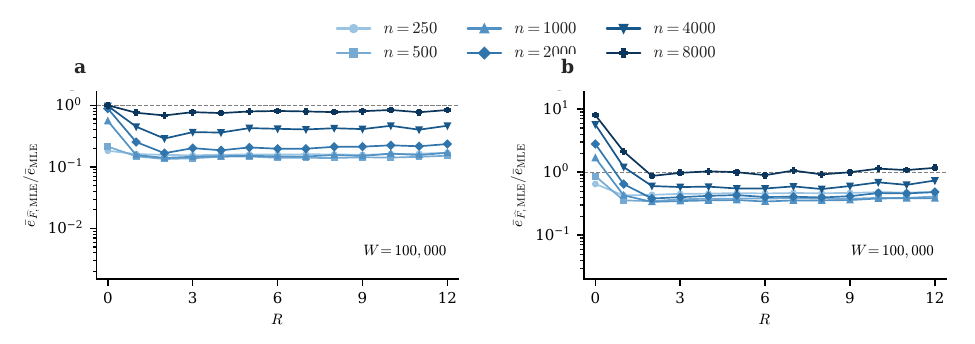}
\caption{Relative distance to the MLE against the number of unshared refinements in linear (a) and logistic (b) regression.}
\label{fig:app-est-unshared}
\end{figure}

Figure~\ref{fig:app-est-unshared} gives the corresponding curves for unshared refinements. At larger sample sizes, their plateaus are higher than those for shared refinements. Since $W$ counts distinct parameters, sharing allows wider blocks at the same number of parameters. The other values of $W$ give the same pattern for both architectures, with higher plateaus at smaller $W$.

\subsection{Bootstrap}\label{app:impl-bootstrap}

\begin{figure}[!htb]
\centering
\includegraphics[width=\textwidth]{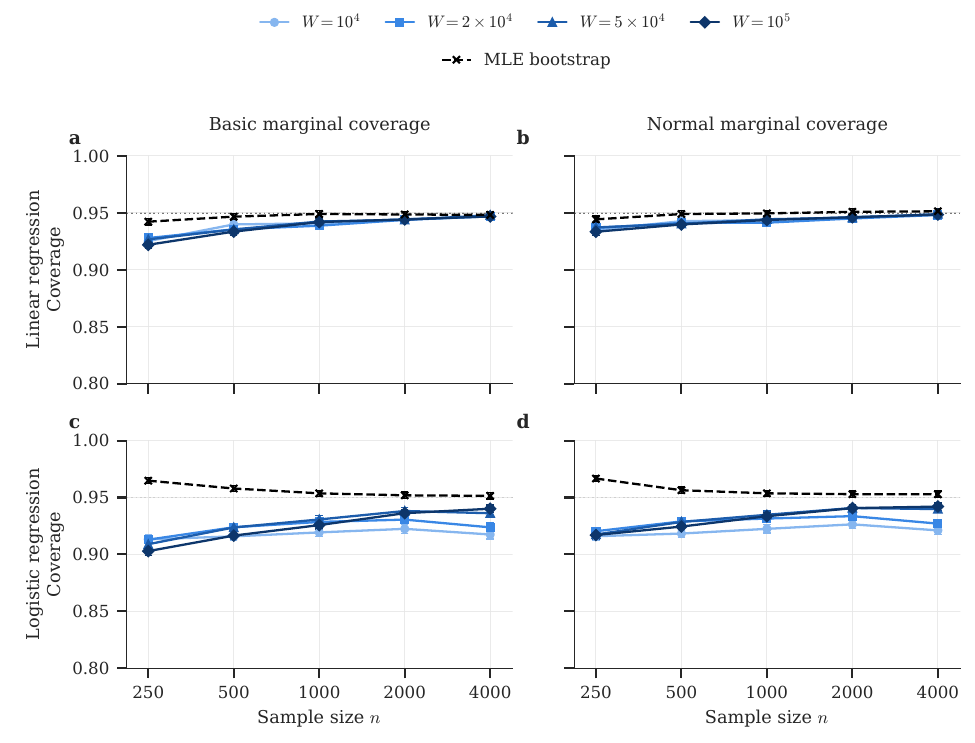}
\caption{Average marginal coverage of basic (left) and normal (right) bootstrap intervals in linear regression (a, b) and logistic regression (c, d).}
\label{fig:boot}
\end{figure}

For each regression model and each $n\in\{250,500,1000,2000,4000\}$, we evaluate the bootstrap on $M_{\mathrm{test}}=5{,}000$ tasks, keeping the networks of Section~\ref{sec:sim-estimation} fixed. For every task, we draw $B=1000$ resamples of the $n$ observed pairs $(Z_i,Y_i)$ and apply the network to each resample~\citep[Chapter~5]{davison1997bootstrap}. We use the normal and basic intervals and the sup-$t$ rectangle defined in Appendix~\ref{app:meta-extensions}. The normal and sup-$t$ constructions use the bootstrap standard deviation $s_j$, computed once from the resamples of the original dataset and held fixed across bootstrap draws. The basic interval uses the empirical quantiles of these same bootstrap estimates. The reference replaces RRDS by the MLE in each construction, using least squares in the linear model. Figure~\ref{fig:boot} reports the marginal coverage of the basic and normal intervals.

In linear regression, the coverage probabilities of the basic and normal intervals based on RRDS approach the nominal level as $n$ increases and remain close to those based on the MLE. In logistic regression, the MLE bootstrap intervals are conservative at small sample sizes, where instability of the MLE leads to wider intervals. Their coverage approaches the nominal level as $n$ increases. The RRDS intervals have lower coverage than the corresponding MLE intervals at all sample sizes considered. Their coverage continues to improve with $n$ for the larger networks, whereas for $W=10^4$ and $2\times10^4$ it initially improves but eventually levels off or decreases.

For $n\geq2000$, increasing the network size improves coverage and reduces the relative discrepancy between RRDS and the MLE. This is consistent with Corollary~\ref{cor:bootstrap}(b), which relates bootstrap consistency to the accuracy of the trained estimator and the stability of its network components. We next compare interval widths, simultaneous coverage, and the variability of the bootstrap estimates.

\begin{figure}[!htb]
\centering
\includegraphics[width=\textwidth]{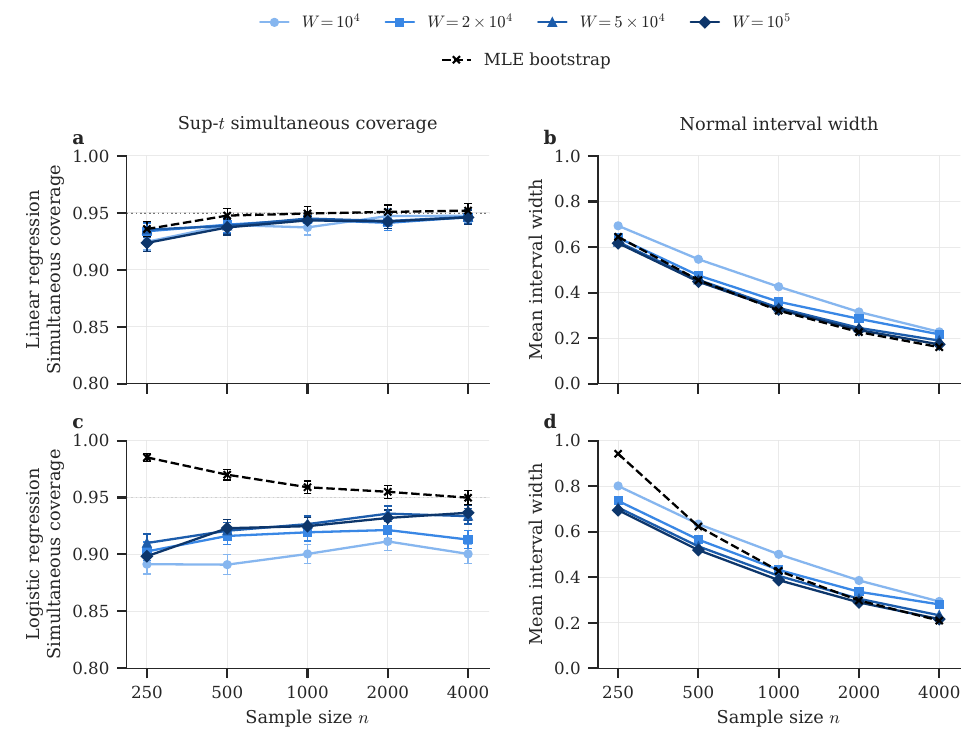}
\caption{Simultaneous coverage of sup-$t$ rectangles (left) and mean width of normal intervals (right) in linear (a, b) and logistic (c, d) regression.}
\label{fig:app-boot}
\end{figure}

The simultaneous coverage in Figure~\ref{fig:app-boot} follows the marginal coverage patterns in Figure~\ref{fig:boot}. Interval widths decrease with $n$ for every method. At small $n$, RRDS intervals are narrower than MLE bootstrap intervals, and the instability of the logistic MLE leads to wider intervals and conservative coverage. At large $n$, the RRDS intervals are wider on the $\sqrt n$ scale yet cover less often. At $n=4000$, the logistic RRDS normal intervals are wider than the corresponding MLE intervals and still have lower coverage.

For logistic RRDS with $W=10^5$, we compare the root mean squared bootstrap standard deviation with the empirical standard deviation of estimation errors across tasks. Write $s_{mj}$ for $s_j$ on task $m$ and $e_{mj}=\widehat\theta_{mj}-\theta_{mj}$. The mean coordinate ratio
\[
\frac{1}{d}\sum_{j=  1}^d \frac{\{M_{\mathrm{test}}^{-1}\sum_{m=  1}^{M_{\mathrm{test}}}s_{mj}^2\}^{1/2}} {\operatorname{SD}_{m}(e_{mj})}
\]
ranges from 0.929 to 0.965 across the five sample sizes. Replacing $s_{mj}$ by the corresponding empirical standard deviation across tasks gives normal intervals with average marginal coverage between 0.9462 and 0.9488. The denominator pools over parameters drawn from $\pi$ and therefore includes variation in conditional bias across tasks.

\subsection{Divide and conquer}\label{app:impl-dc}

We use the networks trained for logistic regression in Section~\ref{sec:sim-estimation} and take block sizes $n_0\in\{250,500,1000,2000\}$. At each $(n_0,J)$, we use 100 evaluation tasks with common data across methods and values of $W$.

For divide-and-conquer estimation, we divide a sample of size $n=Jn_0$ into $J$ disjoint blocks, apply the logistic regression network trained at sample size $n_0$ to each block, and average the resulting estimates. As a benchmark, we use the average of the MLEs computed separately on the $J$ blocks~\citep{zhang2013communication,rosenblatt2016optimality}. For both estimators, we report the ratio $\bar e_{\mathrm{DC},\mathrm{MLE}}/\bar e_{\mathrm{MLE}}$ of their mean squared distances from the full-sample MLE to the MSE of that MLE (Appendix~\ref{app:meta-extensions}). This ratio generally increases with $J$ for both methods. For the larger networks, it remains below one when $J$ is small. It exceeds one for all the methods at $J=n_0$, that is, $n=n_0^2$.

These trends are consistent with the condition $n=o(n_0^2)$ in Corollary~\ref{cor:dc} and its requirement that the error in learning the blockwise estimation rule be negligible relative to the error of the full-sample MLE.

Figure~\ref{fig:dc} plots $\bar e_{\mathrm{DC},\mathrm{MLE}}/\bar e_{\mathrm{MLE}}$, defined in Appendix~\ref{app:meta-extensions}, for all four values of $W$. The horizontal coordinate is $\alpha=1+\log J/\log n_0$, so that $n=n_0^\alpha$. The error bars scale the numerator standard error by the observed mean denominator. We omit the zero discrepancy of the block MLE average at $J=1$ from the logarithmic axis. For errors with respect to $\theta$, both averages approach a plateau as $J$ grows, whereas the full-sample MLE risk continues to decrease.

\begin{figure}[!htb]
\centering
\includegraphics[width=\textwidth]{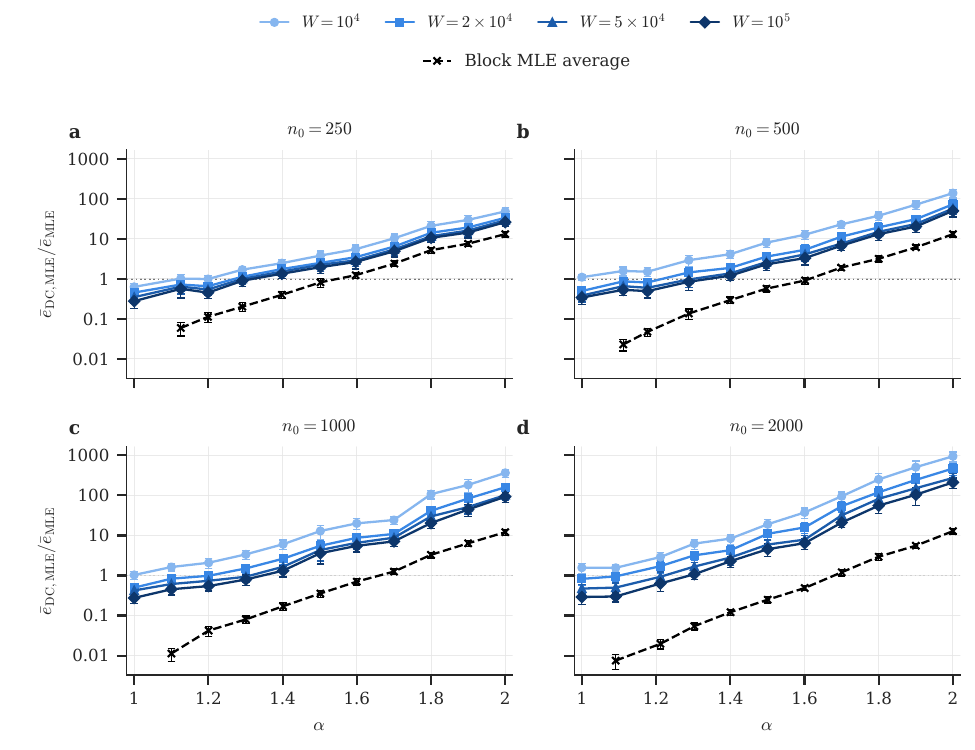}
\caption{Relative distance to the full-sample MLE for averages of RRDS estimates and block MLEs in logistic regression.}
\label{fig:dc}
\end{figure}

\subsection{Model selection}\label{app:impl-model}

The two candidates of Section~\ref{sec:sim-model} are mixtures of the logistic and linear models of Section~\ref{sec:sim-estimation}. Candidate 1 fixes the logistic-component probability at $1/2$ and has five parameters, while candidate 2 has a free mixing probability $\nu\in[0,1]$ and six parameters. The candidates coincide at $\nu=1/2$, where the generating label is unidentifiable and the population accuracy is $1/2$ under equal model probabilities. The task distribution chooses each candidate with probability $1/2$, draws $\theta$ as in Section~\ref{sec:sim-estimation}, and draws $\nu\sim{\rm Uniform}(0,1)$ for candidate 2. We train RRDS with $W\approx10^5$ and $R=12$ on $M=200{,}000$ tasks at each $n\in\{35,1000\}$. We evaluate both RRDS and BIC on 5000 common tasks with equally likely labels at each $\nu$ on a grid in $[0.25,0.75]$.

The BIC rule reduces to a comparison of binary-response counts. The continuous component assigns probability zero to $Y_i\in\{0,1\}$, so the component label is observed almost surely. Let $N_1$ count the binary responses. The log likelihood separates as
$\ell(\theta,\nu)=  \ell_0(\theta)+N_1\log\nu+(n-N_1)\log(1-\nu)$,
where $\ell_0$ is common to the two candidates. Candidate 1 fixes $\nu=1/2$, while candidate 2 has $\widehat\nu=N_1/n$ and one additional parameter. Their maximized log likelihood difference is
\[
\ell(\widehat\theta,\widehat\nu)-\ell(\widehat\theta,1/2) =  nD_{\mathrm{KL}}\{{\rm Bernoulli}(N_1/n)\|{\rm Bernoulli}(1/2)\}.
\]
Thus BIC selects candidate 2 precisely when twice this difference exceeds $\log n$, with equality assigned to candidate 1. The convention $0\log0=0$ covers the cases $N_1=0$ and $N_1=n$.

Cross-entropy training targets the Bayes rule under the task distribution, whereas BIC replaces the log marginal likelihood by its large-sample approximation, with a penalty of $(\log n)/2$ per parameter. In this design, the exact Bayes rule and BIC both depend on the data only through $N_1$ and select candidate 2 when $|N_1-n/2|$ exceeds a cutoff. The Bayes rule selects candidate 2 when $2^n B(N_1+1,n-N_1+1)>1$, where $B$ is the beta function. At $n=1000$, the two cutoffs differ by one count, which carries little probability, so the two reference rules make nearly the same decisions. At $n=35$, they differ at $N_1\in\{12,23\}$, where the Bayes rule selects candidate 2 and BIC does not. These counts are most probable at intermediate values of $\nu$, which is where RRDS improves on BIC. Near $\nu=1/2$, identifiability limits both methods. The large-sample agreement between RRDS and BIC is consistent with Theorems~\ref{thm:ms-approx} and~\ref{thm:ms-gen}, in which the BIC probabilities approximate the posterior model probabilities targeted by training.

\subsection{Variable selection}\label{app:impl-variable}

In the experiment of Section~\ref{sec:sim-variable}, we use a Gaussian linear model without an intercept, with $p=10$ covariates $Z_i\sim\mathcal N(0,I_{10})$, $s_0=3$ active coefficients, noise variance one, and $n=1000$. The support is uniform over subsets of size three. Each active coefficient is $\beta_j=\pm\tau u_j(2\log p/n)^{1/2}$ with equally likely signs and $u_j\sim{\rm Uniform}(1,2)$ independently. For training, we use $\tau\sim{\rm Uniform}(0,4)$, $M=60{,}000$ tasks, and the loss of Section~\ref{sec:meta-formulation} with $\lambda=1/20$. RRDS selects variable $j$ when $q_j>1/2$ and estimates its coefficient by $\widehat\beta_j=q_jb_j$. We tune LASSO and MCP by fivefold cross-validation of the prediction error, with concavity parameter $\gamma=1.5$ for MCP. Both methods select covariate $j$ when $|\widehat\beta_j|>0.05$. We evaluate each value of $\tau$ on 2000 common test tasks.

\subsection{HTRU2}\label{app:impl-htru}

The positive class contains 1,639 of the 17,898 records. We use 2,500 of these records as the development set to determine the feature standardization, calibrate the task distribution of the coefficients, and supply covariates for the simulated logistic training tasks. The remaining 15,398 records form the evaluation pool. We apply the separately trained estimation and joint selection networks to observed subsamples from this pool. In the unrestricted fit to the evaluation pool, the Wald $p$-values for $x_7$ and $x_8$ are 0.336 and 0.081, respectively, whereas those for $x_1,\ldots,x_6$ are below 0.001. The BIC reference retains these first six variables.

For parameter estimation, the task distribution has independent ${\rm Uniform}(c_j-1.5h_j,c_j+1.5h_j)$ coefficients, where $c_j$ and $h_j$ are the midpoint and half-range of development-only pilot envelopes. For variable selection, the support size is uniform on $\{1,\ldots,8\}$ and the support is uniform conditional on its size. Active slopes use the same coordinate intervals, inactive slopes are zero, and the intercept is uniform on $[-7.1114,-3.6150]$. We sample training covariates from the development records, without replacement within estimation tasks and with replacement within selection tasks. We generate responses from the logistic model.

At each $n$, we use $W=49{,}293$ and $M=50{,}000$ for estimation, and $W=97{,}474$ and $M=60{,}000$ for selection. Both networks use 12 shared refinements and the optimization settings in Appendix~\ref{app:impl-protocol}. The estimation network minimizes squared coefficient loss. The selection network adds mean binary cross-entropy over the eight support indicators to $200/9$ times this loss. At evaluation, we set slopes with inclusion probabilities at most $1/2$ to zero. We retain the intercept and do not refit the selected coefficients.

LASSO and ridge regression minimize the mean negative logistic log likelihood plus $\lambda\|\beta\|_1$ and $\lambda\|\beta\|_2^2/2$, respectively, with the intercept unpenalized. Within each task, fivefold stratified cross-validation chooses $\lambda$ from $\{10^{-5},10^{-4.5},\ldots,10\}$ by minimizing held-out binomial deviance. Ties favor the larger penalty. The final Karush--Kuhn--Tucker residuals for the mean objectives are at most $10^{-5}$. BIC searches all $2^8$ supports using $-2\ell+(1+|S|)\log n$ and refits the selected model. Candidate fits must converge, have finite coefficients and likelihood, show no separation, and have total score residual at most $10^{-6}$. We treat BIC values within $10^{-8}$ as tied, preferring smaller supports before lexicographic order.

At $n=500$, the unrestricted MLE is unavailable on 5 of the 1000 tasks following separation or convergence checks. The estimation comparison in Figure~\ref{fig:htru}(a) uses the 995 common available tasks, while the selection procedures use all 1000 tasks. All methods are available on all 1000 tasks at $n=1000$. Restricting both RRDS networks to the same tasks gives mean squared distances of 22.7 for the selection network and 29.6 for the estimator at $n=500$, and 17.2 and 19.4 at $n=1000$.

To examine the large coefficient errors, let $I_m(\widehat\theta_{\rm ref})$ be the total Fisher information on task $m$, evaluated at the empirical reference, and let $v_m$ be a unit eigenvector for its smallest eigenvalue. For $\Delta_m=\widehat\theta_m-\widehat\theta_{\rm ref}$, the decomposition
$\|\Delta_m\|_2^2=  (v_m^\top\Delta_m)^2+ \|(I_9-v_mv_m^\top)\Delta_m\|_2^2$
separates the least informative direction from its orthogonal complement. About 70\% of the aggregate MLE squared distance lies along this least informative direction. The mean of the first component is 172.6 for the MLE and 7.9 for RRDS at $n=500$, and 29.0 and 7.5 at $n=1000$. The information matrices are positive definite in the fixed standardized coordinates. Their smallest and largest eigenvalues, $\lambda_{\min}$ and $\lambda_{\max}$, pass the numerical check $\lambda_{\min}>9\epsilon\lambda_{\max}$, where $\epsilon$ is machine precision.

At $n=500$, 43 finite MLE fits and one LASSO fit have squared distance at least 1000, whereas no ridge fit does. At $n=1000$, all these distances are below 1000. For the largest finite MLE error, the fitted information matrix has a condition number of about $1.05\times10^5$, and the total score residual is $7.4\times10^{-15}$. The extreme LASSO fit has a mean-objective KKT residual of $1.17\times10^{-6}$. Separation checks do not detect complete or quasi-separation in these finite fits. Thus, these large discrepancies occur even though the fits have converged numerically.

\begin{figure}[!htb]
\centering
\includegraphics[width=0.85\textwidth]{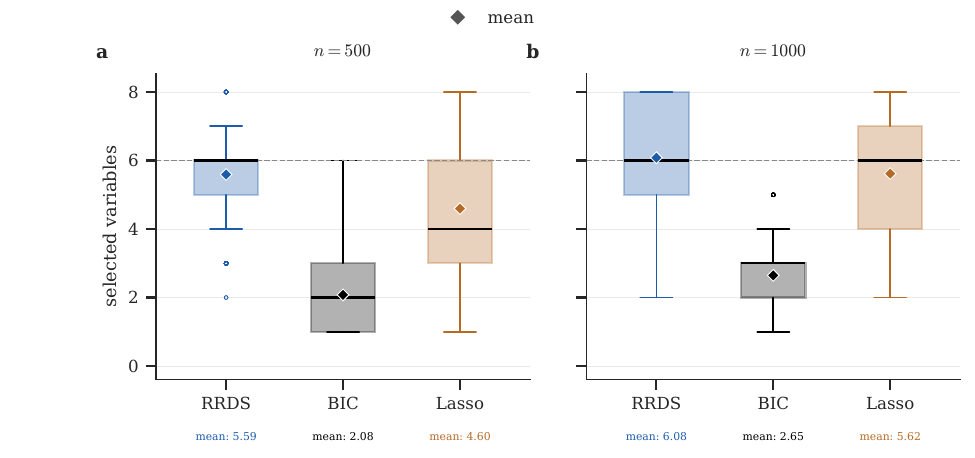}
\caption{Numbers of selected HTRU2 variables. The dashed line marks the reference size of six.}
\label{fig:app-htru-size}
\end{figure}

Figure~\ref{fig:app-htru-size} shows that the RRDS selection network chooses sizes closest to the reference size of six. BIC tends to select fewer variables than the reference.

\section{Proofs of the Main Results}\label{app:proofs}

\subsection{Overview}\label{app:proof-overview}

The proofs are organized by inference task, as in Section~\ref{sec:theory-results}. We state the auxiliary lemmas near the arguments that use them and defer their longer proofs to Appendix~\ref{app:technical}. Appendix~\ref{app:tools} collects the general approximation and probability results. The following overview describes the role of these results so that the reader can follow a proof without first reading all of the technical material.

For parameter estimation (Appendix~\ref{app:proofs-est}), the first comparison is between the RRDS network and the constrained MLE, using the one-step expansion~\eqref{eq:approx-bound} and its mean-square version~\eqref{eq:integrated-module-bound}. The second is between the constrained MLE and the posterior mean (Proposition~\ref{prop:bayes-mle}), which is the population target of squared-error training. Lemma~\ref{lem:fast-rate} then bounds the error due to training on finitely many tasks.

For model and variable selection, we likewise need two comparisons before applying the generalization bound. The first concerns the error incurred when replacing the posterior probabilities by probabilities obtained from an information criterion. The second concerns the error incurred when replacing the criterion by its neural approximation. The proofs follow the three steps of Sections~\ref{sec:theory-model} and~\ref{sec:theory-variable}. We compare the posterior probabilities with the BIC and EBIC probabilities through Laplace bounds (Lemmas~\ref{lem:ms-laplace} and~\ref{lem:vs-laplace}).

We also compare the network logits with the BIC and EBIC logits through the criterion decomposition of Lemma~\ref{lem:criterion}. These comparisons are carried out on the logit scale using the tools of Appendix~\ref{app:logit-tools}. We then compare the trained networks with the constructed networks through Lemmas~\ref{lem:excess-moments} and~\ref{lem:bernstein-erm}.

Variable selection requires additional arguments because the number of candidate supports grows with $p$ and because arbitrarily small nonzero coefficients are allowed under the task distribution. For a support containing the true support, the likelihood gain comes from fitting additional coefficients whose true values are zero; for a support omitting a nonzero coefficient, the loss depends on the size of the omitted signal. Lemma~\ref{lem:sparse-local} gives likelihood expansions that hold simultaneously for all candidate supports containing the true support. Lemma~\ref{lem:vs-omitted-energy} compares the information criterion of any candidate support with that of the true support and explicitly accounts for the energy of the omitted signals. Lemma~\ref{lem:vs-overfit} treats supports with added variables. Lemma~\ref{lem:vs-soft-oracle} gives the risk bound without a signal condition. Appendix~\ref{app:vs-smax} derives the explicit dependence on $s_{\max}$ in the results of Section~\ref{sec:theory-variable}.

For the results under analyticity conditions, the statistical arguments remain unchanged; we replace the finite-smoothness approximation by Lemma~\ref{lem:analytic-approx}. For divide-and-conquer estimation, the key observation is that averaging blockwise score terms gives the full-sample score. For bootstrap inference, we use the stability condition to control the network on resampled datasets and compare the conditional distribution of the resulting estimator with the Gaussian limit.

\medskip
\noindent{\bf Notation.}
Throughout Appendices~\ref{app:proofs}--\ref{app:tools}, $\|\cdot\|_{\rm op}$ and $\|\cdot\|_{\rm F}$ denote the operator and Frobenius norms, $a\vee b=\max\{a,b\}$ and $a\wedge b=\min\{a,b\}$, and ${\rm Lip}_A(f)$ is the Lipschitz constant of $f$ restricted to a set $A$. For a scalar function $f$, a probability law $P$ and $1\leq q<\infty$, $\|f\|_{L^q(P)}=(P|f|^q)^{1/q}$.

For symmetric matrices, $A\preceq B$ means that $B-A$ is positive semidefinite, and $\succeq$ is the reverse order. We write $P_\theta f=\mb E_\theta [f(X_1)]$ and $P_nf=n^{-1}\sum_{i=1}^n f(X_i)$ for the empirical average.

For random vectors $Z_n$ and deterministic $b_n>0$, $Z_n=O_P(b_n)$ means that $\|Z_n\|_2/b_n$ is bounded in $P$-probability, and $Z_n=o_P(b_n)$ means that this ratio converges to zero in $P$-probability, where the subscript identifies the probability law. We use the notation $A\Subset B$ of Appendix~\ref{app:notation}, and write $\xrightarrow{P}$ for convergence in $P$-probability. For a ReLU network $f$, $\|f\|_{\rm coef}$ denotes the largest absolute value of its coefficients, that is, of its weights and biases.

\medskip
\noindent{\bf Constants.}
Throughout Appendices~\ref{app:proofs} and~\ref{app:technical}, $C$ and $c$ denote positive constants whose values may change from line to line. They do not depend on the sample size $n$, the numbers of training tasks, the confidence level $\delta$, the accuracy levels $\varepsilon$ and $\varepsilon_n$, or the optimization error $\varepsilon_{\rm opt}$. The dependence on these quantities is always displayed.

In the proofs for parameter estimation and model selection, the constants may depend on the fixed model or candidate collection, including its dimensions, smoothness indices, working sets and observation space, and they depend on a compact set $\Theta_1$, on the task distributions or on the stability constant $K_{\rm st}$ only as specified in the corresponding statement.

In the proofs for variable selection, the constants depend only on the constants of the setting of Section~\ref{sec:theory-variable} and of Assumption~\ref{ass:sparse}. They do not depend on $p$, the supports or the generating coefficients. The initial variable-selection lemmas allow their constants to depend on $D=1+s_{\max}$. Appendix~\ref{app:vs-smax} and the proofs that invoke its bounds instead track $D$ explicitly. In the latter arguments the constants are independent of $D$, as specified there. Unless fixed in a result or a shared construction, numbered constants are local to each proof and may depend on the same quantities as $C$ and $c$.

\subsection{Proofs for parameter estimation}\label{app:proofs-est}

\subsubsection{Preliminary results for the constrained MLE and moment initialization}\label{app:class}

The dimensions $d,d_X$ are fixed. Fix a compact $\Theta_1\Subset\operatorname{int}\Theta_0$ and write $\ell_n(x^n,\eta)=\sum_{i=1}^n\ell_\eta(x_i)$, $H_n(z)=-P_n\nabla_z^2\log p_z$, and $\lambda_0=\inf_{\theta\in\Theta_0}\lambda_{\min}\{I(\theta)\}>0$, where positivity follows from Lemma~\ref{lem:consequences} and compactness. Since $\ell_n(\cdot,\eta)$ is Borel, $\ell_n(x^n,\cdot)$ is continuous and $\Theta_0$ is nonempty and compact, the measurable maximum theorem~\citep[Theorem~18.19(3), p.~605]{aliprantis2006infinite} gives a Borel selection $\widehat\theta_{0,n}:\mathcal X^n\to\Theta_0$ with $\widehat\theta_{0,n}(x^n)\in\argmax_{\eta\in\Theta_0}\ell_n(x^n,\eta)$, which we fix for all datasets, including resamples.

\medskip
The following lemma summarizes the consequences of Assumption~\ref{ass:regular}. In particular, it establishes the smoothness of $\mu$ and $I$, a quadratic lower bound on the KL divergence, and a first-order expansion of $\widehat\theta_{0,n}$ with a uniform remainder bound.

\begin{lemma}[Regularity and localization]\label{lem:consequences}
Under Assumption~\ref{ass:regular}, $\mu\in C^{r+1}(U)$, $I\in C^r(U;\mathbb S_{++}^d)$, and $\mathbb P_\theta=\mathbb P_\eta$ implies $\theta=\eta$ for $\theta,\eta\in U$. Moreover,
\[
\sup_{x\in\mathcal X,\theta\in\Theta_0} \max_{|\nu|\leq  2}\|D_\theta^\nu s_\theta(x)\|_2<\infty,
\]
$P_\theta s_\theta=0$ and $-P_\theta\nabla_\theta^2\log p_\theta=I(\theta)$. There is $c>0$ such that $D_{\mathrm{KL}}(\mb P_\theta\|\mb P_\eta)\geq c\|\theta-\eta\|_2^2$ for all $\theta,\eta\in\Theta_0$. For each fixed $a>0$ there are $C_a,c_a>0$ such that $\sup_{\theta\in\Theta_0}\mb P_\theta\{\|\widehat\theta_{0,n}-\theta\|_2>a\}\leq C_ae^{-c_an}$. Define the event
\[
\mathcal H_n=  \{ \argmax_{z\in\Theta_0}\ell_n(z)=  \{\widehat\theta_{0,n}\} ,\qquad \widehat\theta_{0,n}\in\operatorname{int}(\Theta_0) ,\qquad H_n(\widehat\theta_{0,n})\succeq\tfrac12\lambda_0I_d\}.
\]
Then $\sup_{\theta\in\Theta_1}\mb P_\theta(\mathcal H_n^c)\leq Ce^{-cn}$. Uniformly over $\theta\in\Theta_1$,
\begin{equation}
 \mb E_\theta^{(n)}[\|\widehat\theta_{0,n}-\theta\|_2^4]\leq Cn^{-2},
 \qquad
 \widehat\theta_{0,n}-\theta
 =I(\theta)^{-1}P_ns_\theta+R_n(\theta),
 \quad \mb E_\theta^{(n)}[\|R_n(\theta)\|_2^2]\leq Cn^{-2}.
 \label{eq:uniform-mle-expansion}
\end{equation}
\end{lemma}

The next two subsections prove the approximation and the meta-generalization results under finite smoothness.

The initialization block $\phi_0$ of the RRDS network in Theorem~\ref{thm:est-approx} computes the moment function, which Assumption~\ref{ass:regular} only requires to be bounded and measurable. The next lemma shows that it can be taken to be a ReLU network without changing the assumption.

\begin{lemma}[A ReLU moment function]\label{lem:relu-moment}
Suppose Assumption~\ref{ass:regular} holds with $(T,U)$. Then there are a ReLU network $\widetilde T:\mb R^{d_X}\to\mb R^d$ and an open set $U'$ with $\Theta_0\subset U'\subset U$ such that Assumption~\ref{ass:regular} holds with $(\widetilde T,U')$ in place of $(T,U)$. The network $\widetilde T$ depends only on the model, $T$ and $\Theta_0$. In particular, its depth and its number of parameters do not depend on $n$ or on any accuracy level.
\end{lemma}

Accordingly, in all constructions the moment function $T$ denotes this ReLU network, the neighborhood $U$ is replaced by $U'$, and the constants depend on the model through $\widetilde T$. The log likelihood $\ell_n$, the information $I$ and $\widehat\theta_{0,n}$ are not affected. In Proposition~\ref{prop:ms-misspec}, the point $\theta_{m,k}$ is defined through the moment function used in the network. For the analytic results of Appendix~\ref{app:analytic}, the extension condition on $g_{\rm MoM}$ likewise refers to the moment function used in the network, which is $T$ itself when $T$ is piecewise linear, for example when $T(x)=x$.

\subsubsection{Proof of Theorem~\ref{thm:est-approx}}\label{app:proof-approx}

The argument has two parts. First, we approximate the inverse moment map, the score and the inverse information on domains that do not depend on the sample size. We then show that the resulting projected one-step estimator differs from the constrained MLE by a sum of the neural approximation error and a second-order statistical error. The projections keep the estimates in the working domains even for datasets on which localization fails.

For $x\in\mathbb R^d$, since $\Theta_0$ is compact and convex and the map $v\mapsto\|x-v\|_2^2$ is continuous and strictly convex, there is a unique minimizer $\Pi_{\Theta_0}(x)$ of $\|x-v\|_2$ over $v\in\Theta_0$. Its first-order optimality condition is
\begin{equation}
 \langle x-\Pi_{\Theta_0}(x),v-\Pi_{\Theta_0}(x)\rangle\leq0,
 \qquad v\in\Theta_0.
 \label{eq:projection-variational}
\end{equation}
For $x,y\in\mathbb R^d$, write $p=\Pi_{\Theta_0}(x)$ and $q=\Pi_{\Theta_0}(y)$. Applying~\eqref{eq:projection-variational} at $x$ with $v=q$ and at $y$ with $v=p$, and adding the inequalities, gives $\|p-q\|_2^2\leq\langle x-y,p-q\rangle\leq\|x-y\|_2\|p-q\|_2$ by Cauchy--Schwarz. Since projection fixes every point of $\Theta_0$, we have
\begin{equation}
 \|\Pi_{\Theta_0}(x)-\Pi_{\Theta_0}(y)\|_2\leq\|x-y\|_2,
 \qquad \Pi_{\Theta_0}(v)=v,\quad v\in\Theta_0.
 \label{eq:projection-nonexpansive}
\end{equation}
With $D_0=\operatorname{diam}(\Theta_0)$, any two points of $\Theta_0$ are at distance at most $D_0$, and by convexity $\int_{\Theta_0}z\,\nu(dz)\in\Theta_0$ for every probability measure $\nu$ on $\Theta_0$.

The output map $\Psi=\Pi_{\Theta_0}$ is a fixed map. For the initial estimate, Lemma~\ref{lem:polytope}(a) gives a polytope $P$ with $\Theta_0\subseteq P\subset U$, and we take $P=\Theta_0$ when $\Theta_0$ is a polytope. Let $\Pi_P$ be the Euclidean projection onto $P$. Then $\Pi_P$ satisfies~\eqref{eq:projection-variational} and~\eqref{eq:projection-nonexpansive} with $P$ in place of $\Theta_0$. By Lemma~\ref{lem:polytope}(b), $\Pi_P$ is computed exactly by a ReLU network of fixed size, which forms part of the block $\rho_0=\Pi_P\circ g_{\rm init}$. Since $P$ is a compact convex subset of $U$, the derivative bounds and the positivity of $I$ in Lemma~\ref{lem:consequences} hold on $P$ as well. By Lemma~\ref{lem:relu-moment}, $T$ is a ReLU network of fixed size, which is the block $\phi_0$. Averaging and the output projection count toward neither $L$ nor $W$. Within the ReLU blocks, fixed internal affine maps count toward $L$, and fixed internal connections count toward the coefficient bounds but not toward $W$.

Lemma~\ref{lem:consequences} and the inverse function theorem give
\[
\mu(U)\text{ open},\qquad g_{\rm MoM}=  \mu^{-1}\in C^{r+1}(\mu(U);U),\qquad \mu(\Theta_0)\subset\operatorname{int}(\mathcal K)\subset\mathcal K\Subset\mu(U).
\]
The compact set $\mathcal K$ does not depend on $n$ or on the accuracies. Each of the maps $g_{\rm MoM}$, $s$ and $I^{-1}$ is of class $C^{r}$ or $C^{r+1}$ on an open set $O$ containing its compact domain $K_0$, namely $\mathcal K$, $\mathcal X\times P$ or $P$. Choose open sets $U_1,V_1$ with $K_0\subset U_1$, $\overline U_1\subset V_1$ and $\overline V_1\subset O$ compact. By \citet[Proposition~2.25, p.~44]{lee2013smooth}, there is $\chi\in C^\infty$ with $\chi=1$ on $\overline U_1$ and $\operatorname{supp}\chi\subset V_1$. Given a function $f$ of class $C^r$ or $C^{r+1}$ on $O$, the product $\chi f$, extended by zero outside $O$, equals $f$ on $U_1$ because $\chi=1$ there, and it is zero outside $\operatorname{supp}\chi$, which is a compact subset of $O$. Hence it has the smoothness of $f$ on the whole space. Applied to each coordinate, this gives fixed extensions
\[
\begin{gathered}
g^{\rm ext}\in C^{r+1}(\mathbb R^d;\mathbb R^d),\qquad g^{\rm ext}|_{\mathcal K}=  g_{\rm MoM},\qquad s^{\rm ext}\in C^r(\mathbb R^{d_X+d};\mathbb R^d),\\
s^{\rm ext}|_{\mathcal X\times P}=  s,\qquad A^{\rm ext}\in C^r(\mathbb R^d;\mathbb R^{d\times d}),\qquad A^{\rm ext}|_{P}=  I^{-1}.
\end{gathered}
\]

Put
$B_s=  1\vee\sup_{x\in\mathcal X,\ z\in P}\|s_z(x)\|_2$, $B_I=  1\vee\sup_{z\in P}\|I(z)^{-1}\|_{\rm op}$,
which are finite by Lemma~\ref{lem:consequences}. The next lemma, which follows from Lemma~\ref{lem:sobolev-approx}, supplies the component networks.

\begin{lemma}[Approximation of the component maps]\label{lem:modules}
For $a,b\in(0,1)$, put $\varepsilon_b=b/(B_s+B_I+2)$. There are ReLU networks $g_{\rm init}$, $\mathsf s$ and a matrix-valued network $A$ such that
\[
 \begin{gathered}
 \sup_{t\in\mathcal K}\|g_{\rm init}(t)-g_{\mathrm{MoM}}(t)\|_2\leq a,\\
 \sup_{x\in\mathcal X,\ z\in P}\|\mathsf s(x,z)-s_z(x)\|_2\leq\varepsilon_b,\qquad
 \sup_{z\in P}\|A(z)-I(z)^{-1}\|_{\rm F}\leq\varepsilon_b.
 \end{gathered}
\]
Their complexity bounds satisfy
\[
\begin{gathered}
\max\{L(g_{\rm init}),L(\mathsf s),L(A)\}\leq   C\log\{e/(a\wedge b)\} ,\qquad W(g_{\rm init})\leq   Ca^{-d/(r+1)}\log(e/a),\\
W(\mathsf s)+W(A)\leq   Cb^{-(d_X+d)/r}\log(e/b) ,\\
\max\{\|g_{\rm init}\|_{\rm coef},\|\mathsf s\|_{\rm coef},\|A\|_{\rm coef}\}\leq   C(a\wedge b)^{-C}.
\end{gathered}
\]
Together, the three networks have at most $C\{a^{-d/(r+1)}\log(e/a)+b^{-(d_X+d)/r}\log(e/b)\}$ coefficient positions. Here $C$ depends only on the fixed domains, dimensions and smoothness bounds.
\end{lemma}

\begin{proof}
In this proof, $C$ depends only on $d$, $d_X$, $r$, the fixed domains and the smoothness bounds. Fix rectangles $R_1\subset\mathbb R^d$, $R_2\subset\mathbb R^{d_X+d}$ and $R_3\subset\mathbb R^d$ that contain $\mathcal K$, $\mathcal X\times P$ and $P$, respectively. The extensions $g^{\rm ext}$, $s^{\rm ext}$ and $A^{\rm ext}$ are of class $C^{r+1}$, $C^r$ and $C^r$ on the whole space, so they have bounded $C^{r+1}$, $C^r$ and $C^r$ norms on these compact rectangles.

We apply Lemma~\ref{lem:sobolev-approx} three times, with the vector assembly at the end of Appendix~\ref{app:relu}. In the notation of that lemma, the input dimension, smoothness order, output dimension and accuracy $(m,s,q,\epsilon)$ are $(d,r+1,d,a)$ for $g^{\rm ext}$ on $R_1$, $(d_X+d,r,d,\varepsilon_b)$ for $s^{\rm ext}$ on $R_2$, and $(d,r,d^2,\varepsilon_b)$ for the vectorization of $A^{\rm ext}$ on $R_3$. Let $g_{\rm init}$, $\mathsf s$ and $A$ be the resulting networks. Since $g^{\rm ext}=g_{\rm MoM}$ on $\mathcal K\subset R_1$,
\[
\sup_{t\in\mathcal K}\|g_{\rm init}(t)-g_{\rm MoM}(t)\|_2\leq\sup_{t\in R_1}\|g_{\rm init}(t)-g^{\rm ext}(t)\|_2\leq a.
\]
The bounds for $\mathsf s$ and $A$ follow in the same way from $s^{\rm ext}=s$ on $\mathcal X\times P$ and $A^{\rm ext}=I^{-1}$ on $P$, where for the matrix output the Euclidean norm of the vectorization is the Frobenius norm. The lemma also gives
\[
\begin{gathered}
L(g_{\rm init})\leq C\log(e/a),\qquad W(g_{\rm init})\leq Ca^{-d/(r+1)}\log(e/a),\\
L(\mathsf s),L(A)\leq C\log(e/\varepsilon_b),\qquad W(\mathsf s)\leq C\varepsilon_b^{-(d_X+d)/r}\log(e/\varepsilon_b),\\
W(A)\leq C\varepsilon_b^{-d/r}\log(e/\varepsilon_b),
\end{gathered}
\]
with coefficients bounded by $Ca^{-C}$ for $g_{\rm init}$ and by $C\varepsilon_b^{-C}$ for $\mathsf s$ and $A$.

Since $\varepsilon_b=b/(B_s+B_I+2)$ and $B_s$, $B_I$ are fixed, we have $\log(e/\varepsilon_b)\leq C\log(e/b)$ and $\varepsilon_b^{-t}=(B_s+B_I+2)^tb^{-t}\leq Cb^{-t}$ for $t=(d_X+d)/r$, for $t=d/r$ and for the exponent of the coefficient bound. Since $b<1$ and $d/r\leq(d_X+d)/r$, this gives $W(\mathsf s)+W(A)\leq Cb^{-(d_X+d)/r}\log(e/b)$. Finally, $\log(e/a)$ and $\log(e/b)$ are at most $\log\{e/(a\wedge b)\}$, and $a^{-C}$ and $b^{-C}$ are at most $(a\wedge b)^{-C}$. These are the claimed complexity bounds. We take the architecture of each network to be the set of its nonzero weights and biases. By the proof of Lemma~\ref{lem:sobolev-approx}, their number is at most the size bound of that lemma, which gives the bound on the number of coefficient positions.
\end{proof}

\begin{proof}[Proof of Theorem~\ref{thm:est-approx}]
In this proof, $C$ and $c$ depend only on $d$, $d_X$, $r$, the maps $\ell$ and $T$ of Assumption~\ref{ass:regular}, $\Theta_0$ and $\Theta_1$, and the constants in the network complexity bounds do not depend on $\Theta_1$. We divide the proof into four steps.

\medskip
\noindent{\bf Step 1: Network construction and localization.} We first bound the error of the approximate Fisher scoring correction uniformly over the working domain. Take the networks $g_{\rm init}$, $\mathsf s$ and $A$ of Lemma~\ref{lem:modules} at $a,b\in(0,1)$, where $\varepsilon_b=b/(B_s+B_I+2)<1$. For $x^n\in\mathcal X^n$ and $z\in P$, set $u=n^{-1}\sum_{i=1}^n\mathsf s(x_i,z)$ and $v=n^{-1}\sum_{i=1}^ns_z(x_i)$. By the triangle inequality and the definition of $B_s$ and Lemma~\ref{lem:modules}, we have $\|v\|_2\leq n^{-1}\sum_{i=1}^n\|s_z(x_i)\|_2\leq B_s$ and
\[
\|u-v\|_2\leq\frac1n\sum_{i=1}^n\|\mathsf s(x_i,z)-s_z(x_i)\|_2\leq\varepsilon_b,
\]
so $\|u\|_2\leq\|v\|_2+\|u-v\|_2\leq B_s+\varepsilon_b\leq B_s+1$. Similarly, since $\|\cdot\|_{\rm op}\leq\|\cdot\|_{\rm F}$, the definition of $B_I$ and Lemma~\ref{lem:modules} give
\[
\|A(z)\|_{\rm op}\leq\|I(z)^{-1}\|_{\rm op}+\|A(z)-I(z)^{-1}\|_{\rm F}\leq B_I+\varepsilon_b\leq B_I+1.
\]
Put $R_\times=1+B_s+B_I$. For $1\leq j,k\leq d$, each coordinate of $u$ satisfies $|u_k|\leq\|u\|_2\leq B_s+1\leq R_\times$ and each entry of $A(z)$ satisfies $|A_{jk}(z)|\leq\|A(z)\|_{\rm op}\leq B_I+1\leq R_\times$, so every coordinate input of the matrix-vector product $A(z)u$ lies in $[-R_\times,R_\times]$. Let $h$ approximate $A(z)u$ with $d^2$ copies of $\mathsf M=\mathsf M_{R_\times,\varepsilon_b/d^{3/2}}$ from Lemma~\ref{lem:multiplication}, with $j$th coordinate $h_j(u,z)=\sum_{k=1}^d\mathsf M\{A_{jk}(z),u_k\}$ for $1\leq j\leq d$. Since each product has error at most $\varepsilon_b/d^{3/2}$, each coordinate satisfies
\[
|h_j(u,z)-\{A(z)u\}_j|\leq\sum_{k=1}^d|\mathsf M\{A_{jk}(z),u_k\}-A_{jk}(z)u_k|\leq d\,\frac{\varepsilon_b}{d^{3/2}},
\]
and hence $\|h(u,z)-A(z)u\|_2\leq\sqrt d\,d\,\varepsilon_b/d^{3/2}=\varepsilon_b$. Writing
\[
h(u,z)-I(z)^{-1}v=\{h(u,z)-A(z)u\}+\{A(z)-I(z)^{-1}\}u+I(z)^{-1}(u-v),
\]
the triangle inequality gives
\begin{equation}
 \begin{aligned}
 \|h(u,z)-I(z)^{-1}v\|_2
 &\leq\varepsilon_b+\|A(z)-I(z)^{-1}\|_{\rm op}\|u\|_2+\|I(z)^{-1}\|_{\rm op}\|u-v\|_2\\
 &\leq\varepsilon_b(B_s+B_I+2)=b.
 \end{aligned}
 \label{eq:separate-module-accuracy}
\end{equation}
Here the first inequality uses $\|h(u,z)-A(z)u\|_2\leq\varepsilon_b$, and the second line uses $\|A(z)-I(z)^{-1}\|_{\rm op}\leq\|A(z)-I(z)^{-1}\|_{\rm F}\leq\varepsilon_b$, $\|u\|_2\leq B_s+1$, $\|I(z)^{-1}\|_{\rm op}\leq B_I$, $\|u-v\|_2\leq\varepsilon_b$ and the definition of $\varepsilon_b$.

The score network $\mathsf s$ is shared across observations, so its parameters are counted once. By Lemma~\ref{lem:multiplication} with $R=R_\times$ and $\zeta=\varepsilon_b/d^{3/2}$, each multiplier $\mathsf M$ has depth and size at most $C\{1+\log(R_\times^2d^{3/2}/\varepsilon_b)\}\leq C\log(e/b)$, since $\varepsilon_b=b/(B_s+B_I+2)$ and $R_\times$ and $d$ are fixed. The proof of Lemma~\ref{lem:modules} gives $L(\mathsf s),L(A)\leq C\log(e/b)$. Hence the $d^2$ multipliers and the identity connections that carry the inputs of one component through the layers of the others have depth and size at most $C\log(e/b)$. Adding the depths and sizes of Lemma~\ref{lem:modules} in series for the depth and in parallel for the size gives
\begin{equation}
  L\leq C\log\{e/(a\wedge b)\},\qquad W\leq C\left\{a^{-d/(r+1)}\log(e/a)
       +b^{-(d_X+d)/r}\log(e/b)\right\}.
 \label{eq:separate-module-complexity}
\end{equation}
Let $\theta_n^{\mathrm m}=g_{\mathrm{MoM}}(\bar T_n)$ when $\bar T_n\in\mathcal K$, with a fixed completion in $\Theta_0$ otherwise, and define
\[
\begin{gathered}
\mathcal E_n=  \{\bar T_n\in\mathcal K, \ \theta_n^{\mathrm m}\in\Theta_0, \ \widehat\theta_{0,n}\in{\rm int}(\Theta_0)\},\\
\rho_n=  \|\theta_n^{\mathrm m}-\widehat\theta_{0,n}\|_2,\qquad
\delta_n=  \|I(\theta_n^{\mathrm m})-H_n(\theta_n^{\mathrm m})\|_{\rm op}.
\end{gathered}
\]

\medskip
\noindent{\bf Step 2: Bound on $\mathcal E_n$.} We next use the local likelihood expansion to compare the approximate update with the MLE. We work on $\mathcal E_n$ and write $z=\Pi_Pg_{\rm init}(\bar T_n)\in P$ and $\widehat\theta=\widehat\theta_{0,n}$. On this event, $\bar T_n\in\mathcal K$ and $\theta_n^{\mathrm m}=g_{\mathrm{MoM}}(\bar T_n)\in\Theta_0\subseteq P$. Since $\Pi_P$ is nonexpansive and fixes $\theta_n^{\mathrm m}$, Lemma~\ref{lem:modules} gives
\[
\|z-\theta_n^{\mathrm m}\|_2
 =\|\Pi_Pg_{\rm init}(\bar T_n)-\Pi_Pg_{\mathrm{MoM}}(\bar T_n)\|_2 \leq\|g_{\rm init}(\bar T_n)-g_{\mathrm{MoM}}(\bar T_n)\|_2\leq a,
\]
so $\|z-\widehat\theta\|_2\leq\|z-\theta_n^{\mathrm m}\|_2+\rho_n\leq a+\rho_n$. Since $\widehat\theta$ is an interior maximizer of $\ell_n$, we have $P_ns_{\widehat\theta}=0$. The segment from $\widehat\theta$ to $z$ lies in the convex set $P$, so the fundamental theorem of calculus gives
\[
 P_ns_z=P_ns_z-P_ns_{\widehat\theta}
 =-\left[\int_0^1 H_n\{\widehat\theta+t(z-\widehat\theta)\}\,dt\right](z-\widehat\theta).
\]
Substituting this into $z+I(z)^{-1}P_ns_z-\widehat\theta$ gives
\[
 z+I(z)^{-1}P_ns_z-\widehat\theta
 =I(z)^{-1}\left[I(z)-\int_0^1
 H_n\{\widehat\theta+t(z-\widehat\theta)\}\,dt\right]
 (z-\widehat\theta).
\]

By Lemma~\ref{lem:consequences}, $I$ and $H_n$ are Lipschitz on $P$ with a constant that does not depend on $n$ or the data. For $t\in[0,1]$, put $z_t=\widehat\theta+t(z-\widehat\theta)\in P$, so that $\|\theta_n^{\mathrm m}-z_t\|_2\leq\rho_n+\|z-\widehat\theta\|_2\leq a+2\rho_n$. We write
\[
I(z)-H_n(z_t)
 =\{I(z)-I(\theta_n^{\mathrm m})\}
 +\{I(\theta_n^{\mathrm m})-H_n(\theta_n^{\mathrm m})\} \quad+\{H_n(\theta_n^{\mathrm m})-H_n(z_t)\}.
\]
The three terms have operator norms at most $C\|z-\theta_n^{\mathrm m}\|_2\leq Ca$, $\delta_n$ and $C\|\theta_n^{\mathrm m}-z_t\|_2\leq C(a+2\rho_n)$. Hence the bracket in the identity is at most $\delta_n+C(a+\rho_n)$ in operator norm.

Let $u=n^{-1}\sum_i\mathsf s(X_i,z)$, so that $\widetilde F=\Pi_{\Theta_0}\{z+h(u,z)\}$. Since the output projection fixes $\widehat\theta$,~\eqref{eq:projection-nonexpansive} gives
\[
\|\widetilde F-\widehat\theta\|_2
 \leq\|z+h(u,z)-\widehat\theta\|_2 \leq\|h(u,z)-I(z)^{-1}P_ns_z\|_2+\|z+I(z)^{-1}P_ns_z-\widehat\theta\|_2.
\]
The first term is at most $b$ by~\eqref{eq:separate-module-accuracy} with $v=P_ns_z$. By the identity, $\|I(z)^{-1}\|_{\rm op}\leq B_I$ and the bounds on the bracket and on $\|z-\widehat\theta\|_2$, the second term is at most $B_I\{\delta_n+C(a+\rho_n)\}(a+\rho_n)$. Hence
\begin{equation}
 \|\widetilde F-\widehat\theta\|_2
 \leq b+\|z+I(z)^{-1}P_ns_z-\widehat\theta\|_2
 \leq b+C(a+\rho_n)(\delta_n+a+\rho_n),
 \label{eq:approx-bound}
\end{equation}
where $C$ depends only on the fixed model, $\Theta_0$ and $P$.

\medskip
\noindent{\bf Step 3: Mean-square approximation error.} Fix $\theta\in\Theta_1$ and let $\mathcal M_n=\{\bar T_n\in\mathcal K,\ \theta_n^{\rm m}\in\Theta_0\}$. Since $\Theta_1$ has a positive distance from $\partial\Theta_0$ and $g_{\rm MoM}$ is Lipschitz near the compact set $\mu(\Theta_1)$, there is $b_T>0$ such that $\|t-\mu(\theta)\|_2\leq b_T$ implies $t\in\mathcal K$ and $g_{\rm MoM}(t)\in\Theta_0$.

Let $B_T=\max_j\sup_x|T_j(x)|$. Since $\|\bar T_n-\mu(\theta)\|_2\leq b_T$ implies $\mathcal M_n$ and $P_\theta T=\mu(\theta)$, on $\mathcal M_n^c$ at least one coordinate satisfies $|\bar T_{n,j}-P_\theta T_j|>b_T/\sqrt d$. Each coordinate $T_j$ takes values in $[-B_T,B_T]$, so Hoeffding's inequality \citep[Proposition~2.5 and (2.11)]{wainwright2019high} gives $\mb P_\theta\{|\bar T_{n,j}-P_\theta T_j|>b_T/\sqrt d\}\leq 2\exp\{-nb_T^2/(2dB_T^2)\}$, and a union bound over the $d$ coordinates yields
\[
\mb P_\theta(\mathcal M_n^c)\leq 2d\exp\Bigl\{-\frac{nb_T^2}{2dB_T^2}\Bigr\}.
\]
Moreover, $\mathcal E_n^c\subseteq\mathcal M_n^c\cup\{\widehat\theta\notin\operatorname{int}(\Theta_0)\}$, and Lemma~\ref{lem:consequences} gives $\mb P_\theta\{\widehat\theta\notin\operatorname{int}(\Theta_0)\}\leq\mb P_\theta(\mathcal H_n^c)\leq Ce^{-cn}$ uniformly over $\theta\in\Theta_1$. Hence $\mb P_\theta(\mathcal E_n^c)\leq\mb P_\theta(\mathcal M_n^c)+\mb P_\theta(\mathcal H_n^c)\leq Ce^{-cn}$.

Next, we bound $\rho_n$ and $\delta_n$ on $\mathcal E_n$. Since $\theta=g_{\rm MoM}(\mu(\theta))$ and $g_{\rm MoM}$ is Lipschitz near $\mu(\Theta_1)$, we have $\|\theta_n^{\rm m}-\theta\|_2\ind_{\mathcal M_n}\leq C\|\bar T_n-P_\theta T\|_2$. On $\mathcal E_n$, the triangle inequality gives $\rho_n=\|\theta_n^{\rm m}-\widehat\theta\|_2\leq\|\theta_n^{\rm m}-\theta\|_2+\|\widehat\theta-\theta\|_2$. Likewise, by the triangle inequality and the Lipschitz bounds on $I$ and $H_n$ from Lemma~\ref{lem:consequences},
\[
\begin{aligned}
\delta_n
&\leq\|I(\theta_n^{\rm m})-I(\theta)\|_{\rm op}+\|I(\theta)-H_n(\theta)\|_{\rm op}+\|H_n(\theta)-H_n(\theta_n^{\rm m})\|_{\rm op}\\
&\leq\|H_n(\theta)-I(\theta)\|_{\rm op}+C\|\theta_n^{\rm m}-\theta\|_2.
\end{aligned}
\]

To bound the fourth moments, note that $\bar T_n-P_\theta T$ and the entries of $H_n(\theta)-I(\theta)$ are averages of $n$ i.i.d. centered bounded terms, since $T$ is bounded and Lemma~\ref{lem:consequences} gives $-P_\theta\nabla_\theta^2\log p_\theta=I(\theta)$ and a uniform bound on $\nabla^2\ell_\theta$. Applying~\eqref{eq:average-fourth-moment} therefore gives
\[
\mb E_\theta^{(n)}[\|\bar T_n-P_\theta T\|_2^4]\leq Cn^{-2},\qquad
\mb E_\theta^{(n)}[\|H_n(\theta)-I(\theta)\|_{\rm op}^4]\leq Cn^{-2},
\]
and hence $\mb E_\theta^{(n)}[\|\theta_n^{\rm m}-\theta\|_2^4\ind_{\mathcal M_n}]\leq Cn^{-2}$. In addition,~\eqref{eq:uniform-mle-expansion} gives $\mb E_\theta^{(n)}[\|\widehat\theta-\theta\|_2^4]\leq Cn^{-2}$. Combining these bounds with the estimates for $\rho_n$ and $\delta_n$ on $\mathcal E_n$ yields $\mb E_\theta^{(n)}[(\rho_n^4+\delta_n^4)\ind_{\mathcal E_n}]\leq Cn^{-2}$.

We now combine the contributions of $\mathcal E_n$ and $\mathcal E_n^c$. On $\mathcal E_n$, squaring~\eqref{eq:approx-bound} gives
\[
\|\widetilde F-\widehat\theta\|_2^2
\leq 2b^2+2C^2(a+\rho_n)^2(\delta_n+a+\rho_n)^2
\leq C(b^2+a^4+\rho_n^4+\delta_n^4).
\]
Taking expectations and using the fourth-moment bound gives $\mb E_\theta^{(n)}[\|\widetilde F-\widehat\theta\|_2^2\ind_{\mathcal E_n}]\leq C(b^2+a^4+n^{-2})$. On $\mathcal E_n^c$, both $\widetilde F$ and $\widehat\theta$ lie in $\Theta_0$, so $\|\widetilde F-\widehat\theta\|_2\leq D_0$ and $\mb E_\theta^{(n)}[\|\widetilde F-\widehat\theta\|_2^2\ind_{\mathcal E_n^c}]\leq D_0^2\mb P_\theta(\mathcal E_n^c)\leq Cn^{-2}$. Adding the two bounds gives, for all $n\geq1$,
\begin{equation}
 \sup_{\theta\in\Theta_1}\mb E_\theta^{(n)}[\|\widetilde F(X^n)-\widehat\theta_{0,n}(X^n)\|_2^2]
 \leq C\{b^2+a^4+n^{-2}\}.
 \label{eq:integrated-module-bound}
\end{equation}
At a fixed $\theta\in\operatorname{int}\Theta_0$, the same bounds give $\mathbb P_\theta(\mathcal E_n)\to1$, $\rho_n=O_{P_\theta}(n^{-1/2})$ and $\delta_n=O_{P_\theta}(n^{-1/2})$.

Take $a=b=\varepsilon$. Since $d/(r+1)<\kappa$,~\eqref{eq:separate-module-complexity} and~\eqref{eq:integrated-module-bound} give $L\leq C\log(e/\varepsilon)$, $W\leq C\varepsilon^{-\kappa}\log(e/\varepsilon)$ and~\eqref{eq:est-approx-mse}, with complexity bound constants that do not depend on $\Theta_1$.

\medskip
\noindent{\bf Step 4: Asymptotic normality.} The mean-square bound yields the first-order comparison directly. Fix $\theta\in{\rm int}(\Theta_0)$ and a compact $\Theta_1\subset{\rm int}(\Theta_0)$ containing it, and let $\varepsilon=\varepsilon_n=o(n^{-1/2})$. For every $t>0$, Markov's inequality and~\eqref{eq:est-approx-mse} give
\[
\mb P_\theta\{\sqrt n\|\widetilde F-\widehat\theta_{0,n}\|_2>t\}
\leq nt^{-2}\mb E_\theta^{(n)}[\|\widetilde F-\widehat\theta_{0,n}\|_2^2] \leq Ct^{-2}(n\varepsilon_n^2+n^{-1})\to0,
\]
since $n\varepsilon_n^2\to0$. Hence $\sqrt n(\widetilde F-\widehat\theta_{0,n})\to0$ in $\mb P_\theta$-probability, which is the first asymptotic claim of the theorem. Writing $\widetilde F-\theta=(\widehat\theta_{0,n}-\theta)+(\widetilde F-\widehat\theta_{0,n})$ and using~\eqref{eq:uniform-mle-expansion} with $P_ns_\theta=n^{-1}\sum_{i=1}^ns_\theta(X_i)$, we have
\[
\sqrt n(\widetilde F-\theta) =  I(\theta)^{-1}\frac1{\sqrt n}\sum_{i=  1}^n s_\theta(X_i) +\sqrt nR_n(\theta) +\sqrt n(\widetilde F-\widehat\theta_{0,n}),
\]
where $\mb E_\theta^{(n)}[\|\sqrt nR_n(\theta)\|_2^2]\leq C/n$. Markov's inequality gives $\sqrt nR_n(\theta)\to0$ in $\mb P_\theta$-probability. The scores $s_\theta(X_i)$ are i.i.d., bounded and centered with covariance $I(\theta)$ under $\mb P_\theta$, so the multivariate central limit theorem \citep[Proposition~2.27]{van2000asymptotic} gives
\[
n^{-1/2}\sum_{i=1}^ns_\theta(X_i)\xrightarrow{d}\mathcal N\{0,I(\theta)\}.
\]
Since $I(\theta)^{-1}I(\theta)I(\theta)^{-1}=I(\theta)^{-1}$, the first term of the decomposition converges in distribution to $\mathcal N\{0,I(\theta)^{-1}\}$. The other two terms tend to zero in $\mb P_\theta$-probability, so Slutsky's lemma gives
$\sqrt n\{\widetilde F(X^n)-\theta\}\xrightarrow{d}\mathcal N\{0,I(\theta)^{-1}\}$.
\end{proof}

\subsubsection{Proofs of Proposition~\ref{prop:bayes-mle} and Theorem~\ref{thm:est-gen}}\label{app:proof-gen}

The approximation result alone does not control the error of a network trained on finitely many tasks. To obtain a generalization bound, we compare the empirical and population excess risks uniformly over the network class. The squared-error loss satisfies a variance bound in terms of its excess risk, which permits a fast rate; see Lemma~\ref{lem:fast-rate}. To apply that lemma, we also need to control the size of a covering of the network class. The following bound on the Lipschitz constant of the loss as a function of the network parameters allows us to construct such a covering from a net of the parameter ball.

\begin{lemma}[Lipschitz continuity in the network parameters]
\label{lem:param-lipschitz}
Consider a fixed architecture with a single refinement of~\eqref{eq:structured-class} with fixed input and state dimensions. Let $N_{\rm pos}$ count its scalar coefficient positions before repeating the shared score module over observations. Assume
\[
\max_{\text{modules }f}L(f)\leq   L_n ,\qquad N_{\rm pos}\leq   CW ,\qquad \dim(\omega)\leq   W ,\qquad \Gamma\geq  1 ,\qquad \|\omega\|_2\leq  \Gamma,
\]
and that every fixed coefficient of the network is bounded by $\Gamma$ in absolute value.
Put $\Xi_n=(1+L_n)\log\{e(W+1)(\Gamma+1)\}$. There is a loss Lipschitz constant $L_{f,n}$ such that, uniformly over all tasks with $\theta\in\Theta_0$,
\[
|\|F_\omega(X^n)-\theta\|_2^2 -\|F_{\omega'}(X^n)-\theta\|_2^2| \leq   L_{f,n}\|\omega-\omega'\|_2 ,\qquad \log(1+\Gamma L_{f,n})\leq   C\Xi_n.
\]
In particular, for the finite-smoothness architecture in Theorem~\ref{thm:est-gen},
\begin{equation}
 \log(1+\Gamma L_{f,n})\leq C\log^2(en).
 \label{eq:parameter-lipschitz-log}
\end{equation}
For the analytic architecture at accuracy $n^{-1}$, with $\Gamma\leq Cn^C$,
\begin{equation}
 \log(1+\Gamma L_{f,n})\leq C\log^3(en).
 \label{eq:analytic-parameter-lipschitz-log}
\end{equation}
In the first display, $C$ depends only on $d,d_X,T,\Theta_0$ and the fixed connections between the component networks. In~\eqref{eq:parameter-lipschitz-log} and~\eqref{eq:analytic-parameter-lipschitz-log}, $C$ depends in addition on the constants of the complexity bounds of the architecture and on the exponent of the polynomial radius bound, and in the analytic case on the extension rectangles, ellipse parameters and amplitude bounds.
\end{lemma}

The next lemma extends Lemma~\ref{lem:param-lipschitz} to a general RRDS network~\eqref{eq:rrds-init}--\eqref{eq:rrds-output}. It is used for model and variable selection, and its constants do not depend on the state and pooled dimensions.

\begin{lemma}[Lipschitz continuity in the parameters of an RRDS network]\label{lem:rrds-lipschitz}
Consider an RRDS architecture with $R$ refinements whose blocks $\phi_0,\rho_0,\phi,\rho$ are ReLU networks, with depth $L$ as in~\eqref{eq:rrds-complexity} and $W$ distinct trainable parameters $\omega$. Suppose that every affine row of every block has at most $N_{\rm pos}$ coefficient positions, counting fixed coefficients and repeated occurrences of a shared parameter, that $\|\omega\|_2\leq\Gamma$ with $\Gamma\geq1$, that all fixed coefficients are bounded by $\Gamma$ in absolute value, and that the observations satisfy $\|x\|_\infty\leq R_X$. Then, for all $\omega,\omega'$ in the parameter ball,
\[
\begin{gathered}
\sup_{x^n}\|z_\omega^{(R)}(x^n)-z_{\omega'}^{(R)}(x^n)\|_\infty\leq L_z\|\omega-\omega'\|_2,\\
L_z=(L+R)\{2(N_{\rm pos}+1)(\Gamma+1)\}^{L+R}(R_X+1).
\end{gathered}
\]
Consequently, if the output map is $\Psi(z)=\operatorname{softmax}\{\operatorname{clip}_B(a)\}$ for $K$ coordinates $a$ of the state, then $|\log q_{\omega,k}(x^n)-\log q_{\omega',k}(x^n)|\leq2L_z\|\omega-\omega'\|_2$ for every $k$ and $x^n$. If $N_{\rm pos}\leq CW$ and $R_X\leq C$, then $\log(1+\Gamma L_z)\leq C(1+L)\log\{e(W+1)(\Gamma+1)\}$, where $C$ depends only on $R$ and the constants in these two bounds.
\end{lemma}

\begin{proof}[Proof of Proposition~\ref{prop:bayes-mle}]

The proof uses the fact that the posterior score has mean zero, after including the prior score. Expanding the likelihood score about the constrained MLE then expresses the difference between the posterior mean and the MLE in terms of the prior score and a quadratic remainder. Averaging over the joint task distribution allows us to control the latter by the fourth moment of the ordinary MLE, without requiring a separate uniform posterior concentration argument.
In this proof, $C$ and $c$ depend only on $d$, $d_X$, $r$, $\ell$, $T$, $\Theta_0$ and $\tau=\operatorname{dist}(\operatorname{supp}\pi,\partial\Theta_0)>0$. Let $\Theta_\tau=\{\theta\in\Theta_0:\operatorname{dist}(\theta,\partial\Theta_0)\geq\tau\}$, so that $\operatorname{supp}\pi\subseteq\Theta_\tau\Subset\operatorname{int}(\Theta_0)$. Write $\ell_n(z)=\sum_i\log p_z(X_i)$, $\widehat\theta=\widehat\theta_{0,n}$ and $\mb E_{\rm post}$ for posterior expectation, so that $F^*=\mb E_{\rm post}[z]$. Since the log density is bounded, $Z_n=\int e^{\ell_n(z)}\pi(z)\,dz\in(0,\infty)$. Since $\pi\in C_c^1(\operatorname{int}(\Theta_0))$, integration by parts has no boundary term:
\[
 0=\int\nabla\{e^{\ell_n(z)}\pi(z)\}\,dz
 =\int e^{\ell_n(z)}\{\pi(z)\nabla\ell_n(z)+\nabla\pi(z)\}\,dz.
\]
Because $\pi\geq0$ is differentiable, $\nabla\pi=0$ on $\{\pi=0\}$, where we set $\nabla\log\pi=0$. Hence $\nabla\pi=\pi\nabla\log\pi$ everywhere, and the identity reads
\[
\int e^{\ell_n(z)}\pi(z)\nabla\ell_n(z)\,dz=-\int e^{\ell_n(z)}\pi(z)\nabla\log\pi(z)\,dz.
\]
Division by $Z_n$ yields
\begin{equation}\label{eq:posterior-score-identity}
 \mb E_{\rm post}[\nabla\ell_n(z)]=-\mb E_{\rm post}[\nabla\log\pi(z)].
\end{equation}

Let $\mathcal H_n^{\rm post}=\{\widehat\theta\in\operatorname{int}(\Theta_0),\ H_n(\widehat\theta)\succeq(\lambda_0/2)I_d\}$, an event that depends only on the data. On this event, $\nabla\ell_n(\widehat\theta)=0$ and the segment from $\widehat\theta$ to any $z\in\Theta_0$ lies in $\Theta_0$, so Taylor's formula and the Lipschitz bound on $H_n$ from Lemma~\ref{lem:consequences} give
\[
\nabla\ell_n(z)=  -nH_n(\widehat\theta)(z-\widehat\theta)+r_n(z),\qquad \|r_n(z)\|_2\leq Cn\|z-\widehat\theta\|_2^2.
\]
Taking posterior expectations and using~\eqref{eq:posterior-score-identity} and $F^*=\mb E_{\rm post}[z]$, we have
\[
\begin{gathered}
-\mb E_{\rm post}[\nabla\log\pi(z)]=-nH_n(\widehat\theta)(F^*-\widehat\theta)+\mb E_{\rm post}[r_n(z)],\\
F^*-\widehat\theta=n^{-1}H_n(\widehat\theta)^{-1}\bigl\{\mb E_{\rm post}[\nabla\log\pi(z)]+\mb E_{\rm post}[r_n(z)]\bigr\}
\end{gathered}
\]
on this event.
Since $\|H_n(\widehat\theta)^{-1}\|_{\rm op}\leq2/\lambda_0$ on $\mathcal H_n^{\rm post}$, the bound on $r_n$ gives
\[
\|F^*-\widehat\theta\|_2\ind_{\mathcal H_n^{\rm post}}
\leq\frac2{\lambda_0}\bigl\{n^{-1}\mb E_{\rm post}[\|\nabla\log\pi(z)\|_2]+C\mb E_{\rm post}[\|z-\widehat\theta\|_2^2]\bigr\}\ind_{\mathcal H_n^{\rm post}}.
\]
Squaring and applying Jensen's inequality to the two posterior expectations,
\[
\|F^*-\widehat\theta\|_2^2\ind_{\mathcal H_n^{\rm post}} \leq   C\{n^{-2}\mb E_{\rm post}[\|\nabla\log\pi(z)\|_2^2] +\mb E_{\rm post}[\|z-\widehat\theta\|_2^4]\}\ind_{\mathcal H_n^{\rm post}}.
\]

For every nonnegative measurable $q(z,x^n)$, the posterior/joint-law identity and Tonelli's theorem give
\[
 \mb E_{Q_n}\mb E_{\rm post}[q(z,X^n)]
 =\int\mb E_\theta^{(n)} q(\theta,X^n)\pi(\theta)\,d\theta.
\]
For $q(z,x^n)=\|\nabla\log\pi(z)\|_2^2$, the right-hand side is $\mathcal J(\pi)$. For $q(z,x^n)=\|z-\widehat\theta(x^n)\|_2^4$, it is $\int\mb E_\theta^{(n)}[\|\theta-\widehat\theta\|_2^4]\pi(\theta)\,d\theta\leq Cn^{-2}$ by~\eqref{eq:uniform-mle-expansion} on $\Theta_\tau$, since $\operatorname{supp}\pi\subseteq\Theta_\tau$. Hence $\mb E_{Q_n}[\|F^*-\widehat\theta\|_2^2\ind_{\mathcal H_n^{\rm post}}]\leq C\{1+\mathcal J(\pi)\}n^{-2}$.

The event $\mathcal H_n^{\rm post}$ contains the event $\mathcal H_n$ of Lemma~\ref{lem:consequences}, so this lemma on $\Theta_\tau$ gives $Q_n\{(\mathcal H_n^{\rm post})^c\}\leq\sup_{\theta\in\Theta_\tau}\mb P_\theta\{(\mathcal H_n^{\rm post})^c\}\leq Ce^{-cn}$. Both $F^*$ and $\widehat\theta$ lie in $\Theta_0$, the former by convexity, so $\|F^*-\widehat\theta\|_2\leq D_0$. Therefore
\[
\mb E_{Q_n}[\|F^*-\widehat\theta\|_2^2] \leq C\{1+\mathcal J(\pi)\}n^{-2}+CD_0^2e^{-cn} \leq C\{1+\mathcal J(\pi)\}n^{-2},
\]
where $C$ depends on the task distribution only through $\tau$.
\end{proof}

\begin{proof}[Proof of Theorem~\ref{thm:est-gen}]
In this proof, $C$ depends only on $d$, $d_X$, $r$, $\ell$, $T$, $\Theta_0$, $\tau=\operatorname{dist}(\operatorname{supp}\pi,\partial\Theta_0)$ and the radius constant $c_\Gamma$, and $\mathcal J(\pi)$ remains explicit.

\medskip
\noindent{\bf Step 1: The reference network and its parameter bounds.} Apply Theorem~\ref{thm:est-approx} at $a=b=\varepsilon_n$, and count the decoder coefficients equal to one among the trainable coordinates, so that the parameter vector $\omega_{0,n}$ of the constructed network $\widetilde F=F_{\omega_{0,n}}$ satisfies $\|\omega_{0,n}\|_2\geq1$. Since $n^{-1}\leq\varepsilon_n<1$, we have $\log(e/\varepsilon_n)\leq\log(en)$ and $\varepsilon_n^{-\kappa}\log(e/\varepsilon_n)\leq n^\kappa\log(en)$. By Theorem~\ref{thm:est-approx} and Lemma~\ref{lem:modules}, we have $W\leq C\varepsilon_n^{-\kappa}\log(e/\varepsilon_n)\leq Cn^\kappa\log(en)$, the module depths are at most $C\log(e/\varepsilon_n)\leq C\log(en)$, and the coefficients are bounded in absolute value by $C\varepsilon_n^{-C}\leq Cn^C$, with constants that do not depend on $\tau$ or $c_\Gamma$. Since $\dim(\omega_{0,n})\leq W$, it follows for $n\geq3$ that
\[
\|\omega_{0,n}\|_2\leq\sqrt W\,\|\omega_{0,n}\|_\infty\leq C\sqrt W\,n^C\leq Cn^C,
\]
again with constants that do not depend on $\tau$ or $c_\Gamma$. Hence there is a fixed $c_{\Gamma,0}\geq1$ such that, for every $c_\Gamma\geq c_{\Gamma,0}$, $n\geq3$ and $\Gamma\in[\|\omega_{0,n}\|_2,c_\Gamma n^{c_\Gamma}]$, the class $\mathcal F=\{F_\omega:\|\omega\|_2\leq\Gamma\}$ contains $\widetilde F$ and~\eqref{eq:parameter-lipschitz-log} holds. The architecture has at most $Cn^\kappa\log(en)$ coefficient positions: those of $g_{\rm init}$, $\mathsf s$ and $A$ by Lemma~\ref{lem:modules}, at most $C\log(en)$ for the multipliers and the identity connections, and a fixed number for $T$ and $\Pi_P$. Hence the assumption $N_{\rm pos}\leq CW$ of Lemma~\ref{lem:param-lipschitz} holds with $W$ equal to the size bound $Cn^\kappa\log(en)$ of the class. Here the fixed coefficients of the architecture are bounded by a constant $C$ that does not depend on $n$, and we apply Lemma~\ref{lem:param-lipschitz} with $\max\{\Gamma,C\}$ in place of $\Gamma$.

\medskip
\noindent{\bf Step 2: Approximation of the posterior mean.} Put $A_{\mathcal F}=\inf_{F\in\mathcal F}\mb E_{Q_n}[\|F-F^*\|_2^2]$. Since $\widetilde F\in\mathcal F$, we have
\[
A_{\mathcal F}\leq\mb E_{Q_n}[\|\widetilde F-F^*\|_2^2] \leq2\mb E_{Q_n}[\|\widetilde F-\widehat\theta_{0,n}\|_2^2]
+2\mb E_{Q_n}[\|\widehat\theta_{0,n}-F^*\|_2^2].
\]
Let $\Theta_\tau=\{\theta\in\Theta_0:\operatorname{dist}(\theta,\partial\Theta_0)\geq\tau\}$, a compact subset of $\operatorname{int}(\Theta_0)$ that contains $\operatorname{supp}\pi$ and depends only on $\Theta_0$ and $\tau$. Integrating~\eqref{eq:est-approx-mse} with $\Theta_1=\Theta_\tau$ over $\theta\sim\pi$ gives
\[
\mb E_{Q_n}[\|\widetilde F-\widehat\theta_{0,n}\|_2^2]
=\int\mb E_\theta^{(n)}[\|\widetilde F(X^n)-\widehat\theta_{0,n}\|_2^2]\,\pi(d\theta)
\leq C(\varepsilon_n^2+n^{-2}).
\]
Proposition~\ref{prop:bayes-mle} gives $\mb E_{Q_n}[\|\widehat\theta_{0,n}-F^*\|_2^2]\leq C\{1+\mathcal J(\pi)\}n^{-2}$, so $A_{\mathcal F}\leq C[\varepsilon_n^2+\{1+\mathcal J(\pi)\}n^{-2}]$.

\medskip
\noindent{\bf Step 3: Excess risk and comparison with the MLE.} Since $\theta\in\operatorname{supp}\pi\subset\Theta_0$ and the output projection keeps $F_\omega(X^n)\in\Theta_0$ for all $\|\omega\|_2\leq\Gamma$, Lemma~\ref{lem:fast-rate} applies with $B_\Theta=1\vee D_0^2$. By~\eqref{eq:parameter-lipschitz-log} and the last claim of Lemma~\ref{lem:fast-rate}, we have $\log\{1+2\Gamma M(1+L_{f,n})\}\leq C\{\log^2(en)+\log(eM)\}$. Substituting this bound into~\eqref{eq:fast-erm-proof} and absorbing $B_\Theta$ into $C$, we obtain an event $\mathcal G$ with $\mb P_{\mathcal D_M}(\mathcal G^c)\leq\delta$ on which every approximate empirical risk minimizer satisfying~\eqref{eq:approx-erm} obeys
\[
\mathcal R(\widehat F)-\mathcal R^*
\leq3A_{\mathcal F}+C\frac{W\{\log^2(en)+\log(eM)\}+\log(2/\delta)}{M}+2\varepsilon_{\rm opt}.
\]

Since the new task has law $Q_n$ and is independent of $\mathcal D_M$,
\[
r_n:=\mb E[\|\widehat F(X^n)-\widehat\theta_{0,n}(X^n)\|_2^2\mid\mathcal D_M]
=\int\|\widehat F(x^n)-\widehat\theta_{0,n}(x^n)\|_2^2\,Q_n(d\theta,dx^n),
\]
and~\eqref{eq:excess-mse} gives $\mb E[\|\widehat F-F^*\|_2^2\mid\mathcal D_M]=\mathcal R(\widehat F)-\mathcal R^*$. Since $\|\widehat F-\widehat\theta_{0,n}\|_2^2\leq2\|\widehat F-F^*\|_2^2+2\|F^*-\widehat\theta_{0,n}\|_2^2$, on $\mathcal G$ we have
\[
\begin{aligned}
r_n
 &\leq2\{\mathcal R(\widehat F)-\mathcal R^*\}
          +2\mb E_{Q_n}[\|F^*-\widehat\theta_{0,n}\|_2^2]\\
&\leq6A_{\mathcal F}+C\frac{W\{\log^2(en)+\log(eM)\}+\log(2/\delta)}M+4\varepsilon_{\rm opt}
          +2\mb E_{Q_n}[\|F^*-\widehat\theta_{0,n}\|_2^2]\\
&\leq C\left[\frac{W\{(\log n)^2+\log(eM)\}+\log(3/\delta)}M
                 +\varepsilon_n^2+\frac{1+\mathcal J(\pi)}{n^2}\right]+4\varepsilon_{\rm opt},
\end{aligned}
\]
where the last inequality uses the bound on $A_{\mathcal F}$, Proposition~\ref{prop:bayes-mle}, $\log^2(en)\leq C(\log n)^2$ for $n\geq3$ and $\log(2/\delta)\leq\log(3/\delta)$. This is~\eqref{eq:est-gen-bound}.

\medskip
\noindent{\bf Step 4: First-order equivalence and asymptotic normality.} To establish the asymptotic results, fix $\delta\in(0,1)$. On $\mathcal G$, $nr_n$ is at most $n$ times the right-hand side of~\eqref{eq:est-gen-bound}. The stated rates give $n\varepsilon_n^2\to0$, $nW\{(\log n)^2+\log(eM)\}/M\to0$ and $n\varepsilon_{\rm opt}\to0$. The second limit gives $n/M\to0$, so $n\log(3/\delta)/M\to0$ for fixed $\delta$, and $\{1+\mathcal J(\pi)\}/n\to0$. Hence $n$ times the right-hand side tends to zero, so $\limsup_n\mb P_{\mathcal D_M}(nr_n>\eta)\leq\delta$ for every $\eta>0$, and letting $\delta\downarrow0$ gives $nr_n\to0$ in training probability. Conditional Markov's inequality yields
\[
\mb P\{\sqrt n\|\widehat F-\widehat\theta_{0,n}\|_2>\gamma\mid\mathcal D_M\} \leq  \min\{1,nr_n/\gamma^2\}\xrightarrow{P_{\mathcal D_M}}0.
\]

To establish asymptotic normality, put $u_n(\theta,\mathcal D_M)=\mb E_\theta^{(n)}[\|\widehat F-\widehat\theta_{0,n}\|_2^2\mid\mathcal D_M]$, so that $\int u_n(\theta,\mathcal D_M)\,\pi(d\theta)=r_n$. For every function $f$ with $|f|\leq1$ and Lipschitz constant at most one, evaluating $\widehat F$ and $\widehat\theta_{0,n}$ on the same sample gives
\[
\begin{aligned}
&\bigl|\mb E_\theta^{(n)}[f\{\sqrt n(\widehat F-\theta)\}-f\{\sqrt n(\widehat\theta_{0,n}-\theta)\}\mid\mathcal D_M]\bigr|\leq\sqrt n\,\mb E_\theta^{(n)}[\|\widehat F-\widehat\theta_{0,n}\|_2\mid\mathcal D_M]
\leq\sqrt{nu_n(\theta,\mathcal D_M)}.
\end{aligned}
\]
Taking the supremum over $f$ gives
\[
b_n:=  d_{\rm BL}[\mathcal L\{\sqrt n(\widehat F-\theta)\mid\theta,\mathcal D_M\}, \mathcal L_\theta\{\sqrt n(\widehat\theta_{0,n}-\theta)\}] \leq  \sqrt{nu_n(\theta,\mathcal D_M)}.
\]
Markov's inequality in $\theta\sim\pi$ given $\mathcal D_M$ then gives
\[
\mb P_{\theta,\mathcal D_M}(b_n>\eta)\leq\mb P_{\theta,\mathcal D_M}\{nu_n(\theta,\mathcal D_M)>\eta^2\} \leq  \mb E_{\mathcal D_M}[\min\{1,nr_n/\eta^2\}]\to0,
\]
where the limit follows because $nr_n\to0$ in training probability and the integrand is bounded by one. For fixed $\theta\in\operatorname{supp}\pi$,~\eqref{eq:uniform-mle-expansion} and the central limit theorem give $a_n(\theta):=  d_{\rm BL}[\mathcal L_\theta\{\sqrt n(\widehat\theta_{0,n}-\theta)\}, \mathcal N\{0,I(\theta)^{-1}\}]\to0$, as in Step~4 of the proof of Theorem~\ref{thm:est-approx}. Since $0\leq a_n\leq2$, bounded convergence gives $\int a_n\,d\pi\to0$, so $a_n(\theta)\to0$ in $\pi$-probability. The triangle inequality for $d_{\rm BL}$ bounds the distance in the theorem by $b_n+a_n(\theta)$, which tends to zero in $P_{\theta,\mathcal D_M}$-probability.
\end{proof}

\subsubsection{Proofs of Lemma~\ref{lem:analytic-density} and Corollary~\ref{cor:analytic-est-gen}}\label{app:proof-analytic}

We first prove Lemma~\ref{lem:analytic-density} and then Corollary~\ref{cor:analytic-est-gen}. The extension condition~\eqref{eq:extension-main}, the affine map $A_{\mathcal R}$ and the regions $E_\varrho$ are defined in Appendix~\ref{app:analytic-extension}.

\begin{proof}[Proof of Lemma~\ref{lem:analytic-density}]
The proof has three parts. The main issue is to obtain a common complex neighborhood on which the relevant functions remain bounded and the information matrix remains invertible. We first extend $\bar\ell$ to a bounded holomorphic function on a complex neighborhood of $\mathcal R_X\times\mathcal R_\Theta$, which gives the claims for the log density and the score. We then extend the Fisher information by integrating the extended log density over $\mathcal X$. Finally, we use the uniform positive definiteness of the real information matrix to show that its complex extension remains invertible in a sufficiently small neighborhood. A Neumann series then gives a bounded holomorphic inverse. Write $m=d_X+d$ and $\mathcal R=\mathcal R_X\times\mathcal R_\Theta$, let $\|\cdot\|_\infty$ denote the maximum norm on $\mb C^m$ and on $\mb C^d$, and for $\delta>0$ put
\[
N_\delta=\{w\in\mb C^m:\operatorname{dist}_\infty(w,\mathcal R)<\delta\},\qquad
N_\delta^\Theta=\{\theta\in\mb C^d:\operatorname{dist}_\infty(\theta,\mathcal R_\Theta)<\delta\}.
\]

Since $\bar\ell$ is real analytic on $O_\ell$, every $a\in\mathcal R$ has a radius $r_a>0$ such that the real cube $\{x\in\mb R^m:\|x-a\|_\infty<r_a\}$ lies in $O_\ell$ and the Taylor series of $\bar\ell$ at $a$ converges absolutely on the polydisc $\Delta_a=\{w\in\mb C^m:\|w-a\|_\infty<r_a\}$ and equals $\bar\ell$ at its real points. Let $L_a$ be the sum of this series on $\Delta_a$, which is holomorphic and takes real values at real points. If $\Delta_a\cap\Delta_{a'}\neq\emptyset$, then this intersection is a product of intersections of two discs centered on the real axis. It is therefore convex, and it contains a nonempty open subset of $\mb R^m$, on which $L_a=\bar\ell=L_{a'}$. All partial derivatives of $L_a-L_{a'}$ vanish at a point of this real open set, so $L_a=L_{a'}$ on the connected set $\Delta_a\cap\Delta_{a'}$ by the identity theorem. The functions $L_a$ therefore define a single holomorphic function $L$ on the open set $\bigcup_{a\in\mathcal R}\Delta_a\supset\mathcal R$. Since $\mathcal R$ is compact, there is $\delta_0>0$ such that the closure of $N_{2\delta_0}$ is a compact subset of this open set. The closure of $N_{\delta_0}$ is then compact as well, and $L$ and its partial derivatives are continuous on it, so there is $B_0<\infty$ with
\[
\sup_{w\in N_{\delta_0}}\max_{|\nu|\leq2}|D^\nu L(w)|\leq B_0,\qquad
L(x,\theta)=\bar\ell(x,\theta)\quad((x,\theta)\in\mathcal R).
\]
For a rectangle with half side lengths at most $\bar h$ and $\varrho=1+t$ with $0<t\leq1$, the calculation in the proof of Lemma~\ref{lem:glm-inverse-extension} shows that every coordinate of a point of $A_{\mathcal R}^{-1}(E_\varrho^m)$ is within distance $\sqrt2\,t\bar h$ of the corresponding coordinate interval of the rectangle. Let $\bar h$ be the largest half side length of $\mathcal R$ and put $\varrho_1=1+\min\{1,\delta_0/(2\bar h)\}$. Then $A_{\mathcal R}^{-1}(E_{\varrho_1}^m)\subset N_{\delta_0}$, and $O=A_{\mathcal R}(N_{\delta_0})$ is an open set containing $E_{\varrho_1}^m$. The map $F_\ell=L\circ A_{\mathcal R}^{-1}$ is holomorphic on $O$, bounded by $B_0$, real on $[-1,1]^m$, and satisfies $F_\ell(A_{\mathcal R}(x,\theta))=\ell_\theta(x)$ for $x\in\mathcal X$ and $\theta\in\Theta_0$ by~\eqref{eq:analytic-log-density}. This is~\eqref{eq:extension-main} for the log density. Both $\bar\ell(x,\cdot)$ and $\ell_\cdot(x)$ are continuously differentiable near $\mathcal R_\Theta$ and agree on $\mathcal R_\Theta$, which is the closure of its interior, so $\nabla_\theta L(x,\theta)=s_\theta(x)$ for $x\in\mathcal X$ and $\theta\in\mathcal R_\Theta$. The same argument with $F_s=(\nabla_\theta L)\circ A_{\mathcal R}^{-1}$ gives~\eqref{eq:extension-main} for the score with constants $(\mathcal R,\varrho_1,B_0)$.

Let $\nu$ be the finite dominating measure on $\mathcal X$ in Section~\ref{sec:theory-estimation}. For $x\in\mathcal X$ and $\theta\in N_{\delta_0}^\Theta$ we have $(x,\theta)\in N_{\delta_0}$ because $\mathcal X\subseteq\mathcal R_X$. Define
\[
\widetilde I(\theta)=-\int_{\mathcal X}\nabla_\theta^2L(x,\theta)\,e^{L(x,\theta)}\,\nu(dx),\qquad\theta\in N_{\delta_0}^\Theta.
\]
For each $x$ the integrand is holomorphic in $\theta$, and its entries are bounded in modulus by $B_0e^{B_0}$ on $\mathcal X\times N_{\delta_0}^\Theta$. Since $\nu(\mathcal X)<\infty$, Fubini's theorem and Morera's theorem show that $\widetilde I$ is holomorphic on $N_{\delta_0}^\Theta$, with entries bounded by $B_0e^{B_0}\nu(\mathcal X)$. For real $\theta\in\mathcal R_\Theta\subset U$ we have $e^{L(x,\theta)}=p_\theta(x)$ on $\mathcal X$, and Lemma~\ref{lem:consequences} gives
\[
\widetilde I(\theta)=-P_\theta\nabla_\theta^2\log p_\theta=I(\theta)\in\mathbb S_{++}^d,\qquad\theta\in\mathcal R_\Theta.
\]

By Lemma~\ref{lem:consequences}, $I$ is continuous on $U$, so $\lambda_\ell=\min_{\theta\in\mathcal R_\Theta}\lambda_{\min}\{I(\theta)\}>0$ by compactness. The map $\widetilde I$ is uniformly continuous on the compact closure of $N_{\delta_0/2}^\Theta$. Hence there is $\delta_1\in(0,\delta_0/2]$ such that
\[
\|\widetilde I(\theta)-I(\theta')\|_{\rm op}\leq\lambda_\ell/2\qquad
\text{whenever }\theta'\in\mathcal R_\Theta\text{ and }\|\theta-\theta'\|_\infty\leq\delta_1.
\]
For $\theta\in N_{\delta_1}^\Theta$, choose such a $\theta'$ and write $Q=I(\theta')^{-1}\{\widetilde I(\theta)-I(\theta')\}$. Then $\|Q\|_{\rm op}\leq1/2$, so
\[
\widetilde I(\theta)^{-1}=\sum_{j=0}^\infty(-Q)^jI(\theta')^{-1},\qquad
\|\widetilde I(\theta)^{-1}\|_{\rm op}\leq2/\lambda_\ell .
\]
The entries of $\widetilde I^{-1}$ are quotients of polynomials in the entries of $\widetilde I$ with nonvanishing denominator $\det\widetilde I$, so $\widetilde I^{-1}$ is holomorphic on $N_{\delta_1}^\Theta$. It is real on $\mathcal R_\Theta$ and equals $I^{-1}$ there. Let $\bar h_\Theta$ be the largest half side length of $\mathcal R_\Theta$ and put $\varrho_2=1+\min\{1,\delta_1/(2\bar h_\Theta)\}$. As in the first part, $F_I=\widetilde I^{-1}\circ A_{\mathcal R_\Theta}^{-1}$ is holomorphic on the open set $A_{\mathcal R_\Theta}(N_{\delta_1}^\Theta)\supset E_{\varrho_2}^d$, real on $[-1,1]^d$, has entries bounded by $2/\lambda_\ell$, and satisfies $F_I(A_{\mathcal R_\Theta}(\theta))=I(\theta)^{-1}$ for $\theta\in\Theta_0$. The lemma follows with $\varrho_\ell=\min\{\varrho_1,\varrho_2\}$ and $\mathcal B_\ell=\max\{B_0,2/\lambda_\ell\}$.
\end{proof}

In the proof of Corollary~\ref{cor:analytic-est-gen}, the extension rectangles, their side lengths and the reciprocals of these lengths, the ellipse parameters and the amplitude bounds of $g_{\rm MoM}$, $s$ and $I^{-1}$ are those of the hypothesis and of Lemma~\ref{lem:analytic-density}. They are fixed independently of $n$ and the accuracy. We write $\varrho$ for the minimum of the ellipse parameters and $\mathcal B$ for the maximum of the amplitude bounds.

\begin{proof}[Proof of Corollary~\ref{cor:analytic-est-gen}]
In part~(a), $C$ depends only on $d$, $d_X$, $r$, $\ell$, $T$, $\Theta_0$ and the fixed extension rectangles, radii and amplitude bounds, and in the risk bound also on $\Theta_1$.

For $a,b\in(0,1)$, set $\varepsilon_b=b/(B_s+B_I+2)$. By~\eqref{eq:extension-main} and Lemma~\ref{lem:analytic-density}, the coordinates of $g_{\rm MoM}$, $s$ and $I^{-1}$ admit holomorphic extensions on $E_\varrho^d$, $E_\varrho^{d_X+d}$ and $E_\varrho^d$ bounded in modulus by $\mathcal B$. Apply Lemma~\ref{lem:analytic-approx} coordinatewise on $[-1,1]^d$, $[-1,1]^{d_X+d}$ and $[-1,1]^d$ with scalar accuracies $a/\sqrt d$ for the $d$ coordinates of $g_{\rm MoM}$, $\varepsilon_b/\sqrt d$ for the $d$ coordinates of $s$ and $\varepsilon_b/d$ for the $d^2$ entries of $I^{-1}$, and fold the affine maps $A_{\mathcal R}$ into the first layers. The resulting Euclidean errors for $g_{\rm MoM}$ and $s$ and Frobenius error for $I^{-1}$ are at most $a$, $\varepsilon_b$ and $\varepsilon_b$, which are the accuracies of Lemma~\ref{lem:modules}. Since the extension rectangles and the reciprocals of their side lengths are fixed, the coefficients remain bounded by $C$.

For each coordinate, Lemma~\ref{lem:analytic-approx} with input dimensions $d$, $d_X+d$ and $d$ gives depths at most $C\log^2(e\sqrt d/a)$, $C\log^2(e\sqrt d/\varepsilon_b)$ and $C\log^2(ed/\varepsilon_b)$, and at most $C\log^{d+1}(e\sqrt d/a)$, $C\log^{d_X+d+1}(e\sqrt d/\varepsilon_b)$ and $C\log^{d+1}(ed/\varepsilon_b)$ parameters. Since $d$, $B_s$ and $B_I$ are fixed, $\log(e\sqrt d/a)\leq C\log(e/a)$ and $\log(e\sqrt d/\varepsilon_b)\leq\log(ed/\varepsilon_b)\leq C\log(e/b)$. Multiplying by the numbers of coordinates $d$, $d$ and $d^2$ and assembling the component networks as at the end of Appendix~\ref{app:relu} gives
\[
W_g\leq Cd\log^{d+1}(e/a),\qquad W_{\mathsf s}\leq Cd\log^{d_X+d+1}(e/b),\qquad W_A\leq Cd^2\log^{d+1}(e/b),
\]
with component depths at most $C\log^2\{e/(a\wedge b)\}$ and coefficients bounded by $C$.

The matrix-vector product and~\eqref{eq:separate-module-accuracy} are as in the proof of Theorem~\ref{thm:est-approx}. The $d^2$ multipliers have depth at most $C\log(e/b)$ and at most $Cd^2\log(e/b)$ coefficients, which are bounded by $CR_\times^2\leq C$ by Lemma~\ref{lem:multiplication}, and carrying the inputs through the components costs at most $Cd\log^2(e/b)$ coefficients. At $a=b=\varepsilon$, since $d+1\leq d_X+d+1$ and $2\leq d_X+d+1$, adding the component sizes and depths gives
\[
W\leq W_g+W_{\mathsf s}+W_A+Cd^2\log(e/\varepsilon)+Cd\log^2(e/\varepsilon)\leq C\log^{d_X+d+1}(e/\varepsilon)
\]
and $L\leq C\log^2(e/\varepsilon)$, with component networks that do not depend on $n$ or $\Theta_1$. The statistical part of the proof of Theorem~\ref{thm:est-approx} uses the networks only through the accuracy of $g_{\rm init}$ on $\mathcal K$ and~\eqref{eq:separate-module-accuracy}, so~\eqref{eq:integrated-module-bound} gives $\sup_{\theta\in\Theta_1}\mb E_\theta^{(n)}[\|\widetilde F_\varepsilon-\widehat\theta_{0,n}\|_2^2]\leq C(\varepsilon^2+n^{-2})$, and the asymptotic statements follow as there. This proves part~(a).

In part~(b), $C$ depends in addition on $c_\Gamma$ and $\tau=\operatorname{dist}(\operatorname{supp}\pi,\partial\Theta_0)$. The proof of Theorem~\ref{thm:est-gen} applies with the following changes. At $a=b=n^{-1}$, part~(a) gives $L_n\leq C\log^2(en)$ and $W\leq C\log^{m+1}(en)$, with $m=d_X+d$. By its proof, the size bound of Lemma~\ref{lem:analytic-approx} counts every coefficient of the network, including the fixed connections. As in the proof of Theorem~\ref{thm:est-gen}, the architecture therefore has at most $C\log^{m+1}(en)$ coefficient positions, and the assumption $N_{\rm pos}\leq CW$ of Lemma~\ref{lem:param-lipschitz} holds with $W$ equal to this size bound. Since the coefficients are bounded by $C$, counting a decoder coefficient equal to one gives
\[
1\leq\|\omega_{0,n}\|_2\leq C\sqrt W\leq C\log^{(m+1)/2}(en)\leq Cn^{(m+1)/2},
\]
with constants that do not depend on $c_\Gamma$ or $\tau$. Hence a fixed sufficiently large $c_\Gamma$ gives $\widetilde F_{n^{-1}}\in\mathcal F$. Integrating the risk bound of part~(a) at $\varepsilon=n^{-1}$ over $\theta\sim\pi$, with $\Theta_1=\Theta_\tau$ as in that proof, and applying Proposition~\ref{prop:bayes-mle} gives
\[
\begin{aligned}
A_{\mathcal F}
&\leq2\mb E_{Q_n}[\|\widetilde F_{n^{-1}}-\widehat\theta_{0,n}\|_2^2]+2\mb E_{Q_n}[\|\widehat\theta_{0,n}-F^*\|_2^2]\\
&\leq Cn^{-2}+C\{1+\mathcal J(\pi)\}n^{-2}\leq C\{1+\mathcal J(\pi)\}n^{-2}.
\end{aligned}
\]
In Lemma~\ref{lem:fast-rate}, \eqref{eq:analytic-parameter-lipschitz-log} replaces~\eqref{eq:parameter-lipschitz-log}, so $\log^2(en)$ becomes $\log^3(en)$ in the covering term. On the event $\mathcal G$ of that lemma, the decomposition of $r_n$ in the proof of Theorem~\ref{thm:est-gen} gives~\eqref{eq:analytic-est-gen-bound}.

The two conditional convergence conclusions follow from $nr_n\to0$ as in the last paragraph of the proof of Theorem~\ref{thm:est-gen}, with the same $u_n$, $b_n$ and $a_n(\theta)$. Finally, take $M=\lceil n\log^{m+5}(en)\rceil$ and $\delta=n^{-2}$. Since $n\geq3$, we have $\log(eM)\leq C\log(en)\leq C\log^3(en)$ and $\log(3n^2)\leq C\log(en)$. Combined with $W\leq C\log^{m+1}(en)$, this gives $W\{\log^3(en)+\log(eM)\}\leq C\log^{m+4}(en)$. Since $M\geq n\log^{m+5}(en)$ and $\log(en)\geq1$, we obtain
\[
\frac{W\{\log^3(en)+\log(eM)\}+\log(3n^2)}{M}
\leq\frac{C\log^{m+4}(en)}{n\log^{m+5}(en)}+\frac{C\log(en)}{n\log^{m+5}(en)}
\leq\frac{C}{n\log(en)},
\]
which proves the sufficient task complexity bound.
\end{proof}

\subsubsection{Proof of Corollary~\ref{cor:dc}}\label{app:proof-dc}

\begin{proof}[Proof of Corollary~\ref{cor:dc}]
In this proof, $C$ depends only on $d$, $d_X$, $r$, $\ell$, $T$, $\Theta_0$, $\tau=\operatorname{dist}(\operatorname{supp}\pi,\partial\Theta_0)$ and $c_\Gamma$. Let $n=Jn_0$, let $P_{n_0}^{(j)}$ be the empirical law of block $j$, and put $\widehat\theta_j=\widehat\theta_{0,n_0}^{(j)}$ and $\bar\theta_0=J^{-1}\sum_{j=1}^J\widehat\theta_j$. We apply~\eqref{eq:uniform-mle-expansion} on the compact set $\Theta_\tau=\{\theta\in\Theta_0:\operatorname{dist}(\theta,\partial\Theta_0)\geq\tau\}$, to each block at sample size $n_0$ and to the full sample:
\[
\widehat\theta_j-\theta=I(\theta)^{-1}P_{n_0}^{(j)}s_\theta+R_{n_0}^{(j)}(\theta),\qquad
\widehat\theta_{0,n}-\theta=I(\theta)^{-1}P_ns_\theta+R_n(\theta).
\]
The leading terms in these expansions are averages of the same score function, and hence combine to give the full-sample score exactly, $J^{-1}\sum_jI(\theta)^{-1}P_{n_0}^{(j)}s_\theta=I(\theta)^{-1}P_ns_\theta$, so averaging the block expansions and subtracting the full-sample expansion gives
\[
\bar\theta_0-\widehat\theta_{0,n}=J^{-1}\sum_jR_{n_0}^{(j)}(\theta)-R_n(\theta).
\]
By Jensen's inequality and the remainder bounds of~\eqref{eq:uniform-mle-expansion}, uniformly over $\theta\in\Theta_\tau$,
\[
\mb E_\theta^{(n)}[\|\bar\theta_0-\widehat\theta_{0,n}\|_2^2]
\leq\frac2J\sum_{j=1}^J\mb E_\theta^{(n)}[\|R_{n_0}^{(j)}(\theta)\|_2^2]+2\mb E_\theta^{(n)}[\|R_n(\theta)\|_2^2]
 \leq C(n_0^{-2}+n^{-2}).
\]

By~\eqref{eq:dc-estimator}, $\widehat\theta_{\rm DC}-\bar\theta_0=J^{-1}\sum_j\{\widehat F(X_{(j)}^{n_0})-\widehat\theta_j\}$. Given $(\theta,\mathcal D_M)$, the blocks $X_{(1)}^{n_0},\ldots,X_{(J)}^{n_0}$ are i.i.d.\ from $\mb P_\theta^{(n_0)}$ and are processed by the same trained network. With $\theta\sim\pi$, each pair $(\theta,X_{(j)}^{n_0})$ therefore has law $Q_{n_0}$ and is independent of $\mathcal D_M$. Jensen's inequality gives
\[
\mb E[\|\widehat\theta_{\rm DC}-\bar\theta_0\|_2^2\mid\mathcal D_M] \leq   J^{-1}\sum_j\mb E[\|\widehat F(X_{(j)}^{n_0})-\widehat\theta_j\|_2^2\mid\mathcal D_M] =  r_{n_0}(\mathcal D_M),
\]
where $r_{n_0}$ is the quantity $r_n$ of the proof of Theorem~\ref{thm:est-gen} at sample size $n_0$.

By~\eqref{eq:est-gen-bound} at sample size $n_0$, on an event of training probability at least $1-\delta$, $nr_{n_0}$ is at most $n$ times the right-hand side of that bound. The stated rates give $nW\{(\log n_0)^2+\log(eM)\}/M\to0$, hence $n/M\to0$, together with $n\varepsilon_{n_0}^2\to0$ and $n\varepsilon_{\rm opt}\to0$, and $n\{1+\mathcal J(\pi)\}/n_0^2\to0$ because $n=o(n_0^2)$. As in the proof of Theorem~\ref{thm:est-gen}, this gives $nr_{n_0}\to0$ in training probability.

Since $\operatorname{supp}\pi\subseteq\Theta_\tau$ and the new task is independent of $\mathcal D_M$, the uniform bound above gives $\mb E[\|\bar\theta_0-\widehat\theta_{0,n}\|_2^2\mid\mathcal D_M]\leq C(n_0^{-2}+n^{-2})$. Since $\|\widehat\theta_{\rm DC}-\widehat\theta_{0,n}\|_2^2\leq2\|\widehat\theta_{\rm DC}-\bar\theta_0\|_2^2+2\|\bar\theta_0-\widehat\theta_{0,n}\|_2^2$ and $n/n_0^2=J/n_0\to0$, we have
\[
n\mb E[\|\widehat\theta_{\rm DC}-\widehat\theta_{0,n}\|_2^2\mid\mathcal D_M] \leq  2nr_{n_0}(\mathcal D_M)+C(n/n_0^2+n^{-1})\xrightarrow{P_{\mathcal D_M}}0.
\]
For each $\gamma>0$, conditional Markov's inequality bounds the probability that $\sqrt n\|\widehat\theta_{\rm DC}-\widehat\theta_{0,n}\|_2>\gamma$ by the minimum of one and $n\mb E[\|\widehat\theta_{\rm DC}-\widehat\theta_{0,n}\|_2^2\mid\mathcal D_M]/\gamma^2$. This bound converges to zero in training probability by the last display. Since it is bounded by one, its expectation also tends to zero, giving $\sqrt n(\widehat\theta_{\rm DC}-\widehat\theta_{0,n})\to0$ in probability. The limit law follows as in the last paragraph of the proof of Theorem~\ref{thm:est-gen}, with $\widehat F$ replaced by $\widehat\theta_{\rm DC}$ and $u_n(\theta,\mathcal D_M)=\mb E_\theta^{(n)}[\|\widehat\theta_{\rm DC}-\widehat\theta_{0,n}\|_2^2\mid\mathcal D_M]$, whose $\pi$-integral is bounded by $n^{-1}$ times the left-hand side of the last display.
\end{proof}

\subsubsection{Proof of Corollary~\ref{cor:bootstrap}}\label{app:proof-boot}

For the constructed estimator, bootstrap validity follows from the one-step comparison on the sample and on the resample. For a trained estimator, the proof writes the scaled error as a uniformly Lipschitz function of finitely many empirical averages, whose sample and resample limits are centered Gaussian laws with covariances $\Sigma$ and $2\Sigma$. A density comparison between these Gaussian laws transfers the vanishing error to the resample. This argument uses the boundedness of the stability constants in Corollary~\ref{cor:bootstrap}(b).

We write $P_n^*$ for the empirical measure of a size-$n$ resample
drawn with replacement from $P_n$, and $\mb P^*$ for its conditional
law. Define $\widehat\theta_{0,n}^*=\widehat\theta_{0,n}(X^{*n})$
by applying the same measurable maximizer rule over $\Theta_0$ to the
resample. Every resample belongs to $\mathcal X^n$, so this statistic
exists even when observations are repeated. A statement
$Y_n=o_{P^*}(1)$ in probability means that
$\mb P^*(\|Y_n\|_2>\epsilon)\to0$ in probability for every
$\epsilon>0$. Let $H_n^*(z)=-P_n^*\nabla_z^2\log p_z$, and let $\mathcal E_n^*$, $\rho_n^*$ and $\delta_n^*$ denote the event $\mathcal E_n$ and the errors $\rho_n$ and $\delta_n$ of Appendix~\ref{app:proof-approx} computed from the resample, that is, with $\bar T_n$, $\widehat\theta_{0,n}$ and $H_n$ replaced by $P_n^*T$, $\widehat\theta_{0,n}^*$ and $H_n^*$.

For background on conditional bootstrap convergence, see \citet[Chapter~23]{van2000asymptotic}. The following lemma establishes the uniform bootstrap expansion needed for $\widehat\theta_{0,n}$. We prove this expansion directly by first localizing the bootstrap estimator.

\begin{lemma}[Bootstrap expansion of the MLE]\label{lem:boot-mle}
Under Assumption~\ref{ass:regular}, at a fixed $\theta\in{\rm int}(\Theta_0)$,
\begin{equation}
 \sqrt n(\widehat\theta_{0,n}^*-\widehat\theta_{0,n})
 =I(\theta)^{-1}\sqrt n(P_n^*-P_n)s_\theta+o_{P^*}(1)
 \label{eq:bootstrap-mle-expansion}
\end{equation}
in $\mathbb P_\theta$-probability, and
\[
b_n^*(\theta):=  d_{\rm BL}\bigl[\mathcal L\{\sqrt n(\widehat\theta_{0,n}^*- \widehat\theta_{0,n})\mid X^n\},\mathcal N\{0,I(\theta)^{-1}\}\bigr] \xrightarrow{\mb P_\theta}0.
\]
For the resampled localization event and initial error, $\mathbb P^*((\mathcal E_n^*)^c)\xrightarrow{\mb P_\theta}0$, $\rho_n^*=O_{P^*}(n^{-1/2})$ and $\delta_n^*=o_{P^*}(1)$ in $\mathbb P_\theta$-probability. More precisely, for every compact $K\Subset\operatorname{int}\Theta_0$ and $\epsilon,\eta>0$,
\[
\begin{gathered}
\sup_{\theta\in K}\mathbb P_\theta\{b_n^*(\theta)>\epsilon\}\to0,\qquad \sup_{\theta\in K}\mathbb P_\theta\{\mathbb P^*((\mathcal E_n^*)^c)>\eta\}\to0,\\
\sup_{\theta\in K}\mathbb P_\theta\{\mathbb P^*(\delta_n^*>\epsilon)>\eta\}\to0,\qquad \lim_{M\to\infty}\limsup_{n\to\infty}\sup_{\theta\in K} \mathbb P_\theta\{\mathbb P^*(\sqrt n\rho_n^*>M)>\eta\}=  0.
\end{gathered}
\]
\end{lemma}

\begin{proof}[Proof of Corollary~\ref{cor:bootstrap}]
In this proof, $C,c$ depend only on $d$, $d_X$, $r$, $\ell$, $T$, $\Theta_0$ and $P$, and in part~(b) also on $\operatorname{dist}(\operatorname{supp}\pi,\partial\Theta_0)$ and $K_{\rm st}$.

For part~(a), fix $\theta\in\operatorname{int}(\Theta_0)$. On $\mathcal E_n^*$,~\eqref{eq:approx-bound} with $a=b=\varepsilon_n$, applied to the resample, gives
\[
\sqrt n\|\widetilde F(X^{*n})-\widehat\theta_{0,n}^*\|_2 \leq  \sqrt n\varepsilon_n+ C\sqrt n(\varepsilon_n+\rho_n^*)(\delta_n^*+\varepsilon_n+\rho_n^*).
\]
By Lemma~\ref{lem:boot-mle}, $\sqrt n(\varepsilon_n+\rho_n^*)=O_{P^*}(1)$ and $\delta_n^*+\varepsilon_n+\rho_n^*=o_{P^*}(1)$ in $\mathbb P_\theta$-probability, and $\sqrt n\varepsilon_n\to0$. Hence the right-hand side is $o_{P^*}(1)$ in $\mathbb P_\theta$-probability. Since also $\mathbb P^*((\mathcal E_n^*)^c)\to0$ in $\mathbb P_\theta$-probability, we have $\sqrt n\{\widetilde F(X^{*n})-\widehat\theta_{0,n}^*\}=o_{P^*}(1)$ in $\mathbb P_\theta$-probability. We write
\[
\sqrt n\{\widetilde F(X^{*n})-\widetilde F(X^n)\}
=\sqrt n(\widehat\theta_{0,n}^*-\widehat\theta_{0,n}) \quad+\sqrt n\{\widetilde F(X^{*n})-\widehat\theta_{0,n}^*\}
-\sqrt n\{\widetilde F(X^n)-\widehat\theta_{0,n}\}.
\]
By Theorem~\ref{thm:est-approx}, the last term tends to zero in $\mathbb P_\theta$-probability, and it does not depend on the resample. This yields
$\sqrt n\{\widetilde F(X^{*n})-\widetilde F(X^n)\} =  \sqrt n(\widehat\theta_{0,n}^*-\widehat\theta_{0,n})+o_{P^*}(1)$ in $\mathbb P_\theta$-probability.

Lemma~\ref{lem:boot-mle} gives $b_n^*(\theta)\to0$ in $\mathbb P_\theta$-probability, and the conditional bounded Lipschitz distance between random vectors $U$ and $U+V$ on the same bootstrap probability space is at most $\mb E^*[\min\{2,\|V\|_2\}]$. This follows by taking the supremum over the test functions defining $d_{\rm BL}$. If $V=o_{P^*}(1)$ in probability, this upper bound tends to zero in probability, since for any $\eta>0$ it is at most $\eta+2\mb P^*(\|V\|_2>\eta)$. This is the stated conditional limit.

For part~(b), let $\mathcal S_n$ be the stated training stability event, so $\mathbb P_{\mathcal D_M}(\mathcal S_n^c)\to0$, and let $V_n=\sqrt n\|\widehat F(X^n)-\widehat\theta_{0,n}\|_2$. Theorem~\ref{thm:est-gen} gives $V_n\to0$ in probability, so $\mathbb E[\min(1,V_n)]\to0$ by bounded convergence. We prove part~(b) in four steps.

\medskip
\noindent{\bf Step 1: Local equicontinuity.} Condition on $\mathcal D_M\in\mathcal S_n$ and on $\theta\in\operatorname{supp}\pi$. Put
\[
\begin{gathered}
G=\Pi_P\circ g_{\rm init},\qquad \mathsf s_v=\mathsf s(\cdot,v),\qquad m(v)=P_\theta \mathsf s_v,\\
a=G(P_\theta T),\qquad z=G(P_nT),\qquad z^*=G(P_n^*T).
\end{gathered}
\]
On $\mathcal S_n$, $G$ is $K_{\rm st}$-Lipschitz, so $\|z-a\|_2\leq K_{\rm st}\|P_nT-P_\theta T\|_2$ and, by the triangle inequality, $\|z^*-a\|_2\leq K_{\rm st}(\|P_n^*T-P_nT\|_2+\|P_nT-P_\theta T\|_2)$. Since $T$ is bounded, we have $\mathbb E[\|P_nT-P_\theta T\|_2^2]\leq C/n$ and, by the conditional resampling variance, $\mathbb E[\|P_n^*T-P_nT\|_2^2\mid X^n]\leq C/n$. Hence $\mathbb E[\|z-a\|_2^2]+\mathbb E[\|z^*-a\|_2^2] \leq C/n$. By Chebyshev's inequality, the event $\mathcal L_n=  \{\|z-a\|_2\leq   n^{-1/4},\ \|z^*-a\|_2\leq   n^{-1/4}\}$ satisfies $\mathbb P(\mathcal L_n^c)\leq n^{1/2}\{\mathbb E[\|z-a\|_2^2]+\mathbb E[\|z^*-a\|_2^2]\}\leq Cn^{-1/2}$.

Since $G$ takes values in $P$, we have $a\in P$. On $\mathcal S_n$, ${\rm Lip}(\mathsf s(x,\cdot))\leq K_{\rm st}$ on $P$ for every $x\in\mathcal X$, so $|\mathsf s_j(x,v)-\mathsf s_j(x,v')|\leq K_{\rm st}\|v-v'\|_2$ for all $v,v'\in P$ and $j\leq d$, and in particular $|\mathsf s_j(x,v)-\mathsf s_j(x,a)|\leq K_{\rm st}n^{-1/4}$ for $v\in P\cap B(a,n^{-1/4})$. Let $t_n=Cn^{-1/4}\sqrt{\log n}$, $h=n^{-2}$ and $u=t_n/\sqrt n$. For $n\geq 2$ and $C$ sufficiently large, we have $2K_{\rm st}n^{-3/2}\leq t_n/2$, so $u>2K_{\rm st}h$ and $(t_n-2K_{\rm st}n^{-3/2})^2\sqrt n\geq t_n^2\sqrt n/4=(C^2/4)\log n$. We apply Lemma~\ref{lem:net-concentration} to the $d$ classes $\{\mathsf s_j(\cdot,v)-\mathsf s_j(\cdot,a):v\in P\cap B(a,n^{-1/4})\}$ with radius $R=n^{-1/4}$, envelope $B=K_{\rm st}n^{-1/4}$, Lipschitz constant $L=K_{\rm st}$, mesh $h=n^{-2}$ and $u=t_n/\sqrt n$. We apply it under $\mathbb P_\theta$, conditionally on $\mathcal D_M$, and, conditionally also on $X^n$, under $\mathbb P^*$. This gives
\[
\begin{gathered}
 \sup_{v\in P\cap B(a,n^{-1/4}),\,j\leq d}
 |\sqrt n(P_n-P_\theta)(\mathsf s_j(\cdot,v)-\mathsf s_j(\cdot,a))|\leq t_n,\\
 \sup_{v\in P\cap B(a,n^{-1/4}),\,j\leq d}
 |\sqrt n(P_n^*-P_n)(\mathsf s_j(\cdot,v)-\mathsf s_j(\cdot,a))|\leq t_n
\end{gathered}
\]
outside events whose probabilities are each at most
\[
\begin{aligned}
&2d\Bigl(1+\frac{2n^{-1/4}}{n^{-2}}\Bigr)^d
 \exp\Bigl\{-\frac{n(t_n/\sqrt n-2K_{\rm st}n^{-2})^2}{2K_{\rm st}^2n^{-1/2}}\Bigr\}\\
&\qquad=2d(1+2n^{7/4})^d
 \exp\Bigl\{-\frac{(t_n-2K_{\rm st}n^{-3/2})^2\sqrt n}{2K_{\rm st}^2}\Bigr\}\\
&\qquad\leq 2d(1+2n^{7/4})^dn^{-C^2/(8K_{\rm st}^2)}\to0,
\end{aligned}
\]
where the limit holds for $C$ large enough that $C^2/(8K_{\rm st}^2)>7d/4$.

On $\mathcal L_n$ and outside both exceptional events, $z$ and $z^*$ lie in $P\cap B(a,n^{-1/4})$ since $G$ takes values in $P$. Evaluating the first bound at $v=z$ gives $|\sqrt n(P_n-P_\theta)(\mathsf s_j(\cdot,z)-\mathsf s_j(\cdot,a))|\leq t_n$ for every $j\leq d$. For the resample, writing $P_n^*-P_\theta=(P_n^*-P_n)+(P_n-P_\theta)$ and evaluating both bounds at $v=z^*$ gives, for every $j\leq d$,
\[
\begin{aligned}
|\sqrt n(P_n^*-P_\theta)(\mathsf s_j(\cdot,z^*)-\mathsf s_j(\cdot,a))|
&\leq|\sqrt n(P_n^*-P_n)(\mathsf s_j(\cdot,z^*)-\mathsf s_j(\cdot,a))|\\
&\quad+|\sqrt n(P_n-P_\theta)(\mathsf s_j(\cdot,z^*)-\mathsf s_j(\cdot,a))|\leq 2t_n.
\end{aligned}
\]
Since $t_n\to0$, $\mathbb P(\mathcal L_n^c)\leq Cn^{-1/2}\to0$ and the probabilities of the two exceptional events tend to $0$, we obtain
\begin{equation}\label{eq:bootstrap-local-equicontinuity}
 \sqrt n(P_n-P_\theta)(\mathsf s_z-\mathsf s_a)=o_P(1),\qquad
 \sqrt n(P_n^*-P_\theta)(\mathsf s_{z^*}-\mathsf s_a)=o_P(1),
\end{equation}
uniformly over $\theta\in\operatorname{supp}\pi$ and $\mathcal D_M\in\mathcal S_n$, because none of the constants or probability bounds depend on $\theta$ or $\mathcal D_M$.

\medskip
\noindent{\bf Step 2: Reduction to averages.} Define
\[
\begin{gathered}
\xi(x)=  (T(x)-P_\theta T,\ \mathsf s_a(x)-m(a),\ I(\theta)^{-1}s_\theta(x))\in\mathbb R^{3d},\qquad P_\theta\xi=  0,\\
\sup_x\|\xi(x)\|_2\leq   C.
\end{gathered}
\]
Here $P_\theta\xi=0$ because $m(a)=P_\theta\mathsf s_a$ and $P_\theta s_\theta=0$. The bound on $\xi$ holds because $T$ is bounded, $\mathsf s_a$ is bounded by the stability bound, and $I(\theta)^{-1}s_\theta$ is bounded uniformly over $\theta\in\operatorname{supp}\pi$. For $u=(u_1,u_2,u_3)$ put
\[
\begin{gathered}
z_n(u_1)=  G(P_\theta T+u_1/\sqrt n),\\
B_n(u)=  \sqrt n[\Pi_{\Theta_0}\{z_n(u_1)+ h(m(z_n(u_1))+u_2/\sqrt n,z_n(u_1))\}-\theta]-u_3.
\end{gathered}
\]
The map $z_n$ is $(K_{\rm st}/\sqrt n)$-Lipschitz, $m$ is $K_{\rm st}$-Lipschitz on $P$ because every $\mathsf s(x,\cdot)$ is, $h$ is $K_{\rm st}$-Lipschitz, and $\Pi_{\Theta_0}$ is nonexpansive. Hence
\[
\begin{aligned}
\|B_n(u)-B_n(v)\|_2
 &\leq\sqrt n\|z_n(u_1)-z_n(v_1)\|_2+K_{\rm st}\bigl\{\sqrt n\|m(z_n(u_1))-m(z_n(v_1))\|_2\\
&\qquad+\|u_2-v_2\|_2+\sqrt n\|z_n(u_1)-z_n(v_1)\|_2\bigr\}+\|u_3-v_3\|_2\\
&\leq(K_{\rm st}+K_{\rm st}^2+K_{\rm st}^3)\|u_1-v_1\|_2+K_{\rm st}\|u_2-v_2\|_2+\|u_3-v_3\|_2.
\end{aligned}
\]
By the Cauchy--Schwarz inequality and $K_{\rm st}^2\leq2(K_{\rm st}+K_{\rm st}^2+K_{\rm st}^3)$, the right-hand side is at most $(1+K_{\rm st}+K_{\rm st}^2+K_{\rm st}^3)\|u-v\|_2$. Since $|\min\{1,\|y\|_2\}-\min\{1,\|y'\|_2\}|\leq\|y-y'\|_2$, the function $f_n(u)=\min\{1,\|B_n(u)\|_2\}$ satisfies
$0\leq   f_n\leq  1$, $\operatorname{Lip}(f_n)\leq   1+K_{\rm st}+K_{\rm st}^2+K_{\rm st}^3$.

For the sample, write $(u_1,u_2,u_3)=\sqrt nP_n\xi$, so that $u_1=\sqrt n(P_nT-P_\theta T)$, $u_2=\sqrt n\{P_n\mathsf s_a-m(a)\}$ and $u_3=I(\theta)^{-1}\sqrt nP_ns_\theta$. Then $z_n(u_1)=G(P_nT)=z$ and
\[
P_n\mathsf s_z=m(z)+(P_n-P_\theta)\mathsf s_a+(P_n-P_\theta)(\mathsf s_z-\mathsf s_a)
=m(z)+u_2/\sqrt n+(P_n-P_\theta)(\mathsf s_z-\mathsf s_a).
\]
Since $\widehat F(X^n)=\Pi_{\Theta_0}\{z+h(P_n\mathsf s_z,z)\}$, the nonexpansiveness of $\Pi_{\Theta_0}$ and the Lipschitz bound on $h$ give
\[
\begin{aligned}
&\bigl\|\sqrt n\{\widehat F(X^n)-\widehat\theta_{0,n}\}-B_n(\sqrt nP_n\xi)\bigr\|_2\\
&\qquad\leq K_{\rm st}\bigl\|\sqrt n(P_n-P_\theta)(\mathsf s_z-\mathsf s_a)\bigr\|_2+\bigl\|\sqrt n(\widehat\theta_{0,n}-\theta)-u_3\bigr\|_2=o_P(1),
\end{aligned}
\]
where the first term is $o_P(1)$ by~\eqref{eq:bootstrap-local-equicontinuity} and the second by~\eqref{eq:uniform-mle-expansion}. For the resample, the same argument applies with $P_n^*$ and $z^*$ in place of $P_n$ and $z$, with the second part of~\eqref{eq:bootstrap-local-equicontinuity}, and with
\[
\sqrt n(\widehat\theta_{0,n}^*-\theta)=\sqrt n(\widehat\theta_{0,n}^*-\widehat\theta_{0,n})+\sqrt n(\widehat\theta_{0,n}-\theta)=I(\theta)^{-1}\sqrt nP_n^*s_\theta+o_P(1),
\]
which follows from~\eqref{eq:bootstrap-mle-expansion} and~\eqref{eq:uniform-mle-expansion}. Hence
$\min(1,V_n)=  f_n(\sqrt nP_n\xi)+o_P(1)$, $\min(1,V_n^*)=  f_n(\sqrt nP_n^*\xi)+o_P(1)$,
where $V_n^*=\sqrt n\|\widehat F(X^{*n})-\widehat\theta_{0,n}^*\|_2$. The remainders also tend to zero in mean, since both compared quantities lie in $[0,1]$.

\medskip
\noindent{\bf Step 3: Gaussian comparison.} By Step~2, $(P_\theta,\xi)$ belongs to the class $\mathfrak C_{q,B}$ of Lemma~\ref{lem:gauss-compare} with $q=3d$ and $B$ the bound on $\|\xi\|_2$, for every $\theta\in\operatorname{supp}\pi$ and $\mathcal D_M\in\mathcal S_n$. The function $f_n$ takes values in $[0,1]$ and has Lipschitz constant at most $\Lambda=1+K_{\rm st}+K_{\rm st}^2+K_{\rm st}^3$, so $|\mathbb Ef_n(U)-\mathbb Ef_n(N)|\leq\Lambda\,d_{\rm BL}\{\mathcal L(U),\mathcal L(N)\}$ for all random vectors $U$ and $N$. With $\Sigma=P_\theta\xi\xi^\top$, the two limits of Lemma~\ref{lem:gauss-compare} and~\eqref{eq:doubled-gaussian-comparison} give uniformly
\[
\begin{aligned}
\mathbb Ef_n(\sqrt nP_n^*\xi)
 &=\mathbb Ef_n\{N(0,2\Sigma)\}+o(1) \leq2^{d/2}[\mathbb Ef_n\{N(0,\Sigma)\}]^{1/3}+o(1)\\
&=2^{d/2}[\mathbb Ef_n(\sqrt nP_n\xi)]^{1/3}+o(1).
\end{aligned}
\]
The factor $2^{d/2}$ is the constant $2^{q/6}$ of~\eqref{eq:doubled-gaussian-comparison} with $q=3d$, the dimension of $\xi$. The last line also uses $|x^{1/3}-y^{1/3}|\leq|x-y|^{1/3}$ for $x,y\geq0$. By Step~2, given $\theta\in\operatorname{supp}\pi$ and $\mathcal D_M\in\mathcal S_n$, the expectations of $f_n(\sqrt nP_n^*\xi)$ and $f_n(\sqrt nP_n\xi)$ differ from those of $\min(1,V_n^*)$ and $\min(1,V_n)$ by $o(1)$, uniformly. We average over the task parameter and training, bound $\min(1,V_n^*)$ by one on $\mathcal S_n^c$, and apply Jensen's inequality to the concave map $x\mapsto x^{1/3}$. This gives
\[
\mathbb E[\min(1,V_n^*)] \leq  2^{d/2}\{\mathbb E[\min(1,V_n)]\}^{1/3} +\mathbb P_{\mathcal D_M}(\mathcal S_n^c)+o(1)\to0.
\]

\medskip
\noindent{\bf Step 4: Bootstrap approximation for the network.} Let $\mathcal C_n=\sigma(X^n,\theta,\mathcal D_M)$. Conditional Markov's inequality gives, for each $\epsilon>0$,
\[
\mathbb P\{\mathbb E[\min(1,V_n^*)\mid\mathcal C_n]>\epsilon\} \leq  \epsilon^{-1}\mathbb E[\min(1,V_n^*)]\to0.
\]
Define the two bootstrap laws, with $\widehat F$ held fixed, by
\[
\mathcal L_n^{\rm net} =  \mathcal L\![\sqrt n\{\widehat F(X^{*n})-\widehat F(X^n)\} \mid X^n,\mathcal D_M] ,\qquad \mathcal L_n^{\rm MLE} =  \mathcal L\![\sqrt n(\widehat\theta_{0,n}^*-\widehat\theta_{0,n}) \mid X^n,\mathcal D_M].
\]
The empirical resampling kernel satisfies
$\mathcal L(X^{*n}\mid X^n,\theta,\mathcal D_M) =  P_n^{\otimes n} =  \mathcal L(X^{*n}\mid X^n,\mathcal D_M)$.
Thus additional conditioning on $\theta$ leaves both laws unchanged. The two statistics are functions of the same resample, and their difference has norm at most $V_n^*+V_n$. For every function $f$ with $|f|\leq1$ and Lipschitz constant at most one, we have $|f(x)-f(y)|\leq\min\{2,\|x-y\|_2\}$, and $\min\{2,V_n+V_n^*\}\leq2\min(1,V_n)+2\min(1,V_n^*)$. Since $V_n$ is $\mathcal C_n$-measurable, taking the conditional expectation given $\mathcal C_n$ and the supremum over $f$ gives
\[
d_{\rm BL}(\mathcal L_n^{\rm net},\mathcal L_n^{\rm MLE}) \leq  2\min(1,V_n)+2\mathbb E[\min(1,V_n^*)\mid\mathcal C_n] \xrightarrow{P_{\theta,X^n,\mathcal D_M}}0.
\]
Here $\min(1,V_n)\to0$ in probability by Theorem~\ref{thm:est-gen}, and the second term tends to zero in probability by the first display of this step. The law $\mathcal L_n^{\rm MLE}$ does not depend on $\mathcal D_M$, so $d_{\rm BL}(\mathcal L_n^{\rm MLE},\mathcal N\{0,I(\theta)^{-1}\})=b_n^*(\theta)$, and the uniform statement of Lemma~\ref{lem:boot-mle} on the compact set $\operatorname{supp}\pi$ gives $b_n^*(\theta)\to0$ in probability when $\theta\sim\pi$. The triangle inequality yields
\[
 d_{\rm BL}(\mathcal L_n^{\rm net},\mathcal N\{0,I(\theta)^{-1}\})
 \leq d_{\rm BL}(\mathcal L_n^{\rm net},\mathcal L_n^{\rm MLE})
       +d_{\rm BL}(\mathcal L_n^{\rm MLE},\mathcal N\{0,I(\theta)^{-1}\})
 \xrightarrow{P_{\theta,X^n,\mathcal D_M}}0.
\]
\end{proof}

\subsubsection{Proof of Theorem~\ref{thm:theory-unbounded}}\label{app:proof-unbounded}

\begin{proof}[Proof of Theorem~\ref{thm:theory-unbounded}]
In this proof, $C$ depends only on $d$, $d_X$, $r$, $\alpha$, $\Theta_0$, $P$, $M_I$, $\mathcal K$ and the fixed ReLU network computing $T$. The construction is that of Theorem~\ref{thm:est-approx}, with inputs clipped to $\mathcal X_n$, with $\phi_0=T\circ\mathsf C_n$ computed exactly, and with accuracies
\[
 a=\varepsilon,\qquad e_s=e_I=\varepsilon/(CB_n),\qquad
 e_\times=\varepsilon/(Cd^{3/2}),\qquad R_\times=2+B_n+M_I.
\]
Each scalar component $f$ of $g_{\rm MoM}^{\rm ext}$, $s^{\rm ext}$ and $(I^{-1})^{\rm ext}$, composed with the affine map from the unit cube onto its approximation rectangle, has $C^\alpha$ or $C^r$ norm at most $CB_n$ by the definition of $B_n$, because the rectangles of $g_{\rm MoM}^{\rm ext}$ and $(I^{-1})^{\rm ext}$ are fixed. For a target accuracy $e$ of $f$, apply Lemma~\ref{lem:sobolev-approx} to $f/(CB_n)$, whose norm is at most one, at accuracy $e/(CB_n)$, and multiply the output by $CB_n$. The constants of Lemma~\ref{lem:sobolev-approx} then do not depend on $n$, and the affine normalizations and the output factor have coefficients bounded by $CB_n$, since $K_{{\rm geom},n}\leq B_n$. The lemma then gives at most $C(CB_n/e)^{m/s}\log(eCB_n/e)$ parameters for a component with input dimension $m$ and smoothness $s$. We take $e=a/\sqrt d$ for the $d$ coordinates of $g_{\rm MoM}^{\rm ext}$, with $(m,s)=(d,\alpha)$, $e=e_s/\sqrt d$ for the $d$ coordinates of $s^{\rm ext}$, with $(m,s)=(d_X+d,r)$, and $e=e_I/d$ for the $d^2$ entries of $(I^{-1})^{\rm ext}$, with $(m,s)=(d,r)$. Since $d$ is fixed, this gives
\[
 \|g_{\rm init}-g_{\rm MoM}\|_{\infty,2}\leq a,\quad
 \|\mathsf s-s\|_{\infty,2}\leq e_s,\quad
 \|A-I^{-1}\|_{\infty,\rm F}\leq e_I,
\]
with sizes of order $(B_n/\varepsilon)^{d/\alpha}$, $(B_n^2/\varepsilon)^{(d_X+d)/r}$ and $(B_n^2/\varepsilon)^{d/r}$, each up to the factor $\log(eB_n/\varepsilon)$.

Since $\|P_n\mathsf s(\cdot,z)\|_2\leq B_n+e_s\leq B_n+1$ and $\|A(z)\|_{\rm op}\leq M_I+e_I\leq M_I+1$, every multiplier input lies in $[-R_\times,R_\times]$. The $d^2$ multipliers add a matrix-vector error of at most $d^{3/2}e_\times$, and as in~\eqref{eq:separate-module-accuracy},
\[
 \|h(P_n\mathsf s,z)-I(z)^{-1}P_ns_z\|_2
 \leq d^{3/2}e_\times+(B_n+1)e_I+M_Ie_s\leq\varepsilon
\]
when $C\geq3+M_I$, because $d^{3/2}e_\times=\varepsilon/C$, $(B_n+1)e_I\leq2\varepsilon/C$ and $M_Ie_s\leq M_I\varepsilon/C$. Adding the complexity bounds gives
\[
\begin{aligned}
L&\leq C\log(eB_n/\varepsilon), W\leq C\bigl\{(B_n/\varepsilon)^{d/\alpha}
            +(B_n^2/\varepsilon)^{(d_X+d)/r}
            +(B_n^2/\varepsilon)^{d/r}+1\bigr\}
                  \log(eB_n/\varepsilon)\\
&\leq C(B_n^2/\varepsilon)^\kappa\log(eB_n/\varepsilon).
\end{aligned}
\]
The last inequality uses $B_n/\varepsilon\leq B_n^2/\varepsilon$ and that each exponent is at most $\kappa$.

Let $\mathcal T_n=\bigcap_{i=1}^n\{X_i\in\mathcal X_n\}$. On $\mathcal E_n\cap\mathcal T_n$, clipping leaves every observation unchanged. We argue as in Step~2 of the proof of Theorem~\ref{thm:est-approx}, where the Lipschitz constants of $I$ and $H_n$ on $\Theta_0$ are now bounded by $B_n$. With $z=\Pi_Pg_{\rm init}(\bar T_n)$, the bracket of the one-step identity is at most $\delta_n+2B_n(\varepsilon+\rho_n)$ in operator norm, so
\[
 \|\widetilde F-\widehat\theta_{0,n}\|_2
 \leq\varepsilon+\|z+I(z)^{-1}P_ns_z-\widehat\theta_{0,n}\|_2
 \leq\varepsilon+M_I(\varepsilon+\rho_n)\{\delta_n+2B_n(\varepsilon+\rho_n)\},
\]
which gives the stated deterministic bound. The union bound gives $\mathbb P_\theta\{(\mathcal E_n\cap\mathcal T_n)^c\}\leq\mathbb P_\theta(\mathcal E_n^c)+n\mathbb P_\theta(X_1\notin\mathcal X_n)\to0$, and on $\mathcal E_n\cap\mathcal T_n$ the stated rates imply
\[
\begin{aligned}
\sqrt n\|\widetilde F-\widehat\theta_{0,n}\|_2
 &\leq\sqrt n\varepsilon+C\sqrt n(\varepsilon+\rho_n)\delta_n
                  +CB_n\sqrt n(\varepsilon+\rho_n)^2\\
&=o_P(1)+O_P(1)o_P(1)+O_P(B_n/\sqrt n)=o_P(1).
\end{aligned}
\]
Here $\sqrt n\varepsilon\to0$, $\sqrt n(\varepsilon+\rho_n)=O_P(1)$, $\delta_n=o_P(1)$ and $B_n\sqrt n(\varepsilon+\rho_n)^2=(B_n/\sqrt n)\{\sqrt n(\varepsilon+\rho_n)\}^2=O_P(B_n/\sqrt n)$. The assumed original-model MLE limit and Slutsky's lemma give the stated Gaussian limit.
\end{proof}

\subsection{Proofs for model selection}\label{app:model-proofs}

We first give a comparison of approximate information criteria that also applies to variable selection. We then specialize to the model-selection network and derive likelihood and marginal likelihood bounds. These results are used to prove Proposition~\ref{prop:ms-bridge}, Theorem~\ref{thm:ms-approx}, Proposition~\ref{prop:ms-misspec}, and Theorem~\ref{thm:ms-gen}, in that order. The proof of the improved bounds for analytic models follows at the end of the subsection.
\subsubsection{Approximate information criteria and likelihood deficits}\label{app:selection-setup}

For now, we use a general finite collection of candidates; the candidates will later be either parametric families or supports in a sparse regression model.

Let $\mathcal I$ be a finite index set, let $\Theta_i$ be the compact convex
working parameter set for candidate $i$, and write $f_{i,\theta}(x)=  \log p_{i,\theta}(x)$, $\ell_{n,i}(\theta)=  \sum_{j=  1}^n f_{i,\theta}(X_j)$ and $R_i(\theta)=  D_{\mathrm{KL}}(\mb P_0\|\mb P_{i,\theta})$.
The candidate densities use the same dominating measure. Fix measurable estimates and nonnegative deterministic penalties satisfying $\widehat\theta_i\in\argmax_{\eta\in\Theta_i}\ell_{n,i}(\eta)$, $\widetilde\theta_i\in\Theta_i$ and $\operatorname{pen}(i)\geq  0$ for $i\in\mathcal I$.
Write $\operatorname{IC}_i=-\ell_{n,i}(\widehat\theta_i)+\operatorname{pen}(i)$ and $D_i=\ell_{n,i}(\widehat\theta_i)-\ell_{n,i}(\widetilde\theta_i)\geq0$. Assume $|f_{i,\theta}(x)|\le B$ for all $i\in\mathcal I$, $x\in\mathcal X$ and $\theta\in\Theta_i$, so that $D_i\leq2Bn$.
Suppose
$\max_{i\in\mathcal I}\sup_{x\in\mathcal X,\theta\in\Theta_i} |\varphi_i(x,\theta)-f_{i,\theta}(x)|\leq  \varepsilon_\ell$,
and define $nc_i=-\sum_{j=1}^n\varphi_i(X_j,\widetilde\theta_i)+\operatorname{pen}(i)$ and $\widetilde i\in\argmin_{i\in\mathcal I}c_i$. All selections use a fixed measurable rule to break ties.

We begin by decomposing an approximate criterion $nc_i$ into the exact information criterion $\operatorname{IC}_i$, the likelihood deficit $D_i$ of the fitted parameter and the likelihood approximation error $e_i$.

\begin{lemma}[Comparison of approximate criteria]\label{lem:criterion}
For every dataset, $nc_i=\operatorname{IC}_i+D_i-e_i$ with $|e_i|\leq n\varepsilon_\ell$. Consequently, for every $i$ and $\theta\in\Theta_i$,
\begin{equation}
-\ell_{n,\widetilde i}(\widetilde\theta_{\widetilde i})
 +\operatorname{pen}(\widetilde i)
\leq-\ell_{n,i}(\theta)+\operatorname{pen}(i)+D_i
 +2n\varepsilon_\ell.                                      \label{eq:criterion-basic}
\end{equation}
For a designated index $i_0$, put $A_{i_0}=D_{i_0}+2n\varepsilon_\ell$.
Then, for every $t\geq0$ and $i\ne i_0$,
$\{\widetilde i=  i,\ A_{i_0}\leq   t\} \subseteq\{\operatorname{IC}_i-\operatorname{IC}_{i_0}\leq   t\}$.
If
$\widehat\theta_i\in\operatorname{int}\Theta_i$, $\sup_{x\in\mathcal X,\theta\in\Theta_i} \|\nabla_\theta^2 f_{i,\theta}(x)\|_{\rm op}\leq  \Lambda$,
then $D_i\leq \frac{n\Lambda}{2}\|\widetilde\theta_i-\widehat\theta_i\|_2^2$.
\end{lemma}

\begin{proof}
\emergencystretch=3em
Let $e_i=\sum_j\{\varphi_i(X_j,\widetilde\theta_i)-f_{i,\widetilde\theta_i}(X_j)\}$. Then $\vert{}e_i\vert{} \le \sum_{j=1}^n \vert{}\varphi_i(X_j, \widetilde\theta_i) - f_{i,\widetilde\theta_i}(X_j)\vert{} \le n\varepsilon_\ell$ and $nc_i=  -\ell_{n,i}(\widetilde\theta_i)+\operatorname{pen}(i)-e_i =  \operatorname{IC}_i+D_i-e_i$.
Since $c_{\widetilde i}\leq c_i$ and $\ell_{n,i}(\widehat\theta_i)\geq\ell_{n,i}(\theta)$,
\[
\begin{aligned}
 -\ell_{n,\widetilde i}(\widetilde\theta_{\widetilde i})+\operatorname{pen}(\widetilde i)&= nc_{\widetilde i} + e_{\widetilde i}\le  nc_i + e_{\widetilde i}
 =-\ell_{n,i}(\widetilde\theta_i) + \operatorname{pen}(i) - e_i + e_{\widetilde i}\\
 &\le-\ell_{n,i}(\widetilde\theta_i)+\operatorname{pen}(i)+2n\varepsilon_\ell
 =-\ell_{n,i}(\widehat\theta_i)+D_i+\operatorname{pen}(i)+2n\varepsilon_\ell,
\end{aligned}
\]
which gives~\eqref{eq:criterion-basic}. Since $nc_{\widetilde i}\leq nc_{i_0}$, the identity $nc_i=\operatorname{IC}_i+D_i-e_i$ and $D_{\widetilde i}\geq0$ give
$\operatorname{IC}_{\widetilde i}-\operatorname{IC}_{i_0} \leq   D_{i_0}-D_{\widetilde i}+e_{\widetilde i}-e_{i_0} \leq   A_{i_0}$,
which gives the event inclusion. For the last bound, let $v=\widetilde\theta_i-\widehat\theta_i$. Convexity and interiority give
$\widehat\theta_i+tv\in\Theta_i$ for $0\leq   t\leq  1$ and $\nabla\ell_{n,i}(\widehat\theta_i)=  0$,
so Taylor's formula yields
\[
 D_i=-\int_0^1(1-t)v^\top\nabla^2\ell_{n,i}(\widehat\theta_i+tv)v\,dt
 \leq\frac{n\Lambda}{2}\|v\|_2^2,
\]
where the last inequality follows from
\[
\left\vert{}v^\top \nabla^2 \ell_{n,i}(\widehat\theta_i + t v) v\right\vert{} \le \sum_{j=1}^n \left\Vert{}\nabla_\theta^2 f_{i,\widehat\theta_i + t v}(X_j)\right\Vert{}_{\rm op} \Vert{}v\Vert{}_2^2 \le n\Lambda \Vert{}v\Vert{}_2^2 .
\]
\end{proof}

The next lemma bounds the likelihood deficit $D_k$ of the one-step fit under misspecification, where the moment and likelihood targets differ by the one-step discrepancy $\tau_k$.

\begin{lemma}[Likelihood deficit under misspecification]\label{lem:ms-deficit}
Suppose Assumption~\ref{ass:models}\textup{(i)} holds, and let $\mb P_0$ and candidate $k$ satisfy conditions \textup{(i)}--\textup{(iii)} preceding Proposition~\ref{prop:ms-misspec}, with $\theta_k^*$, $\theta_{m,k}$ and $\tau_k$ as defined there. Let $\widetilde\theta_k$ be the output of the one-step network of Theorem~\ref{thm:est-approx}, constructed with component accuracies $a=b=\varepsilon\in(0,1)$, and let $D_k$ be its likelihood deficit. Then $\frac1n\mb E_0[D_k]
\leq C\{\tau_k^2+\varepsilon^2+n^{-1}\}$.
\end{lemma}

For the generating model, the deterministic bound~\eqref{eq:approx-bound} makes the deficit of order $n^{-1}$.

\begin{lemma}[Likelihood deficit in the generating model]\label{lem:ms-target}
Suppose $\mb P_0=\mb P_{k_0,\theta_0}$ with
$\theta_0\in{\rm int}(\Theta_{0,k_0})$. For the construction with
$\varepsilon=n^{-1}$, $\mb E_0[D_{k_0}]\leq Cn^{-1}$. For every fixed compact $K_0\subset\operatorname{int}\Theta_{0,k_0}$, there exists $C_{K_0}<\infty$ such that $\sup_{\theta_0\in K_0}\mathbb E_{k_0,\theta_0}[D_{k_0}]\leq   C_{K_0}/n$.
More generally, for the construction with component accuracy $\varepsilon\in(0,1)$,
$\sup_{\theta_0\in K_0}\mathbb E_{k_0,\theta_0}[D_{k_0}]\leq C_{K_0}(n\varepsilon^2+n^{-1})$, and there is a constant $c_{K_0}>0$ such that
\[
n\varepsilon^2\leq c_{K_0}\quad\Longrightarrow\quad
\sup_{\theta_0\in K_0}\mathbb P_{k_0,\theta_0}^{(n)}(D_{k_0}>1)\leq C_{K_0}/n .
\]
Here $C_{K_0}$ and $c_{K_0}$ depend on the fixed candidate bounds and working set, and on $K_0$, including its positive distance from $\partial\Theta_{0,k_0}$.
\end{lemma}

The tail bound is needed for model recovery: the loss of likelihood from using a network estimate must be small relative to the information-criterion penalty. It holds at the accuracy $\varepsilon=n^{-1/2}/\log(en)$, for which the expectation bound alone would only give $\mathbb P(D_{k_0}>1)\leq C/\log^2(en)$.

\subsubsection{Network construction and marginal likelihood comparisons}

We now specialize the preceding criterion comparison to model selection with $K$ and the dimensions $d_k$ fixed.

The remaining lemmas concern the construction of Section~\ref{sec:theory-model} and the comparison of the posterior model probabilities with the BIC probabilities. We use the following notation. The support of $\pi_k$ is the rectangle $\Xi_k$, with $\tau=\min_k\operatorname{dist}(\Xi_k,\partial\Theta_{0,k})>0$, and $0<\underline\pi\leq\bar\pi<\infty$ bound the densities of the $\pi_k$ on their supports. Put $c_w=\max_{j,l}w_j/w_l$ and $K_{j,\rho}=\{\vartheta\in\Theta_{0,j}:\operatorname{dist}(\vartheta,\partial\Theta_{0,j})\geq\rho\}$, where $\rho\in(0,\tau]$ and $A_1$ are as in~\eqref{eq:ms-A1}, and let $\bar\Lambda$ be such that $\|\nabla_\theta^2\ell_{j,\theta}(x)\|_{\rm op}\le \bar\Lambda$ for all $j\in[K]$, $x\in\mathcal X$ and $\theta\in\Theta_{0,j}$.

For a generating pair $(k,\theta)$ and $j\ne k$, let $\operatorname{gap}_j(k,\theta)=\inf_{\vartheta\in\Theta_{0,j}}D_{\rm KL}(\mb P_{k,\theta}\|\mb P_{j,\vartheta})$. For a separated pair, $\mathcal I_+(k,\theta)$ is the set of competitors of the first kind in~\eqref{eq:weak-separation}, with representing parameter $\eta_j=\eta_j(k,\theta)\in\Xi_j$, which is unique by identifiability, and $b_j=d_j-d_k\geq1$. The remaining competitors form $\mathcal I_-(k,\theta)$ and satisfy $\operatorname{gap}_j(k,\theta)\geq A_1\log n/n$. We write $\mathcal J(k,\theta)=\{k\}\cup\mathcal I_+(k,\theta)$ and $\eta_k=\theta$, and for $j\in\mathcal J(k,\theta)$,
\[
\begin{gathered}
\varrho_j=\operatorname{dist}(\eta_j,\partial\Xi_j)\geq\min\{\varrho_k,\rho\},\qquad
U_{n,j}=n\operatorname{dist}(\widehat\theta_j,\Xi_j)^2,\\
\delta_j=\log m_j-\Bigl\{\ell_{n,j}(\widehat\theta_j)-\frac{d_j}2\log n\Bigr\},
\end{gathered}
\]
so that $\log(w_jm_j)=a_j^{\rm BIC}+\delta_j$. The selection setup above is used with $\mathcal I=[K]$, $\Theta_i=\Theta_{0,i}$, $\varepsilon_\ell=n^{-1}$, the one-step estimates $\widetilde\theta_k$ and the penalties $\operatorname{pen}(k)=(d_k/2)\log n+\log(1/w_k)\geq0$, so that $\operatorname{IC}_k=-a_k^{\rm BIC}$ and $nc_k=-\widetilde a_k$.

The first lemma writes the network of Section~\ref{sec:theory-model} in the form~\eqref{eq:rrds-init}--\eqref{eq:rrds-output}. Fix $n\geq3$ and $\varepsilon_n=n^{-1/2}/\log(en)$. For each $k$, let $g_{{\rm init},k}$, $S_k$ and $h_k$ be the networks of Theorem~\ref{thm:est-approx} for candidate $k$ at accuracy $\varepsilon_n$, let $T_k$ be its ReLU moment function (Lemma~\ref{lem:relu-moment}), and let $\varphi_k$ be a ReLU network with $\sup_{\mathcal X\times\Theta_{0,k}}|\varphi_k-\ell_k|\leq n^{-1}$. Write $D_\Sigma=\sum_kd_k$, the state as $z=(\vartheta^0,\theta,a)$ with $\vartheta_k^0,\theta_k\in\mb R^{d_k}$ and $a\in\mb R^K$, and the pooled vector as $u=(u_1^s,\ldots,u_K^s,u^\ell)$ with $u_k^s\in\mb R^{d_k}$ and $u^\ell\in\mb R^K$.

\begin{lemma}[The model-selection network]\label{lem:ms-rrds}
Let $R\geq2$, $d_z=2D_\Sigma+K$ and $d_u=D_\Sigma+K$, and define the blocks
\begin{align*}
\phi_0(x,z)&=\bigl(T_1(x),\ldots,T_K(x),\,0_K\bigr),\\
\rho_0(u,z)&=\bigl(\vartheta^{\rm m},\ \vartheta^{\rm m},\ 0_K\bigr),\qquad\vartheta_k^{\rm m}=\Pi_{\Theta_{0,k}}g_{{\rm init},k}(u_k^s)\quad(k\in[K]),\\
\phi(x,z)&=\bigl(S_1(x,\vartheta_1^0),\ldots,S_K(x,\vartheta_K^0),\ \varphi_1(x,\theta_1),\ldots,\varphi_K(x,\theta_K)\bigr),\\
\rho(u,z)&=\bigl(0_{D_\Sigma},\ \ \Pi_{\Theta_{0,k}}\{\vartheta_k^0+h_k(u_k^s,\vartheta_k^0)\}-\theta_k\ \ (k\in[K]),\ \ V(u^\ell)-a\bigr),
\end{align*}
where
\[
V(u^\ell)_k=\max\Bigl\{v_k-\max_{l\in[K]}v_l,\ -\log n\Bigr\},\qquad v_k=\log w_k+n\,u_k^\ell-\frac{d_k}2\log n .
\]
With the output map $\Psi(z)=\operatorname{softmax}\{\operatorname{clip}_{\log n}(a)\}$, this is an RRDS network $\widetilde F=\widetilde q$ of the form~\eqref{eq:rrds-init}--\eqref{eq:rrds-output}, and the following hold for every dataset.

\textup{(a)} For all $r\geq1$, $\theta_k^{(r)}=\widetilde\theta_k\in\Theta_{0,k}$, the one-step estimate of Theorem~\ref{thm:est-approx} for candidate $k$ at accuracy $\varepsilon_n$.

\textup{(b)} With $\widetilde a_k=\log w_k+\sum_{i=1}^n\varphi_k(X_i,\widetilde\theta_k)-(d_k/2)\log n$, the logits of Section~\ref{sec:theory-model} evaluated at $\widetilde\theta_k$, $a_k^{(r)}=\max\{\widetilde a_k-\max_l\widetilde a_l,-\log n\}$ for all $r\geq2$. In particular, the output is the same for every $R\geq2$, $\argmax_k\widetilde q_k=\argmax_k\widetilde a_k$, and $\widetilde q_k\geq(Kn)^{-1}$.

\textup{(c)} For $R=2$, $\widetilde F\in\mathcal F(L,W)$ with $L\leq C\log n$ and $W\leq CKn^{\kappa_{\max}'}\log n$, all coefficients are bounded by $Cn^C$, and every affine row has at most $CW$ coefficient positions.
\end{lemma}

The next lemma gives upper and lower bounds on the marginal likelihoods in terms of the maximized log likelihoods.

\begin{lemma}[Laplace bounds]\label{lem:ms-laplace}
Under Assumption~\ref{ass:models}, there are constants $C_0,C,c$ and $n_0$ and, for every $j\in[K]$ and $\eta\in K_{j,\rho}$, an event $\mathcal G_n(\eta)$ with $\sup_{\eta\in K_{j,\rho}}\mb P_{j,\eta}^{(n)}\{\mathcal G_n(\eta)^c\}\leq Ce^{-cn}$, such that for $n\geq n_0$, on $\mathcal G_n(\eta)$, $\widehat\theta_j\in\operatorname{int}(\Theta_{0,j})$ and
$-C_0-\bar\Lambda\,U_{n,j}\leq\delta_j\leq C_0$. Moreover, $m_j\leq e^{\ell_{n,j}(\widehat\theta_j)}$ for every dataset, and
\[
\sup_{\eta\in K_{j,\rho}}\mb P_{j,\eta}^{(n)}\bigl\{n\|\widehat\theta_j-\eta\|_2^2>s\bigr\}\leq Ce^{-cs}+Ce^{-cn},\qquad s\geq0 .
\]
\end{lemma}

The network logits $\widetilde a_k$ differ from $a_k^{\rm BIC}$ by the likelihood deficit $D_k$ and by the error of $\varphi_k$.

\begin{lemma}[Network logits and BIC logits]\label{lem:ms-logits}
For every dataset and every $k$, $\widetilde a_k=a_k^{\rm BIC}-D_k+e_k$ with $D_k\geq0$ and $|e_k|\leq1$. Moreover, there is $n_0$ such that, for $n\geq n_0$, every $j\in[K]$ and every $\eta\in K_{j,\rho}$, $\mb P_{j,\eta}^{(n)}(D_j>1)\leq C/n$.
\end{lemma}

\begin{proof}
The identity $-\widetilde a_k=nc_k = \operatorname{IC}_k + D_k - e_k= -a_k^{\rm BIC} + D_k - e_k$ is Lemma~\ref{lem:criterion} with $n\varepsilon_\ell=1$. The tail bound is the last assertion of Lemma~\ref{lem:ms-target} for candidate $j$ with the compact set $K_{j,\rho}$, whose hypothesis $n\varepsilon_n^2=\log^{-2}(en)\leq c_{K_{j,\rho}}$ holds for $n\geq n_0$.
\end{proof}

The next lemma collects the tail bounds for the comparisons between candidates. A candidate with a positive KL gap is selected with exponentially small probability, a larger candidate containing the generating distribution needs a likelihood gain of the order of its extra penalty, and for separated tasks both bounds hold uniformly.

\begin{lemma}[Likelihood comparisons between candidates]\label{lem:ms-compare}
Take the BIC penalties $\operatorname{pen}(j)=(d_j/2)\log n$, $j\in[K]$.

\textup{(a)} Suppose $\mb P_0=\mb P_{k_0,\theta_0}$ with $\theta_0\in\Theta_{0,k_0}$, and let $k$ be a candidate with $\gamma_{{\rm KL},k}=\inf_{\theta\in\Theta_{0,k}} D_{\mathrm{KL}}(\mb P_0\|\mb P_{k,\theta})>0$.
For every fixed $t<\infty$, there are constants $c,C>0$ such that
\[
\mb P_0^{(n)} \{\operatorname{IC}_k-\operatorname{IC}_{k_0}\leq t\} \leq Cn^{d_k}e^{-cn}.
\]
For a collection $\mathcal A$ of generating parameters and incorrect candidates with uniform likelihood envelopes and Lipschitz bounds,
\[
\inf_{(\theta_0,k)\in\mathcal A}\gamma_{{\rm KL},k}(\theta_0)>0
\quad\Longrightarrow\quad
\sup_{(\theta_0,k)\in\mathcal A} n^{-d_k}\mathbb P_{k_0,\theta_0}^{(n)} \{\operatorname{IC}_k-\operatorname{IC}_{k_0}\leq   t\} \leq   Ce^{-cn}.
\]

\textup{(b)} Suppose $\mb P_0=\mb P_{k,\theta_{0,k}}$ with $\theta_{0,k}\in{\rm int}(\Theta_{0,k})$, let $q=d_k$, and put $G_{n,k}=2\{\ell_{n,k}(\widehat\theta_k)-\ell_{n,k}(\theta_{0,k})\}$. For every fixed $a>0$, there are $c,C>0$ such that, for all sufficiently large $n$ and $0<w\leq a\log n$,
\begin{equation}
\mb P_0^{(n)}(G_{n,k}\geq w)\leq C(\log n)^{q/2}e^{-w/2}+Ce^{-cn/(\log n)^4}.
\label{eq:ms-gain-tail}
\end{equation}
For every fixed compact $K_k\subset\operatorname{int}\Theta_{0,k}$, the constants and the lower threshold on $n$ can be chosen so that~\eqref{eq:ms-gain-tail} holds uniformly over $\theta_{0,k}\in K_k$.

\textup{(c)} Under Assumption~\ref{ass:models} there is $n_0$ such that, for $n\geq n_0$, every separated pair $(k,\theta)$ with $\theta\in\Xi_k$ and every $j\in\mathcal I_-(k,\theta)$,
\[
\mb P_{k,\theta}^{(n)}\{a_k^{\rm BIC}-a_j^{\rm BIC}\leq\log n-\log c_w\}\leq C/n .
\]

\textup{(d)} In the setting of \textup{(c)}, for every $j\in\mathcal I_+(k,\theta)$, with $G_{n,j}=2\{\ell_{n,j}(\widehat\theta_j)-\ell_{n,j}(\eta_j)\}$,
\[
a_k^{\rm BIC}-a_j^{\rm BIC}\geq\tfrac12\{b_j\log n-G_{n,j}\}-\log c_w ,
\]
and $G_{n,j}$ satisfies~\eqref{eq:ms-gain-tail} under $\mb P_{k,\theta}^{(n)}$ with $q=d_j$ for $0<w\leq b_j\log n$.
\end{lemma}

The last lemma bounds the contributions from tasks near the boundary of the prior support and from events on which the preceding bounds fail.

\begin{lemma}[Boundary layer and envelope]\label{lem:ms-boundary}
Under Assumption~\ref{ass:models}, there is $n_0$ such that the following hold for $n\geq n_0$.

\textup{(a)} For every separated pair $(k,\theta)$ and $j\in\mathcal J(k,\theta)$, $\mb E_{k,\theta}^{(n)}[U_{n,j}]\leq Ce^{-cn\varrho_j^2}+Cne^{-cn}$, and for every $k$, $\int e^{-cn\operatorname{dist}(\theta,\partial\Xi_k)^2}\,\pi_k(d\theta)\leq\frac C{\sqrt n}$.

\textup{(b)} Let $L_1$ bound the Lipschitz constants of $\vartheta\mapsto\ell_{j,\vartheta}(x)$ on $\Theta_{0,j}$. For every generating pair $(k,\theta)$ with $\theta\in\Xi_k$ and every dataset, $m_k\geq c\,n^{-d_k}e^{\ell_{n,k}(\theta)-L_1}$ and
\[
D_{\rm KL}(q^*\|q^{\rm BIC})\leq C\log n+T_n,\qquad
T_n=\max_{j\in[K]}\sup_{\vartheta\in\Theta_{0,j}}\{\ell_{n,j}(\vartheta)-\ell_{n,k}(\theta)\}\geq0,
\]
where $\mb E_{k,\theta}^{(n)}[T_n^2]\leq C(\log n)^2$.
\end{lemma}

\subsubsection{Proof of Proposition~\ref{prop:ms-bridge}}\label{app:proof-ms-bridge}

\begin{proof}[Proof of Proposition~\ref{prop:ms-bridge}]
In this proof, $C,c$ depend only on $K$, $d_X$, $(d_k,r_k)_{k\in[K]}$, the candidate families and working sets, $c_w$, $\rho$, $\underline\pi$, $\bar\pi$ and the rectangles $\Xi_k$. Write $p=q^*$ and $r=q^{\rm BIC}$, and let $n\geq n_0$, where $n_0$ is at least the thresholds of Lemmas~\ref{lem:ms-laplace}--\ref{lem:ms-boundary} and is increased below. For $3\leq n<n_0$, the proof of Lemma~\ref{lem:ms-boundary}(b), with $c\,n^{-d_k}$ replaced by a positive constant, gives $\mb E_{\mb Q_n}[D_{\rm KL}(p\|r)]\leq C$, and the bound holds after enlarging $C$.

\medskip
\noindent{\bf Step 1: Localization and separated candidates.} Fix a separated pair $(k,\theta)$, write $\mathcal J=\mathcal J(k,\theta)$, $\mathcal I_-=\mathcal I_-(k,\theta)$, $Y_{j,\vartheta}=\log(p_{k,\theta}/p_{j,\vartheta})$ and $g_j=\operatorname{gap}_j(k,\theta)$, and define the event
\[
\mathcal G=\bigcap_{j\in\mathcal J}\mathcal G_n(\eta_j)\ \cap\bigcap_{j\in\mathcal I_-}\Bigl\{\inf_{\vartheta\in\Theta_{0,j}}P_nY_{j,\vartheta}>\frac{g_j}4\Bigr\},
\]
where $\mathcal G_n(\eta_j)$ is the event of Lemma~\ref{lem:ms-laplace}. By the union bound,
\[
\mb P_{k,\theta}^{(n)}(\mathcal G^c)
\leq\sum_{j\in\mathcal J}\mb P_{k,\theta}^{(n)}\bigl\{\mathcal G_n(\eta_j)^c\bigr\}
+\sum_{j\in\mathcal I_-}\mb P_{k,\theta}^{(n)}\Bigl\{\inf_{\vartheta\in\Theta_{0,j}}P_nY_{j,\vartheta}\leq\frac{g_j}4\Bigr\}.
\]
For every $j\in\mathcal J$, we have $\eta_j\in\Xi_j\subseteq K_{j,\rho}$ and $\mb P_{j,\eta_j}=\mb P_{k,\theta}$, so Lemma~\ref{lem:ms-laplace} gives $\mb P_{k,\theta}^{(n)}\{\mathcal G_n(\eta_j)^c\}\leq Ce^{-cn}$. Since $|\mathcal J|\leq K$, the first sum is at most $KCe^{-cn}$.

For the second sum, fix $j\in\mathcal I_-$. For every $\vartheta\in\Theta_{0,j}$, we have $|Y_{j,\vartheta}|=|\ell_{k,\theta}-\ell_{j,\vartheta}|\leq2B\leq B_Y$, the map $\vartheta\mapsto Y_{j,\vartheta}$ is $L_Y$-Lipschitz on $\Theta_{0,j}$, and $\mb E_{k,\theta}[Y_{j,\vartheta}]=D_{\rm KL}(\mb P_{k,\theta}\|\mb P_{j,\vartheta})\geq g_j$ by the definition of $g_j$. Hence Lemma~\ref{lem:lik-process} applies with $\mb P_0=\mb P_{k,\theta}$ and $R_j^*=g_j$. Since the working sets $\Theta_{0,j}$ are bounded and $d_j\leq d_{\max}$, we may take $N_j(n^{-1})\leq Cn^{d_{\max}}$. Substituting $h=n^{-1}$ and $v=g_j/4$ into~\eqref{eq:likelihood-gap} gives
\[
\begin{aligned}
\mb P_{k,\theta}^{(n)}\Bigl\{\inf_{\vartheta\in\Theta_{0,j}}P_nY_{j,\vartheta}\leq\frac{g_j}4\Bigr\}
&\leq N_j(n^{-1})\exp\Bigl[-\lambda_Yn\Bigl\{\frac{g_j}2-\frac{g_j}4-\frac{L_Y}n\Bigr\}\Bigr]\\
&=N_j(n^{-1})\exp\bigl\{-\lambda_Y(ng_j/4-L_Y)\bigr\}\\
&\leq Cn^{d_{\max}}\exp(-\lambda_Yng_j/4)
\leq Cn^{d_{\max}-\lambda_YA_1/4}\leq Cn^{-2},
\end{aligned}
\]
where the last two inequalities use $g_j\geq A_1\log n/n$, which gives $\exp(-\lambda_Yng_j/4)\leq n^{-\lambda_YA_1/4}$, and $A_1\geq4(d_{\max}+2)/\lambda_Y$ from~\eqref{eq:ms-A1}, which gives $d_{\max}-\lambda_YA_1/4\leq-2$. Summing over $j\in\mathcal I_-$, with $|\mathcal I_-|\leq K$, and combining with the bound on the first sum gives
\[
\mb P_{k,\theta}^{(n)}(\mathcal G^c)\leq KCe^{-cn}+KCn^{-2}\leq Cn^{-2},
\]
since $e^{-cn}\leq Cn^{-2}$.

\medskip
\noindent{\bf Step 2: Comparison among correctly specified candidates.} We now bound $D_{\rm KL}(p\|r)$ on $\mathcal G$. Put $t=p(\mathcal J^c)$ and $u=r(\mathcal J^c)$, and let $p^{\mathcal J}=p|_{\mathcal J}/(1-t)$ and $r^{\mathcal J}=r|_{\mathcal J}/(1-u)$ be the restrictions of $p$ and $r$ to $\mathcal J$. Since $p_j=w_jm_j/\sum_{l=1}^Kw_lm_l$, $\log(w_jm_j)=a_j^{\rm BIC}+\delta_j$ and $r=\operatorname{softmax}(a^{\rm BIC})$, we have $p^{\mathcal J}=\operatorname{softmax}(a_j^{\rm BIC}+\delta_j)_{j\in\mathcal J}$ and $r^{\mathcal J}=\operatorname{softmax}(a_j^{\rm BIC})_{j\in\mathcal J}$. Since $\mathcal G\subseteq\mathcal G_n(\eta_j)$ for every $j\in\mathcal J$, Lemma~\ref{lem:ms-laplace} gives $-C_0-\bar\Lambda U_{n,j}\leq\delta_j\leq C_0$ on $\mathcal G$ for all $j\in\mathcal J$. Applying Lemma~\ref{lem:kl-tools}(b) on $\mathcal J$ with $c=C_0$ and $u_j=\bar\Lambda U_{n,j}$ gives
\[
D_{\rm KL}(p^{\mathcal J}\|r^{\mathcal J})
\leq e^{2C_0}\bigl(1-\max_{j\in\mathcal J}p_j^{\mathcal J}\bigr)+\bar\Lambda\max_{j\in\mathcal J}U_{n,j}.
\]
When $\mathcal J=\{k\}$, we have $D_{\rm KL}(p^{\mathcal J}\|r^{\mathcal J})=0$, so the second term can be multiplied by $\ind\{|\mathcal J|\geq2\}$. Multiplying by $1-t=1-p(\mathcal J^c)\leq1$, using the second inequality of Lemma~\ref{lem:kl-tools}(a), $(1-t)\bigl(1-\max_{j\in\mathcal J}p_j^{\mathcal J}\bigr)\leq1-\max_jp_j$, and bounding $\max_{j\in\mathcal J}U_{n,j}\leq\sum_{j\in\mathcal J}U_{n,j}$, we obtain
\begin{equation}
\{1-p(\mathcal J^c)\}\,D_{\rm KL}(p^{\mathcal J}\|r^{\mathcal J})\leq e^{2C_0}\bigl(1-\max_jp_j\bigr)+\bar\Lambda\,\ind\{|\mathcal J|\geq2\}\sum_{j\in\mathcal J}U_{n,j}.
\label{eq:ms-bridge-J}
\end{equation}

\medskip
\noindent{\bf Step 3: Contributions of the remaining candidates.} Next we bound the contributions of $j\in\mathcal I_-=\mathcal J^c$ on $\mathcal G$. For $j\in\mathcal I_-$, by the definition of $\mathcal G$,
\[
\ell_{n,j}(\widehat\theta_j)-\ell_{n,k}(\theta)
=\sup_{\vartheta\in\Theta_{0,j}}\ell_{n,j}(\vartheta)-\ell_{n,k}(\theta) =-n\inf_{\vartheta\in\Theta_{0,j}}P_nY_{j,\vartheta}\leq-\frac{ng_j}4.
\]
By Lemma~\ref{lem:ms-laplace}, $m_j\leq e^{\ell_{n,j}(\widehat\theta_j)}$, and by Lemma~\ref{lem:ms-boundary}(b), $m_k\geq c\,n^{-d_k}e^{\ell_{n,k}(\theta)-L_1}$. Using $w_j/w_k\leq c_w$, $g_j\geq A_1\log n/n$ and $A_1\geq4(d_{\max}+2)$, we obtain
\[
p_j\leq\frac{w_jm_j}{w_km_k}
\leq\frac{c_we^{\ell_{n,j}(\widehat\theta_j)}}{c\,n^{-d_k}e^{\ell_{n,k}(\theta)-L_1}}
\leq Cn^{d_k}e^{-ng_j/4} \leq Cn^{d_{\max}-A_1/4}\leq\frac C{n^2}.
\]
Similarly, by~\eqref{eq:bic-logits} and $\ell_{n,k}(\widehat\theta_k)\geq\ell_{n,k}(\theta)$, together with $w_j/w_k\leq c_w$, $g_j\geq A_1\log n/n$ and $A_1\geq4(d_{\max}+2)$,
\[
\begin{aligned}
\frac{r_j}{r_k}&=e^{a_j^{\rm BIC}-a_k^{\rm BIC}}
=\frac{w_j}{w_k}n^{(d_k-d_j)/2}e^{\ell_{n,j}(\widehat\theta_j)-\ell_{n,k}(\widehat\theta_k)}\\
&\leq c_wn^{d_{\max}/2}e^{\ell_{n,j}(\widehat\theta_j)-\ell_{n,k}(\theta)}
\leq c_wn^{d_{\max}/2}e^{-ng_j/4}\leq\frac C{n^2}.
\end{aligned}
\]
Since $r_j\leq r_j/r_k$, summing over $j\in\mathcal I_-$ gives
\[
u=r(\mathcal J^c)=\sum_{j\in\mathcal I_-}r_j\leq\sum_{j\in\mathcal I_-}\frac{r_j}{r_k}\leq KCn^{-2}.
\]
Increase $n_0$ so that $KCn^{-2}\leq1/2$, which gives $u\leq1/2$ on $\mathcal G$. Moreover, since $|\ell_{l,\vartheta}(x)|\leq B$ for all $l\in[K]$, $x\in\mathcal X$ and $\vartheta\in\Theta_{0,l}$, the definition~\eqref{eq:bic-logits} gives $|a_j^{\rm BIC}-a_l^{\rm BIC}|\leq2Bn+C\log n$ for all $j,l\in[K]$, so
\[
\log\frac1{r_j}=\log\sum_{l=1}^Ke^{a_l^{\rm BIC}-a_j^{\rm BIC}}\leq\log K+2Bn+C\log n\leq Cn
\]
for every dataset. Applying the first inequality of Lemma~\ref{lem:kl-tools}(a) with $J=\mathcal J$, $t=p(\mathcal J^c)$ and $u=r(\mathcal J^c)$ gives
\[
D_{\rm KL}(p\|r)\leq(1-t)\,D_{\rm KL}(p^{\mathcal J}\|r^{\mathcal J})+2u+\sum_{j\notin\mathcal J}p_j\log\frac1{r_j}.
\]
On $\mathcal G$, the bounds on $u$, $p_j$ and $\log(1/r_j)$ give
\[
2u+\sum_{j\notin\mathcal J}p_j\log\frac1{r_j}
\leq2KCn^{-2}+K(Cn^{-2})(Cn)\leq\frac Cn.
\]
Combining this with~\eqref{eq:ms-bridge-J}, we obtain on $\mathcal G$ that
\begin{equation}
D_{\rm KL}(p\|r)\leq e^{2C_0}\bigl(1-\max_jp_j\bigr)+\bar\Lambda\,\ind\{|\mathcal J|\geq2\}\sum_{j\in\mathcal J}U_{n,j}+\frac Cn .
\label{eq:ms-bridge-good}
\end{equation}

\medskip
\noindent{\bf Step 4: Integration over datasets and tasks.} On $\mathcal G^c$, Lemma~\ref{lem:ms-boundary}(b) gives $D_{\rm KL}(p\|r)\leq C\log n+T_n$ for every dataset and $\mb E_{k,\theta}[T_n^2]\leq C(\log n)^2$, so $\mb E_{k,\theta}[(C\log n+T_n)^2]\leq C(\log n)^2$. By the Cauchy--Schwarz inequality and $\mb P_{k,\theta}^{(n)}(\mathcal G^c)\leq Cn^{-2}$,
\[
\mb E_{k,\theta}\bigl[D_{\rm KL}(p\|r)\ind_{\mathcal G^c}\bigr]
\leq\mb E_{k,\theta}\bigl[(C\log n+T_n)^2\bigr]^{1/2}\mb P_{k,\theta}^{(n)}(\mathcal G^c)^{1/2}
\leq\frac{C\log n}n .
\]
For each $j\in\mathcal J$, Lemma~\ref{lem:ms-boundary}(a) and $\varrho_j\geq\min\{\varrho_k,\rho\}$ give
\[
\mb E_{k,\theta}[U_{n,j}]
\leq Ce^{-cn\varrho_j^2}+Cne^{-cn}
\leq C\bigl(e^{-cn\varrho_k^2}+e^{-cn\rho^2}+ne^{-cn}\bigr).
\]
Since $|\mathcal J|\leq K$, summing over $j\in\mathcal J$ gives $\mb E_{k,\theta}\bigl[\sum_{j\in\mathcal J}U_{n,j}\bigr]\leq C\bigl(e^{-cn\varrho_k^2}+e^{-cn\rho^2}+ne^{-cn}\bigr)$. We take expectations in~\eqref{eq:ms-bridge-good} on $\mathcal G$ and add the bound on $\mathcal G^c$. Since $\varrho_k=\operatorname{dist}(\theta,\partial\Xi_k)$ and $e^{-cn\rho^2}+ne^{-cn}+n^{-1}\leq C\log n/n$, this gives, for every separated pair $(k,\theta)$,
\begin{equation}
\mb E_{k,\theta}[D_{\rm KL}(p\|r)]\leq e^{2C_0}\,\mb E_{k,\theta}\bigl[1-\max_jp_j\bigr]+C\,\ind\{|\mathcal J|\geq2\}e^{-cn\operatorname{dist}(\theta,\partial\Xi_k)^2}+\frac{C\log n}n .
\label{eq:ms-bridge-task}
\end{equation}

For a pair $(k,\theta)$ that is not separated, Lemma~\ref{lem:ms-boundary}(b) gives $D_{\rm KL}(p\|r)\leq C\log n+T_n$ and $\mb E_{k,\theta}[T_n]\leq\mb E_{k,\theta}[T_n^2]^{1/2}\leq C\log n$. Hence $\mb E_{k,\theta}[D_{\rm KL}(p\|r)]\leq C\log n$. We now integrate $\mb E_{k,\theta}[D_{\rm KL}(p\|r)]$ over $k\sim w$ and $\theta\sim\pi_k$, using~\eqref{eq:ms-bridge-task} on separated pairs and the bound $C\log n$ on the remaining pairs. Since $1-\max_jp_j=1-\max_jq_j^*(X^n)\geq0$ and ${\rm BE}_n=\mb E_{\mb Q_n}[1-\max_jq_j^*(X^n)]$, the first terms of~\eqref{eq:ms-bridge-task} integrate to at most $e^{2C_0}{\rm BE}_n$. For the second terms, bounding $\ind\{|\mathcal J|\geq2\}\leq1$ and extending the integral to all pairs $(k,\theta)$, the second claim of Lemma~\ref{lem:ms-boundary}(a) and $\sum_{k=1}^Kw_k=1$ give
\[
\sum_{k=1}^Kw_k\int e^{-cn\operatorname{dist}(\theta,\partial\Xi_k)^2}\,\pi_k(d\theta)\leq\frac C{\sqrt n}.
\]
Moreover, these terms vanish when $\mathcal I_+(k,\theta)=\varnothing$ for every separated pair, since then $|\mathcal J|=1$. The third term of~\eqref{eq:ms-bridge-task} integrates to at most $C\log n/n$. The pairs that are not separated have prior probability $\psi_n$ and hence contribute at most $C\psi_n\log n$. Combining these bounds gives
\[
\begin{aligned}
\mb E_{\mb Q_n}\bigl[D_{\rm KL}(q^*\|q^{\rm BIC})\bigr]
&\leq e^{2C_0}{\rm BE}_n+C\psi_n\log n+\frac C{\sqrt n}+\frac{C\log n}n\\
&\leq C\Bigl\{{\rm BE}_n+\psi_n\log n+\frac1{\sqrt n}\Bigr\},
\end{aligned}
\]
since $\log n/n\leq n^{-1/2}$. This proves the first bound, and when $\mathcal I_+(k,\theta)=\varnothing$ for every separated pair the term $C/\sqrt n$ is absent, so $n^{-1/2}$ is replaced by $\log n/n$. For the BIC probabilities $r^{[\log n]}$ with logits centered by their maximum and clipped at $-\log n$, Lemma~\ref{lem:kl-tools}(c) with $B=\log n$ gives $D_{\rm KL}(p\|r^{[\log n]})\leq D_{\rm KL}(p\|r)+Ke^{-\log n}=D_{\rm KL}(p\|r)+K/n$ for every dataset. Taking expectations under $\mb Q_n$ adds $K/n$ to the bound, which is absorbed into $C\log n/n$ and $C/\sqrt n$ for $n\geq3$, so the first bound also holds for $r^{[\log n]}$.

Finally, by Lemma~\ref{lem:ce-selection}(a), ${\rm BE}_n\leq\mb Q_n(\widetilde k\ne k)$ for the selection $\widetilde k$ of the network of Lemma~\ref{lem:ms-rrds}. The bound~\eqref{eq:ms-select-error} below, whose proof does not use the present proposition, then gives ${\rm BE}_n\leq C(\log n)^{d_{\max}/2}/\sqrt n+\psi_n$, which is the second bound of the proposition. If $\mathcal I_+(k,\theta)=\varnothing$ for every separated pair, the proof of~\eqref{eq:ms-select-error} gives $\mb P_{k,\theta}^{(n)}(\widetilde k\ne k)\leq C/n$ for every separated pair, and integrating as there gives ${\rm BE}_n\leq\mb Q_n(\widetilde k\ne k)\leq C/n+\psi_n$.
\end{proof}

\subsubsection{Proofs of Theorem~\ref{thm:ms-approx} and Proposition~\ref{prop:ms-misspec}}\label{app:proof-ms-approx}

\begin{proof}[Proofs of Theorem~\ref{thm:ms-approx} and Proposition~\ref{prop:ms-misspec}]
In this proof, $C,c$ depend only on the quantities listed in the proof of Proposition~\ref{prop:ms-bridge}. In the recovery bound and the oracle inequalities they depend in addition on $\mb P_0$, including the interior margin of $\theta_0$ and the KL gaps of the incorrect candidates, and in~\eqref{eq:ms-oracle-fast} on $B_0$. The constants in the network complexity bound do not depend on $\mb P_0$.

Take the network of Lemma~\ref{lem:ms-rrds} with $R=2$. Part~(c) of that lemma gives the form and the complexity bounds, and by parts~(a) and~(b) its fitted distribution is $\widetilde{\mb P}=\mb P_{\widetilde k,\widetilde\theta_{\widetilde k}}$ with $\widetilde k\in\argmax_k\widetilde a_k=\argmin_kc_k$. We prove the bounds in four steps.

\medskip
\noindent{\bf Step 1: Selection error.} Take $n_0$ large enough for Lemma~\ref{lem:ms-logits} and parts~(c) and~(d) of Lemma~\ref{lem:ms-compare} to hold, and for $\log n_0>6+2\log c_w$. Fix a separated pair $(k,\theta)$, let $n\geq n_0$, and write $\mb P=\mb P_{k,\theta}^{(n)}$. Put $\Delta_j=\widetilde a_j-\widetilde a_k$ for $j\ne k$. Since $\widetilde k$ maximizes $\widetilde a$, we have $\{\widetilde k\ne k\}\subseteq\bigcup_{j\ne k}\{\Delta_j\geq0\}$. By Lemma~\ref{lem:ms-logits}, $\widetilde a_j=a_j^{\rm BIC}-D_j+e_j$ with $D_j\geq0$ and $|e_j|\leq1$ for every $j$. Hence, for each $j\ne k$,
\[
\Delta_j
=(a_j^{\rm BIC}-a_k^{\rm BIC})-D_j+D_k+e_j-e_k \leq-(a_k^{\rm BIC}-a_j^{\rm BIC})+D_k+2,
\]
which gives $\Delta_j\leq3-(a_k^{\rm BIC}-a_j^{\rm BIC})$ on $\{D_k\leq1\}$. Moreover, $\theta\in\Xi_k\subseteq K_{k,\rho}$, so Lemma~\ref{lem:ms-logits} also gives $\mb P(D_k>1)\leq C/n$.

We bound $\mb P(\Delta_j\geq0)$ separately for $j\in\mathcal I_-(k,\theta)$ and $j\in\mathcal I_+(k,\theta)$. For $j\in\mathcal I_-(k,\theta)$, put $E_j=\{a_k^{\rm BIC}-a_j^{\rm BIC}\leq\log n-\log c_w\}$, so that $\mb P(E_j)\leq C/n$ by Lemma~\ref{lem:ms-compare}(c). On $\{D_k\leq1\}\setminus E_j$,
\[
\Delta_j\leq3-(a_k^{\rm BIC}-a_j^{\rm BIC})<3+\log c_w-\log n<0,
\]
where the last inequality uses $\log n\geq\log n_0>6+2\log c_w$ and $c_w\geq1$. Thus $\{\Delta_j\geq0\}\subseteq\{D_k>1\}\cup E_j$, and $\mb P(\Delta_j\geq0)\leq\mb P(D_k>1)+\mb P(E_j)\leq C/n$.
For $j\in\mathcal I_+(k,\theta)$, Lemma~\ref{lem:ms-compare}(d) gives $a_k^{\rm BIC}-a_j^{\rm BIC}\geq\tfrac12\{b_j\log n-G_{n,j}\}-\log c_w$. On $\{\Delta_j\geq0,D_k\leq1\}$,
\[
0\leq\Delta_j
\leq3-(a_k^{\rm BIC}-a_j^{\rm BIC}) \leq3+\log c_w-\tfrac12\{b_j\log n-G_{n,j}\},
\]
so $G_{n,j}\geq b_j\log n-6-2\log c_w$. Put $u_j=b_j\log n-6-2\log c_w$, so that $\{\Delta_j\geq0,D_k\leq1\}\subseteq\{G_{n,j}\geq u_j\}$. Since $b_j\geq1$, $c_w\geq1$ and $\log n\geq\log n_0>6+2\log c_w$, we have $0<u_j\leq b_j\log n$. By Lemma~\ref{lem:ms-compare}(d), $G_{n,j}$ satisfies~\eqref{eq:ms-gain-tail} under $\mb P$ with $q=d_j$ and $w=u_j$. Using $e^{-u_j/2}=e^3c_wn^{-b_j/2}$, we obtain
\[
\mb P(G_{n,j}\geq u_j)
\leq C(\log n)^{d_j/2}e^{-u_j/2}+Ce^{-cn/(\log n)^4} \leq C(\log n)^{d_j/2}n^{-b_j/2}+Ce^{-cn/(\log n)^4}.
\]
Since $\{\Delta_j\geq0\}\subseteq\{D_k>1\}\cup\{G_{n,j}\geq u_j\}$, combining this with $\mb P(D_k>1)\leq C/n$ gives
\[
\begin{aligned}
\mb P(\Delta_j\geq0)
&\leq\mb P(D_k>1)+\mb P(G_{n,j}\geq u_j) \leq\frac Cn+C(\log n)^{d_j/2}n^{-b_j/2}+Ce^{-cn/(\log n)^4}\\
&\leq C\frac{(\log n)^{d_{\max}/2}}{\sqrt n},
\end{aligned}
\]
where the last inequality uses $b_j\geq1$ and $d_j\leq d_{\max}$, together with $n^{-1}\leq n^{-1/2}$ and $e^{-cn/(\log n)^4}\leq C/n$.

The union bound over the $K-1$ indices $j\ne k$ gives $\mb P_{k,\theta}^{(n)}(\widetilde k\ne k)\leq C(\log n)^{d_{\max}/2}/\sqrt n$ for every separated pair $(k,\theta)$, and the bound is $C/n$ if $\mathcal I_+(k,\theta)=\varnothing$. Finally, $\mb Q_n(\widetilde k\ne k)$ is the integral of $\mb P_{k,\theta}^{(n)}(\widetilde k\ne k)$ over $k\sim w$ and $\theta\sim\pi_k$. On separated pairs we apply the bound just proved, and on the remaining pairs we bound the probability by one. Since the remaining pairs have prior probability $\psi_n$, we obtain
\begin{equation}
\mb Q_n(\widetilde k\ne k)\leq C\frac{(\log n)^{d_{\max}/2}}{\sqrt n}+\psi_n ,
\label{eq:ms-select-error}
\end{equation}
after enlarging $C$ to cover $3\leq n<n_0$. This is the second bound in~\eqref{eq:ms-approx-kl}.

\medskip
\noindent{\bf Step 2: Divergence bound.} We prove the first bound in~\eqref{eq:ms-approx-kl}. Unlike Step~1, which only requires correct ranking of the logits, this comparison controls the entire vector of probabilities. Write $p=q^*$, $r=q^{\rm BIC}$ and $s=\operatorname{softmax}(\widetilde a)$. By Lemma~\ref{lem:ms-rrds}(b), $\widetilde q=s^{[\log n]}$ in the notation of Lemma~\ref{lem:kl-tools}(c), and $\widetilde q_j\geq1/(Kn)$ for every $j\in[K]$. Hence $D_{\rm KL}(p\|\widetilde q)\leq\sum_jp_j\log(1/\widetilde q_j)\leq\log(Kn)$ for every dataset.

Let $n\geq n_0$, where $n_0$ is at least the thresholds in the proof of Proposition~\ref{prop:ms-bridge} and Lemma~\ref{lem:ms-logits} and is increased below. Fix a separated pair $(k,\theta)$, let $\mathcal J=\mathcal J(k,\theta)$, $\mathcal I_-=\mathcal I_-(k,\theta)$ and $\mathcal G$ be as in the proof of Proposition~\ref{prop:ms-bridge}, and put $\mathcal G'=\mathcal G\cap\{D_j\leq1\text{ for all }j\in\mathcal J\}$.

Every $j\in\mathcal J$ satisfies $\mb P_{j,\eta_j}=\mb P_{k,\theta}$ with $\eta_j\in\Xi_j\subseteq K_{j,\rho}$, so Lemma~\ref{lem:ms-logits} gives $\mb P_{k,\theta}^{(n)}(D_j>1)=\mb P_{j,\eta_j}^{(n)}(D_j>1)\leq C/n$. Together with $\mb P_{k,\theta}^{(n)}(\mathcal G^c)\leq Cn^{-2}$ from the proof of Proposition~\ref{prop:ms-bridge}, the union bound gives $\mb P_{k,\theta}^{(n)}(\mathcal G'^c)\leq\mb P_{k,\theta}^{(n)}(\mathcal G^c)+\sum_{j\in\mathcal J}\mb P_{k,\theta}^{(n)}(D_j>1)\leq C/n$.

We now bound $D_{\rm KL}(p\|\widetilde q)$ on $\mathcal G'$. The restrictions $p^{\mathcal J}$ and $s^{\mathcal J}$ of $p$ and $s$ to $\mathcal J$ have logits $a_j^{\rm BIC}+\delta_j$ and $\widetilde a_j$. By Lemma~\ref{lem:ms-logits}, $\widetilde a_j=a_j^{\rm BIC}-D_j+e_j$ with $D_j\geq0$ and $|e_j|\leq1$, so the logit difference is $(a_j^{\rm BIC}+\delta_j)-\widetilde a_j=\delta_j+D_j-e_j$. On $\mathcal G'$, for every $j\in\mathcal J$, Lemma~\ref{lem:ms-laplace} gives $-C_0-\bar\Lambda U_{n,j}\leq\delta_j\leq C_0$, while $0\leq D_j\leq1$ and $|e_j|\leq1$, so
\[
-(C_0+2)-\bar\Lambda U_{n,j}
\leq-C_0-1-\bar\Lambda U_{n,j}
\leq\delta_j+D_j-e_j
\leq C_0+2 .
\]
Hence Lemma~\ref{lem:kl-tools}(b) with $c=C_0+2$ and $u_j=\bar\Lambda U_{n,j}$ (with $D_{\rm KL}(p^{\mathcal J}\|s^{\mathcal J})=0$ when $\mathcal J=\{k\}$), together with the second claim of Lemma~\ref{lem:kl-tools}(a), gives
\[
\begin{aligned}
\{1-p(\mathcal J^c)\}\,D_{\rm KL}(p^{\mathcal J}\|s^{\mathcal J})
&\leq\{1-p(\mathcal J^c)\}\Bigl\{e^{2C_0+4}\bigl(1-\max_{j\in\mathcal J}p_j^{\mathcal J}\bigr)+\bar\Lambda\,\ind\{|\mathcal J|\geq2\}\max_{j\in\mathcal J}U_{n,j}\Bigr\}\\
&\leq e^{2C_0+4}\bigl(1-\max_jp_j\bigr)+\bar\Lambda\,\ind\{|\mathcal J|\geq2\}\sum_{j\in\mathcal J}U_{n,j}.
\end{aligned}
\]
Next, let $j\in\mathcal I_-=\mathcal J^c$. On $\mathcal G'$, since $D_j\geq0$, $D_k\leq1$ and $|e_j|,|e_k|\leq1$, we have
\[
\widetilde a_j-\widetilde a_k
=(a_j^{\rm BIC}-a_k^{\rm BIC})-D_j+D_k+e_j-e_k
\leq a_j^{\rm BIC}-a_k^{\rm BIC}+3 .
\]
Using $r_j/r_k\leq Cn^{-2}$ on $\mathcal G$ from the proof of Proposition~\ref{prop:ms-bridge}, we obtain
\[
s_j\leq\frac{s_j}{s_k}=e^{\widetilde a_j-\widetilde a_k}\leq e^3\frac{r_j}{r_k}\leq\frac C{n^2} .
\]
Summing over $j\in\mathcal I_-$ gives $s(\mathcal J^c)\leq KCn^{-2}$. Increase $n_0$ so that $s(\mathcal J^c)\leq KCn^{-2}\leq1/2$ on $\mathcal G'$ for all $n\geq n_0$. Moreover, the network log likelihoods are bounded in magnitude by $(B+1)n$, so $|\widetilde a_l-\widetilde a_j|\leq2(B+1)n+C\log n$ for all $j,l$, which gives $\log(1/s_j)\leq Cn$ for every dataset. Combined with $p_j\leq Cn^{-2}$ on $\mathcal G$ for $j\in\mathcal I_-$ from the proof of Proposition~\ref{prop:ms-bridge}, this gives $\sum_{j\notin\mathcal J}p_j\log(1/s_j)\leq C/n$. Since $\widetilde q=s^{[\log n]}$, applying Lemma~\ref{lem:kl-tools}(c) with clipping threshold $\log n$ and then Lemma~\ref{lem:kl-tools}(a) with $J=\mathcal J$ gives, on $\mathcal G'$,
\[
\begin{aligned}
D_{\rm KL}(p\|\widetilde q)
&\leq D_{\rm KL}(p\|s)+\frac Kn \leq\{1-p(\mathcal J^c)\}\,D_{\rm KL}(p^{\mathcal J}\|s^{\mathcal J})+2s(\mathcal J^c)+\sum_{j\notin\mathcal J}p_j\log\frac1{s_j}+\frac Kn\\
&\leq e^{2C_0+4}\bigl(1-\max_jp_j\bigr)+\bar\Lambda\,\ind\{|\mathcal J|\geq2\}\sum_{j\in\mathcal J}U_{n,j}+\frac Cn .
\end{aligned}
\]

We now take expectations and integrate over $(k,\theta)$. For a separated pair $(k,\theta)$, using $D_{\rm KL}(p\|\widetilde q)\leq\log(Kn)$ on $\mathcal G'^c$ and $\mb P_{k,\theta}^{(n)}(\mathcal G'^c)\leq C/n$ gives $\mb E_{k,\theta}[D_{\rm KL}(p\|\widetilde q)\ind_{\mathcal G'^c}]\leq\log(Kn)\,\mb P_{k,\theta}^{(n)}(\mathcal G'^c)\leq C\log n/n$. The expectation of $\sum_{j\in\mathcal J}U_{n,j}$ is bounded as in the proof of Proposition~\ref{prop:ms-bridge}. Hence~\eqref{eq:ms-bridge-task} holds for every separated pair with $\widetilde q$ in place of $r$ and $e^{2C_0+4}$ in place of $e^{2C_0}$. For a pair that is not separated, the bound $D_{\rm KL}(p\|\widetilde q)\leq\log(Kn)$ gives $\mb E_{k,\theta}[D_{\rm KL}(p\|\widetilde q)]\leq\log(Kn)$. Integrating over $k\sim w$ and $\theta\sim\pi_k$ as in the proof of Proposition~\ref{prop:ms-bridge}, the first terms integrate to at most $e^{2C_0+4}{\rm BE}_n$ because $1-\max_jq_j^*$ is nonnegative, the second terms integrate to at most $Cn^{-1/2}$ by Lemma~\ref{lem:ms-boundary}(a), the third term is at most $Cn^{-1/2}$, and the pairs that are not separated contribute at most $\psi_n\log(Kn)\leq C\psi_n\log n$. Enlarging $C$ to cover $3\leq n<n_0$ (using $D_{\rm KL}(p\|\widetilde q)\leq\log(Kn)$), we obtain, for all $n\geq3$,
\[
\mb E_{\mb Q_n}\bigl[D_{\rm KL}(q^*\|\widetilde q)\bigr]
\leq C\Bigl\{{\rm BE}_n+\psi_n\log n+\frac1{\sqrt n}\Bigr\}.
\]
Finally, Lemma~\ref{lem:ce-selection}(a) and~\eqref{eq:ms-select-error} give ${\rm BE}_n\leq\mb Q_n(\widetilde k\ne k)\leq Cb_n+\psi_n$. Since $n^{-1/2}\leq b_n$ and $\psi_n\leq\psi_n\log n$ for $n\geq3$, we have ${\rm BE}_n+\psi_n\log n+n^{-1/2}\leq C(b_n+\psi_n\log n)$, which gives the second inequality in the first line of~\eqref{eq:ms-approx-kl}.

\medskip
\noindent{\bf Step 3: Recovery.} We prove~\eqref{eq:ms-recovery-rate}. Put
$\mathcal I_+=\{k\ne k_0:\exists\eta\in\Theta_{0,k},\ \mb P_{k,\eta}=\mb P_0\}$ and $\mathcal I_-=[K]\setminus(\{k_0\}\cup\mathcal I_+)$. By the hypothesis of the theorem, every $k\in\mathcal I_+$ has $d_k>d_{k_0}$ and $\mb P_{k,\theta_k^*}=\mb P_0$ for some $\theta_k^*\in\operatorname{int}(\Theta_{0,k})$, and every $k\in\mathcal I_-$ has $\gamma_{{\rm KL},k}=\inf_{\Theta_{0,k}}R_k>0$.

We first reduce the selection error to comparisons of the information criteria. Since $n\varepsilon_\ell=1$, Lemma~\ref{lem:criterion} with $i_0=k_0$ has $A_{k_0}=D_{k_0}+2$, which is at most $3$ on $\{D_{k_0}\leq1\}$. With $t=3$ it gives
$\{\widetilde k=k,D_{k_0}\leq1\}\subseteq\{\operatorname{IC}_k-\operatorname{IC}_{k_0}\leq3\}$ for every $k\ne k_0$, so the union bound gives
\[
\mb P_0(\widetilde k\ne k_0)\leq\mb P_0(D_{k_0}>1)+\sum_{k\ne k_0}\mb P_0\{\operatorname{IC}_k-\operatorname{IC}_{k_0}\leq3\}.
\]
Since $\operatorname{pen}(k)=(d_k/2)\log n+\log(1/w_k)$ and $\log(w_k/w_{k_0})\leq\log c_w$,
\[
\begin{aligned}
\operatorname{IC}_k-\operatorname{IC}_{k_0}
&=\ell_{n,k_0}(\widehat\theta_{k_0})-\ell_{n,k}(\widehat\theta_k)+\frac{d_k-d_{k_0}}2\log n-\log\frac{w_k}{w_{k_0}}\\
&\geq\ell_{n,k_0}(\widehat\theta_{k_0})-\ell_{n,k}(\widehat\theta_k)+\frac{d_k-d_{k_0}}2\log n-\log c_w .
\end{aligned}
\]
For $k\in\mathcal I_+\cup\{k_0\}$, put $G_{n,k}=2\{\ell_{n,k}(\widehat\theta_k)-\log p_0^{(n)}(X^n)\}$. Then $G_{n,k_0}\geq0$ because $\theta_0\in\Theta_{0,k_0}$ and $\widehat\theta_{k_0}$ maximizes $\ell_{n,k_0}$ over $\Theta_{0,k_0}$. Hence, for $k\in\mathcal I_+$,
\[
\begin{aligned}
2(\operatorname{IC}_k-\operatorname{IC}_{k_0})
&=G_{n,k_0}-G_{n,k}+(d_k-d_{k_0})\log n-2\log\frac{w_k}{w_{k_0}}\\
&\geq(d_k-d_{k_0})\log n-G_{n,k}-2\log c_w .
\end{aligned}
\]
We use the first lower bound for $k\in\mathcal I_-$ and the second for $k\in\mathcal I_+$ in the union bound. This gives
\begin{equation}\label{eq:ms-split}
\begin{aligned}
 \mb P_0(\widetilde k\ne k_0)
 &\leq\mb P_0(D_{k_0}>1)
 +\sum_{k\in\mathcal I_-}\mb P_0\Bigl\{\ell_{n,k_0}(\widehat\theta_{k_0})-\ell_{n,k}(\widehat\theta_k)+\frac{d_k-d_{k_0}}2\log n\leq3+\log c_w\Bigr\}\\
 &\quad+\sum_{k\in\mathcal I_+}
 \mb P_0\{G_{n,k}\geq(d_k-d_{k_0})\log n-6-2\log c_w\}.
\end{aligned}
\end{equation}
We bound the three terms in turn. For the first, $n\varepsilon_n^2=\log^{-2}(en)\to0$, so the condition $n\varepsilon_n^2\leq c_{K_0}$ of Lemma~\ref{lem:ms-target} with $K_0=\{\theta_0\}$ holds for all large $n$, and its tail bound gives $\mb P_0(D_{k_0}>1)\leq C/n$ for all large $n$. For the second, Lemma~\ref{lem:ms-compare}(a) with $t=3+\log c_w$ and the finiteness of $\mathcal I_-$ give the bound $C\sum_{k\in\mathcal I_-}n^{d_k}e^{-cn}\leq Ce^{-cn}$. For the third, let $k\in\mathcal I_+$ and put $b_k=d_k-d_{k_0}\geq1$ and $u_k=b_k\log n-6-2\log c_w$. Since $c_w\geq1$, we have $0<u_k\leq b_k\log n$ for all large $n$. Since $\mb P_{k,\theta_k^*}=\mb P_0$, Lemma~\ref{lem:ms-compare}(b) applies with $\theta_{0,k}=\theta_k^*$, $q=d_k$, $a=b_k$ and $w=u_k$, and $e^{-u_k/2}=e^3c_wn^{-b_k/2}$ gives
\[
\begin{aligned}
\mb P_0\{G_{n,k}\geq u_k\}
 &\leq C(\log n)^{d_k/2}e^{-u_k/2}+Ce^{-cn/(\log n)^4}\\
&\leq C(\log n)^{d_k/2}n^{-b_k/2}+Ce^{-cn/(\log n)^4} \leq C(\log n)^{d_{\max}/2}n^{-1/2}+Cn^{-1},
\end{aligned}
\]
where the last inequality uses $b_k\geq1$, $d_k\leq d_{\max}$ and $e^{-cn/(\log n)^4}\leq Cn^{-1}$. Since $e^{-cn}\leq Cn^{-1}$ and $n^{-1}\leq(\log n)^{d_{\max}/2}n^{-1/2}$, combining the three bounds in~\eqref{eq:ms-split}, and covering the finitely many smaller $n$ by enlarging $C$, proves~\eqref{eq:ms-recovery-rate}. The third term is absent when $\mathcal I_+=\varnothing$, and the bound is then at most $C/n$.

\medskip
\noindent{\bf Step 4: Oracle inequalities.} Since $\widetilde{\mb P}=\mb P_{\widetilde k,\widetilde\theta_{\widetilde k}}$, we have $D_{\rm KL}(\mb P_0\|\widetilde{\mb P})=R_{\widetilde k}(\widetilde\theta_{\widetilde k})$, so it suffices to bound $\mb E_0[R_{\widetilde k}(\widetilde\theta_{\widetilde k})]$.

For~\eqref{eq:ms-oracle}, put $\Delta_{n,j}(\theta)=(P_n-P_0)f_{j,\theta}$ and $\mathcal Z_n=\max_{j\in[K]}\sup_{\theta\in\Theta_{0,j}}|\Delta_{n,j}(\theta)|$. Fix $k\in[K]$. Since $n\varepsilon_\ell=1$, $\operatorname{pen}(k)=(d_k/2)\log n+\log(1/w_k)$ and $\operatorname{pen}(\widetilde k)\geq0$, dividing~\eqref{eq:criterion-basic} with comparator $(k,\theta_k^*)$ by $n$, using $\ell_{n,j}(\theta)/n=P_nf_{j,\theta}$ and dropping $\operatorname{pen}(\widetilde k)/n$ gives
\[
-P_nf_{\widetilde k,\widetilde\theta_{\widetilde k}}
\leq-P_nf_{k,\theta_k^*}+\frac{d_k\log n}{2n}
+\frac{\log(1/w_k)}n+\frac{D_k}n+\frac2n .
\]
Because $\mb E_0[|\log p_0(X_1)|]<\infty$ and the candidate log densities $f_{j,\theta}$ are bounded, we have $R_j(\theta)=\mb E_0[\log p_0(X_1)]-P_0f_{j,\theta}$ for all $j\in[K]$ and $\theta\in\Theta_{0,j}$. Since $|\Delta_{n,j}(\theta)|\leq\mathcal Z_n$, we obtain
\[
R_{\widetilde k}(\widetilde\theta_{\widetilde k})
=\mb E_0[\log p_0(X_1)]-P_nf_{\widetilde k,\widetilde\theta_{\widetilde k}}
+\Delta_{n,\widetilde k}(\widetilde\theta_{\widetilde k}) \leq\mb E_0[\log p_0(X_1)]-P_nf_{\widetilde k,\widetilde\theta_{\widetilde k}}+\mathcal Z_n ,
\]
and similarly $-P_nf_{k,\theta_k^*}=-P_0f_{k,\theta_k^*}-\Delta_{n,k}(\theta_k^*)\leq-P_0f_{k,\theta_k^*}+\mathcal Z_n$. Combining these two bounds with the preceding inequality from~\eqref{eq:criterion-basic} and $\mb E_0[\log p_0(X_1)]-P_0f_{k,\theta_k^*}=R_k(\theta_k^*)$ gives
\[
R_{\widetilde k}(\widetilde\theta_{\widetilde k})
\leq R_k(\theta_k^*)+\frac{d_k\log n}{2n}+\frac{\log(1/w_k)}n
+\frac{D_k}n+\frac2n+2\mathcal Z_n .
\]
Since $\varepsilon_n=n^{-1/2}/\log(en)\in(0,1)$ satisfies $\varepsilon_n^2\leq n^{-1}$, Lemma~\ref{lem:ms-deficit} with $\varepsilon=\varepsilon_n$ gives $\mb E_0[D_k]/n\leq C\{\tau_k^2+\varepsilon_n^2+n^{-1}\}\leq C(\tau_k^2+n^{-1})$. Applying~\eqref{eq:likelihood-additive} with $\mathcal I=[K]$ gives $\mb E_0[\mathcal Z_n]\leq C\{(\log K+d_{\max})/n\}^{1/2}$. Because the weights $w_k$ are fixed and positive, $\log(1/w_k)/n\leq C/n$. Taking expectations in the bound on $R_{\widetilde k}(\widetilde\theta_{\widetilde k})$ and using $n^{-1}\leq\{(\log K+d_{\max})/n\}^{1/2}$ gives
\[
\begin{aligned}
\mb E_0[R_{\widetilde k}(\widetilde\theta_{\widetilde k})]
&\leq R_k(\theta_k^*)+\frac{d_k\log n}{2n}+C\tau_k^2+\frac Cn+2\mb E_0[\mathcal Z_n]\\
&\leq R_k(\theta_k^*)+\frac{d_k\log n}{2n}+C\tau_k^2
+C\left(\frac{\log K+d_{\max}}n\right)^{1/2}.
\end{aligned}
\]
Since $R_{\widetilde k}(\widetilde\theta_{\widetilde k})=D_{\rm KL}(\mb P_0\|\widetilde{\mb P})$ and $R_k(\theta_k^*)=D_{\rm KL}(\mb P_0\|\mb P_{k,\theta_k^*})$, minimizing over $k\in[K]$ proves~\eqref{eq:ms-oracle}.

For~\eqref{eq:ms-oracle-fast}, put $Y_{j,\eta}=\log(p_0/p_{j,\eta})$ for $j\in[K]$ and $\eta\in\Theta_{0,j}$. By assumption, $|Y_{j,\eta}|\leq B_0$ holds $\mb P_0$-almost surely for all $j\in[K]$ and $\eta\in\Theta_{0,j}$. Moreover, $Y_{j,\eta}-Y_{j,\eta'}=-(f_{j,\eta}-f_{j,\eta'})$ for all $\eta,\eta'\in\Theta_{0,j}$. Hence the conditions of the second part of Lemma~\ref{lem:lik-process} hold with $\mathcal I=[K]$, $B_Y=\max\{1,B_0\}$ and $L_Y$ a bound on the candidate scores over $\mathcal X\times\Theta_{0,j}$, $j\in[K]$. Each working set $\Theta_{0,j}$ is bounded, so the covering bound in that lemma gives $\log\sum_{j=1}^K N_j(n^{-1})\leq C\log n$. Put $Z=\max_{j\in[K]}\sup_{\eta\in\Theta_{0,j}}\{R_j(\eta)/2-P_nY_{j,\eta}\}$. By the definition of $Z$, we have $\tfrac12 R_{\widetilde k}(\widetilde\theta_{\widetilde k})-Z\leq P_nY_{\widetilde k,\widetilde\theta_{\widetilde k}}$. Fix $k\in[K]$. Since $P_nY_{j,\eta}=P_n\log p_0-P_nf_{j,\eta}$ for all $j\in[K]$ and $\eta\in\Theta_{0,j}$, adding $P_n\log p_0$ to both sides of the inequality obtained above from~\eqref{eq:criterion-basic} (with comparator $(k,\theta_k^*)$, $\varepsilon_\ell=n^{-1}$ and $\operatorname{pen}(\widetilde k)\geq0$ dropped) gives
\[
\tfrac12R_{\widetilde k}(\widetilde\theta_{\widetilde k})-Z
\leq P_nY_{\widetilde k,\widetilde\theta_{\widetilde k}} \leq P_nY_{k,\theta_k^*}+\frac{d_k\log n}{2n}
+\frac{\log(1/w_k)}n+\frac{D_k}n+\frac2n .
\]
Here $\mb E_0[P_nY_{k,\theta_k^*}]=R_k(\theta_k^*)$. Since $\lambda_Y$ depends only on $B_0$ and $n\geq3$, applying~\eqref{eq:likelihood-uniform} with $h=n^{-1}$ gives $\mb E_0[Z]\leq\log\sum_jN_j(n^{-1})/(\lambda_Yn)+3L_Y/(2n)\leq C\log n/n$. We take expectations in this chain and multiply by two. Since $\mb E_0[D_k]/n\leq C(\tau_k^2+n^{-1})$, $\mb E_0[Z]\leq C\log n/n$ and $n^{-1}\leq\log n/n$ for $n\geq3$, we have $2\mb E_0[D_k]/n+4/n+2\mb E_0[Z]\leq C\tau_k^2+C\log n/n$. Hence
\[
\mb E_0[R_{\widetilde k}(\widetilde\theta_{\widetilde k})]
\leq2R_k(\theta_k^*)+\frac{d_k\log n}n+\frac{2\log(1/w_k)}n
+C\tau_k^2+C\frac{\log n}n .
\]
Because the weights $w_k$ are fixed and positive, $2\log(1/w_k)/n\leq C/n\leq C\log n/n$ for all $k\in[K]$. Minimizing over $k\in[K]$ proves~\eqref{eq:ms-oracle-fast}.
\end{proof}

\subsubsection{Proof of Theorem~\ref{thm:ms-gen}}\label{app:proof-ms-gen}

\begin{proof}[Proof of Theorem~\ref{thm:ms-gen}]
In this proof, $C$ depends only on the quantities listed in the proof of Proposition~\ref{prop:ms-bridge} and on the constant in $\Gamma=O(n^C)$.

Every $F_\omega\in\mathcal F(L,W)$ outputs $q_\omega=\operatorname{softmax}\{\operatorname{clip}_{\log n}(a_\omega)\}$. Since the clipped logits lie in $[-\log n,\log n]$, every coordinate satisfies
\[
q_{\omega,k}(X^n)\geq\frac{e^{-\log n}}{Ke^{\log n}}=\frac1{Kn^2}=e^{-b},
\]
where $b=2\log n+\log K$. Since $n\geq3$ and $K\geq1$, we have $1\leq b\leq C\log n$. For a task $(k,X^n)$, put
\[
h_\omega=\log\frac{q_k^*(X^n)}{q_{\omega,k}(X^n)},\qquad a_\omega=\mb Eh_\omega.
\]
Applying Lemma~\ref{lem:excess-moments}(a) with $\mathsf T=k$, $X=X^n$, and $q=q_\omega$ gives $a_\omega=\mb E_{\mb Q_n}[D_{\rm KL}(q^*\|q_\omega)]\geq0$ and $\mb E[|h_\omega|^m]\leq e\,m!\,b^{m-1}a_\omega\qquad(m\geq2)$,
which is the moment condition of Lemma~\ref{lem:bernstein-erm} with $\kappa=e$ and $c=b$. Next, by Lemma~\ref{lem:rrds-lipschitz}, for all $\omega,\omega'$ in the parameter ball,
\[
|h_\omega-h_{\omega'}|=|\log q_{\omega',k}(X^n)-\log q_{\omega,k}(X^n)|\leq2L_z\|\omega-\omega'\|_2,
\]
so the pointwise Lipschitz condition of Lemma~\ref{lem:bernstein-erm} holds with $L_f=2L_z$. Since $L\leq C\log n$, $W\leq CKn^{\kappa_{\max}'}\log n$, and $\Gamma=O(n^C)$, we have $\log\{e(W+1)(\Gamma+1)\}\leq C\log n$. By Lemma~\ref{lem:ms-rrds}(c), every affine row has at most $CW$ coefficient positions, and the observations are bounded. Substituting $N_{\rm pos}\leq CW$ and $R=2$ into the formula for $L_z$ in Lemma~\ref{lem:rrds-lipschitz} gives
\[
\log(1+\Gamma L_z)\leq C(1+L)\log\{e(W+1)(\Gamma+1)\}\leq C(\log n)^2.
\]
Since $1+\Gamma L_f=1+2\Gamma L_z\leq2(1+\Gamma L_z)$, it follows that $\log(1+\Gamma L_f)\leq\log2+\log(1+\Gamma L_z)\leq C(\log n)^2$. Using $b\geq1$ and $M\geq1$, we have $1+2\Gamma M(1+L_f/b)\leq2M(1+\Gamma)(1+\Gamma L_f)$, and hence
\[
\begin{aligned}
\log\{1+2\Gamma M(1+L_f/b)\}
&\leq\log(2M)+\log(1+\Gamma)+\log(1+\Gamma L_f)\\
&\leq C\{(\log n)^2+\log(eM)\},
\end{aligned}
\]
where the last inequality uses $\log(2M)\leq\log(eM)$, $\log(1+\Gamma)\leq C\log n$, and $\log(1+\Gamma L_f)\leq C(\log n)^2$.

The empirical cross-entropy risk of $q_\omega$ is $-\widehat P_M\log q_{\omega,k}$, and
\[
\widehat P_Mh_\omega=\widehat P_M\log q_k^*-\widehat P_M\log q_{\omega,k}.
\]
The first term on the right does not depend on $\omega$. Let $\omega_0$ be the parameter vector of the constructed network $\widetilde q$ from Theorem~\ref{thm:ms-approx}. By Lemma~\ref{lem:ms-rrds}(c), all coefficients of $\widetilde q$ are bounded by $Cn^C$, so $\|\omega_0\|_2\leq\sqrt W\,Cn^C\leq\Gamma$ for an admissible $\Gamma$, and thus $\omega_0$ belongs to the class. Therefore $\widehat q=q_{\widehat\omega}$ satisfies $\widehat P_Mh_{\widehat\omega}\leq\widehat P_Mh_{\omega_0}+\varepsilon_{\rm opt}$. Applying Lemma~\ref{lem:bernstein-erm} with $\kappa=e$, $c=b$, $L_f=2L_z$, and $\varepsilon=\varepsilon_{\rm opt}$, and using $8e+16\leq40$ and $a_{\omega_0}=\mb E_{\mb Q_n}[D_{\rm KL}(q^*\|\widetilde q)]$, we obtain on an event of probability at least $1-\delta$ over $\mathcal D_M$ that
\[
\begin{aligned}
\mb E\bigl[D_{\rm KL}(q^*\|\widehat q)\mid\mathcal D_M\bigr]
&\leq3\,\mb E_{\mb Q_n}\bigl[D_{\rm KL}(q^*\|\widetilde q)\bigr]+40\,b\,\frac{W\log\{1+2\Gamma M(1+L_f/b)\}+\log(2/\delta)}M+2\varepsilon_{\rm opt},
\end{aligned}
\]
where the left side is $a_{\widehat\omega}$ because the new task is independent of $\mathcal D_M$. Using $b\leq C\log n$, the preceding bound on $\log\{1+2\Gamma M(1+L_f/b)\}$, and $\log(2/\delta)\leq\log(3/\delta)$, we have
\[
\begin{aligned}
&40\,b\,\frac{W\log\{1+2\Gamma M(1+L_f/b)\}+\log(2/\delta)}M+2\varepsilon_{\rm opt}\\
&\quad\leq C\frac{\log n\,[W\{(\log n)^2+\log(eM)\}+\log(3/\delta)]}M+2\varepsilon_{\rm opt}\leq C\Delta_{\rm ms}.
\end{aligned}
\]
Combining the last two displays with the bound $\mb E_{\mb Q_n}[D_{\rm KL}(q^*\|\widetilde q)]\leq C(b_n+\psi_n\log n)$ from Theorem~\ref{thm:ms-approx} yields
\[
\mb E_M\bigl[D_{\rm KL}(q^*\|\widehat q)\bigr]
\leq3\,\mb E_{\mb Q_n}\bigl[D_{\rm KL}(q^*\|\widetilde q)\bigr]+C\Delta_{\rm ms}
\leq C(b_n+\psi_n\log n+\Delta_{\rm ms}),
\]
which proves~\eqref{eq:ms-gen-kl}. By Proposition~\ref{prop:ms-bridge}, ${\rm BE}_n\leq Cb_n+\psi_n\leq C(b_n+\psi_n\log n)$ since $n\geq3$. Applying Lemma~\ref{lem:ce-selection}(a) conditionally on $\mathcal D_M$ and substituting this bound together with~\eqref{eq:ms-gen-kl} gives, on the same event of probability at least $1-\delta$,
\[
\mb P_M(\widehat k\ne k)\leq5\,{\rm BE}_n+8\,\mb E_M\bigl[D_{\rm KL}(q^*\|\widehat q)\bigr]\leq C(b_n+\psi_n\log n+\Delta_{\rm ms}),
\]
which proves~\eqref{eq:ms-gen-err}. If $\psi_n\log n\to0$ and $\Delta_{\rm ms}\to0$, take $\delta=\delta_n\to0$ sufficiently slowly to obtain convergence in probability.

\end{proof}

\subsubsection{Proof of Corollary~\ref{cor:analytic-ms}}\label{app:proof-ms-analytic}

\begin{proof}[Proof of Corollary~\ref{cor:analytic-ms}]
In this proof, $C$ depends in addition on the extension rectangles, ellipse parameters and amplitude bounds.

For~(a), we use the blocks of Lemma~\ref{lem:ms-rrds} with the following component networks. For each $k$, we take the networks $g_{{\rm init},k}$, $S_k$ and $h_k$ from Corollary~\ref{cor:analytic-est-gen}(a) at accuracy $\varepsilon=n^{-1}$. By Lemma~\ref{lem:analytic-density}, Lemma~\ref{lem:analytic-approx} applies to $\ell_k$ and gives a likelihood network $\varphi_k$ at accuracy $n^{-1}$. These component networks have depth at most $C\log^2(en)$ and coefficients bounded by a constant, and their sizes satisfy $W(g_{{\rm init},k})+W(S_k)+W(h_k)+W(\varphi_k)\leq C\log^{d_X+d_k+1}(en)$. Parts~(a) and~(b) of Lemma~\ref{lem:ms-rrds} do not depend on how the component networks are obtained. Hence the assembled network has the stated form, with $R=2$, $d_z=2\sum_kd_k+K$ and $d_u=\sum_kd_k+K$. Its unrolled depth satisfies $L(\phi_0)+L(\rho_0)+2\{L(\phi)+L(\rho)\}\leq C\log^2(en)$, giving $L\leq C\log^2(en)$. Its size satisfies $W\leq CK\log^{m_{\rm ms}+1}(en)$, and its coefficients are bounded by $Cn$. The size bound of Lemma~\ref{lem:analytic-approx} counts every coefficient of the network, so the four blocks have at most $CK\log^{m_{\rm ms}+1}(en)$ coefficient occurrences. With $W$ equal to this bound, every affine row has at most $CW$ coefficient positions.

The statistical arguments use the networks only through the component accuracies, the assembly~\eqref{eq:separate-module-accuracy}, the projections and the uniform accuracy $n^{-1}$ of $\varphi_k$. In Lemma~\ref{lem:ms-logits}, for every $j\in[K]$, applying the expectation bound of Lemma~\ref{lem:ms-target} on $K_0=K_{j,\rho}$ at $\varepsilon=n^{-1}$ gives
\[
\sup_{\eta\in K_{j,\rho}}\mb E_{j,\eta}[D_j]\leq C_{K_{j,\rho}}(n\varepsilon^2+n^{-1})\leq\frac Cn,
\]
and Markov's inequality yields $\mb P_{j,\eta}^{(n)}(D_j>1)\leq\mb E_{j,\eta}[D_j]\leq C/n$ for every $\eta\in K_{j,\rho}$. Lemmas~\ref{lem:ms-laplace},~\ref{lem:ms-compare} and~\ref{lem:ms-boundary} do not involve the networks. The proofs of Theorem~\ref{thm:ms-approx} and Proposition~\ref{prop:ms-misspec} use the networks only through Lemmas~\ref{lem:ms-rrds},~\ref{lem:ms-logits},~\ref{lem:ms-deficit} and~\ref{lem:ms-target}. Moreover, Lemma~\ref{lem:ms-deficit} at $\varepsilon=n^{-1}$ gives
\[
\mb E_0[D_k/n]\leq C(\tau_k^2+n^{-2}+n^{-1})\leq C(\tau_k^2+n^{-1}).
\]
Hence all conclusions of Theorem~\ref{thm:ms-approx} and Proposition~\ref{prop:ms-misspec} follow from the same proofs without change.

For~(b), we modify the proof of Theorem~\ref{thm:ms-gen} as follows. Since $W\leq CK\log^{m_{\rm ms}+1}(en)$ and $\Gamma=O(n^C)$, we have $\log\{(W+1)(\Gamma+1)\}\leq C\log(en)$. By part~(a), every affine row has at most $CW$ coefficient positions. Together with $L\leq C\log^2(en)$, substituting $N_{\rm pos}\leq CW$ and $R=2$ into the formula for $L_z$ in Lemma~\ref{lem:rrds-lipschitz} gives
\[
\log(1+\Gamma L_f)\leq C(1+L)\log\{e(W+1)(\Gamma+1)\}\leq C\log^3(en),
\]
so $\log\{1+2\Gamma M(1+L_f/b)\}\leq C\{\log^3(en)+\log(eM)\}$. This replaces $(\log n)^2$ by $\log^3(en)$ in the proof of Theorem~\ref{thm:ms-gen} and in the definition of $\Delta_{\rm ms}$. By~(a), the constructed network has size $W\leq CK\log^{m_{\rm ms}+1}(en)$ and coefficients bounded by $Cn$, so its parameter vector $\omega_0$ satisfies $\|\omega_0\|_2\leq C\sqrt W\,n\leq Cn^2$. Hence $\omega_0$ belongs to the class for an admissible $\Gamma$. The rest of the proof of Theorem~\ref{thm:ms-gen} is unchanged, which gives the first claim of~(b).

For the training error bound, put $\lambda_n=\log(en)$ and take $M=\lceil\sqrt n\,\lambda_n^{m_{\rm ms}+5}\rceil$, $\delta=n^{-1}$ and $\varepsilon_{\rm opt}\leq n^{-1/2}$. Since $W\leq CK\lambda_n^{m_{\rm ms}+1}$ and $\log(eM)\leq C\lambda_n$, we have $W\{\lambda_n^3+\log(eM)\}\leq CK\lambda_n^{m_{\rm ms}+4}$. Together with $\log(3/\delta)=\log(3n)\leq C\lambda_n$, $\log n\leq\lambda_n$ and $M\geq\sqrt n\,\lambda_n^{m_{\rm ms}+5}$, this gives
\[
\frac{\log n\,[W\{\lambda_n^3+\log(eM)\}+\log(3n)]}{M}\leq\frac{C\lambda_n^{m_{\rm ms}+5}}{\sqrt n\,\lambda_n^{m_{\rm ms}+5}}=\frac C{\sqrt n}.
\]
Adding $\varepsilon_{\rm opt}\leq n^{-1/2}$ shows that the training error is at most $C/\sqrt n$.
\end{proof}

\subsection{Proofs for variable selection}\label{app:variable-proofs}

We establish the variable-selection results in three stages. We first develop the likelihood and information bounds for sparse GLMs and construct the network for an individual support. We then analyze the probabilities assigned to the candidate supports, distinguishing supports that contain the true support from those that omit signals. Finally, we use these bounds to prove Theorem~\ref{thm:vs-approx}, Propositions~\ref{prop:vs-bridge} and~\ref{prop:vs-bridge-est}, and Theorem~\ref{thm:vs-gen}. The preliminary bounds are derived for fixed $D=1+s_{\max}$ and are made explicit in $D$ in Appendix~\ref{app:vs-smax}. The intercept is always included and $p$ may grow with $n$.

\subsubsection{Likelihood, information, and regularity of the sparse GLM}\label{app:sparse-regularity}

Throughout this subsection, $X=(Y,Z)$ has the random-design joint law in~\eqref{eq:glm-model}, with $\mathcal L_{S,\theta_S}(Z)=Q_Z$ for $S\in\mathcal S$ and $\theta_S\in\Theta_{0,S}$. With dominating measure $\nu_Y\otimes Q_Z$, the joint log density is $\ell_{S,\theta_S}(y,z)=a^{-1}\big[y\{\icpt+z_S^\top\beta_S\}-b\{\icpt+z_S^\top\beta_S\}\big]+c(y,a)$. Thus the common baseline in the likelihood comparisons can be taken as $a_n(y,z)=c(y,a)$. If instead a design density is included in the dominating-measure convention, its log density is added to this same baseline and cancels from every comparison. Put $v_S(z)=(1,z_S^\top)^\top$. The score and per-observation Fisher information are
\begin{equation}
 \begin{aligned}
 s_{S,\theta_S}(y,z)
   &=a^{-1}\big[y-b'\{\eta_{\theta^{(S)}}(z)\}\big]v_S(z),\\
 I_S(\theta_S)
   &=a^{-1}\mb E_{Z\sim Q_Z}\!\left[
       b''\{\eta_{\theta^{(S)}}(Z)\}v_S(Z)v_S(Z)^\top\right].
 \end{aligned}
 \label{eq:glm-score-information}
\end{equation}
By~\eqref{eq:glm-model},
\[
\mb E[s_{S,\theta_S}\mid Z]=0,\qquad \mb E[s_{S,\theta_S}s_{S,\theta_S}^\top\mid Z]=a^{-1}b''(\eta_{\theta^{(S)}})v_Sv_S^\top=-\nabla^2_{\theta_S}\ell_{S,\theta_S},
\]
so integration over $Q_Z$ gives~\eqref{eq:glm-score-information}.

Under the setting of Section~\ref{sec:theory-variable} and Assumption~\ref{ass:sparse}\textup{(i)}--\textup{(ii)}, we establish regularity bounds uniformly over the candidate supports. Recall that $D=1+s_{\max}$ and $\Lambda_n=\log(en)+\log(ep)$. Every candidate has $2\leq d_S\leq D$. The constants $B,\rho,c_\Theta>0$ are those of the setting of Section~\ref{sec:theory-variable}. The constant $c>0$, fixed below, depends only on $a$ and the constants of Assumption~\ref{ass:sparse}\textup{(i)}--\textup{(ii)}.

As in Section~\ref{sec:theory-variable}, the working rectangles satisfy $\Theta_{0,S}\subseteq[-B,B]^{d_S}$, have side lengths at least $c_\Theta$ and are nested. Setting additional coordinates to zero gives the same density. Write $\iota_{S,S'}$ for the zero extension from $S$ to $S'$, which retains the intercept. Define the allowed generating set
\[
 \mathcal A_{\rho,S}=\left\{\theta_S\in\Theta_{0,S}:
 \min_{\substack{S'\in\mathcal S\\S\subseteq S'}}
 \operatorname{dist}(\iota_{S,S'}\theta_S,\partial\Theta_{0,S'})
 \geq\rho/D\right\}.
\]
This set is closed because the collection of supports is finite and the distance functions are continuous. When Assumption~\ref{ass:sparse-proposal} is imposed, its interior margin implies $\pi_S(\mathcal A_{\rho,S})=1$ and $\operatorname{supp}\pi_S\subseteq\mathcal A_{\rho,S}$. We write $K_S=\operatorname{supp}\pi_S=\mathcal I_0\times\prod_{j\in S}\mathcal I_j$, so that $K_S\subseteq\mathcal A_{\rho,S}$ and $\pi_S\{\operatorname{supp}(\beta)=S\}=1$. For an allowed truth on $S_0$, put $\theta_{0,S}=\iota_{S_0,S}\theta_{0,S_0}$. Then
\[
\theta_{0,S_0}\in K_{S_0},\qquad \operatorname{dist}(\theta_{0,S},\partial\Theta_{0,S})\geq\rho/D,\qquad S\supseteq S_0.
\]
The working rectangles and the topological supports of the task distributions may contain zero coefficients.

As in Section~\ref{sec:theory-variable}, let $\mathcal K_n\subseteq[-B,B]\times\mathbb R^p$ be the common closed support of the observations and $\mathcal Y\subseteq[-B,B]$ the common support of the conditional response laws, with $\nu_Y(\mathcal Y)<\infty$. Thus $\mathbb P_{S,\theta_S}(X\in\mathcal K_n)=1$ for every $S\in\mathcal S$ and $\theta_S\in\Theta_{0,S}$, and $\operatorname{supp}\mathcal L_{S,\theta_S}(Y\mid Z=z)=\mathcal Y$ for $Q_Z$-almost every $z$. We construct the computational rectangle $\mathcal X_n\supseteq\mathcal K_n$ below. The joint dominating measure remains $\nu_Y\otimes Q_Z$. For the common baseline $a_n(x)=c(y,a)$, we use the bounds
\begin{equation}
\begin{gathered}
 \sup_{\substack{S\in\mathcal S,\ x\in\mathcal K_n\\\theta_S\in\Theta_{0,S}}}
       |\log p_{S,\theta_S}(x)-a_n(x)|\leq B,\quad 
 \sup_{\substack{S\in\mathcal S,\ x\in\mathcal K_n\\\theta_S\in\Theta_{0,S}}}
       |\partial^\nu_{\theta_S}\log p_{S,\theta_S}(x)|\leq B,
 \qquad1\leq|\nu|\leq3.
\end{gathered}
 \label{eq:sparse-envelopes}
\end{equation}
These bounds follow from the boundedness of the observations in the setting of Section~\ref{sec:theory-variable} after enlarging the constant $B$, which is the convention stated after Assumption~\ref{ass:sparse}. To verify this, write $B_0$ for the constant that bounds the observations and the working rectangles, and put $Y_0=V_0=1\vee B_0$. Step~1 of the proof of Lemma~\ref{lem:glm-natural-modules} uses only the bound $|Y|\leq Y_0$ on the responses and shows that $b$ is finite and infinitely differentiable on $\mathbb R$, with
\[
\begin{gathered}
|b'(u)|\leq Y_0,\qquad 0\leq b''(u)\leq Y_0^2/a,\qquad |b'''(u)|\leq16Y_0^3/a^2,\\
|b(u)|\leq|b(0)|+Y_0|u|,\qquad u\in\mathbb R .
\end{gathered}
\]
For $\theta_S\in\Theta_{0,S}$ and $(y,z)\in\mathcal K_n$, the linear predictor $u=v_S(z)^\top\theta_S$ lies in the compact interval $\mathcal I_{\rm pred}$ of Assumption~\ref{ass:sparse}\textup{(i)}, because $\mathcal K_n$ is the closed support of the observations and $\mathcal I_{\rm pred}$ is closed. Hence $|u|\leq R_0=\max\{|v|:v\in\mathcal I_{\rm pred}\}$. Since $\log p_{S,\theta_S}(x)-a_n(x)=a^{-1}\{yu-b(u)\}$ and
\[
\partial_j\ell_{S,\theta_S}=\frac{y-b'(u)}a\,v_{S,j},\quad
\partial_{jk}\ell_{S,\theta_S}=-\frac{b''(u)}a\,v_{S,j}v_{S,k},\quad
\partial_{jkl}\ell_{S,\theta_S}=-\frac{b'''(u)}a\,v_{S,j}v_{S,k}v_{S,l},
\]
with $|v_{S,j}|\leq V_0$, the bounds~\eqref{eq:sparse-envelopes} hold with $B$ replaced by
\[
B_{\rm env}=\max\Bigl\{B_0,\ \frac{2Y_0R_0+|b(0)|}a,\ \frac{2Y_0V_0}a,\ \frac{Y_0^2V_0^2}{a^2},\ \frac{16Y_0^3V_0^3}{a^3}\Bigr\},
\]
which depends only on $B_0$, $a$, $b(0)$ and $\mathcal I_{\rm pred}$, and not on $s_{\max}$. The constant $B$ enters the arguments below only through upper bounds, so we write $B$ for $B_{\rm env}$ from now on. In particular, $\Theta_{0,S}\subseteq[-B,B]^{d_S}$ and $\mathcal K_n\subseteq[-B,B]^{1+p}$ remain valid. The log density is infinitely differentiable in $\theta_S$ on $\mathbb R^{d_S}$. Since the derivatives displayed above are bounded on $\mathcal K_n$ uniformly over bounded sets of parameters, differentiation under the integral is valid by dominated convergence. We use the pointwise bounds~\eqref{eq:sparse-envelopes} in three places. The first is the uniform approximation of the scores and log densities by ReLU networks on the compact rectangle $\mathcal X_n$. The second is the Bernstein-type bounds for the log-likelihood ratios and scores that hold simultaneously over the supports (Lemmas~\ref{lem:sparse-local},~\ref{lem:vs-overfit} and~\ref{lem:vs-soft-oracle}). The third is the bound on the training loss in the proof of Theorem~\ref{thm:vs-gen}. Bounds on moments of the derivatives would not suffice for these steps, so we assume bounded observations and not a moment condition. By Assumption~\ref{ass:sparse}\textup{(i)}--\textup{(ii)}, equations~\eqref{eq:glm-score-information} and~\eqref{eq:glm-sparse-design} give
\[
cI_{d_S}\preceq\frac{m_b\underline\lambda_Z}{a}I_{d_S} \preceq I_S(\theta_S) \preceq\frac{M_b\overline\lambda_Z}{a}I_{d_S} \preceq c^{-1}I_{d_S},\qquad \theta_S\in\Theta_{0,S},
\]
after fixing $c>0$ sufficiently small. The common baseline cancels, so
\[
 \sup_{\substack{S\in\mathcal S,\ x\in\mathcal K_n\\\theta_S\in\Theta_{0,S}}}
 \left|\log\frac{p_{\theta_0}(x)}{p_{S,\theta_S}(x)}\right|\leq2B.
\]
We also take $c\leq m_b\underline\lambda_Z/(2a)$. Then~\eqref{eq:glm-kl-parameter} gives $D_{\mathrm{KL}}(\mb P_{\theta_0}\|\mb P_{S,\theta_S})\geq c\|\theta_S-\theta_{0,S}\|_2^2$ for $S\supseteq S_0$, and~\eqref{eq:glm-kl-omitted} gives the corresponding bound for incorrect supports.

Put $Y_*=1\vee B$ and $V_*=1\vee\sqrt{aB/m_b}$. By the bound on the responses, Assumption~\ref{ass:sparse}\textup{(i)} and~\eqref{eq:sparse-envelopes}, on the observation support $\mathcal K_n$ we have
\[
|Y|\leq Y_*,\qquad \frac{m_b}{a}|v_{S,j}|^2\leq|\partial_{jj}\ell_{S,\theta_S}|=a^{-1}b''(\eta)|v_{S,j}|^2\leq B,\qquad |v_{S,j}|\leq V_*.
\]
Every covariate occurs in a candidate support, and the intercept coordinate is at most $V_*$. Hence $Q_Z\!(\max_{j\leq p}|Z_j|\leq V_*)=1$. The network components are approximated on rectangles, whereas the observation support $\mathcal K_n$ need not be one. Since $\mathcal K_n$ is the common closed support, we can therefore choose a rectangle $\mathcal X_n$ with $\mathcal K_n\subseteq\mathcal X_n\subseteq[-Y_*,Y_*]\times[-V_*,V_*]^p$. We call $\mathcal X_n$ the computational rectangle, and all component maps are defined and approximated on it. We write $\mathcal Z_{{\rm cal},n}=\operatorname{proj}_Z(\mathcal X_n)$ for the set of covariate values that occur in $\mathcal X_n$. We use the fixed moment statistic $T_S(Y,Z)=Yv_S(Z)$ and $\mu_S(\theta_S)=\mb E_{Q_Z}[b'(v_S^\top\theta_S)v_S]$. This statistic uses $(Y,Z_S)$, with $T_S:\mathcal X_n\to\mathbb R^{d_S}$ and $\|T_S\|_\infty\leq Y_*V_*$. Moreover, $\operatorname{Cov}_{S,\theta_S}(T_S)\preceq\mb E_{S,\theta_S}[Y^2v_Sv_S^\top]\preceq Y_*^2\overline\lambda_Z I_{d_S}$. Lemma~\ref{lem:glm-natural-modules} gives the real inverse and information properties needed for this choice.

In the sparse analysis below, $\ell_{n,S}$ denotes the centered computational log likelihood. Using the real extension of $b$ in Lemma~\ref{lem:glm-natural-modules}, we define
\[
\ell_{n,S}(\theta_S)=a^{-1}\sum_{i=1}^n\{Y_iv_S(Z_i)^\top\theta_S-b(v_S(Z_i)^\top\theta_S)\},\qquad X^n\in\mathcal X_n^n.
\]
The same formula defines $\ell_{n,U}$ for auxiliary supports, and we write $\ell_n(\theta_0)=\ell_{n,S_0}(\theta_{0,S_0})$. On the observation support,
\[
\sum_{i=1}^n\log p_{S,\theta_S}(X_i)=\ell_{n,S}(\theta_S)+\sum_{i=1}^na_n(X_i),\qquad X^n\in\mathcal K_n^n.
\]
The baseline term $\sum_ia_n(X_i)$ is common to all supports and parameters, so it leaves maximizers, deficits and criterion differences unchanged. Continuity in the parameter, Borel measurability in the dataset and compactness give a measurable selection with
\[
\begin{gathered}
\widehat\theta_S:\mathcal X_n^n\to\Theta_{0,S},\qquad \widehat\theta_S(X^n)\in\argmax_{\theta_S\in\Theta_{0,S}}\ell_{n,S}(\theta_S),\\
D_S(F)=\ell_{n,S}(\widehat\theta_S)-\ell_{n,S}\{F(X^n)\}\geq 0,\qquad X^n\in\mathcal X_n^n.
\end{gathered}
\]
Ties between supports are resolved by a fixed ordering. We write
\[
\operatorname{pen}(S)=\frac{d_S}{2}\log n+\log\binom p{|S|},\qquad
\operatorname{IC}_S=-\ell_{n,S}(\widehat\theta_S)+\operatorname{pen}(S).
\]

The following lemma summarizes the consequences of Assumption~\ref{ass:sparse}\textup{(i)}--\textup{(ii)} for the KL divergence of the random-design GLM.

\begin{lemma}[KL geometry of the random-design GLM]\label{lem:glm-kl}
Suppose the GLM specification~\eqref{eq:glm-model} and Assumption~\ref{ass:sparse}\textup{(i)}--\textup{(ii)} hold. For any $\theta_0,\theta\in\Theta_{\mathcal S}$, put $\Delta\eta(Z)=\eta_\theta(Z)-\eta_{\theta_0}(Z)$. Then
\begin{equation}
 \begin{aligned}
 D_{\mathrm{KL}}(\mb P_{\theta_0}\|\mb P_\theta)
 &=\frac1a\mb E_{Q_Z}\!\left[
 b\{\eta_\theta(Z)\}-b\{\eta_{\theta_0}(Z)\}
       -b'\{\eta_{\theta_0}(Z)\}\Delta\eta(Z)\right],\\
 \frac{m_b}{2a}\mb E_{Q_Z}[(\Delta\eta)^2]
 &\leq D_{\mathrm{KL}}(\mb P_{\theta_0}\|\mb P_\theta)
 \leq\frac{M_b}{2a}\mb E_{Q_Z}[(\Delta\eta)^2] .
 \end{aligned}
 \label{eq:glm-kl-prediction}
\end{equation}
Consequently,
\begin{equation}
 \frac{m_b\underline\lambda_Z}{2a}\|\theta-\theta_0\|_2^2
 \leq D_{\mathrm{KL}}(\mb P_{\theta_0}\|\mb P_\theta)
 \leq\frac{M_b\overline\lambda_Z}{2a}\|\theta-\theta_0\|_2^2 .
 \label{eq:glm-kl-parameter}
\end{equation}
In particular, for a true support $S_0$ and any candidate $S$,
\begin{equation}
 \inf_{\theta_S\in\Theta_{0,S}}
 D_{\mathrm{KL}}(\mb P_{\theta_0}\|\mb P_{S,\theta_S})
 \geq\frac{m_b\underline\lambda_Z}{2a}
                 \|\beta_{0,S_0\setminus S}\|_2^2 .
 \label{eq:glm-kl-omitted}
\end{equation}
The infimum includes arbitrary refitting of the intercept and the retained slopes within their working domains.
\end{lemma}

\begin{proof}
Write $\theta_0=(\icpt_0,\beta_0)$, $\theta=(\icpt,\beta)$ and $\Delta\eta=\eta_\theta-\eta_{\theta_0}$. By~\eqref{eq:glm-model}, $\log\{p_{\theta_0}(y\mid z)/p_\theta(y\mid z)\}=a^{-1}[-y\,\Delta\eta(z)+b\{\eta_\theta(z)\}-b\{\eta_{\theta_0}(z)\}]$, and $\mb P_{\theta_0}$ and $\mb P_\theta$ share the design law $Q_Z$. Conditioning on $Z$ and using $\mb E_{\theta_0}[Y\mid Z]=b'\{\eta_{\theta_0}(Z)\}$ give
\[
\begin{aligned}
D_{\rm KL}(\mb P_{\theta_0}\|\mb P_\theta)
 &=\mb E_{Q_Z}\mb E_{\theta_0}\!\left[
 \frac{-Y\Delta\eta+b(\eta_\theta)-b(\eta_{\theta_0})}{a}\,\middle|\,Z\right]=\frac1a\mb E_{Q_Z}[b(\eta_\theta)-b(\eta_{\theta_0})-b'(\eta_{\theta_0})\Delta\eta],
\end{aligned}
\]
which is the first line of~\eqref{eq:glm-kl-prediction}. In the setting of Section~\ref{sec:theory-variable}, the responses are bounded and, by Assumption~\ref{ass:sparse}\textup{(i)}, both linear predictors lie in the compact interval $\mathcal I_{\rm pred}$ with $Q_Z$-probability one, so all these expectations are finite.

Since $\mathcal I_{\rm pred}$ is an interval, on this full-$Q_Z$-measure event $\eta_{\theta_0}+t\Delta\eta\in\mathcal I_{\rm pred}$ for all $0\leq t\leq 1$, so Taylor's theorem and $m_b\leq b''\leq M_b$ on $\mathcal I_{\rm pred}$ give
\[
\frac{m_b}{2}(\Delta\eta)^2
\leq b(\eta_\theta)-b(\eta_{\theta_0})-b'(\eta_{\theta_0})\Delta\eta
\leq \frac{M_b}{2}(\Delta\eta)^2 .
\]
Taking $\mb E_{Q_Z}$ and dividing by $a$ gives the second line of~\eqref{eq:glm-kl-prediction}.

Since $\theta_0,\theta\in\Theta_{\mathcal S}$ each have at most $s_{\max}$ nonzero slopes, $\beta-\beta_0$ has at most $2s_{\max}$ nonzero entries and $\Delta\eta=(\icpt-\icpt_0)+Z^\top(\beta-\beta_0)$. Applying~\eqref{eq:glm-sparse-design} with $h_{\icpt}=\icpt-\icpt_0$ and $h_\beta=\beta-\beta_0$ gives $\underline\lambda_Z\|\theta-\theta_0\|_2^2\leq\mb E_{Q_Z}[(\Delta\eta)^2]\leq\overline\lambda_Z\|\theta-\theta_0\|_2^2$, which with~\eqref{eq:glm-kl-prediction} proves~\eqref{eq:glm-kl-parameter}.

Finally, for a true support $S_0$, any candidate $S$ and every $\theta_S\in\Theta_{0,S}$, the zero extension $\theta^{(S)}\in\Theta_{\mathcal S}$ has zero coefficients on $S_0\setminus S$, so $\|\theta^{(S)}-\theta_0\|_2^2\geq\|\beta_{0,S_0\setminus S}\|_2^2$. Since $\mb P_{S,\theta_S}=\mb P_{\theta^{(S)}}$, the lower bound in~\eqref{eq:glm-kl-parameter} gives
\[
D_{\rm KL}(\mb P_{\theta_0}\|\mb P_{S,\theta_S})
\geq\frac{m_b\underline\lambda_Z}{2a}\|\theta^{(S)}-\theta_0\|_2^2
\geq\frac{m_b\underline\lambda_Z}{2a}\|\beta_{0,S_0\setminus S}\|_2^2 .
\]
Taking the infimum over $\theta_S\in\Theta_{0,S}$ proves~\eqref{eq:glm-kl-omitted}.
\end{proof}

The next remark verifies the conditions of Lemma~\ref{lem:glm-kl} for logistic and Gaussian regression.

\begin{remark}[Logistic and Gaussian examples]\label{rem:glm-kl-examples}
For Bernoulli logistic regression,
\[
\begin{gathered}
a=  1,\qquad b(u)=  \log(1+e^u),\qquad \frac{e^{R_*}}{(1+e^{R_*})^2}=  :m_b \leq   b''(u)\leq   M_b:=  \frac14,\\
|u|\leq   R_*.
\end{gathered}
\]
A fixed predictor radius $R_*<\infty$ gives uniform curvature constants. For independent symmetric Rademacher covariates,
\[
 \mathbb EZ=0,\qquad\mathbb EZZ^\top=I_p,\qquad
 \mathbb E\begin{pmatrix}1\\Z\end{pmatrix}
             \begin{pmatrix}1&Z^\top\end{pmatrix}=I_{p+1},
 \qquad\underline\lambda_Z=\overline\lambda_Z=1.
\]
For fixed $R>0$, compatible working boxes with
\[
|\beta_j|\leq R/D,\qquad |\icpt|\leq R\quad\Longrightarrow\quad|\eta_\theta(Z)|\leq |\icpt|+\sum_{j\in S}|\beta_jZ_j|\leq2R
\]
verify the predictor condition with $R_*=2R$; omit the intercept term when absent. Correlated designs are also covered whenever~\eqref{eq:glm-sparse-design} holds.

For Gaussian linear regression with known variance,
\[
a=  \sigma^2,\qquad b(u)=  u^2/2,\qquad b''(u)=  1,\qquad (u\in\mathbb R).
\]
If $\mathbb E_{Q_Z}[(\Delta\eta)^2]<\infty$, direct integration gives
\[
D_{\rm KL}(\mathbb P_{\theta_0}\|\mathbb P_\theta) =  \frac1{2\sigma^2}\mathbb E_{Q_Z} [\{(\icpt-\icpt_0)+Z^\top(\beta-\beta_0)\}^2].
\]
The sparse design condition yields~\eqref{eq:glm-kl-omitted}. For logistic curvature on an unrestricted predictor range, $\inf_{u\in\mathbb R}\frac{e^u}{(1+e^u)^2}= 0$.

\end{remark}

The next lemma gives the consequences of Assumption~\ref{ass:sparse}\textup{(i)}--\textup{(ii)} for the natural moment statistic and the inverse information, with constants uniform in $n$, $p$ and $S$.

\begin{lemma}[Natural moments and inverse information in the GLM]
\label{lem:glm-natural-modules}
In the setting of Section~\ref{sec:theory-variable} and under Assumption~\ref{ass:sparse}\textup{(i)}--\textup{(ii)}, define
\[
 \begin{gathered}
 c_\mu=m_b\underline\lambda_Z,\qquad
 h_*=\pi a/(4Y_*),\qquad K_3=16Y_*^3/a^2,\\
 L_\mu=K_3V_*\overline\lambda_Z,\qquad
 r_\mu=\min\{1,h_*/(4V_*),c_\mu/(4L_\mu)\},\qquad
 \rho_\mu=c_\mu r_\mu/4.
 \end{gathered}
\]
These constants are independent of $n,p,S$. The moment map satisfies
\[
D\mu_S(\theta_S)=aI_S(\theta_S),\qquad c_\mu I_{d_S}\preceq D\mu_S(\theta_S)\preceq M_b\overline\lambda_Z I_{d_S},\qquad \theta_S\in\Theta_{0,S}.
\]
Its inverse is defined on
$\mathcal K_{\mu,S}=  \{m\in\mathbb R^{d_S}: \operatorname{dist}(m,\mu_S(\Theta_{0,S}))\leq  \rho_\mu/D\}$,
with
$\|\mu_S^{-1}(m)-\mu_S^{-1}(m')\|_2 \leq  (2/c_\mu)\|m-m'\|_2$, $m,m'\in\mathcal K_{\mu,S}$.
Let $A_S$ be the coordinatewise affine map from $[-1,1]^{d_S}$ onto $\Theta_{0,S}$. Each entry of $I_S^{-1}\circ A_S$ extends holomorphically to an open neighborhood of $E_{\varrho_{I,D}}^{d_S}$, where
\begin{equation}
 \varrho_{I,D}=1+\min\{1,r_\mu/(4(1\vee B)D)\},\qquad
 \sup_{z\in E_{\varrho_{I,D}}^{d_S}}
     \|I_S(A_Sz)^{-1}\|_{\rm op}\leq2a/c_\mu.
 \label{eq:glm-information-ellipse}
\end{equation}
Finally,
$b\in\operatorname{Hol}\{z\in\mathbb C:|\operatorname{Im}z|<\pi a/(2Y_*)\}$,
and
\begin{equation}
 \begin{gathered}
 |b'(z)|\leq\sqrt2Y_*,\qquad |b^{(3)}(z)|\leq K_3,
 \qquad |\operatorname{Im}z|\leq h_*,\\
 |b'(u)|\leq Y_*,\qquad b''(u)\leq Y_*^2/a,\qquad
 |b(u)|\leq|b(0)|+Y_*|u|,\qquad u\in\mathbb R.
 \end{gathered}
 \label{eq:glm-cumulant-strip}
\end{equation}
\end{lemma}

As in Section~\ref{sec:theory-variable}, we take $\rho\leq\rho_\mu$, and $K_{g,S}$ is the set~\eqref{eq:moment-inverse-domain} on which $\mu_S^{-1}$ is used. The next lemma extends the real inverse of Lemma~\ref{lem:glm-natural-modules} to a complex neighborhood and gives conditions under which Assumption~\ref{ass:sparse}\textup{(iii)} follows from Assumption~\ref{ass:sparse}\textup{(i)}--\textup{(ii)}. In it, $\|\cdot\|_2$ on $\mathbb C^{d_S}$ is the Hermitian norm, and distances are taken in this norm.

\begin{lemma}[Analytic extension of the moment inverse]\label{lem:glm-inverse-extension}
Suppose the conditions of Lemma~\ref{lem:glm-natural-modules} hold, and keep its constants $h_*,V_*,L_\mu,c_\mu,\rho_\mu$. Fix $S\in\mathcal S$. Let $\Theta_S^+\subset\mathbb R^{d_S}$ be a compact convex set with $\Theta_{0,S}\subseteq\Theta_S^+\subseteq[-B_+,B_+]^{d_S}$, and let $m_+\in(0,m_b]$ be such that
\[
 b''\{v_S(z)^\top\theta\}\geq m_+
 \qquad\text{for every }\theta\in\Theta_S^+\text{ and }Q_Z\text{-almost every }z.
\]
Put
\[
 c_+=m_+\underline\lambda_Z,\qquad
 r_+=\min\{1,h_*/(4V_*),c_+/(4L_\mu)\},\qquad
 \rho_+=c_+r_+/4 .
\]

\textup{(a)} The map $\mu_S$ is injective on $\mathbb R^{d_S}$. Its inverse on $\mu_S(\Theta_S^+)$ extends to a holomorphic map $G_S^{-1}$ on the open set
\[
 N_S^+=\bigl\{m\in\mathbb C^{d_S}:\operatorname{dist}\{m,\mu_S(\Theta_S^+)\}<\rho_+/\sqrt D\bigr\},
\]
which takes real values at real points, satisfies $\mu_S\{G_S^{-1}(m)\}=m$ for real $m\in N_S^+$, and has every coordinate bounded in modulus by $B_++2$.

\textup{(b)} Let $\mathcal R=\prod_{j=1}^{d_S}[u_j,v_j]$ be a rectangle with $u_j<v_j$ for every $j$ and
\[
 K_{g,S}\subseteq\mathcal R\subseteq
 \bigl\{m\in\mathbb R^{d_S}:\operatorname{dist}\{m,\mu_S(\Theta_S^+)\}\leq\rho_+/(2\sqrt D)\bigr\},
 \qquad \bar h=\max_j(v_j-u_j)/2 .
\]
Then $\mu_S^{-1}$ satisfies~\eqref{eq:extension-main} on $K_{g,S}$ with constants
\[
 (\mathcal R,\varrho_g,\mathcal B_g),\qquad
 \varrho_g=1+\min\{1,\rho_+/(4D\bar h)\},\qquad
 \mathcal B_g=B_++2 .
\]

\textup{(c)} If $\rho\leq\rho_\mu/6$ and $\operatorname{diam}_2(\Theta_{0,S})\leq\rho_\mu/(4DM_b\overline\lambda_Z)$, part~\textup{(b)} applies with $\Theta_S^+=\Theta_{0,S}$, $m_+=m_b$, $B_+=B$ and $\mathcal R$ the smallest rectangle containing $K_{g,S}$. Then $\varrho_g=2$, and every side of $\mathcal R$ has length at least $\rho/D$.
\end{lemma}

Part~\textup{(c)} gives the case discussed in Appendix~\ref{app:vs-setting}, to which the orders stated after Assumption~\ref{ass:sparse}\textup{(iii)} apply. Under its diameter bound, $\varrho_g=2$, $\mathcal B_g=B+2$ and the reciprocal side lengths of $\mathcal R$ are at most $D/\rho$. Part~\textup{(b)} covers larger working rectangles as long as a rectangle containing $K_{g,S}$ stays close to the image of a compact convex set $\Theta_S^+$. Since $b''$ is continuous and positive on $\mathbb R$, such a set always has a curvature bound $m_+>0$, which decreases as the set grows. The restriction in part~\textup{(b)} is on the geometry of the moment image. The range $\mu_S(\mathbb R^{d_S})$ is a bounded open set because $|b'|\leq Y_*$ on $\mathbb R$, the covariates are bounded, and $\mu_S$ is an injective local diffeomorphism. The inverse is unbounded near every boundary point of this range, since a bounded sequence of preimages would have a limit point whose image is that boundary point. A rectangle that contains a large curved image $\mu_S(\Theta_{0,S})$ can leave the range, and then no bounded holomorphic extension exists.

\subsubsection{Component functions of the network for a candidate support}\label{app:proof-vs-approx}

The network for a given support uses only the response and the covariates belonging to that support. This is important for the approximation bounds: the functions evaluated by each network have input dimensions determined by the support size rather than by the total number of covariates. We first describe these functions and then specify the domains and accuracies used in their approximation.

The network for $S$ first uses fixed coordinate extraction
\[
E_S:\mathbb R^{p+1}\to\mathbb R^{|S|+1},\qquad E_S(y,z)=  (y,z_S),\qquad W(E_S)=  |S|+1,\qquad W_{\rm free}(E_S)=  0.
\]
The observation retains its original joint law. Extraction changes only the inputs used by this network. For $x=(y,z)\in\mathcal X_n$, define the maps
\[
\begin{gathered}
\ell^\circ_S(x,\theta_S) =  a^{-1}\{y v_S(z)^\top\theta_S-b(v_S(z)^\top\theta_S)\},\quad
s_S(x,\theta_S) =  a^{-1}\{y-b'(v_S(z)^\top\theta_S)\}v_S(z).
\end{gathered}
\]
On $\mathcal K_n$, these are the centered log density and statistical score: $\ell^\circ_S(x,\theta_S)=\log p_{S,\theta_S}(x)-a_n(x)$ and $s_S(x,\theta_S)=\nabla_{\theta_S}\log p_{S,\theta_S}(x)$. On the computational rectangle, \eqref{eq:glm-cumulant-strip} gives
\[
\begin{gathered}
\sup_{\mathcal X_n\times\Theta_{0,S}}\max_j|s_{S,j}| \leq  2Y_*V_*/a ,\qquad \sup_{\mathcal X_n\times\Theta_{0,S}}\max_{j,k} |\partial_{\theta_k}s_{S,j}| \leq   Y_*^2V_*^2/a^2,\\
\sup_{\mathcal X_n\times\Theta_{0,S}}\max_{j,k,l} |\partial_{\theta_k\theta_l}s_{S,j}| \leq   K_3V_*^3/a.
\end{gathered}
\]
Define the fixed computational derivative envelope
\[
B_{\rm cal}=  \max\{1,B,2Y_*V_*/a,Y_*^2V_*^2/a^2,K_3V_*^3/a\}.
\]
The common baseline $a_n$ is subtracted only in the statistical comparisons on $\mathcal K_n$. The formulas above define the maps on the rest of $\mathcal X_n$ without using it. We compute the inner product by scalar multipliers, evaluate $b$ or $b'$ on its one-dimensional argument, and then form the displayed products. The input dimensions satisfy
\[
d_{\rm raw}=  1+|S|+d_S\leq  2D+1,\qquad d_{b}=  d_{b'}=  1,\qquad d_{\rm multiply}=  2.
\]

The network uses the natural moment statistic $T_S(Y,Z)=Yv_S(Z)$, whose empirical average $\overline T_S=n^{-1}\sum_iT_S(X_i)$ has $d_S$ coordinates. In the RRDS network, a multiplication network computes the products in this statistic (Lemma~\ref{lem:vs-moment}). Write $\Pi_{\mathcal R}$ for the coordinatewise projection onto a rectangle $\mathcal R$. For a moment rectangle $\mathcal R_{g,S}$ and a map $g_S$ on it, both specified below, the network approximates the one-step map
\begin{equation}
 \theta_S^{(0)}=\Pi_{\Theta_{0,S}}g_S\bigl(\Pi_{\mathcal R_{g,S}}\overline T_S\bigr),\qquad
 F_S^{\rm os}=\Pi_{\Theta_{0,S}}\bigl\{\theta_S^{(0)}+I_S(\theta_S^{(0)})^{-1}P_ns_{S,\theta_S^{(0)}}\bigr\}.
 \label{eq:vs-one-step-map}
\end{equation}
All rectangles are fixed before observing the task, and $\dim\operatorname{dom}(g_S)=\dim\operatorname{dom}\{(I_S^{-1})_{jk}\}=d_S$ for $j,k\leq d_S$.

\subsubsection{Analytic approximation of the component functions}\label{app:vs-analytic}
In the notation of Appendix~\ref{app:analytic-extension}, Assumption~\ref{ass:sparse}\textup{(iii)} states that $\mu_S^{-1}$ satisfies~\eqref{eq:extension-main} on the set $K_{g,S}$ of~\eqref{eq:moment-inverse-domain} with constants $(\mathcal R_{g,S},\varrho_{g,n},\mathcal B_{g,n})$. Let $g_S$ denote the resulting real extension to $\mathcal R_{g,S}$. Lemma~\ref{lem:glm-inverse-extension} gives conditions under which Assumption~\ref{ass:sparse}\textup{(iii)} follows from Assumption~\ref{ass:sparse}\textup{(i)}--\textup{(ii)}. For every correct candidate $S\supseteq S_0$, we have $\mu_S(\theta_{0,S})\in\mu_S(\Theta_{0,S})\subseteq K_{g,S}\subseteq\mathcal R_{g,S}$. By Lemma~\ref{lem:glm-natural-modules}, the entries of $I_S^{-1}$ satisfy~\eqref{eq:extension-main} on $\Theta_{0,S}$ with constants $(\Theta_{0,S},\varrho_{I,D},2a/c_\mu)$.

For the response functions, we fix a predictor interval $\mathcal I_n=[-R_{\eta,n},R_{\eta,n}]$ with
\[
1\leq R_{\eta,n}\leq 1+BDV_*,\qquad |v_S(z)^\top\theta_S|\leq R_{\eta,n},\qquad S\in\mathcal S,\qquad z\in\mathcal Z_{{\rm cal},n},\qquad \theta_S\in\Theta_{0,S}.
\]
The value $R_{\eta,n}=1+BDV_*$ is always admissible, and a smaller bound may be used for a specified model. By~\eqref{eq:glm-cumulant-strip}, $b$ and $b'$ satisfy~\eqref{eq:extension-main} on $\mathcal I_n$ with constants $(\mathcal I_n,\varrho_{b,n},\mathcal B_{b,n})$, where $\varrho_{b,n}=1+\min\{1,h_*/(2R_{\eta,n})\}$ and $\mathcal B_{b,n}=1+|b(0)|+2Y_*R_{\eta,n}+\sqrt2Y_*(1+h_*)$. Indeed, for $z\in E_{\varrho_{b,n}}$, we have $|\operatorname{Im}(R_{\eta,n}z)|\leq h_*/2$ and $|\operatorname{Re}(R_{\eta,n}z)|\leq 2R_{\eta,n}$. We use the common constants $\varrho_n=\min\{\varrho_{g,n},\varrho_{I,D},\varrho_{b,n}\}$ and $\mathcal B_n=\max\{1,\mathcal B_{g,n},2a/c_\mu,\mathcal B_{b,n}\}$. Sharper values for a particular model can replace these.

For a common arithmetic range, take
\[
R_n=2+B_{\rm cal}+Y_*+V_*+R_{\eta,n}+|b(0)| +Y_*R_{\eta,n}+Y_*^2/a+a^{-1}+c^{-1}.
\]
Equation~\eqref{eq:glm-cumulant-strip} gives the common range bound
\[
\begin{gathered}
\max\{|y|,\|v_S(z)\|_\infty,\|\theta_S\|_\infty, |u|,|b(u)|,|b'(u)|,|b''(u)|\}\leq   R_n,\\
\max\{a^{-1},B_{\rm cal}+1,c^{-1}+1\}\leq   R_n ,\quad x=  (y,z)\in\mathcal X_n ,\quad \theta_S\in\Theta_{0,S} ,\quad u\in\mathcal I_n,
\quad 1\leq   R_n\leq   C(D+1).
\end{gathered}
\]
Set
\begin{equation}
 \delta_n=(C_0n^3D^4)^{-1},\qquad
 \epsilon_{{\rm c},n}=\frac{\delta_n}{16D R_n^3}.
 \label{eq:glm-component-precision}
\end{equation}
Let $\widetilde g_S$, $\widetilde J_S$, $\widetilde b$ and $\widetilde b'$ approximate $g_S$, $I_S^{-1}$, $b$ and $b'$, respectively, with
\[
\begin{gathered}
\sup_{m\in\mathcal R_{g,S}}\max_{j\leq   d_S} |\widetilde g_{S,j}(m)-g_{S,j}(m)|\leq  \epsilon_{{\rm c},n},\quad 
\sup_{\theta\in\Theta_{0,S}}\max_{j,k\leq   d_S} |\widetilde J_{S,jk}(\theta)-(I_S(\theta)^{-1})_{jk}|\leq  \epsilon_{{\rm c},n},\\
\sup_{u\in\mathcal I_n}|\widetilde b(u)-b(u)| \ \vee\ \sup_{u\in\mathcal I_n}|\widetilde b'(u)-b'(u)|\leq  \epsilon_{{\rm c},n}.
\end{gathered}
\]
Here $\widetilde b'$ denotes the network approximating $b'$.

For each scalar multiplier, Lemma~\ref{lem:multiplication} gives $\sup_{|u|,|v|\leq 3R_n}|\mathsf M_{3R_n,\epsilon_{{\rm c},n}}(u,v)-uv|\leq \epsilon_{{\rm c},n}$.

The affine normalization of each moment, parameter and predictor rectangle is a fixed input map. Put
$K_{{\rm dom},n}=1+\max_{\mathcal R,j}\frac{2}{v_j-u_j}$, $\mathcal R=\prod_j[u_j,v_j]$. Thus $K_{{\rm dom},n}$ bounds the slopes of the affine normalizations of all domains of the analytic maps. The translations of these maps are fixed coefficients, and parameter perturbations through them involve only the inverse side lengths. For every $S$, the analytic scalar maps have input and output dimensions
\[
\begin{gathered}
g_S:\mathcal R_{g,S}\to\mb R^{d_S},\quad I_S^{-1}:\Theta_{0,S}\to\mb R^{d_S\times d_S},\quad
b,b':\mathcal I_n\to\mb R,\quad d_S+d_S^2+2\leq (D+1)^2.
\end{gathered}
\]
The five terms below count products for the score predictor, score coordinates, separate likelihood predictor, response times predictor and matrix-vector update:
\[
N_{\rm mult}(S)=d_S+d_S+d_S+1+d_S^2=d_S^2+3d_S+1\leq 2(D+1)^2.
\]
Affine sums and multiplication by $a^{-1}$ are exact. The score coefficients are shared across observations, and $T_S$ and averaging are fixed operations.

Define
\begin{equation}
 \begin{aligned}
 A_{{\rm an},n}
   &=\frac{2^D\mathcal B_n}{(1-\varrho_n^{-1})^D},
 \quad R_{{\rm an},n}
   =1\vee\frac{2\mathcal B_n}{(1-\varrho_n^{-1})^2},
 \quad k_{{\rm an},n}=\max\left\{1,
       \left\lceil\frac{\log(4DA_{{\rm an},n}/\epsilon_{{\rm c},n})}
                             {\log\varrho_n}\right\rceil\right\},\\
 \zeta_n
   &=\frac{\epsilon_{{\rm c},n}}{8DA_{{\rm an},n}3^{k_{{\rm an},n}}},
 \qquad J_{{\rm an},n}
   =1+\left\lceil\log_4(2R_{{\rm an},n}^2/\zeta_n)\right\rceil.
 \end{aligned}
 \label{eq:sparse-analytic-scales}
\end{equation}
These are the quantities in the proof of Lemma~\ref{lem:analytic-approx}, with $D$ as a common upper bound on the input dimensions and with ellipse parameter $\varrho_n$, modulus bound $\mathcal B_n$ and accuracy $\epsilon_{{\rm c},n}$. After rounding depth and coefficient counts up to integers, the depth, parameter and coefficient bounds for one support are
\begin{equation}
 \begin{aligned}
 L_{{\rm vs},n}
   &=C_L\{k_{{\rm an},n}+1+\log(eD)\}J_{{\rm an},n},\quad
 V_{{\rm vs},n}
   =C_V(D+1)^2D(k_{{\rm an},n}+1)^D J_{{\rm an},n},\\
 P_{{\rm vs},n}
   &=C_P\{1+R_{{\rm an},n}^2+2^D\mathcal B_n+R_n^2\},\quad
 \Gamma_{{\rm vs},n}=C_\Gamma\sqrt{D(D+1)^2(k_{{\rm an},n}+1)^D J_{{\rm an},n}}\,P_{{\rm vs},n}.
 \end{aligned}
 \label{eq:sparse-analytic-complexity}
\end{equation}
The positive constants $C_L,C_V,C_P,C_\Gamma$ do not depend on $n,p,S$. The coefficient bound $P_{{\rm vs},n}$ covers the constructed coefficients, the scalar multipliers and the bounded fixed coefficients within the component blocks. The quantity $\Gamma_{{\rm vs},n}$ bounds the Euclidean norm of the parameter vector of the networks for one support.

Write $\widetilde F_S$ for the parameter estimator for $S$ and $\varphi_S$ for its likelihood network. The proof of Theorem~\ref{thm:vs-approx} shows that we can build these networks within the bounds~\eqref{eq:sparse-analytic-complexity} so that
\begin{equation}
 \begin{gathered}
  \widetilde F_S(X^n)\in\Theta_{0,S},\qquad
 \sup_{X^n\in\mathcal X_n^n}\|\widetilde F_S(X^n)-F_S^{\rm os}(X^n)\|_2\leq n^{-3},\\
 \sup_{x\in\mathcal X_n,\ \theta_S\in\Theta_{0,S}}
 |\varphi_S(x,\theta_S)-\ell_S^\circ(x,\theta_S)|\leq\{6n\log(en)\}^{-1}.
 \end{gathered}
 \label{eq:vs-module-accuracy}
\end{equation}
The lemmas below hold for any estimators and likelihood networks satisfying~\eqref{eq:vs-module-accuracy}.

\begin{remark}[A logistic family satisfying the assumptions]
\label{ex:glm-verified-modules}
Let $Z_1,\ldots,Z_p$ be independent with $\mb P(Z_j=1)=\mb P(Z_j=-1)=1/2$, and let $Y\mid Z\sim\operatorname{Bernoulli}\{\sigma_{\rm log}(\icpt+Z^\top\beta)\}$ with $\sigma_{\rm log}(u)=(1+e^{-u})^{-1}$, so that $a=1$ and $b(u)=\log(1+e^u)$. For a number $r\in(0,1/D]$, take $\Theta_{0,S}=[-r,r]^{d_S}$ for every $S\in\mathcal S$, and let the intercept and every coefficient be uniform on $[-r/2,r/2]$ under the task distribution. The responses and the covariates are bounded by one, so the setting of Section~\ref{sec:theory-variable} applies, and the convention on $B$ is met with a numerical constant. All linear predictors then lie in $[-1,1]$, so Assumption~\ref{ass:sparse}\textup{(i)} holds with $m_b=\sigma_{\rm log}'(1)$ and $M_b=1/4$. Since $\mb E_{Q_Z}[(h_{\icpt}+Z^\top h_\beta)^2]=h_{\icpt}^2+\|h_\beta\|_2^2$, part~(ii) holds with $\underline\lambda_Z=\overline\lambda_Z=1$. Assumption~\ref{ass:sparse-proposal} holds with $c_-=c_+=1/r$, $L_\pi=0$, $a_\pi=r/2$ and margin $r/2$. The constant $\rho_\mu$ of Lemma~\ref{lem:glm-natural-modules} depends only on these bounds and not on $r$. If in addition $2r\sqrt D\leq\rho_\mu/(4DM_b\overline\lambda_Z)$ and $\rho=\min\{r/2,\rho_\mu/6\}$, Lemma~\ref{lem:glm-inverse-extension}\textup{(c)} gives Assumption~\ref{ass:sparse}\textup{(iii)} with $\varrho_{g,n}=2$ and $\mathcal B_{g,n}=B+2$, uniformly in $n$ and $p$. Thus all assumptions of Theorems~\ref{thm:vs-approx} and~\ref{thm:vs-gen} hold for every $p\geq2s_{\max}$, with $L_{{\rm vs},n}=O\{\log^2(en)\}$ and $V_{{\rm vs},n}=O\{\log^{D+1}(en)\}$. The task distribution includes arbitrarily weak signals and parameters arbitrarily close to the boundaries of the slabs.
\end{remark}

The next lemma gives likelihood expansions uniformly over all candidate supports containing the true support.

\begin{lemma}[Simultaneous local expansions]\label{lem:sparse-local}
Suppose $\Lambda_n^3\leq c_0n$, where $c_0$ is sufficiently small.
There is an event $\mathcal E_n$ on which the following bounds hold simultaneously for every $S\in\mathcal S$ with $S\supseteq S_0$:
\begin{align}
 \|\widehat\theta_S-\theta_{0,S}\|_2
 +\|\theta_S^{(0)}-\theta_{0,S}\|_2
 \leq C\sqrt{\Lambda_n/n},&\quad D_S(\widetilde F_S)\leq\frac{1}{4\log(en)},\notag\\
 \left|2\{\ell_{n,S}(\widehat\theta_S)
       -\ell_{n,S}(\theta_{0,S})\}-\|V_S\|_2^2\right|&\leq\frac18,\label{eq:sparse-quadratic}
\end{align}
where $\theta_S^{(0)}$ is the moment initializer in~\eqref{eq:vs-one-step-map}, $\widetilde F_S$ satisfies~\eqref{eq:vs-module-accuracy}, and
\[
V_S=  I_S(\theta_{0,S})^{-1/2}n^{-1/2}\sum_i s_{S,\theta_{0,S}}(X_i).
\]
On $\mathcal E_n$, for every $S\in\mathcal S$ with $S\supseteq S_0$,
\[
\widehat\theta_S\in\operatorname{int}\Theta_{0,S},\qquad \Pi_{\mathcal R_{g,S}}\overline T_S=  \overline T_S,\qquad \mu_S^{-1}(\overline T_S)\in\operatorname{int}\Theta_{0,S},\qquad \theta_S^{(0)}=  \mu_S^{-1}(\overline T_S).
\]
Furthermore,
\begin{equation}
 \|\widetilde F_S-\theta_{0,S}\|_2^2
 \leq \frac Cn\{1+\|V_S\|_2^2\}
 \quad\hbox{on }\mathcal E_n.
 \label{eq:sparse-local-estimation}
\end{equation}
Uniformly over the generating parameters,
$\mathbb P_0(\mathcal E_n^c)\leq   Cn^{-10}$.
Choose a fixed $N\geq3$ large enough for the numerical thresholds in the proof, and then choose $c_0\leq N^{-1}$. The sample-size condition gives
$n\geq\Lambda_n^3/c_0\geq c_0^{-1}\geq N$.
\end{lemma}

For every candidate support $S$, we write
\[
r_-(S)=  |S_0\setminus S|,\qquad r_+(S)=  |S\setminus S_0|,\qquad e_S=  \|\beta_{0,S_0\setminus S}\|_2^2,
\]
and retain the signed EBIC difference
\begin{equation*}
 \Delta_S=\operatorname{pen}(S)-\operatorname{pen}(S_0)
 =\frac{r_+(S)-r_-(S)}{2}\log n+
   \log\frac{\binom p{s_0-r_-(S)+r_+(S)}}{\binom p{s_0}}.
\end{equation*}
Unlike the strict-superset increment, $\Delta_S$ can be negative.

We increase the uniform envelope constant so that $B\geq1$ and use $B_Y=\max\{1,2B\}$ and $\lambda_Y$ from Lemma~\ref{lem:lik-process}. We set
\[
h=  \frac{n^{-2}}{B\sqrt D},\qquad N_{n,d}=  (1+2B^2Dn^2)^d,\qquad N_{n,0}=  1,
\]
and, for $t\geq0$, define the clipped probability bound
\[
q_{n,S}(t)=  \min\{1, N_{n,d_S}\exp[-\lambda_Y \{\frac{nc_K e_S}{2}+\Delta_S-t-\frac1n\}]\}.
\]
Here $c_K=m_b\underline\lambda_Z/(2a)$ is the constant already derived in Lemma~\ref{lem:glm-kl}. All constants in this definition are independent of $S$, $n$, $p$ and $\theta_0$.

The next two lemmas compare the information criterion of a candidate $S$ with that of $S_0$. The first applies to every candidate and explicitly accounts for the energy of the omitted signals. The second bounds the likelihood gain of a strict superset by a standardized efficient score.

\begin{lemma}[Comparison retaining omitted signal energy]
\label{lem:vs-omitted-energy}
For every candidate $S$ and every $t\geq0$,
\begin{equation}
\mb P_0\{\operatorname{IC}_S- \operatorname{IC}_{S_0}\leq t\} \leq q_{n,S}(t).\label{eq:vs-exact-energy-comparison}
\end{equation}
\end{lemma}

\begin{proof}
In this proof, $C$ depends only on the response family, $a$, $m_b$, $M_b$, $\underline\lambda_Z$, $\overline\lambda_Z$, $B$, $\rho$ and $c_\Theta$.
All likelihood ratios are with respect to $\nu_Y\otimes Q_Z$. Subtracting the common baseline from each criterion preserves its minimizer and differences. By definition, since $\widehat\theta_{S_0}$ maximizes $\ell_{n,S_0}$ on $\Theta_{0,S_0}\ni\theta_{0,S_0}$,
\[
\begin{aligned}
\operatorname{IC}_S-\operatorname{IC}_{S_0}
&=\ell_{n,S_0}(\widehat\theta_{S_0})-\ell_{n,S}(\widehat\theta_S)+\Delta_S\\
&\geq\ell_n(\theta_0)-\ell_{n,S}(\widehat\theta_S)+\Delta_S
=\inf_{\theta_S\in\Theta_{0,S}}nP_n\log(p_{\theta_0}/p_{S,\theta_S})+\Delta_S,
\end{aligned}
\]
where the last step uses optimality of $\widehat\theta_S$ and baseline cancellation. Hence
\[
\{\operatorname{IC}_S-\operatorname{IC}_{S_0}\leq t\}\subseteq\Bigl\{\inf_{\theta_S\in\Theta_{0,S}}P_n\log\frac{p_{\theta_0}}{p_{S,\theta_S}}\leq\frac{t-\Delta_S}{n}\Bigr\}.
\]

Put $Y_{S,\theta_S}=\log(p_{\theta_0}/p_{S,\theta_S})$. It has envelope $2B\leq B_Y=\max\{1,2B\}$. By~\eqref{eq:sparse-envelopes}, first derivatives are bounded by $B$, so $d_S\leq D$ gives Lipschitz constant $L_Y=B\sqrt D$. Since $\Theta_{0,S}$ lies in a ball of radius $B\sqrt{d_S}\leq B\sqrt D$, $N_S(h)\leq(1+2B\sqrt D/h)^{d_S}=N_{n,d_S}$. Also $L_Yh=n^{-2}$. By~\eqref{eq:glm-kl-omitted} of Lemma~\ref{lem:glm-kl}, with all retained coefficients refitted, $R_S^*=\inf_{\theta_S\in\Theta_{0,S}}D_{\rm KL}(\mb P_{\theta_0}\|\mb P_{S,\theta_S})\geq c_Ke_S$. Applying~\eqref{eq:likelihood-gap} of Lemma~\ref{lem:lik-process} to $S$ with $v=(t-\Delta_S)/n$ and mesh $h$ gives
\[
\begin{aligned}
\mb P_0\{\operatorname{IC}_S-\operatorname{IC}_{S_0}\leq t\}
&\leq N_{n,d_S}\exp[-\lambda_Yn\{c_Ke_S/2-(t-\Delta_S)/n-n^{-2}\}]\\
&=N_{n,d_S}\exp[-\lambda_Y\{nc_Ke_S/2+\Delta_S-t-n^{-1}\}].
\end{aligned}
\]
Since a probability is at most $1$, this gives $q_{n,S}(t)$ and~\eqref{eq:vs-exact-energy-comparison}.
\end{proof}

\begin{lemma}[Added variables]\label{lem:vs-overfit}
For $S\supsetneq S_0$, put $b=|S\setminus S_0|$ and
\[
 \Delta_b=\operatorname{pen}(S)-\operatorname{pen}(S_0)
 =\frac b2\log n+\log\frac{\binom p{s_0+b}}{\binom p{s_0}}.
\]
Partition the score and information at $\theta_{0,S}$ into the true coordinates $A$ and added coordinates $B$, and define
\[
\begin{gathered}
J_S=  I_{BB}-I_{BA}I_{AA}^{-1}I_{AB},\quad \xi_S=  J_S^{-1/2}(s_B-I_{BA}I_{AA}^{-1}s_A),\quad
U_S=  n^{-1/2}\sum_{i=  1}^n\xi_S(X_i).
\end{gathered}
\]
Then $\mathbb E_0[\xi_S]=0$, $\mathbb E_0[\xi_S\xi_S^\top]=I_b$ and $\|\xi_S\|_2\leq C$, and $\{\operatorname{IC}_S-\operatorname{IC}_{S_0}\leq1\}\cap\mathcal E_n\subseteq\{\|U_S\|_2^2\geq2\Delta_b-9/4\}\cap\mathcal E_n$. For every such support,
\begin{align}
 \mb P_0(\|U_S\|_2^2\geq2\Delta_b-9/4)
 &\leq (C\sqrt{\Lambda_n})^b e^{-\Delta_b},
 \notag\\
 \mb E_0\left[\|U_S\|_2^2
  \ind\{\|U_S\|_2^2\geq2\Delta_b-9/4\}\ind_{\mathcal E_n}\right]
 &\leq Cb\Lambda_n(C\sqrt{\Lambda_n})^b e^{-\Delta_b}.
 \label{eq:sparse-overfit-moment}
\end{align}
\end{lemma}

\subsubsection{Support probabilities and auxiliary bounds for the variable-selection network}\label{app:vs-network}

The component networks give an approximate likelihood and parameter estimate for each support. We now study how errors in these quantities affect the weights assigned to the supports and, ultimately, the inclusion probabilities and coefficient estimates. In the probability comparisons, it is useful to work with the weights on the full collection of supports before taking their marginals.

In the remainder of this subsection, $C,c$ depend on $D$, on the constants of the setting and of Assumptions~\ref{ass:sparse} and~\ref{ass:sparse-proposal}, and on $a$. They do not depend on $n$, $p$, the supports or the generating coefficients. Appendix~\ref{app:vs-smax} derives the dependence on $s_{\max}$ displayed in Section~\ref{sec:theory-variable}. There the constants do not depend on $D$, and the sample-size condition reads $DA_n^3\leq c_0n$. The condition $\Lambda_n^3\leq c_0n$ with a sufficiently small $c_0$ implies all lower thresholds on $n$ used below. We write
\[
N_{\mathcal S}=|\mathcal S|\leq(ep/s_{\max})^{s_{\max}},\quad
A_n=D\log(en)+s_{\max}\log(ep/s_{\max}),\quad
\epsilon_n=(1+np)^{-1},
\]
so that $\Lambda_n\leq A_n\leq D\Lambda_n$, $\log N_{\mathcal S}\leq A_n$, $\operatorname{pen}(S)\leq A_n$ for every $S\in\mathcal S$, and $p\epsilon_n\leq n^{-1}$. A generating task $(S_0,\theta_0)$ is \emph{allowed} if $\theta_{0,S_0}$ belongs to the support $K_{S_0}=\mathcal I_{S_0}=\mathcal I_0\times\prod_{j\in S_0}\mathcal I_j$ of $\pi_{S_0}$. Since $0\in\mathcal I_j$ for $j\in[p]$, its zero extension $\theta_{0,S}$ to a support $S\supseteq S_0$ belongs to $\mathcal I_S$. By Assumption~\ref{ass:sparse-proposal}(iii), $\theta_{0,S}$ therefore lies at least $\rho$ from the boundary of $\Theta_{0,S}$, so the preceding lemmas apply. An allowed task may have zero coefficients in $S_0$. We write $\beta_{\min}=\min_{j\in S_0}|\beta_{0j}|$ and $\mathcal A_+=\{S\in\mathcal S:S\supseteq S_0\}$. For a candidate $S$, we set $e_S=\|\beta_{0,S_0\setminus S}\|_2^2$.

We denote the likelihood networks by $\varphi_S$. We use the selection setup of Appendix~\ref{app:selection-setup} with the supports as candidates, $\varepsilon_\ell=\{6n\log(en)\}^{-1}$, the estimates $\widetilde\theta_S=\widetilde F_S(X^n)$ and the penalties $\operatorname{pen}(S)$. Thus, with $\widetilde a_S$ as in~\eqref{eq:vs-network-logits} evaluated at $\widetilde\theta_S$ and $D_S=D_S(\widetilde F_S)$, Lemma~\ref{lem:criterion} gives, for every dataset,
\begin{equation}
\widetilde a_S=a_S^{\rm E}-D_S+e_S^\ell,\qquad D_S\geq0,\qquad|e_S^\ell|\leq\frac1{6\log(en)},\qquad a_S^{\rm E}=-\operatorname{IC}_S .
\label{eq:vs-logit-identity}
\end{equation}
Let $\widetilde Q_S=e^{\widetilde a_S}/\sum_{S'}e^{\widetilde a_{S'}}$ and $\widetilde q_j^{\rm ex}=\sum_{S\ni j}\widetilde Q_S$. We compute the marginal likelihoods $m_S$ with the centered log likelihood $\ell_{n,S}$, which multiplies all of them by a common factor and leaves the posterior probabilities $Q_S^*$ unchanged. With
\[
\delta_S=\log m_S-\Bigl\{\ell_{n,S}(\widehat\theta_S)-\frac{d_S}2\log n\Bigr\},
\]
Assumption~\ref{ass:sparse-proposal}(i) gives $\log\{w(S)m_S\}=a_S^{\rm E}+\delta_S-\log s_{\max}$, so that
\begin{equation}
Q_S^*\propto Q_S^{\rm E}\,e^{\delta_S},\qquad
Q_S^*\propto\widetilde Q_S\,e^{\delta_S+D_S-e_S^\ell}.
\label{eq:vs-posterior-tilt}
\end{equation}
For $b\geq1$ and $S\supsetneq S_0$ with $|S\setminus S_0|=b$, the penalty difference $\Delta_b$ of Lemma~\ref{lem:vs-overfit} satisfies $b\leq\frac b2\log n\leq\Delta_b\leq b\Lambda_n$ for $n\geq e^2$, and
\begin{equation}\label{eq:ebic-cancellation}
 \binom{p-s_0}{b}e^{-\Delta_b}=\binom{s_0+b}{b}n^{-b/2}.
\end{equation}

In the RRDS network, the products $y z_j$ that form the moment statistic $T_S$ are computed by a multiplication network. Let $G_{g,n}\geq1$ bound the Lipschitz constants of the maps $g_S$ on their rectangles $\mathcal R_{g,S}$, which are finite by the Cauchy estimates for the extensions in Assumption~\ref{ass:sparse}\textup{(iii)}. For $\upsilon\in(0,1)$, let $\mathsf M_{R_n,\upsilon}$ be the multiplication network of Lemma~\ref{lem:multiplication} on $[-R_n,R_n]^2$ with accuracy $\upsilon$, put $T_S^\upsilon(y,z)=(y,(\mathsf M_{R_n,\upsilon}(y,z_j))_{j\in S})$, and let $F_S^{{\rm os},\upsilon}$ be the map~\eqref{eq:vs-one-step-map} with $P_nT_S^\upsilon$ in place of $\overline T_S$.

\begin{lemma}[Moment products]\label{lem:vs-moment}
There is a constant $C_{\rm os}$ such that, for every $S\in\mathcal S$, every $\upsilon\in(0,1)$ and every dataset in $\mathcal X_n^n$, $\|F_S^{{\rm os},\upsilon}(X^n)-F_S^{\rm os}(X^n)\|_2\leq C_{\rm os}G_{g,n}\,\upsilon$. Consequently, with $\upsilon_n=(2C_{\rm os}G_{g,n}n^3)^{-1}$ and the constant $C_0$ in~\eqref{eq:glm-component-precision} doubled, the estimators $\widetilde F_S$ computed from the pooled statistics $P_nT_S^{\upsilon_n}$ satisfy~\eqref{eq:vs-module-accuracy}. The $p$ products are computed once and shared by all supports, with depth and number of parameters at most $CJ_{T,n}$ and $CpJ_{T,n}$, respectively, where $J_{T,n}=1+\log(R_n^2/\upsilon_n)$.
\end{lemma}

We include $J_{T,n}$ in the bounds $L_{{\rm vs},n}$ and $V_{{\rm vs},n}$ of~\eqref{eq:sparse-analytic-complexity} by enlarging $C_L$ and $C_V$ and adding $J_{T,n}$ to the two brackets. Under the conditions stated after Assumption~\ref{ass:sparse}\textup{(iii)}, $\log G_{g,n}=O\{\log(en)\}$ and the stated orders of $L_{{\rm vs},n}$ and $V_{{\rm vs},n}$ are unchanged.

The next lemma writes the network of Section~\ref{sec:theory-variable} in the form~\eqref{eq:rrds-init}--\eqref{eq:rrds-output}. Let $H_{\mathcal S}=\lceil\log_2N_{\mathcal S}\rceil$ and $\eta_n=(np)^{-2}$. For a nonempty set $\mathcal C\subseteq\mathcal S$ and numbers $v_S\in[-B,B]$, $S\in\mathcal C$, put
\[
\operatorname{lse}(\mathcal C)=\log\sum_{S\in\mathcal C}e^{\widetilde a_S},\qquad
\operatorname{avg}(\mathcal C;v)=\frac{\sum_{S\in\mathcal C}e^{\widetilde a_S}v_S}{\sum_{S\in\mathcal C}e^{\widetilde a_S}}.
\]
The proof of Lemma~\ref{lem:vs-rrds} constructs ReLU approximations $\operatorname{LSE}(\mathcal C)$ and $\operatorname{AVG}(\mathcal C;v)$ of these two functions of $(\widetilde a_S,v_S)_{S\in\mathcal C}$. For all real logits and all values in $[-B,B]$, they satisfy
\begin{equation}
|\operatorname{LSE}(\mathcal C)-\operatorname{lse}(\mathcal C)|\leq\frac{\eta_n}2,\qquad
|\operatorname{AVG}(\mathcal C;v)-\operatorname{avg}(\mathcal C;v)|\leq\eta_n,\qquad|\operatorname{AVG}(\mathcal C;v)|\leq B .
\label{eq:vs-tree-accuracy}
\end{equation}
The state is $z=((\vartheta_S^0)_S,(\theta_S)_S,(a_j)_{j\in[p]},(b_j)_{j\in[p]},b_0)$, and the decoder computes
\begin{equation}
\begin{aligned}
a_j&=\operatorname{LSE}\{S:S\ni j\}-\operatorname{LSE}\{S:S\not\ni j\},\\
b_j&=\operatorname{AVG}\bigl(\{S:S\ni j\};(\theta_{S,j})_S\bigr),\qquad
b_0=\operatorname{AVG}\bigl(\mathcal S;(\theta_{S,0})_S\bigr),
\end{aligned}
\label{eq:vs-outputs}
\end{equation}
where $\theta_{S,0}$ is the intercept coordinate of $\theta_S$ and the logits $\widetilde a_S$ are computed from the pooled likelihood values at the current $\theta_S$. Both index sets in the first line are nonempty because $s_{\max}\leq p/2$.

\begin{lemma}[The variable-selection network]\label{lem:vs-rrds}
Let $R\geq2$, $d_z=2\sum_{S\in\mathcal S}d_S+2p+1$ and $d_u=\sum_{S\in\mathcal S}d_S+N_{\mathcal S}$. Define the blocks as in Lemma~\ref{lem:ms-rrds}, with the supports in the role of the candidates. The block $\phi_0$ computes the moment statistics $T_S^{\upsilon_n}$ and $\rho_0$ the projected moment estimates, stored in both $\vartheta_S^0$ and $\theta_S$, and zeros for the remaining coordinates. The block $\phi$ evaluates the score networks at $\vartheta_S^0$ and the likelihood networks $\varphi_S$ at $\theta_S$. The block $\rho$ leaves $\vartheta_S^0$ unchanged, overwrites $\theta_S$ by the projected one-step estimate computed from $\vartheta_S^0$, and overwrites $(a_j,b_j)_{j\in[p]}$ and $b_0$ by~\eqref{eq:vs-outputs}. With the output map $\Psi$ of Section~\ref{sec:theory-variable}, this is an RRDS network $\widetilde F=(\widetilde\theta,\widetilde q)$ of the form~\eqref{eq:rrds-init}--\eqref{eq:rrds-output}, and the following hold for every dataset.

\textup{(a)} For $r\geq1$, $\theta_S^{(r)}=\widetilde\theta_S\in\Theta_{0,S}$, and for $r\geq2$ the coordinates $(a_j,b_j)_j$ and $b_0$ are given by~\eqref{eq:vs-outputs} with $\theta_S=\widetilde\theta_S$. In particular, the output is the same for every $R\geq2$.

\textup{(b)} For every $j\in[p]$, $|a_j-\log\{\widetilde q_j^{\rm ex}/(1-\widetilde q_j^{\rm ex})\}|\leq\eta_n$.

\textup{(c)} Let $c_0^{\rm av}=\sum_{S\in\mathcal S}\widetilde Q_S\widetilde\theta_{S,0}$ and $c_j^{\rm av}=\sum_{S\ni j}\widetilde Q_S\widetilde\theta_{S,j}/\widetilde q_j^{\rm ex}$ for $j\in[p]$. Then $|b_0-c_0^{\rm av}|\leq\eta_n$ and $|b_j-c_j^{\rm av}|\leq\eta_n$, and $b_0,b_j\in[-B,B]$.

\textup{(d)} For $R=2$, $\widetilde F\in\mathcal F(L,W)$ with $L\leq C(L_{{\rm vs},n}+H_{\mathcal S}\Lambda_n^2)$ and $W\leq N_{\mathcal S}\{V_{{\rm vs},n}+Cp\Lambda_n^3+CDH_{\mathcal S}\Lambda_n^2\}$, where $C$ depends only on $B$. All coefficients are bounded by $C(n+A_n+P_{{\rm vs},n}+K_{{\rm dom},n})$, and every affine row has at most $CW$ coefficient positions.
\end{lemma}

The next lemma relates the network output to the weighted average of the fitted parameters. Let $v=(b_0,\widetilde q_1b_1,\ldots,\widetilde q_pb_p)$ be the coefficient vector before the projection $\Pi_{\rm b}$ in $\Psi$, so that $\widetilde\theta=\Pi_{\rm b}v$, and let $\theta^{\rm av}=\sum_{S\in\mathcal S}\widetilde Q_S\,\iota_S\widetilde\theta_S\in\mb R^{1+p}$ be the exact weighted average of the zero extensions of the one-step estimates, whose coordinates are $c_0^{\rm av}$ and $\widetilde q_j^{\rm ex}c_j^{\rm av}$.

\begin{lemma}[Properties of the output]\label{lem:vs-heads}
For every dataset and $n\geq3$, the output of the network of Lemma~\ref{lem:vs-rrds} satisfies
\begin{gather*}
\widetilde q_j\leq e^{\eta_n}\widetilde q_j^{\rm ex}+\epsilon_n,\quad
1-\widetilde q_j\leq e^{\eta_n}(1-\widetilde q_j^{\rm ex})+\epsilon_n,\quad
\sum_{j=1}^p\widetilde q_j^{\rm ex}\leq s_{\max},\quad 
\sum_{j=1}^p\widetilde q_j\leq2s_{\max},\quad
\widetilde\theta=v,\\
\|\widetilde\theta\|_2^2\leq3B^2s_{\max},\quad
|\widetilde S|\leq4s_{\max},\quad
\|\widetilde\theta-\theta^{\rm av}\|_1\leq\frac{3+4B}n,\quad
\sum_{S\in\mathcal S}\widetilde Q_S(\widetilde a_{S_0}-\widetilde a_S)\leq\log N_{\mathcal S}\quad(S_0\in\mathcal S).
\end{gather*}
\end{lemma}

\begin{proof}
Let $t_j$ be the log odds of $\widetilde q_j^{\rm ex}$, that is, $t_j=\log\{\widetilde q_j^{\rm ex}/(1-\widetilde q_j^{\rm ex})\}$. Then $\sigma_{\log}(t_j)=\widetilde q_j^{\rm ex}$, and $|a_j-t_j|\leq\eta_n$ by Lemma~\ref{lem:vs-rrds}(b). For $\gamma\geq0$, $\sigma_{\log}$ is increasing with $\sigma_{\log}(t+\gamma)\leq e^\gamma\sigma_{\log}(t)$ and $1-\sigma_{\log}(t-\gamma)\leq e^\gamma\{1-\sigma_{\log}(t)\}$, so before clipping
\[
\sigma_{\log}(a_j)\leq\sigma_{\log}(t_j+\eta_n)\leq e^{\eta_n}\widetilde q_j^{\rm ex}, 1-\sigma_{\log}(a_j)\leq1-\sigma_{\log}(t_j-\eta_n)\leq e^{\eta_n}(1-\widetilde q_j^{\rm ex}).
\]
Since $\sigma_{\log}\{\log(np)\}=1-\epsilon_n$, clipping $a_j$ to $[-\log(np),\log(np)]$ clips $\sigma_{\log}(a_j)$ to $[\epsilon_n,1-\epsilon_n]$ and changes it by at most $\epsilon_n$, giving the first two bounds.

Next, exchanging sums gives $\sum_j\widetilde q_j^{\rm ex}=\sum_j\sum_{S\ni j}\widetilde Q_S=\sum_S|S|\widetilde Q_S\leq s_{\max}$. Summing the first bound gives $\sum_j\widetilde q_j\leq e^{\eta_n}s_{\max}+p\epsilon_n\leq2s_{\max}$, since $\eta_n\leq1/9$ and $p\epsilon_n\leq1/3$.

By Lemma~\ref{lem:vs-rrds}(c), $|b_0|,|b_j|\leq B$. Since $\widetilde q_j^2\leq\widetilde q_j$,
\[
\|v\|_2^2=b_0^2+\sum_j\widetilde q_j^2b_j^2\leq B^2\Bigl(1+\sum_j\widetilde q_j\Bigr)\leq B^2(1+2s_{\max})\leq3B^2s_{\max}.
\]
Since $3B^2s_{\max}\leq(2B\sqrt{s_{\max}})^2$, $\widetilde\theta=\Pi_{\rm b}v=v$ and $\|\widetilde\theta\|_2^2\leq3B^2s_{\max}$. Also $|\widetilde S|=|\{j:\widetilde q_j>1/2\}|\leq2\sum_j\widetilde q_j\leq4s_{\max}$.

For $\theta^{\rm av}$, the first bound gives $\widetilde q_j-\widetilde q_j^{\rm ex}\leq(e^{\eta_n}-1)\widetilde q_j^{\rm ex}+\epsilon_n$. For the lower bound, the first logistic inequality at $t_j-\eta_n$ gives $\sigma_{\log}(t_j-\eta_n)\geq e^{-\eta_n}\widetilde q_j^{\rm ex}$, so $\widetilde q_j\geq e^{-\eta_n}\widetilde q_j^{\rm ex}-\epsilon_n$ and $\widetilde q_j^{\rm ex}-\widetilde q_j\leq(1-e^{-\eta_n})\widetilde q_j^{\rm ex}+\epsilon_n\leq(e^{\eta_n}-1)\widetilde q_j^{\rm ex}+\epsilon_n$. Hence $|\widetilde q_j-\widetilde q_j^{\rm ex}|\leq2\eta_n\widetilde q_j^{\rm ex}+\epsilon_n$, since $e^{\eta_n}-1\leq2\eta_n$. Also $|c_j^{\rm av}|\leq B$, and $\theta^{\rm av}$ has entries $c_0^{\rm av},\widetilde q_j^{\rm ex}c_j^{\rm av}$. With $\widetilde q_jb_j-\widetilde q_j^{\rm ex}c_j^{\rm av}=\widetilde q_j(b_j-c_j^{\rm av})+c_j^{\rm av}(\widetilde q_j-\widetilde q_j^{\rm ex})$ and Lemma~\ref{lem:vs-rrds}(c),
\[
\begin{aligned}
\|\widetilde\theta-\theta^{\rm av}\|_1
&\leq|b_0-c_0^{\rm av}|+\sum_{j=1}^p\bigl\{\widetilde q_j|b_j-c_j^{\rm av}|+B|\widetilde q_j-\widetilde q_j^{\rm ex}|\bigr\}\leq\eta_n(1+2s_{\max})+B(2\eta_ns_{\max}+p\epsilon_n),
\end{aligned}
\]
which is at most $(3+4B)/n$, since $s_{\max}\eta_n,p\epsilon_n\leq n^{-1}$ give $\eta_n(1+2s_{\max})\leq3s_{\max}\eta_n\leq3/n$ and $B(2\eta_ns_{\max}+p\epsilon_n)\leq3B/n$.

Finally, let $Z=\sum_Se^{\widetilde a_S}$. Then $\widetilde a_S=\log Z+\log\widetilde Q_S$. The entropy of $\widetilde Q$ is at most $\log N_{\mathcal S}$, so
\[
\sum_S\widetilde Q_S\widetilde a_S=\log Z+\sum_S\widetilde Q_S\log\widetilde Q_S\geq\log Z-\log N_{\mathcal S},
\]
and $\widetilde a_{S_0}\leq\log Z$ for $S_0\in\mathcal S$, so $\sum_S\widetilde Q_S(\widetilde a_{S_0}-\widetilde a_S)\leq\log N_{\mathcal S}$.
\end{proof}

The next lemma establishes a deterministic oracle inequality for the output $\widetilde\theta$. It requires no accuracy of the estimates $\widetilde\theta_S$ on the supports that omit active variables, and its constants do not depend on $D$. For $S\in\mathcal S$ and $u\in\Theta_{0,S}$, let $Y_{S,u}=\log(p_{\theta_0}/p_{S,u})$ and $R_S(u)=\mb E_0[Y_{S,u}]$, and put
\[
Z_n=\max_{S\in\mathcal S}\sup_{u\in\Theta_{0,S}}\Bigl\{\frac{R_S(u)}2-P_nY_{S,u}\Bigr\}\geq0,\qquad
k_n=nZ_n+D_{S_0}+\operatorname{pen}(S_0)+1 .
\]

\begin{lemma}[Oracle inequality for the coefficient estimate]\label{lem:vs-soft-oracle}
Let $c_K=m_b\underline\lambda_Z/(2a)$. For every allowed task and every dataset,
\[
\sum_{S\in\mathcal S}\widetilde Q_S\|\iota_S\widetilde\theta_S-\theta_0\|_2^2\leq\frac2{nc_K}\{k_n+\log N_{\mathcal S}\},
\]
and consequently
\[
\begin{gathered}
\|\widetilde\theta-\theta_0\|_2^2\leq\frac4{nc_K}\{k_n+\log N_{\mathcal S}\}+\frac C{n^2},\quad \mb E_{Q_Z}\bigl[\{\eta_{\widetilde\theta}(Z)-\eta_{\theta_0}(Z)\}^2\bigr]\leq\frac{4\overline\lambda_Z}{nc_K}\{k_n+\log N_{\mathcal S}\}+\frac C{n^2},
\end{gathered}
\]
where the expectation is over a new covariate vector $Z$ with $\widetilde\theta$ fixed, and $C$ depends only on $B$. Moreover, if $n\geq D$, then $\mb E_0[nZ_n]\leq CA_n$ and $\mb P_0\{nZ_n>C(A_n+t)\}\leq e^{-t}$ for $t\geq0$, with $C$ depending only on $a$, $B$ and $m_b$.
\end{lemma}

The next lemma bounds the contribution of the supports that strictly contain $S_0$. The vectors $U_S$ are the standardized efficient scores of Lemma~\ref{lem:vs-overfit}.

\begin{lemma}[Supports with added variables]\label{lem:vs-sq-odds}
Let $(S_0,\theta_0)$ be an allowed task and suppose $\Lambda_n^3\leq c_0n$. For $S\supsetneq S_0$ with $b=|S\setminus S_0|$, put $\varpi_S=\min\{1,e^b\exp(\|U_S\|_2^2/2-\Delta_b)\}$. Then
\[
\begin{gathered}
\mb E_0[\varpi_S^2]\leq(C\sqrt{\Lambda_n})^be^{-\Delta_b},\qquad
\mb E_0[\varpi_S]\leq C\Lambda_n(C\sqrt{\Lambda_n})^be^{-\Delta_b},\\
\mb E_0\bigl[(\|U_S\|_2^2/2-\Delta_b+b)_+\ind_{\mathcal E_n}\bigr]\leq(C\sqrt{\Lambda_n})^be^{-\Delta_b}.
\end{gathered}
\]
Consequently,
\[
\begin{gathered}
\sum_{S\supsetneq S_0}\mb E_0[\varpi_S^2]\leq C\sqrt{\frac{\Lambda_n}n},\qquad
\sum_{S\supsetneq S_0:\,b\geq2}b\,\mb E_0[\varpi_S]\leq\frac{C\Lambda_n^2}n,\\
\mb E_0\Bigl[\ind_{\mathcal E_n}\log\Bigl(1+\sum_{S\supsetneq S_0}e^b e^{\|U_S\|_2^2/2-\Delta_b}\Bigr)\Bigr]\leq\frac{C\Lambda_n^{3/2}}{\sqrt n}.
\end{gathered}
\]
\end{lemma}

The remaining lemmas are used for the comparison with the posterior inclusion probabilities. Let $K_0$ and $K_1$ be constants fixed in Lemma~\ref{lem:vs-near}, and put
\[
t_n=\sqrt{K_0A_n/n},\qquad w_n=K_1\sqrt{A_n/n},\qquad
d_\pi(\theta_0)=\min_{j\in S_0\cup\{0\}}\operatorname{dist}(\theta_{0j},\partial\mathcal I_j),
\]
where $\theta_{00}=\icpt_0$. A task is \emph{weak} if $\beta_{\min}\leq t_n$, and it is a \emph{boundary} task if $d_\pi(\theta_0)\leq w_n$. Let $\mathcal T$ and $\mathcal B$ denote the sets of weak and boundary tasks. A task that is neither weak nor a boundary task is called \emph{regular}.

For weak tasks, the supports that omit only weak signals behave as slightly misspecified candidates. Let $\mathcal A_0=\{S\in\mathcal S:e_S\leq t_n^2\}$, which contains $\mathcal A_+$ and equals $\mathcal A_+$ when the task is not weak, and let $U_S^\pi=n\operatorname{dist}(\widehat\theta_S,K_S)^2$.

\begin{lemma}[Supports that omit weak signals]\label{lem:vs-near}
There are constants $K_0$, $K_1$, $m$, $M$ and $c_0$ such that, if $\Lambda_n^3\leq c_0n$, every allowed task admits an event $\mathcal G\subseteq\mathcal E_n$ with $\mb P_0(\mathcal G^c)\leq Cn^{-10}$ on which the following hold.

\textup{(a)} For every $S\in\mathcal S$ and $\vartheta\in\Theta_{0,S}$, $mI_{d_S}\preceq-n^{-1}\nabla^2\ell_{n,S}(\vartheta)\preceq MI_{d_S}$.

\textup{(b)} For every $S\in\mathcal A_0$, $\widehat\theta_S\in\operatorname{int}(\Theta_{0,S})$ and $D_S\leq\{4\log(en)\}^{-1}$.

\textup{(c)} $Q^*(\mathcal A_0^c)$, $Q^{\rm E}(\mathcal A_0^c)$ and $\widetilde Q(\mathcal A_0^c)$ are at most $e^{-8A_n}$.

\textup{(d)} If the task is not a boundary task, then $\widehat\theta_S\in K_S$ for every $S\in\mathcal A_0$.
\end{lemma}

\begin{lemma}[Laplace bounds for sparse candidates]\label{lem:vs-laplace}
There is a constant $C_\pi$, depending only on $c_-$, $c_+$, $m$ and $M$, such that, on the event in Lemma~\ref{lem:vs-near}\textup{(a)} and for $n\geq a_\pi^{-2}$, $\delta_S\leq C_\pi d_S$ for every $S\in\mathcal S$, and $\delta_S\geq-C_\pi d_S-MU_S^\pi$ for every $S$ with $\widehat\theta_S\in\operatorname{int}(\Theta_{0,S})$.
\end{lemma}

\subsubsection{Bounds with explicit dependence on the maximal support size}\label{app:vs-smax}

The statements of Section~\ref{sec:theory-variable} display the dependence on $s_{\max}$, whereas the lemmas above are written with constants that may depend on $D=1+s_{\max}$. This subsection derives that dependence explicitly, so that the constants in the main statistical bounds need not absorb powers of the maximal support size. We first record how $D$ enters the lemmas above. We then prove two results that are specific to this part. The first is a moderate-deviation bound for the score of one added variable that retains the Gaussian tail prefactor (Lemmas~\ref{lem:cramer-prefactor} and~\ref{lem:vs-added-explicit}). The second compares the Laplace corrections of neighboring supports under the Lipschitz condition of Assumption~\ref{ass:sparse-proposal}\textup{(ii)} (Lemma~\ref{lem:vs-laplace-diff}).

In this part, and in the proofs of Propositions~\ref{prop:vs-bridge} and~\ref{prop:vs-bridge-est}, $C$ and $c$ denote constants that depend on $a$, $B$, $m_b$, $M_b$, $\underline\lambda_Z$, $\overline\lambda_Z$, $\rho$, $c_\Theta$ and the constants of Assumption~\ref{ass:sparse-proposal}, and \emph{not} on $D$. We assume
\begin{equation}
DA_n^3\leq c_0n
\label{eq:vs-smax-condition}
\end{equation}
for a sufficiently small such constant $c_0$. Since $D\leq A_n$, this implies $n\geq D^4$. Since $s_{\max}\log(ep/s_{\max})\geq s_{\max}\log(ep)-s_{\max}\log n$ and $\log(en)\geq1$, we have $A_n\geq\log(en)+s_{\max}\log(ep)$, and therefore
\begin{equation}
\Lambda_n\leq A_n\leq D\Lambda_n\leq2s_{\max}\Lambda_n,\qquad D\Lambda_n\leq3A_n .
\label{eq:vs-An-Lambda}
\end{equation}
Hence~\eqref{eq:vs-smax-condition} holds if $s_{\max}^4\Lambda_n^3\leq c_0n/16$, which is the condition of Theorem~\ref{thm:vs-approx} with $c_0/16$ in place of $c_0$. Since $\Lambda_n^3\leq9\{8+\log^3n+(\log p)^3\}$, for bounded $s_{\max}$ this condition holds if $(\log p)^3\leq cn$ for a sufficiently small constant $c$ and $n$ is sufficiently large. We write $r_n=\sqrt{A_n/n}$, so that
\begin{equation}
\sqrt D\,r_n\leq\frac{\sqrt{c_0}}{A_n},\qquad D^{3/2}r_n\leq\sqrt{c_0},\qquad
s_{\max}r_n\leq\sqrt2\,s_{\max}^{3/2}\sqrt{\frac{\Lambda_n}n}.
\label{eq:vs-rn}
\end{equation}

By~\eqref{eq:sparse-envelopes} and Assumption~\ref{ass:sparse}\textup{(i)}, the coordinates of the scores and of the moment statistics are bounded by constants, and $|b''|+|b'''|\leq C$. Hence $\|s_{S,\theta_S}\|_2\leq C\sqrt D$ and $\|\xi_S\|_2\leq C\sqrt D$. The eigenvalues of $I_S(\theta_S)$ on $\Theta_{0,S}$ are bounded above and below by constants. For $U\subseteq[p]$, we write $v_U(z)=(1,z_U^\top)^\top$, so that $\|v_U\|_2\leq C\sqrt D$ when $|U|\leq2s_{\max}$. We also set $H_{n,S}(\theta)=-n^{-1}\nabla^2\ell_{n,S}(\theta)=a^{-1}P_n[b''(v_S^\top\theta)v_Sv_S^\top]$.

\begin{lemma}[Hessians of the sparse candidates]\label{lem:vs-hessian}
Under~\eqref{eq:vs-smax-condition}, there is an event of probability at least $1-Cn^{-12}$ on which, for every $U\subseteq[p]$ with $|U|\leq2s_{\max}$, every $S\in\mathcal S$, and all $\theta,\theta'\in\Theta_{0,S}$,
\[
\begin{gathered}
\bigl\|P_n[v_Uv_U^\top]-\mb E_{Q_Z}[v_Uv_U^\top]\bigr\|_{\rm op}\leq C\sqrt D\,r_n,\qquad
\|H_{n,S}(\vartheta_S)-I_S(\vartheta_S)\|_{\rm op}\leq C\sqrt D\,r_n,\\
\|H_{n,S}(\theta)-H_{n,S}(\theta')\|_{\rm op}\leq C\sqrt D\,\|\theta-\theta'\|_2,
\end{gathered}
\]
where $\vartheta_S\in\Theta_{0,S}$ is any point chosen independently of the data, one for each $S$. The last bound also holds for two supports $S,S'\in\mathcal S$ and the blocks of $H_{n,S}(\theta)$ and $H_{n,S'}(\theta')$ on their common coordinates, with $\|\iota_S\theta-\iota_{S'}\theta'\|_2$ on the right side, and for $I_S$ in place of $H_{n,S}$.
\end{lemma}

\begin{proof}
The argument controls quadratic forms of the empirical Hessian on all candidate supports. A finite net reduces the matrix norm bound to scalar concentration, and the sparse design assumption controls the variance in each direction.
For the first claim, fix $U$ with $|U|\leq2s_{\max}$ and a unit vector $u$, and let $X=(v_U^\top u)^2$. Then $0\leq X\leq\|v_U\|_2^2\leq CD$. Since $u$ has at most $2s_{\max}$ slope coordinates,~\eqref{eq:glm-sparse-design} gives $\mb E X\leq\overline\lambda_Z$, so $\operatorname{Var}(X)\leq\mb E X^2\leq CD\,\mb E X\leq CD$. Take a $1/4$-net of the unit sphere in $\mb R^{1+|U|}$ for each $U$. There are at most $(ep)^{2s_{\max}}$ such sets $U$ and at most $9^{1+2s_{\max}}$ net points per sphere, so $A_n\geq\log(en)+s_{\max}\log(ep)$ and $A_n\geq D$ give at most $e^{CA_n}$ pairs $(U,u)$. Since $\log n\leq A_n$, Bernstein's inequality with variance and range $CD$ at level $CA_n$ gives
\[
|P_nX-\mb E X|\leq C(\sqrt{DA_n/n}+DA_n/n)=C(\sqrt D\,r_n+Dr_n^2)\leq C\sqrt D\,r_n
\]
for all pairs with probability at least $1-n^{-12}$, using $\sqrt D\,r_n\leq1$ from~\eqref{eq:vs-rn}. Applying $\|M\|_{\rm op}\leq2\max_u|u^\top Mu|$ over the net to the symmetric matrix $M=P_n[v_Uv_U^\top]-\mb E_{Q_Z}[v_Uv_U^\top]$ gives the first claim. By~\eqref{eq:glm-sparse-design} and $C\sqrt D\,r_n\leq C\sqrt{c_0}$, the eigenvalues of $P_n[v_Uv_U^\top]$ lie in $[\underline\lambda_Z/2,2\overline\lambda_Z]$ for $c_0$ small.

For the second claim, let $X_u=a^{-1}b''(v_S^\top\vartheta_S)(v_S^\top u)^2$ for $S\in\mathcal S$ and a unit vector $u$. Since $\vartheta_S$ is chosen independently of the data, we argue conditionally on $\vartheta_S$. Given $\vartheta_S$, the $n$ copies of $X_u$ are i.i.d.\ with $\mb E[X_u\mid\vartheta_S]=u^\top I_S(\vartheta_S)u$, and $0\leq b''\leq M_b$ gives $0\leq X_u\leq CD$ and $\operatorname{Var}(X_u\mid\vartheta_S)\leq CD$ as above, uniformly in the value of $\vartheta_S$. The number of pairs $(S,u)$ is again at most $e^{CA_n}$, so for each $S$ and each net point $u$ we apply Bernstein's inequality conditionally on $\vartheta_S$ and integrate the resulting bound over $\vartheta_S$. Summing the failure probabilities over the pairs $(S,u)$ gives the second claim with probability at least $1-n^{-12}$. Both claims thus hold on an event of probability at least $1-2n^{-12}$.

For the third claim, on this event let $\delta=\theta-\theta'$. Then
\[
H_{n,S}(\theta)-H_{n,S}(\theta')=a^{-1}P_n[\{b''(v_S^\top\theta)-b''(v_S^\top\theta')\}v_Sv_S^\top].
\]
Since $|b'''|\leq C$, we have $|b''(v_S^\top\theta)-b''(v_S^\top\theta')|\leq C|v_S^\top\delta|$ and $|v_S^\top\delta|\leq C\sqrt D\,\|\delta\|_2$, so
\[
\begin{aligned}
\|H_{n,S}(\theta)-H_{n,S}(\theta')\|_{\rm op}
&\leq C\sup_{\|u\|_2=1}P_n\bigl[|v_S^\top\delta|(v_S^\top u)^2\bigr]\\
&\leq C\sqrt D\,\|\delta\|_2\sup_{\|u\|_2=1}P_n\bigl[(v_S^\top u)^2\bigr]\leq C\sqrt D\,\|\delta\|_2 ,
\end{aligned}
\]
where the last factor is bounded by the first claim with $U=S$. For two supports the same computation applies with $v_{S\cup S'}$, which has at most $2s_{\max}$ slope coordinates, and for $I_S$ we replace the empirical moments by the moments under $Q_Z$.
\end{proof}

\emph{Lemmas~\ref{lem:sparse-local} and~\ref{lem:vs-near}.} We carry out the proofs with the radius $r_n$ in place of $\sqrt{\Lambda_n/n}$. The union bounds are over at most $e^{CA_n}$ events. We apply the vector Bernstein bound to the scores and the moment statistics with norm bound $C\sqrt D$, where $\sqrt D\,r_n\leq\sqrt{c_0}$. This gives deviations of at most $Cr_n$ in Euclidean norm, with $C$ free of $D$. By Lemma~\ref{lem:vs-hessian}, the Hessians at the generating and pseudo-true parameters deviate from their expectations by at most $C\sqrt D\,r_n$, and the Hessians are Lipschitz with constant $C\sqrt D$. The proofs then require five bounds. First is the interior condition $Cr_n\leq\rho/(2D)$. Second is the Hessian deviation $C\sqrt D\,r_n\leq\underline\lambda_I/2$ on the balls of radius $Cr_n$. Third is the cubic remainder $C\sqrt D\,nr_n^3=C(DA_n^3/n)^{1/2}\leq1/8$ of the quadratic expansion. Fourth is the one-step error $\|\widetilde F_S-\widehat\theta_S\|_2\leq C\sqrt D\,r_n^2+n^{-3}$. Fifth is $D_S(\widetilde F_S)\leq CDA_n^2/n\leq\{4\log(en)\}^{-1}$. All five follow from~\eqref{eq:vs-smax-condition} and~\eqref{eq:vs-rn}, because $\log(en)\leq A_n$. Hence Lemma~\ref{lem:sparse-local} holds with constants free of $D$, except that its first display reads $\|\widehat\theta_S-\theta_{0,S}\|_2+\|\theta_S^{(0)}-\theta_{0,S}\|_2\leq Cr_n$. Moreover, on $\mathcal E_n$ we have $\|V_S\|_2^2\leq CA_n$ for $S\supseteq S_0$.

In the same way, Lemma~\ref{lem:vs-near} holds with constants $K_0$, $K_1$, $m$, $M$ free of $D$. In the proof of part~\textup{(c)}, the lower bound for $m_{S_0}$ needs one change, because the $\pi_{S_0}$-probability of $B(\theta_{0,S_0},1/n)\cap K_{S_0}$ depends on $d_0$. Let $h=(n\sqrt{d_0})^{-1}$. Since $K_{S_0}$ is a rectangle with sides of length at least $2a_\pi$ and $h\leq a_\pi$ for $c_0$ small, $K_{S_0}$ contains a cube $\mathcal C$ with sides of length $h$ and a vertex at $\theta_{0,S_0}$, and $\mathcal C\subseteq B(\theta_{0,S_0},1/n)$. On $K_{S_0}$, the density of $\pi_{S_0}$ is at least $c_-^{d_0}$. On $\mathcal C$, $\ell_{n,S_0}$ varies by at most $C\sqrt D$, because $\|s_{S_0,\theta}\|_2\leq C\sqrt D$ and $\mathcal C$ has diameter $1/n$. Hence
\[
\begin{aligned}
\log m_{S_0}&\geq\ell_{n,S_0}(\theta_{0,S_0})-C\sqrt D-d_0\log n-\tfrac12d_0\log d_0-d_0|\log c_-|\geq\ell_{n,S_0}(\theta_{0,S_0})-CA_n,
\end{aligned}
\]
where the last inequality follows from $d_0\leq D\leq n$ and $D\log(en)\leq A_n$. The constant $C_2$ of that proof is therefore free of $D$.

For $S\in\mathcal A_0$, the pseudo-true parameter $\theta_S^\dagger$ lies within $C\sqrt{e_S}\leq Cr_n$ of the restriction of $\theta_0$ to $S$, and $\|\widehat\theta_S-\theta_S^\dagger\|_2\leq Cr_n$. We may enlarge $K_1$, which only enlarges the set of boundary tasks. For a constant $R$ fixed in Lemma~\ref{lem:vs-laplace-diff}, we take $K_1$ large enough that, on $\mathcal G$ and for every task that is not a boundary task,
\begin{equation}
\{\theta:\|\theta-\widehat\theta_S\|_2\leq Rr_n\}\subseteq K_S\qquad(S\in\mathcal A_0).
\label{eq:vs-ball-interior}
\end{equation}
Indeed, the coordinates of $\widehat\theta_S$ differ from those of $\theta_0$ by at most $Cr_n$. The retained coordinates of $\theta_0$ are at distance more than $K_1r_n$ from the endpoints of their intervals. The other coordinates of $S$ are within $Cr_n$ of zero, while $[-a_\pi,a_\pi]\subseteq\mathcal I_j$ and $(C+R)r_n<a_\pi$.

We next bound $\mb Q_n(\mathcal T)$, $\mb Q_n(\mathcal B)$ and $\mb Q_n(\mathcal T\cap\mathcal B)$. By Assumption~\ref{ass:sparse-proposal}\textup{(ii)}, conditionally on $S$, the coordinates of $\theta_S$ are independent with densities bounded by $c_+$. For each $j\in S$, $\beta_j$ lies in $[-t_n,t_n]$ with probability at most $2c_+t_n$. Each of the $d_S$ coordinates lies within $w_n$ of the boundary of its interval with probability at most $2c_+w_n$. Since $t_n+w_n<a_\pi$ for $c_0$ small, a coordinate with $|\beta_j|\leq t_n$ is at distance at least $a_\pi-t_n>w_n$ from the boundary of $\mathcal I_j$, so the same coordinate cannot be both weak and near the boundary. For two different coordinates, the two events are independent given $S$. The union bound over the coordinates and over the at most $s_{\max}D$ ordered pairs gives
\begin{equation}
\mb Q_n(\mathcal T)\leq2c_+s_{\max}t_n,\qquad\mb Q_n(\mathcal B)\leq2c_+Dw_n,\qquad\mb Q_n(\mathcal T\cap\mathcal B)\leq4c_+^2s_{\max}Dt_nw_n ,
\label{eq:vs-prior-mass}
\end{equation}
and therefore, by~\eqref{eq:vs-rn},
\begin{equation}
\mb Q_n(\mathcal T\cup\mathcal B)\leq Cs_{\max}r_n\leq Cs_{\max}^{3/2}\sqrt{\frac{\Lambda_n}n},\qquad\mb Q_n(\mathcal T\cap\mathcal B)\leq Cs_{\max}^2r_n^2 .
\label{eq:vs-irregular-mass}
\end{equation}

The next lemma gives a moderate deviation bound for a normalized sum that retains the factor $1/x$ of the Gaussian tail, which is absent in Bernstein's inequality.

\begin{lemma}[A tail bound with the Gaussian prefactor]\label{lem:cramer-prefactor}
Let $\xi_1,\ldots,\xi_n$ be i.i.d.\ random variables with $\mb E\xi_1=0$, $\mb E\xi_1^2=1$ and $|\xi_1|\leq K$, where $K\geq1$, and let $U=n^{-1/2}\sum_{i=1}^n\xi_i$. There is a numerical constant $C$ such that, for $1\leq x\leq\sqrt n/(8eK)$, $\mb P(U\geq x)\leq(C/x)\exp(-x^2/2+Kx^3/\sqrt n)$.
\end{lemma}

\begin{proof}
A direct Chernoff bound gives the exponential factor, but does not yield the factor $1/x$. We obtain the latter by changing measure so that the tail threshold is at the mean of the tilted distribution and then applying a normal approximation under that distribution.

Let $\psi(h)=\log\mb E e^{h\xi_1}$ and $h_0=(4eK)^{-1}$, and for $0\leq h\leq h_0$ let $\mb P_h$ (with expectation $\mb E_h$) be the law under which the $\xi_i$ are i.i.d.\ with density $e^{h\xi-\psi(h)}$ with respect to their original law. Since $|h\xi_1|\leq h_0K\leq1$ and $\psi(h)\geq0$ by Jensen's inequality, this density is at most $e$. Hence $\psi''(h)=\mb E_h(\xi_1-\mb E_h\xi_1)^2\leq\mb E_h\xi_1^2\leq e\,\mb E\xi_1^2=e$, and since $|\xi_1-\mb E_h\xi_1|\leq2K$, the third central moment satisfies $|\psi'''(h)|\leq2K\psi''(h)\leq2eK$. With $\psi(0)=\psi'(0)=0$ and $\psi''(0)=1$, integrating gives $|\psi''(h)-1|\leq2eKh\leq2eKh_0=1/2$, then $|\psi'(h)-h|\leq eKh^2\leq h/4$, and finally $\psi(h)\leq h^2/2+eKh^3/3$. Hence, for $0\leq h\leq h_0$,
\[
|\psi''(h)-1|\leq\tfrac12,\qquad\tfrac34h\leq\psi'(h)\leq\tfrac54h,\qquad\psi(h)\leq\tfrac12h^2+\tfrac{eK}3h^3 .
\]

Choose the tilting parameter so that the tilted mean is the threshold per observation. Let $t=x/\sqrt n$, so $0<t\leq h_0/2$. Since $\psi'$ is continuous and increasing on $[0,h_0]$ with $\psi'(0)=0$ and $\psi'(h_0)\geq3h_0/4\geq t$, there is $h\in(0,h_0]$ with $\psi'(h)=t$, and $t=\psi'(h)\leq5h/4$ gives $h\geq4t/5$. This $h$ maximizes the concave function $h'\mapsto h't-\psi(h')$. Since $t\leq h_0$, comparing with $h'=t$ gives $ht-\psi(h)\geq t^2-\psi(t)\geq t^2/2-eKt^3/3$. With $t=x/\sqrt n$, we have $n\{\psi(h)-ht\}\leq-x^2/2+(e/3)Kx^3/\sqrt n\leq-x^2/2+Kx^3/\sqrt n$.

Set $T=\sum_{i=1}^n(\xi_i-t)$, which is centered under $\mb P_h$ because $\mb E_h\xi_1=\psi'(h)=t$. Then $\{U\geq x\}=\{T\geq0\}$ and $d\mb P/d\mb P_h=\exp\{-h\sum_{i=1}^n\xi_i+n\psi(h)\}=e^{n\{\psi(h)-ht\}}e^{-hT}$. Writing $e^{-hT}=\int_T^\infty he^{-hy}\,dy$ and applying Fubini's theorem gives $\mb E_h[e^{-hT}\ind\{T\geq0\}]=\int_0^\infty he^{-hy}\,\mb P_h(0\leq T\leq y)\,dy$. Together with the bound on $n\{\psi(h)-ht\}$, we obtain
\[
\begin{aligned}
\mb P(U\geq x)&=e^{n\{\psi(h)-ht\}}\,\mb E_h\bigl[e^{-hT}\ind\{T\geq0\}\bigr]\\
&\leq\exp\Bigl(-\frac{x^2}2+\frac{Kx^3}{\sqrt n}\Bigr)\int_0^\infty he^{-hy}\,\mb P_h(0\leq T\leq y)\,dy .
\end{aligned}
\]

The remaining integral accounts for the Gaussian prefactor. Below, $C$ denotes a numerical constant that may vary. Under $\mb P_h$, the summands $\xi_i-t$ have mean zero, variance $\sigma^2=\psi''(h)\in[1/2,3/2]$, and third absolute central moment at most $2K\sigma^2$. The ratio of this third moment to $\sigma^3$ is at most $2K/\sigma\leq CK$, so the Berry--Esseen theorem \citep[Section~XVI.5]{feller1971introduction} places the distribution function of $T/(\sigma\sqrt n)$ within $CK/\sqrt n$ of the standard normal one. For $y\geq0$, the standard normal probability of $[0,y/(\sigma\sqrt n)]$ is at most $y/(\sigma\sqrt{2\pi n})\leq y/\sqrt n$, so $\mb P_h(0\leq T\leq y)\leq y/\sqrt n+CK/\sqrt n$. Since $\int_0^\infty he^{-hy}y\,dy=1/h$, the integral is at most $(h\sqrt n)^{-1}+CK/\sqrt n\leq5/(4x)+C/x\leq C/x$, using $h\sqrt n\geq4x/5$ and $K/\sqrt n\leq1/x$ from $x\leq\sqrt n/(8eK)$. Substituting this into the display above gives the claimed bound.
\end{proof}

\begin{lemma}[Supports with added variables, explicit form]\label{lem:vs-added-explicit}
Let $(S_0,\theta_0)$ be an allowed task and suppose~\eqref{eq:vs-smax-condition} holds. For a constant $\kappa\geq1$ and $S\supsetneq S_0$ with $b=|S\setminus S_0|$, put $\varpi_S^\kappa=\min\{1,e^{\kappa(b+1)}\exp(\|U_S\|_2^2/2-\Delta_b)\}$, so that $\varpi_S\leq\varpi_S^\kappa$ for the weights $\varpi_S$ of Lemma~\ref{lem:vs-sq-odds}. Then, with constants that may depend on $\kappa$ but not on $D$,
\begin{gather}
\sum_{S\supsetneq S_0}\mb E_0\bigl[(\varpi_S^\kappa)^2\bigr]\leq Cs_{\max}\sqrt{\frac{\Lambda_n}n},
\qquad\sum_{S\supsetneq S_0:\,b\geq2}b^2\,\mb E_0[\varpi_S^\kappa]\leq\frac{Cs_{\max}^2\Lambda_n^2}n,
\label{eq:vs-added-sums}\\
\begin{gathered}
\mb E_0\Bigl[\ind_{\mathcal E_n}\log\Bigl(1+\sum_{S\supsetneq S_0}e^be^{\|U_S\|_2^2/2-\Delta_b}\Bigr)\Bigr]\leq\frac{Cs_{\max}\Lambda_n^{3/2}}{\sqrt n},\\
\sum_{S\supsetneq S_0}\mb E_0\bigl[\varpi_S\|U_S\|_2^2\ind_{\mathcal E_n}\bigr]\leq C .
\end{gathered}
\label{eq:vs-added-weighted}
\end{gather}
\end{lemma}

\begin{proof}
We first consider a single support. For a unit vector $u$, the variable $u^\top\xi_S$ has mean zero and unit variance and satisfies $|u^\top\xi_S|\leq\|\xi_S\|_2\leq C\sqrt D$, so $\mb E_0|u^\top\xi_S|^3\leq C\sqrt D\,\mb E_0(u^\top\xi_S)^2=C\sqrt D$ and $\log\mb E_0e^{hu^\top\xi_S}\leq h^2/2+C\sqrt D\,|h|^3$ for $|h|\leq(C\sqrt D)^{-1}$. Hence the tail bound of the score statistic in the proof of Lemma~\ref{lem:vs-overfit}, which is used for $4b\leq w\leq CA_n$, has the correction term $C\sqrt D\,w^{3/2}/\sqrt n\leq C(DA_n^3/n)^{1/2}\leq C$, since the admissible range of $h$ contains $\sqrt{w/n}$. Thus the bounds of Lemmas~\ref{lem:vs-overfit} and~\ref{lem:vs-sq-odds} for a single support hold with constants free of $D$, and replacing $e^b$ by $e^{\kappa(b+1)}$ in the weights only changes these constants, because $\Delta_b\geq(b/2)\log n$.

For the sums, there are $\binom{p-s_0}b$ supports with $b$ added variables. Since $s_0+b\leq s_{\max}$, we have $\binom{s_0+b}b\leq(s_0+b)^b/b!\leq s_{\max}^b/b!$, and~\eqref{eq:ebic-cancellation} gives
\[
\binom{p-s_0}b(C\sqrt{\Lambda_n})^be^{-\Delta_b}=\binom{s_0+b}b\Bigl(C\sqrt{\frac{\Lambda_n}n}\Bigr)^b\leq\frac1{b!}\Bigl(Cs_{\max}\sqrt{\frac{\Lambda_n}n}\Bigr)^b.
\]
With $x=s_{\max}\sqrt{\Lambda_n/n}\leq1$, we have, for $k\leq2$ and $b_0\in\{1,2\}$,
\[
\sum_{b\geq b_0}\frac{b^k(Cx)^b}{b!}\leq x^{b_0}\sum_{b\geq b_0}\frac{b^kC^b}{b!}\leq Cx^{b_0}.
\]
We combine this with the single-support bounds of Lemma~\ref{lem:vs-sq-odds} (with $\varpi_S^\kappa$ in place of $\varpi_S$ for the first two). The bound on $\mb E_0[(\varpi_S^\kappa)^2]$ with $b_0=1$ and $k=0$ gives the first bound of~\eqref{eq:vs-added-sums}. The bound on $\mb E_0[\varpi_S^\kappa]$, which has the extra factor $C\Lambda_n$, with $b_0=2$ and $k=2$ gives the second bound of~\eqref{eq:vs-added-sums}, of order $\Lambda_n(s_{\max}\sqrt{\Lambda_n/n})^2=s_{\max}^2\Lambda_n^2/n$. Finally, write $r_S=e^b\exp(\|U_S\|_2^2/2-\Delta_b)$. As in the proof of Lemma~\ref{lem:vs-sq-odds}, $\log(1+\sum_Sr_S)\leq\sum_S\log(1+r_S)\leq\sum_S\{\varpi_S+(\log r_S)_+\}$, so the second and third single-support bounds with $b_0=1$ and $k=0$ give the first bound of~\eqref{eq:vs-added-weighted}, of order $\Lambda_ns_{\max}\sqrt{\Lambda_n/n}$.

It remains to prove the second bound of~\eqref{eq:vs-added-weighted}. For $b\geq2$, we have $\|U_S\|_2^2\leq\|V_S\|_2^2\leq CA_n$ on $\mathcal E_n$, so $\varpi_S\|U_S\|_2^2\ind_{\mathcal E_n}\leq CA_n\varpi_S$. Since $\varpi_S\leq\varpi_S^\kappa$ and $b\geq2$, we have $\sum_{S\supsetneq S_0:\,b\geq2}\mb E_0[\varpi_S]\leq\sum_{S\supsetneq S_0:\,b\geq2}b^2\,\mb E_0[\varpi_S^\kappa]$, and the second bound of~\eqref{eq:vs-added-sums} gives
\[
\sum_{S\supsetneq S_0:\,b\geq2}\mb E_0\bigl[\varpi_S\|U_S\|_2^2\ind_{\mathcal E_n}\bigr]\leq\frac{CA_ns_{\max}^2\Lambda_n^2}n\leq\frac{9CA_n^3}n\leq C,
\]
where the penultimate inequality uses $s_{\max}\Lambda_n\leq D\Lambda_n\leq3A_n$ from~\eqref{eq:vs-An-Lambda} and the last uses~\eqref{eq:vs-smax-condition}.

Now let $b=1$, so that $U=U_S$ is a normalized sum of i.i.d.\ centered variables with unit variance, bounded by $K=C\sqrt D$. Put $t=\Delta_1-1$ and $x_t=\sqrt{2t}$, and let $G(x)=\mb P_0(|U|\geq x)$. Since $\Delta_1\geq(\log n)/2\geq2$ for $n\geq e^4$ and $\Delta_1\leq\Lambda_n$, we have $1\leq t\leq\Lambda_n$. For $1\leq x\leq2x_t$ we have $x\leq\sqrt n/(8eK)$ and $Kx^3/\sqrt n\leq C\sqrt D\Lambda_n^{3/2}/\sqrt n\leq C$, so Lemma~\ref{lem:cramer-prefactor}, applied to $\pm U$, gives $G(x)\leq(C/x)e^{-x^2/2}$. Since $\varpi_S=\min\{1,e^{U^2/2-t}\}$ and $U^2/2\geq t$ holds exactly when $|U|\geq x_t$, we have
\[
\varpi_SU^2=U^2\ind\{|U|\geq x_t\}+e^{-t}U^2e^{U^2/2}\ind\{|U|<x_t\}.
\]
For the second term, let $f(x)=x^2e^{x^2/2}$, so that $f(0)=0$ and $f'(x)=(2x+x^3)e^{x^2/2}\geq0$ for $x\geq0$. Hence
\[
f(|U|)\ind\{|U|<x_t\}=\int_0^{x_t}f'(x)\ind\{x\leq|U|<x_t\}\,dx\leq\int_0^{x_t}f'(x)\ind\{|U|\geq x\}\,dx .
\]
For $1\leq x\leq x_t$, the bound $G(x)\leq(C/x)e^{-x^2/2}$ gives $f'(x)G(x)\leq C(2+x^2)$. Taking expectations and using $G(x)\leq1$ on $[0,1]$, together with $\int_1^{x_t}(2+x^2)\,dx\leq Cx_t^3=Ct^{3/2}$ and $t\geq1$, we obtain
\[
\mb E_0\bigl[f(|U|)\ind\{|U|<x_t\}\bigr]\leq\int_0^{x_t}f'(x)G(x)\,dx\leq\int_0^1f'(x)\,dx+C\int_1^{x_t}(2+x^2)\,dx\leq Ct^{3/2}.
\]
For the first term, on $\mathcal E_n$ we have $|U|\leq x_n:=\sqrt{CA_n}$, and
\[
\mb E_0\bigl[U^2\ind\{|U|\geq x_t\}\ind_{\mathcal E_n}\bigr]\leq x_t^2G(x_t)+\int_{x_t}^{x_n}2xG(x)\,dx .
\]
For $2x_t\leq x\leq x_n$, we have $Kx/\sqrt n\leq C\sqrt{DA_n/n}\leq1$, so Bernstein's inequality gives
\[
G(x)\leq2\exp\Bigl[-\frac{x^2}{2\{1+Kx/(3\sqrt n)\}}\Bigr]\leq2e^{-3x^2/8}.
\]
The part of the integral over $x\geq2x_t$ is therefore at most
\[
\int_{2x_t}^\infty4xe^{-3x^2/8}\,dx=\frac{16}3e^{-3x_t^2/2}\leq\frac{16}3e^{-t},
\]
since $x_t^2=2t\geq2$. On $[x_t,2x_t]$, the bound on $G(x)$ gives $2xG(x)\leq Ce^{-x^2/2}$, so the part of the integral over $[x_t,2x_t]$ is at most
\[
C\int_{x_t}^\infty e^{-x^2/2}\,dx\leq\frac C{x_t}e^{-x_t^2/2}\leq Ce^{-t},
\]
since $x_t=\sqrt{2t}\geq1$. Since $x_t^2G(x_t)\leq Cx_te^{-t}$, combining the two terms gives
\[
\mb E_0[\varpi_SU_S^2\ind_{\mathcal E_n}]\leq Ct^{3/2}e^{-t}=Ce\,t^{3/2}e^{-\Delta_1}\leq C\Lambda_n^{3/2}e^{-\Delta_1},
\]
because $e^{-t}=e\cdot e^{-\Delta_1}$ and $t\leq\Lambda_n$. There are $p-s_0$ supports with one added variable, and $(p-s_0)e^{-\Delta_1}=(s_0+1)n^{-1/2}\leq s_{\max}n^{-1/2}$ by~\eqref{eq:ebic-cancellation}, so their contribution is at most $Cs_{\max}\Lambda_n^{3/2}/\sqrt n\leq C(27A_n^3/n)^{1/2}\leq C$ by~\eqref{eq:vs-smax-condition} and~\eqref{eq:vs-An-Lambda}. Combining the bounds for $b\geq2$ and $b=1$ proves the second bound of~\eqref{eq:vs-added-weighted}.
\end{proof}

\emph{Lemma~\ref{lem:vs-soft-oracle} and the risk bounds of Theorem~\ref{thm:vs-approx}.} The constants of Lemma~\ref{lem:vs-soft-oracle} do not depend on $D$. With the lemmas in the form recorded above, the constants in Steps~3 and~4 of the proof of Theorem~\ref{thm:vs-approx} below do not depend on $D$ either. Hence~\eqref{eq:vs-uniform-risk} and~\eqref{eq:vs-prior-average} hold with the constants of this part, and $A_n\leq2s_{\max}\Lambda_n$ from~\eqref{eq:vs-An-Lambda} gives the risk bounds of parts~\textup{(b)} and~\textup{(a)}.

The last ingredient concerns the Laplace corrections $\delta_S$ of Appendix~\ref{app:vs-network}.

\begin{lemma}[Laplace corrections of neighboring supports]\label{lem:vs-laplace-diff}
Suppose~\eqref{eq:vs-smax-condition} holds, and let $\mathcal G$ be the event of Lemma~\ref{lem:vs-near}, intersected with the event of Lemma~\ref{lem:vs-hessian}. There are constants $c_1$ and $C_1$ such that, on $\mathcal G$ and for every regular task, $|\delta_S-\delta_{S_0}|\leq c_1|S\setminus S_0|+C_1$ for every $S\in\mathcal A_+$.
\end{lemma}

\begin{proof}
Fix a regular task and $S\in\mathcal A_+$, and write $d=d_S$, $\widehat\theta=\widehat\theta_S$ and $H=H_{n,S}(\widehat\theta)$. By~\eqref{eq:vs-ball-interior}, the ball $B(\widehat\theta,Rr_n)$ lies in $K_S$, where $R$ is chosen below. Since the task is regular, $\mathcal A_0=\mathcal A_+$, so $\widehat\theta\in\operatorname{int}(\Theta_{0,S})$ by Lemma~\ref{lem:vs-near}\textup{(b)} and $\nabla\ell_{n,S}(\widehat\theta)=0$. The Hessian $-nH_{n,S}$ of $\ell_{n,S}$ is Lipschitz with constant $C\sqrt D\,n$ by Lemma~\ref{lem:vs-hessian}, so for $\|h\|_2\leq Rr_n$,
\[
\Bigl|\ell_{n,S}(\widehat\theta+h)-\ell_{n,S}(\widehat\theta)+\frac n2h^\top Hh\Bigr|
\leq C\sqrt D\,n\|h\|_2^3\leq C\sqrt D\,nR^3r_n^3 =CR^3\Bigl(\frac{DA_n^3}n\Bigr)^{1/2}\leq1,
\]
for $c_0$ small by~\eqref{eq:vs-smax-condition}.

Let $\mathsf Z\sim N(\widehat\theta,(nH)^{-1})$ and $\bar g_S=\mb E[\pi_S(\mathsf Z)\ind\{\mathsf Z\in K_S\}]$. Since $\int e^{-(n/2)h^\top Hh}\,dh=(2\pi/n)^{d/2}(\det H)^{-1/2}$, the integral of $\pi_S$ against the quadratic approximation over $K_S$, written as an expectation under $\mathsf Z$, equals $e^{\ell_{n,S}(\widehat\theta)}(2\pi/n)^{d/2}(\det H)^{-1/2}\bar g_S$. By Lemma~\ref{lem:vs-near}\textup{(a)}, $mI_d\preceq-n^{-1}\nabla^2\ell_{n,S}\preceq MI_d$ on $\Theta_{0,S}$, so $mI_d\preceq H\preceq MI_d$. Since $\nabla\ell_{n,S}(\widehat\theta)=0$, outside $B(\widehat\theta,Rr_n)$ both the likelihood and its quadratic approximation are at most $e^{\ell_{n,S}(\widehat\theta)}e^{-nm\|h\|_2^2/2}$, where $h=\theta-\widehat\theta$. Since $c_-^d\leq\pi_S\leq c_+^d$ on $K_S$ by Assumption~\ref{ass:sparse-proposal}\textup{(ii)} and $nr_n^2=A_n$, each integral outside the ball is at most $e^{\ell_{n,S}(\widehat\theta)}$ times
\[
c_+^d\int_{\|h\|_2>Rr_n}e^{-nm\|h\|_2^2/2}\,dh\leq c_+^d\Bigl(\frac{4\pi}{nm}\Bigr)^{d/2}e^{-mR^2A_n/4}.
\]
On $B(\widehat\theta,Rr_n)\subseteq K_S$, we have $\pi_S\geq c_-^d$ and $\det H\leq M^d$. Since $H\succeq mI_d$, $\mb E\|\sqrt n(\mathsf Z-\widehat\theta)\|_2^2\leq d/m$. By Chebyshev's inequality, $\mb P(\|\mathsf Z-\widehat\theta\|_2>Rr_n)\leq d/(mR^2A_n)\leq1/2$ when $R^2A_n\geq2d/m$, which holds for $R^2\geq2/m$ because $A_n\geq d$. Hence the integral of the quadratic approximation over the ball is at least $\frac12c_-^d\{2\pi/(nM)\}^{d/2}e^{\ell_{n,S}(\widehat\theta)}$. Since $A_n\geq d$, we can fix $R\geq\sqrt{2/m}$, depending only on $m$, $M$, $c_-$ and $c_+$, such that the ratio of these two bounds is at most $1/4$. On the ball the likelihood and its quadratic approximation differ by a factor in $[e^{-1},e]$, while outside it each integral is at most one quarter of the integral of the quadratic approximation over the ball. Hence $m_S$ and $e^{\ell_{n,S}(\widehat\theta)}(2\pi/n)^{d/2}(\det H)^{-1/2}\bar g_S$ differ by a factor in $[e^{-2},e^2]$. Taking logarithms and using the definition of $\delta_S$ gives
\begin{equation}
\delta_S=\frac d2\log(2\pi)-\frac12\log\det H_{n,S}(\widehat\theta_S)+\log\bar g_S+\epsilon_S,\qquad|\epsilon_S|\leq2 .
\label{eq:vs-laplace-exact}
\end{equation}

The same holds for $S_0$. We compare the three terms for $S$ and $S_0$, with $b=|S\setminus S_0|$. The first terms differ by $(b/2)\log(2\pi)$. For the determinants, partition $H_{n,S}(\widehat\theta_S)$ into the coordinates of $S_0\cup\{0\}$ and the $b$ added ones, with first diagonal block $H_0$. Its determinant is the product of $\det H_0$ and that of a $b\times b$ Schur complement, whose eigenvalues lie in $[m,M]$ by Lemma~\ref{lem:vs-near}\textup{(a)}. The matrices $H_0$ and $H_{n,S_0}(\widehat\theta_{S_0})$ are the blocks of the two Hessians on their common coordinates, and have eigenvalues at least $m$. Since both estimates are within $Cr_n$ of $\theta_0$, $\|\widehat\theta_S-\iota_{S_0,S}\widehat\theta_{S_0}\|_2\leq Cr_n$, so Lemma~\ref{lem:vs-hessian} gives $\|H_0-H_{n,S_0}(\widehat\theta_{S_0})\|_{\rm op}\leq C\sqrt D\,\|\widehat\theta_S-\iota_{S_0,S}\widehat\theta_{S_0}\|_2\leq C\sqrt D\,r_n$. Since $|\log\det A_1-\log\det A_2|\leq d_0\|A_1-A_2\|_{\rm op}/m$ for symmetric $d_0\times d_0$ matrices $A_1,A_2$ with eigenvalues at least $m$,
\[
\begin{aligned}
\bigl|\log\det H_{n,S}(\widehat\theta_S)-\log\det H_{n,S_0}(\widehat\theta_{S_0})\bigr|
&\leq b\max\{|\log m|,|\log M|\}+\frac{Cd_0\sqrt D\,r_n}m\leq Cb+C,
\end{aligned}
\]
because $d_0\sqrt D\,r_n\leq D^{3/2}r_n\leq\sqrt{c_0}$ by~\eqref{eq:vs-rn}.

For the prior terms, Assumption~\ref{ass:sparse-proposal}\textup{(ii)} gives $|\log\pi_S(\theta)-\log\pi_S(\widehat\theta)|\leq L_\pi\|\theta-\widehat\theta\|_1\leq L_\pi\sqrt d\,\|\theta-\widehat\theta\|_2\leq L_\pi R\sqrt d\,r_n\leq C$ on $B(\widehat\theta,Rr_n)$, since $\sqrt d\,r_n\leq1$. Moreover, $\mb P(\|\mathsf Z-\widehat\theta\|_2\leq Rr_n)\geq1/2$, while $\det H\leq M^d$ and the choice of $R$ give $c_+^d\,\mb P(\|\mathsf Z-\widehat\theta\|_2>Rr_n)\leq c_-^d\leq\pi_S(\widehat\theta)$. Hence $|\log\bar g_S-\log\pi_S(\widehat\theta_S)|\leq C$, and likewise for $S_0$. Finally,
\[
\log\pi_S(\widehat\theta_S)-\log\pi_{S_0}(\widehat\theta_{S_0})
={}\sum_{j\in S_0\cup\{0\}}\bigl\{\log g_j(\widehat\theta_{S,j})-\log g_j(\widehat\theta_{S_0,j})\bigr\} +\sum_{j\in S\setminus S_0}\log g_j(\widehat\theta_{S,j}),
\]
where the first sum is at most $L_\pi\sqrt{d_0}\,Cr_n\leq C$ in absolute value and the second is at most $b\max\{|\log c_-|,|\log c_+|\}$. Collecting the three comparisons in~\eqref{eq:vs-laplace-exact} gives $|\delta_S-\delta_{S_0}|\leq c_1|S\setminus S_0|+C_1$.
\end{proof}

Without the Lipschitz condition, the first sum in the last display can only be bounded by $d_0\log(c_+/c_-)$, which is of order $s_{\max}$. This is why we assume the condition. For the tasks that are weak or boundary tasks, but not both, the proof of Proposition~\ref{prop:vs-bridge} uses only Lemma~\ref{lem:vs-laplace}, which bounds the Laplace corrections by a multiple of $s_{\max}$, up to the boundary term $MU_S^\pi$ in the lower bound. For the weak boundary tasks it uses the bound~\eqref{eq:vs-envelope}, which holds for every dataset. We write that proof, in Appendix~\ref{app:proof-vs-bridge}, with constants free of $D$. It gives the first bound of Theorem~\ref{thm:vs-approx}(a).

\emph{Network size and Theorem~\ref{thm:vs-gen}.} In Lemma~\ref{lem:vs-rrds}(d), $H_{\mathcal S}\leq1+2A_n\leq5s_{\max}\Lambda_n$, so $L\leq C(L_{{\rm vs},n}+s_{\max}\Lambda_n^3)$ and $W\leq N_{\mathcal S}\{V_{{\rm vs},n}+C(p+s_{\max}^2)\Lambda_n^3\}$. In the proof of Theorem~\ref{thm:vs-gen}, $b_1=(R_b+r_0)^2\leq(2+\sqrt2)^2B^2s_{\max}$, so the bound on the loss is $c_\lambda\leq C(s_{\max}+\lambda\Lambda_n)$, which is the prefactor in~\eqref{eq:sparse-complexity-training}. The other constants of that proof do not involve $D$. With $\mathcal R(\widetilde F)-\mathcal R^*\leq C\{s_{\max}/n+(\lambda/p)J_n\}$ this gives the four displays of the theorem.

\subsubsection{Proof of Theorem~\ref{thm:vs-approx}}\label{app:proof-vs-thm}

\begin{proof}[Proof of Theorem~\ref{thm:vs-approx}]
In this proof, $C,c$ depend only on the response family, $a$, $D$, and the constants of the setting and of Assumptions~\ref{ass:sparse} and~\ref{ass:sparse-proposal}. They are independent of $n$, $p$, $S_0$ and the generating coefficients. We divide the proof into five steps. Steps~1 and~2 construct, for every support $S$, the networks that compute $F_S^{\rm os}$ and $\ell_S^\circ$ and verify~\eqref{eq:vs-module-accuracy}, Steps~3 and~4 prove parts~(b) and~(a), and Step~5 bounds the Hamming loss of the selected support.

\medskip
\noindent{\bf Step 1: Score and likelihood components.} Fix $x=(y,z)\in\mathcal X_n$ and $\theta_S\in\Theta_{0,S}$. Set $\eta=v_S(z)^\top\theta_S\in\mathcal I_n$ and $\widetilde\eta=\Pi_{\mathcal I_n}\sum_j\mathsf M_{3R_n,\epsilon_{{\rm c},n}}(v_{S,j},\theta_{S,j})$. Since $|v_{S,j}|,|\theta_{S,j}|\leq R_n$, Lemma~\ref{lem:multiplication} bounds each of the $d_S$ multiplier errors by $\epsilon_{{\rm c},n}$. Since $\eta\in\mathcal I_n$ and the interval projection $\Pi_{\mathcal I_n}$ is nonexpansive and fixes $\eta$,
\[
|\widetilde\eta-\eta|\leq\sum_j|\mathsf M_{3R_n,\epsilon_{{\rm c},n}}(v_{S,j},\theta_{S,j})-v_{S,j}\theta_{S,j}|\leq d_S\epsilon_{{\rm c},n}\leq D\epsilon_{{\rm c},n}.
\]
Clip the response networks to the real ranges of $b,b'$ on $\mathcal I_n$, which does not increase their errors and gives $|\widetilde b'(\widetilde\eta)|\leq R_n$. Since $|b'|\leq R_n$ on $\mathcal I_n$ by the range bound from~\eqref{eq:glm-cumulant-strip},
\[
|\widetilde b(\widetilde\eta)-b(\eta)|\leq|\widetilde b(\widetilde\eta)-b(\widetilde\eta)|+|b(\widetilde\eta)-b(\eta)|\leq\epsilon_{{\rm c},n}+R_nD\epsilon_{{\rm c},n}.
\]
The same holds for $b'$ with $|b''|\leq R_n$, so both errors are at most $(1+R_nD)\epsilon_{{\rm c},n}$. Since $|y|,|\widetilde b'(\widetilde\eta)|,|v_{S,j}|,|\widetilde\eta|\leq R_n$, we have $|y-\widetilde b'(\widetilde\eta)|\leq2R_n$ and the multipliers stay in their domains. Here $s_{S,\theta_S,j}(x)=a^{-1}\{y-b'(\eta)\}v_{S,j}$, $\ell_S^\circ(x,\theta_S)=a^{-1}\{y\eta-b(\eta)\}$ and $\varphi_S(x,\theta_S)=a^{-1}\{\mathsf M_{3R_n,\epsilon_{{\rm c},n}}(y,\widetilde\eta)-\widetilde b(\widetilde\eta)\}$.

For the score, the multiplier error is at most $\epsilon_{{\rm c},n}$ and replacing $b'(\eta)$ by $\widetilde b'(\widetilde\eta)$ adds at most $|v_{S,j}|(1+R_nD)\epsilon_{{\rm c},n}\leq R_n(1+R_nD)\epsilon_{{\rm c},n}$. For the likelihood, $|\mathsf M_{3R_n,\epsilon_{{\rm c},n}}(y,\widetilde\eta)-y\eta|\leq\epsilon_{{\rm c},n}+|y|\,|\widetilde\eta-\eta|\leq\epsilon_{{\rm c},n}+R_nD\epsilon_{{\rm c},n}$. With $a^{-1}\leq R_n$ and $D,R_n\geq1$, this gives
\[
\begin{aligned}
|a^{-1}\mathsf M_{3R_n,\epsilon_{{\rm c},n}}(y-\widetilde b'(\widetilde\eta),v_{S,j})-s_{S,\theta_S,j}(x)|
 &\leq R_n\{\epsilon_{{\rm c},n}+R_n(1+R_nD)\epsilon_{{\rm c},n}\}\leq3DR_n^3\epsilon_{{\rm c},n},\\
|\varphi_S(x,\theta_S)-\ell_S^\circ(x,\theta_S)|
 &\leq R_n(2\epsilon_{{\rm c},n}+2R_nD\epsilon_{{\rm c},n})\leq4DR_n^2\epsilon_{{\rm c},n}.
\end{aligned}
\]
By~\eqref{eq:glm-component-precision} and $R_n\geq1$, we have $3DR_n^3\epsilon_{{\rm c},n}=3\delta_n/16\leq\delta_n$ and $4DR_n^2\epsilon_{{\rm c},n}=\delta_n/(4R_n)\leq\delta_n$. When $C_0\geq6$, using $\log(en)\leq n^2$ and $D\geq1$ gives $\delta_n=(C_0n^3D^4)^{-1}\leq\{6n\log(en)\}^{-1}$, so $\max_j\|\widetilde s_{S,j}-s_{S,j}\|_\infty\leq\delta_n$ and $\|\varphi_S-\ell_S^\circ\|_\infty\leq\delta_n\leq\{6n\log(en)\}^{-1}$.

There are $d_S+d_S^2+2\leq(D+1)^2$ scalar analytic maps and at most $d_S^2+3d_S+1\leq2(D+1)^2$ scalar products. By~\eqref{eq:analytic-explicit-complexity}, with the common scales in~\eqref{eq:sparse-analytic-scales},
\[
\begin{gathered}
L_{\rm map}\leq C\{k_{{\rm an},n}+1+\log(eD)\}J_{{\rm an},n},\qquad W_{\rm maps}\leq C(D+1)^2D(k_{{\rm an},n}+1)^DJ_{{\rm an},n},\\
W_{\rm prod}\leq C(D+1)^2J_{{\rm an},n},\qquad \|\mathsf M_{3R_n,\epsilon_{{\rm c},n}}\|_{\rm coef}\leq CR_n^2,
\end{gathered}
\]
where the last line uses $1+\log(R_n^2/\epsilon_{{\rm c},n})\leq CJ_{{\rm an},n}$, since $\epsilon_{{\rm c},n}\leq(16DR_n^3)^{-1}$. Parallel products and affine sums require only a fixed number of serial stages, and the update sends the inverse-information entries, computed after the score average, directly to their products. Hence the fixed signed coordinate carries satisfy
\[
W_{\rm carry}\leq CD\{k_{{\rm an},n}+1+\log(eD)\}J_{{\rm an},n}\leq C(D+1)^2D(k_{{\rm an},n}+1)^DJ_{{\rm an},n}.
\]
Let $\mathcal C_{{\rm comp},S}$ contain the constructed free coefficients and the fixed coefficients within the component blocks, excluding the prescribed normalization and projection interfaces. Choosing the structural constants initially to dominate these counts and bounds gives
\[
L\leq L_{{\rm vs},n},\qquad W\leq V_{{\rm vs},n},\qquad \max_{c\in\mathcal C_{{\rm comp},S}}|c|\leq P_{{\rm vs},n}.
\]
The coordinates of the parameter network satisfy
\[
\begin{gathered}
W_{{\rm free},S}=\dim(\omega_S)\leq C_\Gamma^2D(D+1)^2(k_{{\rm an},n}+1)^DJ_{{\rm an},n},\\
\|\omega_S\|_2\leq\sqrt{W_{{\rm free},S}}\,\|\omega_S\|_\infty\leq\sqrt{W_{{\rm free},S}}\,P_{{\rm vs},n}\leq\Gamma_{{\rm vs},n}.
\end{gathered}
\]
The likelihood network is included in $L,V$. The total number of parameters of the networks for all supports is at most $|\mathcal S|V_{{\rm vs},n}$, including fixed extraction and zero-extension connections.

\medskip
\noindent{\bf Step 2: One-step update and network construction.} For $\widetilde\theta_S^{(0)}=\Pi_{\Theta_{0,S}}\widetilde g_S(\Pi_{\mathcal R_{g,S}}\overline T_S)$ and $\theta_S^{(0)}=\Pi_{\Theta_{0,S}}g_S(\Pi_{\mathcal R_{g,S}}\overline T_S)$, both in $\Theta_{0,S}$, the coordinatewise projection onto a rectangle is nonexpansive in the sup norm and $\|\widetilde g_S-g_S\|_\infty\leq\epsilon_{{\rm c},n}$ on $\mathcal R_{g,S}$, so $\|\widetilde\theta_S^{(0)}-\theta_S^{(0)}\|_\infty\leq\epsilon_{{\rm c},n}\leq\delta_n$ by~\eqref{eq:glm-component-precision}. For each $x\in\mathcal X_n$ and $j\leq d_S$,
\[
\begin{aligned}
|\widetilde s_{S,\widetilde\theta_S^{(0)},j}(x)-s_{S,\theta_S^{(0)},j}(x)|
&\leq|\widetilde s_{S,\widetilde\theta_S^{(0)},j}(x)-s_{S,\widetilde\theta_S^{(0)},j}(x)|+|s_{S,\widetilde\theta_S^{(0)},j}(x)-s_{S,\theta_S^{(0)},j}(x)|\\
&\leq\delta_n+B_{\rm cal}D\delta_n ,
\end{aligned}
\]
by Step~1 and because on $\mathcal X_n\times\Theta_{0,S}$ the parameter derivatives of the score are bounded by $B_{\rm cal}$ and at most $d_S\leq D$ coordinates move by at most $\delta_n$ each. Averaging gives $\max_j|P_n\{\widetilde s_{S,\widetilde\theta_S^{(0)},j}-s_{S,\theta_S^{(0)},j}\}|\leq(B_{\rm cal}D+1)\delta_n$ for every $X^n\in\mathcal X_n^n$.

Differentiation under the bounded integral gives
\[
\partial_cI_{ab}(\theta)=-\mathbb E_\theta[\partial_{abc}\ell_\theta(X)+\partial_{ab}\ell_\theta(X)s_{\theta,c}(X)],
\]
so the entrywise envelopes in~\eqref{eq:sparse-envelopes} and the lower bound $I_S(\theta_S)\succeq cI_{d_S}$ of Appendix~\ref{app:sparse-regularity} give
\[
\|\partial_cI_S\|_{\rm op}\leq CD,\qquad \|\partial_cI_S^{-1}\|_{\rm op}=\|I_S^{-1}(\partial_cI_S)I_S^{-1}\|_{\rm op}\leq CD.
\]
Write $\widetilde A=\widetilde J_S(\widetilde\theta_S^{(0)})$ and $\widetilde u=P_n\widetilde s_{S,\widetilde\theta_S^{(0)}}$ for the approximate inverse and averaged score, and $A=I_S(\theta_S^{(0)})^{-1}$ and $u=P_ns_{S,\theta_S^{(0)}}$ for their exact values at $\theta_S^{(0)}$. Since each partial derivative of $(I_S^{-1})_{jk}$ is at most $CD$ and each of the $d_S\leq D$ coordinates moves by at most $\delta_n$, the mean value theorem along the segment in the convex set $\Theta_{0,S}$ gives $|(I_S(\widetilde\theta_S^{(0)})^{-1})_{jk}-A_{jk}|\leq CD^2\delta_n$. With the accuracy $\epsilon_{{\rm c},n}\leq\delta_n$ of $\widetilde J_S$, the triangle inequality gives $|\widetilde A_{jk}-A_{jk}|\leq CD^2\delta_n$. By the definition of $R_n$, $|\widetilde u_j|\leq B_{\rm cal}+1\leq R_n$ and $|\widetilde A_{jk}|\leq c^{-1}+1\leq R_n$, so the multipliers $\mathsf M_{3R_n,\epsilon_{{\rm c},n}}$ are used inside their domain $[-3R_n,3R_n]^2$. Applying the decomposition in~\eqref{eq:separate-module-accuracy} to each coordinate gives
\[
\begin{aligned}
\left|\sum_k\mathsf M_{3R_n,\epsilon_{{\rm c},n}}(\widetilde A_{jk},\widetilde u_k)-(Au)_j\right|
&\leq D\epsilon_{{\rm c},n}+\sum_k\{|\widetilde A_{jk}-A_{jk}|\,|\widetilde u_k|+|A_{jk}|\,|\widetilde u_k-u_k|\}\leq CD^3\delta_n ,
\end{aligned}
\]
where the $d_S\leq D$ multiplier errors contribute at most $D\epsilon_{{\rm c},n}$, while each of the $d_S\leq D$ terms satisfies $|\widetilde A_{jk}-A_{jk}|\,|\widetilde u_k|\leq CD^2\delta_n(B_{\rm cal}+1)$ and $|A_{jk}|\,|\widetilde u_k-u_k|\leq c^{-1}(B_{\rm cal}D+1)\delta_n$. Adding the initialization error $\delta_n$ in each of the $d_S\leq D$ coordinates, applying the nonexpansive final projection and passing to the Euclidean norm with the factor $\sqrt D$ give $\sup_{X^n\in\mathcal X_n^n}\|\widetilde F_S-F_S^{\rm os}\|_2\leq CD^{7/2}\delta_n\leq n^{-3}$ once $C_0\geq C$, since $\delta_n=(C_0n^3D^4)^{-1}$. The architectures and coefficients do not depend on $S_0$ or $\theta_0$, and $\widetilde F_S(X^n)\in\Theta_{0,S}$ by the final projection.

By Lemma~\ref{lem:vs-moment}, the same accuracies hold when $\overline T_S$ is computed with the shared products $\mathsf M_{R_n,\upsilon_n}(y,z_j)$, after doubling $C_0$, so the estimators $\widetilde F_S$ and the likelihood networks $\varphi_S$ used in the network satisfy~\eqref{eq:vs-module-accuracy}. Lemma~\ref{lem:vs-rrds} assembles them into an RRDS network $\widetilde F=(\widetilde\theta,\widetilde q)$ with $R=2$ and the stated $d_z$, $d_u$, $L$ and $W$. In the remaining steps, $(S_0,\theta_0)$ is an allowed task.

\medskip
\noindent{\bf Step 3: Estimation and prediction at a fixed parameter (part~(b)).} By Lemma~\ref{lem:sparse-local} for $S=S_0$, $D_{S_0}\leq\{4\log(en)\}^{-1}$ on $\mathcal E_n$ and $\mb P_0(\mathcal E_n^c)\leq Cn^{-10}$. With $0\leq D_{S_0}\leq4B^2n/(am_b)$, splitting on $\mathcal E_n$ and $\mathcal E_n^c$ gives
\[
\mb E_0[D_{S_0}]\leq\frac1{4\log(en)}+\frac{4B^2n}{am_b}\,\mb P_0(\mathcal E_n^c)\leq\frac1{4\log(en)}+Cn^{-9}.
\]
Since $n\geq D$ under the sample-size condition, Lemma~\ref{lem:vs-soft-oracle} gives $\mb E_0[nZ_n]\leq CA_n$. With $\operatorname{pen}(S_0)\leq A_n$, we have $\mb E_0[k_n]=\mb E_0[nZ_n]+\mb E_0[D_{S_0}]+\operatorname{pen}(S_0)+1\leq CA_n$, and since $\log N_{\mathcal S}\leq A_n$,
\[
\frac4{nc_K}\bigl\{\mb E_0[k_n]+\log N_{\mathcal S}\bigr\}+\frac C{n^2}\leq\frac{CA_n}n .
\]
Taking expectations in Lemma~\ref{lem:vs-soft-oracle} (absorbing $\overline\lambda_Z$ into $C$) therefore gives
\begin{equation}
\mb E_0[\|\widetilde\theta-\theta_0\|_2^2]\leq\frac{CA_n}n,\qquad
\mb E_0\Bigl[\mb E_{Q_Z}\bigl[\{\eta_{\widetilde\theta}(Z)-\eta_{\theta_0}(Z)\}^2\bigr]\Bigr]\leq\frac{CA_n}n .
\label{eq:vs-uniform-risk}
\end{equation}
For the prediction risk, the exponential family~\eqref{eq:glm-model} is defined for every real natural parameter because the responses are bounded, and $b''\leq Y_*^2/a$ on $\mb R$ by~\eqref{eq:glm-cumulant-strip}. The first identity in~\eqref{eq:glm-kl-prediction} holds for every $\theta\in\mb R^{1+p}$, so Taylor's theorem gives
\[
\begin{aligned}
D_{\rm KL}(\mb P_{\theta_0}\|\mb P_{\widetilde\theta})
&=\frac1a\,\mb E_{Q_Z}\bigl[b\{\eta_{\widetilde\theta}(Z)\}-b\{\eta_{\theta_0}(Z)\} -b'\{\eta_{\theta_0}(Z)\}\{\eta_{\widetilde\theta}(Z)-\eta_{\theta_0}(Z)\}\bigr]\\
&\leq\frac{Y_*^2}{2a^2}\,\mb E_{Q_Z}\bigl[\{\eta_{\widetilde\theta}(Z)-\eta_{\theta_0}(Z)\}^2\bigr],
\end{aligned}
\]
where the expectation is over a new covariate vector $Z$ with $\widetilde\theta$ fixed. Taking expectations and using~\eqref{eq:vs-uniform-risk} and $A_n\leq D\Lambda_n$ proves part~(b).

\medskip
\noindent{\bf Step 4: Estimation risk averaged over tasks (part~(a)).} To bound the task-averaged risk, split the tasks into $\mathcal T$ and $\mathcal T^c$, where $\mathcal T$ is defined in Appendix~\ref{app:vs-network}. For a task in $\mathcal T^c$, $\mathcal A_0=\mathcal A_+$. Work on the event $\mathcal G\subseteq\mathcal E_n$ of Lemma~\ref{lem:vs-near}, with $\mb P_0(\mathcal G^c)\leq Cn^{-10}$. For $S\in\mathcal A_+$,~\eqref{eq:sparse-local-estimation} and $\|V_S\|_2^2=\|V_{S_0}\|_2^2+\|U_S\|_2^2$ (Lemma~\ref{lem:vs-overfit}, with $U_{S_0}=0$) give $\|\iota_S\widetilde\theta_S-\theta_0\|_2^2\leq(C/n)\{1+\|V_{S_0}\|_2^2+\|U_S\|_2^2\}$. For every $S\in\mathcal S$, $\iota_S\widetilde\theta_S-\theta_0$ has at most $2D$ nonzero coordinates, each bounded by $2B$, so $\|\iota_S\widetilde\theta_S-\theta_0\|_2^2\leq8B^2D$. Lemma~\ref{lem:vs-near}(c) gives $\widetilde Q(\mathcal A_+^c)\leq e^{-8A_n}$.

For $S\supsetneq S_0$ with $b=|S\setminus S_0|$,~\eqref{eq:vs-logit-identity} and $D_{S_0}\leq\{4\log(en)\}^{-1}$ (Lemma~\ref{lem:vs-near}(b)) give
\[
\begin{aligned}
\widetilde a_S-\widetilde a_{S_0}
&=(a_S^{\rm E}-a_{S_0}^{\rm E})-D_S+D_{S_0}+e_S^\ell-e_{S_0}^\ell\\
&\leq\ell_{n,S}(\widehat\theta_S)-\ell_{n,S_0}(\widehat\theta_{S_0})-\Delta_b+\frac1{4\log(en)}+\frac1{3\log(en)}.
\end{aligned}
\]
By~\eqref{eq:sparse-quadratic} for $S$ and $S_0$ and $\|V_S\|_2^2=\|V_{S_0}\|_2^2+\|U_S\|_2^2$, we have $\ell_{n,S}(\widehat\theta_S)-\ell_{n,S_0}(\widehat\theta_{S_0})\leq\|U_S\|_2^2/2+1/8$. Since $\log(en)\geq1$ and $1/8+1/4+1/3<1\leq b$, this gives $\widetilde a_S-\widetilde a_{S_0}\leq\|U_S\|_2^2/2-\Delta_b+b$. Since $\widetilde Q_S\leq\min\{1,\widetilde Q_S/\widetilde Q_{S_0}\}$,
\[
\widetilde Q_S\leq\min\{1,e^{\widetilde a_S-\widetilde a_{S_0}}\}\leq\min\{1,e^b\exp(\|U_S\|_2^2/2-\Delta_b)\}=\varpi_S,
\]
where $\varpi_S$ is defined in Lemma~\ref{lem:vs-sq-odds}.

By the convexity of the squared norm, $\theta^{\rm av}$ of Lemma~\ref{lem:vs-heads} satisfies
\[
\begin{aligned}
\|\theta^{\rm av}-\theta_0\|_2^2
&\leq\sum_{S\in\mathcal S}\widetilde Q_S\|\iota_S\widetilde\theta_S-\theta_0\|_2^2\leq\frac Cn\Bigl\{\widetilde Q(\mathcal A_+)(1+\|V_{S_0}\|_2^2)+\sum_{S\supsetneq S_0}\widetilde Q_S\|U_S\|_2^2\Bigr\}+8B^2D\,\widetilde Q(\mathcal A_+^c)\\
&\leq\frac Cn\Bigl\{1+\|V_{S_0}\|_2^2+\sum_{S\supsetneq S_0}\varpi_S\|U_S\|_2^2\Bigr\}+8B^2De^{-8A_n},
\end{aligned}
\]
where we split the sum over $S=S_0$, $S\supsetneq S_0$ and $S\in\mathcal A_+^c$, and used $\widetilde Q(\mathcal A_+)=\widetilde Q_{S_0}+\sum_{S\supsetneq S_0}\widetilde Q_S\leq1$, $\widetilde Q_S\leq\varpi_S$ and $\widetilde Q(\mathcal A_+^c)\leq e^{-8A_n}$. Since $V_{S_0}$ has identity covariance, $\mb E_0[\|V_{S_0}\|_2^2]=d_0$. Since $\mathcal G\subseteq\mathcal E_n$, the last bound of~\eqref{eq:vs-added-weighted} gives $\sum_{S\supsetneq S_0}\mb E_0[\varpi_S\|U_S\|_2^2\ind_{\mathcal G}]\leq C$, and $e^{-8A_n}\leq n^{-8}$ with $D\leq n$ gives $8B^2De^{-8A_n}\leq Cd_0/n$. Hence $\mb E_0[\|\theta^{\rm av}-\theta_0\|_2^2\ind_{\mathcal G}]\leq Cd_0/n$.

By Lemma~\ref{lem:vs-heads}, $\|\widetilde\theta-\theta^{\rm av}\|_2\leq\|\widetilde\theta-\theta^{\rm av}\|_1\leq(3+4B)/n$ for every dataset, so $\|\widetilde\theta-\theta_0\|_2^2\leq2\|\theta^{\rm av}-\theta_0\|_2^2+2\|\widetilde\theta-\theta^{\rm av}\|_2^2$ gives $\mb E_0[\|\widetilde\theta-\theta_0\|_2^2\ind_{\mathcal G}]\leq Cd_0/n+C/n^2\leq CD/n$. On $\mathcal G^c$, Lemma~\ref{lem:vs-heads} gives $\|\widetilde\theta-\theta_0\|_2^2\leq8B^2D$, contributing at most $8B^2D\cdot Cn^{-10}\leq CD/n$. Integrating over $\mathcal T^c$ thus contributes at most $CD/n$. On $\mathcal T$,~\eqref{eq:vs-uniform-risk} contributes at most $(CA_n/n)\,\mb Q_n(\mathcal T)$, with $\mb Q_n(\mathcal T)\leq Cs_{\max}r_n$ from~\eqref{eq:vs-prior-mass}. Together these give
\begin{equation}
\mb E_{\mb Q_n}[\|\widetilde\theta-\theta\|_2^2]\leq\frac{CD}n+\frac{CA_n}n Cs_{\max}r_n\leq\frac{Cs_{\max}}n,
\label{eq:vs-prior-average}
\end{equation}
since $A_nr_n=A_n^{3/2}/\sqrt n\leq\sqrt{c_0}$ and $D\leq2s_{\max}$. The first claim of part~(a) is~\eqref{eq:vs-bridge-general} below with $Q=\widetilde Q$, and $J_n\leq Ch_n$ by the bound on $H_n^*$ in Proposition~\ref{prop:vs-bridge}.

\medskip
\noindent{\bf Step 5: Hamming loss of the selected support.} We bound the Hamming loss of $\widetilde S=\{j:\widetilde q_j>1/2\}$, which gives the bound on $H_n^*$ in Proposition~\ref{prop:vs-bridge}. Fix a task in $\mathcal T^c$, so that $\mathcal A_0=\mathcal A_+$, and work on the event $\mathcal G$ of Step~4, on which $\widetilde Q(\mathcal A_+^c)\leq e^{-8A_n}$ and $\widetilde Q_S\leq\varpi_S$ for $S\supsetneq S_0$.

For $j\in S_0$, a support that omits $j$ does not contain $S_0$ and lies in $\mathcal A_+^c$, so $1-\widetilde q_j^{\rm ex}=\sum_{S\not\ni j}\widetilde Q_S\leq\widetilde Q(\mathcal A_+^c)\leq e^{-8A_n}$. Since $\eta_n\leq1/9$, $\epsilon_n\leq1/7$ and $A_n\geq1$, Lemma~\ref{lem:vs-heads} gives $1-\widetilde q_j\leq e^{\eta_n}(1-\widetilde q_j^{\rm ex})+\epsilon_n\leq e^{\eta_n}e^{-8A_n}+\epsilon_n<1/2$. Hence $S_0\subseteq\widetilde S$.

For $j\notin S_0$, Lemma~\ref{lem:vs-heads} gives $1/2<\widetilde q_j\leq e^{\eta_n}\widetilde q_j^{\rm ex}+\epsilon_n$ when $j\in\widetilde S$, so $\widetilde q_j^{\rm ex}\geq e^{-\eta_n}(1/2-\epsilon_n)\geq1/4$. A support $S$ containing $j\notin S_0$ belongs to $\mathcal A_+$ or $\mathcal A_+^c$. If $S\in\mathcal A_+$, then $S\supsetneq S_0$, so $S=S_0\cup\{j\}$ (one added variable) or $b=|S\setminus S_0|\geq2$, and $\widetilde Q_S\leq\varpi_S$. Since $\widetilde q_j^{\rm ex}\leq1$,
\[
\widetilde q_j^{\rm ex}\leq\min\Bigl\{1,\ \varpi_{S_0\cup\{j\}}+\sum_{S\supsetneq S_0:\,S\ni j,\,b\geq2}\varpi_S+\sum_{S\in\mathcal A_+^c:\,S\ni j}\widetilde Q_S\Bigr\},
\]
where the first term is absent if $s_0=s_{\max}$. We use $\min\{1,x+y+z\}^2\leq3\{\min(1,x)^2+y+z\}$ for $x,y,z\geq0$, since if $x\geq1$ the right side is at least $3$, if $x+y+z\geq1$ and $x<1$ then $x^2+y+z\geq x^2+1-x\geq3/4$, and if $x+y+z<1$ then $(x+y+z)^2\leq3(x^2+y^2+z^2)\leq3(x^2+y+z)$. When we apply this bound with $\varpi_{S_0\cup\{j\}}\leq1$ and sum over $j\notin S_0$, each support with $b\geq2$ added variables is counted $b$ times and each support in $\mathcal A_+^c$ at most $s_{\max}$ times, so the last sum is at most $s_{\max}\widetilde Q(\mathcal A_+^c)\leq s_{\max}e^{-8A_n}$. Since $S_0\subseteq\widetilde S$, this gives
\[
\begin{aligned}
|\widetilde S\mathbin{\triangle}S_0|&=|\{j\notin S_0:\widetilde q_j>1/2\}|
\leq\sum_{j\notin S_0}\ind\{\widetilde q_j^{\rm ex}\geq1/4\}
\leq16\sum_{j\notin S_0}(\widetilde q_j^{\rm ex})^2\\
&\leq48\Bigl\{\sum_{S\supsetneq S_0:\,b=1}\varpi_S^2+\sum_{S\supsetneq S_0:\,b\geq2}b\,\varpi_S+s_{\max}e^{-8A_n}\Bigr\}.
\end{aligned}
\]
By~\eqref{eq:vs-added-sums}, since $\varpi_S\leq\varpi_S^\kappa$ (with $\kappa=1$), $b\leq b^2$ and $e^{-8A_n}\leq n^{-8}$, the expectation of the right side is at most
\[
C\Bigl\{s_{\max}\sqrt{\frac{\Lambda_n}n}+\frac{s_{\max}^2\Lambda_n^2}n\Bigr\}+\frac{Cs_{\max}}{n^8}\leq Cs_{\max}\sqrt{\frac{\Lambda_n}n},
\]
because $s_{\max}^2\Lambda_n^3\leq27A_n^3\leq27c_0n$ by~\eqref{eq:vs-An-Lambda} and~\eqref{eq:vs-smax-condition}. On $\mathcal G^c$, $|\widetilde S\mathbin{\triangle}S_0|\leq|\widetilde S|+s_0\leq5s_{\max}$ by Lemma~\ref{lem:vs-heads}, and $\mb P_0(\mathcal G^c)\leq Cn^{-10}$. Thus $\mb E_0[|\widetilde S\mathbin{\triangle}S_0|]\leq Cs_{\max}\sqrt{\Lambda_n/n}$ for every task in $\mathcal T^c$. On $\mathcal T$, use $|\widetilde S\mathbin{\triangle}S_0|\leq5s_{\max}$ and $\mb Q_n(\mathcal T)\leq Cs_{\max}r_n\leq Cs_{\max}^{3/2}\sqrt{\Lambda_n/n}$ from~\eqref{eq:vs-prior-mass} and~\eqref{eq:vs-rn}. Together these give
\begin{equation}
\mb E_{\mb Q_n}[|\widetilde S\mathbin{\triangle}S|]\leq Cs_{\max}\sqrt{\frac{\Lambda_n}n}+5s_{\max}\cdot Cs_{\max}^{3/2}\sqrt{\frac{\Lambda_n}n}\leq Cs_{\max}^{5/2}\sqrt{\frac{\Lambda_n}n}.
\label{eq:vs-network-hamming}
\end{equation}
\end{proof}

\subsubsection{Comparison with posterior inclusion probabilities}\label{app:proof-vs-bridge}

\begin{proof}[Proof of Proposition~\ref{prop:vs-bridge} and of Theorem~\ref{thm:vs-approx}\textup{(a)}]

The aim is to bound the sum of the binary KL divergences, rather than the KL divergence between the full support distributions. We first discard supports with negligible probability under both distributions and then compare the marginal probabilities on the remaining supports. The final averaging step separates tasks with weak signals or parameters near a prior boundary from the other tasks, because the accuracy of the Laplace approximation differs in these cases.
In this proof, $C,c$ are as in Appendix~\ref{app:vs-smax}, so they do not depend on $n$, $p$, $D$, the supports or $\theta_0$. We assume~\eqref{eq:vs-smax-condition} and use the lemmas of Appendix~\ref{app:vs-network} in the form recorded in Appendix~\ref{app:vs-smax}.
The two comparisons differ only in the additional approximation of the marginal logits by the network. We therefore treat $Q^{\rm E}$ and $\widetilde Q$ together. Let $Q$ be either $Q^{\rm E}$ or $\widetilde Q$, with marginals $q_j^Q$. In the first case put $e_j=0$, and in the second $e_j=a_j-\log\{\widetilde q_j^{\rm ex}/(1-\widetilde q_j^{\rm ex})\}$, so that $|e_j|\leq\eta_n\leq1$ by Lemma~\ref{lem:vs-rrds}(b). In the notation of Lemma~\ref{lem:bern-tilt} with $\gamma=1$ and $\epsilon=\epsilon_n$, the probabilities to be compared with $q_j^*$ are $q_j=q_j^{Q,e}$, which are the clipped EBIC inclusion probabilities in the first case and $\widetilde q_j$ in the second. Let $\mathcal G$ be the event of Lemma~\ref{lem:vs-laplace-diff}. We prove that
\begin{equation}
\begin{gathered}
\sum_{j=1}^p\mb E_{\mb Q_n}\bigl[\ind_{(\mathcal T\cup\mathcal B)^c}D_{\rm KL}(q_j^*\|q_j)\bigr]\leq C\Bigl\{H_n^*+\frac{s_{\max}^2\Lambda_n^2}n+\frac1n\Bigr\},\\
\sum_{j=1}^p\mb E_{\mb Q_n}\bigl[\ind_{\mathcal T\cup\mathcal B}D_{\rm KL}(q_j^*\|q_j)\bigr]\leq Cs_{\max}^3r_n .
\end{gathered}
\label{eq:vs-bridge-general}
\end{equation}
Since $s_{\max}^2\Lambda_n^2/n\leq s_{\max}\sqrt{\Lambda_n/n}$ and $s_{\max}^3r_n\leq\sqrt2\,s_{\max}^{7/2}\sqrt{\Lambda_n/n}$ by~\eqref{eq:vs-smax-condition} and~\eqref{eq:vs-rn}, this proves the proposition and the first claim of Theorem~\ref{thm:vs-approx}(a).

Fix an allowed task. For every dataset, since $\log(1/\epsilon_n)=\log(1+np)\leq\Lambda_n$ and $p\epsilon_n\leq n^{-1}$, Lemma~\ref{lem:bern-tilt}(a) with $P=Q^*$, the vector $Q$, $s=s_{\max}$, $\gamma=1$ and $\epsilon=\epsilon_n$ gives
\begin{equation}
\sum_{j=1}^pD_{\rm KL}(q_j^*\|q_j)\leq Cs_{\max}\{1+\log(1/\epsilon_n)\}+Cp\epsilon_n\leq Cs_{\max}\Lambda_n ,
\label{eq:vs-envelope}
\end{equation}
so the complement of $\mathcal G$ contributes at most $Cs_{\max}\Lambda_nn^{-10}\leq Cn^{-1}$ to~\eqref{eq:vs-bridge-general}, since $\mb P_0(\mathcal G^c)\leq Cn^{-10}$ and $s_{\max}\Lambda_n\leq n$ by~\eqref{eq:vs-smax-condition}. On $\mathcal G$, Lemma~\ref{lem:vs-near}(c) gives $\alpha=Q^*(\mathcal A_0^c)\leq e^{-8A_n}$ and $\beta=Q(\mathcal A_0^c)\leq e^{-8A_n}$, so Lemma~\ref{lem:bern-tilt}(b) with $\mathcal A=\mathcal A_0$, $s=s_{\max}$, $\gamma=1$ and $\epsilon=\epsilon_n$ gives
\begin{equation}
\sum_{j=1}^pD_{\rm KL}(q_j^*\|q_j)\leq\sum_{j=1}^pD_{\rm KL}\bigl(p_j^{\mathcal A_0}\|q_j^{\mathcal A_0,e}\bigr)+\frac C{n^5},
\label{eq:vs-removal}
\end{equation}
where $p_j^{\mathcal A_0}$ and $q_j^{\mathcal A_0,e}$ are computed from the conditional vectors $P$ and $Q^{\mathcal A_0}$ of $Q^*$ and $Q$ on $\mathcal A_0$. Indeed, Lemma~\ref{lem:bern-tilt}(b) adds the three terms
\[
2s_{\max}\alpha\log(1/\epsilon_n),\qquad p\,h(\min\{2s_{\max}\alpha/p,1/2\})\qquad\text{and}\qquad2s_{\max}e\beta/\epsilon_n.
\]
Since $\log(1/\epsilon_n)\leq\Lambda_n\leq A_n$, the first term is at most $2s_{\max}A_ne^{-8A_n}$. For the second, since $2s_{\max}\alpha/p\leq2s_{\max}e^{-8A_n}/p\leq1/2$, $h(x)\leq x\log(e/x)$ with $x\mapsto x\log(e/x)$ increasing on $(0,1]$, and $\log(ep)\leq A_n$, we have
\[
\begin{aligned}
p\,h(2s_{\max}\alpha/p)
&\leq2s_{\max}\alpha\log\{ep/(2s_{\max}\alpha)\}\\
&\leq2s_{\max}e^{-8A_n}\{\log(ep)+8A_n\}\leq18s_{\max}A_ne^{-8A_n}.
\end{aligned}
\]
For the third, $1/\epsilon_n=1+np\leq e^{\Lambda_n}\leq e^{A_n}$ gives $\beta/\epsilon_n\leq e^{-7A_n}\leq e^{-6A_n}$, so $2s_{\max}e\beta/\epsilon_n\leq2es_{\max}e^{-6A_n}$. Since $A_n\geq2\log n$ and $s_{\max}\leq A_n$, all three terms are at most $CA_n^2e^{-6A_n}\leq C/n^5$. By~\eqref{eq:vs-posterior-tilt}, $P_S\propto Q_S^{\mathcal A_0}e^{\delta_S'}$ with $\delta_S'=\delta_S$ for $Q=Q^{\rm E}$ and $\delta_S'=\delta_S+D_S-e_S^\ell$ for $Q=\widetilde Q$, where $0\leq D_S\leq1/4$ and $|e_S^\ell|\leq1/6$ on $\mathcal G$ for $S\in\mathcal A_0$. We consider four cases. For tasks with neither weak signals nor boundary effects, only supports containing the true support need to be retained. The penalty for adding variables then controls the error. If either condition fails, the remaining supports or the Laplace corrections require separate bounds.

\medskip
\noindent{\bf Case 1: Neither weak signals nor boundary effects.} Then $\mathcal A_0=\mathcal A_+$. Set $b_S=|S\setminus S_0|$. On $\mathcal G\subseteq\mathcal E_n$, Lemma~\ref{lem:vs-laplace-diff} gives $|\delta_S-\delta_{S_0}|\leq c_1b_S+C_1$, and $\delta_S'=\delta_S$ for $Q=Q^{\rm E}$. For $Q=\widetilde Q$, Lemma~\ref{lem:vs-near}(b) and~\eqref{eq:vs-logit-identity} give $0\leq D_S\leq\{4\log(en)\}^{-1}$ and $|e_S^\ell|\leq\{6\log(en)\}^{-1}$, so $|(D_S-e_S^\ell)-(D_{S_0}-e_{S_0}^\ell)|\leq1/2$ since $\log(en)\geq2$ for $n\geq3$. Since $b_S\geq1$ for $S\supsetneq S_0$, $|\delta_S'-\delta_{S_0}'|\leq c_1b_S+C_1+1/2\leq cb_S$ with $c=c_1+C_1+1/2$. The two bounds of Lemma~\ref{lem:added-tilt} for $P,Q^{\mathcal A_+}$, tilts $\delta_S'$ and $\gamma=1$ give
\[
\sum_{j=1}^pD_{\rm KL}\bigl(p_j^{\mathcal A_+}\|q_j^{\mathcal A_+,e}\bigr)\leq C\Bigl\{(1-P_{S_0})+\sum_{S\supsetneq S_0:\,b_S\geq2}b_S^2\bigl(P_S+Q_S^{\mathcal A_+}\bigr)\Bigr\}+2p\epsilon_n ,
\]
where $C$ depends on $c$. Here $P_{S_0}=Q_{S_0}^*/Q^*(\mathcal A_+)\geq Q_{S_0}^*$, so $1-P_{S_0}\leq1-Q_{S_0}^*$. For $S\supsetneq S_0$, $Q_S^{\mathcal A_+}/Q_{S_0}^{\mathcal A_+}=Q_S/Q_{S_0}$ is $e^{a_S^{\rm E}-a_{S_0}^{\rm E}}$ or $e^{\widetilde a_S-\widetilde a_{S_0}}$. By~\eqref{eq:sparse-quadratic} for $S,S_0$, $\|V_S\|_2^2=\|V_{S_0}\|_2^2+\|U_S\|_2^2$ and~\eqref{eq:vs-logit-identity}, $Q_S/Q_{S_0}\leq e^{b_S}\exp(\|U_S\|_2^2/2-\Delta_{b_S})$ since $1/8+1/2\leq b_S$. Hence $Q_S^{\mathcal A_+}\leq Q_S/Q_{S_0}$ and $P_S\leq P_S/P_{S_0}=(Q_S/Q_{S_0})e^{\delta_S'-\delta_{S_0}'}\leq e^{(c+1)b_S}\exp(\|U_S\|_2^2/2-\Delta_{b_S})$. Since $P_S,Q_S^{\mathcal A_+}\leq1$ and $e^{b_S}\leq e^{(c+1)b_S}\leq e^{\kappa(b_S+1)}$ with $\kappa=c+1$, we have $P_S+Q_S^{\mathcal A_+}\leq2\varpi_S^\kappa$ (Lemma~\ref{lem:vs-added-explicit}). By~\eqref{eq:vs-removal}, the first display, $2p\epsilon_n\leq2/n$ and the second bound of~\eqref{eq:vs-added-sums}, every regular task satisfies
\[
\sum_{j=1}^p\mb E_0\bigl[\ind_{\mathcal G}D_{\rm KL}(q_j^*\|q_j)\bigr]\leq C\,\mb E_0[1-Q_{S_0}^*]+\frac{Cs_{\max}^2\Lambda_n^2}n+\frac Cn .
\]
Integrating over regular tasks and adding $C/n$ for $\mathcal G^c$ gives the first bound of~\eqref{eq:vs-bridge-general}, since $\mb E_{\mb Q_n}[1-Q_S^*]\leq2H_n^*$ for the generating support $S$ by Lemma~\ref{lem:ce-selection}(b).

\medskip
\noindent{\bf Case 2: Weak signals, away from the boundary.} On $\mathcal G$, Lemma~\ref{lem:vs-near}(b) and~(d) show that, for every $S\in\mathcal A_0$, the MLE $\widehat\theta_S$ is an interior point of $\Theta_{0,S}$ and belongs to $K_S$, so $U_S^\pi=0$ and $|\delta_S|\leq CD$ by Lemma~\ref{lem:vs-laplace}. Hence $\max_S\delta_S'-\min_S\delta_S'\leq CD$ on $\mathcal A_0$, and
\[
D_{\rm KL}(P\|Q^{\mathcal A_0})=\sum_{S\in\mathcal A_0}P_S\delta_S'-\log\sum_{S\in\mathcal A_0}Q_S^{\mathcal A_0}e^{\delta_S'}\leq CD .
\]
Lemma~\ref{lem:sparse-contraction} bounds the sum of the divergences between the marginals by $4s_{\max}$ times this quantity, and Lemma~\ref{lem:bern-tilt}(c) bounds the effect of the errors $e_j$ and of the clipping by $2e\,\eta_ns_{\max}+2p\epsilon_n$. For a coordinate whose marginal under $Q^{\mathcal A_0}$ equals zero or one, the marginal under $P$ is the same, and only the clipping contributes, by at most $2\epsilon_n$. With~\eqref{eq:vs-removal}, $\sum_jD_{\rm KL}(q_j^*\|q_j)\leq Cs_{\max}^2$ on $\mathcal G$.

\medskip
\noindent{\bf Case 3: Boundary effects, without weak signals.} In this case $\mathcal A_0=\mathcal A_+$. Write $P^+$ and $Q^+$ for the conditional vectors of $Q^*$ and $Q$ on $\mathcal A_+$. On $\mathcal G$, Lemma~\ref{lem:vs-near}(b) and Lemma~\ref{lem:vs-laplace} give $\delta_S'\leq CD$ for every $S\in\mathcal A_+$ and $\delta_{S_0}'\geq-CD-MU_{S_0}^\pi$. Only $S_0$ needs a lower bound, because
\[
\begin{aligned}
D_{\rm KL}(P^+\|Q^+)&=\sum_{S\in\mathcal A_+}P_S^+(\delta_S'-\delta_{S_0}')-\log\sum_{S\in\mathcal A_+}Q_S^+e^{\delta_S'-\delta_{S_0}'}\\
&\leq(1-P_{S_0}^+)\bigl(CD+MU_{S_0}^\pi\bigr)-\log Q_{S_0}^+ .
\end{aligned}
\]
Lemma~\ref{lem:sparse-contraction} bounds the sum of the divergences between the marginals of $P^+$ and $Q^+$ by $4s_{\max}$ times this quantity, and Lemma~\ref{lem:bern-tilt}(c) bounds the effect of the errors $e_j$ and of the clipping as in Case~2. With~\eqref{eq:vs-removal}, on $\mathcal G$,
\begin{equation}
\sum_{j=1}^pD_{\rm KL}(q_j^*\|q_j)\leq Cs_{\max}\bigl\{D+U_{S_0}^\pi-\log Q_{S_0}^+\bigr\}+\frac Cn .
\label{eq:vs-boundary-task}
\end{equation}
Since $\theta_{0,S_0}\in K_{S_0}$, $U_{S_0}^\pi\leq n\|\widehat\theta_{S_0}-\theta_{0,S_0}\|_2^2$, and on $\mathcal G$ the strong concavity of $\ell_{n,S_0}$ and~\eqref{eq:sparse-quadratic} give $n\|\widehat\theta_{S_0}-\theta_{0,S_0}\|_2^2\leq(\|V_{S_0}\|_2^2+1/8)/m$, so $\mb E_0[\ind_{\mathcal G}U_{S_0}^\pi]\leq CD$. As in Case~1, $Q_S^+/Q_{S_0}^+\leq e^{b_S}\exp(\|U_S\|_2^2/2-\Delta_{b_S})$ for $S\supsetneq S_0$, so the first bound of~\eqref{eq:vs-added-weighted} gives $\mb E_0[\ind_{\mathcal G}(-\log Q_{S_0}^+)]\leq Cs_{\max}\Lambda_n^{3/2}/\sqrt n\leq C$. Hence $\sum_j\mb E_0[\ind_{\mathcal G}D_{\rm KL}(q_j^*\|q_j)]\leq Cs_{\max}^2$ for every task of this case.

\medskip
\noindent{\bf Case 4: Both weak signals and boundary effects.} We use the envelope~\eqref{eq:vs-envelope}.

By~\eqref{eq:vs-irregular-mass}, the tasks of Cases~2 and~3 have probability at most $Cs_{\max}r_n$ and those of Case~4 at most $Cs_{\max}^2r_n^2$, so together they contribute at most
\[
Cs_{\max}^2\cdot s_{\max}r_n+Cs_{\max}\Lambda_n\cdot s_{\max}^2r_n^2+\frac Cn\leq Cs_{\max}^3r_n
\]
to~\eqref{eq:vs-bridge-general}, because $\Lambda_nr_n\leq A_n^{3/2}/\sqrt n\leq1$ and $n^{-1}\leq r_n$. The four cases cover all tasks, which proves the second bound of~\eqref{eq:vs-bridge-general}.

Finally, by Lemma~\ref{lem:ce-selection}(b), $H_n^*\leq\mb E_{\mb Q_n}[|\widetilde S\mathbin{\triangle}S|]$ for the selection $\widetilde S$ of the network of Lemma~\ref{lem:vs-rrds}, and the bound~\eqref{eq:vs-network-hamming}, whose proof does not use the present proposition, gives the bound on $H_n^*$.
\end{proof}

\subsubsection{Proof of Proposition~\ref{prop:vs-bridge-est}}\label{app:proof-vs-bridge-est}

\begin{proof}[Proof of Proposition~\ref{prop:vs-bridge-est}]
In this proof, $C,c$ depend only on the constants of the setting of Appendix~\ref{app:vs-setting} and of Assumptions~\ref{ass:sparse} and~\ref{ass:sparse-proposal}, and they do not depend on $n$, $p$, $D$, the supports or $\theta_0$. We assume~\eqref{eq:vs-smax-condition} and use the lemmas of Appendix~\ref{app:vs-network} in the form recorded in Appendix~\ref{app:vs-smax}, in which the dependence on $D$ is explicit. We use the notation of Appendix~\ref{app:vs-network}. In particular, $a_S^{\rm E}=\ell_{n,S}(\widehat\theta_S)-\operatorname{pen}(S)=-\operatorname{IC}_S$, the log likelihoods are the centered ones, and $\iota_S\widehat\theta_S=\widehat\theta^{(S)}$ denotes zero extension.

\emph{Part~(a).} Fix an allowed task and a dataset, and let $c_K=m_b\underline\lambda_Z/(2a)$. For every $S\in\mathcal S$, both $\iota_S\widehat\theta_S$ and $\theta_0$ belong to $\Theta_{\mathcal S}$, so~\eqref{eq:glm-kl-parameter} gives $R_S(\widehat\theta_S)\geq c_K\|\iota_S\widehat\theta_S-\theta_0\|_2^2$. By the definition of $Z_n$ and since $nP_nY_{S,u}=\ell_{n,S_0}(\theta_{0,S_0})-\ell_{n,S}(u)$,
\[
\frac{nc_K}2\|\iota_S\widehat\theta_S-\theta_0\|_2^2
\leq nZ_n+\ell_{n,S_0}(\theta_{0,S_0})-\ell_{n,S}(\widehat\theta_S)
\leq nZ_n+\ell_{n,S_0}(\widehat\theta_{S_0})-\ell_{n,S}(\widehat\theta_S),
\]
where the second inequality is the optimality of $\widehat\theta_{S_0}$ on $\Theta_{0,S_0}\ni\theta_{0,S_0}$. The last difference equals $a_{S_0}^{\rm E}-a_S^{\rm E}+\operatorname{pen}(S_0)-\operatorname{pen}(S)$, and $\operatorname{pen}(S)\geq0$, so
\begin{equation}
\frac{nc_K}2\|\iota_S\widehat\theta_S-\theta_0\|_2^2\leq nZ_n+\operatorname{pen}(S_0)+\bigl(a_{S_0}^{\rm E}-a_S^{\rm E}\bigr)_+ ,\qquad S\in\mathcal S .
\label{eq:vs-mle-support-distance}
\end{equation}
Since $Q^{\rm E}$ is a probability vector and $\theta^{\rm E}-\theta_0=\sum_SQ_S^{\rm E}(\iota_S\widehat\theta_S-\theta_0)$, the convexity of the squared norm and~\eqref{eq:vs-mle-support-distance} give
\[
\|\theta^{\rm E}-\theta_0\|_2^2\leq\sum_{S\in\mathcal S}Q_S^{\rm E}\|\iota_S\widehat\theta_S-\theta_0\|_2^2
\leq\frac2{nc_K}\Bigl\{nZ_n+\operatorname{pen}(S_0)+\sum_{S\in\mathcal S}Q_S^{\rm E}\bigl(a_{S_0}^{\rm E}-a_S^{\rm E}\bigr)_+\Bigr\}.
\]
The last sum is at most $\log N_{\mathcal S}$ by the entropy argument in the proof of Lemma~\ref{lem:vs-heads}, which applies to the softmax weights of any logits: with $M=\max_Sa_S^{\rm E}$, $\sum_SQ_S^{\rm E}a_S^{\rm E}=\log\sum_Se^{a_S^{\rm E}}+\sum_SQ_S^{\rm E}\log Q_S^{\rm E}\geq M-\log N_{\mathcal S}$. This is the displayed bound, since $2/c_K=4a/(m_b\underline\lambda_Z)$. For the expectation, $\operatorname{pen}(S_0)\leq A_n$ and $\log N_{\mathcal S}\leq A_n$, and the proof of Lemma~\ref{lem:vs-soft-oracle} gives $\mb E_0[nZ_n]\leq\lambda_Y^{-1}\log\sum_SN_S(h)+2n^{-1}\leq C_1'A_n$ for $n\geq D$, where $\lambda_Y$ and the constant in the covering bound depend only on $a$, $B$ and $m_b$. Hence $\mb E_0[\|\theta^{\rm E}-\theta_0\|_2^2]\leq C_1A_n/n$ with $C_1$ depending only on $a$, $B$, $m_b$ and $\underline\lambda_Z$. This part does not use the sample-size condition.

\emph{Part~(b).} The first inequality follows from Lemma~\ref{lem:ce-selection}(c), which gives $\mb E_{\mb Q_n}[\|\theta^{\rm E}-\theta\|_2^2]=\mb E_{\mb Q_n}[\|\theta^{\rm E}-\theta^*\|_2^2]+\mb E_{\mb Q_n}[\|\theta^*-\theta\|_2^2]$. For the second, first consider an allowed task outside $\mathcal T$, so that $\mathcal A_0=\mathcal A_+$. Let $\mathcal G\subseteq\mathcal E_n$ be the event of Lemma~\ref{lem:vs-near}, with $\mb P_0(\mathcal G^c)\leq Cn^{-10}$. We use three facts that hold on $\mathcal G$.

First, $Q^{\rm E}(\mathcal A_+^c)\leq e^{-8A_n}$ by Lemma~\ref{lem:vs-near}(c). Second, let $S\in\mathcal A_+$. By Lemma~\ref{lem:vs-near}(b), $\widehat\theta_S$ is an interior maximizer, so $\nabla\ell_{n,S}(\widehat\theta_S)=0$, and a second-order expansion at $\widehat\theta_S$ along the segment to $\theta_{0,S}$, which lies in the convex set $\Theta_{0,S}$, together with the curvature bound of Lemma~\ref{lem:vs-near}(a), gives $\ell_{n,S}(\widehat\theta_S)-\ell_{n,S}(\theta_{0,S})\geq(nm/2)\|\widehat\theta_S-\theta_{0,S}\|_2^2$. With~\eqref{eq:sparse-quadratic} and the identity $\|V_S\|_2^2=\|V_{S_0}\|_2^2+\|U_S\|_2^2$ from the proof of Lemma~\ref{lem:vs-overfit}, where we put $U_{S_0}=0$,
\[
\|\iota_S\widehat\theta_S-\theta_0\|_2^2=\|\widehat\theta_S-\theta_{0,S}\|_2^2\leq\frac{\|V_{S_0}\|_2^2+\|U_S\|_2^2+1/8}{nm},\qquad S\in\mathcal A_+ .
\]
Third, let $S\supsetneq S_0$ with $b=|S\setminus S_0|$. By~\eqref{eq:sparse-quadratic} and the efficient score identity, $\operatorname{IC}_S-\operatorname{IC}_{S_0}\geq\Delta_b-\|U_S\|_2^2/2-1/8$, so
\[
Q_S^{\rm E}\leq\min\Bigl\{1,\frac{Q_S^{\rm E}}{Q_{S_0}^{\rm E}}\Bigr\}=\min\bigl\{1,e^{\operatorname{IC}_{S_0}-\operatorname{IC}_S}\bigr\}\leq\min\bigl\{1,e^b\exp(\|U_S\|_2^2/2-\Delta_b)\bigr\}=\varpi_S ,
\]
with $\varpi_S$ as in Lemma~\ref{lem:vs-sq-odds}. For every $S$, the vector $\iota_S\widehat\theta_S-\theta_0$ has at most $2D$ nonzero coordinates, each bounded by $2B$ in absolute value, so $\|\iota_S\widehat\theta_S-\theta_0\|_2^2\leq8B^2D$. Combining the three facts with the convexity of the squared norm, on $\mathcal G$,
\begin{equation}
\|\theta^{\rm E}-\theta_0\|_2^2\leq\sum_{S\in\mathcal S}Q_S^{\rm E}\|\iota_S\widehat\theta_S-\theta_0\|_2^2
\leq\frac{\|V_{S_0}\|_2^2+1/8}{nm}+\frac1{nm}\sum_{S\supsetneq S_0}\varpi_S\|U_S\|_2^2+8B^2De^{-8A_n}.
\label{eq:vs-mle-average}
\end{equation}

By the last bound of~\eqref{eq:vs-added-weighted}, the expectation of the sum in the middle term on $\mathcal G\subseteq\mathcal E_n$ is at most $C$. Taking expectations in~\eqref{eq:vs-mle-average}, with $\mb E_0[\|V_{S_0}\|_2^2]=d_0\leq D$, $A_n\geq2\log n$, and the bound $\|\theta^{\rm E}-\theta_0\|_2^2\leq8B^2D$ on $\mathcal G^c$, we obtain, for every allowed task outside $\mathcal T$,
\begin{equation}
\mb E_0\bigl[\|\theta^{\rm E}-\theta_0\|_2^2\bigr]\leq\frac{CD}n .
\label{eq:vs-mle-strong}
\end{equation}

Finally we integrate over the task distribution as in Step~4 of the proof of Theorem~\ref{thm:vs-approx}. Under $\mb Q_n$ the generating parameter is an allowed task almost surely. By~\eqref{eq:vs-prior-mass}, $\mb Q_n(\mathcal T)\leq Cs_{\max}r_n$, and on these tasks part~(a) gives $\mb E_0[\|\theta^{\rm E}-\theta_0\|_2^2]\leq CA_n/n$. On the remaining tasks we use~\eqref{eq:vs-mle-strong}. Hence
\[
\mb E_{\mb Q_n}\bigl[\|\theta^{\rm E}-\theta\|_2^2\bigr]\leq\frac{CD}n+\frac{Cs_{\max}A_nr_n}n\leq\frac{Cs_{\max}}n,
\]
because $A_nr_n=A_n^{3/2}/\sqrt n\leq\sqrt{c_0}$.
\end{proof}

\subsubsection{Proof of Theorem~\ref{thm:vs-gen}}\label{app:proof-vs-gen}

\begin{proof}[Proof of Theorem~\ref{thm:vs-gen}]
In this proof, $C$ does not depend on $\Gamma$ or $\lambda$. The radius $\Gamma\geq1$ is at least the Euclidean norm of the parameter vector of the constructed network and at least the absolute value of every fixed coefficient of the architecture; by Lemma~\ref{lem:vs-rrds}(d), such a $\Gamma$ is finite.

Put $r_0=B\sqrt D$ and $R_b=2B\sqrt{s_{\max}}$, so that $r_0\leq R_b$. For a parameter vector $\omega$, the output of the network is $F_\omega=(\Pi_{\rm b}v_\omega,q_\omega)$, with
\[
q_{\omega,j}\in[\epsilon_n,1-\epsilon_n]\qquad\text{and}\qquad v_\omega=(\operatorname{clip}_B(b_0),(q_{\omega,j}\operatorname{clip}_B(b_j))_j),
\]
where $\Pi_{\rm b}$ is the projection onto the ball of radius $R_b$ in the output map $\Psi$. Thus $\|\Pi_{\rm b}v_\omega\|_2\leq R_b$ for every $\omega$, and $\widehat F=F_{\widehat\omega}$. Every label satisfies $\|\theta\|_2\leq r_0$. Let $\widetilde\omega$ be the parameter vector of the constructed network of Theorem~\ref{thm:vs-approx}, which belongs to the class by Lemma~\ref{lem:vs-rrds}(d) for an admissible $\Gamma$, so that $F_{\widetilde\omega}=\widetilde F$. By Lemma~\ref{lem:vs-heads}, the projection leaves the output of $\widetilde\omega$ unchanged for every dataset. With $\widehat P_M$ the empirical measure of the training tasks and $\ell_{\rm tr}$ the loss of Section~\ref{sec:meta-formulation}, the definition~\eqref{eq:approx-erm} of the approximate empirical risk minimizer gives
\begin{equation}
\widehat P_M\ell_{\rm tr}(F_{\widehat\omega})\leq\widehat P_M\ell_{\rm tr}(F_{\widetilde\omega})+\varepsilon_{\rm opt}.
\label{eq:vs-empirical-comparison}
\end{equation}

For a task $(\theta,S,X^n)$, let $h_\omega=\ell_{\rm tr}\{F_\omega(X^n),(\theta,S)\}-\ell_{\rm tr}\{(\theta^*,q^*)(X^n),(\theta,S)\}$, so $a_\omega=\mb Eh_\omega=\mathcal R(F_\omega)-\mathcal R^*$ by~\eqref{eq:vs-excess} and~\eqref{eq:vs-empirical-comparison} reads $\widehat P_Mh_{\widehat\omega}\leq\widehat P_Mh_{\widetilde\omega}+\varepsilon_{\rm opt}$. Write $h_\omega=G+K$ as in Lemma~\ref{lem:excess-moments}(b) with $v=\Pi_{\rm b}v_\omega$. Since $\|\Pi_{\rm b}v_\omega-\theta\|_2\leq R_b+r_0$ and $\|\theta^*-\theta\|_2\leq2r_0\leq R_b+r_0$, both squared distances are at most $b_1=(R_b+r_0)^2$. Moreover, $q_{\omega,j}\in[\epsilon_n,1-\epsilon_n]=[e^{-b},1-e^{-b}]$ with $b=\log(1+np)$, where $1\leq b\leq\Lambda_n$. Lemma~\ref{lem:excess-moments}(b) therefore gives
\[
\mb E[|h_\omega|^m]\leq5\,m!\,c_\lambda^{m-1}a_\omega\quad(m\geq2),\qquad c_\lambda=2\max\{b_1,\lambda b\}\leq C(D+\lambda\Lambda_n),
\]
where $c_\lambda\leq C(s_{\max}+\lambda\Lambda_n)$ with $C$ free of $D$, because $b_1\leq(2+\sqrt2)^2B^2s_{\max}$.

Let $L_z$ be the constant of Lemma~\ref{lem:rrds-lipschitz} for this architecture. Since the logistic function is $1/4$-Lipschitz and clipping is nonexpansive, coordinate $0$ of $v_\omega$ is $L_z$-Lipschitz in $\omega$, and for $j\in[p]$, with $b_j,b_j'$ the state coordinates at $\omega,\omega'$, we have $|q_{\omega,j}\operatorname{clip}_B(b_j)-q_{\omega',j}\operatorname{clip}_B(b_j')|\leq|b_j-b_j'|+B|q_{\omega,j}-q_{\omega',j}|\leq(1+B/4)L_z\|\omega-\omega'\|_2$. Taking the Euclidean norm over the $p+1$ coordinates and using that $\Pi_{\rm b}$ is nonexpansive gives $\|\Pi_{\rm b}v_\omega-\Pi_{\rm b}v_{\omega'}\|_2\leq(1+B/4)\sqrt{p+1}\,L_z\|\omega-\omega'\|_2$. In $h_\omega$, the term of the Bayes predictor does not depend on $\omega$, the squared-error part changes by at most $2(R_b+r_0)\|\Pi_{\rm b}v_\omega-\Pi_{\rm b}v_{\omega'}\|_2$ because both vectors lie within $R_b+r_0$ of $\theta$, and the cross-entropy part is $\lambda/p$ times a sum of $p$ terms, each $1$-Lipschitz in a logit that is $L_z$-Lipschitz in $\omega$. Hence $|h_\omega-h_{\omega'}|\leq L_f\|\omega-\omega'\|_2$ with
\[
\begin{gathered}
L_f=\bigl\{2(R_b+r_0)(1+B/4)\sqrt{p+1}+\lambda\bigr\}L_z,\\
\frac{L_f}{c_\lambda}\leq\Bigl\{\frac{(1+B/4)\sqrt{p+1}}{R_b+r_0}+\frac1{2b}\Bigr\}L_z\leq C\sqrt p\,L_z .
\end{gathered}
\]
Here the first bound on $L_f/c_\lambda$ uses $c_\lambda\geq2b_1=2(R_b+r_0)^2$ for the first part and $c_\lambda\geq2\lambda b$ for the second, and the last inequality uses $R_b+r_0\geq2B$ and $b\geq1$.

By Lemma~\ref{lem:rrds-lipschitz} and Lemma~\ref{lem:vs-rrds}(d), $\log(1+\Gamma L_z)\leq C(1+L)\log\{e(W+1)(\Gamma+1)\}\leq CL\log(eW\Gamma)$, since the observations are bounded, every affine row has at most $CW$ coefficient positions, and $L,W,\Gamma\geq1$. Since $L_z\geq1$, we have $1+2\Gamma M(1+L_f/c_\lambda)\leq2eM(1+C\sqrt p)(1+\Gamma L_z)$, and $W\geq p$ then gives $\log\{1+2\Gamma M(1+L_f/c_\lambda)\}\leq\log(2eM)+\log(1+C\sqrt p)+\log(1+\Gamma L_z)\leq C\{L\log(eW\Gamma)+\log(eM)\}$. Lemma~\ref{lem:bernstein-erm} with $\kappa=5$ (so $8\kappa+16=56$), $c=c_\lambda$ and $\omega_0=\widetilde\omega$ gives, on an event of probability at least $1-\delta$,
\[
\mathcal R(\widehat F)-\mathcal R^*\leq3\{\mathcal R(\widetilde F)-\mathcal R^*\}+56\,c_\lambda\,\frac{W\log\{1+2\Gamma M(1+L_f/c_\lambda)\}+\log(2/\delta)}M+2\varepsilon_{\rm opt},
\]
which is at most $3\{\mathcal R(\widetilde F)-\mathcal R^*\}+C\Delta_{\rm vs}$ by~\eqref{eq:sparse-complexity-training}, using $c_\lambda\leq C(s_{\max}+\lambda\Lambda_n)$, the bound on the logarithm and $\log(2/\delta)\leq\log(3/\delta)$. Here $\mathcal R(\widehat F)-\mathcal R^*=a_{\widehat\omega}$ is the excess risk conditional on $\mathcal D_M$, because the new task is independent of $\mathcal D_M$.

By~\eqref{eq:vs-excess}, $\mathcal R(\widetilde F)-\mathcal R^*=\mb E_{\mb Q_n}[\|\widetilde\theta-\theta^*\|_2^2]+(\lambda/p)\sum_{j=1}^p\mb E_{\mb Q_n}[D_{\rm KL}(q_j^*\|\widetilde q_j)]$. By Theorem~\ref{thm:vs-approx}(a) and Lemma~\ref{lem:ce-selection}(c), $\mb E_{\mb Q_n}[\|\widetilde\theta-\theta^*\|_2^2]\leq\mb E_{\mb Q_n}[\|\widetilde\theta-\theta\|_2^2]\leq Cs_{\max}/n$, and the sum is at most $CJ_n$ by~\eqref{eq:vs-bridge-general} with $Q=\widetilde Q$, so $\mathcal R(\widetilde F)-\mathcal R^*\leq C\{s_{\max}n^{-1}+(\lambda/p)J_n\}$. Since $J_n\leq Ch_n$ by Proposition~\ref{prop:vs-bridge}, this proves~\eqref{eq:vs-trained-excess}.

The remaining bounds hold on the same event. By Lemma~\ref{lem:ce-selection}(c), conditionally on $\mathcal D_M$,
\[
\mb E\bigl[\|\widehat\theta-\theta\|_2^2\mid\mathcal D_M\bigr]=\mb E\bigl[\|\widehat\theta-\theta^*\|_2^2\mid\mathcal D_M\bigr]+\mb E_{\mb Q_n}[\|\theta^*-\theta\|_2^2]\leq\{\mathcal R(\widehat F)-\mathcal R^*\}+\mb E_{\mb Q_n}[\|\widetilde\theta-\theta\|_2^2],
\]
using $\mb E[\|\widehat\theta-\theta^*\|_2^2\mid\mathcal D_M]\leq\mathcal R(\widehat F)-\mathcal R^*$ from~\eqref{eq:vs-excess} and $\mb E_{\mb Q_n}[\|\theta^*-\theta\|_2^2]\leq\mb E_{\mb Q_n}[\|\widetilde\theta-\theta\|_2^2]$ from Lemma~\ref{lem:ce-selection}(c). Together with~\eqref{eq:vs-trained-excess}, this gives~\eqref{eq:vs-trained-mse}. The second term of~\eqref{eq:vs-excess} gives the additional bound
\begin{equation}
\sum_j\mb E_M[D_{\rm KL}(q_j^*\|\widehat q_j)]\leq(p/\lambda)\{\mathcal R(\widehat F)-\mathcal R^*\}\leq C\{J_n+s_{\max}p/(\lambda n)+p\Delta_{\rm vs}/\lambda\}.
\label{eq:vs-trained-kl}
\end{equation}
Finally, Lemma~\ref{lem:ce-selection}(b), applied conditionally on $\mathcal D_M$, gives
\[
\mb E[H\mid\mathcal D_M]\leq5H_n^*+(8p/\lambda)\{\mathcal R(\widehat F)-\mathcal R^*\},
\]
and $H_n^*\leq J_n{}\leq Ch_n$ by the definition of $J_n$. This proves~\eqref{eq:vs-trained-hamming}.

\end{proof}

\section{Proofs of the Technical Lemmas}\label{app:technical}

This appendix gives the proofs of the technical lemmas stated in Appendix~\ref{app:proofs}, in the same order of inference tasks. These results justify the likelihood expansions, probability comparisons, and network bounds used in the proofs of the main theorems.

\subsection{Proofs of the lemmas for parameter estimation}\label{app:proof-est-lemmas}

The proofs use the setting of Appendix~\ref{app:proofs-est} and the tools of Appendix~\ref{app:tools}.

\subsubsection{Regularity and localization (Lemma~\ref{lem:consequences})}

\begin{proof}
In this proof, $C$ and $c$ depend only on $d$, $d_X$, $r$, $\ell$, $T$, $\Theta_0$ and, for the statements at interior parameters, $\Theta_1$. Constants that depend in addition on a compact set $K$ or a radius $a$ carry it as an index. We divide the proof into five steps.

\medskip
\noindent{\bf Step 1: Regularity and identifiability.} Fix $K\Subset\operatorname{int}(K^+)\subset K^+\Subset U$, and let $\nu_{\mathcal X}$ be the finite dominating measure. Since $(x,\theta)\mapsto\ell_\theta(x)$ has a $C^{r+1}$ extension to a neighborhood of $\mathcal X\times U$, each $(x,\theta)\mapsto D_\theta^\nu p_\theta(x)$ with $|\nu|\leq\lfloor r+1\rfloor$ is continuous on the compact set $\mathcal X\times K^+$. Since $T$ is bounded, the envelope $\sup_{\theta\in K^+}\max_{|\nu|\leq\lfloor r+1\rfloor}\{1+\|T(x)\|_2\}|D_\theta^\nu p_\theta(x)|$ is bounded, and hence integrable because $\nu_{\mathcal X}$ is finite. Differentiation under the integral sign gives $D^\nu\mu(\theta)=\int TD_\theta^\nu p_\theta\,d\nu_{\mathcal X}$ for $|\nu|\leq\lfloor r+1\rfloor$. If $r+1$ is not an integer, the H\"older bound on the top derivatives of $p_\theta$ on $K$ integrates to the same bound for $\mu$. If $r+1$ is an integer, dominated convergence gives continuity of the top derivatives. Hence $\mu\in C^{r+1}(U)$. Since $s_\theta=\nabla_\theta\ell_\theta$ has a joint $C^r$ extension, the same argument applied to $I(\theta)=\int s_\theta s_\theta^\top p_\theta\,d\nu_{\mathcal X}$ gives $I\in C^r(U)$. The bounds on the derivatives of $s_\theta$ follow from this joint extension on the compact set $\mathcal X\times\Theta_0$. Since $\nabla_\theta p_\theta=s_\theta p_\theta$ and $\nabla_\theta^2 p_\theta=(\nabla_\theta^2\ell_\theta+s_\theta s_\theta^\top)p_\theta$, differentiating $\int p_\theta\,d\nu_{\mathcal X}=1$ once gives $\int\nabla_\theta p_\theta\,d\nu_{\mathcal X}=P_\theta s_\theta=0$, and differentiating it twice gives $P_\theta\{\nabla_\theta^2\ell_\theta+s_\theta s_\theta^\top\}=0$. Using $P_\theta s_\theta=0$ in $D\mu(\theta)=\int T(\nabla_\theta p_\theta)^\top\,d\nu_{\mathcal X}=P_\theta[Ts_\theta^\top]$, we obtain
\[
P_\theta s_\theta=  0,\qquad -P_\theta\nabla^2\ell_\theta=  P_\theta s_\theta s_\theta^\top=  I(\theta),\qquad D\mu(\theta)=  P_\theta[(T-\mu(\theta))s_\theta^\top].
\]
If $v^\top I(\theta)v=P_\theta(s_\theta^\top v)^2=0$, then $s_\theta^\top v=0$ $\mb P_\theta$-almost surely, so $D\mu(\theta)v=P_\theta[(T-\mu(\theta))s_\theta^\top v]=0$, and hence $v=0$ because $D\mu(\theta)$ is nonsingular. Thus $I(\theta)\in\mathbb S_{++}^d$ on $U$, and $\lambda_0=\inf_{\Theta_0}\lambda_{\min}(I)>0$ by compactness and continuity of $I$. If $\mb P_\theta=\mb P_\eta$ for $\theta,\eta\in U$, then $\mu(\theta)=\mu(\eta)$ and hence $\theta=\eta$ because $\mu$ is one-to-one on $U$.

\medskip
\noindent{\bf Step 2: Quadratic KL bound.} Fix $\theta\in\Theta_0$ and put $k(\eta)=D_{\rm KL}(\mb P_\theta\|\mb P_\eta)=P_\theta(\ell_\theta-\ell_\eta)$ for $\eta\in\Theta_0$. By the uniform bound on $s_\eta$ and its derivatives up to order two on $\mathcal X\times\Theta_0$ from Step~1, differentiating under $P_\theta$ gives
\[
k(\theta)=0,\qquad \nabla k(\theta)=-P_\theta s_\theta=0,\qquad \nabla^2 k(\theta)=-P_\theta\nabla^2\ell_\theta=I(\theta),
\]
where the last two equalities also follow from Step~1. By the same bound on the second derivatives of the score, the third derivatives of $k$ are bounded on $\Theta_0$ uniformly in $\theta$. Since $\Theta_0$ is convex, Taylor's formula around $\theta$ gives a constant $C>0$, independent of $\theta$ and $\eta$, such that for all $\eta\in\Theta_0$,
\[
|k(\eta)-\tfrac12(\eta-\theta)^\top I(\theta)(\eta-\theta)|\leq C\|\eta-\theta\|_2^3.
\]
Put $a_0=\min\{\lambda_0/(4C),D_0\}>0$. Since $I(\theta)\succeq\lambda_0I_d$ on $\Theta_0$ by Step~1, whenever $\theta,\eta\in\Theta_0$ satisfy $\|\eta-\theta\|_2\leq a_0$ we have
\[
\begin{aligned}
D_{\rm KL}(\mb P_\theta\|\mb P_\eta)
&\geq\tfrac12(\eta-\theta)^\top I(\theta)(\eta-\theta)-C\|\eta-\theta\|_2^3\\
&\geq\tfrac12\lambda_0\|\eta-\theta\|_2^2-C\|\eta-\theta\|_2^3
\geq\tfrac14\lambda_0\|\eta-\theta\|_2^2.
\end{aligned}
\]
On the compact set $\{(\theta,\eta)\in\Theta_0\times\Theta_0:\|\eta-\theta\|_2\geq a_0\}$, the map $(\theta,\eta)\mapsto D_{\rm KL}(\mb P_\theta\|\mb P_\eta)$ is continuous by dominated convergence and positive because $\mb P_\theta\neq\mb P_\eta$ by the identifiability in Step~1, so it has a minimum $c_1>0$. Since $\|\eta-\theta\|_2\leq D_0$, for all $\theta,\eta\in\Theta_0$ with $\|\eta-\theta\|_2\geq a_0$ we have
\[
D_{\rm KL}(\mb P_\theta\|\mb P_\eta)\geq c_1\geq\frac{c_1}{D_0^2}\|\eta-\theta\|_2^2.
\]
Hence $D_{\rm KL}(\mb P_\theta\|\mb P_\eta)\geq c\|\eta-\theta\|_2^2$ for all $\theta,\eta\in\Theta_0$, with $c=\min\{\lambda_0/4,c_1/D_0^2\}>0$.

\medskip
\noindent{\bf Step 3: Localization.} Fix $a>0$ and $\theta\in\Theta_0$. For every $\eta\in\argmax_{\Theta_0}\ell_n$ with $\|\eta-\theta\|_2\geq a$, the maximality of $\eta$ over $\Theta_0$ gives $P_n\ell_\eta\geq P_n\ell_\theta$, so $P_n(\ell_\eta-\ell_\theta)\geq 0$. By the triangle inequality and Step~2,
\[
\begin{aligned}
2\sup_{z\in\Theta_0}|(P_n-P_\theta)\ell_z|
&\geq |(P_n-P_\theta)\ell_\eta|+|(P_n-P_\theta)\ell_\theta|\geq (P_n-P_\theta)(\ell_\eta-\ell_\theta)
\geq -P_\theta(\ell_\eta-\ell_\theta)\\
&= D_{\rm KL}(\mb P_\theta\|\mb P_\eta)
\geq c\|\eta-\theta\|_2^2\geq ca^2.
\end{aligned}
\]
Hence the event $\{\sup_{\eta\in\argmax_{\Theta_0}\ell_n}\|\eta-\theta\|_2>a\}$ is contained in $\{\sup_{z\in\Theta_0}|(P_n-P_\theta)\ell_z|\geq ca^2/2\}$, which is contained in $\{\sup_{z\in\Theta_0}|(P_n-P_\theta)\ell_z|>ca^2/4\}$. We have $\Theta_0\subseteq B_2(\theta,D_0)$. Let $B_\ell=\sup_{z\in\Theta_0,\,x\in\mathcal X}|\ell_z(x)|$, and let $L_\ell$ be the Lipschitz constant of $z\mapsto\ell_z(x)$ on $\Theta_0$, uniformly in $x\in\mathcal X$. Take $u=ca^2/4$ and a fixed mesh $0<h<ca^2/(16L_\ell)$, so that $u-2L_\ell h>ca^2/8$. Lemma~\ref{lem:net-concentration}, applied to the single class $\{\ell_z:z\in\Theta_0\}$ with radius $D_0$, envelope $B_\ell$ and Lipschitz constant $L_\ell$, then gives
\[
\mb P_\theta\Bigl\{\sup_{z\in\Theta_0}|(P_n-P_\theta)\ell_z|>\tfrac14ca^2\Bigr\} \leq 2\Bigl(1+\frac{2D_0}{h}\Bigr)^d\exp\Bigl\{-\frac{nc^2a^4}{128B_\ell^2}\Bigr\}=C_ae^{-c_an},
\]
where $C_a$ and $c_a$ do not depend on $\theta$. By the event inclusion above, taking the supremum over $\theta\in\Theta_0$ yields
\[
\sup_{\theta\in\Theta_0}\mb P_\theta\Bigl\{\sup_{\eta\in\argmax_{\Theta_0}\ell_n}\|\eta-\theta\|_2>a\Bigr\}\leq C_ae^{-c_an}.
\]

\medskip
\noindent{\bf Step 4: First-order expansion.} All bounds in this step are uniform over $\theta\in\Theta_1$. Choose $b>0$ such that $B(\theta,2b)\subset\operatorname{int}(\Theta_0)$ for $\theta\in\Theta_1$ and $\sup_{z\in B(\theta,2b)}\|H_n(z)-H_n(\theta)\|_{\rm op}\leq\lambda_0/4$, which is possible because $H_n$ is Lipschitz on $\Theta_0$ uniformly in $n$ and the data. Let
$\mathcal H_n^{\rm loc}(\theta)=  \{ \argmax_{\Theta_0}\ell_n\subset B(\theta,b)$, $\|H_n(\theta)-I(\theta)\|_{\rm op}\leq  \lambda_0/4\}$.
On this event, every $z\in B(\theta,2b)$ satisfies
\[
\begin{aligned}
\lambda_{\min}\{H_n(z)\}
&\geq\lambda_{\min}\{I(\theta)\}-\|H_n(\theta)-I(\theta)\|_{\rm op}-\|H_n(z)-H_n(\theta)\|_{\rm op}\\
&\geq\lambda_0-\tfrac14\lambda_0-\tfrac14\lambda_0=\tfrac12\lambda_0.
\end{aligned}
\]
Hence $\ell_n$ is strictly concave on $B(\theta,2b)$, which contains all maximizers. The maximizer is therefore unique and interior, $P_ns_{\widehat\theta_{0,n}}=0$ and $H_n(\widehat\theta_{0,n})\succeq\tfrac12\lambda_0I_d$, so $\mathcal H_n^{\rm loc}(\theta)\subseteq\mathcal H_n$. With $e_n=\widehat\theta_{0,n}-\theta$, Taylor's formula gives $0=P_ns_{\widehat\theta_{0,n}}=P_ns_\theta-A_ne_n$, that is,
\[
 A_ne_n=P_ns_\theta,\qquad
 A_n=\int_0^1H_n(\theta+te_n)\,dt\succeq\tfrac12\lambda_0I_d,
 \qquad\|e_n\|_2\leq2\lambda_0^{-1}\|P_ns_\theta\|_2.
\]

The complement of $\mathcal H_n^{\rm loc}(\theta)$ is contained in the union of the events $\{\argmax_{\Theta_0}\ell_n\not\subset B(\theta,b)\}$ and $\{\|H_n(\theta)-I(\theta)\|_{\rm op}>\lambda_0/4\}$. The first event has probability at most $C_be^{-c_bn}$ by Step~3 with $a=b$. For the second, the entries of $H_n(\theta)-I(\theta)$ are averages of $n$ i.i.d.\ centered bounded terms and $\|A\|_{\rm op}\leq d\max_{jk}|A_{jk}|$, so Hoeffding's inequality \citep[Proposition~2.5 and (2.11)]{wainwright2019high}, applied to the $d^2$ entries at threshold $\lambda_0/(4d)$, gives an exponential bound. Hence $\sup_{\theta\in\Theta_1}\mb P_\theta\{\mathcal H_n^{\rm loc}(\theta)^c\}\leq Ce^{-cn}$, which proves the bound on $\mb P_\theta(\mathcal H_n^c)$. We have $\|e_n\|_2^4\leq16\lambda_0^{-4}\|P_ns_\theta\|_2^4$ on $\mathcal H_n^{\rm loc}(\theta)$ and $\|e_n\|_2\leq D_0$ always, and $P_ns_\theta$ is an average of i.i.d.\ centered bounded vectors. By~\eqref{eq:average-fourth-moment},
$\mb E_\theta^{(n)}[\|e_n\|_2^4] \leq  16\lambda_0^{-4}\mb E_\theta^{(n)}[\|P_ns_\theta\|_2^4] +CD_0^4e^{-cn}\leq   Cn^{-2}$.

Put $R_n=e_n-I(\theta)^{-1}P_ns_\theta$ for every dataset. On $\mathcal H_n^{\rm loc}(\theta)$, $e_n=A_n^{-1}P_ns_\theta$, so
\[
R_n=\{A_n^{-1}-I(\theta)^{-1}\}P_ns_\theta=A_n^{-1}\{I(\theta)-A_n\}I(\theta)^{-1}P_ns_\theta.
\]
The Lipschitz bound on $H_n$ and the bound on $\|e_n\|_2$ give $\|I(\theta)-A_n\|_{\rm op}\leq\|I(\theta)-H_n(\theta)\|_{\rm op}+C\|e_n\|_2\leq\|H_n(\theta)-I(\theta)\|_{\rm op}+C\|P_ns_\theta\|_2$. Since $\|A_n^{-1}\|_{\rm op}\leq2/\lambda_0$ and $\|I(\theta)^{-1}\|_{\rm op}\leq1/\lambda_0$, we have $\|R_n\|_2\leq C\{\|H_n(\theta)-I(\theta)\|_{\rm op}\|P_ns_\theta\|_2+\|P_ns_\theta\|_2^2\}$ on $\mathcal H_n^{\rm loc}(\theta)$. On the complement, $\|R_n\|_2\leq D_0+\lambda_0^{-1}B_s$. By the Cauchy--Schwarz inequality and~\eqref{eq:average-fourth-moment}, applied to $s_\theta$ and to the entries of $\nabla^2\ell_\theta$,
\[
\begin{aligned}
&\mb E_\theta^{(n)}\bigl[\|H_n(\theta)-I(\theta)\|_{\rm op}^2\|P_ns_\theta\|_2^2\bigr]\leq\bigl(\mb E_\theta^{(n)}[\|H_n(\theta)-I(\theta)\|_{\rm op}^4]\bigr)^{1/2}
\bigl(\mb E_\theta^{(n)}[\|P_ns_\theta\|_2^4]\bigr)^{1/2}\leq Cn^{-2},
\end{aligned}
\]
and the complement contributes at most $Ce^{-cn}$. Hence $\mb E_\theta^{(n)}[\|R_n\|_2^2]\leq Cn^{-2}$, which completes the proof of~\eqref{eq:uniform-mle-expansion}. In particular, $\sqrt nR_n\to0$ in $\mb P_\theta$-probability, and the central limit theorem \citep[Proposition~2.27]{van2000asymptotic} for the bounded score gives
\[
\sqrt n(\widehat\theta_{0,n}-\theta) =  I(\theta)^{-1}n^{-1/2}\sum_i s_\theta(X_i)+o_{\mb P_\theta}(1) \xrightarrow{d}\mathcal N\{0,I(\theta)^{-1}\}.
\]

\medskip
\noindent{\bf Step 5: Equivalence to the unrestricted MLE.} Let $\widehat\theta_{\rm MLE}$ be a measurable estimator completed by a fixed $\theta_{\rm fix}\in\Theta$ outside a measurable event $G_n$. Suppose
\[
\begin{gathered}
\widehat\theta_{\rm MLE}\in\argmax_{\eta\in\Theta}\ell_n(\eta)\quad\text{on }G_n,\quad \widehat\theta_{\rm MLE}=\theta_{\rm fix}\quad\text{on }G_n^c,\quad \mathbb P_\theta(G_n)\longrightarrow1,\quad
\widehat\theta_{\rm MLE}\xrightarrow{\mb P_\theta}\theta.
\end{gathered}
\]
On $G_n$, $\widehat\theta_{\rm MLE}$ maximizes $\ell_n$ over $\Theta$. If also $\widehat\theta_{\rm MLE}\in B(\theta,b)\subset\Theta_0\subset\Theta$, then $\widehat\theta_{\rm MLE}$ maximizes $\ell_n$ over $\Theta_0$. On $\mathcal H_n^{\rm loc}(\theta)$ this maximizer is unique, so $\widehat\theta_{\rm MLE}=\widehat\theta_{0,n}$. Hence
\[
\mathcal H_n^{\rm loc}(\theta)\cap G_n\cap\{\widehat\theta_{\rm MLE}\in B(\theta,b)\}\subseteq\{\widehat\theta_{\rm MLE}=\widehat\theta_{0,n}\}.
\]
Taking complements and applying the union bound gives
\[
\begin{aligned}
\mb P_\theta(\widehat\theta_{\rm MLE}\ne\widehat\theta_{0,n})
&\leq\mb P_\theta\{\mathcal H_n^{\rm loc}(\theta)^c\}+\mb P_\theta(G_n^c)+\mb P_\theta\{G_n,\|\widehat\theta_{\rm MLE}-\theta\|_2\geq b\}\\
&\leq Ce^{-cn}+\mb P_\theta(G_n^c)+\mb P_\theta\{G_n,\|\widehat\theta_{\rm MLE}-\theta\|_2\geq b\}\to0,
\end{aligned}
\]
where the second inequality uses Step~4 and the last two terms tend to zero since $\mb P_\theta(G_n)\to1$ and $\widehat\theta_{\rm MLE}\xrightarrow{\mb P_\theta}\theta$.
\end{proof}

\subsubsection{Lipschitz continuity in the network parameters (Lemma~\ref{lem:param-lipschitz})}

\begin{proof}
Let $C\geq2$ bound the number of times a coordinate is reused by the fixed connections and the magnitude of the input coordinates. It depends only on $d$, $d_X$, $T$, $\Theta_0$ and the fixed connections. The number $q_n$ of retained coefficient positions in each module then satisfies $q_n\leq N_{\rm pos}\leq CW$. Put $A=C(q_n+1)(\Gamma+1)\geq2$ and $e=\|\omega-\omega'\|_2$, and let $B_j$ and $E_j$ bound, uniformly over admissible inputs and over all $\omega,\omega'$ in the parameter ball, the largest absolute coordinate at layer $j$ and the largest coordinate difference between $\omega$ and $\omega'$. Write an affine row at layer $j+1$ as $x\mapsto\sum_k w_k x_k+w_0$, which has at most $q_n$ retained positions and therefore at most $Cq_n$ terms in the sum. Every coefficient satisfies $|w_k|,|w_0|\leq\Gamma$, for trainable coordinates because $\|\omega\|_2\leq\Gamma$ and for fixed ones by assumption. Since ReLU is nonexpansive and vanishes at $0$, we have $B_{j+1}\leq Cq_n\Gamma B_j+\Gamma$, so
\[
B_{j+1}+1\leq Cq_n\Gamma B_j+\Gamma+1\leq C(q_n+1)(\Gamma+1)(B_j+1)=A(B_j+1).
\]
For $\omega$ and $\omega'$, with layer-$j$ inputs $x,x'$ and row coefficients $(w_k,w_0)$ and $(w_k',w_0')$, nonexpansiveness of ReLU and the triangle inequality give
\[
\begin{aligned}
&\Bigl|\sum_k w_k x_k+w_0-\Bigl(\sum_k w_k' x_k'+w_0'\Bigr)\Bigr|\\
&\qquad\leq\sum_k |w_k||x_k-x_k'|+\sum_k |w_k-w_k'||x_k'|+|w_0-w_0'| \leq Cq_n\Gamma E_j+Cq_n e B_j+e,
\end{aligned}
\]
since each coefficient difference, including repeated occurrences of a shared coefficient, is at most $e$. Because $Cq_n\Gamma\leq A$ and $Cq_n B_j+1\leq C(q_n+1)(B_j+1)\leq A(B_j+1)$, this yields $E_{j+1}\leq AE_j+A(B_j+1)e$. By induction from $B_0$ and $E_0=0$, we obtain $B_l+1\leq A^l(B_0+1)$ and $E_l\leq lA^l(B_0+1)e$. Indeed, if $E_l\leq lA^l(B_0+1)e$, then
\[
E_{l+1}\leq A\bigl\{lA^l(B_0+1)e\bigr\}+A\bigl\{A^l(B_0+1)\bigr\}e=(l+1)A^{l+1}(B_0+1)e.
\]

With unchanged coefficients ($e=0$), the same recurrence gives $\Delta_{j+1}\leq A\Delta_j$ for the largest coordinate difference $\Delta_j$ at layer $j$, so each module of depth at most $L_n$ has sup-norm input Lipschitz constant at most $A^{L_n}$. We apply these bounds successively to the initialization, score and update modules, using the triangle inequality to separate input and parameter perturbations. The initial projection $\Pi_P$ has a fixed bounded range and is nonexpansive, and the input bound of the update module includes the magnitude bound of the score module at the given parameters. Averaging the shared score module over the observations increases neither the sup norm of its output nor that of its parameter perturbation. Since the dimensions and the number of serial modules are fixed, there are fixed integers $a_1,a_2\geq0$ such that
\[
\sup_{x_{1:n}}\|F_\omega(x_{1:n})-F_{\omega'}(x_{1:n})\|_2\leq C(L_n+1)^{a_1}A^{a_2L_n}e,
\]
where $C$, $a_1$ and $a_2$ do not depend on $n$, $W$, $\Gamma$ or the parameters.

The final projection and $\theta\in\Theta_0$ give $|\|F_\omega-\theta\|_2^2-\|F_{\omega'}-\theta\|_2^2|\leq2D_0\|F_\omega-F_{\omega'}\|_2$, so we may take $L_{f,n}=2D_0C(L_n+1)^{a_1}A^{a_2L_n}$. Since $q_n\leq CW$ and $\Gamma\geq1$, we have $(W+1)(\Gamma+1)\geq2$, so $\log A\leq C\log\{(W+1)(\Gamma+1)\}$ after enlarging $C$. Since $1+\Gamma L_{f,n}\leq(\Gamma+1)(1+L_{f,n})$ and $(L_n+1)^{a_1}A^{a_2L_n}\geq1$, enlarging $C$ again gives
\[
\begin{aligned}
\log(1+\Gamma L_{f,n})
&\leq\log(\Gamma+1)+\log(1+L_{f,n})\\
&\leq C\bigl\{1+\log(\Gamma+1)+\log(L_n+1)+L_n\log A\bigr\}\\
&\leq C\bigl\{1+\log(L_n+1)+(L_n+1)\log[(W+1)(\Gamma+1)]\bigr\}\leq C\Xi_n,
\end{aligned}
\]
where the last inequality uses $1+\log(L_n+1)\leq 1+L_n$ and $1+\log[(W+1)(\Gamma+1)]=\log\{e(W+1)(\Gamma+1)\}$.

For the finite-smoothness architecture in Theorem~\ref{thm:est-gen} at accuracy $\varepsilon_n\geq n^{-1}$, we have $L_n\leq C\log(en)$, $W\leq Cn^\kappa\log(en)$ and $\Gamma\leq c_\Gamma n^{c_\Gamma}$, so $\log\{e(W+1)(\Gamma+1)\}\leq C\log(en)$ and $\Xi_n\leq C\log^2(en)$, which gives~\eqref{eq:parameter-lipschitz-log}. For the analytic architecture at accuracy $n^{-1}$ with $\Gamma\leq Cn^C$, we have $L_n\leq C\log^2(en)$ and $W\leq C\log^{d_X+d+1}(en)$, so $\log\{e(W+1)(\Gamma+1)\}\leq C\log(en)$ and $\Xi_n\leq C\log^3(en)$, which gives~\eqref{eq:analytic-parameter-lipschitz-log}. In these two bounds, $C$ depends in addition on the constants of the complexity bounds, the radius bound and, in the analytic case, the extension constants.
\end{proof}

\subsubsection{Bootstrap expansion of the MLE (Lemma~\ref{lem:boot-mle})}

\begin{proof}
Fix a compact $\Theta_1\Subset\operatorname{int}(\Theta_0)$. All convergence and order statements below are uniform over $\theta\in\Theta_1$, and we write $P_\theta H(z)=-P_\theta\nabla_z^2\log p_z$ for the population version of $H_n(z)$ and $H_n^*(z)$. For each bounded Lipschitz family $\{f_z:z\in\Theta_0\}$ with envelope $B_f$ and Lipschitz constant $L_f$, used for likelihoods or entries of $-\nabla_z^2\log p_z$, the conditional form of Lemma~\ref{lem:net-concentration} given $X^n$ gives, with $N(h)=(1+2D_0/h)^d$,
$P^*\!\{\sup_z|(P_n^*-P_n)f_z|>t+2L_fh\} \leq 2N(h)\exp\{-nt^2/(2B_f^2)\}$.
Taking $h$ and then $t$ small gives $\sup_z|(P_n^*-P_n)f_z|=o_{P^*}(1)$, while Lemma~\ref{lem:net-concentration} for the sample gives $\sup_z|(P_n-P_\theta)f_z|=o_P(1)$. Since $\sup_z|(P_n^*-P_\theta)f_z|\leq\sup_z|(P_n^*-P_n)f_z|+\sup_z|(P_n-P_\theta)f_z|$, applying this to $\ell_z$ and each entry of $-\nabla_z^2\log p_z$ yields
$\sup_z|(P_n^*-P_\theta)\ell_z|=o_{P^*}(1)$, $\sup_z\|H_n^*(z)-P_\theta H(z)\|_{\rm op}=o_{P^*}(1)$
in probability.

By Lemma~\ref{lem:consequences}, $P_\theta H(\theta)=I(\theta)\succeq\lambda_0I_d$ and $z\mapsto P_\theta H(z)$ is Lipschitz on $\Theta_0$ uniformly in $\theta$. Choose a common $b>0$ with
\[
B(\theta,2b)\subset\operatorname{int}\Theta_0,\qquad \inf_{z\in B(\theta,2b)}\lambda_{\min}(P_\theta H(z))\geq 3\lambda_0/4,\qquad \theta\in\Theta_1.
\]
As in Step~3 of the proof of Lemma~\ref{lem:consequences}, if $\|\widehat\theta_{0,n}^*-\theta\|_2\geq b$, then optimality of $\widehat\theta_{0,n}^*$ and the quadratic KL bound of Lemma~\ref{lem:consequences} give
\[
2\sup_z|(P_n^*-P_\theta)\ell_z|
\geq(P_n^*-P_\theta)(\ell_{\widehat\theta_{0,n}^*}-\ell_\theta) \geq D_{\rm KL}(\mb P_\theta\|\mb P_{\widehat\theta_{0,n}^*})\geq cb^2.
\]
Hence $P^*(\|\widehat\theta_{0,n}^*-\theta\|_2\geq b)\to 0$ in probability. On $\{\sup_z\|H_n^*(z)-P_\theta H(z)\|_{\rm op}\leq\lambda_0/4\}$, the smallest eigenvalue of $H_n^*(z)$ is at least $\lambda_0/2$ on $B(\theta,2b)$, so
\[
P^*\!\{\widehat\theta_{0,n}^*\in B(\theta,b),\ \inf_{z\in B(\theta,2b)}\lambda_{\min}(H_n^*(z))\geq \lambda_0/2\}\longrightarrow 1
\]
in probability. Intersecting with the event $\mathcal H_n$ of Lemma~\ref{lem:consequences}, both maximizers are interior on an event of joint sample/resample probability tending to one, giving
$P_n^*s_{\widehat\theta_{0,n}^*}=0$, $P_ns_{\widehat\theta_{0,n}}=0$.

Boundedness and centering of the score (Lemma~\ref{lem:consequences}) give
\[
\begin{gathered}
E^*P_n^*s_\theta=P_ns_\theta=O_P(n^{-1/2}),\\
E^*\|(P_n^*-P_n)s_\theta\|_2^2=\frac{P_n\|s_\theta\|_2^2-\|P_ns_\theta\|_2^2}{n}=O_P(n^{-1}).
\end{gathered}
\]
Thus $E^*\|P_n^*s_\theta\|_2^2=\|P_ns_\theta\|_2^2+E^*\|(P_n^*-P_n)s_\theta\|_2^2=O_P(n^{-1})$, so $\|P_n^*s_\theta\|_2=O_{P^*}(n^{-1/2})$ in probability. On the localization event above, Taylor's formula gives $0=P_n^*s_{\widehat\theta_{0,n}^*}=P_n^*s_\theta-A_n^*(\widehat\theta_{0,n}^*-\theta)$ with $A_n^*=\int_0^1H_n^*\{\theta+t(\widehat\theta_{0,n}^*-\theta)\}\,dt\succeq(\lambda_0/2)I_d$. Hence $\|(A_n^*)^{-1}\|_{\rm op}\leq 2/\lambda_0$ and $\|\widehat\theta_{0,n}^*-\theta\|_2\leq(2/\lambda_0)\|P_n^*s_\theta\|_2=O_{P^*}(n^{-1/2})$ in probability. By $P_\theta H(\theta)=I(\theta)$ and the uniform Lipschitz bound on $P_\theta H$,
\[
\|A_n^*-I(\theta)\|_{\rm op}\leq\sup_z\|H_n^*(z)-P_\theta H(z)\|_{\rm op}+C\|\widehat\theta_{0,n}^*-\theta\|_2=o_{P^*}(1)
\]
in probability. On the same event, $\widehat\theta_{0,n}^*-\theta=(A_n^*)^{-1}P_n^*s_\theta=I(\theta)^{-1}P_n^*s_\theta+(A_n^*)^{-1}\{I(\theta)-A_n^*\}I(\theta)^{-1}P_n^*s_\theta$, so $\widehat\theta_{0,n}^*-\theta=I(\theta)^{-1}P_n^*s_\theta+o_{P^*}(n^{-1/2})$ in probability. By~\eqref{eq:uniform-mle-expansion}, $\widehat\theta_{0,n}-\theta=I(\theta)^{-1}P_ns_\theta+o_P(n^{-1/2})$.
Subtracting proves~\eqref{eq:bootstrap-mle-expansion}. Given $X^n$, the resampled scores $s_\theta(X_i^*)$ are i.i.d.\ with mean $P_ns_\theta$ and covariance $P_n(s_\theta s_\theta^\top)-(P_ns_\theta)(P_ns_\theta)^\top$. For every $\epsilon>0$, since $\|s_\theta-P_ns_\theta\|_2\leq 2B_s$, the indicator vanishes once $\epsilon\sqrt n>2B_s$, so $P_n\![\|s_\theta-P_ns_\theta\|_2^2 \ind\{\|s_\theta-P_ns_\theta\|_2>\epsilon\sqrt n\}]=0$ for all sufficiently large $n$. Since $P_\theta(s_\theta s_\theta^\top)=I(\theta)$ and $P_\theta s_\theta=0$, the law of large numbers gives $P_n(s_\theta s_\theta^\top)-(P_ns_\theta)(P_ns_\theta)^\top \xrightarrow{P_\theta}I(\theta)$. By the Lindeberg--Feller central limit theorem \citep[Proposition~2.27]{van2000asymptotic}, applied conditionally on $X^n$, the conditional limit of $\sqrt n(P_n^*-P_n)s_\theta$ is $\mathcal N\{0,I(\theta)\}$, so applying $I(\theta)^{-1}$ gives $\mathcal N\{0,I(\theta)^{-1}\}$. By Slutsky's lemma, the $o_{P^*}(1)$ term in~\eqref{eq:bootstrap-mle-expansion} does not change the bounded Lipschitz limit, so $b_n^*(\theta)\xrightarrow{P_\theta}0$.

For the initialization, the same conditional variance calculation gives $E^*\|P_n^*T-P_nT\|_2^2\leq n^{-1}P_n\|T\|_2^2=O_P(n^{-1})$, while boundedness of $T$ and $P_\theta T=\mu(\theta)$ give $P_nT-\mu(\theta)=O_P(n^{-1/2})$. By the triangle inequality,
\[
\|P_n^*T-\mu(\theta)\|_2\leq\|P_n^*T-P_nT\|_2+\|P_nT-\mu(\theta)\|_2=O_{P^*}(n^{-1/2})
\]
in probability. Hence $P_n^*T$ lies with conditional probability tending to one in the neighborhood $\|t-\mu(\theta)\|_2\leq b_T$ of Appendix~\ref{app:proof-approx}, where $t\in\mathcal K$, $g_{\rm MoM}(t)\in\Theta_0$ and $g_{\rm MoM}$ is Lipschitz. Together with $P^*(\widehat\theta_{0,n}^*\in B(\theta,b))\to1$ in probability, this yields $P^*((\mathcal E_n^*)^c)\to0$ in probability. Writing $\theta_m^*=g_{\rm MoM}(P_n^*T)$ on this neighborhood, the Lipschitz bound for $g_{\rm MoM}$ at $\mu(\theta)$, the expansion of $\widehat\theta_{0,n}^*-\theta$ and $P^*((\mathcal E_n^*)^c)\to0$ give
\[
\begin{gathered}
\|\theta_m^*-\theta\|_2=  O_{P^*}(n^{-1/2}),\qquad \rho_n^*\leq  \|\theta_m^*-\theta\|_2+ \|\widehat\theta_{0,n}^*-\theta\|_2=  O_{P^*}(n^{-1/2}),\\
\delta_n^* \leq\|H_n^*(\theta_m^*)-P_\theta H(\theta_m^*)\|_{\rm op} +\|P_\theta H(\theta_m^*)-I(\theta)\|_{\rm op} +\|I(\theta)-I(\theta_m^*)\|_{\rm op}=o_{P^*}(1)
\end{gathered}
\]
in probability. Here the first term in the bound for $\delta_n^*$ is $o_{P^*}(1)$ by the conditional uniform Hessian law, and the other two are $o_{P^*}(1)$ by continuity of $P_\theta H$ and $I$ (with $P_\theta H(\theta)=I(\theta)$) together with $\theta_m^*\to\theta$ in conditional probability.

Finally, the fixed distance from $\Theta_1$ to $\partial\Theta_0$ makes every neighborhood and envelope uniform on $\Theta_1$. The empirical covariance of the score converges to $I(\theta)$ uniformly on $\Theta_1$ by Hoeffding's inequality \citep[Proposition~2.5 and (2.11)]{wainwright2019high} for its $d^2$ entries. If the uniform statement for $b_n^*$ failed along some $\theta_n\in\Theta_1$, a subsequence with $\theta_n\to\theta_\infty$ and almost surely convergent empirical covariances would contradict the conditional central limit theorem along that subsequence, since $I$ is continuous.
\end{proof}

\subsubsection{A ReLU moment function (Lemma~\ref{lem:relu-moment})}

\begin{proof}
Let $\nu$ be the finite dominating measure on $\mathcal X$ and choose $a_0>0$ such that $K^+=\{\theta:\operatorname{dist}(\theta,\Theta_0)\leq a_0\}\subset U$. The set $K^+$ is compact and convex. By the joint $C^{r+1}$ extension of the log density, $C_p=\sup_{x\in\mathcal X,\theta\in K^+}\{p_\theta(x)+\|\nabla_\theta p_\theta(x)\|_2\}<\infty$. For a bounded measurable $G:\mathcal X\to\mb R^d$ put $\mu_G(\theta)=\int G\,p_\theta\,d\nu$. As in the proof of Lemma~\ref{lem:consequences}, $\mu_G\in C^{r+1}$ on a neighborhood of $K^+$ and $D\mu_G(\theta)=\int G\,(\nabla_\theta p_\theta)^\top d\nu$, so that
\[
\sup_{\theta\in K^+}\bigl\{\|\mu_G(\theta)-\mu(\theta)\|_2+\|D\mu_G(\theta)-D\mu(\theta)\|_{\rm op}\bigr\}\leq C_p\,\|G-T\|_{L^1(\nu)}.
\]
Since $\mu$ is one-to-one with a nonsingular Jacobian on $U\supset K^+$ and $K^+$ is compact and convex, there are $\sigma_0,c_0>0$ such that the smallest singular value of $D\mu(\theta)$ is at least $\sigma_0$ and $\|\mu(\theta)-\mu(\theta')\|_2\geq c_0\|\theta-\theta'\|_2$ for $\theta,\theta'\in K^+$. Indeed, the ratio $\|\mu(\theta)-\mu(\theta')\|_2/\|\theta-\theta'\|_2$ is positive and continuous off the diagonal and is at least $\sigma_0/2$ near the diagonal by the uniform continuity of $D\mu$. Let $\epsilon_0=\min\{\sigma_0,c_0\}/(2C_p)$. If $\|G-T\|_{L^1(\nu)}\leq\epsilon_0$, then the displayed bound and the mean value inequality on the convex set $K^+$, applied to $\mu_G-\mu$, show that the smallest singular value of $D\mu_G(\theta)$ is at least $\sigma_0/2$ and
\[
\|\mu_G(\theta)-\mu_G(\theta')\|_2\geq(c_0-C_p\epsilon_0)\|\theta-\theta'\|_2\geq\tfrac12c_0\|\theta-\theta'\|_2,\qquad\theta,\theta'\in K^+,
\]
so $\mu_G$ is one-to-one with a nonsingular Jacobian on $U'=\operatorname{int}(K^+)\supset\Theta_0$.

It remains to find a ReLU network $G=\widetilde T$ with $\|\widetilde T-T\|_{L^1(\nu)}\leq\epsilon_0$. Continuous functions are dense in $L^1(\nu)$ for a finite Borel measure on the compact set $\mathcal X$~\citep[Theorem~3.14, p.~69]{rudin1987real}, so there is a continuous map $T_1$ with $\|T_1-T\|_{L^1(\nu)}\leq\epsilon_0/4$. The restrictions of real polynomials to $\mathcal X$ form an algebra that contains the constants and separates points, so the Stone--Weierstrass theorem \citep[Theorem~7.32, p.~162]{rudin1976principles}, applied to each coordinate, gives a polynomial map $T_2$ with $\sup_{\mathcal X}\|T_2-T_1\|_2\leq\epsilon_0/\{4\nu(\mathcal X)\}$. Hence $\|T_2-T\|_{L^1(\nu)}\leq\epsilon_0/2$. Lemma~\ref{lem:sobolev-approx}, applied to $T_2$ on a rectangle containing $\mathcal X$ with accuracy $\min[1/2,\epsilon_0/\{2\nu(\mathcal X)\}]$, gives a ReLU network $\widetilde T$ with $\|\widetilde T-T_2\|_{L^1(\nu)}\leq\epsilon_0/2$. The map $\widetilde T$ is bounded on $\mathcal X$, and the smoothness part of Assumption~\ref{ass:regular} does not involve the moment function.
\end{proof}

\subsubsection{Lipschitz continuity in the parameters of an RRDS network (Lemma~\ref{lem:rrds-lipschitz})}

\begin{proof}
We follow the proof of Lemma~\ref{lem:param-lipschitz}. Unroll the network for one dataset $x^n$: $\phi_0$ is applied to each $(x_i,0)$, its outputs are averaged, $\rho_0$ is applied, and each refinement applies $\phi$ to each $(x_i,z^{(r-1)})$, averages, applies $\rho$ to $(u^{(r)},z^{(r-1)})$ and adds $z^{(r-1)}$. Let $A=2(N_{\rm pos}+1)(\Gamma+1)$ and $e=\|\omega-\omega'\|_2$. Counting each affine layer and residual addition as one step, let $B_j$ and $E_j$ bound the largest absolute coordinate and coordinate difference between $\omega$ and $\omega'$ produced through step $j$, uniformly over the inputs, the observation index and the parameter ball, so $B_j$ and $E_j$ also bound the carried-forward $x_i$ and $z^{(r-1)}$. Every affine row has at most $N_{\rm pos}$ coefficient positions with magnitudes at most $\Gamma$ and differences at most $e$ (including repeated occurrences of a shared parameter across observations and refinements), and ReLU is nonexpansive and fixes $0$, so splitting each affine difference into weight and input differences bounds the output coordinates in magnitude by $N_{\rm pos}\Gamma B_j+\Gamma$ and in difference by $N_{\rm pos}\Gamma E_j+(N_{\rm pos}+1)(B_j+1)e$, giving
\[
B_{j+1}+1\leq A(B_j+1),\qquad E_{j+1}\leq AE_j+A(B_j+1)e .
\]

Averaging over observations increases neither bound. In each residual addition $z^{(r)}=z^{(r-1)}+\rho(u^{(r)},z^{(r-1)})$, both summands have magnitude at most $B_j$ and difference at most $E_j$, so the new state has magnitude at most $2B_j$ and difference at most $2E_j$. Since $A\geq2$, we have $2B_j+1\leq A(B_j+1)$ and $2E_j\leq AE_j+A(B_j+1)e$, so each of the $R$ residual additions counts as one more step of the same recurrence. By induction from $E_0=0$, as in the proof of Lemma~\ref{lem:param-lipschitz}, $B_j+1\leq A^j(B_0+1)$ and $E_j\leq jA^j(B_0+1)e$. By~\eqref{eq:rrds-complexity}, there are $L$ affine layers and $R$ residual additions, for $L+R$ steps in total, with $B_0\leq R_X$ and $E_0=0$ on the initial inputs $(x_i,0)$. Evaluating at $j=L+R$ gives
\[
\sup_{x^n}\|z_\omega^{(R)}(x^n)-z_{\omega'}^{(R)}(x^n)\|_\infty\leq E_{L+R}\leq(L+R)A^{L+R}(R_X+1)e=L_z\|\omega-\omega'\|_2,
\]
which is the first claim.

For the second claim, for any $a,a'\in\mathbb R^K$ and $k$, the triangle inequality, $\exp(a_m)\leq\exp(a'_m+\|a-a'\|_\infty)$ for all $m$, and symmetry give
\[
\begin{aligned}
&|\log\operatorname{softmax}(a)_k-\log\operatorname{softmax}(a')_k|\\
&\qquad\leq|a_k-a'_k|+\Bigl|\log\sum_{m=1}^K\exp(a_m)-\log\sum_{m=1}^K\exp(a'_m)\Bigr|\leq2\|a-a'\|_\infty .
\end{aligned}
\]
Since $\operatorname{clip}_B$ is $1$-Lipschitz in each coordinate, applying this bound to the clipped versions of the $K$ coordinates of $z_\omega^{(R)}(x^n)$ and $z_{\omega'}^{(R)}(x^n)$ with the first claim gives $|\log q_{\omega,k}(x^n)-\log q_{\omega',k}(x^n)|\leq2L_z\|\omega-\omega'\|_2$ for every $k$ and $x^n$.

For the last claim, since $L_z\geq1$ we have $1+\Gamma L_z\leq(1+\Gamma)L_z$, so
\[
\begin{aligned}
\log(1+\Gamma L_z)
&\leq\log(1+\Gamma)+\log(L+R)+(L+R)\log\{2(N_{\rm pos}+1)(\Gamma+1)\}+\log(R_X+1).
\end{aligned}
\]
When $N_{\rm pos}\leq CW$ and $R_X\leq C$, since $R$ is fixed, each of the four terms is at most a constant multiple of $(1+L)[1+\log\{(W+1)(\Gamma+1)\}]$. This is the last claim, where $C$ depends only on $R$ and the constants in these two bounds.
\end{proof}

\subsection{Proofs of the lemmas for model selection}\label{app:proof-ms-lemmas}

\subsubsection{Likelihood deficit under misspecification (Lemma~\ref{lem:ms-deficit})}

\begin{proof}
The proof of Lemma~\ref{lem:ms-deficit} uses the same one-step comparison as the proof of Theorem~\ref{thm:est-approx}, with the population one-step map at $\theta_{m,k}$ in place of $\widehat\theta_{0,n}$.

In this proof, $C,c$ depend only on the fixed candidate family, its working set and $\mb P_0$, including the interior targets $\theta_{m,k},\theta_k^*$ and the curvature at $\theta_k^*$. They do not depend on $n$ or $\varepsilon$. We suppress $k$ and write $q=d_k$, $\theta_m=\theta_{m,k}$, $\theta^*=\theta_k^*$ and $z=z^{(0)}$.

Let $\mathcal K$ and $P$ be the compact sets of the proof of Theorem~\ref{thm:est-approx} for candidate $k$, so that $\Theta_{0,k}\subseteq P$ and $z=\Pi_Pg_{\rm init}(\bar T_n)\in P$ with $\bar T_n=P_nT$. Since $P_0T=\mu_k(\theta_m)\in\mu_k(\Theta_{0,k})\subset\operatorname{int}(\mathcal K)$, there is $b_T>0$ with $B(P_0T,b_T)\subset\mathcal K$. Put $\mathcal M_n=\{\|P_nT-P_0T\|_2\leq b_T\}$. If $\|P_nT-P_0T\|_2>b_T$, then some coordinate of $P_nT-P_0T$ exceeds $b_T/\sqrt q$ in absolute value. Since the coordinates of $T$ are bounded, Hoeffding's inequality \citep[Proposition~2.5 and (2.11)]{wainwright2019high}, applied to the $q$ coordinates at threshold $b_T/\sqrt q$, gives $P_0(\mathcal M_n^c)\leq Ce^{-cn}$. On $\mathcal M_n$, we have $\bar T_n\in\mathcal K$. Since $\theta_m\in\Theta_{0,k}\subseteq P$, the projection $\Pi_P$ is nonexpansive and fixes $\theta_m$ by~\eqref{eq:projection-nonexpansive}. By Lemma~\ref{lem:modules} with $a=\varepsilon$, we have $\|g_{\rm init}(\bar T_n)-\mu_k^{-1}(\bar T_n)\|_2\leq\varepsilon$, and since $\mu_k^{-1}$ is Lipschitz on $\mathcal K$ with $\mu_k^{-1}(P_0T)=\theta_m$, we have $\|\mu_k^{-1}(\bar T_n)-\theta_m\|_2\leq C\|P_nT-P_0T\|_2$. Hence, on $\mathcal M_n$,
\[
\begin{aligned}
\|z-\theta_m\|_2
&=\|\Pi_Pg_{\rm init}(\bar T_n)-\Pi_P\theta_m\|_2
\leq\|g_{\rm init}(\bar T_n)-\theta_m\|_2\\
&\leq\|g_{\rm init}(\bar T_n)-\mu_k^{-1}(\bar T_n)\|_2+\|\mu_k^{-1}(\bar T_n)-\theta_m\|_2
\leq\varepsilon+C\|P_nT-P_0T\|_2.
\end{aligned}
\]
On $\mathcal M_n^c$, both $z$ and $\theta_m$ lie in the compact set $P$, so $\|z-\theta_m\|_2\leq C$. Squaring these bounds and taking expectations under $P_0$, we obtain
\[
E_0\|z-\theta_m\|_2^2
\leq C\varepsilon^2+CE_0\|P_nT-P_0T\|_2^2+CP_0(\mathcal M_n^c)
\leq C(\varepsilon^2+n^{-1}),
\]
where the last inequality uses $E_0\|P_nT-P_0T\|_2^2\leq C/n$, since $T$ is bounded, and $P_0(\mathcal M_n^c)\leq Ce^{-cn}$.

For the unprojected update $v_n$ and the population update $u_m$, write
\[
v_n=z+I_k(z)^{-1}P_ns_{k,z}+e_n,\qquad
u_m=\theta_m+I_k(\theta_m)^{-1}P_0s_{k,\theta_m},
\]
where~\eqref{eq:separate-module-accuracy} with $b=\varepsilon$ gives $\|e_n\|_2\leq\varepsilon$. Here $P_0s_{k,\theta_m}$ need not vanish. Splitting $P_ns_{k,z}-P_0s_{k,\theta_m}=P_n(s_{k,z}-s_{k,\theta_m})+(P_n-P_0)s_{k,\theta_m}$, we decompose
\[
\begin{aligned}
v_n-u_m
&=(z-\theta_m)+\bigl\{I_k(z)^{-1}-I_k(\theta_m)^{-1}\bigr\}P_ns_{k,z}\\
&\qquad+I_k(\theta_m)^{-1}P_n(s_{k,z}-s_{k,\theta_m})+I_k(\theta_m)^{-1}(P_n-P_0)s_{k,\theta_m}+e_n.
\end{aligned}
\]
By Lemma~\ref{lem:consequences}, the score is bounded on $\mathcal X\times P$, and the maps $\theta\mapsto I_k(\theta)^{-1}$ and $\theta\mapsto s_{k,\theta}(x)$ are bounded and Lipschitz on $P$, uniformly in $x\in\mathcal X$. Hence the second and third terms are at most $C\|z-\theta_m\|_2$ in norm, and
\[
\|v_n-u_m\|_2\leq C\bigl\{\varepsilon+\|z-\theta_m\|_2+\|(P_n-P_0)s_{k,\theta_m}\|_2\bigr\}.
\]
Since $s_{k,\theta_m}$ is bounded, $E_0\|(P_n-P_0)s_{k,\theta_m}\|_2^2\leq C/n$. Together with the bound on $E_0\|z-\theta_m\|_2^2$, this gives $E_0\|v_n-u_m\|_2^2\leq C(\varepsilon^2+n^{-1})$.

Finally, $\widetilde\theta_k=\Pi_{\Theta_{0,k}}(v_n)$, and since $\theta^*\in\Theta_{0,k}$, the output projection $\Pi_{\Theta_{0,k}}$ is nonexpansive and fixes $\theta^*$ by~\eqref{eq:projection-nonexpansive}. Together with $\|u_m-\theta^*\|_2=\tau_k$, the triangle inequality gives
\[
\|\widetilde\theta_k-\theta^*\|_2
\leq\|v_n-\theta^*\|_2
\leq\|v_n-u_m\|_2+\|u_m-\theta^*\|_2
=\|v_n-u_m\|_2+\tau_k.
\]
Since $(x+y)^2\leq2x^2+2y^2$, taking expectations gives
\[
E_0\|\widetilde\theta_k-\theta^*\|_2^2
\leq2E_0\|v_n-u_m\|_2^2+2\tau_k^2
\leq C(\varepsilon^2+n^{-1})+2\tau_k^2.
\]

Let $H_*=-P_0\nabla^2\ell_{k,\theta^*}$. By condition~\textup{(i)} preceding Proposition~\ref{prop:ms-misspec}, $\theta^*\in\operatorname{int}(\Theta_{0,k})$ minimizes $D_{\rm KL}(\mb P_0\|\mb P_{k,\theta})$ over $\Theta_{0,k}$, so it maximizes $z\mapsto P_0\ell_{k,z}$ over $\Theta_{0,k}$ and therefore satisfies $P_0s_{k,\theta^*}=0$. By condition~\textup{(ii)} preceding Proposition~\ref{prop:ms-misspec}, $H_*\in\mathbb S_{++}^q$, so $\lambda_*=\lambda_{\min}(H_*)>0$. Since $\theta^*\in\operatorname{int}(\Theta_{0,k})$ and $z\mapsto -P_0\nabla^2\ell_{k,z}$ is continuous, we can choose $b_*>0$ with $B(\theta^*,2b_*)\subset\operatorname{int}(\Theta_{0,k})$ and $\lambda_{\min}(-P_0\nabla^2\ell_{k,z})\geq 3\lambda_*/4$ for all $z\in B(\theta^*,2b_*)$. Since $\theta^*$ is the unique maximizer of $z\mapsto P_0\ell_{k,z}$ on the compact set $\Theta_{0,k}$, we may set
\[
g_*=\inf_{\substack{z\in\Theta_{0,k}\\\|z-\theta^*\|_2\geq b_*}}
P_0(\ell_{k,\theta^*}-\ell_{k,z})>0.
\]
Define
\[
\mathcal G_n=\Bigl\{\argmax_{\Theta_{0,k}}\ell_{n,k}\subset B(\theta^*,b_*),\ \ \inf_{z\in B(\theta^*,2b_*)}\lambda_{\min}(H_{n,k}(z))\geq\lambda_*/2\Bigr\}.
\]

We first bound $P_0(\mathcal G_n^c)$. Suppose that $\widehat z\in\argmax_{\Theta_{0,k}}\ell_{n,k}$ satisfies $\|\widehat z-\theta^*\|_2\geq b_*$. Since $P_n\ell_{k,\widehat z}\geq P_n\ell_{k,\theta^*}$, the definition of $g_*$ gives
\[
g_*\leq P_0(\ell_{k,\theta^*}-\ell_{k,\widehat z})
\leq(P_n-P_0)(\ell_{k,\widehat z}-\ell_{k,\theta^*})
\leq2\sup_{z\in\Theta_{0,k}}|(P_n-P_0)\ell_{k,z}|.
\]
Next, $\|A\|_{\rm op}\leq q\max_{j,l}|A_{jl}|$ for every $q\times q$ matrix $A$. If every entry of $(P_n-P_0)\nabla^2\ell_{k,z}$ is at most $\lambda_*/(4q)$ in absolute value for all $z\in B(\theta^*,2b_*)$, then $\|H_{n,k}(z)+P_0\nabla^2\ell_{k,z}\|_{\rm op}\leq\lambda_*/4$, and the choice of $b_*$ gives
\[
\lambda_{\min}(H_{n,k}(z))\geq\tfrac34\lambda_*-\tfrac14\lambda_*=\tfrac12\lambda_*,\qquad z\in B(\theta^*,2b_*).
\]
Hence, on $\mathcal G_n^c$, either $\sup_{z\in\Theta_{0,k}}|(P_n-P_0)\ell_{k,z}|\geq g_*/2$ or some entry of $(P_n-P_0)\nabla^2\ell_{k,z}$ exceeds $\lambda_*/(4q)$ in absolute value at some $z\in B(\theta^*,2b_*)$. Lemma~\ref{lem:net-concentration}, applied under $P_0$ to the class $\{\ell_{k,z}:z\in\Theta_{0,k}\}$ with $u=g_*/4$ and to the $q^2$ classes of Hessian entries over $B(\theta^*,2b_*)$ with $u=\lambda_*/(4q)$, each with a fixed mesh $h$ satisfying $2Lh<u$ for the corresponding Lipschitz constant $L$, gives $P_0(\mathcal G_n^c)\leq Ce^{-cn}$.

On $\mathcal G_n$, the function $\ell_{n,k}$ is strictly concave on $B(\theta^*,2b_*)$ and every maximizer over $\Theta_{0,k}$ lies in $B(\theta^*,b_*)\subset\operatorname{int}(\Theta_{0,k})$. Thus the maximizer $\widehat\theta_k$ is unique and interior, and $P_ns_{k,\widehat\theta_k}=0$. Let $A_n$ be the average of $H_{n,k}$ over the segment from $\theta^*$ to $\widehat\theta_k$, which lies in $B(\theta^*,2b_*)$. Then $\lambda_{\min}(A_n)\geq\lambda_*/2$ and
\[
0=P_ns_{k,\widehat\theta_k}=P_ns_{k,\theta^*}-A_n(\widehat\theta_k-\theta^*),
\qquad\|\widehat\theta_k-\theta^*\|_2\leq2\lambda_*^{-1}\|P_ns_{k,\theta^*}\|_2.
\]
On $\mathcal G_n^c$, both $\widehat\theta_k$ and $\theta^*$ lie in the bounded set $\Theta_{0,k}$. Since $P_0s_{k,\theta^*}=0$ and the score is bounded, $E_0\|P_ns_{k,\theta^*}\|_2^2\leq C/n$. Hence
\[
E_0\|\widehat\theta_k-\theta^*\|_2^2
\leq4\lambda_*^{-2}E_0\|P_ns_{k,\theta^*}\|_2^2+CP_0(\mathcal G_n^c)\leq\frac Cn.
\]

Let $\Lambda$ be a fixed Hessian envelope. On $\mathcal G_n$, $\widehat\theta_k\in\operatorname{int}(\Theta_{0,k})$, so Lemma~\ref{lem:criterion} gives $D_k\leq(n\Lambda/2)\|\widetilde\theta_k-\widehat\theta_k\|_2^2$. On $\mathcal G_n^c$, we have $D_k\leq 2Bn$. Taking expectations, dividing by $n$, and using $\|\widetilde\theta_k-\widehat\theta_k\|_2^2\leq 2\|\widetilde\theta_k-\theta^*\|_2^2+2\|\widehat\theta_k-\theta^*\|_2^2$, we obtain
\[
\begin{aligned}
n^{-1}E_0D_k
&\leq\tfrac12\Lambda E_0\|\widetilde\theta_k-\widehat\theta_k\|_2^2+2BP_0(\mathcal G_n^c)\\
&\leq\Lambda\bigl(E_0\|\widetilde\theta_k-\theta^*\|_2^2+E_0\|\widehat\theta_k-\theta^*\|_2^2\bigr)+2BP_0(\mathcal G_n^c) \leq C(\tau_k^2+\varepsilon^2+n^{-1}),
\end{aligned}
\]
where the last inequality uses $E_0\|\widetilde\theta_k-\theta^*\|_2^2\leq C(\varepsilon^2+n^{-1})+2\tau_k^2$, $E_0\|\widehat\theta_k-\theta^*\|_2^2\leq C/n$, and $P_0(\mathcal G_n^c)\leq Ce^{-cn}$.
\end{proof}

\subsubsection{Likelihood deficit in the generating model (Lemma~\ref{lem:ms-target})}

\begin{proof}
In this proof, $C,c$ depend only on the generating candidate $k_0$, its model bounds and working set $\Theta_{0,k_0}$, and the compact set $K_0\subset\operatorname{int}(\Theta_{0,k_0})$, and they are uniform over $\theta_0\in K_0$. Fix $\theta_0\in K_0$, and apply the proof of Theorem~\ref{thm:est-approx} to candidate $k_0$ with $a=b=\varepsilon\in(0,1)$ and with $\Theta_1$ a compact subset of $\operatorname{int}(\Theta_{0,k_0})$ containing $K_0$. Let $\mathcal H_n^{\rm int}=\{\widehat\theta_{k_0}\in\operatorname{int}(\Theta_{0,k_0})\}$. By Lemma~\ref{lem:consequences}, $\mb P_0((\mathcal H_n^{\rm int})^c)\leq Ce^{-cn}$. Let $\Lambda$ be a uniform bound on $\|\nabla_\theta^2 f_{k_0,\theta}(x)\|_{\rm op}$ over $x\in\mathcal X$ and $\theta\in\Theta_{0,k_0}$, which is finite by Lemma~\ref{lem:consequences}, and put $C_0=\Lambda/2$. On $\mathcal H_n^{\rm int}$, the constrained MLE $\widehat\theta_{k_0}$ is interior, so Lemma~\ref{lem:criterion} gives
\[
D_{k_0}\leq\frac{n\Lambda}{2}\|\widetilde\theta_{k_0}-\widehat\theta_{k_0}\|_2^2
=C_0n\|\widetilde\theta_{k_0}-\widehat\theta_{k_0}\|_2^2,
\]
whereas $D_{k_0}\leq 2Bn$ holds for every dataset. Substituting $a=b=\varepsilon$ into~\eqref{eq:integrated-module-bound} yields
\[
\mb E_0\bigl[\|\widetilde\theta_{k_0}-\widehat\theta_{k_0}\|_2^2\bigr]
\leq C(\varepsilon^2+\varepsilon^4+n^{-2})
\leq C(\varepsilon^2+n^{-2}),
\]
since $\varepsilon<1$. Splitting the expectation on $\mathcal H_n^{\rm int}$ and its complement, we obtain
\[
\begin{aligned}
\mb E_0[D_{k_0}]
&\leq C_0n\,\mb E_0\bigl[\|\widetilde\theta_{k_0}-\widehat\theta_{k_0}\|_2^2\bigr]
+2Bn\,\mb P_0\bigl((\mathcal H_n^{\rm int})^c\bigr)\\
&\leq C(n\varepsilon^2+n^{-1})+2BCne^{-cn}
\leq C(n\varepsilon^2+n^{-1}),
\end{aligned}
\]
where the last inequality uses $ne^{-cn}\leq C/n$. At $\varepsilon=n^{-1}$, this gives $\mb E_0[D_{k_0}]\leq C/n$.

For the tail bound, let $\mathcal E_n$, $\rho_n$ and $\delta_n$ be as in the proof of Theorem~\ref{thm:est-approx}, for candidate $k_0$. By definition of $\mathcal E_n$, we have $\mathcal E_n\subseteq\mathcal H_n^{\rm int}$, and Step~3 of that proof gives $\mb P_0(\mathcal E_n^c)\leq Ce^{-cn}$. Substituting $a=b=\varepsilon$ into~\eqref{eq:approx-bound} gives
\[
\|\widetilde\theta_{k_0}-\widehat\theta_{k_0}\|_2
\leq\varepsilon+C_1(\varepsilon+\rho_n)(\delta_n+\varepsilon+\rho_n)
=\varepsilon+V_n
\qquad\text{on }\mathcal E_n,
\]
where $C_1$ is the constant in~\eqref{eq:approx-bound} and $V_n=C_1(\varepsilon+\rho_n)(\delta_n+\varepsilon+\rho_n)$. Since $V_n^2\leq C(\varepsilon^4+\rho_n^4+\delta_n^4)$ and Step~3 of the proof of Theorem~\ref{thm:est-approx} gives $\mb E_0[(\rho_n^4+\delta_n^4)\ind_{\mathcal E_n}]\leq Cn^{-2}$, we have
\[
\mb E_0[V_n^2\ind_{\mathcal E_n}]
\leq C\,\mb E_0\bigl[(\varepsilon^4+\rho_n^4+\delta_n^4)\ind_{\mathcal E_n}\bigr]
\leq C(\varepsilon^4+n^{-2}).
\]
Since $\mathcal E_n\subseteq\mathcal H_n^{\rm int}$ and $(\varepsilon+V_n)^2\leq 2\varepsilon^2+2V_n^2$, on $\mathcal E_n$ we have
\[
D_{k_0}\leq C_0n\|\widetilde\theta_{k_0}-\widehat\theta_{k_0}\|_2^2
\leq C_0n(\varepsilon+V_n)^2
\leq 2C_0n\varepsilon^2+2C_0nV_n^2.
\]
Put $c_0=1/(4C_0)$ and suppose $n\varepsilon^2\leq c_0$. Then $2C_0n\varepsilon^2\leq1/2$, so on $\mathcal E_n$ the event $\{D_{k_0}>1\}$ implies $2C_0nV_n^2>1/2$, that is, $V_n^2>1/(4C_0n)$. Splitting on $\mathcal E_n$ and $\mathcal E_n^c$, applying Markov's inequality to $V_n^2\ind_{\mathcal E_n}$, and using $n\varepsilon^4=(n\varepsilon^2)^2/n\leq c_0^2/n$ and $e^{-cn}\leq C/n$, we obtain
\[
\begin{aligned}
\mb P_0(D_{k_0}>1)
&\leq\mb P_0(\mathcal E_n^c)+\mb P_0\bigl(V_n^2\ind_{\mathcal E_n}>1/(4C_0n)\bigr) \leq\mb P_0(\mathcal E_n^c)+4C_0n\,\mb E_0[V_n^2\ind_{\mathcal E_n}]\\
&\leq Ce^{-cn}+C(n\varepsilon^4+n^{-1})
\leq Ce^{-cn}+C(c_0^2+1)/n\leq C/n.
\end{aligned}
\]
Since the proof of Theorem~\ref{thm:est-approx} is applied with a compact set $\Theta_1$ containing $K_0$, all constants are uniform over $\theta_0\in K_0$. This proves the uniform statements with $c_{K_0}=c_0$.
\end{proof}

\subsubsection{Weak-separation probability for nested and separated candidates (Lemma~\ref{lem:psi-nested})}

\begin{proof}
Let $A_1\log n/n\leq\gamma$. By the definition of $\psi_n$ and the union bound over $j\ne k$,
\[
\psi_n\leq\sum_{k=1}^Kw_k\sum_{j\ne k}\pi_k\bigl\{\theta\in\Xi_k:(k,\theta)\text{ violates~\eqref{eq:weak-separation} for }j\bigr\}.
\]
For a pair $(k,j)$ of type (S), $\inf_{\theta'\in\Theta_{0,j}}D_{\rm KL}(\mb P_{k,\theta}\|\mb P_{j,\theta'})\geq\gamma\geq A_1\log n/n$ for every $\theta\in\Xi_k$, so the second alternative in~\eqref{eq:weak-separation} holds. For a pair of type (U), the matching parameter $\theta'$ gives the first alternative for every $\theta\in\Xi_k$. Hence only the pairs of type (D) contribute to the sum, and $\psi_n=0$ if there is none.

Let $(k,j)$ be of type (D), $N=\iota(\Theta_{0,j})\subset\Theta_{0,k}$ and $b=d_k-d_j$. Since $\mb P_{j,\vartheta}=\mb P_{k,\iota(\vartheta)}$, the quadratic lower bound of Lemma~\ref{lem:consequences} for candidate $k$ gives
\[
\inf_{\vartheta\in\Theta_{0,j}}D_{\rm KL}(\mb P_{k,\theta}\|\mb P_{j,\vartheta})=\inf_{\vartheta\in\Theta_{0,j}}D_{\rm KL}(\mb P_{k,\theta}\|\mb P_{k,\iota(\vartheta)})\geq c_k\inf_{\vartheta\in\Theta_{0,j}}\|\theta-\iota(\vartheta)\|_2^2=c_k\operatorname{dist}(\theta,N)^2 .
\]
Put $r_n=\{A_1\log n/(c_kn)\}^{1/2}$. If $\operatorname{dist}(\theta,N)\geq r_n$, the right side is at least $A_1\log n/n$ and the second alternative in~\eqref{eq:weak-separation} holds. So $(k,\theta)$ can violate~\eqref{eq:weak-separation} for $j$ only if $\operatorname{dist}(\theta,N)<r_n$.

To bound the probability of this set under $\pi_k$, let $L_\iota$ be the Lipschitz constant of $\iota$ and let $r\leq1$. By the covering bound in the proof of Lemma~\ref{lem:net-concentration}, the bounded set $\Theta_{0,j}\subset\mb R^{d_j}$ has an $r/L_\iota$-net $\vartheta_1,\ldots,\vartheta_M$ with $M\leq Cr^{-d_j}$. Every point $\iota(\vartheta)$ of $N$ is within distance $L_\iota(r/L_\iota)=r$ of some $\iota(\vartheta_i)$, so every $\theta$ with $\operatorname{dist}(\theta,N)<r$ is within distance $2r$ of some $\iota(\vartheta_i)$. Hence the balls of radius $2r$ centered at the points $\iota(\vartheta_i)$ cover the $r$-neighborhood of $N$, and their total Lebesgue measure is at most $Cr^{-d_j}(2r)^{d_k}=Cr^b$. Since the density of $\pi_k$ is at most $\bar\pi$,
\[
\pi_k\bigl\{\theta\in\Xi_k:\operatorname{dist}(\theta,N)<r_n\bigr\}\leq C\bar\pi r_n^b\leq C\Bigl(\frac{\log n}n\Bigr)^{(d_k-d_j)/2}
\]
when $r_n\leq1$, and the bound is trivial otherwise after enlarging $C$. Summing over the pairs of type (D) with the weights $w_k$ gives the first bound of the lemma. The second bound follows because $d_k-d_j\geq1$, $\log n/n\leq1$ and $K$ is fixed.
\end{proof}

\subsubsection{The model-selection network (Lemma~\ref{lem:ms-rrds})}

\begin{proof}
In this proof, $C,c$ depend only on $K$, $d_X$, $(d_k,r_k)_{k\in[K]}$, the candidate families and the working sets.

We first verify that all four blocks are ReLU networks. By Lemma~\ref{lem:polytope}(b) with $P=\Theta_{0,k}$, the projection $\Pi_{\Theta_{0,k}}$ onto each polytope $\Theta_{0,k}$ is computed exactly by a ReLU network of fixed size (for rectangles, it is coordinatewise clipping). Let $\sigma$ denote the ReLU function. The vector $v$ is an affine function of $u^\ell$, and the maximum of $K$ numbers is computed exactly by $\lceil\log_2K\rceil$ layers of $\max(x,y)=x+\sigma(y-x)$. The lower clipping uses $\max\{t,-\log n\}=\sigma(t+\log n)-\log n$, and the identity map is $t=\sigma(t)-\sigma(-t)$. Since $T_k$, $g_{{\rm init},k}$, $S_k$, $h_k$ and $\varphi_k$ are ReLU networks, the blocks $\phi_0$, $\rho_0$, $\phi$ and $\rho$ are ReLU networks, so $\widetilde F=\widetilde q$ is an RRDS network of the form~\eqref{eq:rrds-init}--\eqref{eq:rrds-output}.

For (a), fix any dataset $X_1,\ldots,X_n$ and set $\bar T_{k,n}=n^{-1}\sum_{i=1}^nT_k(X_i)$. The initialization~\eqref{eq:rrds-init} gives $u_k^{s,(0)}=\bar T_{k,n}$ and $u^{\ell,(0)}=0_K$, so $a^{(0)}=0_K$ and
\[
\vartheta_k^{0,(0)}=\theta_k^{(0)}=\vartheta_k^{\rm m}=\Pi_{\Theta_{0,k}}g_{{\rm init},k}(\bar T_{k,n})\qquad(k\in[K]).
\]
Since the first component of $\rho$ is $0_{D_\Sigma}$, the update~\eqref{eq:rrds-update} gives $\vartheta^{0,(r)}=\vartheta^{\rm m}$ for all $r\geq0$. Therefore, for every $r\geq1$, the pooling step in~\eqref{eq:rrds-update} gives
\[
u_k^{s,(r)}=\frac1n\sum_{i=1}^nS_k(X_i,\vartheta_k^{\rm m}),\qquad
u_k^{\ell,(r)}=\frac1n\sum_{i=1}^n\varphi_k\bigl(X_i,\theta_k^{(r-1)}\bigr),
\]
and the residual updates in~\eqref{eq:rrds-update} simplify to
\[
\begin{aligned}
\theta_k^{(r)}&=\theta_k^{(r-1)}+\Pi_{\Theta_{0,k}}\bigl\{\vartheta_k^{\rm m}+h_k(u_k^{s,(r)},\vartheta_k^{\rm m})\bigr\}-\theta_k^{(r-1)} =\Pi_{\Theta_{0,k}}\bigl\{\vartheta_k^{\rm m}+h_k(u_k^{s,(r)},\vartheta_k^{\rm m})\bigr\},\\
a^{(r)}&=a^{(r-1)}+V\bigl(u^{\ell,(r)}\bigr)-a^{(r-1)}=V\bigl(u^{\ell,(r)}\bigr).
\end{aligned}
\]
Because $u_k^{s,(r)}$ does not depend on $r$, the iterate $\theta_k^{(r)}$ does not depend on $r$, lies in $\Theta_{0,k}$, and coincides with the one-step estimate $\widetilde\theta_k$ of~\eqref{eq:structured-class} for candidate $k$ at accuracy $\varepsilon_n$ from Theorem~\ref{thm:est-approx}. This proves (a).

For (b), let $r\geq2$. By (a), $\theta_k^{(r-1)}=\widetilde\theta_k$ for all $k$. Hence $n\,u_k^{\ell,(r)}=\sum_i\varphi_k(X_i,\widetilde\theta_k)$ and
\[
v_k=\log w_k+n\,u_k^{\ell,(r)}-\frac{d_k}2\log n =\log w_k+\sum_{i=1}^n\varphi_k(X_i,\widetilde\theta_k)-\frac{d_k}2\log n=\widetilde a_k.
\]
Since $a^{(r)}=V(u^{\ell,(r)})$, we obtain
\[
a_k^{(r)}=\max\Bigl\{\widetilde a_k-\max_{l\in[K]}\widetilde a_l,\ -\log n\Bigr\}\qquad(k\in[K]).
\]
In particular, $a^{(r)}$ does not depend on $r\geq2$, so the output $\widetilde q=\Psi(z^{(R)})$ is the same for every $R\geq2$. Since $n\geq3$, we have $-\log n<0$ and $a_k^{(R)}\in[-\log n,0]$ for all $k\in[K]$, so the clipping in $\Psi$ leaves $a^{(R)}$ unchanged. Moreover, $a_k^{(R)}\leq0$ with equality if and only if $\widetilde a_k=\max_{l\in[K]}\widetilde a_l$. Hence the largest logit is zero and is attained exactly at the maximizers of $\widetilde a$, which gives $\argmax_k\widetilde q_k=\argmax_k\widetilde a_k$. Finally, the softmax of $K$ logits in $[-\log n,0]$ has every coordinate at least $e^{-\log n}/K$, so $\widetilde q_k\geq(Kn)^{-1}$. This proves (b).

For (c), set $\kappa_k=(d_X+d_k)/r_k$ and $\kappa_k'=(d_X+d_k)/(r_k+1)$ for each $k\in[K]$. By Theorem~\ref{thm:est-approx} at accuracy $\varepsilon_n=n^{-1/2}/\log(en)$, the networks $g_{{\rm init},k}$, $S_k$ and $h_k$ have depth at most $C\log(e/\varepsilon_n)\leq C\log n$ and total parameter count at most $C\varepsilon_n^{-\kappa_k}\log(e/\varepsilon_n)$. Since $r_k\geq2$, we have $\kappa_k'-\kappa_k/2=(d_X+d_k)(r_k-1)/\{2r_k(r_k+1)\}>0$. Hence the polynomial factor $n^{\kappa_k'-\kappa_k/2}$ absorbs $\log^{\kappa_k}(en)$, so $\log^{\kappa_k}(en)\leq Cn^{\kappa_k'-\kappa_k/2}$. Using $\varepsilon_n^{-\kappa_k}=n^{\kappa_k/2}\log^{\kappa_k}(en)$ and $\log(e/\varepsilon_n)\leq C\log n$ (as $n\geq3$), we obtain
\[
\begin{aligned}
W(g_{{\rm init},k})+W(S_k)+W(h_k)&\leq C\varepsilon_n^{-\kappa_k}\log(e/\varepsilon_n)\\
&=Cn^{\kappa_k/2}\log^{\kappa_k}(en)\log(e/\varepsilon_n)\leq Cn^{\kappa_k'}\log n.
\end{aligned}
\]
For $\varphi_k$, Assumption~\ref{ass:regular} gives that $\ell_k$ has smoothness $r_k+1$ in its $d_X+d_k$ arguments, and the cutoff construction in the proof of Theorem~\ref{thm:est-approx} extends $\ell_k$ to a rectangle containing $\mathcal X\times\Theta_{0,k}$. We take $\varphi_k$ to be the network that Lemma~\ref{lem:sobolev-approx} gives with input dimension $d_X+d_k$, smoothness $r_k+1$ and accuracy $n^{-1}$. It has depth $L(\varphi_k)\leq C\log(en)\leq C\log n$ and parameter count
\[
W(\varphi_k)\leq Cn^{(d_X+d_k)/(r_k+1)}\log(en)\leq Cn^{\kappa_k'}\log n.
\]
The moment networks $T_k$ (Lemma~\ref{lem:relu-moment}) and projections $\Pi_{\Theta_{0,k}}$ (Lemma~\ref{lem:polytope}(b)) have fixed size, and the clippings and $V$ add $O(K)$ parameters and $O(\log K)$ layers. Placing the $K$ candidates in parallel inside each block gives $L(\phi_0),L(\rho_0),L(\phi),L(\rho)\leq C\log n$ and, since $\kappa_k'\leq\kappa_{\max}'$ for all $k\in[K]$, $W(\phi_0),W(\rho_0),W(\phi),W(\rho)\leq CKn^{\kappa_{\max}'}\log n$. By~\eqref{eq:rrds-complexity} with $R=2$, this gives $L\leq C\log n$ and $W\leq CKn^{\kappa_{\max}'}\log n$, so $\widetilde F\in\mathcal F(L,W)$ with the stated bounds. The same bound holds for the total number of coefficient occurrences in the four blocks, fixed or trainable. For $g_{{\rm init},k}$, $S_k$ and $h_k$ this follows from Lemma~\ref{lem:modules} and Step~1 of the proof of Theorem~\ref{thm:est-approx}, for $\varphi_k$ from the proof of Lemma~\ref{lem:sobolev-approx}, and for the fixed networks, the clippings and $V$ from the counts above. We take $W$ equal to this bound $CKn^{\kappa_{\max}'}\log n$. An affine row has at most as many positions as the blocks have coefficient occurrences, which gives the last claim. Finally, the coefficient bounds are those of Lemmas~\ref{lem:modules} and~\ref{lem:sobolev-approx} at the accuracies $\varepsilon_n$ and $n^{-1}$, where $\varepsilon_n^{-1}\leq Cn$, together with the coefficient $n$ in $v$. Hence all coefficients are bounded by $Cn^C$, where for the bias $\log w_k-(d_k/2)\log n$ of $v_k$ the constant also depends on the weight $w_k$. This proves (c).
\end{proof}

\subsubsection{Laplace bounds (Lemma~\ref{lem:ms-laplace})}

\begin{proof}
In this proof, $C,c$ depend only on $K$, $d_X$, $(d_k,r_k)_{k\in[K]}$, the candidate families and working sets, $\rho$, $\underline\pi$, $\bar\pi$ and the rectangles $\Xi_k$.

Since $\pi_j$ is a probability measure on $\Theta_{0,j}$ and $\ell_{n,j}(\vartheta)\leq\ell_{n,j}(\widehat\theta_j)$ for all $\vartheta\in\Theta_{0,j}$, we have $m_j=\int e^{\ell_{n,j}(\vartheta)}\,\pi_j(d\vartheta)\leq e^{\ell_{n,j}(\widehat\theta_j)}$ for every dataset.

For the remaining claims, we suppress $j$ and write $d=d_j$, $\Theta_0=\Theta_{0,j}$ and $\Xi=\Xi_j$. Let $D_0=\operatorname{diam}(\Theta_0)$, and let $B_s$ be such that $\|s_\vartheta(x)\|_2\leq B_s$ for all $x\in\mathcal X$ and $\vartheta\in\Theta_0$, which is possible by Lemma~\ref{lem:consequences}. Let $c_1$ be the constant of the quadratic lower bound $D_{\rm KL}(\mb P_\eta\|\mb P_\vartheta)\geq c_1\|\eta-\vartheta\|_2^2$ of Lemma~\ref{lem:consequences}, and let $b>0$, $\lambda_0>0$ and $\mathcal H_n^{\rm loc}(\eta)$ be as in the proof of that lemma with $\Theta_1=K_{j,\rho}$. On $\mathcal H_n^{\rm loc}(\eta)$ we have $\widehat\theta\in B(\eta,b)$, $B(\eta,2b)\subset\operatorname{int}(\Theta_0)$, and the negative Hessian of $n^{-1}\ell_n$ is at least $\frac12\lambda_0I_d$ on $B(\eta,2b)$. Put
\[
\mathcal G_n(\eta)=\mathcal H_n^{\rm loc}(\eta)\cap\Bigl\{\sup_{\vartheta\in\Theta_0}|(P_n-P_\eta)(\ell_\vartheta-\ell_\eta)|\leq\tfrac12c_1b^2\Bigr\}.
\]
Lemma~\ref{lem:consequences} gives $\sup_{\eta\in K_{j,\rho}}\mb P_\eta\{\mathcal H_n^{\rm loc}(\eta)^c\}\leq Ce^{-cn}$. For the second event, the convexity of $\Theta_0$ and the bound on the score give $|\ell_\vartheta(x)-\ell_{\vartheta'}(x)|\leq B_s\|\vartheta-\vartheta'\|_2$ for all $\vartheta,\vartheta'\in\Theta_0$ and $x\in\mathcal X$, so $|\ell_\vartheta(x)-\ell_\eta(x)|\leq B_sD_0$, and $\Theta_0\subseteq B(\eta,D_0)$. Lemma~\ref{lem:net-concentration}, applied to the class $\{\ell_\vartheta-\ell_\eta:\vartheta\in\Theta_0\}$ with $u=c_1b^2/2$ and the fixed mesh $h=c_1b^2/(8B_s)$, gives
\[
\begin{aligned}
&\mb P_\eta\Bigl\{\sup_{\vartheta\in\Theta_0}|(P_n-P_\eta)(\ell_\vartheta-\ell_\eta)|>\tfrac12c_1b^2\Bigr\}\leq2\Bigl(1+\frac{16B_sD_0}{c_1b^2}\Bigr)^d\exp\Bigl\{-\frac{nc_1^2b^4}{32B_s^2D_0^2}\Bigr\}\leq Ce^{-cn}.
\end{aligned}
\]
Hence $\sup_{\eta\in K_{j,\rho}}\mb P_\eta\{\mathcal G_n(\eta)^c\}\leq Ce^{-cn}$. On $\mathcal G_n(\eta)$, $\widehat\theta\in B(\eta,b)\subset\operatorname{int}(\Theta_0)$, so $\nabla\ell_n(\widehat\theta)=0$.

For the upper bound on $\delta$, we work on $\mathcal G_n(\eta)$ and bound $\ell_n(\vartheta)-\ell_n(\widehat\theta)$ in two regions. If $\vartheta\in B(\eta,2b)$, the segment from $\widehat\theta$ to $\vartheta$ lies in $B(\eta,2b)$. Since $\nabla\ell_n(\widehat\theta)=0$ and the negative Hessian of $\ell_n$ is at least $\frac12\lambda_0nI_d$ on this ball, Taylor's formula gives
\[
\ell_n(\vartheta)-\ell_n(\widehat\theta)\leq-\tfrac14\lambda_0n\|\vartheta-\widehat\theta\|_2^2 .
\]
If $\vartheta\in\Theta_0$ and $\|\vartheta-\eta\|_2\geq2b$, then
\[
\begin{aligned}
\ell_n(\vartheta)-\ell_n(\widehat\theta)
&\leq\ell_n(\vartheta)-\ell_n(\eta)=nP_n(\ell_\vartheta-\ell_\eta)=n\bigl\{-D_{\rm KL}(\mb P_\eta\|\mb P_\vartheta)+(P_n-P_\eta)(\ell_\vartheta-\ell_\eta)\bigr\}\\
&\leq n\bigl\{-4c_1b^2+\tfrac12c_1b^2\bigr\}\leq-3c_1b^2n
\leq-\frac{3c_1b^2}{D_0^2}\,n\|\vartheta-\widehat\theta\|_2^2 ,
\end{aligned}
\]
where the first inequality holds because $\widehat\theta$ maximizes $\ell_n$ over $\Theta_0$, the second uses $D_{\rm KL}(\mb P_\eta\|\mb P_\vartheta)\geq c_1\|\eta-\vartheta\|_2^2\geq4c_1b^2$ and the definition of $\mathcal G_n(\eta)$, and the last uses $\|\vartheta-\widehat\theta\|_2\leq D_0$. Hence, with $c_2=\min\{\lambda_0/4,3c_1b^2/D_0^2\}$, we have $\ell_n(\vartheta)\leq\ell_n(\widehat\theta)-c_2n\|\vartheta-\widehat\theta\|_2^2$ for all $\vartheta\in\Theta_0$. Since $\pi$ is carried by $\Theta_0$ and its density is at most $\bar\pi$,
\[
m=\int e^{\ell_n(\vartheta)}\,\pi(d\vartheta)
\leq\bar\pi\,e^{\ell_n(\widehat\theta)}\int_{\mb R^d}e^{-c_2n\|\vartheta-\widehat\theta\|_2^2}\,d\vartheta
=\bar\pi\,e^{\ell_n(\widehat\theta)}\Bigl(\frac\pi{c_2n}\Bigr)^{d/2}.
\]
Since $\delta=\log m-\ell_n(\widehat\theta)+\frac d2\log n$, this gives $\delta\leq\log\bar\pi+\frac d2\log(\pi/c_2)$.

For the lower bound on $\delta$, the density of $\pi$ is at least $\underline\pi$ on the rectangle $\Xi$, so there are $c_\pi,r_0>0$ with $\pi\{B(\vartheta,r)\}\geq c_\pi r^d$ for all $\vartheta\in\Xi$ and $0<r\leq r_0$. Let $n\geq r_0^{-2}$, let $\vartheta_0$ be the point of $\Xi$ nearest to $\widehat\theta$ and put $t=\operatorname{dist}(\widehat\theta,\Xi)$, so that $U_n=nt^2$. For $\vartheta\in B(\vartheta_0,n^{-1/2})\cap\Theta_0$, the segment from $\widehat\theta$ to $\vartheta$ lies in the convex set $\Theta_0$. Since $\nabla\ell_n(\widehat\theta)=0$ on $\mathcal G_n(\eta)$ and $\|\nabla^2\ell_n\|_{\rm op}\leq n\bar\Lambda$ on $\Theta_0$, Taylor's formula and $\|\vartheta-\widehat\theta\|_2^2\leq2\|\vartheta-\vartheta_0\|_2^2+2t^2\leq2n^{-1}+2t^2$ give
\[
\ell_n(\vartheta)\geq\ell_n(\widehat\theta)-\tfrac12n\bar\Lambda\|\vartheta-\widehat\theta\|_2^2
\geq\ell_n(\widehat\theta)-\bar\Lambda-\bar\Lambda nt^2 .
\]
Since $\pi$ is carried by $\Theta_0$ and $n^{-1/2}\leq r_0$,
\[
m\geq e^{\ell_n(\widehat\theta)-\bar\Lambda-\bar\Lambda nt^2}\,\pi\{B(\vartheta_0,n^{-1/2})\}
\geq c_\pi n^{-d/2}e^{\ell_n(\widehat\theta)-\bar\Lambda-\bar\Lambda nt^2},
\]
which gives $\delta\geq\log c_\pi-\bar\Lambda-\bar\Lambda U_n$. Hence $-C_0-\bar\Lambda U_n\leq\delta\leq C_0$ on $\mathcal G_n(\eta)$ for $n\geq n_0$, with $C_0=\max\{\log\bar\pi+\frac d2\log(\pi/c_2),\bar\Lambda-\log c_\pi\}$ and $n_0\geq r_0^{-2}$, where we take the maximum over the finitely many $j\in[K]$.

For the tail bound, fix $s\geq0$. By the proof of Lemma~\ref{lem:consequences}, on $\mathcal H_n^{\rm loc}(\eta)$ we have $\|\widehat\theta-\eta\|_2\leq2\lambda_0^{-1}\|P_ns_\eta\|_2$, so $n\|\widehat\theta-\eta\|_2^2>s$ implies $n\|P_ns_\eta\|_2^2>\lambda_0^2s/4$. Hence
\[
\mb P_\eta\bigl\{n\|\widehat\theta-\eta\|_2^2>s\bigr\}
\leq\mb P_\eta\{\mathcal H_n^{\rm loc}(\eta)^c\}
+\mb P_\eta\bigl\{n\|P_ns_\eta\|_2^2>\tfrac14\lambda_0^2s\bigr\}.
\]
The first term is at most $Ce^{-cn}$. For the second, let $s'\geq0$. If $n\|P_ns_\eta\|_2^2>s'$, then some coordinate of $P_ns_\eta$ exceeds $\sqrt{s'/(dn)}$ in absolute value. By Lemma~\ref{lem:consequences}, each coordinate of the score $s_\eta$ has mean zero under $\mb P_\eta$, and it takes values in $[-B_s,B_s]$. Hoeffding's inequality \citep[Proposition~2.5 and (2.11)]{wainwright2019high}, applied to the $d$ coordinates at threshold $\sqrt{s'/(dn)}$, therefore gives $\mb P_\eta\{n\|P_ns_\eta\|_2^2>s'\}\leq2d\exp\{-s'/(2dB_s^2)\}$. Taking $s'=\lambda_0^2s/4$, we obtain
\[
\sup_{\eta\in K_{j,\rho}}\mb P_\eta\bigl\{n\|\widehat\theta-\eta\|_2^2>s\bigr\}
\leq Ce^{-cn}+2d\exp\Bigl\{-\frac{\lambda_0^2s}{8dB_s^2}\Bigr\}
\leq Ce^{-cs}+Ce^{-cn}.
\]
\end{proof}

\subsubsection{Likelihood comparisons between candidates (Lemma~\ref{lem:ms-compare})}

\begin{proof}
For (a), $C,c$ depend only on the fixed candidate families, working sets, $\mb P_0$, $t$ and $\gamma_{{\rm KL},k}$. By definition of $\operatorname{IC}_j$ with the BIC penalties $\operatorname{pen}(j)=(d_j/2)\log n$,
\[
\begin{aligned}
\operatorname{IC}_k-\operatorname{IC}_{k_0}
&=\ell_{n,k_0}(\widehat\theta_{k_0})-\ell_{n,k}(\widehat\theta_k)+\frac{d_k-d_{k_0}}{2}\log n\geq\ell_{n,k_0}(\theta_0)-\sup_{\theta\in\Theta_{0,k}}\ell_{n,k}(\theta)+\frac{d_k-d_{k_0}}{2}\log n\\
&=n\inf_{\theta\in\Theta_{0,k}}P_n\log\frac{p_0}{p_{k,\theta}}+\frac{d_k-d_{k_0}}{2}\log n,
\end{aligned}
\]
where the inequality holds because $\theta_0\in\Theta_{0,k_0}$, $\widehat\theta_{k_0}$ maximizes $\ell_{n,k_0}$ over $\Theta_{0,k_0}$ and $\ell_{n,k}(\widehat\theta_k)=\sup_{\theta\in\Theta_{0,k}}\ell_{n,k}(\theta)$, and the last equality uses $\mb P_0=\mb P_{k_0,\theta_0}$. Put $v_n=\{t-(d_k-d_{k_0})\log n/2\}/n$, which may be negative, and $Y_{k,\theta}=\log(p_0/p_{k,\theta})$. Then
\[
\{\operatorname{IC}_k-\operatorname{IC}_{k_0}\leq t\}
\subseteq\Bigl\{\inf_{\theta\in\Theta_{0,k}}P_nY_{k,\theta}\leq v_n\Bigr\}.
\]
Since the candidate log densities are bounded and Lipschitz in the parameter, $Y_{k,\theta}$ satisfies the conditions of the second part of Lemma~\ref{lem:lik-process} with constants $B_Y$, $L_Y$ and $\lambda_Y$, and its covering bound gives $N_k(n^{-1})\leq Cn^{d_k}$. Under $\mb P_0$, the mean of $Y_{k,\theta}$ is $R_k(\theta)=D_{\mathrm{KL}}(\mb P_0\|\mb P_{k,\theta})$, so $R_k^*=\inf_{\theta\in\Theta_{0,k}}R_k(\theta)=\gamma_{{\rm KL},k}$. Since $v_n+L_Y/n\to0$ as $n\to\infty$ and $\gamma_{{\rm KL},k}>0$, we have $v_n+L_Y/n\leq\gamma_{{\rm KL},k}/4$ for all sufficiently large $n$, so $\gamma_{{\rm KL},k}/2-v_n-L_Y/n\geq\gamma_{{\rm KL},k}/4$. Put $c=\lambda_Y\gamma_{{\rm KL},k}/4$. For all such $n$, applying~\eqref{eq:likelihood-gap} with $h=n^{-1}$ and $v=v_n$ and using $N_k(n^{-1})\leq Cn^{d_k}$ gives
\[
\begin{aligned}
&\mb P_0^{(n)}\{\operatorname{IC}_k-\operatorname{IC}_{k_0}\leq t\}
\leq\mb P_0^{(n)}\Bigl\{\inf_{\theta\in\Theta_{0,k}}P_nY_{k,\theta}\leq v_n\Bigr\}\\
&\qquad\leq N_k(n^{-1})\exp\bigl[-\lambda_Yn\{\gamma_{{\rm KL},k}/2-v_n-L_Y/n\}\bigr]
\leq Cn^{d_k}e^{-cn}.
\end{aligned}
\]
The finitely many smaller $n$ are covered by enlarging $C$, which proves the first bound in (a). Uniform likelihood envelopes, Lipschitz bounds and a uniform positive lower bound on $\gamma_{{\rm KL},k}(\theta_0)$ make $B_Y$, $L_Y$, $\lambda_Y$, the threshold on $n$, and hence $C$ and $c$, uniform over $\mathcal A$, giving the second bound in (a).

For (b), $C,c$ depend only on the fixed correct candidate, its working set and a compact interior set containing $\theta_{0,k}$, over which they are uniform, and $C$ may also depend on $a$. The argument localizes $\widehat\theta_k$ at the scale $(\log n)^{-1}$, bounds the likelihood gain by a quadratic form in the standardized score, and applies Lemma~\ref{lem:sphere-bernstein}, which keeps the factor $1/2$ in the exponent. Write $q=d_k$, $I_0=I_k(\theta_{0,k})$ and
\[
\begin{gathered}
U_j=I_0^{-1/2}s_{k,\theta_{0,k}}(X_j),\quad V_n=n^{-1/2}\sum_{j=1}^n U_j,\quad 
E_0U_j=0,\quad E_0U_jU_j^\top=I_q,\quad \|U_j\|_2\leq B_0,
\end{gathered}
\]
where $B_0$ is a fixed score bound for the correct candidate. Set $a_n=(\log n)^{-1}$. The quadratic lower bound on the KL divergence in the proof of Lemma~\ref{lem:consequences} gives a fixed $c_1>0$ such that
\[
\inf_{\substack{\theta\in\Theta_{0,k}\\\|\theta-\theta_{0,k}\|_2\geq a_n}}
D_{\rm KL}(P_0\|P_{k,\theta})\geq c_1a_n^2.
\]
Let $L_\ell\geq 1$ be a Lipschitz constant of $\theta\mapsto\ell_{k,\theta}(x)$ on $\Theta_{0,k}$, uniform in $x$, and $L_H$ one of $z\mapsto H_{n,k}(z)$ in operator norm. Define
\[
\mathcal H_n^{\rm lik}=\Bigl\{\sup_{\theta\in\Theta_{0,k}}|(P_n-P_0)(\ell_{k,\theta}-\ell_{k,\theta_{0,k}})|\leq\tfrac12c_1a_n^2,\ \|H_{n,k}(\theta_{0,k})-I_0\|_{\rm op}\leq a_n\Bigr\}.
\]

On $\mathcal H_n^{\rm lik}$, for every $\theta\in\Theta_{0,k}$ with $\|\theta-\theta_{0,k}\|_2\geq a_n$, the KL bound and the definition of $\mathcal H_n^{\rm lik}$ give
\[
P_n(\ell_{k,\theta}-\ell_{k,\theta_{0,k}})
=-D_{\rm KL}(P_0\|P_{k,\theta})+(P_n-P_0)(\ell_{k,\theta}-\ell_{k,\theta_{0,k}}) \leq -c_1a_n^2+\tfrac12c_1a_n^2<0,
\]
so no such $\theta$ can maximize $\ell_{n,k}$ over $\Theta_{0,k}$. Hence $\|\widehat\theta_k-\theta_{0,k}\|_2<a_n$, so $\widehat\theta_k\in B(\theta_{0,k},a_n)\subset\operatorname{int}(\Theta_{0,k})$ for all large $n$. Moreover, for $z\in B(\theta_{0,k},a_n)$, the Lipschitz bound on $H_{n,k}$ and the second inequality in $\mathcal H_n^{\rm lik}$ give
\[
H_{n,k}(z)\succeq H_{n,k}(\theta_{0,k})-L_Ha_nI_q\succeq I_0-(1+L_H)a_nI_q\succeq(1-C_1a_n)I_0,
\]
where $C_1\geq\|I_0^{-1}\|_{\rm op}(1+L_H)$ and the last step uses $I_q\preceq\|I_0^{-1}\|_{\rm op}I_0$.

To bound $P_0((\mathcal H_n^{\rm lik})^c)$, note first that $|\ell_{k,\theta}(x)|\leq B$ gives $|\ell_{k,\theta}(x)-\ell_{k,\theta_{0,k}}(x)|\leq 2B$, and $\theta\mapsto\ell_{k,\theta}(x)-\ell_{k,\theta_{0,k}}(x)$ is $L_\ell$-Lipschitz on the bounded set $\Theta_{0,k}$ uniformly in $x$. Lemma~\ref{lem:net-concentration}, applied to $\{\ell_{k,\theta}-\ell_{k,\theta_{0,k}}:\theta\in\Theta_{0,k}\}$ with $u=c_1a_n^2/2$ and mesh $h=c_1a_n^2/(8L_\ell)$, has $u-2L_\ell h=c_1a_n^2/4$ and a net of at most $Ca_n^{-2q}$ points, so it bounds the probability that the first inequality in $\mathcal H_n^{\rm lik}$ fails by $Ca_n^{-2q}e^{-cna_n^4}$.

For the second event in $(\mathcal H_n^{\rm lik})^c$, we have $\|A\|_{\rm op}\leq q\max_{1\leq j,l\leq q}|A_{jl}|$ for any $q\times q$ matrix $A$. By Lemma~\ref{lem:consequences}, the entries of $H_{n,k}(\theta_{0,k})$ are averages of independent bounded variables with mean $(I_0)_{jl}$, so Hoeffding's inequality \citep[Proposition~2.5 and (2.11)]{wainwright2019high} at threshold $a_n/q$ and a union bound over the $q^2$ entries give $P_0(\|H_{n,k}(\theta_{0,k})-I_0\|_{\rm op}>a_n)\leq 2q^2e^{-cna_n^2}$. With $a_n=(\log n)^{-1}$, combining the two bounds gives
\[
P_0((\mathcal H_n^{\rm lik})^c)\leq Ca_n^{-2q}e^{-cna_n^4}+2q^2e^{-cna_n^2}\leq Ce^{-cn/(\log n)^4}.
\]

Next, take $n$ large enough that $B(\theta_{0,k},a_n)\subset\operatorname{int}(\Theta_{0,k})$ and $C_1a_n<1$, and work on $\mathcal H_n^{\rm lik}$. Put $\Delta=\widehat\theta_k-\theta_{0,k}$. Since $\nabla\ell_{n,k}(\theta_{0,k})=nP_ns_{k,\theta_{0,k}}=\sqrt n\,I_0^{1/2}V_n$ and the segment from $\theta_{0,k}$ to $\widehat\theta_k$ lies in $B(\theta_{0,k},a_n)$, Taylor's formula applied to $G_{n,k}=2\{\ell_{n,k}(\widehat\theta_k)-\ell_{n,k}(\theta_{0,k})\}$ yields an average $\bar H$ of $H_{n,k}$ over the segment, which satisfies $\bar H\succeq(1-C_1a_n)I_0$, such that
\[
\begin{aligned}
G_{n,k}&=2\Delta^\top nP_ns_{k,\theta_{0,k}}-n\Delta^\top\bar H\Delta \leq 2\sqrt n\,\Delta^\top I_0^{1/2}V_n-n(1-C_1a_n)\Delta^\top I_0\Delta\\
&\leq 2\sup_{v\in\mathbb R^q}\Bigl\{\sqrt n\,v^\top I_0^{1/2}V_n-\frac{n(1-C_1a_n)}{2}v^\top I_0v\Bigr\}
=\frac{\|V_n\|_2^2}{1-C_1a_n},
\end{aligned}
\]
where the last equality follows by substituting $u=\sqrt n\,I_0^{1/2}v$ and maximizing $2u^\top V_n-(1-C_1a_n)\|u\|_2^2$ over $u\in\mathbb R^q$.

Fix $0<w\leq a\log n$ and set $r=\sqrt{(1-C_1a_n)w}$. On $\mathcal H_n^{\rm lik}$, $G_{n,k}\geq w$ implies $\|V_n\|_2\geq r$, so
\[
P_0(G_{n,k}\geq w)\leq P_0((\mathcal H_n^{\rm lik})^c)+P_0(\|V_n\|_2\geq r).
\]
Applying~\eqref{eq:sphere-bernstein} of Lemma~\ref{lem:sphere-bernstein} to $U_1,\ldots,U_n$ with its constant $K$ equal to $B_0$, $\sigma^2=1$ and $\alpha_n=(\log n)^{-1/2}$ gives
\[
P_0(\|V_n\|_2\geq r)\leq\Bigl(1+\frac{2}{\alpha_n}\Bigr)^q\exp\Bigl\{-\frac{(1-\alpha_n^2/2)^2r^2}{2\{1+B_0r/(3\sqrt n)\}}\Bigr\}.
\]
For $n\geq 3$, we have $\alpha_n\leq 1$, so $(1+2/\alpha_n)^q\leq 3^q(\log n)^{q/2}$. For the exponent, $(1-\alpha_n^2/2)^2\geq 1-\alpha_n^2$, $r^2=(1-C_1a_n)w\leq w$, and $(1+t)^{-1}\geq 1-t$ for $t\geq 0$. Hence, for all large $n$ and $0<w\leq a\log n$,
\[
\begin{aligned}
\frac{(1-\alpha_n^2/2)^2r^2}{2\{1+B_0r/(3\sqrt n)\}}
&\geq\frac{(1-\alpha_n^2)(1-C_1a_n)w}{2}\Bigl(1-\frac{B_0\sqrt w}{3\sqrt n}\Bigr)\geq\frac w2\Bigl(1-C_1a_n-\alpha_n^2-\frac{B_0\sqrt w}{3\sqrt n}\Bigr)\\
&\geq\frac w2-C\Bigl\{w(a_n+\alpha_n^2)+\frac{w^{3/2}}{\sqrt n}\Bigr\}\geq\frac w2-C,
\end{aligned}
\]
where the last inequality uses $wa_n=w\alpha_n^2\leq a$ and $\sup_{n\geq 3}(\log n)^{3/2}/\sqrt n<\infty$. Thus
\[
P_0(\|V_n\|_2\geq r)\leq 3^q(\log n)^{q/2}e^{-w/2+C}\leq C(\log n)^{q/2}e^{-w/2}.
\]
Combining this with the bound on $P_0((\mathcal H_n^{\rm lik})^c)$ yields $P_0(G_{n,k}\geq w)\leq C(\log n)^{q/2}e^{-w/2}+Ce^{-cn/(\log n)^4}$ for all large $n$ and $0<w\leq a\log n$, with constants and lower threshold on $n$ uniform over the compact interior generating set.

For (c), $C$ depends only on the quantities of Assumption~\ref{ass:models} and on $A_1$ and $\rho$. Let $Y_{j,\vartheta}=\log(p_{k,\theta}/p_{j,\vartheta})$, so that $|Y_{j,\vartheta}|\leq2B\leq B_Y$ and $\vartheta\mapsto Y_{j,\vartheta}$ is Lipschitz with a constant $L_Y$. By~\eqref{eq:bic-logits}, since $\log(w_k/w_j)\geq-\log c_w$, $\ell_{n,k}(\widehat\theta_k)\geq\ell_{n,k}(\theta)$, $\ell_{n,j}(\widehat\theta_j)=\sup_{\vartheta\in\Theta_{0,j}}\ell_{n,j}(\vartheta)$ and $d_k-d_j\leq d_{\max}$, we have
\[
\begin{aligned}
a_k^{\rm BIC}-a_j^{\rm BIC}
&=\log(w_k/w_j)+\ell_{n,k}(\widehat\theta_k)-\ell_{n,j}(\widehat\theta_j)-\tfrac12(d_k-d_j)\log n\\
&\geq-\log c_w+\ell_{n,k}(\theta)-\sup_{\vartheta\in\Theta_{0,j}}\ell_{n,j}(\vartheta)-\tfrac12d_{\max}\log n\\
&=n\inf_{\vartheta\in\Theta_{0,j}}P_nY_{j,\vartheta}-\tfrac12d_{\max}\log n-\log c_w .
\end{aligned}
\]
On the event $\{a_k^{\rm BIC}-a_j^{\rm BIC}\leq\log n-\log c_w\}$, since $A_1\geq8$ by~\eqref{eq:ms-A1}, this gives
\[
n\inf_{\vartheta\in\Theta_{0,j}}P_nY_{j,\vartheta}
\leq\Bigl(1+\frac{d_{\max}}2\Bigr)\log n
\leq\Bigl(\frac{A_1}8+\frac{d_{\max}}2\Bigr)\log n ,
\]
so the event in (c) is contained in $\{\inf_{\vartheta\in\Theta_{0,j}}P_nY_{j,\vartheta}\leq v\}$ with $nv=(A_1/8+d_{\max}/2)\log n$. Under $\mb P_{k,\theta}$, we have $\mb E_{k,\theta}[Y_{j,\vartheta}]=D_{\rm KL}(\mb P_{k,\theta}\|\mb P_{j,\vartheta})$, so for $j\in\mathcal I_-(k,\theta)$,
\[
R_j^*=\inf_{\vartheta\in\Theta_{0,j}}D_{\rm KL}(\mb P_{k,\theta}\|\mb P_{j,\vartheta})=\operatorname{gap}_j(k,\theta)\geq\frac{A_1\log n}n .
\]
Applying~\eqref{eq:likelihood-gap} with $h=n^{-1}$ and a net of cardinality $N_j(n^{-1})\leq Cn^{d_{\max}}$, we obtain
\[
\begin{aligned}
\mb P_{k,\theta}^{(n)}\Bigl\{\inf_{\vartheta\in\Theta_{0,j}}P_nY_{j,\vartheta}\leq v\Bigr\}
&\leq N_j(n^{-1})\exp\Bigl[-\lambda_Yn\Bigl\{\frac{R_j^*}2-v-\frac{L_Y}n\Bigr\}\Bigr]\\
&\leq Cn^{d_{\max}}\exp\Bigl[-\lambda_Y\Bigl\{\Bigl(\frac{A_1}2-\frac{A_1}8-\frac{d_{\max}}2\Bigr)\log n-L_Y\Bigr\}\Bigr]\\
&\leq Cn^{d_{\max}-\lambda_Y(3A_1/8-d_{\max}/2)}\leq\frac Cn .
\end{aligned}
\]
Here the last inequality holds because the second term of~\eqref{eq:ms-A1} gives $A_1\geq\frac83\{(d_{\max}+1)/\lambda_Y+d_{\max}/2\}$, that is, $\lambda_Y(3A_1/8-d_{\max}/2)\geq d_{\max}+1$. The envelopes $B_Y$ and $L_Y$ are uniform over the generating pairs.

For (d), since $\mb P_{j,\eta_j}=\mb P_{k,\theta}$, we have $\ell_{n,j}(\eta_j)=\ell_{n,k}(\theta)$ almost surely, and $\ell_{n,k}(\theta)\leq\ell_{n,k}(\widehat\theta_k)$. Hence
\[
2\{\ell_{n,k}(\widehat\theta_k)-\ell_{n,j}(\widehat\theta_j)\}
\geq2\{\ell_{n,j}(\eta_j)-\ell_{n,j}(\widehat\theta_j)\}
=-G_{n,j} .
\]
Using~\eqref{eq:bic-logits}, $d_j-d_k=b_j$ and $\log(w_k/w_j)\geq-\log c_w$, we obtain
\[
\begin{aligned}
a_k^{\rm BIC}-a_j^{\rm BIC}
&=\log(w_k/w_j)+\{\ell_{n,k}(\widehat\theta_k)-\ell_{n,j}(\widehat\theta_j)\}+\tfrac12(d_j-d_k)\log n\\
&\geq-\log c_w-\tfrac12G_{n,j}+\tfrac12b_j\log n
=\tfrac12\{b_j\log n-G_{n,j}\}-\log c_w .
\end{aligned}
\]
For the tail bound, since $\mb P_{k,\theta}=\mb P_{j,\eta_j}$ and $\eta_j\in\Xi_j\subset K_{j,\rho}\subset\operatorname{int}(\Theta_{0,j})$, the uniform assertion of (b), applied to model $j$ on the compact set $K_{j,\rho}$ with $a=b_j$ and $q=d_j$, shows that $G_{n,j}$ satisfies~\eqref{eq:ms-gain-tail} under $\mb P_{k,\theta}^{(n)}$ for $0<w\leq b_j\log n$.
\end{proof}

\subsubsection{Boundary layer and envelope (Lemma~\ref{lem:ms-boundary})}

\begin{proof}
For (a), fix a separated pair $(k,\theta)$ and $j\in\mathcal J(k,\theta)$. Since $\eta_j\in\Xi_j$ and $\varrho_j=\operatorname{dist}(\eta_j,\partial\Xi_j)$, the ball $B(\eta_j,\varrho_j)$ is contained in $\Xi_j$, so $\operatorname{dist}(\widehat\theta_j,\Xi_j)\leq(\|\widehat\theta_j-\eta_j\|_2-\varrho_j)_+$. Moreover, the data have law $\mb P_{k,\theta}^{(n)}=\mb P_{j,\eta_j}^{(n)}$ with $\eta_j\in\Xi_j\subset K_{j,\rho}$. Fix $t\geq0$. If $U_{n,j}=n\operatorname{dist}(\widehat\theta_j,\Xi_j)^2>t$, then $\operatorname{dist}(\widehat\theta_j,\Xi_j)>\sqrt{t/n}$, so $\|\widehat\theta_j-\eta_j\|_2>\varrho_j+\sqrt{t/n}$, and hence $n\|\widehat\theta_j-\eta_j\|_2^2>(\sqrt n\varrho_j+\sqrt t)^2$. Since $(\sqrt n\varrho_j+\sqrt t)^2\geq n\varrho_j^2+t$, the tail bound of Lemma~\ref{lem:ms-laplace} with $s=(\sqrt n\varrho_j+\sqrt t)^2$ gives
\[
\begin{aligned}
\mb P_{j,\eta_j}^{(n)}(U_{n,j}>t)
&\leq\mb P_{j,\eta_j}^{(n)}\bigl\{n\|\widehat\theta_j-\eta_j\|_2^2>(\sqrt n\varrho_j+\sqrt t)^2\bigr\}\\
&\leq Ce^{-c(\sqrt n\varrho_j+\sqrt t)^2}+Ce^{-cn}
\leq Ce^{-cn\varrho_j^2}e^{-ct}+Ce^{-cn}.
\end{aligned}
\]
Let $D_0=\max_{l\in[K]}\operatorname{diam}(\Theta_{0,l})$. Since $\widehat\theta_j\in\Theta_{0,j}$ and $\Xi_j\subset\Theta_{0,j}$, we have $0\leq U_{n,j}\leq nD_0^2$. Integrating the tail bound over $t\in[0,nD_0^2]$ gives
\[
\begin{aligned}
\mb E_{k,\theta}^{(n)}[U_{n,j}]
&=\int_0^{nD_0^2}\mb P_{j,\eta_j}^{(n)}(U_{n,j}>t)\,dt\\
&\leq Ce^{-cn\varrho_j^2}\int_0^\infty e^{-ct}\,dt+CnD_0^2e^{-cn}
=\frac Cc\,e^{-cn\varrho_j^2}+CD_0^2ne^{-cn},
\end{aligned}
\]
which gives $\mb E_{k,\theta}^{(n)}[U_{n,j}]\leq Ce^{-cn\varrho_j^2}+Cne^{-cn}$ after adjusting $C$.

For the second bound, fix $k\in[K]$ and put $F(h)=\pi_k\{\operatorname{dist}(\theta,\partial\Xi_k)<h\}$ for $h>0$. Since $\Xi_k$ is a rectangle, the set of points in $\Xi_k$ within distance $h$ of $\partial\Xi_k$ has Lebesgue measure at most $Ch$. Because $\pi_k$ is supported on $\Xi_k$ with density at most $\bar\pi$, we have $F(h)\leq C\bar\pi h$. For every $\theta\in\Xi_k$, writing $r=\operatorname{dist}(\theta,\partial\Xi_k)$, we have $e^{-cnr^2}=\int_r^\infty 2cnh\,e^{-cnh^2}\,dh$. By Fubini's theorem and the substitution $u=\sqrt{cn}\,h$,
\[
\begin{aligned}
\int e^{-cn\operatorname{dist}(\theta,\partial\Xi_k)^2}\,\pi_k(d\theta)
&=\int_0^\infty 2cnh\,e^{-cnh^2}F(h)\,dh\\
&\leq C\bar\pi\int_0^\infty 2cnh^2e^{-cnh^2}\,dh
=\frac{C\bar\pi}{\sqrt{cn}}\int_0^\infty 2u^2e^{-u^2}\,du.
\end{aligned}
\]
Since $\int_0^\infty 2u^2e^{-u^2}\,du<\infty$, this gives $\int e^{-cn\operatorname{dist}(\theta,\partial\Xi_k)^2}\,\pi_k(d\theta)\leq C/\sqrt n$ after adjusting $C$.

For (b), fix a generating pair $(k,\theta)$ with $\theta\in\Xi_k$, and let $d_{\max}=\max_{j\in[K]}d_j$. Since $\vartheta\mapsto\ell_{k,\vartheta}(x)$ has Lipschitz constant at most $L_1$ on $\Theta_{0,k}$ for every $x\in\mathcal X$, summing over the $n$ observations gives $|\ell_{n,k}(\vartheta)-\ell_{n,k}(\theta)|\leq nL_1\|\vartheta-\theta\|_2\leq L_1$, and hence $\ell_{n,k}(\vartheta)\geq\ell_{n,k}(\theta)-L_1$, for all $\vartheta\in B(\theta,1/n)\cap\Xi_k$. Because the density of $\pi_k$ is at least $\underline\pi$ on the rectangle $\Xi_k$, we have $\pi_k\{B(\theta,1/n)\cap\Xi_k\}\geq c\,n^{-d_k}$ for all $\theta\in\Xi_k$ and $n\geq n_0$. Therefore, for every dataset,
\[
m_k=\int e^{\ell_{n,k}(\vartheta)}\,\pi_k(d\vartheta)
\geq e^{\ell_{n,k}(\theta)-L_1}\,\pi_k\{B(\theta,1/n)\cap\Xi_k\}
\geq c\,n^{-d_k}e^{\ell_{n,k}(\theta)-L_1}.
\]
Next, note that $T_n\geq0$ since $\theta\in\Xi_k\subset\Theta_{0,k}$. By~\eqref{eq:bic-logits}, $e^{a_l^{\rm BIC}}=w_ln^{-d_l/2}e^{\ell_{n,l}(\widehat\theta_l)}\leq w_le^{\ell_{n,l}(\widehat\theta_l)}$ for each $l\in[K]$. Since $\sum_lw_l=1$ and $\ell_{n,l}(\widehat\theta_l)\leq\ell_{n,k}(\theta)+T_n$ by the definition of $T_n$, we have $\sum_le^{a_l^{\rm BIC}}\leq\max_le^{\ell_{n,l}(\widehat\theta_l)}\leq e^{\ell_{n,k}(\theta)+T_n}$. For every $j\in[K]$, we have $q_j^*=w_jm_j/\sum_lw_lm_l$ and $q_j^{\rm BIC}=e^{a_j^{\rm BIC}}/\sum_le^{a_l^{\rm BIC}}$, with $e^{a_j^{\rm BIC}}=w_jn^{-d_j/2}e^{\ell_{n,j}(\widehat\theta_j)}$. By Lemma~\ref{lem:ms-laplace}, $m_j\leq e^{\ell_{n,j}(\widehat\theta_j)}$ for every dataset. Combining these relations with $n^{d_j/2}\leq n^{d_{\max}/2}$, $\sum_lw_lm_l\geq w_km_k$ and the lower bound on $m_k$ gives
\[
\frac{q_j^*}{q_j^{\rm BIC}}
=\frac{m_j}{n^{-d_j/2}e^{\ell_{n,j}(\widehat\theta_j)}}\cdot\frac{\sum_le^{a_l^{\rm BIC}}}{\sum_lw_lm_l} \leq n^{d_{\max}/2}\,\frac{e^{\ell_{n,k}(\theta)+T_n}}{w_k\,c\,n^{-d_k}e^{\ell_{n,k}(\theta)-L_1}}
\leq Cn^{3d_{\max}/2}e^{T_n}.
\]
Since $\sum_{j=1}^Kq_j^*=1$, we obtain, for every dataset,
\[
D_{\rm KL}(q^*\|q^{\rm BIC})
=\sum_{j=1}^Kq_j^*\log\frac{q_j^*}{q_j^{\rm BIC}}
\leq\log\bigl(Cn^{3d_{\max}/2}\bigr)+T_n
\leq C\log n+T_n.
\]

For the moment bound, take a $(1/n)$-net $\mathcal N_j$ in each working set $\Theta_{0,j}$. By the covering bound in the proof of Lemma~\ref{lem:net-concentration}, the total number of net points satisfies $N=\sum_{j=1}^K|\mathcal N_j|\leq Cn^{d_{\max}}$. For a fixed $j\in[K]$ and $\vartheta\in\mathcal N_j$, $\exp\{\ell_{n,j}(\vartheta)-\ell_{n,k}(\theta)\}$ is the likelihood ratio of $\mb P_{j,\vartheta}^{(n)}$ to $\mb P_{k,\theta}^{(n)}$, so $\mb E_{k,\theta}^{(n)}[\exp\{\ell_{n,j}(\vartheta)-\ell_{n,k}(\theta)\}]\leq1$. Put $M_n=\max_{j\in[K]}\max_{\vartheta\in\mathcal N_j}\{\ell_{n,j}(\vartheta)-\ell_{n,k}(\theta)\}$. For $v\geq0$, Markov's inequality and the union bound over the $N$ net points give $\mb P_{k,\theta}^{(n)}(M_n>\log N+v)\leq Ne^{-\log N-v}=e^{-v}$. For any $j\in[K]$ and $\vartheta\in\Theta_{0,j}$, choosing $\vartheta'\in\mathcal N_j$ with $\|\vartheta-\vartheta'\|_2\leq1/n$ gives $|\ell_{n,j}(\vartheta)-\ell_{n,j}(\vartheta')|\leq nL_1\|\vartheta-\vartheta'\|_2\leq L_1$, so $\ell_{n,j}(\vartheta)-\ell_{n,k}(\theta)\leq M_n+L_1$. Taking the supremum over $\vartheta\in\Theta_{0,j}$ and the maximum over $j\in[K]$ yields $T_n\leq M_n+L_1$, and hence $\mb P_{k,\theta}^{(n)}(T_n>L_1+\log N+v)\leq\mb P_{k,\theta}^{(n)}(M_n>\log N+v)\leq e^{-v}$ for all $v\geq0$. Since $T_n\geq0$ and $N\leq Cn^{d_{\max}}$, integrating this tail bound gives $\mb E_{k,\theta}^{(n)}[T_n^2]\leq C(\log n)^2$.
\end{proof}

\subsection{Proofs of the lemmas for variable selection}\label{app:proof-vs-lemmas}

\subsubsection{Natural moments and inverse information in the GLM (Lemma~\ref{lem:glm-natural-modules})}

The natural moments for a canonical GLM play the same role as the moment function in the general estimation problem. Here we exploit their relationship to the Fisher information to obtain an invertible moment map and uniform bounds on its derivatives.

\begin{proof}
The constants $c_\mu,h_*,K_3,L_\mu,r_\mu,\rho_\mu$ depend only on the response family, $a$, $m_b$, $M_b$, $\underline\lambda_Z$, $\overline\lambda_Z$, $B$, $\rho$ and $c_\Theta$. In particular, they are independent of $n$, $p$, $S$ and $\theta_0$. We divide the proof into three steps.

\medskip
\noindent{\bf Step 1: Cumulant function.} Let $z_0\in\mathbb R$ be a point where the partition integral is finite and define
\[
\mathfrak Z(z)=\int e^{yz/a+c(y,a)}\,\nu_Y(dy),\qquad
m_k(z)=\frac{\int y^ke^{yz/a+c(y,a)}\,\nu_Y(dy)}{\mathfrak Z(z)}.
\]
Since $\nu_Y$ is finite and supported on $[-Y_*,Y_*]$, $|e^{y(z-z_0)/a}|\leq\exp\{Y_*|z-z_0|/a\}$ there. As $\mathfrak Z(z_0)<\infty$, the integral converges for all $z\in\mathbb C$ and the exponential series integrates term by term, locally uniformly, so $\mathfrak Z$ is entire. For $z=u+it$ with $|t|<\pi a/(2Y_*)$ and $|y|\leq Y_*$, $|yt/a|\leq Y_*|t|/a<\pi/2$, so $\operatorname{Re}e^{y(u+it)/a}=e^{yu/a}\cos(yt/a)\geq e^{yu/a}\cos(Y_*|t|/a)$. Integrating gives
\[
\operatorname{Re}\mathfrak Z(u+it)\geq\cos(Y_*|t|/a)\mathfrak Z(u)>0.
\]
Since $\operatorname{Re}\mathfrak Z>0$, the principal logarithm gives $b(z)=a\log\mathfrak Z(z)\in\operatorname{Hol}\{z\in\mathbb C:|\operatorname{Im}z|<\pi a/(2Y_*)\}$. Density normalization gives $e^{b(u)/a}=\mathfrak Z(u)$ for real $u$, so $b$ extends the real cumulant. Since $|e^{yz/a}|=e^{yu/a}$, the numerator of $m_k(z)$ has modulus at most $Y_*^k\mathfrak Z(u)$, and $|\mathfrak Z(z)|\geq\operatorname{Re}\mathfrak Z(z)$ gives
\[
|m_k(z)|\leq\frac{Y_*^k\mathfrak Z(u)}{\operatorname{Re}\mathfrak Z(z)}\leq\frac{Y_*^k}{\cos(Y_*|t|/a)}.
\]
On $|t|\leq h_*=\pi a/(4Y_*)$, $\cos(Y_*|t|/a)\geq\cos(\pi/4)=1/\sqrt2$, so $|m_k(z)|\leq\sqrt2Y_*^k$. Differentiating $b=a\log\mathfrak Z$ gives $b'=m_1$, $b''=a^{-1}(m_2-m_1^2)$ and $b'''=a^{-2}(m_3-3m_1m_2+2m_1^3)$. Hence $|b'(z)|\leq\sqrt2Y_*$ and
\[
|b'''(z)|\leq a^{-2}(\sqrt2+3\cdot2+2\cdot2\sqrt2)Y_*^3=(5\sqrt2+6)Y_*^3/a^2\leq K_3,
\]
since $5\sqrt2+6<16$. For real $u$, let $P_u$ be the probability measure on $[-Y_*,Y_*]$ with $P_u(dy)\propto e^{yu/a+c(y,a)}\nu_Y(dy)$. Then $m_k(u)$ is its $k$th moment. Thus $|b'(u)|=|m_1(u)|\leq Y_*$. The variance of $P_u$ is $ab''(u)\in[0,Y_*^2]$, so $0\leq b''(u)\leq Y_*^2/a$. By the mean value theorem, $|b(u)-b(0)|\leq Y_*|u|$, so $|b(u)|\leq|b(0)|+Y_*|u|$. This proves~\eqref{eq:glm-cumulant-strip}.

\medskip
\noindent{\bf Step 2: Moment map and its inverse.} Differentiation under the bounded integral and~\eqref{eq:glm-score-information} give $D\mu_S(\theta)=E[b''(v_S^\top\theta)v_Sv_S^\top]=aI_S(\theta)$. For $\theta\in\Theta_{0,S}$, the predictors lie in $\mathcal I_{\rm pred}$, where $m_b\leq b''\leq M_b$, so~\eqref{eq:glm-sparse-design} gives $c_\mu I_{d_S}\preceq D\mu_S(\theta)\preceq M_b\overline\lambda_ZI_{d_S}$, which is the eigenvalue claim of the lemma. Whenever every predictor on the segment between real $\theta,\theta'$ lies in $\mathcal I_{\rm pred}$, we have $D\mu_S\succeq c_\mu I_{d_S}$ along it, so
\[
\langle\mu_S(\theta)-\mu_S(\theta'),\theta-\theta'\rangle
=\int_0^1(\theta-\theta')^\top D\mu_S(\theta'+t(\theta-\theta'))(\theta-\theta')\,dt \geq c_\mu\|\theta-\theta'\|_2^2.
\]
Since $\Theta_{0,S}$ is convex, this holds on $\Theta_{0,S}$ and $\mu_S$ is injective there.

Set $r_{\mu,D}=r_\mu/\sqrt D$. If every predictor on the complex segment from $\theta$ to $z$ lies in $|\operatorname{Im}\eta|\leq h_*$, then $D\mu_S(z)-D\mu_S(\theta)=E[\{b''(v_S^\top z)-b''(v_S^\top\theta)\}v_Sv_S^\top]$, and $|b'''|\leq K_3$ on this strip by Step~1 gives $|b''(v_S^\top z)-b''(v_S^\top\theta)|\leq K_3|v_S^\top(z-\theta)|$. Hence
\begin{equation}\label{eq:glm-moment-derivative-lipschitz}
\begin{aligned}
 \|D\mu_S(z)-D\mu_S(\theta)\|_{\rm op}
 &\leq K_3V_*\sqrt D\|z-\theta\|_2
 \sup_{\|u\|_2=\|w\|_2=1}E|u^*v_S||v_S^\top w|\leq L_\mu\sqrt D\|z-\theta\|_2.
\end{aligned}
\end{equation}
The first inequality uses $|v_S^\top(z-\theta)|\leq V_*\sqrt D\|z-\theta\|_2$. For the second,~\eqref{eq:glm-sparse-design} on the real and imaginary parts of $u$ gives $E|u^*v_S|^2\leq\overline\lambda_Z\|u\|_2^2$ (similarly for $w$), so complex Cauchy--Schwarz and $L_\mu=K_3V_*\overline\lambda_Z$ give the bound.

The set $U_S=\Theta_{0,S}+B_2(0,2r_{\mu,D})$ is convex and open, and the canonical formula extends to $U_S$ via the entire partition function. For real $z\in U_S$ and its nearest point $\theta\in\Theta_{0,S}$, we have $\|z-\theta\|_2<2r_{\mu,D}$ and real predictors on the segment. Then~\eqref{eq:glm-moment-derivative-lipschitz}, $\sqrt Dr_{\mu,D}=r_\mu$ and $r_\mu\leq c_\mu/(4L_\mu)$ give $\|D\mu_S(z)-D\mu_S(\theta)\|_{\rm op}\leq 2L_\mu r_\mu\leq c_\mu/2$, and $\lambda_{\min}\{D\mu_S(\theta)\}\geq c_\mu$ gives $\lambda_{\min}\{D\mu_S(z)\}\geq c_\mu/2$. On the convex set $U_S$, the segment calculation with $c_\mu/2$ gives $\langle\mu_S(z)-\mu_S(z'),z-z'\rangle\geq(c_\mu/2)\|z-z'\|_2^2$ for $z,z'\in U_S$. By the Cauchy--Schwarz inequality, $\|\mu_S(z)-\mu_S(z')\|_2\geq(c_\mu/2)\|z-z'\|_2$, so $\mu_S$ is injective on $U_S$ and $\|\mu_S^{-1}(w)-\mu_S^{-1}(w')\|_2\leq 2c_\mu^{-1}\|w-w'\|_2$ on $\mu_S(U_S)$.

Fix $\theta\in\Theta_{0,S}$, $A=D\mu_S(\theta)$ and $\|w-\mu_S(\theta)\|_2\leq c_\mu r_{\mu,D}/2$. On $\|q\|_2\leq r_{\mu,D}$ set $\Phi_w(q)=q+A^{-1}\{w-\mu_S(\theta+q)\}$, with derivative $D\Phi_w(q)=A^{-1}\{A-D\mu_S(\theta+q)\}$. Here $\|A^{-1}\|_{\rm op}\leq c_\mu^{-1}$, and~\eqref{eq:glm-moment-derivative-lipschitz} gives $\|A-D\mu_S(\theta+q)\|_{\rm op}\leq L_\mu\sqrt D\|q\|_2\leq L_\mu r_\mu\leq c_\mu/4$, so $\|D\Phi_w(q)\|_{\rm op}\leq 1/4$ and $\|\Phi_w(q)-\Phi_w(q')\|_2\leq\tfrac14\|q-q'\|_2$. Since $\|\Phi_w(0)\|_2=\|A^{-1}\{w-\mu_S(\theta)\}\|_2\leq r_{\mu,D}/2$,
\[
\|\Phi_w(q)\|_2\leq\|\Phi_w(0)\|_2+\|\Phi_w(q)-\Phi_w(0)\|_2\leq r_{\mu,D}/2+r_{\mu,D}/4=3r_{\mu,D}/4.
\]
Thus $\Phi_w$ is a contraction mapping the closed ball $\|q\|_2\leq r_{\mu,D}$ into itself, and its fixed point $q$ solves $\mu_S(\theta+q)=w$ with $\theta+q\in U_S$.

Hence $\overline B_2(\mu_S(\theta),c_\mu r_{\mu,D}/2)\subset\mu_S(U_S)$, and $\rho_\mu/D=c_\mu r_\mu/(4D)\leq c_\mu r_\mu/(2\sqrt D)=c_\mu r_{\mu,D}/2$. Since $\mu_S(\Theta_{0,S})$ is compact, every $w\in\mathcal K_{\mu,S}$ lies within $\rho_\mu/D$ of some $\mu_S(\theta)$. Thus $\mathcal K_{\mu,S}\subset\mu_S(U_S)$, so $\mu_S^{-1}$ is defined on $\mathcal K_{\mu,S}$ with Lipschitz constant $2/c_\mu$, as stated in the lemma. The same contraction on sufficiently small complex neighborhoods converges locally uniformly. Each iterate is holomorphic, so the inverse is holomorphic there and smooth on the real image.

\medskip
\noindent{\bf Step 3: Inverse information on the ellipse.} Let $A_S$ be the affine map in the lemma and put $t=\min\{1,r_\mu/(4(1\vee B)D)\}$, so $\varrho=\varrho_{I,D}=1+t$ in~\eqref{eq:glm-information-ellipse}. Take $\zeta\in E_\varrho^{d_S}$ and $z=A_S\zeta$. The semi-axes of $E_\varrho$ satisfy $(\varrho+\varrho^{-1})/2-1\leq(\varrho-\varrho^{-1})/2\leq t$, so each $\zeta_j$ lies within $\sqrt2\,t\leq2t$ of its nearest point $\zeta_j^\circ\in[-1,1]$. Since the sides of $\Theta_{0,S}\subseteq[-B,B]^{d_S}$ have half-length at most $B$, the point $x=A_S\zeta^\circ\in\Theta_{0,S}$ has $|z_j-x_j|\leq2(1\vee B)t$ for each of the $d_S\leq D$ coordinates, and the definition of $t$ gives
\[
\operatorname{dist}_2(z,\Theta_{0,S})\leq\|z-x\|_2\leq2(1\vee B)\sqrt D\,t\leq r_\mu/(2\sqrt D)=r_{\mu,D}/2.
\]
For a nearest $\theta\in\Theta_{0,S}$, both $\theta$ and $v_S(Z)$ are real, so the predictor on the segment has imaginary part $u\operatorname{Im}\{v_S(Z)^\top(z-\theta)\}$, whose modulus is at most $\|v_S(Z)\|_2\|z-\theta\|_2$. Since $\|v_S(Z)\|_2\leq V_*\sqrt D$ $Q_Z$-almost surely, $\|z-\theta\|_2\leq r_{\mu,D}/2$ and $r_\mu\leq h_*/(4V_*)$, we have, $Q_Z$-almost surely,
\[
\sup_{0\leq u\leq1}|\operatorname{Im}\{v_S(Z)^\top[\theta+u(z-\theta)]\}|\leq V_*\sqrt D\|z-\theta\|_2\leq V_*r_\mu/2\leq h_*/8.
\]
Here $I_S(z)=a^{-1}E[b''(v_S^\top z)v_Sv_S^\top]=a^{-1}D\mu_S(z)$ uses the extension of $b$ from Step~1. Hence~\eqref{eq:glm-moment-derivative-lipschitz} applies, and $r_\mu\leq c_\mu/(4L_\mu)$ gives
\[
\|I_S(z)-I_S(\theta)\|_{\rm op}\leq a^{-1}L_\mu\sqrt D\|z-\theta\|_2\leq L_\mu r_\mu/(2a)\leq c_\mu/(8a).
\]
By Step~2, $aI_S(\theta)=D\mu_S(\theta)\succeq c_\mu I_{d_S}$, so $\|I_S(\theta)^{-1}\|_{\rm op}\leq a/c_\mu$. Put $Q=I_S(\theta)^{-1}\{I_S(z)-I_S(\theta)\}$, so $I_S(z)=I_S(\theta)(I_{d_S}+Q)$ and $\|Q\|_{\rm op}\leq(a/c_\mu)\{c_\mu/(8a)\}=1/8<1$. Hence the Neumann series $I_S(z)^{-1}=\sum_{j=0}^{\infty}(-Q)^jI_S(\theta)^{-1}$ converges. Since $\sum_{j=0}^{\infty}8^{-j}=8/7$,
\[
\|I_S(z)^{-1}\|_{\rm op}\leq\frac{a/c_\mu}{1-1/8}=\frac{8a}{7c_\mu}<\frac{2a}{c_\mu},
\]
which gives the bound in~\eqref{eq:glm-information-ellipse}. The entries of $z\mapsto I_S(z)$ are holomorphic near $z$ and the Neumann series converges locally uniformly, so the entries of $I_S(z)^{-1}$ are holomorphic there. The strict margins and compactness extend $\zeta\mapsto I_S(A_S\zeta)^{-1}$ to an open neighborhood of the closed normalized product ellipse, as required in Lemma~\ref{lem:analytic-approx}.
\end{proof}

\subsubsection{Analytic extension of the moment inverse (Lemma~\ref{lem:glm-inverse-extension})}

The goal is to extend the real inverse moment map to a complex neighborhood of a rectangle on which it will be approximated. Local invertibility at each real parameter does not immediately give such an extension on a common rectangle. The proof constructs the inverse on a neighborhood of the working set with quantitative control of the radius.

\begin{proof}
We prove each part separately.

Write $d=d_S$, $v=v_S(Z)$ and $E$ for expectation under $Q_Z$. On $\mathbb C^d$ let $\langle x,y\rangle=\sum_jx_j\overline{y_j}$, so that $\|x\|_2^2=\langle x,x\rangle$, and put $\delta_+=2r_+/\sqrt D\leq2$. Since every coordinate of $v$ is bounded by $V_*$ and $d\leq D$, we have $\|v\|_2\leq V_*\sqrt D$ almost surely.

We first show that $\mu_S$ is injective on $\mathbb R^d$. By the proof of Lemma~\ref{lem:glm-natural-modules}, for real $u$ the quantity $ab''(u)=m_2(u)-m_1(u)^2$ is the variance of the probability measure $P_u(dy)\propto e^{yu/a+c(y,a)}\nu_Y(dy)$. These measures are mutually absolutely continuous. If $b''(u)=0$ for some real $u$, then $P_u$ is a point mass, so every $P_{u'}$ is the same point mass and $b''$ vanishes identically, which contradicts $b''\geq m_b>0$ on $\mathcal I_{\rm pred}$. Hence $b''>0$ on $\mathbb R$. For real $\theta\ne\theta'$, write $\eta=v^\top\theta$ and $\eta'=v^\top\theta'$. Then
\[
\begin{aligned}
 \langle\mu_S(\theta)-\mu_S(\theta'),\theta-\theta'\rangle
 &=E\bigl[\{b'(\eta)-b'(\eta')\}(\eta-\eta')\bigr]=E\Bigl[(\eta-\eta')^2\int_0^1b''\{\eta'+t(\eta-\eta')\}\,dt\Bigr].
\end{aligned}
\]
The integrand is nonnegative, and it is positive whenever $\eta\ne\eta'$. By~\eqref{eq:glm-sparse-design}, $E(\eta-\eta')^2\geq\underline\lambda_Z\|\theta-\theta'\|_2^2>0$, so $\eta\ne\eta'$ with positive probability and the display is positive. Thus $\mu_S(\theta)\ne\mu_S(\theta')$.

We next extend the moment map to a complex domain on which its derivative has uniformly positive curvature. Let $\Omega=\bigl\{w\in\mathbb C^d:\operatorname{dist}(w,\Theta_S^+)<\delta_+\bigr\}$,
which is open, convex, and invariant under complex conjugation. For $w\in\Omega$ choose $\theta\in\Theta_S^+$ with $\|w-\theta\|_2<\delta_+$. Almost surely, for $0\leq t\leq1$,
\[
 \bigl|\operatorname{Im}v^\top\{\theta+t(w-\theta)\}\bigr|
 \leq\|v\|_2\|w-\theta\|_2
 \leq V_*\sqrt D\,\delta_+=2V_*r_+\leq h_*/2 .
\]
By~\eqref{eq:glm-cumulant-strip}, $b'$ is holomorphic and bounded by $\sqrt2Y_*$ on the strip $|\operatorname{Im}z|\leq h_*$. Hence $G(w)=E[b'(v^\top w)v]$ is well defined and bounded on $\Omega$, and differentiation under the bounded integral shows that $G$ is holomorphic with $DG(w)=E[b''(v^\top w)vv^\top]$. It equals $\mu_S$ at real points, and $G(\overline w)=\overline{G(w)}$ because $b'$ is real on $\mathbb R$. The segment from $\theta$ to $w$ satisfies the hypothesis of~\eqref{eq:glm-moment-derivative-lipschitz}, so
\[
 \|DG(w)-D\mu_S(\theta)\|_{\rm op}
 \leq L_\mu\sqrt D\|w-\theta\|_2<2L_\mu r_+\leq c_+/2 .
\]
For $h\in\mathbb C^d$, the curvature bound on $\Theta_S^+$ and~\eqref{eq:glm-sparse-design}, applied to the real and imaginary parts of $h$, give
\[
 \langle D\mu_S(\theta)h,h\rangle
 =E\bigl[b''(v^\top\theta)|v^\top h|^2\bigr]
 \geq m_+\bigl\{E(v^\top\operatorname{Re}h)^2+E(v^\top\operatorname{Im}h)^2\bigr\}
 \geq c_+\|h\|_2^2 .
\]
Combining the last two displays,
\begin{equation}
 \operatorname{Re}\langle DG(w)h,h\rangle\geq\frac{c_+}2\|h\|_2^2,
 \qquad w\in\Omega,\ h\in\mathbb C^d .
 \label{eq:glm-complex-curvature}
\end{equation}
In particular $DG(w)$ is invertible for every $w\in\Omega$.

We now show that $G$ is injective on $\Omega$ and has a holomorphic inverse. For $w_1,w_2\in\Omega$ put $h=w_1-w_2$. The segment between them lies in $\Omega$, and $G(w_1)-G(w_2)=\int_0^1DG(w_2+th)h\,dt$. By~\eqref{eq:glm-complex-curvature} and the Cauchy--Schwarz inequality,
\[
 \|G(w_1)-G(w_2)\|_2\|h\|_2
 \geq\operatorname{Re}\langle G(w_1)-G(w_2),h\rangle
 \geq\frac{c_+}2\|h\|_2^2,
\]
and therefore
\begin{equation}
 \|G(w_1)-G(w_2)\|_2\geq\frac{c_+}2\|w_1-w_2\|_2,
 \qquad w_1,w_2\in\Omega .
 \label{eq:glm-complex-lower-lipschitz}
\end{equation}
Thus $G$ is injective on $\Omega$. Since its derivative is invertible, the holomorphic inverse function theorem shows that $G$ is an open map, that $G(\Omega)$ is open, and that $G^{-1}:G(\Omega)\to\Omega$ is holomorphic.

It remains to show that $N_S^+\subseteq G(\Omega)$. Fix $\theta\in\Theta_S^+$ and $m\in\mathbb C^d$ with $\|m-\mu_S(\theta)\|_2<\rho_+/\sqrt D=c_+\delta_+/8$. Choose $\delta'<\delta_+$ with $\|m-\mu_S(\theta)\|_2<c_+\delta'/8$. The closed ball $\overline B=\{w:\|w-\theta\|_2\leq\delta'\}$ is a compact subset of $\Omega$, so $w\mapsto\|G(w)-m\|_2$ attains its minimum over $\overline B$ at some $w_*$. If $\|w_*-\theta\|_2=\delta'$, then~\eqref{eq:glm-complex-lower-lipschitz} gives
\[
 \|G(w_*)-m\|_2\geq\|G(w_*)-G(\theta)\|_2-\|G(\theta)-m\|_2
 >\frac{c_+\delta'}2-\frac{c_+\delta'}8>\|G(\theta)-m\|_2,
\]
which contradicts minimality because $\theta\in\overline B$. Hence $w_*$ is an interior point of $\overline B$. Since $G$ is open, $G(\overline B)$ contains a neighborhood of $G(w_*)$. If $G(w_*)\ne m$, this neighborhood contains a point closer to $m$ than $G(w_*)$, again contradicting minimality. Therefore $G(w_*)=m$, and $N_S^+\subseteq G(\Omega)$. The set $N_S^+$ is open because $\mu_S(\Theta_S^+)$ is compact.

Let $G_S^{-1}$ be the restriction of $G^{-1}$ to $N_S^+$. If $m\in N_S^+$ is real and $w=G_S^{-1}(m)$, then $\overline w\in\Omega$ and $G(\overline w)=\overline m=m$, so $\overline w=w$ by injectivity. Thus $w$ is real and $\mu_S(w)=m$, and by the injectivity of $\mu_S$ on $\mathbb R^d$ it is the only real solution of this equation. Hence $G_S^{-1}$ agrees at real points with the inverse of $\mu_S$, and in particular with the map $\mu_S^{-1}$ of Lemma~\ref{lem:glm-natural-modules} wherever both are defined. Finally, $G_S^{-1}$ takes values in $\Omega$, whose points have coordinates bounded in modulus by $B_++\delta_+\leq B_++2$. This proves part~\textup{(a)}.

We turn to part~\textup{(b)}. Write $\varrho=\varrho_g=1+t$ with $t=\min\{1,\rho_+/(4D\bar h)\}$. The region $E_\varrho$ is bounded by the ellipse with semi-axes $\alpha=(\varrho+\varrho^{-1})/2$ along the real axis and $\beta=(\varrho-\varrho^{-1})/2$ along the imaginary axis, so every $\zeta\in E_\varrho$ has $|\operatorname{Re}\zeta|\leq\alpha$ and $|\operatorname{Im}\zeta|\leq\beta$. Its nearest point $\zeta^\circ$ in $[-1,1]$ therefore satisfies
\[
 |\zeta-\zeta^\circ|^2\leq(\alpha-1)^2+\beta^2\leq2\beta^2,
 \qquad
 \beta=\frac{t(2+t)}{2(1+t)}\leq t,
\]
where $\alpha-1=(\varrho-1)^2/(2\varrho)\leq\beta$. Let $A_{\mathcal R}$ be the affine normalization of $\mathcal R$ in Appendix~\ref{app:analytic-extension}, extended to $\mathbb C^d$. For $\zeta\in E_\varrho^d$, the point $z=A_{\mathcal R}^{-1}\zeta$ has coordinates $z_j=(u_j+v_j)/2+\zeta_j(v_j-u_j)/2$. Replacing each $\zeta_j$ by $\zeta_j^\circ$ gives a point $x\in\mathcal R$ with
\[
 \|z-x\|_2\leq\sqrt d\,\bar h\,\sqrt2\,t
 \leq\sqrt{2D}\,\bar h\,\frac{\rho_+}{4D\bar h}
 =\frac{\sqrt2}4\frac{\rho_+}{\sqrt D}.
\]
Since $\operatorname{dist}\{x,\mu_S(\Theta_S^+)\}\leq\rho_+/(2\sqrt D)$,
\[
 \operatorname{dist}\{z,\mu_S(\Theta_S^+)\}
 \leq\Bigl(\frac{\sqrt2}4+\frac12\Bigr)\frac{\rho_+}{\sqrt D}<\frac{\rho_+}{\sqrt D},
\]
so $A_{\mathcal R}^{-1}(E_\varrho^d)\subseteq N_S^+$. Put $O=A_{\mathcal R}(N_S^+)$ and $F=G_S^{-1}\circ A_{\mathcal R}^{-1}$. Then $O$ is open and contains $E_\varrho^d$, and $F$ is holomorphic on $O$. For $x\in K_{g,S}\subseteq\mathcal R$, part~\textup{(a)} gives $F\{A_{\mathcal R}(x)\}=G_S^{-1}(x)=\mu_S^{-1}(x)$. Moreover $F([-1,1]^d)=G_S^{-1}(\mathcal R)\subseteq\mathbb R^d$, and every coordinate of $F$ is bounded in modulus by $B_++2$ on $O$. This is~\eqref{eq:extension-main} with $K=K_{g,S}$ and constants $(\mathcal R,\varrho_g,\mathcal B_g)$.

Finally, we prove part~\textup{(c)}. Here $\Theta_S^+=\Theta_{0,S}$ and $m_+=m_b$, which is admissible by Assumption~\ref{ass:sparse}\textup{(i)}, and then $c_+=c_\mu$, $r_+=r_\mu$ and $\rho_+=\rho_\mu$. By Lemma~\ref{lem:glm-natural-modules}, $\mu_S$ is Lipschitz on the convex set $\Theta_{0,S}$ with constant $M_b\overline\lambda_Z$.

Let $\mathcal R$ be the smallest rectangle containing $K_{g,S}$. Since
\[
 \operatorname{diam}_2(K_{g,S})\leq M_b\overline\lambda_Z\operatorname{diam}_2(\Theta_{0,S})+\frac\rho D
 \leq\frac{\rho_\mu}{4D}+\frac\rho D,
\]
every side of $\mathcal R$ has length at most this bound, and every $x\in\mathcal R$ is within $\sqrt d\operatorname{diam}_2(K_{g,S})$ of a point of $K_{g,S}$. With $d\leq D$ and $\rho\leq\rho_\mu/6$,
\[
 \operatorname{dist}\{x,\mu_S(\Theta_{0,S})\}
 \leq\sqrt D\Bigl(\frac{\rho_\mu}{4D}+\frac\rho D\Bigr)+\frac\rho{2D}
 \leq\frac{\rho_\mu}{4\sqrt D}+\frac{3\rho}{2\sqrt D}
 \leq\frac{\rho_\mu}{2\sqrt D}.
\]
Thus part~\textup{(b)} applies. The half side lengths satisfy $\bar h\leq\rho_\mu/(8D)+\rho/(2D)<\rho_\mu/(4D)$, so $\rho_\mu/(4D\bar h)>1$ and $\varrho_g=2$. Finally, $K_{g,S}$ contains a Euclidean ball of radius $\rho/(2D)$, so every side of $\mathcal R$ has length at least $\rho/D$.
\end{proof}

\subsubsection{Simultaneous local expansions (Lemma~\ref{lem:sparse-local})}

For every support containing the true support, the model is correctly specified. We need the usual root-$n$ localization and one-step likelihood expansion to hold simultaneously over these supports, since their number grows with $p$.

\begin{proof}
In this proof, $C,c$ depend only on the response family, $a$, $m_b$, $M_b$, $\underline\lambda_Z$, $\overline\lambda_Z$, $B$, $\rho$ and $c_\Theta$, and they are independent of $n,p,S$ and $\theta_0$. The radius constants $C_2,C_5$ below are fixed in the same way. We divide the proof into two steps.

\medskip
\noindent{\bf Step 1: The event $\mathcal E_n$.} Put $t_n=\sqrt{D\Lambda_n/n}$, $\mathcal S_+=\{S\in\mathcal S:S\supseteq S_0\}$, $a_S=P_ns_{S,\theta_{0,S}}$ and $H_{n,S}=-n^{-1}\nabla^2\ell_{n,S}$. The mesh $h=n^{-2}/(B\sqrt D)$ gives $B\sqrt D\,h=n^{-2}$. Since $\Theta_{0,S}\subseteq[-B,B]^{d_S}$ lies in a ball of radius $B\sqrt D$, Lemma~\ref{lem:lik-process} gives an $h$-net of $\Theta_{0,S}$ of size $N_S(h)\leq(1+2B\sqrt D/h)^{d_S}=(1+2B^2Dn^2)^{d_S}$. With $d_S\leq D$ and $\log|\mathcal S|\leq s_{\max}\log(ep)$, we have
\[
\log\sum_{S\in\mathcal S}N_S(h)\leq\log|\mathcal S|+D\log(1+2B^2Dn^2)\leq C\Lambda_n.
\]
For $C_2$ fixed below, define
\[
\begin{aligned}
\mathcal A_n&=\Bigl\{\max_{S\in\mathcal S_+}\|\widehat\theta_S-\theta_{0,S}\|_2\leq C_2t_n\Bigr\}, \mathcal B_n=\Bigl\{\max_{S\in\mathcal S_+}\bigl(\|a_S\|_2+\|\overline T_S-\mu_S(\theta_{0,S})\|_2\bigr)\leq C_2t_n\Bigr\},\\
\mathcal C_n&=\Bigl\{\max_{S\in\mathcal S_+}\|H_{n,S}(\theta_{0,S})-I_S(\theta_{0,S})\|_{\rm op}\leq C_2Dt_n\Bigr\},
\end{aligned}
\]
and $\mathcal E_n=\mathcal A_n\cap\mathcal B_n\cap\mathcal C_n$. Since $\Lambda_n\geq1$ and $\Lambda_n^3\leq c_0n$, we have $t_n^2=D\Lambda_n/n\leq D\Lambda_n^3/n\leq Dc_0$. Also $n\geq c_0^{-1}$ gives $\Lambda_n\geq\log(e/c_0)$, so small $c_0$ ensures $\Lambda_n^2\geq D$ and $t_n\leq\sqrt{c_0}$. Decrease $c_0$ so that $C_2t_n\leq C_2\sqrt{c_0}\leq1$.

For $\mathcal A_n$, put $c_K=m_b\underline\lambda_Z/(2a)$ from~\eqref{eq:glm-kl-parameter} and $Y_{S,z}=\log(p_{\theta_0}/p_{S,z})$. For $S\in\mathcal S_+$, the true density is $p_{S,\theta_{0,S}}$, so $Y_{S,\theta_{0,S}}=0$ and the maximizer $\widehat\theta_S$ satisfies $P_nY_{S,\widehat\theta_S}\leq0$. Hence, if $\|\widehat\theta_S-\theta_{0,S}\|_2>C_2t_n$, then the infimum of $P_nY_{S,z}$ over $\{z\in\Theta_{0,S}:\|z-\theta_{0,S}\|_2>C_2t_n\}$ is at most $0$, giving
\[
\mathcal A_n^c\subseteq\bigcup_{S\in\mathcal S_+}
\Bigl\{\inf_{\substack{z\in\Theta_{0,S}\\\|z-\theta_{0,S}\|_2>C_2t_n}}P_nY_{S,z}\leq0\Bigr\}.
\]
On this restricted set, \eqref{eq:glm-kl-parameter} gives $P_0Y_{S,z}=D_{\mathrm{KL}}(\mb P_{\theta_0}\|\mb P_{S,z})\geq c_K\|z-\theta_{0,S}\|_2^2\geq c_KC_2^2t_n^2$. By~\eqref{eq:sparse-envelopes}, $|Y_{S,z}|\leq2B$ and $Y_{S,z}$ is Lipschitz in $z$ with constant $L_Y\leq B\sqrt D$, so $L_Yh\leq B\sqrt D\,h=n^{-2}\leq c_KC_2^2t_n^2/4$ for $C_2$ large. Since $c_KC_2^2t_n^2/2-L_Yh\geq c_KC_2^2t_n^2/4$, \eqref{eq:likelihood-gap} applied with $v=0$ to each restricted set, with nets taken inside them, whose sizes are again at most $(1+2B^2Dn^2)^{d_S}$ because the restricted sets lie in the same ball, gives the bound
\[
(1+2B^2Dn^2)^{d_S}\exp\Bigl\{-\frac{\lambda_Ync_KC_2^2t_n^2}{4}\Bigr\}
=(1+2B^2Dn^2)^{d_S}\exp\Bigl\{-\frac{\lambda_Yc_KC_2^2D\Lambda_n}{4}\Bigr\}
\]
for each $S\in\mathcal S_+$, because $nt_n^2=D\Lambda_n$. By a union bound over $\mathcal S_+$ and $\sum_{S\in\mathcal S}(1+2B^2Dn^2)^{d_S}\leq e^{C\Lambda_n}$,
\[
P_0(\mathcal A_n^c)\leq\exp\{C\Lambda_n-cC_2^2D\Lambda_n\}\leq n^{-12}
\]
for $C_2$ large, because $\Lambda_n\geq\log n$.

For $\mathcal B_n$, fix $S\in\mathcal S_+$. Under $P_0$, $s_{S,\theta_{0,S}}$ and $T_S-\mu_S(\theta_{0,S})$ have mean zero, and by~\eqref{eq:sparse-envelopes} and the bounds on $I_S$ and $T_S$ above, their norms and directional variances are bounded by $C$. For unit $u$,
\[
E_0(u^\top s_{S,\theta_{0,S}})^2+\mathbb E_0\{u^\top(T_S-\mathbb E_0[T_S])\}^2\leq C, |u^\top s_{S,\theta_{0,S}}|+|u^\top(T_S-\mathbb E_0[T_S])|\leq C.
\]
We apply Lemma~\ref{lem:sphere-bernstein} with $\alpha=1/2$ and $r=C_2t_n\sqrt n/2$ to both centered vectors. With $\alpha=1/2$ the prefactor in~\eqref{eq:sphere-bernstein} is $5^{d_S}$, while $r^2=C_2^2D\Lambda_n/4$ and the norm bound $K\leq C$ gives $Kr/(3\sqrt n)=KC_2t_n/6\leq C$ because $C_2t_n\leq1$. Since $\|a_S\|_2\geq C_2t_n/2$ is the event $\|n^{-1/2}\sum_{i=1}^ns_{S,\theta_{0,S}}(X_i)\|_2\geq r$, and similarly for $\overline T_S-\mu_S(\theta_{0,S})$, \eqref{eq:sphere-bernstein} bounds $\|a_S\|_2$ and $\|\overline T_S-\mu_S(\theta_{0,S})\|_2$ by $C_2t_n/2$ each outside an event of probability at most $5^{d_S}\exp(-cC_2^2D\Lambda_n)$ for each $S\in\mathcal S_+$. Hence $P_0(\|a_S\|_2+\|\overline T_S-\mu_S(\theta_{0,S})\|_2>C_2t_n)\leq2\cdot5^{d_S}\exp(-cC_2^2D\Lambda_n)$.

For $\mathcal C_n$, fix $S\in\mathcal S_+$. By~\eqref{eq:glm-score-information} and~\eqref{eq:sparse-envelopes}, the $d_S^2$ entries of $H_{n,S}(\theta_{0,S})$ are averages of $n$ independent variables bounded in absolute value by $B$ with means equal to the entries of $I_S(\theta_{0,S})$. Since the operator norm is at most $d_S$ times the largest entry in absolute value and $d_S\leq D$, Hoeffding's inequality \citep[Proposition~2.5 and (2.11)]{wainwright2019high}, applied to the $d_S^2$ entries at threshold $C_2t_n$, gives
\[
\begin{aligned}
&P_0\bigl(\|H_{n,S}(\theta_{0,S})-I_S(\theta_{0,S})\|_{\rm op}>C_2Dt_n\bigr)\leq2d_S^2\exp\Bigl\{-\frac{nC_2^2t_n^2}{2B^2}\Bigr\}
=2d_S^2\exp\Bigl\{-\frac{C_2^2D\Lambda_n}{2B^2}\Bigr\}.
\end{aligned}
\]
Taking a union bound over $S\in\mathcal S_+$, the union factors $|\mathcal S|5^D$ and $|\mathcal S|D^2$ are at most $e^{C\Lambda_n}$. After fixing $C_2$ large enough, we obtain $P_0(\mathcal B_n^c)+P_0(\mathcal C_n^c)\leq Cn^{-12}$, and hence $P_0(\mathcal E_n^c)\leq P_0(\mathcal A_n^c)+P_0(\mathcal B_n^c)+P_0(\mathcal C_n^c)\leq Cn^{-12}$.

\medskip
\noindent{\bf Step 2: Expansions on $\mathcal E_n$.} We work on $\mathcal E_n$ and fix $S\in\mathcal S_+$. By Lemma~\ref{lem:glm-natural-modules}, $\mu_S^{-1}$ has Lipschitz constant $L_g=2/c_\mu$ on $\mathcal K_{\mu,S}\supseteq K_{g,S}$. Choose
\[
C_5\geq C_2\max\{1,L_g,\sup_S\|I_S(\theta_{0,S})^{-1}\|_{\rm op}\}
\]
and decrease the fixed $c_0$ so that $C_5t_n\leq\rho/(2D)=\rho_g$. On $\mathcal B_n$, we have $\|\overline T_S-\mu_S(\theta_{0,S})\|_2\leq C_2t_n\leq C_5t_n\leq\rho_g$, so $\overline T_S\in K_{g,S}\subseteq\mathcal R_{g,S}$ and $\Pi_{\mathcal R_{g,S}}\overline T_S=\overline T_S$. The Lipschitz bound on $\mu_S^{-1}$ then gives $\|\mu_S^{-1}(\overline T_S)-\theta_{0,S}\|_2\leq L_gC_2t_n\leq C_5t_n\leq\rho/(2D)$. On $\mathcal A_n$, we also have $\|\widehat\theta_S-\theta_{0,S}\|_2\leq C_2t_n\leq\rho/(2D)$. Since $\operatorname{dist}(\theta_{0,S},\partial\Theta_{0,S})\geq\rho/D$, both $\widehat\theta_S$ and $\mu_S^{-1}(\overline T_S)$ lie in $\operatorname{int}(\Theta_{0,S})$, so the two projections defining $\theta_S^{(0)}$ in~\eqref{eq:vs-one-step-map} are inactive and
\[
\widehat\theta_S\in\operatorname{int}(\Theta_{0,S}),\qquad \Pi_{\mathcal R_{g,S}}\overline T_S=\overline T_S,\qquad \theta_S^{(0)}=\Pi_{\Theta_{0,S}}\mu_S^{-1}(\overline T_S)=\mu_S^{-1}(\overline T_S).
\]

Let $r_n=2C_5t_n$, and decrease $c_0$ so that $r_n\leq\rho/(2D)$, which gives $B_2(\theta_{0,S},r_n)\subseteq\Theta_{0,S}$. By~\eqref{eq:sparse-envelopes}, every third partial derivative of $\ell_{S,\theta}(X)$ is bounded by $B$ on $\Theta_{0,S}$, so $\sup_{\theta\in\Theta_{0,S}}\|\nabla^3\ell_{S,\theta}(X)\|_{\rm op}\leq C$ and $H_{n,S}$ is operator-norm Lipschitz on $\Theta_{0,S}$ with constant $CD^{3/2}$. Comparing $H_{n,S}(\theta_{0,S}+u)$ with $H_{n,S}(\theta_{0,S})$ and using $\mathcal C_n$ gives, by the triangle inequality,
\[
\sup_{\|u\|_2\leq r_n}\|H_{n,S}(\theta_{0,S}+u)-I_S(\theta_{0,S})\|_{\rm op}\leq C_2Dt_n+CD^{3/2}r_n\leq C\sqrt{\Lambda_n/n}=:\varsigma_n,
\]
since $D$ is fixed. Write $\underline\lambda_I=m_b\underline\lambda_Z/a$, and decrease $c_0$ so that $\varsigma_n\leq\underline\lambda_I/2$. Since $\underline\lambda_I I_{d_S}\preceq I_S(\theta_{0,S})\preceq(M_b\overline\lambda_Z/a)I_{d_S}$ by Assumption~\ref{ass:sparse}\textup{(i)}--\textup{(ii)}, we have, for $\|u\|_2\leq r_n$,
\[
\frac{\underline\lambda_I}{2}I_{d_S}\preceq H_{n,S}(\theta_{0,S}+u)\preceq\Bigl(\frac{M_b\overline\lambda_Z}{a}+\frac{\underline\lambda_I}{2}\Bigr)I_{d_S}=:\overline\lambda_H I_{d_S}.
\]
Since $n^{-1}\nabla\ell_{n,S}(\theta_{0,S})=a_S$ and $-n^{-1}\nabla^2\ell_{n,S}=H_{n,S}$, Taylor's formula around $\theta_{0,S}$ and the bound $\|H_{n,S}(\theta_{0,S}+u)-I_S(\theta_{0,S})\|_{\rm op}\leq\varsigma_n$ yield, for $\|u\|_2\leq r_n$,
\[
n^{-1}\{\ell_{n,S}(\theta_{0,S}+u)-\ell_{n,S}(\theta_{0,S})\}=a_S^\top u-\tfrac12u^\top I_S(\theta_{0,S})u+R_S(u),\qquad |R_S(u)|\leq\tfrac12\varsigma_n\|u\|_2^2.
\]

Put $q_S(u)=a_S^\top u-\tfrac12u^\top I_S(\theta_{0,S})u$, $\widehat u=\widehat\theta_S-\theta_{0,S}$ and $u^*=I_S(\theta_{0,S})^{-1}a_S$. The maximum of $q_S$ is $q_S(u^*)=\tfrac12a_S^\top I_S(\theta_{0,S})^{-1}a_S=\|V_S\|_2^2/(2n)$, attained at $u^*$. On $\mathcal E_n$, we have $\|\widehat u\|_2\leq C_2t_n\leq C_5t_n$ and $\|u^*\|_2\leq\|I_S(\theta_{0,S})^{-1}\|_{\rm op}\|a_S\|_2\leq C_5t_n=r_n/2$, so $\theta_{0,S}+u^*\in\Theta_{0,S}$. Applying the expansion at $\widehat u$ gives $n^{-1}\{\ell_{n,S}(\widehat\theta_S)-\ell_{n,S}(\theta_{0,S})\}=q_S(\widehat u)+R_S(\widehat u)\leq q_S(u^*)+\tfrac12\varsigma_nC_5^2t_n^2$. Since $\widehat\theta_S$ maximizes $\ell_{n,S}$ over $\Theta_{0,S}$, the expansion at $u^*$ gives $n^{-1}\{\ell_{n,S}(\widehat\theta_S)-\ell_{n,S}(\theta_{0,S})\}\geq q_S(u^*)+R_S(u^*)\geq q_S(u^*)-\tfrac12\varsigma_nC_5^2t_n^2$. Multiplying by $2n$ gives
\[
|2\{\ell_{n,S}(\widehat\theta_S)-\ell_{n,S}(\theta_{0,S})\}-\|V_S\|_2^2|\leq 2n\varsigma_n C_5^2t_n^2\leq C\Lambda_n^{3/2}/\sqrt n\leq 1/8
\]
after decreasing $c_0$, since $\Lambda_n^{3/2}/\sqrt n\leq\sqrt{c_0}$. This proves~\eqref{eq:sparse-quadratic}.

Since the design law $Q_Z$ does not depend on $\theta$ and~\eqref{eq:sparse-envelopes} bounds the third partial derivatives, differentiating~\eqref{eq:glm-score-information} under the integral gives
\[
\begin{gathered}
\partial_l[I_S(\theta)]_{jk}=a^{-1}E_{Q_Z}[b'''(v_S^\top\theta)v_{S,j}v_{S,k}v_{S,l}]=-E_\theta\partial_{jkl}\ell_{S,\theta}(X),\\
\operatorname{Lip}_{\Theta_{0,S}}(I_S)\vee\operatorname{Lip}_{\Theta_{0,S}}(H_{n,S})\leq C,
\end{gathered}
\]
with both operator-norm Lipschitz constants at most $CD^{3/2}$. Put $\rho_{n,S}=\|\theta_S^{(0)}-\widehat\theta_S\|_2$ and $\delta_{n,S}=\|I_S(\theta_S^{(0)})-H_{n,S}(\theta_S^{(0)})\|_{\rm op}$. By the triangle inequality, $\rho_{n,S}\leq\|\theta_S^{(0)}-\theta_{0,S}\|_2+\|\widehat\theta_S-\theta_{0,S}\|_2\leq 2C_5t_n$. Comparing $I_S(\theta_S^{(0)})$ with $I_S(\theta_{0,S})$, $I_S(\theta_{0,S})$ with $H_{n,S}(\theta_{0,S})$ on $\mathcal C_n$, and $H_{n,S}(\theta_{0,S})$ with $H_{n,S}(\theta_S^{(0)})$ gives $\delta_{n,S}\leq C_2Dt_n+CD^{3/2}t_n$. We apply the exact-component case of the one-step comparison~\eqref{eq:approx-bound} with inverse-information bound $\underline\lambda_I^{-1}$ and Lipschitz bound $C$. On $\mathcal E_n$, $\widehat\theta_S$ is an interior maximizer, the segment from $\theta_S^{(0)}$ to $\widehat\theta_S$ lies in the convex rectangle $\Theta_{0,S}$, and the nonexpansive projection $\Pi_{\Theta_{0,S}}$ fixes $\widehat\theta_S$. Adding the module error $\|\widetilde F_S-F_S^{\rm os}\|_2\leq n^{-3}$ from~\eqref{eq:vs-module-accuracy} by the triangle inequality gives
\[
\|\widetilde F_S-\widehat\theta_S\|_2\leq n^{-3}+\underline\lambda_I^{-1}\{\delta_{n,S}+C\rho_{n,S}\}\rho_{n,S}\leq C\Lambda_n/n,
\]
where the last inequality uses $t_n^2=D\Lambda_n/n$.

By the triangle inequality, $\|\widetilde F_S-\theta_{0,S}\|_2\leq\|\widehat\theta_S-\theta_{0,S}\|_2+\|\widetilde F_S-\widehat\theta_S\|_2\leq C_5t_n+C\Lambda_n/n\leq r_n$ after decreasing $c_0$. Hence the segment $[\widehat\theta_S,\widetilde F_S]$ lies in $B_2(\theta_{0,S},r_n)$, and $\nabla\ell_{n,S}(\widehat\theta_S)=0$ because $\widehat\theta_S\in\operatorname{int}(\Theta_{0,S})$ maximizes $\ell_{n,S}$. Taylor's formula around $\widehat\theta_S$ with the curvature bound $H_{n,S}\preceq\overline\lambda_H I_{d_S}$ on $B_2(\theta_{0,S},r_n)$ and $\Lambda_n^2\log(en)/n\leq\Lambda_n^3/n\leq c_0$ gives
\[
D_S(\widetilde F_S)\leq\tfrac12n\overline\lambda_H\|\widetilde F_S-\widehat\theta_S\|_2^2\leq C\Lambda_n^2/n\leq\{4\log(en)\}^{-1}
\]
for $c_0$ small, where $c_0$ may depend on the fixed $D$.

For~\eqref{eq:sparse-local-estimation}, expanding $0=n^{-1}\nabla\ell_{n,S}(\widehat\theta_S)$ around $\theta_{0,S}$ gives $0=a_S-\bar H(\widehat\theta_S-\theta_{0,S})$, where $\bar H$ is the average of $H_{n,S}$ over the segment from $\theta_{0,S}$ to $\widehat\theta_S$. Since $H_{n,S}\succeq(\underline\lambda_I/2)I_{d_S}$ on $B_2(\theta_{0,S},r_n)$, we have $\bar H\succeq(\underline\lambda_I/2)I_{d_S}$, and hence $\|\widehat\theta_S-\theta_{0,S}\|_2\leq 2\underline\lambda_I^{-1}\|a_S\|_2$. By the definition of $V_S$, $\|a_S\|_2^2\leq\|I_S(\theta_{0,S})\|_{\rm op}\|V_S\|_2^2/n\leq C\|V_S\|_2^2/n$. Using the triangle inequality and $(x+y)^2\leq 2x^2+2y^2$ with $\|\widetilde F_S-\widehat\theta_S\|_2\leq C\Lambda_n/n$ and $\Lambda_n^2/n^2\leq c_0/n$ gives
\[
\|\widetilde F_S-\theta_{0,S}\|_2^2\leq 8\underline\lambda_I^{-2}\|a_S\|_2^2+C\Lambda_n^2/n^2\leq\frac{C}{n}(1+\|V_S\|_2^2),
\]
which proves~\eqref{eq:sparse-local-estimation}. Finally, $\|\widehat\theta_S-\theta_{0,S}\|_2+\|\theta_S^{(0)}-\theta_{0,S}\|_2\leq(C_2+C_5)t_n\leq C\sqrt{\Lambda_n/n}$, giving the first bound of the lemma. All conclusions hold simultaneously on $\mathcal E_n$, and $P_0(\mathcal E_n^c)\leq Cn^{-12}\leq Cn^{-10}$.
\end{proof}

\subsubsection{Added variables (Lemma~\ref{lem:vs-overfit})}

For a support containing the true support, we decompose the score into the score for the true model and the additional score after removing its projection onto the true-model score. The latter accounts for the likelihood gain from the extra variables.

\begin{proof}
In this proof, $C$ and the named constants $C_2,C_4$ depend only on the response family, $a$, $m_b$, $M_b$, $\underline\lambda_Z$, $\overline\lambda_Z$, $B$, $\rho$ and $c_\Theta$, uniformly over $0<c_0\leq1$. They are independent of $n,p,S$ and $\theta_0$.

Order the coordinates so that the first $d_0$ are those of $S_0$ and write $I_S$ in blocks $I_{AA},I_{AB},I_{BA},I_{BB}$. Compatibility identifies $(I_{AA},s_A)$ with $(I_{S_0}(\theta_{0,S_0}),s_{S_0,\theta_{0,S_0}})$. Put $J=I_{BB}-I_{BA}I_{AA}^{-1}I_{AB}$, $\xi_S=J^{-1/2}(s_B-I_{BA}I_{AA}^{-1}s_A)$ and $U_S=n^{-1/2}\sum_i\xi_S(X_i)$. Under $P_0$, $s_{S,\theta_{0,S}}$ has mean zero and covariance $I_S(\theta_{0,S})$, so $s_B-I_{BA}I_{AA}^{-1}s_A$ has mean zero and covariance $I_{BB}-I_{BA}I_{AA}^{-1}I_{AB}=J$. Thus $E_0[\xi_S]=0$ and $E_0[\xi_S\xi_S^\top]=I_b$. Since $cI_{d_S}\preceq I_S(\theta_{0,S})\preceq c^{-1}I_{d_S}$, we have $\lambda_{\min}(J)\geq\lambda_{\min}\{I_S(\theta_{0,S})\}\geq c$, and~\eqref{eq:sparse-envelopes} with fixed $D$ gives $\|\xi_S\|_2\leq C$. By the Schur complement decomposition of $I_S(\theta_{0,S})^{-1}$ and this identification,
\[
\|V_S\|_2^2=\Bigl\|I_{AA}^{-1/2}n^{-1/2}\sum_i s_A(X_i)\Bigr\|_2^2+\|U_S\|_2^2=\|V_{S_0}\|_2^2+\|U_S\|_2^2.
\]

Since $\ell_{n,S}(\theta_{0,S})=\ell_{n,S_0}(\theta_{0,S_0})$ and $\|V_S\|_2^2-\|V_{S_0}\|_2^2=\|U_S\|_2^2$, applying~\eqref{eq:sparse-quadratic} to $S$ and $S_0$ on $\mathcal E_n$ and subtracting gives
\[
\bigl|2\{\ell_{n,S}(\widehat\theta_S)-\ell_{n,S_0}(\widehat\theta_{S_0})\}-\|U_S\|_2^2\bigr|\leq\frac18+\frac18=\frac14.
\]
By definition, $\operatorname{IC}_S-\operatorname{IC}_{S_0}=\Delta_b-\{\ell_{n,S}(\widehat\theta_S)-\ell_{n,S_0}(\widehat\theta_{S_0})\}$. Thus $\operatorname{IC}_S-\operatorname{IC}_{S_0}\leq1$ implies $2\{\ell_{n,S}(\widehat\theta_S)-\ell_{n,S_0}(\widehat\theta_{S_0})\}\geq2\Delta_b-2$, so $\|U_S\|_2^2\geq2\Delta_b-2-1/4$ on $\mathcal E_n$. Hence $\{\operatorname{IC}_S-\operatorname{IC}_{S_0}\leq1\}\cap\mathcal E_n\subseteq\{\|U_S\|_2^2\geq2\Delta_b-9/4\}\cap\mathcal E_n$.

Since $\|U_S\|_2^2\leq\|V_S\|_2^2\leq c^{-1}\|n^{-1/2}\sum_i s_{S,\theta_{0,S}}(X_i)\|_2^2$, choose a fixed $C_2\geq2$ so that the simultaneous score bound $\max_{S'\supseteq S_0}\|P_ns_{S',\theta_{0,S'}}\|_2\leq C\sqrt{D\Lambda_n/n}$, which holds on $\mathcal E_n$ by the definition of this event in Step~1 of the proof of Lemma~\ref{lem:sparse-local}, implies $\|U_S\|_2^2\leq C_2D\Lambda_n$ on $\mathcal E_n$. For $4b\leq w\leq C_2D\Lambda_n$, applying~\eqref{eq:sphere-bernstein-adaptive} of Lemma~\ref{lem:sphere-bernstein} to $\xi_S(X_1),\ldots,\xi_S(X_n)$ with $\sigma^2=1$ and $K=C$ gives
\[
P_0(\|U_S\|_2^2\geq w)\leq3^b\Bigl(\frac wb\Bigr)^{b/2}\exp\Bigl\{-\frac w2+\frac b2+\frac{Cw^{3/2}}{6\sqrt n}\Bigr\}.
\]
Since $\Lambda_n^3\leq c_0n$, $w^{3/2}/\sqrt n\leq(C_2D)^{3/2}(\Lambda_n^3/n)^{1/2}\leq(C_2D)^{3/2}\sqrt{c_0}$. As $b,D\geq1$ and $c_0\leq1$, the last two terms in the exponent are at most $(C+CC_2^{3/2})b$, where $C$ may depend on the fixed $D$. Thus, for a tail factor $C_4$ chosen uniformly for $0<c_0\leq1$, $P_0(\|U_S\|_2^2\geq w)\leq C_4^b(w/b)^{b/2}e^{-w/2}$ for $4b\leq w\leq C_2D\Lambda_n$.

Next, $2\Delta_b=b\log n+2\log\{\binom p{s_0+b}/\binom p{s_0}\}$, where $\binom p{s_0+b}/\binom p{s_0}=\prod_{i=1}^b(p-s_0-i+1)/(s_0+i)\in[1,p^b]$ because $s_0+b\leq s_{\max}\leq p/2$. Hence
\[
b\log n\leq2\Delta_b\leq b(\log n+2\log p)\leq2b\Lambda_n.
\]
Put $w_b=2\Delta_b-9/4$. For $n\geq N$ with the fixed sufficiently large $N$ in Lemma~\ref{lem:sparse-local}, $w_b\geq b\log n-9/4\geq4b$. Since $b\leq s_{\max}<D$ and $C_2\geq2$, we also have $w_b\leq2b\Lambda_n\leq2D\Lambda_n\leq C_2D\Lambda_n$, so the tail bound applies at $w=w_b$. With $(w_b/b)^{b/2}\leq(2\Lambda_n)^{b/2}$ and $e^{-w_b/2}=e^{9/8}e^{-\Delta_b}$,
\[
P_0(\|U_S\|_2^2\geq w_b)\leq C_4^b(w_b/b)^{b/2}e^{-w_b/2} \leq e^{9/8}(C_4\sqrt{2\Lambda_n})^be^{-\Delta_b}\leq(C\sqrt{\Lambda_n})^be^{-\Delta_b}.
\]

Let $Q=\|U_S\|_2^2$. Since $Q\leq C_2D\Lambda_n$ on $\mathcal E_n$, tail integration gives
\[
\begin{aligned}
E_0[Q\ind\{Q\geq w_b\}\ind_{\mathcal E_n}]
 &\leq w_bP_0(Q\geq w_b)+\int_{w_b}^{C_2D\Lambda_n}P_0(Q\geq w)\,dw\\
&\leq C_4^bb^{-b/2}(w_b+8/3)w_b^{b/2}e^{-w_b/2} \leq Cb\Lambda_n(C\sqrt{\Lambda_n})^be^{-\Delta_b}.
\end{aligned}
\]
For the second line, we apply the tail bound and note that
\[
\frac{d}{dw}\log\bigl\{(w/b)^{b/2}e^{-w/2}\bigr\}=\frac{b}{2w}-\frac12\leq-\frac38,\qquad w\geq4b,
\]
so the integral of $(w/b)^{b/2}e^{-w/2}$ over $[w_b,\infty)$ is at most $8/3$ times its value at $w_b$. For the third line, $w_b+8/3\leq Cb\Lambda_n$ and the bound at $w_b$ give~\eqref{eq:sparse-overfit-moment}.
\end{proof}

\subsubsection{Moment products (Lemma~\ref{lem:vs-moment})}

\begin{proof}
The first coordinate of $T_S^\upsilon$ equals that of $T_S$, and each of the other coordinates differs from the corresponding product by at most $\upsilon$ for every observation in $\mathcal X_n$, so $\|P_nT_S^\upsilon-\overline T_S\|_2\leq\sqrt D\,\upsilon$. The projections onto rectangles are nonexpansive, so the two initial estimates in~\eqref{eq:vs-one-step-map} differ by at most $G_{g,n}\sqrt D\,\upsilon$. On the computational rectangle, the averaged score $\vartheta\mapsto P_ns_{S,\vartheta}$ is bounded by $\sqrt DB_{\rm cal}$ and Lipschitz with constant $DB_{\rm cal}$, and $\vartheta\mapsto I_S(\vartheta)^{-1}$ is bounded in operator norm by $c^{-1}$ and Lipschitz with constant $CD^{3/2}$, by the identity $A(\vartheta)-A(\vartheta')=A(\vartheta)\{I_S(\vartheta')-I_S(\vartheta)\}A(\vartheta')$ for $A=I_S^{-1}$ and the derivative bounds in the proof of Theorem~\ref{thm:vs-approx}. Hence $\vartheta\mapsto\vartheta+I_S(\vartheta)^{-1}P_ns_{S,\vartheta}$ is Lipschitz on $\Theta_{0,S}$ with a constant depending only on $D$ and the constants of the setting, and the final projection is nonexpansive. This proves the first claim.

In the proof of Theorem~\ref{thm:vs-approx}, the bound $\sup_{X^n}\|\widetilde F_S-F_S^{\rm os}\|_2\leq CD^{7/2}\delta_n$ is obtained from uniform approximations of $g_S$ on $\mathcal R_{g,S}$, after the projection onto this rectangle, and of $s_{S,\theta}$ and $I_S^{-1}$ on the computational rectangle. It does not use the form of the pooled vector that enters the initialization, and therefore holds for the pooled vector $P_nT_S^{\upsilon_n}$, with $F_S^{{\rm os},\upsilon_n}$ in place of $F_S^{\rm os}$. After doubling $C_0$, this bound is at most $n^{-3}/2$, and the first claim with $\upsilon=\upsilon_n$ adds at most $n^{-3}/2$. The complexity of the multiplication networks is that of Lemma~\ref{lem:multiplication}, and the $p$ products are placed in parallel.
\end{proof}

\subsubsection{The variable-selection network (Lemma~\ref{lem:vs-rrds})}

The main computational issue is to aggregate the weights of many supports without evaluating exponentially large likelihoods. The proof uses pairwise combinations of log sums and weighted averages, which are functions of differences of log weights.

\begin{proof}
In this proof, $C$ depends only on $B$. We first construct the approximations $\operatorname{LSE}$ and $\operatorname{AVG}$ in~\eqref{eq:vs-tree-accuracy}.

\emph{A node.} Let $\zeta\in(0,1/8]$, $T=\lceil\log(16/\zeta)\rceil$ and $J=1+\log\{16(B+1)/\zeta\}$. We construct a ReLU network $\mathcal N_\zeta$ that maps two pairs $(v_1,a_1),(v_2,a_2)\in[-B,B]\times\mb R$ to a pair $(\widehat v,\widehat a)$ satisfying
\begin{equation}
\bigl|\widehat a-\log(e^{a_1}+e^{a_2})\bigr|\leq\zeta,\qquad
\bigl|\widehat v-\{v_2+\sigma_{\log}(a_1-a_2)(v_1-v_2)\}\bigr|\leq(2B+1)\zeta,\qquad|\widehat v|\leq B,
\label{eq:vs-node}
\end{equation}
with depth at most $CJ^2$, at most $CJ^3$ parameters and coefficients bounded by $CJ$.

Write $d=a_1-a_2$ and ${\rm sp}(t)=\log(1+e^t)$, so that $\log(e^{a_1}+e^{a_2})=\max(a_1,a_2)+{\rm sp}(-|d|)$. The maximum, the absolute value and the clipping of a number to an interval are exact ReLU maps. Clipping $d$ to $[-T,T]$ changes $\sigma_{\log}(d)$ by at most $e^{-T}\leq\zeta/16$, and clipping $-|d|$ to $[-T,0]$ changes ${\rm sp}(-|d|)$ by at most $e^{-T}$.

For an integer $k$, let $w_k(t)=(1-|t-k|)_+$, which is an exact ReLU map. For $t\in[-T,T]$, these functions sum to one and at most two of them are positive.

For $|x|\leq1$, the functions $x\mapsto\sigma_{\log}(k+x)$ and, for $k\leq0$, $x\mapsto{\rm sp}(k+x)$ extend holomorphically to a neighborhood of the Bernstein ellipse $E_{3/2}$. Every point $x\in E_{3/2}$ satisfies $|\operatorname{Im}x|\leq5/12$ and $\operatorname{Re}x\leq13/12$, and these extensions are bounded on $E_{3/2}$ by $4$, uniformly in $k$. Indeed, the real part of $1+e^{k+x}$ is positive and $|e^{k+x}/(1+e^{k+x})|\leq1/\cos(5/12)<2$. Moreover, $|\log(1+w)|\leq|\log|1+w||+\pi/2$, with $|1+e^{k+x}|\leq1+e^{13/12}$ and $|1+e^{k+x}|\geq1$ for $k\leq0$.

By Lemma~\ref{lem:analytic-approx} with $m=1$, each of these functions is approximated on $[-1,1]$ within $\zeta/32$ by a ReLU network of depth and size at most $CJ^2$ with bounded coefficients, where the constants do not depend on $k$. Let $\zeta'=\zeta/\{32(2T+3)\}$, and let $\mathsf M_{4,\zeta'}$ be the multiplication network of Lemma~\ref{lem:multiplication}. We apply the $k$th local network to the clipped argument $t-k$, clip its output to $[0,4]$, multiply it by $w_k(t)$ using $\mathsf M_{4,\zeta'}$, and sum over $|k|\leq T$ for $\sigma_{\log}$ and over $-T\leq k\leq0$ for ${\rm sp}$. Since the weights $w_k(t)$ sum to one and vanish where the $k$th local approximation is not valid, the resulting sums approximate $\sigma_{\log}$ on $[-T,T]$ and ${\rm sp}$ on $[-T,0]$ within $\zeta/16$. We clip the sum for $\sigma_{\log}$ to $[0,1]$, which does not increase its error. With the two clipping errors, this gives $\widehat\sigma\in[0,1]$ satisfying $|\widehat\sigma-\sigma_{\log}(d)|\leq\zeta/4$ for every real $d$, together with the first claim of~\eqref{eq:vs-node}.

Finally, we obtain $\widehat v$ by clipping $v_2+\mathsf M_{2B+1,\zeta/2}(\widehat\sigma,v_1-v_2)$ to $[-B,B]$. The exact value is a convex combination of $v_1$ and $v_2$, so clipping does not increase the error, which is at most $2B\zeta/4+\zeta/2$.

We place the $2T+1$ local networks in parallel, which gives the stated depth and size. The translations $k$ and the scalings are bounded by $CJ$.

\emph{A tree.} Given leaves $(v_S,\widetilde a_S)$, $S\in\mathcal C$, we merge them by $\mathcal N_\zeta$ along a balanced binary tree of height $h\leq H_{\mathcal S}$. Any leaf without a partner is passed on unchanged. With exact nodes, the root is $(\operatorname{avg}(\mathcal C;v),\operatorname{lse}(\mathcal C))$, because the weighted average of two weighted averages with their total weights is the weighted average of all leaves.

Let $E_a(h)$ and $E_v(h)$ be the largest errors of the two coordinates at height $h$. The function $(a_1,a_2)\mapsto\log(e^{a_1}+e^{a_2})$ is $1$-Lipschitz for the maximum norm, so $E_a(h)\leq E_a(h-1)+\zeta$. For the value coordinate, the exact node is a convex combination of the two values, with $|v_1-v_2|\leq2B$. Since $\sigma_{\log}$ is $1/4$-Lipschitz and the difference of the two logits has error at most $2E_a(h-1)$, we have $E_v(h)\leq E_v(h-1)+BE_a(h-1)+(2B+1)\zeta$. Hence
\[
E_a(h)\leq h\zeta,\qquad E_v(h)\leq\Bigl\{\frac B2h(h-1)+(2B+1)h\Bigr\}\zeta .
\]

We take $\zeta=\zeta_{\rm tr}=\eta_n/\{16(B+1)(H_{\mathcal S}+1)^2\}$ and define $\operatorname{LSE}(\mathcal C)$ and $\operatorname{AVG}(\mathcal C;v)$ as the two coordinates of the root. With this choice, the errors in the logit and value coordinates at the root are at most $\eta_n/2$ and $\eta_n$, respectively. This proves~\eqref{eq:vs-tree-accuracy}. The corresponding $J_{\rm tr}=1+\log\{16(B+1)/\zeta_{\rm tr}\}$ satisfies $J_{\rm tr}\leq C\Lambda_n$, because $H_{\mathcal S}\leq1+2A_n\leq3D\Lambda_n$ and $D\leq p$.

Part~(a) follows from the residual update as in the proof of Lemma~\ref{lem:ms-rrds}. The coordinates $\vartheta_S^0$ are constant, and each refinement overwrites the coordinates $\theta_S$, $a_j$, $b_j$ and $b_0$ by functions of the pooled vector and of the state before the update. The logits $\widetilde a_S$ are affine functions of the pooled likelihood values.

Part~(b) follows from~\eqref{eq:vs-tree-accuracy}, since the log odds of $\widetilde q_j^{\rm ex}$ equal $\operatorname{lse}\{S\ni j\}-\operatorname{lse}\{S\not\ni j\}$.

For part~(c), $\operatorname{avg}(\{S\ni j\};(\widetilde\theta_{S,j})_S)=c_j^{\rm av}$ and $\operatorname{avg}(\mathcal S;(\widetilde\theta_{S,0})_S)=c_0^{\rm av}$, because $\widetilde Q_S\propto e^{\widetilde a_S}$. The network does not divide by $\widetilde q_j^{\rm ex}$.

For part~(d), the networks for each support have depth at most $L_{{\rm vs},n}$ and at most $V_{{\rm vs},n}$ parameters, and they are placed in parallel. Since $N_{\mathcal S}\geq p$, the shared products of Lemma~\ref{lem:vs-moment} use at most $CpJ_{T,n}\leq N_{\mathcal S}V_{{\rm vs},n}$ parameters.

For each $j$, the trees over $\{S\ni j\}$ and $\{S\not\ni j\}$ have $N_{\mathcal S}$ leaves in total, and the tree for $b_0$ has $N_{\mathcal S}$ leaves. Hence there are at most $(p+1)N_{\mathcal S}$ nodes, each with at most $CJ_{\rm tr}^3$ parameters. Nodes of the same height are placed in parallel, so the trees have depth at most $C(H_{\mathcal S}+1)J_{\rm tr}^2$. Carrying the states $\vartheta_S^0$ and $\theta_S$ through these layers uses at most $CN_{\mathcal S}D(H_{\mathcal S}+1)J_{\rm tr}^2$ further connections. This gives the bounds on $L$ and $W$ through~\eqref{eq:rrds-complexity}.

The coefficients of the networks for the supports are bounded by $P_{{\rm vs},n}$, the normalization slopes are bounded by $K_{{\rm dom},n}$, and the coefficients of the nodes are at most $CJ_{\rm tr}$. The other coefficients are the factor $n$ multiplying the pooled likelihood values and the penalties $\operatorname{pen}(S)\leq A_n$. An affine row has at most as many positions as the network has coefficients.
\end{proof}

\subsubsection{Oracle inequality for the coefficient estimate (Lemma~\ref{lem:vs-soft-oracle})}

\begin{proof}
Write $\iota_S\widetilde\theta_S\in\mb R^{1+p}$ for the zero extension of $\widetilde\theta_S$. Both $\iota_S\widetilde\theta_S$ and $\theta_0$ belong to $\Theta_{\mathcal S}$, so~\eqref{eq:glm-kl-parameter} gives $R_S(\widetilde\theta_S)\geq c_K\|\iota_S\widetilde\theta_S-\theta_0\|_2^2$. On all of $\mathcal X_n$ we use the version $Y_{S,u}(y,z)=a^{-1}[y\{v_{S_0}(z)^\top\theta_{0,S_0}-v_S(z)^\top u\}-b\{v_{S_0}(z)^\top\theta_{0,S_0}\}+b\{v_S(z)^\top u\}]$ of the log likelihood ratio, so that $nP_nY_{S,u}=\ell_{n,S_0}(\theta_{0,S_0})-\ell_{n,S}(u)$ for every dataset. By the definition of $Z_n$, the optimality of $\widehat\theta_{S_0}$, the accuracy $n\varepsilon_\ell\leq1/6$ of the likelihood networks and $\operatorname{pen}(S)\geq0$,
\begin{equation}
\begin{aligned}
\frac{nc_K}2\|\iota_S\widetilde\theta_S-\theta_0\|_2^2
&\leq nZ_n+\ell_{n,S_0}(\theta_{0,S_0})-\ell_{n,S}(\widetilde\theta_S)\\
&\leq nZ_n+D_{S_0}+2n\varepsilon_\ell+\operatorname{pen}(S_0)-\operatorname{pen}(S)+\widetilde a_{S_0}-\widetilde a_S\\
&\leq k_n-\tfrac23+\widetilde a_{S_0}-\widetilde a_S
\end{aligned}
\label{eq:vs-support-distance}
\end{equation}
for every $S\in\mathcal S$ and every dataset. Averaging over $\widetilde Q$ and using the last bound of Lemma~\ref{lem:vs-heads} gives the first claim. Since $\theta^{\rm av}-\theta_0=\sum_S\widetilde Q_S(\iota_S\widetilde\theta_S-\theta_0)$ and the squared norm is convex, $\|\theta^{\rm av}-\theta_0\|_2^2\leq\sum_S\widetilde Q_S\|\iota_S\widetilde\theta_S-\theta_0\|_2^2$. Each vector $\iota_S\widetilde\theta_S-\theta_0$ has at most $2s_{\max}$ nonzero slope coordinates, so~\eqref{eq:glm-sparse-design} gives $\mb E_{Q_Z}[\{v(Z)^\top(\iota_S\widetilde\theta_S-\theta_0)\}^2]\leq\overline\lambda_Z\|\iota_S\widetilde\theta_S-\theta_0\|_2^2$ with $v(Z)=(1,Z^\top)^\top$, and the function $u\mapsto\mb E_{Q_Z}[\{v(Z)^\top u\}^2]$ is convex. Hence
\[
\mb E_{Q_Z}\bigl[\{\eta_{\theta^{\rm av}}(Z)-\eta_{\theta_0}(Z)\}^2\bigr]\leq\overline\lambda_Z\sum_{S\in\mathcal S}\widetilde Q_S\|\iota_S\widetilde\theta_S-\theta_0\|_2^2 .
\]
We apply the upper bound in~\eqref{eq:glm-sparse-design} to each sparse vector before averaging, and not to the average $\theta^{\rm av}-\theta_0$, which need not be sparse. Finally, $\|\widetilde\theta-\theta^{\rm av}\|_2\leq\|\widetilde\theta-\theta^{\rm av}\|_1\leq(3+4B)/n$ by Lemma~\ref{lem:vs-heads}, $|\eta_{\widetilde\theta}(Z)-\eta_{\theta^{\rm av}}(Z)|\leq(1\vee B)\|\widetilde\theta-\theta^{\rm av}\|_1$, and $(x+y)^2\leq2x^2+2y^2$, which gives the two bounds for $\widetilde\theta$.

For the bounds on $Z_n$, both linear predictors of a log likelihood ratio $Y_{S,u}$ lie in $\mathcal I_{\rm pred}$, whose length is at most $2B/m_b$ because $b'$ takes values in $[-B,B]$ and $b''\geq m_b$ on $\mathcal I_{\rm pred}$. Hence $|Y_{S,u}|\leq B_Y=4B^2/(am_b)$, a bound that does not depend on $D$, and $Y_{S,u}$ is Lipschitz in $u$ with constant $L_Y\leq2B(1\vee B)\sqrt D/a$. Take $h=n^{-2}/L_Y$, for which the nets of the working rectangles satisfy $\log\sum_SN_S(h)\leq CA_n$ when $n\geq D$. Then~\eqref{eq:likelihood-uniform} gives $\mb E_0[nZ_n]\leq CA_n$, and~\eqref{eq:likelihood-tail}, with $3L_Yh/2\leq2n^{-2}$, gives $\mb P_0\{nZ_n>C(A_n+t)\}\leq e^{-t}$.
\end{proof}

\subsubsection{Supports with added variables (Lemma~\ref{lem:vs-sq-odds})}

\begin{proof}
In this proof, $C,c$ depend only on the constants of the setting of Appendix~\ref{app:vs-setting} and of Assumptions~\ref{ass:sparse} and~\ref{ass:sparse-proposal}, and they do not depend on $n$, $p$, the supports or $\theta_0$.
Write $Q=\|U_S\|_2^2$ and $t=\Delta_b-b$, so that $\varpi_S=\min\{1,e^{Q/2-t}\}$ and $2b\leq t\leq b\Lambda_n$ for $n\geq e^6$. We first bound the tail of $Q$. Let $4b\leq w\leq2s_{\max}\Lambda_n$. By the proof of Lemma~\ref{lem:vs-overfit}, the vectors $\xi_S(X_i)$ are independent and centered with identity covariance and norms bounded by $C$, so Lemma~\ref{lem:sphere-bernstein} gives
\[
\mb P_0(Q\geq w)\leq3^b\Bigl(\frac wb\Bigr)^{b/2}\exp\Bigl\{-\frac w2+\frac b2+\frac{Cw^{3/2}}{6\sqrt n}\Bigr\}\leq C^b\Bigl(\frac wb\Bigr)^{b/2}e^{-w/2},
\]
because $w^{3/2}/\sqrt n\leq C\Lambda_n^{3/2}/\sqrt n\leq C$.

Since $\varpi_S^2=e^{-2t}e^{\min(Q,2t)}$,
\begin{align*}
\mb E_0[\varpi_S^2]&\leq e^{-2t}\Bigl\{1+\int_0^{2t}e^w\,\mb P_0(Q>w)\,dw\Bigr\}
\leq e^{-2t}\Bigl\{e^{4b}+C^b\Bigl(\frac{2t}b\Bigr)^{b/2}\int_0^{2t}e^{w/2}dw\Bigr\}\\
&\leq e^{-t}\Bigl\{e^{2b}+2C^b(2\Lambda_n)^{b/2}\Bigr\},
\end{align*}
where the second inequality uses the tail bound for $w\geq4b$, and the last uses $t\geq2b$. Since $e^{-t}=e^be^{-\Delta_b}$, this is the first bound. Since $\varpi_S\leq\ind\{Q\geq2t\}+e^{-t}e^{\min(Q,2t)/2}$ and $\mb E_0[e^{\min(Q,2t)/2}]=1+\frac12\int_0^{2t}e^{w/2}\mb P_0(Q>w)\,dw$,
\[
\begin{aligned}
\mb E_0[\varpi_S]&\leq\mb P_0(Q\geq2t)+e^{-t}\Bigl\{1+\frac12\int_0^{2t}e^{w/2}\,\mb P_0(Q>w)\,dw\Bigr\}\leq e^{-t}\Bigl\{C^b(2\Lambda_n)^{b/2}(1+t)+e^{2b}\Bigr\},
\end{aligned}
\]
which gives the second bound because $t\leq b\Lambda_n$. For the third, Lemma~\ref{lem:vs-overfit} and the quadratic and local estimation bounds of Lemma~\ref{lem:sparse-local} give $Q\leq\|V_S\|_2^2\leq C_1\Lambda_n$ on $\mathcal E_n$, and we may take $C_1\geq2s_{\max}$. The tail bound holds for $4b\leq w\leq C_1\Lambda_n$ with a constant depending on $C_1$, and $w\mapsto(w/b)^{b/2}e^{-w/2}$ has logarithmic derivative at most $-3/8$ for $w\geq4b$. Hence
\[
\mb E_0\bigl[(Q/2-t)_+\ind_{\mathcal E_n}\bigr]=\frac12\int_{2t}^{C_1\Lambda_n}\mb P_0(Q>w,\mathcal E_n)\,dw\leq C^b\Bigl(\frac{2t}b\Bigr)^{b/2}e^{-t}\leq(C\sqrt{\Lambda_n})^be^{-\Delta_b}.
\]

For the consequences, there are $\binom{p-s_0}b$ supports $S\supsetneq S_0$ with $b$ added variables, and~\eqref{eq:ebic-cancellation} gives $\binom{p-s_0}b(C\sqrt{\Lambda_n})^be^{-\Delta_b}=\binom{s_0+b}b(C\sqrt{\Lambda_n/n})^b$. Since $s_{\max}$ is fixed and $\Lambda_n/n\leq c_0$, the sum over $b\geq1$ is at most $C\sqrt{\Lambda_n/n}$ and the sum over $b\geq2$ is at most $C\Lambda_n/n$. This gives the first two consequences. For the third, $\log(1+\sum_Sr_S)\leq\sum_S\log(1+r_S)$ and $\log(1+e^x)\leq\min\{1,e^x\}+x_+$, applied with $x=Q/2-t$, reduce the claim to the second and third bounds.
\end{proof}

\subsubsection{Supports that omit weak signals (Lemma~\ref{lem:vs-near})}

\begin{proof}
We prove each part separately.

In this proof, $C,c$ depend only on the constants of the setting of Appendix~\ref{app:vs-setting} and of Assumptions~\ref{ass:sparse} and~\ref{ass:sparse-proposal}, and they do not depend on $n$, $p$, the supports or $\theta_0$.
Put $r_n=\sqrt{A_n/n}$. The constants are chosen in the order $L_0$, $K_0$, $K_1$, $c_0$, where $L_0$ is fixed in the proof of (c).

For (a), the negative Hessian is $-n^{-1}\nabla^2\ell_{n,S}(\vartheta)=a^{-1}P_n[b''\{v_S(Z)^\top\vartheta\}v_S(Z)v_S(Z)^\top]$, with $m_b\leq b''\leq M_b$ by Assumption~\ref{ass:sparse}\textup{(i)}. The entries of $v_Sv_S^\top$ are bounded by $V_*^2$, and by~\eqref{eq:glm-sparse-design} the eigenvalues of $\mb E_{Q_Z}[v_Sv_S^\top]$ lie in $[\underline\lambda_Z,\overline\lambda_Z]$. Hoeffding's inequality \citep[Proposition~2.5 and (2.11)]{wainwright2019high}, applied to the $(p+1)^2$ entries at threshold $\underline\lambda_Z/(2D)$, and the union bound show that, except on an event of probability $2(p+1)^2\exp\{-n\underline\lambda_Z^2/(8D^2V_*^4)\}\leq Cn^{-12}$, all empirical second moments are within $\underline\lambda_Z/(2D)$ of their expectations. On the complement, $\|P_n[v_Sv_S^\top]-\mb E_{Q_Z}[v_Sv_S^\top]\|_{\rm op}\leq\underline\lambda_Z/2$ for every $S$, which gives (a) with $m=m_b\underline\lambda_Z/(2a)$ and $M=2M_b\overline\lambda_Z/a$.

For (b), let $S\in\mathcal A_0$, and let $\theta_S^{\rm r}$ keep the intercept and the coordinates of $\theta_0$ in $S\cap S_0$ and set the other coordinates of $S$ to zero. By Assumption~\ref{ass:sparse-proposal}(ii) and~(iii), $\theta_S^{\rm r}\in K_S$ and $\operatorname{dist}(\theta_S^{\rm r},\partial\Theta_{0,S})\geq\rho$. The map $\vartheta\mapsto D_{\rm KL}(\mb P_{\theta_0}\|\mb P_{S,\vartheta})$ has Hessian $I_S(\vartheta)$ on $\Theta_{0,S}$, because the Hessian of the log density of a canonical GLM does not involve the response and the covariate law is common, so it is strictly convex with a unique minimizer $\theta_S^\dagger$. By~\eqref{eq:glm-kl-parameter}, its value at $\theta_S^{\rm r}$ is at most $Ce_S$, and at $\theta_S^\dagger$ it is at least $c_K\|\iota_S\theta_S^\dagger-\theta_0\|_2^2$. Hence $\|\theta_S^\dagger-\theta_S^{\rm r}\|_2\leq C\sqrt{e_S}\leq C\sqrt{K_0}\,r_n\leq\rho/2$ for $c_0$ small, so $\theta_S^\dagger$ is an interior point of $\Theta_{0,S}$ and
\begin{equation}
\mb E_0[s_{S,\theta_S^\dagger}(X_1)]=0,\qquad\mb E_0[T_S(X_1)]=\mu_S(\theta_S^\dagger),
\label{eq:vs-pseudo-true}
\end{equation}
where the second identity is the first for the score~\eqref{eq:glm-score-information}. The coordinates of $s_{S,\theta_S^\dagger}$, of $T_S$ and of the Hessian at $\theta_S^\dagger$ are bounded functions that do not depend on the data. Hoeffding's inequality \citep[Proposition~2.5 and (2.11)]{wainwright2019high} and the union bound over the at most $e^{A_n}$ supports and their coordinates give, for a sufficiently large $C_1$, an event of probability at least $1-Cn^{-12}$ on which, for every $S\in\mathcal A_0$,
\[
\|P_ns_{S,\theta_S^\dagger}\|_2+\|\overline T_S-\mu_S(\theta_S^\dagger)\|_2+\|n^{-1}\nabla^2\ell_{n,S}(\theta_S^\dagger)+I_S(\theta_S^\dagger)\|_{\rm op}\leq C_1r_n .
\]
On this event and the event of (a), the strong concavity of $\ell_{n,S}$ and the optimality of $\widehat\theta_S$ on the convex set $\Theta_{0,S}$ give $\|\widehat\theta_S-\theta_S^\dagger\|_2\leq(2/m)\|P_ns_{S,\theta_S^\dagger}\|_2\leq Cr_n$, so $\widehat\theta_S$ is an interior point for $c_0$ small. By Lemma~\ref{lem:glm-natural-modules}, the pooled moment $\overline T_S$ lies in $K_{g,S}$, where $g_S=\mu_S^{-1}$ is Lipschitz, so the initial estimate in~\eqref{eq:vs-one-step-map} satisfies $\|\theta_S^{(0)}-\theta_S^\dagger\|_2\leq Cr_n$ and both projections are inactive. Since $P_ns_{S,\widehat\theta_S}=0$,
\begin{align*}
&\theta_S^{(0)}+I_S(\theta_S^{(0)})^{-1}P_ns_{S,\theta_S^{(0)}}-\widehat\theta_S\\
&\qquad=I_S(\theta_S^{(0)})^{-1}\Bigl\{I_S(\theta_S^{(0)})-\int_0^1H_{n,S}\bigl(\widehat\theta_S+u(\theta_S^{(0)}-\widehat\theta_S)\bigr)du\Bigr\}(\theta_S^{(0)}-\widehat\theta_S),
\end{align*}
with $H_{n,S}=-n^{-1}\nabla^2\ell_{n,S}$. The third derivatives are bounded, so $H_{n,S}$ and $I_S$ are Lipschitz, and the term in braces is at most $Cr_n$ in operator norm. Hence $\|F_S^{\rm os}-\widehat\theta_S\|_2\leq Cr_n^2$, where the final projection is inactive, and~\eqref{eq:vs-module-accuracy} gives $\|\widetilde\theta_S-\widehat\theta_S\|_2\leq CA_n/n$. Taylor's formula at the interior maximizer and (a) give $D_S\leq(nM/2)\|\widetilde\theta_S-\widehat\theta_S\|_2^2\leq CA_n^2/n\leq\{4\log(en)\}^{-1}$, because $A_n^2\log(en)/n\leq C\Lambda_n^3/n\leq Cc_0$.

For (c), let $S\notin\mathcal A_0$, so that $e_S>K_0A_n/n$. Lemma~\ref{lem:vs-omitted-energy} with $t=L_0A_n$ and the union bound give
\[
\mb P_0\{\exists S\notin\mathcal A_0:\operatorname{IC}_S-\operatorname{IC}_{S_0}\leq L_0A_n\}\leq\exp\{CA_n+\lambda_YL_0A_n-\lambda_Yc_KK_0A_n/2\}\leq n^{-12}
\]
for $K_0$ large after $L_0$. On the complement, $Q_S^{\rm E}/Q_{S_0}^{\rm E}=e^{-(\operatorname{IC}_S-\operatorname{IC}_{S_0})}\leq e^{-L_0A_n}$, and~\eqref{eq:vs-logit-identity}, $D_S\geq0$, and $D_{S_0}\leq\{4\log(en)\}^{-1}$ by (b) give $\widetilde Q_S/\widetilde Q_{S_0}\leq e^{1-L_0A_n}$. For the posterior, $m_S\leq e^{\ell_{n,S}(\widehat\theta_S)}$. The set $B(\theta_{0,S_0},1/n)\cap K_{S_0}$ has $\pi_{S_0}$-probability at least $c\,n^{-d_0}$, because $K_{S_0}$ is a rectangle with sides of length at least $2a_\pi$ on which the density is at least $c_-^{d_0}$, and $\ell_{n,S_0}$ varies by at most $B\sqrt D$ on it. Hence $m_{S_0}\geq c\,n^{-d_0}e^{\ell_{n,S_0}(\theta_{0,S_0})-C}$ for every dataset. By the concavity of $\ell_{n,S_0}$ and (b), $\ell_{n,S_0}(\widehat\theta_{S_0})-\ell_{n,S_0}(\theta_{0,S_0})\leq n\|P_ns_{S_0,\theta_{0,S_0}}\|_2\|\widehat\theta_{S_0}-\theta_{0,S_0}\|_2\leq CA_n$. Therefore
\[
\log\frac{Q_S^*}{Q_{S_0}^*}=\log\frac{w(S)m_S}{w(S_0)m_{S_0}}\leq-(\operatorname{IC}_S-\operatorname{IC}_{S_0})+C_2A_n\leq-(L_0-C_2)A_n .
\]
Summing over at most $e^{A_n}$ supports and taking $L_0=C_2+10$ gives (c) for the three probability vectors. The constant $C_2$ does not depend on $K_0$: for $S=S_0$ we have $e_{S_0}=0$ and $\theta_{S_0}^\dagger=\theta_{0,S_0}$, so the bounds of (b) used here involve only $C_1$ and $m$. The order $L_0$, $K_0$, $K_1$, $c_0$ is therefore consistent.

For (d), let $S\in\mathcal A_0$. For $j\in(S\cap S_0)\cup\{0\}$, $|\widehat\theta_{S,j}-\theta_{0j}|\leq\|\widehat\theta_S-\theta_S^\dagger\|_2+\|\theta_S^\dagger-\theta_S^{\rm r}\|_2\leq C(1+\sqrt{K_0})r_n$, and for $j\in S\setminus S_0$ the same bound holds for $|\widehat\theta_{S,j}|$. Take $K_1>C(1+\sqrt{K_0})$ and $c_0$ so small that $w_n<a_\pi$. If $d_\pi(\theta_0)>w_n$, the first group of coordinates lies in the intervals $\mathcal I_j$, and the second lies in $[-a_\pi,a_\pi]\subseteq\mathcal I_j$. The event $\mathcal G$ is the intersection of $\mathcal E_n$ with the three events above.
\end{proof}

\subsubsection{Laplace bounds for sparse candidates (Lemma~\ref{lem:vs-laplace})}

\begin{proof}
Let $d=d_S$. At a maximizer over the convex set $\Theta_{0,S}$, the directional derivative of $\ell_{n,S}$ toward any point of the set is nonpositive, so Lemma~\ref{lem:vs-near}(a) gives $\ell_{n,S}(\vartheta)\leq\ell_{n,S}(\widehat\theta_S)-(nm/2)\|\vartheta-\widehat\theta_S\|_2^2$ on $\Theta_{0,S}$. The density of $\pi_S$ is at most $c_+^d$, and integrating over $\mb R^d$ gives $m_S\leq c_+^de^{\ell_{n,S}(\widehat\theta_S)}\{2\pi/(nm)\}^{d/2}$, that is, $\delta_S\leq d\{\log c_++\frac12\log(2\pi/m)\}$.

For the lower bound, let $z$ be the projection of $\widehat\theta_S$ onto the rectangle $K_S$. Every interval $\mathcal I_j$ has length at least $2a_\pi\geq2n^{-1/2}$, so $K_S$ contains a box with a corner at $z$ and sides of length $n^{-1/2}$. For $\vartheta$ in this box, $\|\vartheta-\widehat\theta_S\|_2^2\leq2U_S^\pi/n+2d/n$. If $\widehat\theta_S$ is an interior maximizer, the gradient vanishes there, so
\[
\ell_{n,S}(\vartheta)\geq\ell_{n,S}(\widehat\theta_S)-\frac{nM}2\|\vartheta-\widehat\theta_S\|_2^2\geq\ell_{n,S}(\widehat\theta_S)-M(U_S^\pi+d),
\]
and integrating the lower bound $c_-^d$ of the density over the box, whose volume is $n^{-d/2}$, gives $\delta_S\geq d(\log c_--M)-MU_S^\pi$. We use a box and not a ball because the volume of a ball of radius $n^{-1/2}$ would contribute a term of order $d\log d$.
\end{proof}

\section{General Tools and Standard Results}\label{app:tools}

This appendix collects general results used in the proofs. Appendix~\ref{app:concentration} gives concentration, covering and empirical-process bounds and the empirical risk minimization bound, Appendix~\ref{app:logit-tools} gives facts on probability vectors and logits, and Appendix~\ref{app:relu} gives ReLU approximation bounds. Constants in the statements are numerical unless their dependence is specified, and within a proof $C$ may change from line to line.

\subsection{Concentration, covering and empirical processes}\label{app:concentration}

\subsubsection{Uniform concentration and likelihood comparisons}

We first give a concentration bound for bounded parametric families using finite nets.

\begin{lemma}[Concentration over a parametric family]\label{lem:net-concentration}
Let $X_1,\ldots,X_n\stackrel{\rm iid}{\sim}P$ on $\mathcal X$, and let $\mathcal I$ be a finite set. For $i\in\mathcal I$, let $\mathcal Z_i\subseteq B_2(z_i,R_i)\subset\mathbb R^{d_i}$ and let $f_{i,z}:\mathcal X\to\mb R$, $z\in\mathcal Z_i$, be measurable with $\sup_{z\in\mathcal Z_i,\,x\in\mathcal X}|f_{i,z}(x)|\leq B_i$ and $|f_{i,z}(x)-f_{i,z'}(x)|\leq L_i\|z-z'\|_2$ for all $z,z'\in\mathcal Z_i$ and $x\in\mathcal X$. Then, for $h_i>0$ and $u_i>2L_ih_i$,
\begin{equation}
 \mb P\Bigl\{\exists i\in\mathcal I:\sup_{z\in\mathcal Z_i}|(P_n-P)f_{i,z}|>u_i\Bigr\}
 \leq 2\sum_{i\in\mathcal I}\Bigl(1+\frac{2R_i}{h_i}\Bigr)^{d_i}
       \exp\Bigl\{-\frac{n(u_i-2L_ih_i)^2}{2B_i^2}\Bigr\}.
 \label{eq:bounded-parametric-concentration}
\end{equation}
The same bound holds conditionally on a $\sigma$-field $\mathcal H$ if, given $\mathcal H$, the observations are i.i.d.\ from a law $P_{\mathcal H}$, the sets $\mathcal Z_i$, the functions $f_{i,z}$ and the constants are $\mathcal H$-measurable, and $P$ is replaced by $P_{\mathcal H}$. In particular, for a resample $X^{*n}$ drawn from the empirical law $P_n$ of $X^n$, it holds conditionally on $X^n$ with $(P_n^*,P_n)$ in place of $(P_n,P)$.
\end{lemma}

\begin{proof}
Fix $i\in\mathcal I$. A maximal $h_i$-separated subset $\mathcal N_i$ of $\mathcal Z_i$ is an $h_i$-net of $\mathcal Z_i$. The balls of radius $h_i/2$ around its points are disjoint and contained in $B_2(z_i,R_i+h_i/2)$, so the volume argument in~\citet[Lemma~5.7 and Example~5.8, pp.~125--126]{wainwright2019high} gives $|\mathcal N_i|\leq(1+2R_i/h_i)^{d_i}$. For any $z\in\mathcal Z_i$, choose $v\in\mathcal N_i$ with $\|z-v\|_2\leq h_i$. By the Lipschitz bound, $|f_{i,z}(x)-f_{i,v}(x)|\leq L_ih_i$ for all $x\in\mathcal X$, so $|P_n(f_{i,z}-f_{i,v})|\leq L_ih_i$ and $|P(f_{i,z}-f_{i,v})|\leq L_ih_i$. Hence $|(P_n-P)(f_{i,z}-f_{i,v})|\leq2L_ih_i$, and the triangle inequality gives
\[
\sup_{z\in\mathcal Z_i}|(P_n-P)f_{i,z}|\leq\max_{v\in\mathcal N_i}|(P_n-P)f_{i,v}|+2L_ih_i .
\]
For fixed $v\in\mathcal N_i$, the variables $f_{i,v}(X_j)$ are independent with values in $[-B_i,B_i]$. Since $u_i-2L_ih_i>0$, Hoeffding's inequality \citep[Proposition~2.5 and (2.11), pp.~24--25]{wainwright2019high} at threshold $u_i-2L_ih_i$ gives $\mb P\{|(P_n-P)f_{i,v}|>u_i-2L_ih_i\}\leq2\exp\{-n(u_i-2L_ih_i)^2/(2B_i^2)\}$. Since $|\mathcal N_i|\leq(1+2R_i/h_i)^{d_i}$, the bound on the supremum and the union bound over $i\in\mathcal I$ and $v\in\mathcal N_i$ prove~\eqref{eq:bounded-parametric-concentration}. The supremum is measurable because $z\mapsto(P_n-P)f_{i,z}$ is continuous on the separable set $\mathcal Z_i$. Given $\mathcal H$, the same argument applies with $P_{\mathcal H}$ in place of $P$.
\end{proof}

The second lemma is Bernstein's inequality on a net of the sphere. The adaptive mesh in~\eqref{eq:sphere-bernstein-adaptive} keeps the factor $1/2$ in the Gaussian exponent, which the likelihood-ratio tails in Appendix~\ref{app:technical} require.

\begin{lemma}[Norm of a bounded random vector]\label{lem:sphere-bernstein}
Let $\xi_1,\ldots,\xi_n\in\mb R^b$ be independent with $E\xi_i=0$, $\|\xi_i\|_2\leq K$ and $\sup_{\|v\|_2=1}E(v^\top\xi_i)^2\leq\sigma^2$, and put $U=n^{-1/2}\sum_{i=1}^n\xi_i$. For $\alpha\in(0,1)$ and $r>0$,
\begin{equation}
\mb P(\|U\|_2\geq r)\leq\Bigl(1+\frac2\alpha\Bigr)^b\exp\Bigl\{-\frac{(1-\alpha^2/2)^2r^2}{2\{\sigma^2+Kr/(3\sqrt n)\}}\Bigr\}.
\label{eq:sphere-bernstein}
\end{equation}
If $\sigma^2=1$ and $w\geq4b$, then
\begin{equation}
\mb P(\|U\|_2^2\geq w)\leq3^b\Bigl(\frac wb\Bigr)^{b/2}\exp\Bigl\{-\frac w2+\frac b2+\frac{Kw^{3/2}}{6\sqrt n}\Bigr\}.
\label{eq:sphere-bernstein-adaptive}
\end{equation}
\end{lemma}

\begin{proof}
Let $\mathcal V$ be a maximal $\alpha$-separated subset of the unit sphere in $\mb R^b$. As in the proof of Lemma~\ref{lem:net-concentration}, $\mathcal V$ is an $\alpha$-net with $|\mathcal V|\leq(1+2/\alpha)^b$. For $U\ne0$, put $t=U/\|U\|_2$ and choose $v\in\mathcal V$ with $\|v-t\|_2\leq\alpha$. Since $v$ and $t$ are unit vectors, $v^\top t=1-\|v-t\|_2^2/2\geq1-\alpha^2/2$, so $v^\top U\geq(1-\alpha^2/2)\|U\|_2$. Put $s=(1-\alpha^2/2)r$. Then
\[
\{\|U\|_2\geq r\}\subseteq\bigcup_{v\in\mathcal V}\{v^\top U\geq s\}.
\]
Fix $v\in\mathcal V$ and put $X_i=v^\top\xi_i$. Since $\|v\|_2=1$, the variables $X_1,\ldots,X_n$ are independent with $EX_i=0$, $|X_i|\leq\|\xi_i\|_2\leq K$ and $\sum_{i=1}^nEX_i^2\leq n\sigma^2$. For every integer $q\geq3$, we have $(X_i)_+^q\leq K^{q-2}X_i^2$, and $3^{q-2}\leq q!/2$ gives $K^{q-2}\leq(q!/2)(K/3)^{q-2}$. Hence
\[
\sum_{i=1}^nE(X_i)_+^q\leq K^{q-2}\sum_{i=1}^nEX_i^2\leq\frac{q!}2\,n\sigma^2\Bigl(\frac K3\Bigr)^{q-2},\qquad q\geq3.
\]
Bernstein's inequality \citep[Theorem~2.10 and Corollary~2.11, p.~37]{boucheron2013concentration} with $v=n\sigma^2$ and $c=K/3$ therefore gives
\[
\mb P(v^\top U\geq s)=\mb P\Bigl(\sum_{i=1}^nX_i\geq s\sqrt n\Bigr)\leq\exp\Bigl\{-\frac{s^2}{2\{\sigma^2+Ks/(3\sqrt n)\}}\Bigr\}.
\]
Since $s=(1-\alpha^2/2)r\leq r$, we have $Ks/(3\sqrt n)\leq Kr/(3\sqrt n)$. Replacing $s$ by $r$ in the denominator and taking a union bound over $v\in\mathcal V$ with $|\mathcal V|\leq(1+2/\alpha)^b$ yields~\eqref{eq:sphere-bernstein}.

For~\eqref{eq:sphere-bernstein-adaptive}, suppose $\sigma^2=1$ and $w\geq4b$, and put $r=\sqrt w$ and $\alpha=\sqrt{b/w}$. Since $w\geq4b$, we have $\alpha\leq1/2$, so $1+2/\alpha\leq3/\alpha=3(w/b)^{1/2}$ and $(1+2/\alpha)^b\leq3^b(w/b)^{b/2}$. We also have $(1-\alpha^2/2)^2w=(1-b/(2w))^2w\geq w-b$. Put $x=K\sqrt w/(3\sqrt n)$. Since $x\geq0$ and $w\geq b$, the inequality $1/(1+x)\geq1-x$ gives
\[
\frac{(1-\alpha^2/2)^2w}{2(1+x)}\geq\frac{(w-b)(1-x)}2\geq\frac w2-\frac b2-\frac{Kw^{3/2}}{6\sqrt n}.
\]
Applying~\eqref{eq:sphere-bernstein} with these values of $r$ and $\alpha$ and substituting the bounds above gives~\eqref{eq:sphere-bernstein-adaptive}.
\end{proof}

For independent vectors $Y_i\in\mathbb R^q$ with $EY_i=0$ and $\sup_iE\|Y_i\|_2^4<\infty$, expanding the fourth power and applying the Cauchy--Schwarz inequality to the cross terms gives $\mb E[\|\sum_iY_i\|_2^4]\leq\{n+3n(n-1)\}\sup_i\mb E[\|Y_i\|_2^4]$, and hence
\begin{equation}
 \mb E\left\|n^{-1}\sum_{i=1}^nY_i\right\|_2^4
 \leq 3n^{-2}\sup_i\mb E[\|Y_i\|_2^4].
 \label{eq:average-fourth-moment}
\end{equation}

The next lemma gives two uniform bounds for log-likelihood processes over a finite collection of parametric families. The additive bound is used for misspecified candidates, and the likelihood-ratio bound for correctly specified comparisons. Its notation is that of Appendix~\ref{app:selection-setup}.

\begin{lemma}[Likelihood-process bounds]\label{lem:lik-process}
Suppose that, for every $i\in\mathcal I$,
\[
 \begin{gathered}
 \Theta_i\subseteq B_2(\theta_i,R)\subset\mathbb R^{d_i},\qquad
 \sup_{x\in\mathcal X,\,\theta\in\Theta_i}|f_{i,\theta}(x)|\leq B,\\
 |f_{i,\theta}(x)-f_{i,\theta'}(x)|
 \leq L_{\rm lik}\|\theta-\theta'\|_2
 \quad(x\in\mathcal X,\ \theta,\theta'\in\Theta_i).
 \end{gathered}
\]
Put $d_{\max}=\max_i d_i$ and
$\mathcal Z_n=  \max_{i\in\mathcal I}\sup_{\theta\in\Theta_i} |(P_n-P_0)f_{i,\theta}|$.
There is a numerical constant $C$ such that
\begin{equation}
\mb E_0[\mathcal Z_n]
\leq C(B+L_{\rm lik}R)
       \left\{\frac{\log(2|\mathcal I|)+d_{\max}}{n}\right\}^{1/2}.
\label{eq:likelihood-additive}
\end{equation}
For the remaining assertions, suppose instead that, $\mathbb P_0$-almost surely, $Y_{i,\theta}=\log(p_0/p_{i,\theta})$ satisfies $\sup_{i,\theta\in\Theta_i}|Y_{i,\theta}|\leq B_Y$, where $B_Y\geq1$, and $|Y_{i,\theta}-Y_{i,\theta'}|\leq L_Y\|\theta-\theta'\|_2$ for all $i\in\mathcal I$ and $\theta,\theta'\in\Theta_i$.
Put $C_Y=B_Y^2/(e^{-B_Y}-1+B_Y)$ and $\lambda_Y=(C_Y+2B_Y/3)^{-1}$. Then $\mb E_0[Y_{i,\theta}^2]\leq C_YR_i(\theta)$ and $\mb E_0\exp[\lambda_Yn\{R_i(\theta)/2-P_nY_{i,\theta}\}]\leq1$.
Let $N_i(h)$ be the size of an $h$-net of $\Theta_i$.  For every $h>0$,
\begin{equation}
\mb E_0\max_i\sup_{\theta\in\Theta_i}
\left\{\frac{R_i(\theta)}2-P_nY_{i,\theta}\right\}
\leq\frac{\log\sum_iN_i(h)}{\lambda_Yn}+\frac{3L_Yh}{2},
\label{eq:likelihood-uniform}
\end{equation}
and, for every $t>0$,
\begin{equation}
\mb P_0\Bigl[\max_i\sup_{\theta\in\Theta_i}
\Bigl\{\frac{R_i(\theta)}2-P_nY_{i,\theta}\Bigr\}
>\frac{\log\sum_iN_i(h)+t}{\lambda_Yn}+\frac{3L_Yh}{2}\Bigr]\leq e^{-t}.
\label{eq:likelihood-tail}
\end{equation}
Also, with $R_i^*=\inf_{\theta\in\Theta_i}R_i(\theta)$,
\begin{equation}
\mb P_0^{(n)}
\left\{\inf_{\theta\in\Theta_i}P_nY_{i,\theta}\leq v\right\}
\leq N_i(h)\exp\left[
-\lambda_Yn\left\{\frac{R_i^*}{2}-v-L_Yh\right\}
\right].                                                 \label{eq:likelihood-gap}
\end{equation}
One may take $N_i(h)\leq(1+2R/h)^{d_i}$.
\end{lemma}
\begin{proof}
For the additive bound we use the sub-Gaussian increments of the centered empirical process. For the likelihood-ratio bounds, we instead exploit the fact that the second moment of the log likelihood ratio is controlled by its KL divergence. This yields a bound that is particularly useful when the divergence is small.

\medskip
\noindent{\bf Additive bound.} Let $\mathcal F=\{0\}\cup\{\pm f_{i,\theta}:i\in\mathcal I,\ \theta\in\Theta_i\}$ and $X_f=(P_n-P_0)f$, so $\mathcal Z_n=\sup_{f\in\mathcal F}X_f\leq\sup_{f,g\in\mathcal F}(X_f-X_g)$. For $f,g\in\mathcal F$, the increment $X_f-X_g=(P_n-P_0)(f-g)$ is an average of $n$ independent centered variables, each with values in an interval of length $2\|f-g\|_\infty$. By the sub-Gaussian bound for bounded variables \citep[Proposition~2.5 and (2.11)]{wainwright2019high}, $\{X_f:f\in\mathcal F\}$ is a zero-mean sub-Gaussian process with respect to the metric $\rho(f,g)=\|f-g\|_\infty/\sqrt n$, in the sense of~\citet[Definition~5.16]{wainwright2019high}. Since $\sup_{f\in\mathcal F}\|f\|_\infty\leq B$, its $\rho$-diameter is at most $2B/\sqrt n$. Moreover, the Lipschitz condition and the covering bound in the proof of Lemma~\ref{lem:net-concentration} give, for $u>0$,
\[
N(u;\mathcal F,\rho)\leq 1+2\sum_{i\in\mathcal I}\left(1+\frac{2L_{\rm lik}R}{u\sqrt n}\right)^{d_i}
\leq(1+2|\mathcal I|)\left(1+\frac{2L_{\rm lik}R}{u\sqrt n}\right)^{d_{\max}}.
\]
Since $\sqrt{a+b}\leq\sqrt a+\sqrt b$ and $\log(1+x)\leq x$, we have
\[
\begin{aligned}
\sqrt{\log N(u;\mathcal F,\rho)}
&\leq\sqrt{\log(1+2|\mathcal I|)}+\sqrt{d_{\max}\log\left(1+\frac{2L_{\rm lik}R}{u\sqrt n}\right)}\\
&\leq\sqrt{\log(1+2|\mathcal I|)}+\sqrt{d_{\max}}\sqrt{\frac{2L_{\rm lik}R}{u\sqrt n}}.
\end{aligned}
\]
Dudley's entropy integral bound \citep[Theorem~5.22, p.~140]{wainwright2019high} with $\delta=0$, combined with the preceding bound over $u\in[0,2B/\sqrt n]$ and the substitution $v=u\sqrt n$, gives
\[
\begin{aligned}
\mb E_0[\mathcal Z_n]
&\leq32\int_0^{2B/\sqrt n}\sqrt{\log N(u;\mathcal F,\rho)}\,du\leq\frac{32}{\sqrt n}\left\{2B\sqrt{\log(1+2|\mathcal I|)}+\sqrt{d_{\max}}\int_0^{2B}\sqrt{\frac{2L_{\rm lik}R}{v}}\,dv\right\}.
\end{aligned}
\]
The last integral satisfies $\int_0^{2B}\sqrt{2L_{\rm lik}R/v}\,dv=4\sqrt{BL_{\rm lik}R}\leq2(B+L_{\rm lik}R)$. Since $\log(1+2|\mathcal I|)\leq2\log(2|\mathcal I|)$, $B\leq B+L_{\rm lik}R$ and $\sqrt a+\sqrt b\leq\sqrt{2(a+b)}$, we obtain $\mb E_0[\mathcal Z_n]\leq C(B+L_{\rm lik}R)\{(\log(2|\mathcal I|)+d_{\max})/n\}^{1/2}$ for a numerical constant $C$. This proves~\eqref{eq:likelihood-additive}.

\medskip
\noindent{\bf Likelihood-ratio bounds.} For $|y|\leq B_Y$, we have
\[
e^{-y}-1+y=y^2\int_0^1(1-t)e^{-ty}\,dt\geq y^2\int_0^1(1-t)e^{-tB_Y}\,dt=C_Y^{-1}y^2.
\]
Since $|Y_{i,\theta}|\leq B_Y$ $\mathbb P_0$-almost surely, $E_0e^{-Y_{i,\theta}}\leq1$ and $E_0Y_{i,\theta}=R_i(\theta)$, we obtain $E_0[Y_{i,\theta}^2]\leq C_YE_0[e^{-Y_{i,\theta}}-1+Y_{i,\theta}]\leq C_YR_i(\theta)$. Put $\xi_j=-Y_{i,\theta}(X_j)$. Then $\sum_jE_0\xi_j^2\leq nC_YR_i(\theta)$ and $(\xi_j)_+\leq B_Y$, so $(\xi_j)_+^q\leq B_Y^{q-2}\xi_j^2$ for every $q\geq3$. Since $(3/2)^{q-2}\leq q!/2$ for $q\geq3$, we have $B_Y^{q-2}\leq(q!/2)(2B_Y/3)^{q-2}$, and hence
\[
\sum_jE_0(\xi_j)_+^q\leq B_Y^{q-2}\sum_jE_0\xi_j^2\leq\frac{q!}2\,nC_YR_i(\theta)\Bigl(\frac{2B_Y}3\Bigr)^{q-2}.
\]
Theorem~2.10 in~\citet[p.~37]{boucheron2013concentration}, applied to $\sum_j(\xi_j-E_0\xi_j)=n\{R_i(\theta)-P_nY_{i,\theta}\}$ with $v=nC_YR_i(\theta)$ and $c=2B_Y/3$, yields
\[
\log E_0\exp\bigl[\lambda n\{R_i(\theta)-P_nY_{i,\theta}\}\bigr]\leq\frac{n\lambda^2C_YR_i(\theta)}{2(1-2B_Y\lambda/3)},\qquad 0<\lambda<\frac3{2B_Y}.
\]
At $\lambda=\lambda_Y=(C_Y+2B_Y/3)^{-1}$, we have $0<\lambda_Y<3/(2B_Y)$ and $\lambda_YC_Y=1-2B_Y\lambda_Y/3$, so the right side equals $n\lambda_YR_i(\theta)/2$. Subtracting $n\lambda_YR_i(\theta)/2$ from both sides gives $E_0\exp[\lambda_Yn\{R_i(\theta)/2-P_nY_{i,\theta}\}]\leq1$.

Now let $Z_{i,\theta}=R_i(\theta)/2-P_nY_{i,\theta}$ and let $\mathcal N_i\subset\Theta_i$ be an $h$-net of size $N_i(h)$. Since $|Y_{i,\theta}-Y_{i,\theta'}|\leq L_Y\|\theta-\theta'\|_2$, both $R_i(\theta)=E_0Y_{i,\theta}$ and $P_nY_{i,\theta}$ are $L_Y$-Lipschitz in $\theta$. When $\|\theta-\theta'\|_2\leq h$, the triangle inequality gives
\[
|Z_{i,\theta}-Z_{i,\theta'}|\leq\frac12|R_i(\theta)-R_i(\theta')|+|P_nY_{i,\theta}-P_nY_{i,\theta'}|\leq\frac{3L_Yh}2,
\]
so the supremum of $Z_{i,\theta}$ over $i$ and $\theta\in\Theta_i$ exceeds the maximum over the nets by at most $3L_Yh/2$. Bounding the maximum of $e^{\lambda_YnZ_{i,\theta'}}$ by the sum, applying Jensen's inequality and using $E_0e^{\lambda_YnZ_{i,\theta'}}\leq1$ for every $i$ and $\theta'\in\mathcal N_i$, we have
\[
\begin{aligned}
E_0\max_i\max_{\theta'\in\mathcal N_i}Z_{i,\theta'}
&\leq\frac1{\lambda_Yn}\log E_0\sum_i\sum_{\theta'\in\mathcal N_i}e^{\lambda_YnZ_{i,\theta'}}\\
&=\frac1{\lambda_Yn}\log\sum_i\sum_{\theta'\in\mathcal N_i}E_0e^{\lambda_YnZ_{i,\theta'}}
\leq\frac{\log\sum_iN_i(h)}{\lambda_Yn}.
\end{aligned}
\]
Adding $3L_Yh/2$ gives $E_0\max_i\sup_{\theta\in\Theta_i}Z_{i,\theta}\leq(\lambda_Yn)^{-1}\log\sum_iN_i(h)+3L_Yh/2$, which is~\eqref{eq:likelihood-uniform}.

For~\eqref{eq:likelihood-tail}, fix $t>0$. Using the same net approximation, the union bound over the nets, Markov's inequality applied to $e^{\lambda_YnZ_{i,\theta'}}$ and $E_0e^{\lambda_YnZ_{i,\theta'}}\leq1$, we obtain
\[
\begin{aligned}
&P_0\Bigl\{\max_i\sup_{\theta\in\Theta_i}Z_{i,\theta}>\frac{\log\sum_iN_i(h)+t}{\lambda_Yn}+\frac{3L_Yh}2\Bigr\}\\
&\leq P_0\Bigl\{\max_i\max_{\theta'\in\mathcal N_i}Z_{i,\theta'}>\frac{\log\sum_iN_i(h)+t}{\lambda_Yn}\Bigr\} \leq\sum_i\sum_{\theta'\in\mathcal N_i}\frac{E_0e^{\lambda_YnZ_{i,\theta'}}}{e^t\sum_iN_i(h)}\leq e^{-t}.
\end{aligned}
\]
For~\eqref{eq:likelihood-gap}, $P_nY_{i,\theta}$ is $L_Y$-Lipschitz in $\theta$ and $\mathcal N_i$ is an $h$-net of $\Theta_i$, so
\[
\Bigl\{\inf_{\theta\in\Theta_i}P_nY_{i,\theta}\leq v\Bigr\}\subseteq\bigcup_{\theta'\in\mathcal N_i}\{P_nY_{i,\theta'}\leq v+L_Yh\}.
\]
For each $\theta'\in\mathcal N_i$, Markov's inequality applied to $e^{\lambda_YnZ_{i,\theta'}}$ and $E_0e^{\lambda_YnZ_{i,\theta'}}\leq1$ give $P_0\{P_nY_{i,\theta'}\leq v+L_Yh\}=P_0\{Z_{i,\theta'}\geq R_i(\theta')/2-v-L_Yh\}\leq e^{-\lambda_Yn\{R_i(\theta')/2-v-L_Yh\}}$. By the union bound over $\theta'\in\mathcal N_i$ and the inequality $R_i(\theta')\geq R_i^*$,
\[
P_0\bigl\{\inf_{\theta\in\Theta_i}P_nY_{i,\theta}\leq v\bigr\}
\leq\sum_{\theta'\in\mathcal N_i}e^{-\lambda_Yn\{R_i(\theta')/2-v-L_Yh\}}
\leq N_i(h)e^{-\lambda_Yn\{R_i^*/2-v-L_Yh\}},
\]
which is~\eqref{eq:likelihood-gap}. The bound on $N_i(h)$ is the covering bound in the proof of Lemma~\ref{lem:net-concentration}.
\end{proof}

The bootstrap proof compares the law of a scaled empirical average with that of its resampled version, whose limiting covariance is doubled.

\subsubsection{Gaussian comparison for bootstrap inference}

The next result is used for the trained estimator in Appendix~\ref{app:proof-boot}. It compares empirical averages centered at their population mean with averages from a resample, also centered at the population mean. The second centering accounts for the additional source of variation in the resample.

\begin{lemma}[Gaussian comparison for a sample and its resample]\label{lem:gauss-compare}
Fix $q$ and $B<\infty$, and let $\mathfrak C_{q,B}$ be the set of pairs $(P,\xi)$ such that $\xi:\mathcal X\to\mathbb R^q$ is measurable, $P\xi=0$ and $\|\xi\|_2\leq B$ $P$-a.s. For an i.i.d.\ sample from $P$ and its size-$n$ empirical resample, put $U_n=\sqrt nP_n\xi$, $U_n^*=\sqrt nP_n^*\xi$ and $\Sigma=P\xi\xi^\top$. Then
\[
\begin{gathered}
\sup_{(P,\xi)\in\mathfrak C_{q,B}} d_{\rm BL}\{\mathcal L(U_n),\mathcal N(0,\Sigma)\}\to0, \quad \sup_{(P,\xi)\in\mathfrak C_{q,B}} d_{\rm BL}\{\mathcal L(U_n^*),\mathcal N(0,2\Sigma)\}\to0,
\end{gathered}
\]
where the second law integrates over the sample and resample. For every $\Sigma\in\mathbb S_+^q$ and measurable $f:\mathbb R^q\to[0,1]$,
\begin{equation}
 \mathbb Ef\{N(0,2\Sigma)\}
 \leq2^{q/6}\left[\mathbb Ef\{N(0,\Sigma)\}\right]^{1/3}.
 \label{eq:doubled-gaussian-comparison}
\end{equation}
\end{lemma}
\begin{proof}
The set $\{\Sigma\succeq0:\operatorname{tr}\Sigma\leq B^2\}$ is compact. If the first claim failed, there would be $\epsilon>0$, $n_j\to\infty$ and $(P_j,\xi_j)\in\mathfrak C_{q,B}$ with $d_{\rm BL}\{\mathcal L(U_{n_j}),\mathcal N(0,\Sigma_j)\}\geq\epsilon$. Passing to a subsequence, we may assume $\Sigma_j\to\Sigma_\infty$. The vectors $Y_{j,i}=\xi_j(X_i)/\sqrt{n_j}$, $i\leq n_j$, form a triangular array of independent centered vectors with $\sum_i\operatorname{Cov}Y_{j,i}=\Sigma_j\to\Sigma_\infty$. For any $\eta>0$, we have $\sum_iE\|Y_{j,i}\|_2^2\ind\{\|Y_{j,i}\|_2>\eta\}=0$ once $B/\sqrt{n_j}\leq\eta$. The Lindeberg--Feller theorem \citep[Proposition~2.27, pp.~20--21]{van2000asymptotic} gives $\mathcal L(U_{n_j})\rightsquigarrow\mathcal N(0,\Sigma_\infty)$, which may be singular, and $\Sigma_j\to\Sigma_\infty$ gives $\mathcal N(0,\Sigma_j)\rightsquigarrow\mathcal N(0,\Sigma_\infty)$. Since $d_{\rm BL}$ metrizes weak convergence, this contradicts $d_{\rm BL}\{\mathcal L(U_{n_j}),\mathcal N(0,\Sigma_j)\}\geq\epsilon$, proving the first claim.

For the resample, let $\widehat\Sigma_n=P_n(\xi\xi^\top)-(P_n\xi)(P_n\xi)^\top$. Since the sample is i.i.d. and $\|\xi\|_2\leq B$, we have $\|\xi\xi^\top\|_{\rm F}\leq B^2$, so $E\|P_n(\xi\xi^\top)-\Sigma\|_{\rm F}^2\leq B^4/n$ and $E\|P_n\xi\|_2^2\leq B^2/n$. Hence
\[
E\|\widehat\Sigma_n-\Sigma\|_{\rm F}\leq\{E\|P_n(\xi\xi^\top)-\Sigma\|_{\rm F}^2\}^{1/2}+E\|P_n\xi\|_2^2\leq B^2n^{-1/2}+B^2n^{-1}.
\]
Conditionally on the sample, the resampled values of $\xi-P_n\xi$ are i.i.d., centered, bounded by $2B$ in norm, and have covariance $\widehat\Sigma_n$. Fix $t\in\mb R^q$, and let $\varphi_n^*$ be the conditional characteristic function of one resampled value at $t/\sqrt n$. Since the resampled values are bounded by $2B$ in norm, the third-order Taylor remainder of the characteristic function at $t/\sqrt n$ is at most $(2B\|t\|_2)^3/(6n^{3/2})$, so $|\varphi_n^*-1+t^\top\widehat\Sigma_nt/(2n)|\leq(2B\|t\|_2)^3/(6n^{3/2})$. We also have $|e^{-x}-1+x|\leq x^2/2$ for $x\geq0$. Since $|a^n-b^n|\leq n|a-b|$ for $|a|,|b|\leq1$, we obtain
\[
\bigl|E^*e^{it^\top\sqrt n(P_n^*-P_n)\xi}-e^{-t^\top\widehat\Sigma_nt/2}\bigr|\leq\frac{(2B\|t\|_2)^3}{6\sqrt n}+\frac{(2B\|t\|_2)^4}{8n},
\]
uniformly over $\mathfrak C_{q,B}$ and the sample. Since $U_n^*=U_n+\sqrt n(P_n^*-P_n)\xi$ and $|e^{-x}-e^{-y}|\leq|x-y|$ for $x,y\geq0$, the previous two bounds give
\[
Ee^{it^\top U_n^*}
=E[e^{it^\top U_n}e^{-t^\top\widehat\Sigma_nt/2}]+o(1) =e^{-t^\top\Sigma t/2}Ee^{it^\top U_n}+o(1)
=e^{-t^\top\Sigma t}+o(1)
\]
uniformly over $\mathfrak C_{q,B}$, where the last equality uses $Ee^{it^\top U_n}=e^{-t^\top\Sigma t/2}+o(1)$ from the first claim. The compactness argument of the first claim, with L\'evy's continuity theorem in place of the Lindeberg--Feller theorem, proves the second claim.

For~\eqref{eq:doubled-gaussian-comparison}, put $r=\operatorname{rank}\Sigma$. If $r=0$, both laws equal $\delta_0$, and since $f(0)\in[0,1]$, we have $f(0)\leq f(0)^{1/3}\leq2^{q/6}f(0)^{1/3}$. If $r>0$, restrict to the range of $\Sigma$ and whiten, so that it suffices to compare $\mathcal N(0,2I_r)$ with $\mathcal N(0,I_r)$. Let $Z\sim\mathcal N(0,I_r)$ and let $r_{\rm G}(z)=2^{-r/2}e^{\|z\|_2^2/4}$ be the density of $\mathcal N(0,2I_r)$ with respect to $\mathcal N(0,I_r)$. Then
\[
Er_{\rm G}(Z)^{3/2}=2^{-3r/4}Ee^{3\|Z\|_2^2/8}=2^{-3r/4}(1-3/4)^{-r/2}=2^{r/4},
\]
so $\|r_{\rm G}\|_{L^{3/2}}=2^{r/6}$. Since $f$ takes values in $[0,1]$, we have $f^3\leq f$. Using H\"older's inequality and $r\leq q$, we obtain
\[
E[r_{\rm G}(Z)f(Z)]
\leq\|r_{\rm G}\|_{L^{3/2}}(Ef(Z)^3)^{1/3}
\leq2^{r/6}(Ef(Z))^{1/3}\leq2^{q/6}(Ef(Z))^{1/3},
\]
which gives~\eqref{eq:doubled-gaussian-comparison}.
\end{proof}

\subsubsection{Generalization over independent training tasks}

Here the sample consists of $M$ independent statistical tasks, not the $n$ observations within a task. We use a Bernstein moment condition for the excess loss, as in the local analysis of~\citet[Section~3.1 and Theorem~3.3]{bartlett2005local}, but use a finite net of the parameter ball instead of localized complexity. This yields a bound of order $M^{-1}$ up to complexity and confidence factors. The key distinction from a direct uniform bound on the risk is that the variance of the excess loss is controlled by its expectation, so the concentration error decreases as the network approaches the population optimum.

\begin{lemma}[Empirical risk minimization under a Bernstein condition]\label{lem:bernstein-erm}
Let $\{h_\omega:\omega\in\Omega\}$, $\Omega\subseteq\{\omega\in\mb R^W:\|\omega\|_2\leq\Gamma\}$, be measurable functions of a task with $a_\omega=\mb Eh_\omega\geq0$, such that for constants $\kappa\geq1$, $c>0$ and $L_f\geq0$, $\mb E[|h_\omega|^m]\leq\kappa\,m!\,c^{m-1}a_\omega$ for all $m\geq2$ and $|h_\omega-h_{\omega'}|\leq L_f\|\omega-\omega'\|_2$ pointwise.
Let $\widehat P_M$ be the empirical measure of $M\geq1$ independent tasks and let $\delta\in(0,1)$. With probability at least $1-\delta$, simultaneously for all $\widehat\omega,\omega_0\in\Omega$ with $\widehat P_Mh_{\widehat\omega}\leq\widehat P_Mh_{\omega_0}+\varepsilon$,
\[
a_{\widehat\omega}\leq3a_{\omega_0}+(8\kappa+16)\,c\,\frac{W\log\{1+2\Gamma M(1+L_f/c)\}+\log(2/\delta)}M+2\varepsilon.
\]
\end{lemma}

\begin{proof}
Fix $\omega\in\Omega$ and let $X_j=h_\omega(\mathsf t_j)$ for the $M$ independent training tasks $\mathsf t_j$ with distribution $P$, so that $Ph_\omega=a_\omega$. The moment condition gives $\mb EX_j^2\leq2\kappa ca_\omega$ at $m=2$ and $\mb E|X_j|^q\leq\kappa q!c^{q-1}a_\omega=(q!/2)(2\kappa ca_\omega)c^{q-2}$ at $m=q\geq3$. Summing over $j$ gives $\sum_{j=1}^M\mb EX_j^2\leq2M\kappa ca_\omega$ and $\sum_{j=1}^M\mb E|X_j|^q\leq(q!/2)(2M\kappa ca_\omega)c^{q-2}$ for $q\geq3$, and the same bounds hold for $-X_j$. Theorem~2.10 in~\citet[p.~37]{boucheron2013concentration}, applied to $\pm X_j$ with $v=2M\kappa ca_\omega$, shows that for every $t>0$, except on an event of probability at most $2e^{-t}$,
\[
|(\widehat P_M-P)h_\omega|\leq\sqrt{\frac{4\kappa a_\omega ct}M}+\frac{ct}M\leq\frac{a_\omega}2+r,\qquad r=\frac{(2\kappa+1)ct}M,
\]
where the second inequality uses $\sqrt{4\kappa a_\omega ct/M}\leq a_\omega/2+2\kappa ct/M$.

Take a net of $\Omega$ of mesh $\{M(1+L_f/c)\}^{-1}$ and cardinality $N\leq\{1+2\Gamma M(1+L_f/c)\}^W$, and set $t=\log(2N/\delta)$. By a union bound over the $N$ net points, the bound above fails for at least one net point with probability at most $2Ne^{-t}=\delta$, so it holds simultaneously on the net with probability at least $1-\delta$. For any $\omega\in\Omega$ and its nearest net point $\nu$, we have $\|\omega-\nu\|_2\leq\{M(1+L_f/c)\}^{-1}$, so the Lipschitz condition gives $|h_\omega-h_\nu|\leq L_f/\{M(1+L_f/c)\}\leq c/M$ pointwise. Hence $|a_\omega-a_\nu|\leq c/M$ and $|\widehat P_Mh_\omega-\widehat P_Mh_\nu|\leq c/M$. On this event, since $\frac12a_\nu-r\leq\widehat P_Mh_\nu\leq\frac32a_\nu+r$, we have for every $\omega\in\Omega$
\[
\begin{aligned}
\widehat P_Mh_\omega&\geq\widehat P_Mh_\nu-\frac cM\geq\frac12a_\nu-r-\frac cM\geq\frac12a_\omega-r-\frac{3c}{2M},\\
\widehat P_Mh_\omega&\leq\widehat P_Mh_\nu+\frac cM\leq\frac32a_\nu+r+\frac cM\leq\frac32a_\omega+r+\frac{5c}{2M}.
\end{aligned}
\]

On the same event, for all $\widehat\omega,\omega_0\in\Omega$ with $\widehat P_Mh_{\widehat\omega}\leq\widehat P_Mh_{\omega_0}+\varepsilon$, applying the lower bound to $\widehat\omega$ and the upper bound to $\omega_0$ gives
\[
\frac12a_{\widehat\omega}-r-\frac{3c}{2M}\leq\widehat P_Mh_{\widehat\omega}\leq\widehat P_Mh_{\omega_0}+\varepsilon\leq\frac32a_{\omega_0}+r+\frac{5c}{2M}+\varepsilon,
\]
so $a_{\widehat\omega}\leq3a_{\omega_0}+4r+8c/M+2\varepsilon$. Since $N\geq1$ and $\delta\in(0,1)$, we have $t=\log(2N/\delta)\geq\log2$, and hence
\[
4r+\frac{8c}M\leq\{4(2\kappa+1)+8/\log2\}\frac{ct}M\leq(8\kappa+16)\frac{ct}M.
\]
Since $N\leq\{1+2\Gamma M(1+L_f/c)\}^W$, we also have $t=\log(2N/\delta)\leq W\log\{1+2\Gamma M(1+L_f/c)\}+\log(2/\delta)$.  Substituting these bounds proves the claim.
\end{proof}

For the squared-error loss of parameter estimation, the moment condition is verified directly.

\begin{lemma}[A fast bound for squared-error training]\label{lem:fast-rate}
Let $\Theta_0$ be compact and convex, and suppose $\Omega\subseteq B_2(0,\Gamma)\subset\mathbb R^W$ and $\theta,F_\omega(X^n)\in\Theta_0$ for $\omega\in\Omega$.
For $f_\omega(X^n,\theta)=\|F_\omega(X^n)-\theta\|_2^2$, assume
$\sup_{X^n,\theta}|f_\omega(X^n,\theta)-f_{\omega'}(X^n,\theta)| \leq L_{f,n}\|\omega-\omega'\|_2$ for $\omega,\omega'\in\Omega$.
Put $B_\Theta=1\vee\operatorname{diam}(\Theta_0)^2$ and $A_{\mathcal F}=\inf_{F\in\mathcal F}\mathbb E_{\mathbb Q_n}[\|F-F^*\|_2^2]$.
With probability at least $1-\delta$, every approximate empirical risk
minimizer satisfying~\eqref{eq:approx-erm} obeys
\begin{equation}
 \mathcal R(\widehat F)-\mathcal R^*
 \leq3A_{\mathcal F}
 +C B_\Theta\frac{W\log\{1+2\Gamma M(1+L_{f,n})\}
                         +\log(2/\delta)}{M}
 +2\varepsilon_{\rm opt}.
 \label{eq:fast-erm-proof}
\end{equation}
Here $C$ is numerical. For the architectures in Lemma~\ref{lem:param-lipschitz},
$\log\{1+2\Gamma M(1+L_{f,n})\}\leq   C\{\Xi_n+\log(eM)\}$,
where this $C$ depends only on the fixed architecture bounds.
\end{lemma}

\begin{proof}
Since $\Theta_0$ is compact and convex and $\theta\in\Theta_0$, we have $F^*(X^n)=E[\theta\mid X^n]\in\Theta_0$. Put $D_0=\operatorname{diam}(\Theta_0)$ and define
\[
h_\omega=\|F_\omega(X^n)-\theta\|_2^2-\|F^*(X^n)-\theta\|_2^2,\qquad a_\omega=Eh_\omega=E\|F_\omega-F^*\|_2^2,
\]
where the last equality follows from Lemma~\ref{lem:ce-selection}(c). Since $\theta,F_\omega(X^n),F^*(X^n)\in\Theta_0$, both squared losses lie in $[0,D_0^2]$, so $|h_\omega|\leq D_0^2$. Moreover, $\|F_\omega+F^*-2\theta\|_2\leq\|F_\omega-\theta\|_2+\|F^*-\theta\|_2\leq2D_0$, so the identity $h_\omega=\langle F_\omega-F^*,F_\omega+F^*-2\theta\rangle$ gives $|h_\omega|\leq2D_0\|F_\omega-F^*\|_2$. Hence $Eh_\omega^2\leq4D_0^2E\|F_\omega-F^*\|_2^2=4D_0^2a_\omega$. For every $m\geq2$, the bound $|h_\omega|\leq D_0^2$ implies $|h_\omega|^m\leq D_0^{2(m-2)}h_\omega^2$, so
\[
E|h_\omega|^m\leq D_0^{2(m-2)}Eh_\omega^2\leq4D_0^{2(m-1)}a_\omega\leq2\,m!\,B_\Theta^{m-1}a_\omega,
\]
where the last inequality uses $m!\geq2$ and $D_0^2\leq B_\Theta$. Furthermore, $|h_\omega-h_{\omega'}|=|f_\omega-f_{\omega'}|\leq L_{f,n}\|\omega-\omega'\|_2$ pointwise for all $\omega,\omega'\in\Omega$. By~\eqref{eq:approx-erm}, every approximate empirical risk minimizer $\widehat F=F_{\widehat\omega}$ satisfies $\widehat P_Mh_{\widehat\omega}\leq\widehat P_Mh_{\omega_0}+\varepsilon_{\rm opt}$ for every $\omega_0\in\Omega$. Since $8\kappa+16=32$ at $\kappa=2$, applying Lemma~\ref{lem:bernstein-erm} with $\kappa=2$, $c=B_\Theta$, $L_f=L_{f,n}$ and $\varepsilon=\varepsilon_{\rm opt}$ gives, with probability at least $1-\delta$ simultaneously for all such $\widehat\omega$ and all $\omega_0\in\Omega$,
\[
\begin{aligned}
a_{\widehat\omega}
&\leq3a_{\omega_0}+32B_\Theta\frac{W\log\{1+2\Gamma M(1+L_{f,n}/B_\Theta)\}+\log(2/\delta)}M+2\varepsilon_{\rm opt}\\
&\leq3a_{\omega_0}+32B_\Theta\frac{W\log\{1+2\Gamma M(1+L_{f,n})\}+\log(2/\delta)}M+2\varepsilon_{\rm opt},
\end{aligned}
\]
where the second inequality uses $1+L_{f,n}/B_\Theta\leq1+L_{f,n}$ because $B_\Theta\geq1$. Taking the infimum over $\omega_0\in\Omega$ and using $\inf_{\omega_0\in\Omega}a_{\omega_0}=A_{\mathcal F}$ and $a_{\widehat\omega}=\mathcal R(\widehat F)-\mathcal R^*$ yields~\eqref{eq:fast-erm-proof}. Finally, $\log\{1+2\Gamma M(1+L_{f,n})\}\leq\log(2eM)+\log(1+\Gamma)+\log(1+\Gamma L_{f,n})$. By Lemma~\ref{lem:param-lipschitz}, the last two terms are bounded by a multiple of $\Xi_n$, so $\log\{1+2\Gamma M(1+L_{f,n})\}\leq C\{\Xi_n+\log(eM)\}$ with $C$ depending only on the fixed architecture bounds.
\end{proof}

\subsection{Probability vectors, logits and excess risks}\label{app:logit-tools}

For model and variable selection, the population targets are probabilities, whereas the network constructions approximate log likelihoods and information criteria. The natural comparison is therefore between the unnormalized log probabilities (logits), rather than directly between probabilities. This subsection explains how errors in these quantities translate into KL divergence, selection errors, and bounds for the cross-entropy and joint training losses. Throughout, $\Delta_K$ is the probability simplex, $\psi(x)=e^x-1-x$, and for $u,v\in[0,1]$ we write $D_{\rm KL}(u\|v)$ for the KL divergence between the Bernoulli distributions with success probabilities $u$ and $v$. The KL divergence satisfies no triangle inequality, so all comparisons below are made through the logits.

\subsubsection{Perturbation of probability vectors}

\begin{lemma}[Perturbation of the logits]\label{lem:logit-kl}
Let $q^*\in\Delta_K$, $\delta\in\mb R^K$ and $q_k=q_k^*e^{\delta_k}/\sum_lq_l^*e^{\delta_l}$. For every index $k_1\in[K]$,
\[
D_{\rm KL}(q^*\|q)\leq\sum_{k\ne k_1}q_k^*\,\psi(\delta_k-\delta_{k_1}).
\]
In particular, if $\|\delta\|_\infty\leq c$, then $D_{\rm KL}(q^*\|q)\leq e^{2c}(1-\max_kq_k^*)$. For $K=2$, with $q^*=(u,1-u)$ and $x=\delta_1-\delta_2$, this reads $D_{\rm KL}(u\|v)\leq\min\{u\,\psi(x),(1-u)\psi(-x)\}$, where $v$ has log odds equal to those of $u$ plus $x$.
\end{lemma}

\begin{proof}
Put $\delta_k'=\delta_k-\delta_{k_1}$, which leaves $q$ unchanged because it multiplies the numerator and the denominator by $e^{-\delta_{k_1}}$. Expanding the KL divergence and substituting $q_k=q_k^*e^{\delta_k'}/\sum_lq_l^*e^{\delta_l'}$ gives
\[
\begin{aligned}
D_{\rm KL}(q^* \Vert{} q) &= \sum_{k=1}^K q_k^* \log \frac{q_k^*}{q_k} = \sum_{k=1}^K q_k^* \left[ \log\left(\sum_{l=1}^K q_l^* e^{\delta_l'}\right) - \delta_k' \right]= \log\left(\sum_{l=1}^K q_l^* e^{\delta_l'}\right) - \sum_{k=1}^K q_k^* \delta_k'.
\end{aligned}
\]
Since $q_{k_1}^* = 1 - \sum_{k \ne k_1} q_k^*$, $\delta_{k_1}'=0$ and $\log(1+y)\leq y$,
\[
\begin{aligned}
D_{\rm KL}(q^*\|q)&= \log\left( q_{k_1}^* \cdot 1 + \sum_{k \ne k_1} q_k^* e^{\delta_k'}\right) - \sum_{k=1}^K q_k^* \delta_k'=\log\left( 1 + \sum_{k \ne k_1} q_k^*(e^{\delta_k'} - 1)\right) - \sum_{k=1}^K q_k^* \delta_k'\\
&\le \sum_{k \ne k_1} q_k^*(e^{\delta_k'} - 1) - \sum_{k=1}^K q_k^* \delta_k'=\sum_{k\ne k_1}q_k^*\psi(\delta_k').
\end{aligned}
\]
If $\|\delta\|_\infty\leq c$, take $k_1\in\argmax_kq_k^*$, so $\sum_{k \ne k_1} q_k^*= 1 - \max_k q_k^*$. Since $\vert{}\delta_k'\vert{} = \vert{}\delta_k - \delta_{k_1}\vert{} \le \vert{}\delta_k\vert{} + \vert{}\delta_{k_1}\vert{} \le 2c$, we have $\psi(\delta_k') \le \psi(\vert{}\delta_k'\vert{}) \le \psi(2c) = e^{2c} - 1 - 2c \le e^{2c}$, so $D_{\rm KL}(q^*\|q)\le e^{2c}\left( 1 - \max_k q_k^*\right)$. The case $K=2$ follows by taking $k_1=2$ and $k_1=1$.
\end{proof}

\begin{lemma}[Restriction, one-sided errors and clipping]\label{lem:kl-tools}
Let $p,r\in\Delta_K$ with $r_j>0$ for all $j$.

\textup{(a)} Let $J\subseteq[K]$ with $p(J)>0$, put $t=p(J^c)$ and $u=r(J^c)$, and let $p^J=p|_J/(1-t)$ and $r^J=r|_J/(1-u)$. If $u\leq1/2$, then
\[
\begin{gathered}
D_{\rm KL}(p\|r)\leq(1-t)\,D_{\rm KL}(p^J\|r^J)+2u+\sum_{j\notin J}p_j\log\frac1{r_j},\\
(1-t)\bigl(1-\max_{j\in J}p_j^J\bigr)\leq1-\max_{j\in[K]}p_j .
\end{gathered}
\]

\textup{(b)} Suppose $r=\operatorname{softmax}(a)$ and $p=\operatorname{softmax}(a+\delta)$ with $-c-u_j\leq\delta_j\leq c$ for constants $c\geq0$ and $u_j\geq0$. Then $D_{\rm KL}(p\|r)\leq e^{2c}(1-\max_jp_j)+\max_ju_j$.

\textup{(c)} For $r=\operatorname{softmax}(a)$ and $B>0$, let $r^{[B]}=\operatorname{softmax}(\max\{a_j-\max_la_l,-B\})_{j\in[K]}$. Then $r_j^{[B]}\geq e^{-B}/K$ and $D_{\rm KL}(p\|r^{[B]})\leq D_{\rm KL}(p\|r)+Ke^{-B}$.
\end{lemma}

\begin{proof}
\emergencystretch=3em
For (a), split $D_{\rm KL}(p\|r)$ into two parts, $D_{\rm KL}(p \Vert{} r) = \sum_{j \in J} p_j \log \frac{p_j}{r_j} + \sum_{j \notin J} p_j \log \frac{p_j}{r_j}$. For the terms $j\in J$, since $p^J_j = \frac{p_j}{1-t}$ and $r^J_j = \frac{r_j}{1-u}$,
\[
\begin{aligned} \sum_{j \in J} p_j \log \frac{p_j}{r_j} &= \sum_{j \in J} (1-t)p^J_j \log \left( \frac{1-t}{1-u} \cdot \frac{p^J_j}{r^J_j} \right)= (1-t)\sum_{j \in J} p^J_j \log \frac{p^J_j}{r^J_j} + (1-t)\left(\sum_{j \in J} p^J_j\right) \log \frac{1-t}{1-u} \\ &= (1-t)D_{\rm KL}(p^J \Vert{} r^J) + (1-t)\log \frac{1-t}{1-u}\le (1-t)D_{\rm KL}(p^J \Vert{} r^J) - \log(1-u),
\end{aligned}
\]
where $-\log(1-u)\leq2u$ because $u\leq1/2$. For $j\notin J$, $p_j\log(p_j/r_j)\leq p_j\log(1/r_j)$. The second claim is $(1-t)\bigl(1-\max_{j\in J}p_j^J\bigr)= 1-t-\max_{j\in J}p_j\leq1-\max_jp_j$, which holds because $\max_jp_j\leq\max_{j\in J}p_j+t$.

For (b), write $\delta_j=\delta_j^0+\epsilon_j$ with $\delta_j^0=\max\{\delta_j,-c\}\in[-c,c]$ and $\epsilon_j\in[-u_j,0]$, and let $r'=\operatorname{softmax}(a+\epsilon)$. Since $r_j' = \frac{r_j e^{\epsilon_j}}{\sum_l r_l e^{\epsilon_l}}$,
\[
\begin{aligned}
D_{\rm KL}(p \Vert{} r) - D_{\rm KL}(p \Vert{} r') &= \sum_{j=1}^K p_j \log \frac{p_j}{r_j} - \sum_{j=1}^K p_j \log \frac{p_j}{r_j'} = \sum_{j=1}^K p_j \log \frac{r_j'}{r_j}=\sum_{j=1}^K p_j \left( \epsilon_j - \log \sum_{l=1}^K r_l e^{\epsilon_l} \right)\\
&= \sum_{j=1}^K p_j \epsilon_j - \log \sum_{l=1}^K r_l e^{\epsilon_l}\le - \log \sum_{l=1}^K r_l e^{\epsilon_l}\le  - \log\left[ \left(\sum_{l=1}^K r_l\right)  e^{\min_m \epsilon_m} \right]\\
&= -\min_m \epsilon_m\le \max_j u_j.
\end{aligned}
\]
Moreover, because $p_j\propto r_j'e^{\delta_j^0}$ and $\Vert{}\delta^0\Vert{}_\infty \le c$, Lemma~\ref{lem:logit-kl} gives $D_{\rm KL}(p\|r')\leq e^{2c}(1-\max_jp_j)$.

For (c), put $v_j=e^{a_j-\max_la_l}$, $v_j'=\max\{v_j,e^{-B}\}$, $Z=\sum_jv_j\geq1$ and $Z'=\sum_jv_j'\leq Z+Ke^{-B}$. Then $r_j/r_j^{[B]}=(v_j/v_j')(Z'/Z)\le Z'/Z\leq1+Ke^{-B}$, so $D_{\rm KL}(p\|r^{[B]})-D_{\rm KL}(p\|r)=\sum_jp_j\log(r_j/r_j^{[B]})\leq Ke^{-B}$, and $r_j^{[B]}\geq e^{-B}/Z'\geq e^{-B}/K$.
\end{proof}

\subsubsection{From probability errors to selection errors}

The training losses measure the accuracy of probabilities, whereas the final decisions are obtained by maximizing or thresholding them. The next lemma relates these two notions of accuracy.

\begin{lemma}[Selection error, Bayes optimality]\label{lem:ce-selection}
Let $\mathsf T\in[K]$ be a label with posterior probabilities $q^*(X^n)$, let $q:\mathcal X^n\to\Delta_K$ be measurable with $k_q\in\argmax_kq_k$, and put ${\rm BE}_n=\mb E[1-\max_kq_k^*(X^n)]$.

\textup{(a)} $\mb P(k_q\ne\mathsf T)\leq5\,{\rm BE}_n+8\,\mb E[D_{\rm KL}(q^*\|q)]$, and ${\rm BE}_n\leq\mb P(\widetilde k\ne\mathsf T)$ for every measurable rule $\widetilde k$.

\textup{(b)} Let $S\subseteq[p]$ be a random set with $q_j^*(X^n)=\mb P(j\in S\mid X^n)$, let $q_j:\mathcal X^n\to[0,1]$ be measurable, $\widehat S_q=\{j:q_j>1/2\}$ and $H_n^*=\sum_j\mb E[\min\{q_j^*,1-q_j^*\}]$. Then $\mb E[|\widehat S_q\mathbin{\triangle}S|]\leq5H_n^*+8\sum_{j=1}^p\mb E[D_{\rm KL}(q_j^*\|q_j)]$, and $H_n^*\leq\mb E[|\widetilde S\mathbin{\triangle}S|]$ for every measurable $\widetilde S$. Moreover, with $Q_S^*=\mb P(S\mid X^n)$ and $S$ the generating set, $\mb E[1-Q_S^*]\leq2H_n^*$.

\textup{(c)} If $\mb E[\|\theta\|_2^2]<\infty$ and $\theta^*=\mb E[\theta\mid X^n]$, then $\mb E[\|\widetilde\theta-\theta\|_2^2]=\mb E[\|\widetilde\theta-\theta^*\|_2^2]+\mb E[\|\theta^*-\theta\|_2^2]$ for every measurable function $\widetilde\theta$ of $X^n$ with finite second moment.

\noindent The same statements hold conditionally on training data that are independent of the task.
\end{lemma}

\begin{proof}
For (a), fix $x^n$, write $m=\max_kq_k^*$ and let $k^*\in\argmax_kq_k^*$. Conditionally on $X^n=x^n$,
\[
\mathbb{P}(k_q \ne \mathsf T \mid X^n = x^n) = 1 - \mathbb{P}(\mathsf T = k_q \mid X^n = x^n) = 1 - q_{k_q}^* \leq(1-m)+\ind\{k_q\ne k^*\}.
\]
If $m>3/4$ and $k_q\ne k^*$, then $q_{k^*}\leq q_{k_q}$ and $q_{k^*}+q_{k_q}\leq1$, so $q_{k^*}\leq1/2$. The total variation distance between $q^*$ and $q$ then satisfies
\[
\begin{aligned} \Vert{}q^* - q\Vert{}_{\rm TV}& = \frac{1}{2} \sum_{k=1}^K \vert{}q_k^* - q_k\vert{} = \frac{1}{2}\vert{}q_{k^*}^* - q_{k^*}\vert{} + \frac{1}{2}\sum_{k \ne k^*} \vert{}q_k^* - q_k\vert{}\ge \frac{1}{2}\vert{}q_{k^*}^* - q_{k^*}\vert{} + \frac{1}{2}\left\vert{} \sum_{k \ne k^*} (q_k^* - q_k) \right\vert{} \\ &= \frac{1}{2}\vert{}q_{k^*}^* - q_{k^*}\vert{} + \frac{1}{2}\vert{}(1 - q_{k^*}^*) - (1 - q_{k^*})\vert{} = \vert{}q_{k^*}^* - q_{k^*}\vert{}=m-q_{k^*} > \frac{1}{4}, \end{aligned}
\]
and Pinsker's inequality \citep[(3.59), p.~78]{wainwright2019high} gives $D_{\rm KL}(q^*\|q)\ge 2\Vert{}q^* - q\Vert{}_{\rm TV}^2>1/8$. Hence $\ind\{k_q\ne k^*\}\leq4(1-m)+8D_{\rm KL}(q^*\|q)$, and taking expectations proves the first claim. The second holds because $\mb P(\widetilde k\ne\mathsf T\mid X^n)=1-q_{\widetilde k}^*\geq1-m$.

For (b), fix $j\in[p]$. Applying part~(a) with $K=2$ to the label $\ind\{j\in S\}$, the two-point law $(q_j^*,1-q_j^*)$ and the prediction $(q_j,1-q_j)$ gives
\[
\mb P\bigl(\ind\{j\in\widehat S_q\}\ne\ind\{j\in S\}\bigr)
\leq5\,\mb E\bigl[\min\{q_j^*,1-q_j^*\}\bigr]+8\,\mb E\bigl[D_{\rm KL}(q_j^*\|q_j)\bigr],
\]
since $1-\max\{q_j^*,1-q_j^*\}=\min\{q_j^*,1-q_j^*\}$. Since $|\widehat S_q\mathbin{\triangle}S|=\sum_{j=1}^p\ind\{\ind\{j\in\widehat S_q\}\ne\ind\{j\in S\}\}$, summing over $j\in[p]$ gives the first claim. Applying the second claim of (a) to each $\ind\{j\in\widetilde S\}$ and summing over $j\in[p]$ gives $H_n^*\leq\mb E[|\widetilde S\mathbin{\triangle}S|]$ for every measurable $\widetilde S$. For the last claim, let $S'$ be drawn from $Q^*$ independently of $S$ given $X^n$. Since $S$ and $S'$ are independent with conditional law $Q^*$ given $X^n$, we have $\mb P(S'\ne S\mid X^n)=\sum_{s\subseteq[p]}Q_s^*(1-Q_s^*)=\mb E[1-Q_S^*\mid X^n]$. Using $\ind\{S'\ne S\}\leq|S\mathbin{\triangle}S'|$ and $u(1-u)\leq\min\{u,1-u\}$ for $u\in[0,1]$, we obtain
\[
\begin{aligned}
\mb E[1-Q_S^*\mid X^n]
&=\mb P(S'\ne S\mid X^n)\leq\mb E[|S\mathbin{\triangle}S'|\mid X^n]=\sum_{j=1}^p\mb P\bigl(\ind\{j\in S\}\ne\ind\{j\in S'\}\mid X^n\bigr)\\
&=2\sum_{j=1}^p q_j^*(1-q_j^*)\leq2\sum_{j=1}^p\min\{q_j^*,1-q_j^*\},
\end{aligned}
\]
where the third equality uses the conditional independence of $S$ and $S'$ given $X^n$ and $\mb P(j\in S\mid X^n)=q_j^*$. Taking expectations gives $\mb E[1-Q_S^*]\leq2H_n^*$.

For (c), $\widetilde\theta-\theta^*$ is a function of $X^n$ and $\mb E[\theta^*-\theta\mid X^n]=0$, so conditioning on $X^n$ gives $\mb E[\langle\widetilde\theta-\theta^*,\theta^*-\theta\rangle]=0$. Expanding $\|\widetilde\theta-\theta\|_2^2=\|(\widetilde\theta-\theta^*)+(\theta^*-\theta)\|_2^2$ and taking expectations gives the identity. Since the training data are independent of the task, the same arguments apply conditionally on the training data and give the conditional versions.
\end{proof}

The next two lemmas concern probability vectors on a family of sparse subsets and their marginals, as in variable selection. Let $\mathcal S$ be a finite family of subsets of $[p]$, each of size at most $s$. For a probability vector $P$ on $\mathcal S$, write $p_j=P\{S\ni j\}$, so that $\sum_jp_j\leq s$.

\subsubsection{Marginal inclusion probabilities}

For variable selection, the candidate space consists of entire supports, but the training loss involves marginal inclusion probabilities. Directly bounding the sum of marginal divergences by $p$ times the divergence on supports would lose the sparsity structure. The following bounds use the fact that each support contains at most $s$ variables.

\begin{lemma}[From supports to marginals]\label{lem:sparse-contraction}
If $P$ and $Q$ are probability vectors on $\mathcal S$ with $Q_S>0$ for all $S$, then $\sum_{j=1}^pD_{\rm KL}(p_j\|q_j)\leq4s\,D_{\rm KL}(P\|Q)$.
\end{lemma}

\begin{proof}
Suppose first that $P_S>0$ for all $S$. For $u\in[0,1]$, put $R_u=(1-u)Q+uP$, with marginals $r_j(u)=(1-u)q_j+up_j$, and let $f_u(S)=(P_S-Q_S)/R_{u,S}$. Since $P$ and $Q$ are probability vectors, we have $\mb E_{R_u}[f_u]=\sum_S R_{u,S}f_u(S)=\sum_S(P_S-Q_S)=0$ and $\mb E_{R_u}[f_u\ind\{j\in S\}]=\sum_{S\ni j}R_{u,S}f_u(S)=p_j-q_j$. Since $P_S>0$ and $Q_S>0$ for all $S$, any coordinate with $q_j\in\{0,1\}$ satisfies $p_j=q_j$ and contributes zero. We omit such coordinates, so $r_j(u)\in(0,1)$ for all $u\in[0,1]$ and all remaining $j$.

The maps $u\mapsto D_{\rm KL}(R_u\|Q)$ and $u\mapsto D_{\rm KL}\{r_j(u)\|q_j\}$ vanish together with their first derivatives at $u=0$. Their second derivatives are
\[
\frac{d^2}{du^2}D_{\rm KL}(R_u\|Q)=\sum_S\frac{(P_S-Q_S)^2}{R_{u,S}}=\mb E_{R_u}[f_u^2], \frac{d^2}{du^2}D_{\rm KL}\{r_j(u)\|q_j\}=\frac{(p_j-q_j)^2}{r_j(u)\{1-r_j(u)\}}.
\]
Taylor's formula with integral remainder therefore gives
\[
D_{\rm KL}(P\|Q)=\int_0^1(1-u)\,\mb E_{R_u}[f_u^2]\,du,\qquad
D_{\rm KL}(p_j\|q_j)=\int_0^1(1-u)\frac{(p_j-q_j)^2}{r_j(u)\{1-r_j(u)\}}\,du.
\]

Fix $u\in[0,1]$ and write $r_j=r_j(u)$. For each $j$ with $r_j\leq1/2$, since $p_j-q_j=\mb E_{R_u}[f_u\ind\{j\in S\}]$ and $\mb E_{R_u}[\ind\{j\in S\}]=r_j$, the Cauchy--Schwarz inequality gives $(p_j-q_j)^2\leq r_j\,\mb E_{R_u}[f_u^2\ind\{j\in S\}]$. Since $1/(1-r_j)\leq2$ for $r_j\leq1/2$ and $\sum_{j=1}^p\ind\{j\in S\}=|S|\leq s$ for every $S\in\mathcal S$, we have
\[
\sum_{j:\,r_j\leq1/2}\frac{(p_j-q_j)^2}{r_j(1-r_j)}
\leq2\sum_{j:\,r_j\leq1/2}\mb E_{R_u}[f_u^2\ind\{j\in S\}] \leq2\,\mb E_{R_u}[f_u^2|S|]\leq2s\,\mb E_{R_u}[f_u^2].
\]
For the remaining coordinates, since $\sum_{j=1}^p r_j\leq s$, at most $2s$ coordinates satisfy $r_j>1/2$. For each of them, since $\mb E_{R_u}[f_u]=0$, we have $p_j-q_j=\mb E_{R_u}[f_u(\ind\{j\in S\}-r_j)]$. Since $\mb E_{R_u}[(\ind\{j\in S\}-r_j)^2]=r_j(1-r_j)$, the Cauchy--Schwarz inequality gives
\[
(p_j-q_j)^2\leq\mb E_{R_u}[(\ind\{j\in S\}-r_j)^2]\,\mb E_{R_u}[f_u^2]=r_j(1-r_j)\,\mb E_{R_u}[f_u^2],
\]
so $(p_j-q_j)^2/\{r_j(1-r_j)\}\leq\mb E_{R_u}[f_u^2]$. Summing over these at most $2s$ coordinates and adding the two groups gives
\[
\sum_j\frac{(p_j-q_j)^2}{r_j(1-r_j)}\leq(2s+2s)\,\mb E_{R_u}[f_u^2]=4s\,\mb E_{R_u}[f_u^2].
\]
Integrating against $(1-u)\,du$ over $[0,1]$ yields $\sum_{j=1}^pD_{\rm KL}(p_j\|q_j)\leq4s\,D_{\rm KL}(P\|Q)$. For general $P$, apply this bound to $(1-\epsilon)P+\epsilon Q$ for $\epsilon\in(0,1)$ and let $\epsilon\to0$.
\end{proof}

For $\epsilon\in(0,1/2)$ and $q\in[0,1]$, let $[q]_\epsilon=\min\{\max(q,\epsilon),1-\epsilon\}$. For a probability vector $Q$ on $\mathcal S$ with marginals $q_j$ and numbers $e_j\in[-\gamma,\gamma]$, let $q_j^{e}$ be obtained by adding $e_j$ to the log odds of $q_j$ and then applying $[\,\cdot\,]_\epsilon$, where a marginal equal to zero or one is left unchanged before clipping.

\begin{lemma}[Marginals under clipping, removal and logit errors]\label{lem:bern-tilt}
Let $P,Q$ be probability vectors on $\mathcal S$ with positive entries, and let $\gamma\geq0$ be fixed. There is a constant $C_\gamma$ such that the following hold.

\textup{(a)} $\sum_{j=1}^pD_{\rm KL}(p_j\|q_j^{e})\leq C_\gamma s\{1+\log(1/\epsilon)\}+C_\gamma p\epsilon$.

\textup{(b)} Let $\mathcal A\subseteq\mathcal S$ with $\alpha=P(\mathcal A^c)<1$ and $\beta=Q(\mathcal A^c)<1$, let $P^{\mathcal A}$, $Q^{\mathcal A}$ be the conditional vectors on $\mathcal A$, and let $q_j^{\mathcal A,e}$ be defined from the marginals of $Q^{\mathcal A}$ with the same $e_j$. Then
\[
\sum_{j=1}^pD_{\rm KL}(p_j\|q_j^{e})\leq\sum_{j=1}^pD_{\rm KL}(p_j^{\mathcal A}\|q_j^{\mathcal A,e})+2s\alpha\log\frac1\epsilon+p\,h\Bigl(\min\Bigl\{\frac{2s\alpha}p,\frac12\Bigr\}\Bigr)+\frac{2se^\gamma\beta}\epsilon,
\]
where $h$ is the binary entropy, and $\sum_j|p_j-p_j^{\mathcal A}|\leq2s\alpha$.

\textup{(c)} For $u\in[0,1]$, $v\in(0,1)$ and $|e|\leq\gamma$, let $v^e$ have log odds equal to those of $v$ plus $e$. Then $D_{\rm KL}(u\|[v^e]_\epsilon)\leq D_{\rm KL}(u\|v)+\gamma e^\gamma\min\{u+v,\,2-u-v\}+2\epsilon$.
\end{lemma}

\begin{proof}
We first record that $D_{\rm KL}(u\|[v]_\epsilon)\leq D_{\rm KL}(u\|v)+2\epsilon$ for all $u\in[0,1]$ and $v\in(0,1)$. If $\epsilon\leq v\leq1-\epsilon$, then $[v]_\epsilon=v$. If $v<\epsilon$, then $[v]_\epsilon=\epsilon$, so replacing $v$ by $\epsilon$ decreases $u\log(u/v)$ and increases $(1-u)\log\{(1-u)/(1-v)\}$ by $(1-u)\log\{(1-v)/(1-\epsilon)\}\leq-\log(1-\epsilon)\leq2\epsilon$, where the last inequality uses $\epsilon\in(0,1/2)$. If $v>1-\epsilon$, then $[v]_\epsilon=1-\epsilon$, so replacing $v$ by $1-\epsilon$ decreases $(1-u)\log\{(1-u)/(1-v)\}$ and increases $u\log(u/v)$ by $u\log\{v/(1-\epsilon)\}\leq-\log(1-\epsilon)\leq2\epsilon$.

For (a), the divergence is at most the cross-entropy $p_j\log(1/q_j^e)+(1-p_j)\log\{1/(1-q_j^e)\}$. Since $q_j^e\geq\epsilon$ and $\sum_jp_j\leq s$, the first terms sum to at most $s\log(1/\epsilon)$. Before clipping, adding $e_j$ to the log odds of $q_j$ gives at most $e^\gamma q_j$. Since $\sum_{j=1}^p q_j\leq s$, the number of coordinates with $e^\gamma q_j>1/2$ is at most $2e^\gamma\sum_{j=1}^p q_j\leq2e^\gamma s$. On each such coordinate, $1-q_j^e\geq\epsilon$, so the second term is at most $\log(1/\epsilon)$. These coordinates therefore contribute at most $2e^\gamma s\log(1/\epsilon)$. On every other coordinate, $e^\gamma q_j\leq1/2$, so $1-q_j^e\geq1-\max\{e^\gamma q_j,\epsilon\}$. Since $\max\{e^\gamma q_j,\epsilon\}\leq1/2$ and $-\log(1-x)\leq2x$ for $x\in[0,1/2]$,
\[
(1-p_j)\log\frac1{1-q_j^e}\leq-\log(1-q_j^e)\leq2\max\{e^\gamma q_j,\epsilon\}\leq2e^\gamma q_j+2\epsilon.
\]
Summing over these coordinates and using $\sum_{j=1}^p q_j\leq s$ gives at most $2e^\gamma s+2p\epsilon$. Combining the three sums yields (a).

For (b), if $\mathcal A=\mathcal S$, then $P^{\mathcal A}=P$, $Q^{\mathcal A}=Q$ and $\alpha=\beta=0$, and there is nothing to prove. Otherwise, $p_j=(1-\alpha)p_j^{\mathcal A}+\alpha p_j^c$ with $p_j^c$ the marginal of the conditional vector on $\mathcal A^c$, so $|p_j-p_j^{\mathcal A}|\leq\alpha(p_j^c+p_j^{\mathcal A})$. Summing over $j$ gives $\sum_{j=1}^p|p_j-p_j^{\mathcal A}|\leq2s\alpha$. The decomposition $q_j=(1-\beta)q_j^{\mathcal A}+\beta q_j^c$ gives $\sum_{j=1}^p|q_j-q_j^{\mathcal A}|\leq2s\beta$ in the same way. Since the map that adds $e_j$ to the log odds is $e^\gamma$-Lipschitz on $[0,1]$ and clipping is nonexpansive, we have $\sum_{j=1}^p|q_j^e-q_j^{\mathcal A,e}|\leq2se^\gamma\beta$. Write $D_{\rm KL}(u\|v)={\rm CE}(u,v)-h(u)$ with ${\rm CE}(u,v)=u\log(1/v)+(1-u)\log\{1/(1-v)\}$. For $v\in[\epsilon,1-\epsilon]$, ${\rm CE}(u,v)$ is $\log(1/\epsilon)$-Lipschitz in $u$ and $\epsilon^{-1}$-Lipschitz in $v$. Moreover, $|h(x)-h(y)|\leq h(\min\{|x-y|,1/2\})$ for $x,y\in[0,1]$. Setting $d_j=|p_j-p_j^{\mathcal A}|$ and using $q_j^e,q_j^{\mathcal A,e}\in[\epsilon,1-\epsilon]$, we obtain
\[
\begin{aligned}
&D_{\rm KL}(p_j\|q_j^e)-D_{\rm KL}(p_j^{\mathcal A}\|q_j^{\mathcal A,e})\\
&\quad=\{{\rm CE}(p_j,q_j^e)-{\rm CE}(p_j^{\mathcal A},q_j^e)\}+\{{\rm CE}(p_j^{\mathcal A},q_j^e)-{\rm CE}(p_j^{\mathcal A},q_j^{\mathcal A,e})\}-\{h(p_j)-h(p_j^{\mathcal A})\}\\
&\quad\leq\log\frac1\epsilon\,d_j+\frac1\epsilon|q_j^e-q_j^{\mathcal A,e}|+h\bigl(\min\{d_j,1/2\}\bigr).
\end{aligned}
\]
Since $h$ is concave and nondecreasing on $[0,1/2]$ and $\sum_{j=1}^p d_j\leq2s\alpha$,
\[
\sum_{j=1}^p h\bigl(\min\{d_j,1/2\}\bigr)\leq p\,h\Bigl(\min\Bigl\{\frac1p\sum_{j=1}^p d_j,\frac12\Bigr\}\Bigr)\leq p\,h\Bigl(\min\Bigl\{\frac{2s\alpha}p,\frac12\Bigr\}\Bigr).
\]
Summing the divergence bound over $j$ and using $\sum_{j=1}^p d_j\leq2s\alpha$ and $\sum_{j=1}^p|q_j^e-q_j^{\mathcal A,e}|\leq2se^\gamma\beta$ gives the bound in (b).

For (c), let $a$ be the log odds of $v$ and put $f(t)=D_{\rm KL}\{u\|\sigma_{\log}(a+t)\}$, where $\sigma_{\log}(t)=(1+e^{-t})^{-1}$. Then $f'(t)=\sigma_{\log}(a+t)-u$. Since $\sigma_{\log}(a+t)\leq e^{|t|}v$ and $1-\sigma_{\log}(a+t)\leq e^{|t|}(1-v)$, we have $|f'(t)|\leq e^\gamma\min\{u+v,\,2-u-v\}$ for $|t|\leq\gamma$. Since $f(0)=D_{\rm KL}(u\|v)$, $f(e)=D_{\rm KL}(u\|v^e)$ and $|e|\leq\gamma$,
\[
D_{\rm KL}(u\|v^e)=f(e)=f(0)+\int_0^e f'(t)\,dt\leq D_{\rm KL}(u\|v)+\gamma e^\gamma\min\{u+v,\,2-u-v\}.
\]
Applying the preliminary bound to $[v^e]_\epsilon$ adds $2\epsilon$ and gives (c).
\end{proof}

The next lemma bounds the divergences between the marginals of two probability vectors that differ by a tilt growing at most linearly with the number of added variables.

\begin{lemma}[Marginals under tilts that grow with the number of added variables]\label{lem:added-tilt}
Let $S_0\in\mathcal S$ and $\mathcal A_+=\{S\in\mathcal S:S\supseteq S_0\}$, and write $b_S=|S\setminus S_0|$. Let $P,Q$ be probability vectors on $\mathcal A_+$ with positive entries such that $P_S\propto Q_Se^{\delta_S}$ and $|\delta_S-\delta_{S_0}|\leq cb_S$ for every $S\in\mathcal A_+$, for a constant $c\geq0$. Let $p_j,q_j$ be their marginals, and let $q_j^e$ be defined as in Lemma~\ref{lem:bern-tilt} with $|e_j|\leq\gamma\leq1$. Then, with $M_P=\sum_SP_Sb_S^2$ and $M_Q=\sum_SQ_Sb_S^2$,
\[
\begin{gathered}
\sum_{j=1}^pD_{\rm KL}(p_j\|q_j^e)\leq(4c+5)(M_P+M_Q)+2p\epsilon,\\
M_P+M_Q\leq(1+2e^c)(1-P_{S_0})+\sum_{S:\,b_S\geq2}b_S^2(P_S+Q_S).
\end{gathered}
\]
\end{lemma}

\begin{proof}
For $j\in S_0$, we have $p_j=q_j=1$, so $D_{\rm KL}(p_j\|q_j)=0$. Let $\mu_P=\sum_SP_Sb_S$ and $\mu_Q=\sum_SQ_Sb_S$. Then $\sum_{j\notin S_0}p_j=\sum_SP_S|S\setminus S_0|=\mu_P$ and $\sum_{j\notin S_0}q_j=\mu_Q$. Since $b_S$ is a nonnegative integer, $b_S\leq b_S^2$, so $\mu_P\leq M_P$ and $\mu_Q\leq M_Q$. By Jensen's inequality, $\mu_P^2\leq M_P$ and $\mu_Q^2\leq M_Q$. Put $\Delta_S=\delta_S-\delta_{S_0}$, so that $|\Delta_S|\leq cb_S$. Since $P_S\propto Q_Se^{\delta_S}$, cancelling the common factor $e^{\delta_{S_0}}$ gives $P_S/Q_S=e^{\Delta_S}/\sum_{S'}Q_{S'}e^{\Delta_{S'}}$. By Jensen's inequality and $|\Delta_S|\leq cb_S$,
\begin{equation}
\log\frac{P_S}{Q_S}=\Delta_S-\log\sum_{S'}Q_{S'}e^{\Delta_{S'}}\leq\Delta_S-\sum_{S'}Q_{S'}\Delta_{S'}\leq cb_S+c\mu_Q,
\label{eq:added-tilt-ratio}
\end{equation}
and in the same way $D_{\rm KL}(P\|Q)=\sum_SP_S\Delta_S-\log\sum_SQ_Se^{\Delta_S}\leq c(\mu_P+\mu_Q)$.

Consider first the indices $j\notin S_0$ with $q_j\leq1/2$. Since $\log x\leq x-1$ and $0\leq1-p_j\leq1$, we have $(1-p_j)\log\{(1-p_j)/(1-q_j)\}\leq(1-p_j)(q_j-p_j)/(1-q_j)\leq(q_j-p_j)_+/(1-q_j)$. Since $1-q_j\geq1/2$, we obtain
\[
D_{\rm KL}(p_j\|q_j)
=p_j\log\frac{p_j}{q_j}+(1-p_j)\log\frac{1-p_j}{1-q_j} \leq p_j\log\frac{p_j}{q_j}+\frac{(q_j-p_j)_+}{1-q_j}\leq p_j\log\frac{p_j}{q_j}+2q_j.
\]
By the log-sum inequality, $p_j\log(p_j/q_j)\leq\sum_{S\ni j}P_S\log(P_S/Q_S)$. Since the right side of~\eqref{eq:added-tilt-ratio} is nonnegative, we may insert it and sum over all $j\notin S_0$. Using $\mu_P\mu_Q\leq(\mu_P^2+\mu_Q^2)/2\leq(M_P+M_Q)/2$ and $\mu_Q\leq M_Q$, this gives
\[
\begin{aligned}
\sum_{j\notin S_0:\,q_j\leq1/2}D_{\rm KL}(p_j\|q_j)
&\leq\sum_SP_Sb_S(cb_S+c\mu_Q)+2\mu_Q=cM_P+c\mu_P\mu_Q+2\mu_Q\\
&\leq cM_P+\tfrac12c(M_P+M_Q)+2M_Q=\tfrac32cM_P+\bigl(\tfrac12c+2\bigr)M_Q.
\end{aligned}
\]
Next, since $\sum_{j\notin S_0}q_j=\mu_Q$, at most $2\mu_Q$ indices $j\notin S_0$ satisfy $q_j>1/2$. For each of them, the data-processing inequality gives $D_{\rm KL}(p_j\|q_j)\leq D_{\rm KL}(P\|Q)\leq c(\mu_P+\mu_Q)$. Using $2\mu_P\mu_Q\leq M_P+M_Q$ and $\mu_Q^2\leq M_Q$, their total contribution is at most $2c\mu_Q(\mu_P+\mu_Q)\leq cM_P+3cM_Q$. Hence $\sum_jD_{\rm KL}(p_j\|q_j)\leq(4c+2)(M_P+M_Q)$. For $j$ with $q_j\in\{0,1\}$, we have $p_j=q_j$, the marginal is left unchanged before clipping, and $D_{\rm KL}(p_j\|q_j^e)=-\log(1-\epsilon)\leq2\epsilon$. For every other $j$, Lemma~\ref{lem:bern-tilt}(c) gives
\[
D_{\rm KL}(p_j\|q_j^e)\leq D_{\rm KL}(p_j\|q_j)+\gamma e^\gamma\min\{p_j+q_j,\,2-p_j-q_j\}+2\epsilon,
\]
and such a $j$ is not in $S_0$, so the minimum is at most $p_j+q_j$. Since $\gamma\leq1$, we have $\gamma e^\gamma\leq e\leq3$. Summing over $j\in[p]$ and using $\sum_{j\notin S_0}(p_j+q_j)=\mu_P+\mu_Q\leq M_P+M_Q$, we see that the errors $e_j$ and the clipping add at most $3(M_P+M_Q)+2p\epsilon$, which gives $\sum_{j=1}^pD_{\rm KL}(p_j\|q_j^e)\leq(4c+5)(M_P+M_Q)+2p\epsilon$ and proves the first claim.

For the second claim, the supports with $b_S=1$ contribute $P(b_S=1)\leq1-P_{S_0}$ to $M_P$ and $Q(b_S=1)$ to $M_Q$, while $S_0$ contributes $0$. For each $S$ with $b_S=1$, we have $|\Delta_S|\leq c$ and $Q_{S_0}\leq1$, so $Q_S=Q_{S_0}(P_S/P_{S_0})e^{-\Delta_S}\leq e^cP_S/P_{S_0}$. Summing over $S$ with $b_S=1$ gives $Q(b_S=1)\leq e^c(1-P_{S_0})/P_{S_0}$. Thus $Q(b_S=1)\leq2e^c(1-P_{S_0})$ if $P_{S_0}\geq1/2$, while if $P_{S_0}<1/2$ then $Q(b_S=1)\leq1\leq2(1-P_{S_0})\leq2e^c(1-P_{S_0})$ since $c\geq0$. Hence $P(b_S=1)+Q(b_S=1)\leq(1+2e^c)(1-P_{S_0})$, and adding $\sum_{S:\,b_S\geq2}b_S^2(P_S+Q_S)$ gives the second claim.
\end{proof}

The next lemma controls all moments of the excess cross-entropy loss and of the excess joint loss by their means, which is the moment condition of Lemma~\ref{lem:bernstein-erm}.

\subsubsection{Moment bounds for the excess training loss}

To apply the empirical risk minimization result above, we must control the second moment of the excess loss for each training objective. The boundedness of the logits or the clipping of the probabilities provides the required control for cross-entropy.

\begin{lemma}[Moments of the excess loss]\label{lem:excess-moments}
\textup{(a)} Let $\mathsf T$ take values in a finite set with conditional law $q^*(X)$ given $X$, let $q(X)$ be a probability vector with $q_k\geq e^{-b}$ for all $k$, where $b\geq1$, and put $H=\log\{q_{\mathsf T}^*(X)/q_{\mathsf T}(X)\}$. Then $\mb E[H]=\mb E[D_{\rm KL}\{q^*(X)\|q(X)\}]$ and, for every integer $m\geq2$, $\mb E[|H|^m]\leq e\,m!\,b^{m-1}\,\mb EH$.

\textup{(b)} Let $Y_1,\ldots,Y_p\in\{0,1\}$ with $q_j^*(X)=\mb P(Y_j=1\mid X)$, let $q_j(X)\in[e^{-b},1-e^{-b}]$ and $\lambda>0$, let $H_j$ be the excess log loss of coordinate $j$ as in \textup{(a)}, and $K=(\lambda/p)\sum_jH_j$. Let $\theta$ be a random vector with $\theta^*=\mb E[\theta\mid X]$, let $v(X)$ be such that $\|v-\theta\|_2^2\leq b_1$ and $\|\theta^*-\theta\|_2^2\leq b_1$ almost surely, and $G=\|v-\theta\|_2^2-\|\theta^*-\theta\|_2^2$. Then $\mb EG=\mb E[\|v-\theta^*\|_2^2]$ and, with $c=2\max\{b_1,\lambda b\}$, $\mb E[|G+K|^m]\leq5\,m!\,c^{m-1}\,\mb E[G+K]$ for every $m\geq2$.
\end{lemma}

\begin{proof}
For (a), conditionally on $X$,
\[
\mathbb{E}[H \mid X] = \sum_{k} \mathbb{P}(\mathsf T = k \mid X) \log \frac{q_k^*(X)}{q_k(X)} = \sum_{k} q_k^*(X) \log \frac{q_k^*(X)}{q_k(X)} = D_{\rm KL}\{q^*(X) \,\Vert{}\, q(X)\},
\]
and taking expectations gives $\mathbb{E}[H] = \mathbb{E}[D_{\rm KL}\{q^*(X) \,\Vert{}\, q(X)\}]$. Let $\Phi(y)=e^{-y}-1+y$. Since $H\leq \log(e^b)=b$ and $\mathbb{E}[e^{-H} \mid X] = \sum_{k :\, q_k^*(X) > 0} q_k^*(X) \cdot \frac{q_k(X)}{q_k^*(X)} = \sum_{k :\, q_k^*(X) > 0} q_k(X) \le 1$, we have $\mathbb{E}[\Phi(H)] = \mathbb{E}[e^{-H}] - 1 + \mathbb{E}[H] \le \mathbb{E}[H]$. We claim that $|y|^m\leq e\,m!\,b^{m-1}\Phi(y)$ for $y\leq b$. Indeed, for $y=-t\leq0$, $\Phi(y)=e^t-1-t\geq t^m/m!$. For $0\leq y\leq b$, $\Phi(y)/y^2=\int_0^1(1-u)e^{-uy}du$ is decreasing in $y$ and $\Phi(b)\geq b/e$ for $b\geq1$, so $\Phi(y)\geq y^2/(eb)$ and $y^m\leq b^{m-2}y^2\leq eb^{m-1}\Phi(y)$. Hence setting $y = H$ and taking expectations gives $\mathbb{E}[\vert{}H\vert{}^m] \le e \cdot m! \cdot b^{m-1} \, \mathbb{E}[\Phi(H)] \le e \cdot m! \cdot b^{m-1} \, \mathbb{E}[H]$.

For (b), part (a) gives $\mb E[|H_j|^m]\leq e\,m!\,b^{m-1}\mb EH_j$ for every $j$, and convexity gives $|K|^m=\lambda^m|p^{-1}\sum_jH_j|^m\leq\lambda^mp^{-1}\sum_j|H_j|^m$. Taking expectations yields
\[
\mb E[|K|^m]\leq\lambda^mp^{-1}\sum_j e\,m!\,b^{m-1}\mb EH_j=e\,m!\,(\lambda b)^{m-1}\mb EK.
\]
Next, expanding $\|v-\theta\|_2^2=\|(v-\theta^*)+(\theta^*-\theta)\|_2^2$ gives $G=\|v-\theta^*\|_2^2+2\langle v-\theta^*,\theta^*-\theta\rangle$. Since $\theta^*=\mb E[\theta\mid X]$, the cross term has conditional mean zero given $X$, so $\mb E[G\mid X]=\|v-\theta^*\|_2^2$ and $\mb EG=\mb E[\|v-\theta^*\|_2^2]$. Also, since $\|v-\theta\|_2^2\leq b_1$ and $\|\theta^*-\theta\|_2^2\leq b_1$ almost surely, we have $|G|\leq b_1$ almost surely. Put $g=\|v-\theta^*\|_2^2$, so that $g\leq\mb E[\|v-\theta\|_2^2\mid X]\leq b_1$ almost surely. By $(x+y)^2\leq2x^2+2y^2$ and the Cauchy--Schwarz inequality, together with $g\leq b_1$ and $\|\theta^*-\theta\|_2^2\leq b_1$ almost surely, we have
\[
\mb E[G^2\mid X]\leq2g^2+8g\,\mb E[\|\theta^*-\theta\|_2^2\mid X]\leq10b_1g.
\]
Since $|G|\leq b_1$ and $\mb Eg=\mb EG$, this gives
\[
\mb E[|G|^m]\leq b_1^{m-2}\mb E[G^2]\leq10\,b_1^{m-1}\mb EG.
\]
Finally, with $c=2\max\{b_1,\lambda b\}$, we have $10\leq5m!$ for $m\geq2$ and $e<5$, so $2^{m-1}\cdot10\,b_1^{m-1}=10(2b_1)^{m-1}\leq5\,m!\,c^{m-1}$ and $2^{m-1}\,e\,m!\,(\lambda b)^{m-1}=e\,m!\,(2\lambda b)^{m-1}\leq5\,m!\,c^{m-1}$. Since $|G+K|^m\leq2^{m-1}(|G|^m+|K|^m)$ and $\mb EG,\mb EK\geq0$, combining the bounds on $\mb E[|G|^m]$ and $\mb E[|K|^m]$ gives
\[
\mb E[|G+K|^m]\leq2^{m-1}(\mb E[|G|^m]+\mb E[|K|^m])\leq5\,m!\,c^{m-1}\mb E[G+K].
\]
\end{proof}

\subsection{ReLU approximation}\label{app:relu}

The constructions in Appendix~\ref{app:theory-discussion} require only a few basic network operations: projection onto a parameter set, multiplication, and approximation of smooth functions. We give the bounds for these operations here. Keeping track of the number of coefficients, including fixed coefficients used to pass intermediate quantities between layers, is important for the subsequent covering arguments. We first give an exact ReLU representation of the projection used in the initialization of the parameter estimation network. A polytope is the convex hull of finitely many points.

\subsubsection{Projection and multiplication}

\begin{lemma}[Projection onto a polytope]\label{lem:polytope}
\textup{(a)} Let $\Theta_0\subset\mb R^d$ be compact and convex and let $U\supset\Theta_0$ be open. Then there is a polytope $P$ with $\Theta_0\subseteq P\subset U$.

\textup{(b)} Let $P\subset\mb R^d$ be a nonempty polytope. The Euclidean projection $\Pi_P:\mb R^d\to P$ is nonexpansive, fixes every point of $P$, and is computed exactly by a ReLU network whose depth and number of parameters depend only on $P$.
\end{lemma}

\begin{proof}
For (a), since $\Theta_0$ is compact and $U$ is open, the distance $\operatorname{dist}(\Theta_0,U^c)$ is positive. Choose $a_0>0$ smaller than $\operatorname{dist}(\Theta_0,U^c)$ and set $K^+=\{\theta:\operatorname{dist}(\theta,\Theta_0)\leq a_0\}$. Then $K^+$ is convex because $\Theta_0$ is convex, and $K^+\subset U$. Consider the closed cubes with vertices in $(a_0/\sqrt d)\mathbb Z^d$ and side length $a_0/\sqrt d$, each of which has diameter $a_0$. Since $\Theta_0$ is compact, finitely many of these cubes meet $\Theta_0$, and they cover $\Theta_0$. Any cube that meets $\Theta_0$ contains a point of $\Theta_0$, so each of its points is within distance $a_0$ of $\Theta_0$ and the cube lies in $K^+$. Let $P$ be the convex hull of the vertices of these cubes. Then $P$ is a polytope. It contains every such cube, so $\Theta_0\subseteq P$, and it lies in the convex set $K^+$ because the vertices of these cubes lie in $K^+$. Hence $\Theta_0\subseteq P\subseteq K^+\subset U$.

For (b), nonexpansiveness and the identity $\Pi_P(v)=v$ for $v\in P$ are~\eqref{eq:projection-nonexpansive}, whose proof applies to every compact convex set. For a face $F$ of $P$, let $N_F$ be the normal cone of $P$ at a point of the relative interior of $F$, which does not depend on the point and is contained in the normal cone at every point of $F$. If $x=v+w$ with $v\in F$ and $w\in N_F$, then $w$ lies in the normal cone of $P$ at $v$, so $\langle x-v,v'-v\rangle=\langle w,v'-v\rangle\leq0$ for all $v'\in P$. For every $v'\in P$, we have
\[
\|x-v'\|_2^2=\|x-v\|_2^2+\|v-v'\|_2^2-2\langle x-v,v'-v\rangle\geq\|x-v\|_2^2,
\]
so $\Pi_P(x)=v$. Since $w$ is orthogonal to the directions of the affine hull of $F$, $v$ is also the orthogonal projection of $x$ onto this affine hull, which is an affine map of $x$. Hence $\Pi_P$ coincides with an affine map on the polyhedron $F+N_F$. Conversely, for every $x\in\mb R^d$, the point $\Pi_P(x)$ lies in the relative interior of a face $F$ of $P$, and $x-\Pi_P(x)\in N_F$ by~\eqref{eq:projection-variational}, so $x\in F+N_F$. Since a polytope has finitely many faces, $\Pi_P$ is continuous and affine on each of finitely many polyhedra that cover $\mb R^d$. Each coordinate of such a map is computed exactly by a ReLU network with at most $\lceil\log_2(d+1)\rceil+1$ affine layers~\citep[Theorem~2.1, p.~3]{arora2018understanding}. Placing these $d$ networks in parallel gives a ReLU network that computes $\Pi_P$ exactly, with depth and number of parameters depending only on $P$.
\end{proof}

For a rectangle $P=\prod_{j=1}^d[l_j,u_j]$, the network in part~(b) is given explicitly by $[\Pi_P(x)]_j=l_j+\sigma(x_j-l_j)-\sigma(x_j-u_j)$ with $\sigma(t)=\max\{t,0\}$. It uses $2d$ ReLU units and two affine layers.

\begin{lemma}[Multiplication network]\label{lem:multiplication}
For $R\geq1$ and $0<\zeta<1$, there is a ReLU network $\mathsf M_{R,\zeta}:\mb R^2\to\mb R$ with $\sup_{|u|,|v|\leq R}|\mathsf M_{R,\zeta}(u,v)-uv|\leq\zeta$, $L(\mathsf M_{R,\zeta}),W(\mathsf M_{R,\zeta})\leq C\{1+\log(R^2/\zeta)\}$ and $\|\mathsf M_{R,\zeta}\|_{\rm coef}\leq CR^2$.
Here $C$ is numerical.
\end{lemma}

This follows from the square construction and the polarization argument in~\citet[Propositions~2 and~3, pp.~105--106]{yarotsky2017error}. Scaling the square approximation to $[-2R,2R]$ uses bounded tent-map coefficients and an output factor $4R^2$, giving the coefficient bound $CR^2$.

The next lemma follows from the approximation theorem for H\"older functions in~\citet[Theorem~5]{schmidthieber2020nonparametric}, which treats real smoothness orders and bounds the network coefficients.

\subsubsection{Approximation under finite smoothness}

The next approximation result is used for the component functions in the one-step construction. The input dimension in this result is the dimension of the component function, not that of the full dataset.

\begin{lemma}[Approximation of a H\"older-smooth function]\label{lem:sobolev-approx}
Let $f:R\to\mathbb R^q$ have bounded $C^s$ norm on a fixed
$m$-dimensional rectangle, where $s>0$ is real and the integer
convention is that in the notation paragraph. For every
$\epsilon\in(0,1)$, there is a ReLU network $\phi$ with
\[
\sup_R\|\phi-f\|_2\leq  \epsilon,\qquad L\leq   C\log(e/\epsilon),\qquad W\leq   C\epsilon^{-m/s}\log(e/\epsilon).
\]
Its coefficients satisfy
$\|\phi\|_{\rm coef}\leq   C\epsilon^{-C}$.
Here the constants depend only on the fixed dimensions, smoothness, rectangle and $C^s$ norm bound.
\end{lemma}

\begin{proof}
Write $R=\prod_{j=1}^m[u_j,v_j]$ and first let $f$ be scalar. The function $g(t)=f\{u_1+(v_1-u_1)t_1,\ldots,u_m+(v_m-u_m)t_m\}$, $t\in[0,1]^m$, belongs to the H\"older ball of smoothness $s$ and radius $K$ on $[0,1]^m$ defined in \citet[Section~3, p.~1880]{schmidthieber2020nonparametric}, for some $K$ depending only on $m$, $s$, $R$ and the $C^s$ norm bound. For integer $s$, this holds because bounded derivatives of order $s$ make the derivatives of order $s-1$ Lipschitz on the convex set $[0,1]^m$. By \citet[Theorem~5, p.~1889]{schmidthieber2020nonparametric}, for all integers $a\geq1$ and $N\geq(s+1)^m\vee(K+1)e^m$ there is a ReLU network $\psi$ with $8+(a+5)(1+\lceil\log_2(m\vee s)\rceil)$ hidden layers, at most $141(m+s+1)^{3+m}N(a+6)$ nonzero parameters, and all weights and shifts in $[-1,1]$ \citep[see (3)--(4), pp.~1878--1879]{schmidthieber2020nonparametric}. It satisfies
\[
\sup_{t\in[0,1]^m}|\psi(t)-g(t)|\leq(2K+1)(1+m^2+s^2)6^mN2^{-a}+K3^sN^{-s/m}.
\]
Choose
\[
\begin{gathered}
N=\Bigl\lceil\max\bigl\{(s+1)^m,\ (K+1)e^m,\ (2K3^s/\epsilon)^{m/s}\bigr\}\Bigr\rceil,\\
a=\Bigl\lceil\log_2\bigl\{2(2K+1)(1+m^2+s^2)6^mN/\epsilon\bigr\}\Bigr\rceil.
\end{gathered}
\]
Then $N\geq(2K3^s/\epsilon)^{m/s}$ gives $K3^sN^{-s/m}\leq\epsilon/2$, and the choice of $a$ gives $(2K+1)(1+m^2+s^2)6^mN2^{-a}\leq\epsilon/2$, so $\sup_{t\in[0,1]^m}|\psi(t)-g(t)|\leq\epsilon$. Since $\epsilon<1$, we have $N\leq C\epsilon^{-m/s}$ and $a\leq C\log(e/\epsilon)$. Let $\phi(x)=\psi\{(x_1-u_1)/(v_1-u_1),\ldots,(x_m-u_m)/(v_m-u_m)\}$, obtained by folding the affine change of variables into the first layer of $\psi$. This changes only the weights of the first layer and the shifts of its units with a nonzero incoming weight, so it at most doubles the number of nonzero parameters. Moreover, the new coefficients are bounded by a constant depending only on $R$. Hence $\sup_R|\phi-f|\leq\epsilon$, $L\leq C\log(e/\epsilon)$, $W\leq C\epsilon^{-m/s}\log(e/\epsilon)$, and the coefficients of $\phi$ are bounded by a constant.

For $q$ output coordinates, construct $\phi_l$ in this way for each coordinate $f_l$ at accuracy $\epsilon/\sqrt q$, using the same $N$ and $a$ for every coordinate so that all $\phi_l$ have the same depth, and place them in parallel to form $\phi$. Then $\sup_R\|\phi-f\|_2^2\leq\sum_{l=1}^q\sup_R|\phi_l-f_l|^2\leq\epsilon^2$, $L(\phi)=L(\phi_l)$ and $W(\phi)\leq\sum_{l=1}^qW(\phi_l)$, which gives the stated bounds.
\end{proof}

\subsubsection{Approximation under analyticity}

For functions satisfying the analytic extension condition in Appendix~\ref{app:analytic-extension}, the polynomial dependence of the preceding approximation bound on the inverse accuracy can be replaced by a polynomial dependence on the logarithm of the inverse accuracy. The key is to truncate a Chebyshev expansion and then approximate the products needed to evaluate the resulting polynomial. We use Chebyshev polynomials because they are uniformly bounded on the normalized real interval and their coefficients decay geometrically for functions admitting a bounded analytic extension.

The following lemma is a version of the exponential ReLU approximation of holomorphic maps in \citet[Theorem~3.6]{opschoor2022exponential}. We make the dependence of the constants on $m$, the ellipse parameter and the amplitude bound explicit for the variable-selection complexity bounds in Appendix~\ref{app:vs-analytic}. The region $E_\rho^m$ is defined in Appendix~\ref{app:analytic-extension}. The rates are consistent with \citet[Theorem~3.6, p.~565]{opschoor2022exponential} and \citet[Theorem~2.6, p.~1736]{e2018exponential}.

The proof truncates a multivariate expansion at a degree of order $\log(1/\epsilon)$ in each coordinate. For fixed dimension this leaves a number of polynomial terms that is a power of $\log(1/\epsilon)$. The construction below also bounds every network coefficient uniformly in the accuracy and counts the fixed connections needed to carry intermediate values through the network. We prove the Chebyshev coefficient bound by a contour argument. See also \citet[Lemma~5.1, p.~4841]{trefethen2017multivariate} for the decay of multivariate coefficients.

\begin{lemma}[Analytic approximation]\label{lem:analytic-approx}
Let $m\geq1$, $\rho>1$, and suppose that $f:[-1,1]^m\to\mathbb R$ and $F\in\operatorname{Hol}(O)$, where $O\subset\mathbb C^m$ is open and $E_\rho^m\subset O$, satisfy $F|_{[-1,1]^m}=f$ and $\sup_{z\in E_\rho^m}|F(z)|\leq B$.
For every $\epsilon\in(0,1)$ there is a ReLU network $\phi$ such that
\[
\sup_{x\in[-1,1]^m}|\phi(x)-f(x)|\leq  \epsilon,\qquad L\leq   C_{m,\rho,B}\log^2(e/\epsilon),\qquad W\leq   C_{m,\rho,B}\log^{m+1}(e/\epsilon).
\]
Its coefficients satisfy $\|\phi\|_{\rm coef}\leq C_{m,\rho,B}$.
The complexity bound includes fixed connections carrying intermediate outputs and inputs.
\end{lemma}

\begin{proof}
In this proof, $C$ denotes a numerical constant. We approximate $f$ by a tensor Chebyshev polynomial and evaluate it with approximate multiplications. We divide the proof into five steps. For $m=1$, Step~4 is replaced by Step~4$'$, a backward recurrence, which avoids storing all polynomial values.

\medskip
\noindent{\bf Step 1: Chebyshev expansion.} Let $T_0(x)=1$, $T_1(x)=x$ and $T_{j+1}(x)=2xT_j(x)-T_{j-1}(x)$ for $j\geq1$, so that $T_j(\cos t)=\cos(jt)$ and $\sup_{x\in[-1,1]}|T_j(x)|\leq1$. For $\nu=(\nu_1,\ldots,\nu_m)\in\mathbb N_0^m$, put
\[
 |\nu|_1=\sum_{r=1}^m\nu_r,\qquad
 q(\nu)=|\{r:\nu_r>0\}|,\qquad
 T_\nu(x)=\prod_{r=1}^mT_{\nu_r}(x_r),
\]
and define the real coefficients
\[
 a_\nu=\frac{2^{q(\nu)}}{\pi^m}
 \int_{[0,\pi]^m} f(\cos t_1,\ldots,\cos t_m)
       \prod_{r=1}^m\cos(\nu_rt_r)\,dt_1\cdots dt_m.
\]
Set $G(w)=F\{(w_1+w_1^{-1})/2,\ldots,(w_m+w_m^{-1})/2\}$. Since $w\mapsto(w+w^{-1})/2$ maps the annulus $\rho^{-1}\leq|w|\leq\rho$ onto $E_\rho$ and $E_\rho^m\subset O$, $G$ is holomorphic on an open set containing the closed product annulus $\{w:\rho^{-1}\leq|w_r|\leq\rho\ (r\in[m])\}$ and satisfies $|G(w)|\leq B$ on this closed product annulus. This map is invariant under $w\mapsto w^{-1}$, so $G$ is invariant under $w_r\mapsto w_r^{-1}$ in each coordinate, and its Laurent coefficients $c_\nu$, $\nu\in\mathbb Z^m$, satisfy $c_\nu=c_{(|\nu_1|,\ldots,|\nu_m|)}$. On the unit circles $w_r=e^{it_r}$ we have $G(e^{it_1},\ldots,e^{it_m})=f(\cos t_1,\ldots,\cos t_m)$, so
\[
\begin{aligned}
c_\nu
 &=\frac{1}{(2\pi)^m}\int_{[-\pi,\pi]^m}
   f(\cos t_1,\ldots,\cos t_m)e^{-i\sum_{r=1}^m\nu_rt_r}\,dt_1\cdots dt_m\\
&=\frac{1}{\pi^m}\int_{[0,\pi]^m}
   f(\cos t_1,\ldots,\cos t_m)\prod_{r=1}^m\cos(\nu_rt_r)\,dt_1\cdots dt_m,
\end{aligned}
\]
where the second equality uses the evenness of $f(\cos t_1,\ldots,\cos t_m)$ in each $t_r$. Hence $a_\nu=2^{q(\nu)}c_\nu$ for $\nu\in\mathbb N_0^m$. By Cauchy's theorem, moving each contour to $|w_r|=\rho$ gives
\[
 c_\nu=\frac{1}{(2\pi i)^m}
 \int_{|w_1|=\rho}\!\cdots\!\int_{|w_m|=\rho}
 G(w)\prod_{r=1}^m w_r^{-\nu_r-1}\,dw_1\cdots dw_m,
 \qquad
 |a_\nu|\leq2^{q(\nu)}B\rho^{-|\nu|_1}.
\]
Here $|G(w)|\leq B$ gives $|c_\nu|\leq B\rho^{-|\nu|_1}$ for $\nu\in\mathbb N_0^m$, yielding the bound on $|a_\nu|$. Hence $\sum_{\nu\in\mathbb Z^m}|c_\nu|=\sum_{\nu\in\mathbb N_0^m}2^{q(\nu)}|c_\nu|\leq B(1+2\sum_{j\geq1}\rho^{-j})^m<\infty$, so the Laurent series of $G$ converges absolutely and uniformly on the unit torus. For $\nu\in\mathbb N_0^m$, the $2^{q(\nu)}$ indices with coordinatewise absolute values $\nu$ contribute
\[
 c_\nu\prod_{r:\nu_r>0}(w_r^{\nu_r}+w_r^{-\nu_r})
 =2^{q(\nu)}c_\nu\prod_{r=1}^m\cos(\nu_rt_r)
 =a_\nu T_\nu(\cos t_1,\ldots,\cos t_m)
\]
at $w_r=e^{it_r}$. Summing over $\nu\in\mathbb N_0^m$ gives $f=\sum_{\nu\in\mathbb N_0^m}a_\nu T_\nu$ uniformly on $[-1,1]^m$.

\medskip
\noindent{\bf Step 2: Truncation.} Put $B_*=1\vee B$ and define
\[
A_m=  \frac{2^mB_*}{(1-\rho^{-1})^m},\qquad H_\rho=  1\vee\frac{2B_*}{(1-\rho^{-1})^2},\qquad k=  \max\{1, \left\lceil\frac{\log(4mA_m/\epsilon)}{\log\rho}\right\rceil\}.
\]
Let $\mathcal A_k=\{0,\ldots,k\}^m$, $N_k=|\mathcal A_k|=(k+1)^m$ and $P_k=\sum_{\nu\in\mathcal A_k}a_\nu T_\nu$. By Step~1, $|a_\nu|\leq2^{q(\nu)}B\rho^{-|\nu|_1}\leq2^mB\rho^{-|\nu|_1}$ and $|T_\nu|\leq1$ on $[-1,1]^m$. Since $B\leq B_*$, summing the geometric series over each coordinate gives $\sum_\nu|a_\nu|\leq2^mB(\sum_{j\geq0}\rho^{-j})^m=2^mB(1-\rho^{-1})^{-m}\leq A_m$. Every index $\nu\notin\mathcal A_k$ has $\nu_r\geq k+1$ for some $r$, and for each fixed $r$ the geometric series yields $\sum_{\nu:\nu_r\geq k+1}2^mB\rho^{-|\nu|_1}=2^mB\rho^{-(k+1)}(1-\rho^{-1})^{-m}\leq A_m\rho^{-(k+1)}$. Since $\rho>1$, the choice of $k$ implies $\rho^{-k}\leq\epsilon/(4mA_m)$, hence $mA_m\rho^{-(k+1)}\leq\epsilon/4$. Combining these bounds with $f-P_k=\sum_{\nu\notin\mathcal A_k}a_\nu T_\nu$ and a union over the coordinates, we obtain
\[
\sum_\nu|a_\nu|\leq   A_m,\qquad \|f-P_k\|_\infty \leq  \sum_{r=  1}^m \sum_{\nu:\nu_r\geq   k+1}2^mB\rho^{-|\nu|_1} \leq   mA_m\rho^{-(k+1)}\leq  \epsilon/4.
\]

\medskip
\noindent{\bf Step 3: Approximate multiplication and clipping.} We use the multiplication networks of Lemma~\ref{lem:multiplication}. A retained signed value uses the representation $z=\sigma(z)-\sigma(-z)$, with two carried coordinates and at most four nonzero coefficients per layer. Take
\[
\zeta= \frac{\epsilon}{8mA_m3^k},\qquad J_\times= 1+\left\lceil\log_4(2H_\rho^2/\zeta)\right\rceil.
\]
Every multiplier used below then satisfies
\[
L(\mathsf M_{R,\zeta}),W(\mathsf M_{R,\zeta})\leq   CJ_\times,\qquad \|\mathsf M_{R,\zeta}\|_{\rm coef}\leq   CH_\rho^2,\qquad R\in\{1,H_\rho\}.
\]
Write $\mathsf C_R(t)=-R+\sigma(t+R)-\sigma(t-R)$ for exact clipping to $[-R,R]$, which does not increase the distance to a point of this interval.

\medskip
\noindent{\bf Step 4: Evaluation for $m\geq2$.} For each input coordinate, compute
\[
\widetilde T_0(x)=1,\qquad \widetilde T_1(x)=x,\qquad \widetilde T_{j+1}(x)=\mathsf C_1\{2\mathsf M_{1,\zeta}(x,\widetilde T_j(x))-\widetilde T_{j-1}(x)\}.
\]
Clipping places every multiplier input in $[-1,1]^2$ and gives $\sup_{[-1,1]}|\widetilde T_j|\leq1$. Put $e_j=\sup_{[-1,1]}|\widetilde T_j-T_j|$, so $e_0=e_1=0$. Since $|T_{j+1}|\leq1$ on $[-1,1]$, clipping does not increase the distance to $T_{j+1}$. For $j\geq1$ and $x\in[-1,1]$, the accuracy $\zeta$ of $\mathsf M_{1,\zeta}$ and $|x|\leq1$ give $|2\mathsf M_{1,\zeta}(x,\widetilde T_j(x))-2xT_j(x)|\leq2\zeta+2|\widetilde T_j(x)-T_j(x)|$. With $T_{j+1}(x)=2xT_j(x)-T_{j-1}(x)$, the triangle inequality then gives $e_{j+1}\leq2e_j+e_{j-1}+2\zeta$. Since $2\cdot3^j+3^{j-1}+2\leq3^{j+1}$ for $j\geq1$, induction yields $e_j\leq3^j\zeta$.

For each $\nu\in\mathcal A_k$, compute $\widetilde T_\nu(x)$ by multiplying the $m$ values $\widetilde T_{\nu_r}(x_r)$ along a balanced binary tree, padding the leaves with ones to a power of two $m'$ with $m\leq m'<2m$ and applying $\mathsf C_1\circ\mathsf M_{1,\zeta}$ at each internal node. Every exact subtree product and every computed output lies in $[-1,1]$. At an internal node with exact inputs $a,b$ and computed inputs $\widetilde a,\widetilde b$, we have
\[
|\mathsf C_1\mathsf M_{1,\zeta}(\widetilde a,\widetilde b)-ab|\leq\zeta+|\widetilde a-a||\widetilde b|+|a||\widetilde b-b|\leq\zeta+|\widetilde a-a|+|\widetilde b-b|,
\]
where the first inequality uses nonexpansiveness of $\mathsf C_1$, the accuracy $\zeta$ of $\mathsf M_{1,\zeta}$ and $\widetilde a\widetilde b-ab=(\widetilde a-a)\widetilde b+a(\widetilde b-b)$, and the second uses $|\widetilde b|\leq1$ and $|a|\leq1$. Summing this bound over the tree shows that the root error $|\widetilde T_\nu(x)-T_\nu(x)|$ is at most the sum of the leaf errors plus $\zeta$ for each of the $m'-1$ internal nodes. Since $\nu_r\leq k$, each of the $m$ true leaves has error at most $3^k\zeta$, and each padded leaf has error zero. Since $m'<2m$ and $k\geq1$, this gives $\|\widetilde T_\nu-T_\nu\|_\infty\leq m3^k\zeta+(m'-1)\zeta\leq2m3^k\zeta$.

For the output $\phi=\sum_{\nu\in\mathcal A_k}a_\nu\widetilde T_\nu$, the bound $\sum_\nu|a_\nu|\leq A_m$ of Step~2 and the choice of $\zeta$ in Step~3 give $\|\phi-P_k\|_\infty\leq\sum_{\nu\in\mathcal A_k}|a_\nu|\cdot2m3^k\zeta\leq2mA_m3^k\zeta=\epsilon/4$. Combining this with the truncation bound $\|f-P_k\|_\infty\leq\epsilon/4$ of Step~2 gives $\|\phi-f\|_\infty\leq\epsilon/4+\epsilon/4\leq\epsilon$.

The $m$ univariate recurrences run in parallel for $k$ stages and, together with the retained values, have depth at most $CkJ_\times$ and at most $CmJ_\times\sum_{j=0}^k(j+2)\leq Cm(k+1)^2J_\times$ parameters. The $N_k$ parallel trees have depth at most $C\log(em)J_\times$ and at most $CmN_kJ_\times$ parameters. Copying and the final affine sum use at most $CmN_k$ parameters. These counts include the signed identity connections and all fixed fan-out coefficients. Since $N_k=(k+1)^m$ and $(k+1)^2\leq(k+1)^m$ for $m\geq2$, the total number of parameters is at most $Cm(k+1)^mJ_\times$.

\medskip
\noindent{\bf Step 4$'$: Evaluation for $m=1$.} A backward recurrence avoids the storage cost $(k+1)^2$. Define
\[
b_{k+1}(x)=b_{k+2}(x)=0,\qquad b_j(x)=a_j+2xb_{j+1}(x)-b_{j+2}(x),\qquad j=k,\ldots,1.
\]
Let $U_0(x)=1$, $U_1(x)=2x$ and $U_{r+1}(x)=2xU_r(x)-U_{r-1}(x)$, so that $U_r(\cos t)=\sum_{h=0}^re^{i(r-2h)t}$ and $\sup_{x\in[-1,1]}|U_r(x)|\leq r+1$. Substitution into the recurrence gives
\[
b_j(x)=\sum_{l=j}^k a_lU_{l-j}(x),\qquad |b_j(x)|\leq 2B\sum_{l=j}^k\rho^{-l}(l-j+1)\leq\frac{2B\rho^{-j}}{(1-\rho^{-1})^2}\leq H_\rho,
\]
where the first inequality uses $|a_l|\leq2B\rho^{-l}$ for $m=1$ and $|U_{l-j}(x)|\leq l-j+1$, the second uses $\sum_{h\geq0}(h+1)\rho^{-h}=(1-\rho^{-1})^{-2}$, and the last uses the definition of $H_\rho$ and $\rho^{-j}\leq1$. With $U_{-1}=0$, the identity $xU_{l-1}(x)-U_{l-2}(x)=T_l(x)$ for $l\geq1$ gives $P_k(x)=a_0+xb_1(x)-b_2(x)$. Starting with $\widetilde b_{k+1}=\widetilde b_{k+2}=0$, compute
$\widetilde b_j=\mathsf C_{H_\rho}\{a_j+2\mathsf M_{H_\rho,\zeta}(x,\widetilde b_{j+1})-\widetilde b_{j+2}\}$ for $j=k,\ldots,1$,
and return
$\phi(x)=a_0+\mathsf M_{H_\rho,\zeta}(x,\widetilde b_1)-\widetilde b_2$.
Since $|x|\leq1\leq H_\rho$ and $|\widetilde b_j|\leq H_\rho$, all multiplier inputs lie in $[-H_\rho,H_\rho]^2$. Put $d_j=\sup_{[-1,1]}|\widetilde b_j-b_j|$, so that $d_{k+1}=d_{k+2}=0$. Since $|b_j|\leq H_\rho$, clipping does not increase the distance to $b_j$. By Lemma~\ref{lem:multiplication}, $|x|\leq1$ and the triangle inequality, $|\mathsf M_{H_\rho,\zeta}(x,\widetilde b_{j+1})-xb_{j+1}|\leq\zeta+|\widetilde b_{j+1}-b_{j+1}|$ for $0\leq j\leq k$. Hence $d_j\leq2d_{j+1}+d_{j+2}+2\zeta$, and backward induction gives $d_j\leq(3^{k-j+1}-1)\zeta$ for $1\leq j\leq k$. Consequently, $\|\phi-P_k\|_\infty\leq\zeta+d_1+d_2\leq2\cdot3^k\zeta\leq\epsilon/4$, where the last inequality uses the choice of $\zeta$ with $m=1$ and $A_1\geq1$. With the truncation bound $\|f-P_k\|_\infty\leq\epsilon/4$, we obtain $\|\phi-f\|_\infty\leq\epsilon$. Each stage carries $x$ and two recurrence values, so $L,W\leq C(k+1)J_\times$.

\medskip
\noindent{\bf Step 5: Complexity.} In both cases,
\begin{equation}
 \begin{aligned}
 L\leq C\{k+1+\log(em)\}J_\times,\quad W\leq Cm(k+1)^mJ_\times,\quad
 \|\phi\|_{\rm coef}\leq C\{1+H_\rho^2+2^mB\}.
 \end{aligned}
 \label{eq:analytic-explicit-complexity}
\end{equation}
The coefficient bound follows from the multiplier construction, the clipping levels $1$ and $H_\rho$, and $|a_\nu|\leq2^mB$; retaining and copying values uses only coefficients of magnitude one. The explicit choices of $k$ and $J_\times$ give $k+1\leq C_{m,\rho,B}\log(e/\epsilon)$ and $J_\times\leq C_{m,\rho,B}\log(e/\epsilon)$, and substitution into~\eqref{eq:analytic-explicit-complexity} proves the lemma.
\end{proof}

For a rectangle $R=\prod_{j=1}^m[u_j,v_j]$ with $u_j<v_j$, we define
\[
(A_Rx)_j=\frac{2x_j-u_j-v_j}{v_j-u_j},\qquad K_R=1+\sum_{j=1}^m\{\frac2{v_j-u_j}+\frac{|u_j+v_j|}{v_j-u_j}\}.
\]
Folding $A_R$ into the first affine layer preserves depth and gives
\[
W(\phi\circ A_R)\leq 2W(\phi),\qquad \|\phi\circ A_R\|_{\rm coef}\leq K_R\|\phi\|_{\rm coef}.
\]
For $q$ real output coordinates, we construct $\phi_l$ at accuracy $\epsilon/\sqrt q$ and assemble them in parallel. Then
\[
\sup_R\|\phi-f\|_2^2\leq\sum_{l=1}^q\sup_R|\phi_l-f_l|^2\leq\epsilon^2,\qquad W\leq C_q\sum_{l=1}^qW(\phi_l),\qquad L\leq\max_lL(\phi_l)+2.
\]
Here $C_q$ accounts for the fixed input copying, depth alignment and output assembly, and depends only on $q$ and the fixed input dimension.

When dimensions, rectangles or models vary with $n$, the bounds hold with the explicit quantities $A_m$, $H_\rho$, $k$, $J_\times$ and the affine scales.

\end{document}